\documentclass[3p,times,11pt]{elsarticle}
\journal{Physics Reports}
\usepackage[english]{babel}
\usepackage[T1]{fontenc}
\usepackage{graphics}
\usepackage{graphicx}
\usepackage{tabularx}
\usepackage{fancyheadings}
\usepackage{amsmath}
\usepackage{amsbsy}
\usepackage{pifont}
\usepackage{float}
\usepackage{amssymb}
\usepackage{delarray}
\usepackage{rotating}
\usepackage{makeidx}
\usepackage{color}
\usepackage[multiple]{footmisc}
\usepackage{xspace}
\usepackage{subfig}
\DeclareGraphicsExtensions{.png,.pdf,.eps}

\DeclareMathAlphabet{\mathbfit}{T1}{cmr}{bx}{it}
\DeclareMathAlphabet{\bbox}{T1}{cmr}{bx}{it}
\DeclareMathAlphabet{\bbbox}{T1}{cmr}{bx}{n}

\graphicspath{{figures/}{./figures/}}

\usepackage{ragged2e}
\newcolumntype{L}[1]{>{\hsize=#1\hsize\RaggedRight} X}

\def\alphaS{\alpha_S}

\def\MG5aMC{{\sc \small MadGraph5\_aMC@NLO}}
\def\MadOnia{{\sc \small MadOnia}}

\def\MadGraph{{\sc \small MadGraph}}

\def\HELACOnia{ {\sc \small HELAC-Onia}}

\newcommand{\ie}{{\it i.e.}}
\newcommand{\eg}{{\it e.g.}}
\newcommand{\etal}{{\it et al.}}
\newcommand{\Br}{{\cal B}}
\renewcommand{\bar}[1]{\overline{#1}}
\definecolor{glaucous}{rgb}{0.38, 0.51, 0.71}
\definecolor{indigo(dye)}{rgb}{0.0, 0.25, 0.42}
\definecolor{mediumelectricblue}{rgb}{0.01, 0.31, 0.59}

\newcommand{\bqa}{\begin{eqnarray}}
\newcommand{\eqa}{\end{eqnarray}}
\newcommand{\bq}{\begin{equation}}
\newcommand{\eq}{\end{equation}}
\renewcommand{\P}{{\cal P}}

\newcommand{\Q}{{\cal Q}}

\providecommand{\pp}{$pp$}
\providecommand{\pp}{$pd$}

\newcommand{\ce}[1]{Eq.~(\ref{#1})}

\newcommand{\cf}[1]{{Fig.~\ref{#1}}}
\newcommand{\ct}[1]{{Tab.~\ref{#1}}}
\newcommand{\cfs}[1]{{Figs.~\ref{#1}}}
\def\be{\begin{equation}}
\def\ee{\end{equation}}
\def\bea{\begin{eqnarray}}
\def\eea{\end{eqnarray}}

\newcommand{\eqs}[1]{\begin{equation} \begin{split} #1\end{split} \end{equation} }

\newcommand{\ep}{\varepsilon}

\newcommand{\vect}[1]{\vec{#1}}

\newcommand{\pt}{\ensuremath{P_{{T}}}\xspace}

\newcommand{\jpsi}{\ensuremath{J/\psi}\xspace}

\newcommand{\chiQ}{\ensuremath{\chi_Q}\xspace}

\newcommand{\ups}{\ensuremath{\Upsilon}\xspace}
\newcommand{\upsi}{\ensuremath{\Upsilon}\xspace}
\newcommand{\upsa}{\ensuremath{\Upsilon{(1S)}}\xspace}
\newcommand{\upsb}{\ensuremath{\Upsilon{(2S)}}\xspace}
\newcommand{\upsc}{\ensuremath{\Upsilon{(3S)}}\xspace}

\newcommand{\QQbar}{\ensuremath{{Q\overline{Q}}}\xspace}

\newcommand{\etac}{\ensuremath{\eta_c}\xspace}

\newcommand{\epem}{\ensuremath{e^+e^-}\xspace}

\def\soge{{\bigl.^{2s+1}\hspace{-1mm}S^{[8]}_J}}

\def\sps{{\bigl.^1\hspace{-0.5mm}S^{[8]}_0}}
\def\so{{\bigl.^3\hspace{-0.5mm}S^{[8]}_1}}
\def\pjs{{\bigl.^3\hspace{-0.5mm}P^{[1]}_J}}
\def\p0s{{\bigl.^3\hspace{-0.5mm}P^{[1]}_0}}
\def\pones{{\bigl.^3\hspace{-0.5mm}P^{[1]}_1}}
\def\ptwos{{\bigl.^3\hspace{-0.5mm}P^{[1]}_2}}

\def\pj{{\bigl.^3\hspace{-0.5mm}P^{[8]}_J}}
\def\p0{{\bigl.^3\hspace{-0.5mm}P^{[8]}_0}}

\def\sa{{\bigl.^1\hspace{-0.5mm}S^{[8]}_0}}
\def\sb{{\bigl.^3\hspace{-0.5mm}S^{[8]}_1}}
\def\mspb{{\langle\mathcal{O}(\bigl.^3\hspace{-0.5mm}S_1^{[1]})\rangle}}
\def\mopb{{\langle\mathcal{O}(\bigl.^3\hspace{-0.5mm}S_1^{[8]})\rangle}}
\def\mopj{{\langle\mathcal{O}(\bigl.^3\hspace{-0.5mm}P_J^{[8]})\rangle}}
\def\mop0{{\langle\mathcal{O}(\bigl.^3\hspace{-0.5mm}P_0^{[8]})\rangle}}

\def\msp0{{\langle\mathcal{O}(\bigl.^3\hspace{-0.5mm}P_0^{[1]})\rangle}}
\def\mops{{\langle\mathcal{O}(\bigl.^1\hspace{-0.5mm}S_0^{[8]})\rangle}}

\def\mopbchiQ{{\langle\mathcal{O}^{\chiQ}(\bigl.^3\hspace{-0.5mm}S_1^{[8]})\rangle}}

\def\mspzerochiQ{{\langle\mathcal{O}^{\chiQ}(\bigl.^3\hspace{-0.5mm}P_0^{[1]})\rangle}}

\def\pseudos{{\bigl.^1\hspace{-0.5mm}S^{[1]}_0}}
\def\ssnew{{\bigl.^3\hspace{-0.5mm}S^{[1]}_1}}
\def\psip{\psi'}
\def\etac{\eta_c}
\def\etacp{\eta_c'}

\def\coetacp{{\langle\mathcal{O}^{\etacp}(\bigl.^3\! S_1^{[8]})\rangle}}
\def\lsim{\raise0.3ex\hbox{$<$\kern-0.75em\raise-1.1ex\hbox{$\sim$}}}
\def\gsim{\raise0.3ex\hbox{$>$\kern-0.75em\raise-1.1ex\hbox{$\sim$}}}

\renewcommand{\footnote}[1]{\footnotemark\footnotetext{\ #1}}

\usepackage{bm}

\relax
\def\frac#1#2{{#1\over#2}}
\def\<{\langle}\def\>{\rangle}
\newcount\parenthesis \parenthesis=0 \newcount\n
\def\({\global\advance\parenthesis by1\left(}
\def\){\global\advance\parenthesis by-1\right)}
\def\{{\global\advance\parenthesis by1\left\lbrace}
\def\}{\global\advance\parenthesis by-1\right\rbrace}
\def\[{\relax} 
\def\]{\relax} 
\def\Loop#1\Repeat{\global\n=0\global\let\body=#1\iterate}
\def\iterate{\body\let\next=\iterate\else\let\next=\relax\fi\next}
\def\ldd{\ifnum\n<\parenthesis\global\advance\n by1
\left.\nulldelimiterspace=0pt\mathsurround=0pt}
\def\rdd{\ifnum\n<\parenthesis\global\advance\n by1
\right.\nulldelimiterspace=0pt\mathsurround=0pt}

\newcount\exacount\exacount=0\font\caps=cmcsc10
\def\Istrut{\vrule height11pt depth4pt width0pt}
\def\TRIexa#1#2#3#4{\global\advance\exacount by1\par\filbreak
{\offinterlineskip
  \vbox{\hrule\hbox to\hsize{\Istrut\vrule
      \hbox to 8mm{\hfil\caps\the\exacount\hfil}\vrule
      \quad\rm#1\hfill\vrule
      \hbox to 32mm{\hfill{\caps Mode: }{\tt #2}\hfill}\vrule
      \hbox to 32mm{\hfill{\caps Tolerance: }{\tt #3}\hfill}\vrule}
    \hrule\hbox to\hsize{\Istrut\vrule\hfill#4\hfill\vrule}\hrule}
}\nobreak}

\usepackage{comment}
\usepackage{array}
\usepackage{longtable}
\usepackage{calc}
\usepackage{multirow}
\usepackage{hhline}
\usepackage{ifthen}
\usepackage{sectsty}

\newif\ifshowcitations\showcitationsfalse%
\newif\ifshowlinks\showlinksfalse%

\ifshowlinks%
  \usepackage[
         colorlinks=true,
         urlcolor=blue,       
         ]{hyperref}
  \newcommand*{\inspireurl}[1]{\\\href{#1}{INSPIRE-HEP entry}}
\else
  \makeatletter
  \newcommand*{\inspireurl}[1]{\@bsphack\@esphack}
  \makeatother
\fi
\ifshowcitations%
  \newcommand*{\citations}[1]{\\* #1}
\else
  \makeatletter
  \newcommand*{\citations}[1]{\@bsphack\@esphack}
  \makeatother
\fi

\newcommand{\clearemptydoublepage}{\newpage{\pagestyle{empty}\cleardoublepage}}

\usepackage{scalerel}
\usepackage{tikz}
\usetikzlibrary{svg.path}

\definecolor{orcidlogocol}{HTML}{A6CE39}
\tikzset{
  orcidlogo/.pic={
    \fill[orcidlogocol] svg{M256,128c0,70.7-57.3,128-128,128C57.3,256,0,198.7,0,128C0,57.3,57.3,0,128,0C198.7,0,256,57.3,256,128z};
    \fill[white] svg{M86.3,186.2H70.9V79.1h15.4v48.4V186.2z}
                 svg{M108.9,79.1h41.6c39.6,0,57,28.3,57,53.6c0,27.5-21.5,53.6-56.8,53.6h-41.8V79.1z M124.3,172.4h24.5c34.9,0,42.9-26.5,42.9-39.7c0-21.5-13.7-39.7-43.7-39.7h-23.7V172.4z}
                 svg{M88.7,56.8c0,5.5-4.5,10.1-10.1,10.1c-5.6,0-10.1-4.6-10.1-10.1c0-5.6,4.5-10.1,10.1-10.1C84.2,46.7,88.7,51.3,88.7,56.8z};
  }
}

\newcommand\orcidicon[1]{\href{https://orcid.org/#1}{\mbox{\scalerel*{
\begin{tikzpicture}[yscale=-1,transform shape]
\pic{orcidlogo};
\end{tikzpicture}
}{|}}}}

\usepackage[backref=page]{hyperref}

    \renewcommand*{\backrefalt}[4]{
       \ifcase #1 
          No cited.
       \or
          Cited on page #2.
       \else
          Cited on pages #2.
       \fi} 
\hypersetup{
   pdftitle={New Observables in Inclusive Production of Quarkonia},%
    pdfauthor={Jean-Philippe Lansberg},%
    pdfsubject={},%
    pdfkeywords={},%
    pdfstartview={},%
    bookmarksopen=true, breaklinks=true, debug=true, %
    colorlinks=true, linkcolor=mediumelectricblue, citecolor=blue, urlcolor=blue,hyperfootnotes = false
  }

\begin{document}

\begin{frontmatter}

\title{New Observables in Inclusive Production of Quarkonia}

\date{\today}
\author[IJCLab]{Jean-Philippe~Lansberg\orcidicon{0000-0003-2746-5986}}
\address[IJCLab]{Universit\'e Paris-Saclay, CNRS, IJCLab, 91405, Orsay, France}

\begin{abstract}
After an introduction motivating the study of quarkonium production, we review the recent developments in the phenomenology of quarkonium production in inclusive scatterings of hadrons and leptons. We naturally address data and predictions relevant for the LHC, the Tevatron, RHIC, HERA, LEP, $B$ factories and EIC. An up-to-date discussion of the contributions from feed downs within the charmonium and bottomonium families as well as from $b$ hadrons to charmonia is also provided. This contextualises an exhaustive overview of new observables such as the associated production along with a Standard Model boson ($\gamma$, $W$ and $Z$), with another quarkonium, with another heavy quark as well as with light hadrons or jets. 
We address the relevance of these reactions in order to improve our understanding of the mechanisms underlying quarkonium production as well as the physics of multi-parton interactions, in particular the double parton scatterings.
An outlook towards future studies and facilities concludes this review. 
\end{abstract}

\end{frontmatter}

\tableofcontents 
\clearemptydoublepage

\section{Introduction to inclusive quarkonium production}
\label{ch:introduction}

The discovery of the first quarkonia, the $J/\psi$, in 1974 is probably the last  {\it meson} discovery
which was a fundamental milestone for our understanding of the elementary-particle realms. It indeed identified with that of the fourth quark, the charm quark, and thus with the first tangible observation
that quarks were real particles and not only the mathematical entities invented by M. Gell-Mann. Unsurprisingly,
it was rewarded by a Nobel prize to B.~Richter  and S.~Ting only 2 years afterwards. It is still remembered as the ``November Revolution".
 
Since then, many other such mesons, made of a quark and an antiquark of the same (heavy) flavour, have been discovered.
The first radial excitation of the $J/\psi$, the $\psi(2S)$, discovered at SLAC along with the $J/\psi$,
already indicated that the {\it strong} interaction was indeed {\it weak} at short distances resulting in a near Coulombic binding
potential with a confinement part. Quarkonia were indeed one of the first playgrounds where theorists
applied~\cite{Appelquist:1974yr,Eichten:1974af} the asymptotic freedom of Quantum Chromodynamics --the theory of the Strong Interaction within the Standard Model
of particle physics. The discovery of the charmonium $P$ waves, the $\chi_c$, and much later of the
pseudoscalar ground state, the $\eta_c$, confirmed their non-relativistic nature and their rather simple
behaviour. Beside this {\it charmonium} family\footnote{This name directly follows
from the analogy with the positronium in QED~\cite{Appelquist:1974yr}.}, quarkonia made of the heavier bottom quark were 
found thereafter.
In fact, the first ``particle" discovered at the LHC was the $3 ^3P_J$ triplet state~\cite{Aad:2011ih}
of the {\it bottomonium} family.

However, as soon as theorists started to try to predict their {\it production} rates, the enthusiasm created
by their discovery swiftly gave way to a controversy. As early as in the 1980's, animated discussions started when different approaches were proposed: \eg\ one~\cite{Fritzsch:1977ay,Halzen:1977rs} where
gluon emissions from the heavy quarks are so numerous that the hadronisation is essentially
decorrelated from the heavy-quark-pair production; another~\cite{Chang:1979nn,Berger:1980ni,Baier:1981uk} where each gluon emission from the heavy
quarks occurs at short distances (less than a tenth of a femtometer) and costs each time
one power of the strong coupling constant, $\alphaS$, thus roughly 0.2  
at the scale of the $c$-quark mass. 

Nearly 4 decades
later, it is still not clear whether the leading (colour-singlet) Fock state contribution\footnote{Whose size 
does not depend on unconstrained parameters, but on the Schr\"odinger wave function at the origin (for the $S$ waves).} to $\jpsi$ production is dominant or not, because of soft gluon emissions. 
Yet, quarkonia are now ubiquitous in high-energy physics. Here, they signal
the presence of a phase transition by disappearing~\cite{Matsui:1986dk} or, by being more abundanly produced, 
they hint at collective heavy-quark effects~\cite{Thews:2000rj}.
There, they are decay products of newly discovered exotic states~\cite{Choi:2003ue,Acosta:2003zx,Aubert:2005rm,Choi:2007wga,Abe:2004zs,Liu:2013dau,Aaij:2015tga},
of the $H^0$ boson, potentially telling us
about its coupling to heavy quarks~\cite{Bodwin:2013gca,Aad:2015sda}, or of a $B$ meson, allowing us to measure $CP$ symmetry violation~\cite{Aubert:2001nu,Abe:2001xe}.  When produced with other particles, in a ``back-to-back" configuration, 
they give us ways to see how gluons are polarised inside unpolarised protons~\cite{Dunnen:2014eta,Lansberg:2017dzg}, whereas, in
a ``near" configuration, they can give us insights about multi-parton dynamics and correlations of gluons inside the proton~\cite{Kom:2011bd,Lansberg:2014swa}, etc.

For all these reasons, we need to solve the on-going debates and 
advance the theory of quarkonium production into a precision era, at the LHC and elsewhere. Experimentalists
measure them, use them, rely on them and currently perform upgrades (ALICE-MFT) or new experimental projects (AFTER@LHC, EIC, PANDA, SoLID) to study them even more precisely.
In this overview, we will discuss several attempts  aiming at providing better tools --\ie\ more discriminant observables-- and significantly improved precision in the corresponding theory studies. Along these discussion, we will
systematically highlight to which extent these can --or not-- make data-theory comparisons
more reliable and more informative, but also to test the coherence of our theoretical models and ideas applied to the quarkonium realms.

\subsection{Our current understanding of the quarkonium-production mechanisms}

Quarkonium production is intrinsically a two-scale problem, that of the heavy-quark pair production and that of their binding. 
The former is {\it a priori} tractable with perturbative methods, \ie\ using Feynman graphs and perturbative QCD (pQCD), the latter is
non-perturbative and sensitive to the confinement properties of the strong interaction and, as such, cannot be computed {\it ab initio}.
Currently, nearly all the models 
of quarkonium production  rely on a factorisation between the heavy-quark-pair production and its binding --or hadronisation. 
Different approaches  essentially differ 
in the treatment of the hadronisation. In addition, depending on the production regime --proton-proton ($pp$), lepton-proton ($\ell p$), lepton-lepton ($\ell\ell$), proton-nucleus ($pA$), nucleus-nucleus ($AA$) collisions--, different approaches may be adopted to describe how the quarks, gluons, leptons or photons from the initial state effectively collide. The vast majority of the existing theoretical computations
 are nevertheless based on the collinear factorisation~\cite{Brock:1993sz}. Some also used the $k_T$ factorisation~\cite{Catani:1990xk,Catani:1990eg,Collins:1991ty} 
 in order to address specific effects in the high-energy regime whereas a number of recent studies relied on the Transverse-Momentum-Dependent (TMD) factorisation~\cite{Collins:2011zzd,Aybat:2011zv,GarciaEchevarria:2011rb,Angeles-Martinez:2015sea} which can  deal with spin-dependent objects, both for the partons and the hadrons.

Based on an effective theory, nonrelativistic QCD (NRQCD)~\cite{Bodwin:1994jh}, one can express in a rigorous way the hadronisation probability
of a heavy-quark pair into a quarkonium via long-distance matrix elements (LDMEs). In addition to the usual
expansion in powers of the strong coupling constant, $\alphaS$, NRQCD further introduces an expansion in the heavy-quark velocity, $v$, and, 
through the consideration of a couple of higher Fock states, the {\it colour-octet mechanism} (COM) which results in a non-negligible probability 
for a coloured $Q\bar Q$ pair to bind into a quarkonium after low-energy --non-perturbative-- gluon emissions. Whether the COM is dominant, or 
not, matters much for the expected cross section, the quarkonium polarisation\footnote{Irrespective of the reference frame used (see~\cite{Faccioli:2010kd} about details about usual frames).} and for the applications to heavy-ion and spin studies (see \eg\ \cite{Kharzeev:1993qd,Qiu:1998rz,Yuan:2008vn,Ma:2015sia,Dunnen:2014eta,Boer:2016bfj}). On the contrary, if the leading Fock state dominates, NRQCD 
identifies\footnote{To be precise, this only holds for the $S$ waves.} with the {\it colour-singlet model} (CSM)~\cite{Chang:1979nn,Berger:1980ni,Baier:1981uk} for which the $Q\bar Q$ pair is produced at short distances, 
${\cal O}(1/2m_Q)$, already in the very same quantum numbers than the quarkonium in which it will hadronise. 
On the other end of the spectrum, the {\it colour-evaporation model}~\cite{Fritzsch:1977ay,Halzen:1977rs}, which is the emanation of the
principle of quark-hadron duality to quarkonium production, assumes that so many gluons can
be emitted during the hadronisation that the quantum number of the pair is randomised and simple statistical 
rules apply. More details on these 3 approaches which will follow us throughout this overview will be given below. 
Let us however guide the reader to a few relevant reviews:
 \cite{Andronic:2015wma} is the most recent one with comparisons to LHC data, \cite{Brambilla:2014jmp} contains some discussions about the first LHC results, \cite{Brambilla:2010cs} is a pre-LHC one providing an overview of the emergence of Next-to-Leading-Order (NLO) computations, \cite{Lansberg:2006dh} is a slightly older overview of the different existing approaches as of early 2000's, \cite{Kramer:2001hh} offers a complete
overview of (LO) NRQCD successes and failures in the context of the Tevatron and HERA data, \cite{Braaten:1996pv}
is the first review discussing the relevance of the COM contributions to explain the early Tevatron data,
\cite{Schuler:1994hy} is the last pre-NRQCD review likely giving the most complete picture of quarkonium production in the early 1990's.

\subsubsection{The Colour-Evaporation Model (CEM)}

This approach is in line with the principle of quark-hadron duality~\cite{Fritzsch:1977ay,Halzen:1977rs}. As such,
the production cross section of quarkonia is expected to be directly connected to that to  produce a \QQbar pair
in an invariant-mass region where its hadronisation into a quarkonium is possible, that is
between  the kinematical threshold to produce a quark pair, $2m_Q$, and that 
to create the lightest open-heavy-flavour hadron pair, $2m_{H}$. 

The cross section to produce a given quarkonium state is then supposed to be obtained after
a multiplication by a phenomenological factor $\P_{\cal Q}$ related to a process-independent probability that the
pair eventually hadronises into this state. One assumes that a number of non-perturbative-gluon emissions
occur once the $Q \overline Q$ pair is produced and that the quantum state of the pair at its hadronisation
is essentially decorrelated --at least colour-wise-- with that at its production. From the reasonable
assumption~\cite{Amundson:1995em} that one ninth --one colour-{\it singlet} \QQbar 
configuration out of 9 possible-- of the pairs in the suitable kinematical region 
hadronises in a quarkonium,  a simple statistical counting~\cite{Amundson:1995em} was proposed
based on the spin $J_{\cal Q}$ of the quarkonium ${\cal Q}$,
$\P_{\cal Q}= {1}/{9} \times {(2 J_{\cal Q} +1)}/{\sum_i (2 J_i +1)}$,
where the sum over $i$ runs  over all the charmonium states below the open heavy-flavour threshold. It was shown to 
reasonably account for existing \jpsi hadroproduction data of the late 1990's and, in fact, is comparable to 
the fit value in~\cite{Bedjidian:2004gd}. Let us however note that one should in principle treat both direct and prompt yields with 
correspondingly different hadronisation probabilities. The difference between both amounts to the inclusion in the latter of the yield from the decays of higher excited states, in general referred to as feed downs (FDs). Section \ref{sec:FD} is dedicated to their discussion. Since the kinematics of the FDs is not explicitly taken into account, treating a quarkonium with a prompt yield may not give the same result as deriving 
the same prompt yield from a set of direct yields with the relevant decays.

Mathematically, one has
\eqs{d\sigma^{\rm (N)LO,\ direct/prompt}[{\cal Q}+X]= \P^{\rm direct/prompt}_{\cal Q}\int_{2m_Q}^{2m_H} 
\frac{d\sigma^{\rm (N)LO}[Q\overline Q+X]}{d m_{Q\bar Q}}d m_{Q\overline Q},
\label{eq:sigma_CEM}}
where $d\sigma^{\rm (N)LO}[Q\overline Q+X]$ is the hadronic differential cross section to produce
a \QQbar pair at (N)LO accuracy.
In the latter formula, a factorisation between the short-distance \QQbar-pair production and its hadronisation in the 
quarkonium state is of course implied although it does not rely on any factorisation proof. In spite of this, 
this model benefits from a successful phenomenology except for some discrepancies in some transverse momentum spectra and in $e^+e^-$ annihilation. Recent advances in the CEM phenomenology which we will review in section~\ref{subsec:CEM_updates} comprise the first NLO computation~\cite{Lansberg:2016rcx,Lansberg:2020rft} of the $P_T$ differential spectrum, an improved treatment~\cite{Ma:2016exq} 
of the kinematical effects related to the invariant mass of the pair and first studies of the polarisation~\cite{Cheung:2018upe,Cheung:2018tvq,Cheung:2017osx,Cheung:2017loo}.

\subsubsection{The Colour-Singlet Model (CSM)}

The second simplest model to describe quarkonium production relies on the rather opposite assumption 
that the quantum state of the pair does {\it not} evolve between its production and its hadronisation, neither
in spin, nor in colour~\cite{Chang:1979nn,Baier:1981uk,Baier:1983va} -- gluon emissions from the heavy-quark are
suppressed by powers of $\alphaS(m_Q)$. In principle, they are taken into account in the pQCD corrections 
to the hard-scattering part account for the \QQbar-pair production.
If one further assumes that the quarkonia are non-relativistic bound states
with a highly peaked wave function in the momentum space, it can be shown that the
partonic cross section for quarkonium production should then be expressed as that for the production of a heavy-quark 
pair with zero relative velocity, $v$, in a colour-singlet state and 
in the same angular-momentum and spin state as that of the to-be produced quarkonium, 
and the square of the Schr\"odinger wave function at the origin in the position space. In the case of hadroproduction,  
one should further account for the parton $i,j$ densities in the 
colliding hadrons, $f_{i,j}(x,\mu_F)$, in order to get the following hadronic cross section
\begin{equation}
d\sigma[{\cal Q}+X]
=\sum_{i,j}\!\int\! d x_{i} \,d x_{j} \,f_{i}(x_i,\mu_F) \,f_{j}(x_j,\mu_F) d\hat{\sigma}_{i+j\rightarrow (Q\overline{Q})+X} (\mu_R,\mu_F) |R(0)|^2.
\label{eq:sigma_CSM}
\end{equation}
The above formula holds for the collinear factorisation. The corresponding ones
for $k_T$ factorisation or TMD factorisation follow the same logic. 
In the case of $P$-waves, $|R(0)|^2$ vanishes and one should consider its derivative and that of 
the hard scattering. In the CSM, $|R(0)|^2$ or $|R'(0)|^2$ also appear in decay processes and can be extracted
from decay-width measurements. The model then becomes  fully predictive but for the usual unknown values of the unphysical
factorisation and renormalisation scales and of the heavy-quark mass entering the hard-scattering coefficient.

Slightly less than ten years ago, the first evaluations of the QCD 
corrections~\cite{Campbell:2007ws,Artoisenet:2007xi,Gong:2008sn,Gong:2008hk,Artoisenet:2008fc}
to the yields of \jpsi and \ups (also commonly denoted $\cal Q$) in 
hadron  collisions in the CSM appeared. It is now widely accepted~\cite{Lansberg:2008gk,ConesadelValle:2011fw,Brambilla:2010cs} 
that $\alpha^4_s$ and $\alpha^5_s$ corrections to the CSM are significantly larger than the LO  contributions at $\alpha^3_s$
at mid and large \pt and that they should systematically be accounted for in any study of their \pt spectrum. 
This will be discussed in more details and contrasted with the case of the other approaches in section~\ref{ch:dvlpts}.

Possibly due to its high predictive power, the CSM has faced several phenomenological issues although
it accounts reasonably well for the bulk of hadroproduction data from RHIC to LHC energies~\cite{Brodsky:2009cf,Lansberg:2010cn,Feng:2015cba}, 
\epem data at $B$ factories~\cite{Ma:2008gq,Gong:2009kp,He:2009uf} and photoproduction data at HERA~\cite{Aaron:2010gz}. 
Taking into account NLO --one loop-- corrections and approximate NNLO contributions (dubbed as NNLO$^\star$ in the following) 
has reduced the most patent discrepancies in particular for \pt up to a couple of $m_{\cal Q}$~\cite{Lansberg:2010vq,Lansberg:2011hi,Lansberg:2012ta,Lansberg:2013iya}.
A full NNLO computation (\ie~at $\alpha^5_s$) is however needed to confirm this trend, in particular another approximate NNLO computation (dubbed nnLO) seems~\cite{Shao:2018adj} to hint at different conclusions.

It is however true that the CSM is affected by infrared divergences in the case of $P$-wave decay at NLO, which were 
earlier regulated by an ad-hoc binding energy~\cite{Barbieri:1976fp}. These can nevertheless 
be rigorously cured~\cite{Bodwin:1992ye} in the more general framework of NRQCD which we discuss now and which
introduce the concept of colour-octet mechanism.

\subsubsection{The Colour-Octet Mechanism (COM) within Nonrelativistic QCD (NRQCD)}

In NRQCD~\cite{Bodwin:1994jh}, the information on the hadronisation 
of a heavy-quark pair into a quarkonium is encapsulated in long-distance matrix elements (LDMEs). In addition to the usual
expansion in powers of $\alphaS$, NRQCD further introduces an expansion in $v$. It is then natural to account
for the effect of higher-Fock states (in $v$) where the \QQbar pair is in an octet state with a different 
angular-momentum and spin states --the sole consideration of the {\it leading} Fock state (in $v$) amounts\footnote{As aforementioned, such a statement only holds for $S$ waves.} to the CSM, which is thus
{\it a priori} the {\it leading} NRQCD contribution (in $v$). However, this opens the possibility for non-perturbative transitions between
these coloured states and the physical meson. One of the virtues of this is the consideration of $^3S_1^{[8]}$ states
in $P$-wave quarkonium productions, whose contributions cancel the aforementioned divergences in the CSM. The necessity for 
such a cancellation does not however fix the relative importance of these contributions. In this precise case, 
it depends on an unphysical scale $\mu_\Lambda$, replacing an ad-hoc binding energy used as IR regulator
in one-loop CS based studies in the early 1980's~\cite{Barbieri:1980yp,Barbieri:1981xz}.

As compared to the \ce{eq:sigma_CSM}, one has to further
consider additional quantum numbers (angular momentum, spin and colour), generically denoted $n$, involved in the production mechanism: 
\begin{equation}
d\sigma[{\cal Q}+X]
=\sum_{i,j,n}\!\int\! d x_{i} \,d x_{j} \,f_{i}(x_i,\mu_F) \,f_{j}(x_j,\mu_F) d\hat{\sigma}_{i+j\rightarrow (Q\overline{Q})_{n}+X} (\mu_R,\mu_F,\mu_\Lambda) \langle{\cal O}_{\cal Q}^{n} \rangle.
\label{eq:sigma_NRQCD}
\end{equation}

Instead of the Schr\"odinger wave function at the origin squared, the former equation involves 
the aforementioned LDMEs, $\langle{\cal O}_{\cal Q}^{n} \rangle$, which {\it cannot} 
be fixed by decay-width measurements nor lattice studies -- but the leading CSM ones of course. Only 
relations based on Heavy-Quark Spin Symmetry (HQSS) can relate some of them.

It is well-known 
that the first measurements by the CDF Collaboration of the {\it direct}
production of $J/\psi$ and $\psi'$ at $\sqrt{s}=1.8$ 
TeV~\cite{Abe:1997jz,Abe:1997yz} brought to light a striking discrepancies with
the CSM predictions. Via the introduction of the COM, 
NRQCD provided a very appealing solution to this puzzle, hence its subsequent popularity. 
However, as we will see in this review, new puzzles have showed up and many 
issues remain unsolved which motivates
the study of new observables. 

\subsubsection{The advent of the NLO era}\label{subsec:intro_NLO_era}
More than twenty years ago now, the very first NLO calculation of a $P_T$-differential 
quarkonium-production cross section was performed by Kr\"amer. It dealt with unpolarised 
photoproduction of $\psi$~\cite{Kramer:1995nb} via a CS
transition. He found that real-emission $\alpha\alphaS^3$ corrections associated
to new topologies were large, bringing the CS yield close to the experimental data
(see~\cite{Kramer:2001hh} for a nice review). Strictly speaking, this is
however not the first NLO studies of quarkonium production since Kuhn 
and Mirkes~\cite{Kuhn:1992qw} had studied, in 1992,  that on the $P_T$-integrated cross section
for a pseudo-scalar quarkonium, more precisely for a possible toponium.
In 1994, Sch\"uler also studied~\cite{Schuler:1994hy} the same process and 
applied it to $\eta_c$ production.

It took nearly 10 years for the next NLO computations of a $P_T$-differential cross section 
to be done, \ie\ for direct $\gamma\gamma$
collisions~\cite{Klasen:2004az,Klasen:2004tz}, with the aim to address 
the discrepancy~\cite{Klasen:2001cu} between 
the LO CS predictions and the measured rates by DELPHI~\cite{Abdallah:2003du}. 
These ones were based on NRQCD with the consideration of the COM. They followed
the important study of Petrelli \etal~\cite{Petrelli:1997ge} in 1997 where the techniques
to deal with loop corrections in NRQCD for production processes were outlined in
the context of the $P_T$-integrated cross section for hadroproduction. In 1997, 
another study for photoproduction at the end point~\cite{Maltoni:1997pt} was also carried out at NLO. 

NLO corrections were then computed for the integrated cross 
section of $J/\psi$-production observables at the $B$-factories: 
$J/\psi + c \bar c$~\cite{Zhang:2006ay,Gong:2009ng} and its 
inclusive counterpart~\cite{Gong:2009kp,Zhang:2009ym},
accounting for both CS and CO channels following tensions uncovered at $B$ factories~\cite{Abe:2001za,Abe:2002rb,Pakhlov:2009nj}. The account of these NLO corrections and of relativistic corrections solved these tensions with the  data which were then found to be well accounted for by the CS contributions.

At hadron colliders (Tevatron, RHIC, LHC), $\psi$ and $\Upsilon$ production most uniquely  proceeds
via gluon-fusion processes. The corresponding cross
section at NLO ($\alphaS^4$ for hadroproduction processes) are
significantly more complicated to compute than the former ones 
and became only available in 2007~\cite{Campbell:2007ws,Artoisenet:2007xi,Gong:2008sn}. 
Like for photoproduction, NLO corrections were found to be very large, with a different
$P_T$ scaling, significantly altering the predicted polarisation~\cite{Artoisenet:2008fc,Gong:2008sn}
of the produced quarkonia and reducing the gap with the Tevatron data which we mentioned earlier.

One year later, we managed to achieve a partial NNLO computation~\cite{Artoisenet:2008fc}
incorporating what we identified to be the dominant $\alphaS^5$ contributions. 
We found that the difference between the NLO and the partial NNLO results was significant and just filled
the gap with the Tevatron data, especially at large \pt . 
It is now widely accepted --and understood--~\cite{Lansberg:2008gk,Ma:2014svb} 
that the NLO ($\alphaS^4$) and NNLO ($\alphaS^5$) corrections to the production
of spin triplet vector quarkonia in the CSM are 
significantly larger than the LO contributions at mid and large \pt 
and that they should systematically be accounted for in any study of their \pt 
spectrum, although this has not always been the case in the literature. Yet, 
being partial, our NNLO computations exhibit larger uncertainties than 
complete ones. In any case, since the large \pt yield is expected to be dominated
by Born order contributions to $\Q$+3 jets at $\alphaS^5$, the theoretical uncertainties
are necessarily significant. Another approximate NNLO computation (dubbed nnLO) also holding
at large $P_T$ was carried out by Shao in 2018 \cite{Shao:2018adj} and hints at 
a smaller impact of these new NNLO topologies. In both cases, the conclusions may
depend on the infrared treatment and a full NNLO computation is eagerly awaited for.

Owing to these large uncertainties, there is indeed  no consensus whether the CSM
is indeed the dominant mechanism at work for mid and large-\pt $\Upsilon$ and $J/\psi$ 
production. Even at low \pt, where it accounts reasonably well for the bulk 
of hadroproduction data from RHIC to LHC energies as we have shown through NLO studies~\cite{Brodsky:2009cf,Lansberg:2010vq,Feng:2015cba}\footnote{As we will discuss later, such NLO computations
at Tevatron and LHC energies seem to be extremely sensitive to the PDF behaviour and a great care
should then taken when interpreting them.},
some recent studies~\cite{Ma:2014mri,Kang:2013hta} left the door open 
for dominant COM contributions. These apparently opposite conclusions
can be traced back to the large theoretical uncertainties in both approaches.

As we discussed for the $\psi$ and $\Upsilon$, the Born-order cross section
for hadroproduction are at $\alphaS^3$ and for photoproduction at $\alpha\alphaS^2$. 
On the contrary, the CO contributions appears at $\alphaS^2$ for hadroproduction and 
at $\alpha \alphaS$ for photoproduction, which renders the
computation of their loop corrections sometimes easier. As we mentioned above, 
these NLO corrections to the $P_T$-integrated yield in NRQCD had been 
computed~\cite{Petrelli:1997ge} in 1997. However, the first phenomenological 
study~\cite{Maltoni:2006yp} only appeared in 2006.

As what regards the NLO corrections to the COM contribution to the $P_T$-differential 
cross section, the first partial one considering only the $S$-wave 
octet states was released in 2008~\cite{Gong:2008ft} by the IHEP group.
It was completed by 3 groups (Hamburg~\cite{Butenschoen:2012px}, again IHEP~\cite{Gong:2012ug} and PKU~\cite{Chao:2012iv}) which
have carried out a number of NLO studies of cross-section fits to determine the NRQCD CO LDMEs, including
the photoproduction case~\cite{Butenschoen:2009zy} and that of the $\chi_c$~\cite{Ma:2010vd} as well
as many more.
Their conclusions nevertheless differ both qualitatively and quantitatively (see also~\cite{Bodwin:2014gia,Faccioli:2014cqa}). 
The situation is obviously not satisfactory, to the extent that the 
CEM, which had gone out of fashion, refocused our attention~\cite{Lansberg:2016rcx,Ma:2016exq}. This motivated us
to perform the first NLO ($\alphaS^4$) analyses~\cite{Lansberg:2016rcx,Lansberg:2020rft} of the \pt  spectrum in the CEM. 
We will review all these recent updates of NLO computations in section~\ref{ch:dvlpts}.

\subsubsection{What's next ?}\label{subsec:QCD_corrections}

In principle, the recent LHC data should help in settling the situation. For instance, 
the recent $\eta_c$ study by LHCb~\cite{Aaij:2014bga} already significantly reduced, by virtue of heavy-quark-spin symmetry (HQSS), the allowed
region for some $J/\psi$ LDMEs (uncovering significant tensions with data for the Hamburg and IHEP fits).

Along the same lines, we have recently proposed~\cite{Lansberg:2017ozx} to analyse $\eta_c'$ production which should
significantly constraint the $\psi'$ LDMEs which are less constrained than
the $\jpsi$ ones. Indeed, neither $e^+e^-$ nor $ep$ data exist.

$\chi_{Q}$ studies at very low $P_T$ where the magnitude of Landau-Yang suppression of the 
$\chi_{Q1}$ should offer complementary information about the relevant of the CS physics at low $P_T$
and whether the typical off-shellness of the initial gluons predicted to increase with the collision 
energy~\cite{Hagler:2000dd} is large or not. The newly observed~\cite{Aaij:2017vck} $\chi_c$ decay into $J/\psi+\mu^+\mu^-$ should allow LHCb to look further into this physics.

As it will be clear from our detailed discussion of the NLO corrections to the COM, 
more precision for the yield and the polarisation is not what we need. The 
different LDMEs on which the debate bears nearly always appear inside the same linear combinations
in hadroproduction observables.

Much hope is then put in associated-production channels where the LDMEs sometimes appear
 in different linear combinations or where the CSM is expected to be dominant
and can be tested.
So far, data exist for 
$\jpsi+W$ by ATLAS~\cite{Aad:2014rua},
$\jpsi+Z$ by ATLAS~\cite{Aad:2014kba},
$J/\psi+J/\psi$ by LHCb~\cite{Aaij:2011yc,Aaij:2016bqq}, D$0$~\cite{Abazov:2014qba}, CMS~\cite{Khachatryan:2014iia}, ATLAS~\cite{Aaboud:2016fzt}
$\jpsi + \hbox{charm}$ by LHCb~\cite{Aaij:2012dz}, 
$\Upsilon + \hbox{charm}$ by LHCb\cite{Aaij:2015wpa}, 
$J/\psi+\Upsilon$ by D$0$~\cite{Abazov:2015fbl},
$\Upsilon+\Upsilon$ by CMS~\cite{Khachatryan:2016ydm}. More will be available with the forthcoming projected LHC luminosities. Yet, by lack of complete NLO NRQCD computations, the current fits cannot include
them. A vigorous theoretical effort is therefore needed to allow for a drastic reduction
of the uncertainties on the NRQCD LDMEs and finally test NRQCD.
We will review all these observables in section \ref{ch:associated}.

A last aspect not to be overlooked, and recently uncovered by our 
detailed studies of associated hadroproduction~\cite{Lansberg:2015lva,Lansberg:2016rcx,Shao:2016wor,Lansberg:2016muq,Lansberg:2017chq}, is that of
double parton scatterings (DPS), usually assumed to come from two independent 
single parton scatterings (SPS) --\ie\ all the reactions discussed above. 
On the one side, it means that associated-production of quarkonia can help us to gain
more insights into the multiple particle dynamics of the proton-proton collisions.
On the other, this may signify that   the use of  such associated production
channels to study the quarkonium production mechanism or to use them to study the initial state
of a --single-parton-- collision may significantly be complicated.
Along these lines, let us add that triple $J/\psi$ production has lately been theoretically studied~\cite{dEnterria:2016ids,Shao:2019qob} and may also bring in complementary information on 
both the multi parton dynamics involved in such hard scatterings as well as on 
the quarkonium-production mechanisms themselves.

As what concerns the short-distance physics, it is obvious that the completion of our partial
NNLO computations is absolutely essential. This is a clear challenge for it involves the computation
of thousands of real-virtual graphs as well as as double virtual graphs, which has never been 
done before. Such a complete computation would probably significantly decrease the uncertainties
on the short-distance physics and allow us to extend our partial computations to low $P_T$. 
Recent progress in QCD-correction computations make this challenge at reach.

\subsection{Quarkonia as tools: a further motivation to study their production}

Despite all the above drawbacks, quarkonia --especially the vector ones-- 
remain easy to experimentally study in different colliding systems with reasonably high yields 
compared to other hard processes. Except maybe for anti-proton- and meson-induced collisions
at limited energies, quarkonia are dominantly produced by gluon fusion. As such, 
they are at least indirect gluon luminometers. They can probably also
tell us much about the spin of the gluons in polarised and unpolarised protons, about their
momentum distribution in nuclei as well as about the parton correlations inside the proton
as we just discussed.

Beside being probes of the initial state of collisions, quarkonia can also give us information on
the dense medium resulting from ultra  relativistic heavy-ion collisions. Different
effects certainly come into play and call for careful and well-thought analyses
bearing in mind our current understanding of the quarkonium-production
mechanism in the vacuum.

\subsubsection{Tools to study novel aspects of the nucleon spin}

With the constant improvements of the experimental techniques providing larger 
luminosities, higher trigger rates and larger data acquisition bandwidths, but
also with the advent of the first collider of polarised protons, RHIC, 
the study of the nucleon spin in the high-energy domain has definitely 
taken off. The exploration of the contribution of the gluons --dominant
at high energies-- to the nucleon spin is emerging as a crucial aspect for the understanding of the proton spin, earlier referred to as the resolution of the spin {\it crisis}, and now of the spin {\it puzzle}, with the consideration of the role of orbital angular momentum. Clearly a new era will also open with the proposed US EIC
where quarkonia will certainly help further probe the role of the gluon angular momentum in the nucleon spin.

Theory has also progressed much. Two complementary frameworks
are now widely accepted: the aforementioned TMD factorisation, relying on TMD distributions (TMDs)
which extend the usual PDFs with a transverse momentum for the partons and additional spin effects;
the Collinear Twist-3 (CT3) formalism~\cite{Efremov:1981sh,Efremov:1984ip,Qiu:1991pp}, 
relying on 3-parton-correlation functions  which generate spin asymmetries.

In this context, the quarkonia --natural gluonometers as we argued above-- find their place in the
play. Yet, spin studies usually rely on asymmetry measurements which are easily
diluted by unpolarised collisions or backgrounds. Looking at rare probes --as compared to
pions or kaons for instance-- is challenging and only begins to be done. The first 
$J/\psi$ double-spin-asymmetry measurement was only released in 2016 by the
PHENIX collaboration~\cite{Adare:2016cqe}. In this case, one can access the gluon-spin
contribution to the proton spin~\cite{Teryaev:1996sr,Klasen:2003zn}. The first NLO
computation of the corresponding helicity-dependent cross sections was just carried out in 2018 by Feng \etal~\cite{Feng:2018cai}.

Single --transverse-- spin asymmetries (SSA) 
are more accessible and are possible with a sole target polarisation at fixed-target experiments. 
The first SSA study in $J/\psi$ production was carried
out by the PHENIX collaboration in 2010~\cite{Adare:2010bd}. It is possible at the CERN-SPS 
thanks to pion beams with the COMPASS detector~\cite{Quintans:2011zz}, at Fermilab with that of E1039-SpinQuest~\cite{Klein:zoa,Chen:2019hhx} and then at a possible US EIC~\cite{Accardi:2012qut}. Extremely promising
prospects are also offered by the use of the LHC beams in the fixed-target mode (referred to as AFTER@LHC in the following) to study quarkonium SSAs --namely of $J/\psi$, $\psi'$, $\chi_c$, $\eta_c$, $\Upsilon(nS)$, $\chi_b$ and even $\psi+\psi$--  with unprecedented precision~\cite{Brodsky:2012vg,Lansberg:2012kf,Rakotozafindrabe:2013au,Lansberg:2014myg,Massacrier:2015nsm,Lansberg:2016urh,Kikola:2017hnp,Hadjidakis:2018ifr}.

Such SSAs should give us unique information
on the gluon Sivers effect which can be studied in the TMD factorisation for $\eta_c$ or $J/\psi$-pair production.  
Let us recall that the absence of a Sivers effect implies the absence of gluon orbital angular momentum in the proton. In recent years, a number of theoretical
works promoted such spin-related quarkonium studies~\cite{Anselmino:2016fhz,Mukherjee:2016qxa,Schafer:2013wca,Godbole:2014tha,Godbole:2013bca,Godbole:2012bx,Yuan:2008vn}. Yet, all of them were limited to LO and none could go in the details of the FD pattern effects.

With the advent of the TMD factorisation, the spin realm can also be studied 
without polarised hadrons at the LHC. Indeed, the correlations between the parton transverse
momentum inside an unpolarised hadron and its own spin suffice to generate observable
effects --azimuthal and $P_T$-spectrum modulations. In the quark sector, this effect is referred
to as the Boer-Mulders effect~\cite{Boer:1997nt}.
In the gluon sector, it is connected to the linear polarisation of the gluons inside
unpolarised protons~\cite{Mulders:2000sh}. TMD factorisation was first applied to Higgs
and di-photon hadroproduction~\cite{Boer:2011kf,Qiu:2011ai,Boer:2013fca}, then 
to low $P_T$ $C$-even quarkonia~\cite{Boer:2012bt} -- which can in principle 
be produced without any additional gluons, but which is challenging to 
measure\footnote{The first ever measurement~\cite{Aaij:2014bga} of inclusive $\eta_c$ production which we mentioned above 
has only been carried out in 2014 by the LHCb collaboration but with a
\pt cut at 6 GeV. Below 6 GeV, the background quickly rises and trigger rates are difficult
to handle.}. 
Ma \etal\ then successfully advanced such an idea for $\eta_c$ 
to NLO~\cite{Ma:2012hh}\footnote{In~\cite{Ma:2014oha}, they however pointed at possible difficulties with the $\chi_{c,b}$'s
owing to IR divergences specific to $P$-wave production.}. 
This confirmed that the computation of the sole virtual 
corrections is sufficient to perform a complete NLO computations in the TMD factorisation.
Very recently a TMD-factorisation proof has been established for quarkonium production \cite{Echevarria:2019ynx,Fleming:2019pzj}, which pointed out that in fact new soft hadronic matrix elements are involved for quarkonium production at low $p_T$, in addition to the TMDs. These encode the soft physics of the process.
We then proposed~\cite{Dunnen:2014eta} to study this effect in the production of an  $\Upsilon$ 
or $J/\psi$ associated with a photon and showed that sufficient LHC data are already
on tape to tell if this linear polarisation indeed exists. The similar process with a Drell-Yan
pair instead of a photon unfortunately likely exhibits~\cite{Lansberg:2017tlc} asymmetries which would be very challenging to measure. Yet, we recently showed~\cite{Lansberg:2017dzg,Scarpa:2019fol} that di-$J/\psi$ 
production is currently the best process to measure this linearly polarisation 
of the gluon with azimuthal asymmetries as large as 10-40\%. This further illustrates the interest to study
associated quarkonium production.

\subsubsection{Tools to study a possible deconfined state of matter at extreme conditions}

Numerical lattice QCD computations predict~\cite{Boyd:1996bx} that the nuclear matter at extremely high temperature and density behaves 
as a deconfined Plasma of Quarks and Gluons (QGP). This phase of matter is expected to be created 
in ultra-relativistic nucleus-nucleus collisions for a short time before the fireball cools down and the 
process of hadronisation takes place. Quarkonia are sensitive probes of the properties 
of the QGP since they are created in hard scatterings during the early stage of the collisions and since
they subsequently interact with the hot and dense medium out of which they escape. In particular, 
quarkonia are predicted to be melted due to the screening effects induced by the high density of colour charges~\cite{Matsui:1986dk}.
As such, they should provide information on the medium and, in particular, on its temperature.
Moreover, at LHC collider energies, thus in the TeV range, the number of charm (anti) quarks produced by unit of volume 
becomes so large\footnote{About 80 pairs are produced on average in each PbPb collision at 5 TeV} that
it was suggested~\cite{BraunMunzinger:2000px,Thews:2000rj}  that heavy quarks 
produced in the same {\it nucleus-nucleus} collision, but from different {\it nucleon-nucleon}
collisions, could bind into quarkonia. Such a recombination would {\it de facto} signal the production
of a deconfined medium as well as hint at a collective behaviour of heavy quarks.
Among all the quarkonia, the $J/\psi$ and the $\Upsilon$ are the most studied since
their decay into muons offer a tractable background in collisions where more than one thousand
of particles are produced at once.

An entire community of theorists and experimentalists is dedicated to these studies
in proton-nucleus and nucleus-nucleus collisions to predict and respectively measure a relative suppression or enhancement 
of their production (see the following reviews~\cite{Albacete:2016veq,Andronic:2015wma,Brambilla:2010cs}). 
The above picture is however an idealised one. Many competing effects come in fact
into play. Since the EMC measurement~\cite{Aubert:1983xm}, we know that the parton flux inside nuclei --as encoded in nuclear PDF (nPDF)\cite{Eskola:2016oht,Kovarik:2015cma,deFlorian:2011fp,Frankfurt:2011cs,Eskola:2009uj,Hirai:2007sx,Kulagin:2004ie}--  differs from that of in the nucleon
used as references. Quarkonia can simply be broken up~\cite{Gerschel:1988wn,Vogt:1999cu}, suffer multiple scatterings~\cite{Kopeliovich:2001ee,Fujii:2013gxa,Qiu:2013qka,Ducloue:2015gfa,Ma:2015sia} or loose their energy by radiation~\cite{Gavin:1991qk,Brodsky:1992nq,Arleo:2012hn}, in their way out of the 
nucleus overlapping region, even in absence of deconfinement. They can also be broken
by {\it comovers} ~\cite{Capella:2000zp,Ferreiro:2012rq,Ferreiro:2014bia,Ferreiro:2018wbd}.
Besides the possible ambiguity in the nuclear effects at work, the mechanism(s) via which a quarkonium
is created also matters. Indeed, an octet state admittedly does not propagate in a nuclear environment as a colourless state. As such, quarkonia remain a potential source of interesting information once 
the different nuclear phenomena can be disentangled. 

Among all the above effects, the simplest to account for and which can then be 
subtracted is that of the nPDFs which are extracted from other data sets. 
In fact, the conception of a unique code, {\sc Jin}~\cite{Ferreiro:2008wc}, to compute 
the effect of the nPDF and of the nuclear break-up using a correct kinematics 
for the quarkonium production allowed us to show that most of the high 
energy data (from the LHC and RHIC) are compatible with a scenario~\cite{Ferreiro:2013pua,Ferreiro:2011xy} where the 
modification of the gluon densities is the dominant effect at the LHC  along the lines of Vogt's works~\cite{Vogt:2015uba,Vogt:2010aa,Vogt:2004dh}.  If this observation 
can be confirmed, this would significantly simplify the interpretation of quarkonium
data in $AA$ collisions.

We also recently 
proposed a simple method~\cite{Lansberg:2016deg} which can easily be interfaced 
to automated tools to account for the kinematics in a model-independent way 
and which allowed us to systematically account for the nPDF uncertainty as 
encoded in the most recent nPDF set and to study the factorisation scale 
uncertainty -- which we found to be significant. The compatibility of the data 
with the dominance of nPDF effects does not rule out the presence of other 
effects, in particular in the backward regions at RHIC where the quarkonia are 
formed while traversing the nuclear matter. To further assess this compatibility, 
we performed~\cite{Kusina:2017gkz,Kusina:2018pbp} in 2017 a Bayesian reweighting  analysis 
to quantify the constraints which can be set by the heavy-flavour $pA$ data, on
both open (charmed and beauty hadrons) and closed (quarkonia). We found that the
$D$ and $J/\psi$ meson data point at a similar depletion in the gluon densities
at $x < 10^{-3}$. Given that different nuclear effects are expected to impact
these particles, this is a non-trivial indication that the above picture may indeed
be correct.

To go even further, we do hope that the new observables that we will discuss in section~\ref{ch:associated} can clarify
in the near future the production mechanisms of the quarkonia in the vacuum. In turn, 
we should be able to improve the aforementioned description of quarkonium
production in proton-nucleus and nucleus-nucleus collisions. A better understanding of quarkonium production in the vacuum also has very practical implications with regards the possible FDs from the excited states
to the observed lower-lying states as it can induce different values of the nuclear modification factors. We will
provide an up-to-date discussion of these in section \ref{sec:FD}.
Quarkonium
production in proton-nucleus and nucleus-nucleus collisions are by themselves 
research fields worth a dedicated review which is outside the scope of
this one. Yet, a specific effort towards the definition and the promotion of new observables for quarkonium
production in proton-nucleus and nucleus-nucleus collisions should also be accomplished 
in the near future in order to propose new measurements at the LHC, RHIC, SPS and AFTER@LHC.

\section{Recent developments in the phenomenology of inclusive quarkonium production }
\label{ch:dvlpts}
In this section, we summarise the recent developments --most of them related 
with the realisations of NLO computations-- in the application of the 3 aforementioned
theoretical approaches to the realms of hadron colliders (the Tevatron, RHIC and
the LHC), electron-proton colliders (HERA and EIC) and electron-positron colliders (LEP and
$B$ factories). 

Its main purpose is to propose to the reader an up-to-date context to 
the discussion of the associated-production observables in section~\ref{ch:associated}. We have however
decided not to cover the recent updates~\cite{Baranov:2015laa,Baranov:2015yea,Baranov:2016clx,Baranov:2019joi,Baranov:2019lhm,He:2019qqr,Abdulov:2019uyx,	Karpishkov:2019vyt,Babiarz:2019mag,Babiarz:2020jkh} as what regards the $k_T$ factorisation~\cite{Catani:1990xk,Catani:1990eg,Collins:1991ty}
--which is however still being applied at LO in $\alpha_s$-- and the recent
progress in addressing the high-$P_T$ regime in the fragmentation-function approach~\cite{Kang:2011mg,Ma:2013yla,Kang:2014tta,Ma:2014eja,Ma:2014svb,Kang:2014pya,Ma:2015yka}
-- whose predictions for the associated-production channels discussed in section~\ref{ch:associated} are still lacking.
As such, and unless stated otherwise, all the discussion will focus on computations made in the
collinear factorisation~\cite{Brock:1993sz} at fixed-order in $\alpha_s$. We will also not address the attempts to account for the off-shellness of the heavy-quark in the bound state using the Bethe-Salpeter vertex functions~\cite{Haberzettl:2007kj,Lansberg:2005pc,Lansberg:2005aw}. These have not been pursued lately as they require the introduction of poorly
 known 4-point functions and they have not yet been applied to any associated-production studies. The same applies for approaches including low-$x$ effects~\cite{Khoze:2004eu} or higher-twist effects~\cite{Alonso:1989pz,Motyka:2015kta,Schmidt:2018gep,Levin:2018qxa} or finally a newly proposed factorisation scheme~\cite{Ma:2017xno}.

As we noted in the previous section, some of these  specific reactions~\cite{Dunnen:2014eta,Zhang:2014vmh,Lansberg:2017tlc,Lansberg:2017dzg} have recently been studied in the framework 
of the Transverse-Momentum-Dependent (TMD) factorisation~\cite{Collins:2011zzd,Aybat:2011zv,GarciaEchevarria:2011rb,Angeles-Martinez:2015sea} since in general quarkonia are thought~\cite{Boer:2012bt,Schafer:2013wca,Lansberg:2014myg,Lansberg:2015hla,Mukherjee:2015smo,Mukherjee:2016qxa,Signori:2016jwo,Signori:2016lvd,Boer:2016bfj,Rajesh:2018qks,Kishore:2018ugo,Bacchetta:2018ivt} to give a unique handle on the tri-dimensional momentum distribution of the gluons inside nucleons. We do not cover these aspects here as it certainly deserves a dedicated review.
 
 In the hadroproduction case, Born-order cross sections are systematically dominated by gluon fusion and the majority of the data sets focus on events where the quarkonium has a finite, nonzero, $P_T$. This is not only justified by experimental requirements but also because it is believed that pQCD computations are more reliable when $P_T$ is getting larger. Yet, hadroproduction data also exist for $P_T$-integrated cross sections which we will also discuss in a dedicated section owing to its relevance for heavy-ion and spin applications. Photoproduction and leptoproduction have extensively been studied at HERA and provide a complementary playground to hadroproduction, in a more restricted $P_T$ range though. At HERA, quarkonium production is usually dominated by photon-gluon fusion. We will also report on 
$P_T$-integrated cross sections in this case. LEP studies of inclusive quarkonium production mostly cover a region where photon-photon fusion occurs in a $P_T$ range similar to that of HERA. At $B$ factories, imposing a nonzero $P_T$ seems less crucial because of the clean initial state.

Besides restrictions on $P_T$, most of the available recent datasets are also restricted in rapidity, usually towards values close to zero in the hadron centre-of-momentum system (c.m.s.). The notable exceptions are the LHCb detector and the ALICE and PHENIX muon arms which cover more forward rapidities. Since we will not systematically review the whole set of existing experimental data, we naturally guide the reader to the HEPData data base (\href{https://www.hepdata.net/}{\tt https://www.hepdata.net/}) and to a dedicated repository of quarkonium measurements up to 2012 (\href{http://hepdata.cedar.ac.uk/review/quarkonii/}{\tt http://hepdata.cedar.ac.uk/review/quarkonii/}) and documented in~\cite{Andronic:2013zim}.  A  review of the existing RHIC data was also recently published~\cite{Tang:2020ame}.

\subsection{Digression about the feed downs}
\label{sec:FD}

In order to properly address the phenomenology of quarkonium inclusive production, it is
fundamental to bear in mind that a given quarkonium can be produced through the decays of other particles.
The mechanisms underlying these reactions can be very different than the production picture presented above, according to which
a heavy-quark pair is produced in a hard scattering and then hadronises in the
observed quarkonium states. Such a production mode is in fact referred to as the ``direct'' one.

There exist two other frequent modes of production, that from a (strong or electromagnetic) decay of an excited states of the same family and, for the charmonium case, that from a (weak) decay of a $b$-hadrons. Of course, $H^0$, $Z^0$, $W^\pm$ bosons can also decay into a quarkonium
but the corresponding rates are usually so small that their contributions to the measured yield can safely be neglected. The same applies to the decay of a bottomonium into a charmonium. These decays however are interesting production processes on their own, but they need not
to be corrected for or subtracted to interpret other inclusive measurements. 

In principle, with the appropriate detectors, one can disentangle or separately measure the FD yields from the direct one. One usually define $F^{\rm (non)prompt}_\Q$ as the fraction of (non)prompt $\Q$ ($\Q$ being here a charmonium), $F^{\Q'}_\Q$ as the fraction of $\Q$ from $\Q'$ ($\Q'$ being a higher state of the same family) and 
$F^{\rm direct}_\Q$ as the (prompt) fraction of $\Q$ which are produced without transiting by any other quarkonium states.

In the case of $b$-hadron decays to charmonia, owing to the $b$-hadron lifetime, the tracks left by the charmonium are displaced in a way which is measurable by a ``vertex'' detector. Thanks to topological cuts regarding this displacement, this ``nonprompt`` component can be isolated.  Just as for the other aforementioned decays, charmonium production in $b$-hadron decay can also be considered as a production mode by itself. As what concerns the decay from excited states, the FD yield is obtained either by measuring the production rates of the excited states which are then multiplied\footnote{This simple picture can in fact be complicated by efficiency, (polarisation-dependent) acceptance  and kinematical effects.} by the branching ratio to the measured quarkonium or by looking at a specific decay channel. Well known examples of measured decay channels are $\chi_b \to \Upsilon \, \gamma $,  $\chi_c \to J/\psi \, \gamma $ and $\psi(2S) \to J/\psi \,\pi\, \pi$.   

From this discussion, it clearly appears that, whereas dealing with direct yields is significantly simpler theory wise, it
brings in nontrivial experimental complications. It even gets  more complicated when it comes to differential-yield measurements since, in principle, the FDs should also be differentially determined. 

Finally, we note that the $\psi(2S)$ has a special status since the possible FD it can receive is limited to the nonprompt yield which is rather easy to subtract. The $\chi_b(3P)$ are currently the only states with a negligible FD of any kind but $P$-wave quarkonia are notably more complicated to measure via
their electromagnetic decays of $S$ waves and a photon. As such, the data sample of $\chi_b(3P)$ remains very limited.

\subsubsection{$b$-hadron FD}

Disentangling the prompt from the nonprompt yield is thus usually straightforward and does not generate any specific ambiguities. In fact, it can even be done on a event-by-event basis such that the isolation of the nonprompt yield can also be made in associated-production processes which we discuss in section~\ref{ch:associated}. However, some important past measurements on which we will report later were performed without a vertex detector or without such a subtraction and some muon detectors are still running
without any (the ALICE muon arm, the COMPASS detector). It is therefore still relevant to remember a few point about such a FD. 

First, it is generally believed that the nonprompt fractions for hadroproduction is negligible at RHIC energies ($\sqrt{s}=200$ GeV) and below. Yet, no direct determination exists. A recent study by PHENIX~\cite{Aidala:2017iad}, but at $\sqrt{s}=500$~GeV, showed
that the nonprompt fraction of the $P_T$-integrated yield is $8.1\% \pm 2.3\%\text{(stat)} \pm 1.9\%\text{(syst)}$ at forward rapidities. Such a value is a little smaller than the LHC measurements, \ie\  around 12\%~(see~\cite{Andronic:2015wma}). It would be interesting to meaasure such a nonprompt fraction in the LHC fixed-target mode in $p$H collisions at $\sqrt{s}=115$~GeV~\cite{Brodsky:2012vg,Hadjidakis:2018ifr}.
It should however be kept in mind that as soon as $P_T$ cuts are imposed such a FD is expected to steadily grow even up to more than 20\% at $P_T \simeq 6$~GeV in the range covered by the RHIC experiments~\cite{Adamczyk:2012ey} . For the interpretation of most of the current $pp$ data, it is however not problematic.
At TeV energies, the nonprompt fraction is about 12\% at low $P_T$ and can reach 70\% at $P_T$ on the order of 50-100 GeV 
(see~\cite{Andronic:2015wma}). Clearly, with such a high nonprompt fraction, one should always be careful in interpreting more specific aspects of the data than the yield. We guide the reader to~\cite{Andronic:2015wma} for a review on these aspects.

\begin{figure}[hbt!]
\centering
\subfloat[]{\includegraphics[width=0.5\columnwidth]{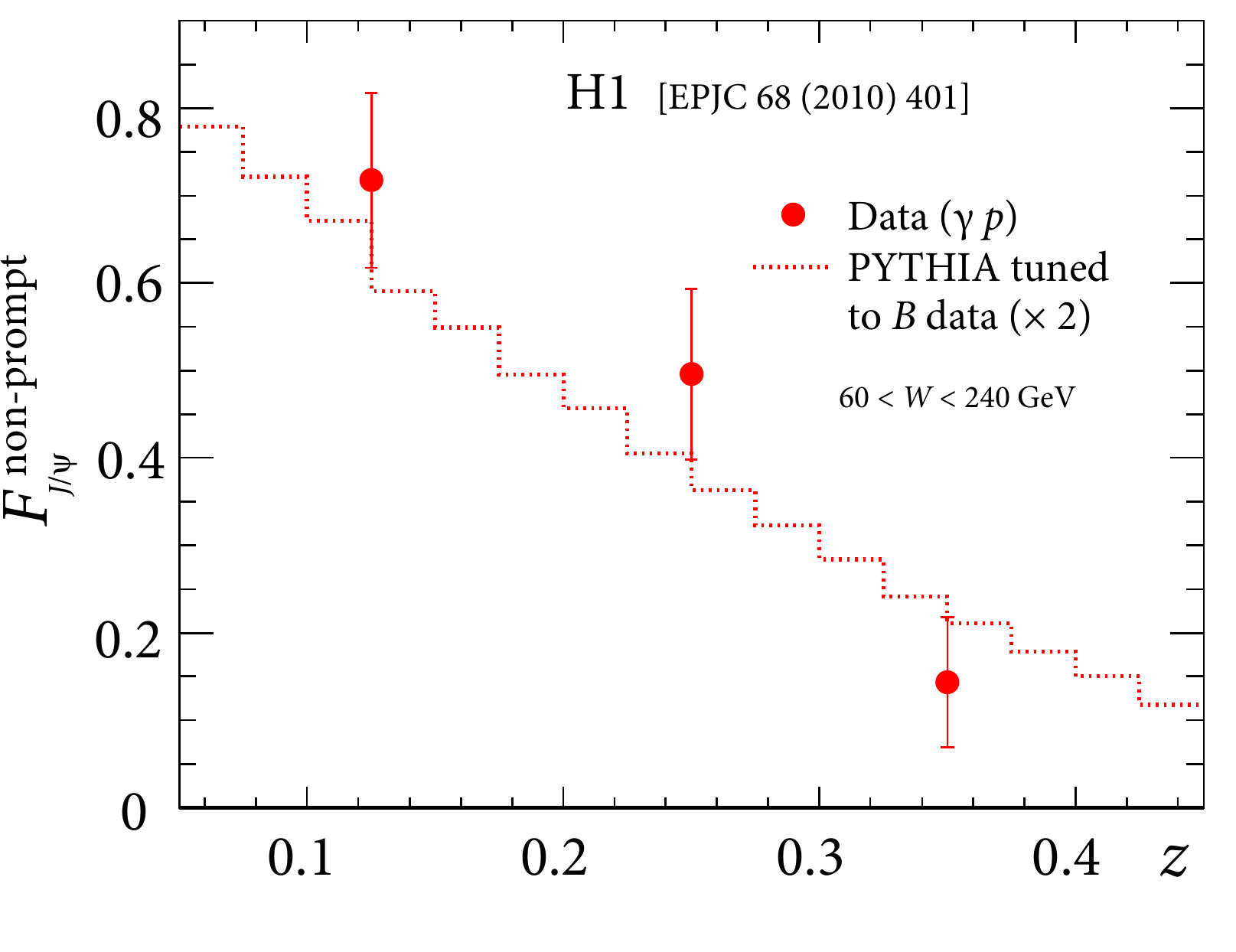}\label{fig:EPJC-68-401-H1-fig4c}}
\subfloat[]{\includegraphics[width=0.5\columnwidth]{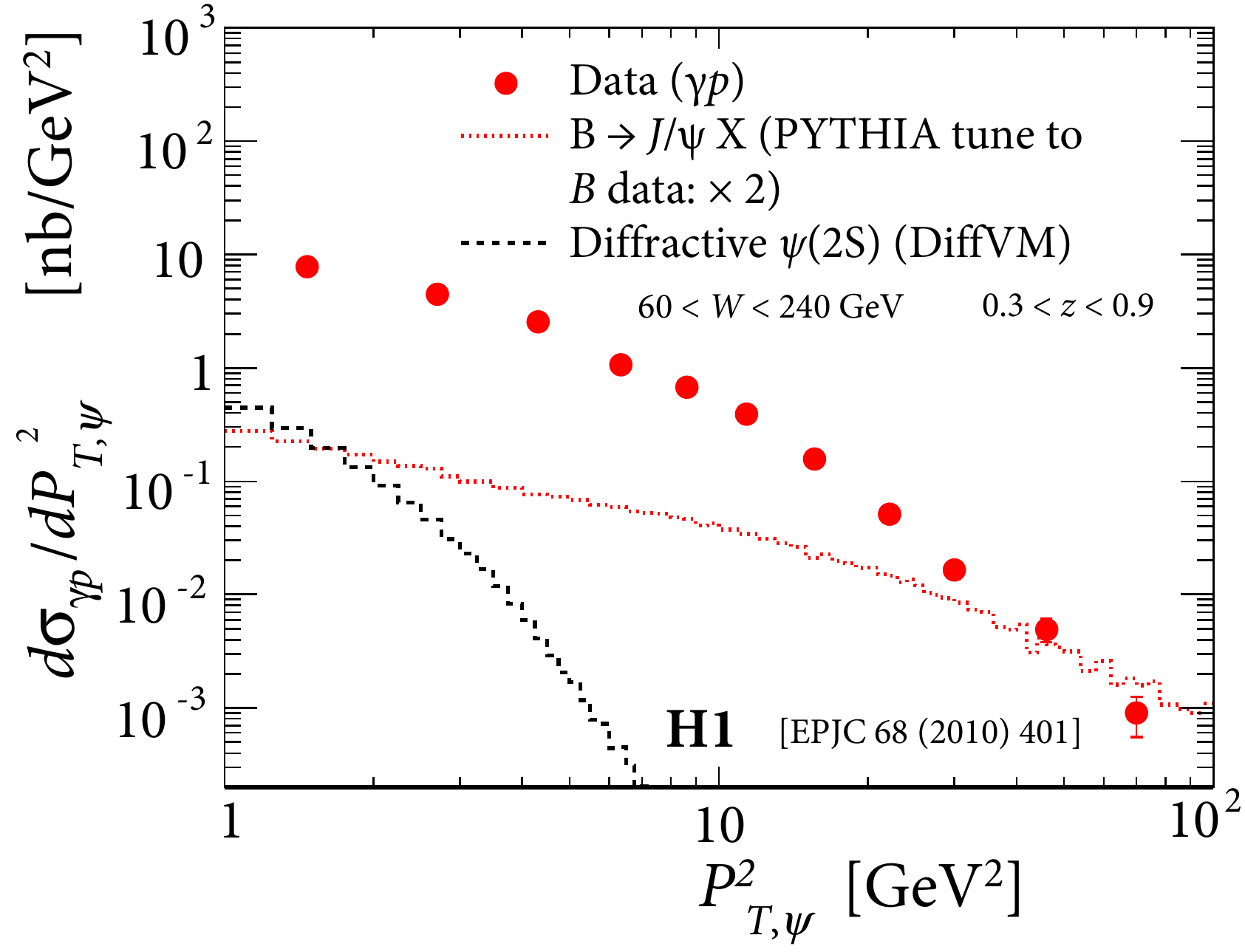}\label{fig:EPJC-68-401-H1-fig5c}}
\caption{(a) nonprompt fraction of photoproduced $J/\psi$ for $60 < W < 240$~ GeV. 
(b) $P^2_T$ differential cross section of photoproduced $J/\psi$ for $60 < W < 240$~ GeV and $0.3 < z < 0.9$ : comparison between H1 data, simulation of the $B\to J/\psi$ FD with {\sc Pythia} tuned to $B$ data and of diffractive $\psi(2S)$ FD with DiffVM. 
Adapted from ~\cite{Aaron:2010gz}.}
\label{fig:photoproduction-b-FD}
\end{figure}

The nonprompt fraction of photoproduced $J/\psi$ was measured by H1 in 2010 and is shown on \cf{fig:EPJC-68-401-H1-fig4c}. One clearly sees that $F^{\rm nonprompt}_{J/\psi}$ becomes very large at small $z$ and is about 20\% at $z\simeq 0.35$. According to the simulations shown on~\cf{fig:EPJC-68-401-H1-fig5c}, one should expect $F^{\rm nonprompt}_{J/\psi}$ also to become very large at large $P_T$ (compare the red dotted curve with the data points).
In DIS, $F^{\rm nonprompt}_{J/\psi}$ seems to be smaller and barely reaches 20\% for $0.3 < z < 0.45$ according to tuned simulation done by H1 in 2002~\cite{Adloff:2002ey}.

\subsubsection{Excited-quarkonium FD}
\label{sec:FD-onium_to_onium}
Let us now quickly review what is known about the FD of higher-mass quarkonium to the most studied
states. According to the above definition, one has \eg\ $F^{\chi_b(1P)}_{\Upsilon(1S)} = \frac{\sigma_{\chi_b(1P)} \, {\cal B}^{\chi_b(1P)}_{\Upsilon(1S)}}{\sigma_{\Upsilon(1S)}}$ with the obvious notation ${\cal B}^{\chi_b(1P)}_{\Upsilon(1S)}$ as the branching fraction
of the $\chi_b(1P)$ into $\Upsilon(1S)$.

Some knowledge on the branching ratio is thus necessary to anticipate whether or not a higher-mass quarkonium can provide a relevant FD fraction to a lower-lying state. Yet, 
one also needs to know the production yields. It has been seen in several examples that predictions should always be taken with a grain of salt before measurements of both yields are made. Along these lines, care should be taken when discussing the production in collision systems or associated-production channels where yields are only theoretically known. That being said, there is no known cases of significant measured FD fractions with branching ratios below the 10 \% level.

\paragraph{Charmonium-to-charmonium FD.}
\ct{tab:charmonium-branching} gathers our current knowledge~\cite{Tanabashi:2018oca} on the branching ratios 
for charmonium-to-charmonium decays.  Without any surprise, the branching fractions from and to the spin-triplet $S$ wave are the best known. For some decays, only upper limits are known and, for some, the decay has even never been seen. Among those, that of $\eta_c(2S) \to \eta_c$ is probably that which would be the most important to learn about for phenomenology.

\begin{table}[hbt!]
\centering\renewcommand{\arraystretch}{1.3}
\begin{tabular}{c|c|c|c|c|c|c|c}
             & $J/\psi$ & $\chi_{c0}$ & $\chi_{c1}$ & $h_c$       & $\chi_{c2}$  & $\eta_c(2S)$ & $\psi(2S)$ \\
\hline \hline
$\eta_c$           &    $1.7 \pm 0.4$        &  $< 0.07$           &    $< 0.32$         & $51 \pm 6$ & $< 0.32$ & $< 25$ & $0.34 \pm 0.05$ \\
$J/\psi$              &   --     &    $1.40 \pm 0.05$         &   $34.3 \pm 1.0$          &  $< 18$            & $19.0 \pm 0.5$  & $< 1.4$ & $63.4\pm 0.6$         \\
$\chi_{c0}$           &          &     --      &     ?        &      ?       &      ?        &  ?& $9.79 \pm 0.20$  \\
$\chi_{c1}$           &          &             &     --      &       ?      &       ?       & ? & $9.75 \pm 0.24$  \\
$h_c$                 &          &             &             &      --     &       ?       &  ? & $0.086 \pm 0.013$ \\
$\chi_{c2}$          &          &             &             &             &      --      &  ?& $9.52 \pm 0.20$     \\
$\eta_c(2S)$          &          &             &             &             &            & -- &$0.07 \pm 0.05$ \\
\end{tabular}
\caption{Known values of or limits on the branching ratios for charmonium to charmonium decays [in per cent]~\cite{Tanabashi:2018oca}.}\label{tab:charmonium-branching}
\end{table}

\begin{figure}[!htb]
\begin{center}
\subfloat[]{
        \label{fig:chic-FD-LHCb}
        \includegraphics[width=0.35\columnwidth]{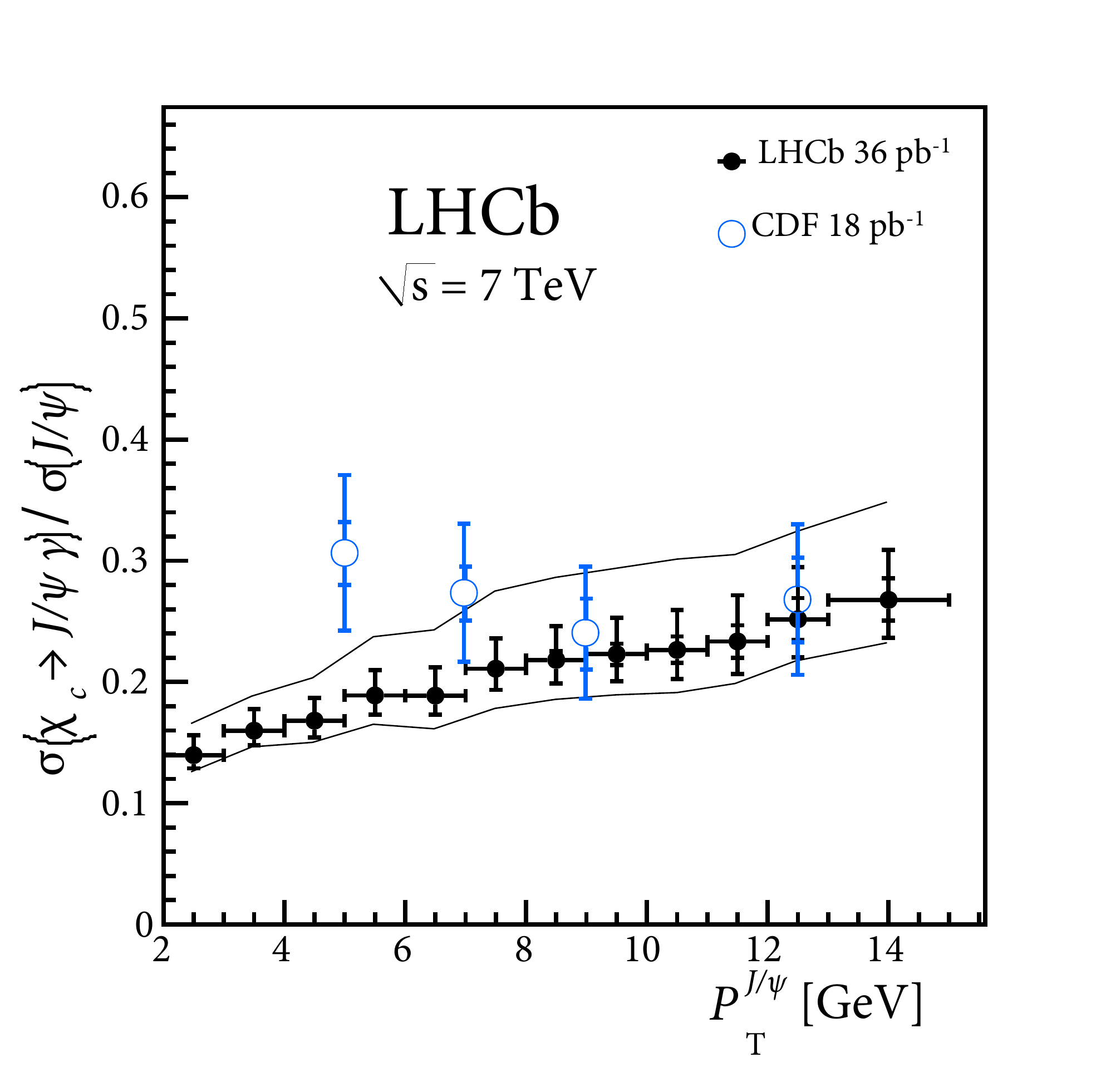}
        }
\subfloat[Low $P_T$ $J/\psi$]{\raisebox{1.8cm}{\includegraphics[width=0.3\columnwidth]{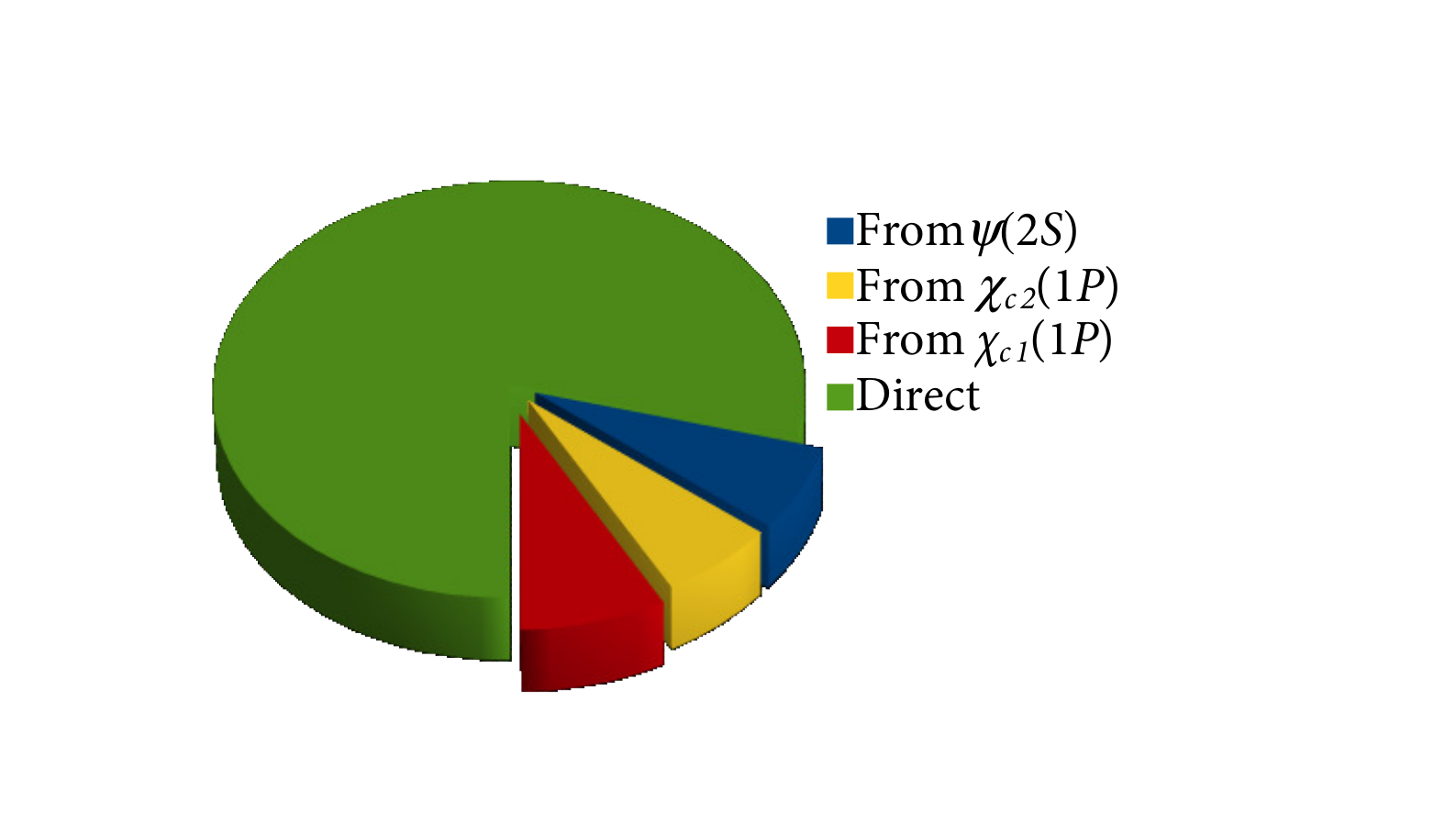}}\label{fig:pp:FeedDownFraction-psi-lowPT}}
\subfloat[High $P_T$ $J/\psi$]{\raisebox{1.8cm}{\includegraphics[width=0.3\columnwidth]{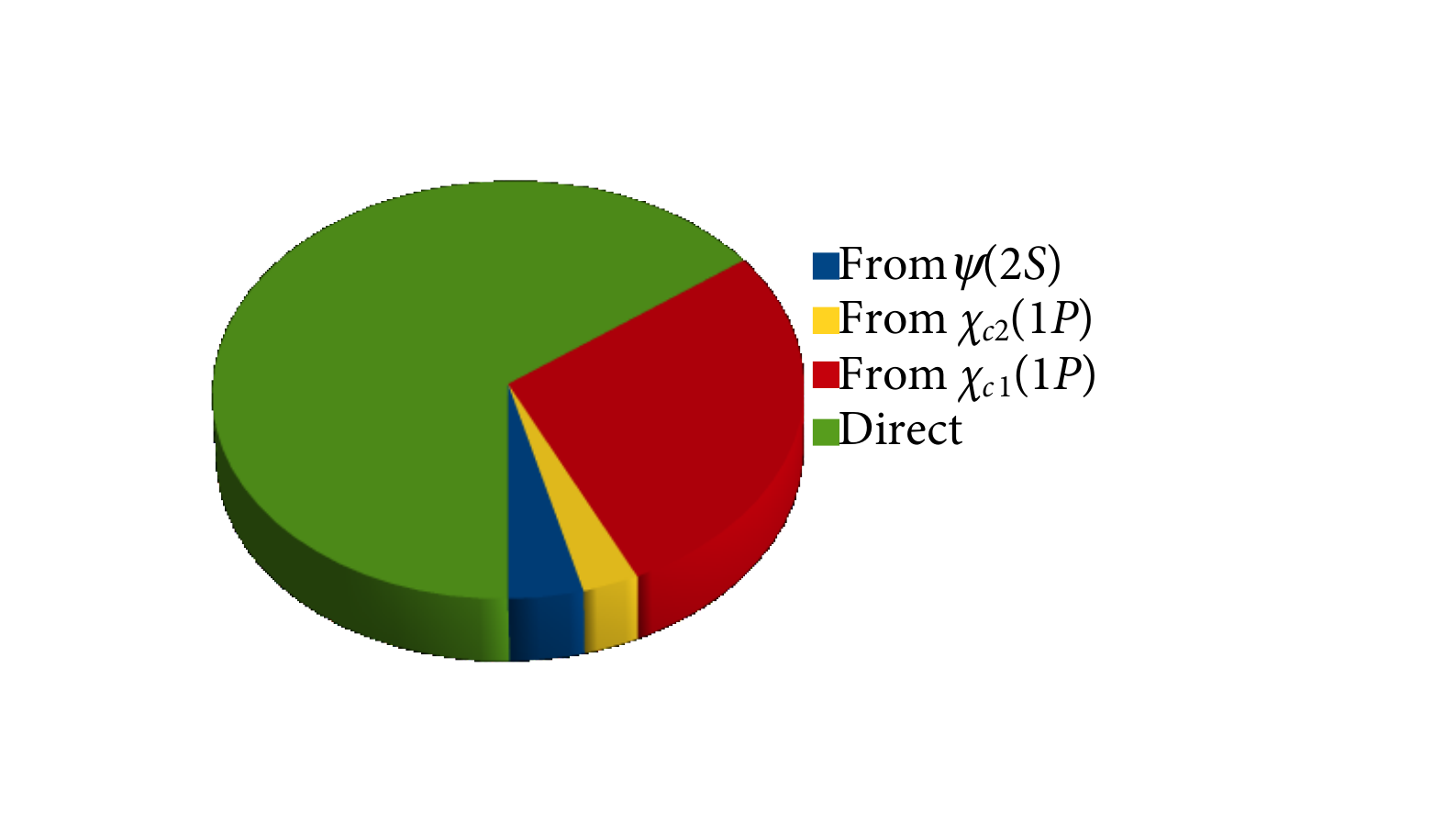}}\label{fig:pp:FeedDownFraction-psi-highPT}}
\caption{(a) $F^{\chi_{c}}_{J/\psi}$ as function of $P_T^{J/\psi}$ from CDF~\cite{Abe:1997yz} and LHCb~\cite{LHCb:2012af}. (b) Typical sources of hadroproduced $J/\psi$ at low and high \pt.  These numbers are mostly derived from LHC measurements~\cite{Sirunyan:2017qdw,Abelev:2014qha,Aaij:2013yaa,Aaij:2013dja,Chatrchyan:2012ub,LHCb:2012af,LHCb:2012ac} assuming  an absence of a significant rapidity dependence. (a) Adapted from~\cite{LHCb:2012af}.
}
\end{center}
\end{figure}

Since hadroproduction is the most studied case, we start by reviewing it and we then only complement the discussion by specific statements for
lepto/photoproduction and $e^+e^-$ annihilation.
Let us start with the $J/\psi$ case. From~\ct{tab:charmonium-branching}, one naturally expects the FDs from the $\chi_{c1}$,
$\chi_{c2}$ and $\psi(2S)$ to be the largest. In fact, historically, based a simplistic $\alpha_s$ counting, it was believed that
$\chi_c$ FD would be the main source of (prompt) $J/\psi$. This was contradicted by the pioneering measurement of the 
CDF collaboration released in 1997~\cite{Abe:1997yz} with a FD fraction on the order of 30~\% (see the open blue circle on \cf{fig:chic-FD-LHCb}) . Measurements were then performed at the LHC by LHCb and then ATLAS~\cite{LHCb:2012af,ATLAS:2014ala}. We now know that this fraction increases between $\pt=2$~GeV and 15~GeV and that the $\chi_c$ FD fraction is rather $10\div15 \%$ at low $P_T$.
Measurements of the $\chi_{c2}/\chi_{c1}$ yield ratio~\cite{LHCb:2012ac,Chatrchyan:2012ub,Aaij:2013dja,ATLAS:2014ala} then allow one to derive the sharing of this FD between both states, namely $F^{\chi_{c1}}_{J/\psi}$ and $F^{\chi_{c2}}_{J/\psi}$. In 2013, LHCb reported~\cite{Aaij:2013dja} the first evidence for $\chi_{c0}$ hadroproduction and derived
that $\sigma(\chi_{c0}) \simeq \sigma(\chi_{c2})$ in the kinematical region cover by their acceptance. As such $F^{\chi_{c0}}_{J/\psi}$ is expected to be about half a per cent.
 Overall, the situation can be summarised as on \cf{fig:pp:FeedDownFraction-psi-lowPT} \& \ref{fig:pp:FeedDownFraction-psi-highPT} and \ct{tab:FD-fractions-psi},

\begin{table}[!htb]
\begin{center}\renewcommand{\arraystretch}{1.3}
\begin{tabular}{c|cccc}
                        & direct & from $\chi_{c1}$ &  from $\chi_{c2}$ & from $\psi(2S)$ \\\hline\hline
``low'' $P_T$ $J/\psi$  & $79.5 \pm 4$ \% & $ 8 \pm 2$ \% & $ 6 \pm 1.5$ \% &$ 6.5 \pm 1.5$ \%\\
``high'' $P_T$ $J/\psi$ & $ 64.5 \pm 5$ \%&$ 23 \pm 5$ \% &$ 5 \pm 2$ \% &$ 7.5 \pm 0.5$ \%
\end{tabular}
\caption{$J/\psi$ FD fraction in hadroproduction at Tevatron and LHC energies. \label{tab:FD-fractions-psi} }
\end{center}
\end{table}

As what concerns the $\eta_c$, there is no experimental FD determination. Existing phenomenological studies  indicate that the sole $h_c$ --whose inclusive production has never been measured-- may contribute in a relevant way to the $\eta_c$. However, they also tend to agree~\cite{Han:2014jya,Butenschoen:2014dra,Zhang:2014ybe} that the $h_c$ cross section should be slightly suppressed compared to the other quarkonium states and thus that its FD can be neglected in a first approximation. According to \ct{tab:charmonium-branching}, another possible relevant FD source may be that of the $\eta_c(2S)$. Its prompt cross section  is not yet measured and its branching fraction to  $\eta_c$ is essentially unknown --only a loose upper bound exists. If this fraction is found to be large and the $\eta_c(2S)$ cross section is particularly large (see~\cite{Lansberg:2017ozx} for some expectations and section~\ref{subsubsec:etaQ-CSM}), the $\eta_c(2S)$ could be a visible source of FD to $\eta_c$ --unlikely larger than 10~\% though. 
From~\ct{tab:charmonium-branching}, one might expect some $\psi(2S)$ FD to the $\chi_c$ states. This has so far often been ignored in the literature. Indeed, using the $\psi(2S)$ and $\chi_c$ FD to $J/\psi$ of \cf{fig:chic-FD-LHCb}, one easily derives that the $\psi(2S)$ FD to the $\chi_{c1}$ is on the order of $2\div 4 \%$ depending on $\pt$ and to the $\chi_{c2}$ on the order of 2 \%. However, there may be situations (like lepto/photoproduction) where vector-quarkonium production
would highly be favoured compared to the $C$-even $P$-wave production and where such a FD would become relevant.   
The same may happen for the $\eta_c(2S)$ FD to the $h_c$. In this case, none are measured in any inclusive reaction and thus the size of such a FD remains an academical topic. 

We end this short survey on the FD in hadroproduction by the $\eta_c(2S)$ case which has recently been measured in inclusive $b$-hadron decays by LHCb~\cite{Aaij:2016kxn} and whose
production should be measurable with data currently on tapes~\cite{Lansberg:2017ozx}. The only possible decay is obviously
from the $\psi(2S)$ and its branching is at the per-mil level, so that it is unlikely to affect the $\eta_c(2S)$ phenomenology. 

FD can in principle also affect the phenomenology of photo- and electro-production. We have just seen that $b$-hadron decay can be a significant source of $J/\psi$ in some part of the phase space. The $\psi(2S)/J/\psi$ cross-section ratio has been measured by ZEUS~\cite{Chekanov:2002at} in photoproduction for $50 < W < 180$~GeV and $0.55 < z < 0.9$. From this ratio, ZEUS derived $F^{\psi(2S)}_{J/\psi} =15.5^{+3.8}_{-4.2}\%$. 

Simulations of the contribution of $\chi_c$ FD in photoproduction at HERA were reported in~\cite{Adloff:2002ex} and $F^{\chi_{c}}_{J/\psi}$ was found negligibly small in the region, $0.3 < z<0.9$. So far no published result are available; the observation of some counts with H1 has been reported in a Master thesis in 2001~\cite{Kuckens:2001} but this was not followed by any publication. Future measurements both in photo- and electro-production at a possible EIC would be invaluable. 

As what concerns $e^+e^-$ annihilation, $F^{\psi(2S)}_{J/\psi}$ is expected to be about 25\% if the CS contributions dominate. In principle, the measurement of $\sigma(\psi(2S)+X)$ by Belle~\cite{Abe:2001za} in 2001 should allow one to derive such a FD fraction. However, as we will see in section~\ref{sec:CSM_eeproduction}, it is very important to distinguish the production with or without another charm pair since they arise from different mechanisms. Such a separation could not be done for the $\psi(2S)$.

\begin{figure}[!htb]
\begin{center}
             \includegraphics[width=0.55\columnwidth]{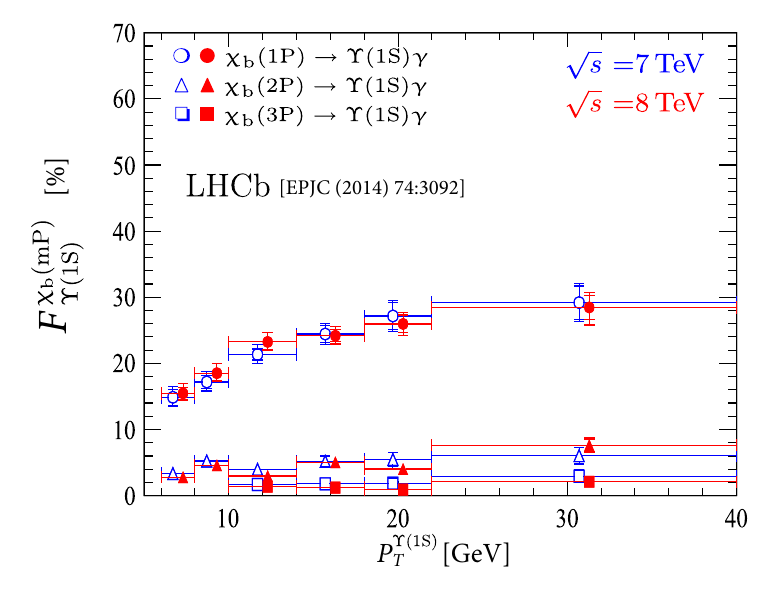}
   	
\caption{
\label{fig:pp:FD-fractions-LHCb-ChibtoUpsi1S}
Measured FD fraction of $\chi_b(nP)$ to $\Upsilon(1S)$. Adapted from~\cite{Aaij:2014caa}.
}
\end{center}
\end{figure}

\paragraph{Bottomonium-to-bottomonium FD.}

The discussion of the $\Upsilon(nS)$ case is the most complex, yet the most relevant since the FD fraction are significant. The first quantitative information were provided by a study of the CDF collaboration~\cite{Affolder:1999wm} in the early 2000's, which indicated that, for $P_T > 8$ GeV, half of the $\Upsilon(1S)$ 
yield was from FD. We note here that deriving the FD fractions of the $\Upsilon(mS)$ into $\Upsilon(nS)$ (with $n<m$) is easy if one has the corresponding yields and branching fractions. For quickly varying differential cross section, it is however important to take into account the mass difference between the states\footnote{The same of course 
applies for the charmonia.}. This was carefully done by W\"ohri in~\cite{Woehri:QWG2014}. The following discussion
 relies on her work and LHC $\chi_b$ experimental measurements.

 Since the discovery of the $\chi_b(3P)$ by ATLAS~\cite{Aad:2011ih}, we know that the 3 $\Upsilon(nS)$ receive FD from $\chi_b(nP)$. A crucial study by LHCb in 2014~\cite{Aaij:2014caa} in fact indicated that, contrary to initial beliefs, the FD fraction $F^{\chi_b(3P)}_{\Upsilon(3S)}$ was significant. For this precise transition, the measurement was however carried out  at rather large $P_T$, on the order of $20\div30$~GeV. Owing to the large mass splitting between the
$\chi_b(3P)$ and the $\Upsilon(1S)$, $F^{\chi_b(3P)}_{\Upsilon(1S)}$ could however be measured down to lower $P_T$. This can  be used to derive information on the $\chi_b(3P)$ yield\footnote{In 2018, CMS performed~\cite{Sirunyan:2018dff} the first separation of the 
$\chi_{b1}(3P)$ and $\chi_{b2}(3P)$ peaks but cross section ratios have not yet been measured} and extrapolate the LHCb measurement of $\chi_b(3P)$ FDs to lower $P_T$.

Such an extrapolation is necessary. Indeed, $F^{\chi_b(1P)}_{\Upsilon(1S)}$ decreases by a factor of 2 between $P_T=30$ GeV and $P_T=7$ GeV (see~\cf{fig:pp:FD-fractions-LHCb-ChibtoUpsi1S})\footnote{In view of \cf{fig:pp:FD-fractions-LHCb-ChibtoUpsi1S}, it is even legitimate to wonder whether $F^{\chi_b(1P)}_{\Upsilon(1S)}$  further decreases below 7 GeV or saturates when $P_T \leq M_\Upsilon$. Only a dedicated low-$P_T$ measurement could clarify this.}. 
One can thus wonder whether $F^{\chi_b(2P)}_{\Upsilon(2S)}$, $F^{\chi_b(3P)}_{\Upsilon(3S)}$ and $F^{\chi_b(3P)}_{\Upsilon(2S)}$ also strongly vary with $P_T$.
One first notes that $ \frac{F^{\chi_b(nP)}_{{\Upsilon(mS)}}  F^{{\Upsilon(mS)}}_{\Upsilon(1S)}}{F^{\chi_b(nP)}_{\Upsilon(1S)}} =  \frac{{\cal B}^{\chi_b(nP)}_{\Upsilon(mS)}  {\cal B}^{\Upsilon(mS)}_{1S}}{{\cal B}^{nP}_{1S}}$ and is thus constant. 
The $\Upsilon(nS)$ (differential) yields and the $\Upsilon(nS) \to \Upsilon(mS)$ branching ratios are reliably known such that $F^{\Upsilon(mS)}_{\Upsilon(1S)}$ are well determined and allow one to evaluate $F^{\Upsilon(nS)}_{\Upsilon(1S)}/F^{\chi_b(nP)}_{\Upsilon(1S)}$ ($n=2,3$).
Combining existing experimental information, one finds that $F^{\Upsilon(mS)}_{\Upsilon(1S)}/F^{\chi_b(nP)}_{\Upsilon(1S)}$ ($n=m=2,3$ and $m=2,n=3$) are mostly constant vs. $P_T$ (down to $P_T=(7,12)$ GeV for $n=(2,3)$). One can thus reasonably infer that $F^{\chi_b(2P)}_{\Upsilon(2S)}$,  $F^{\chi_b(3P)}_{\Upsilon(3S)}$ and $F^{\chi_b(3P)}_{\Upsilon(2S)}$ are mostly constant in the corresponding ranges, at variance with $F^{\chi_b(1P)}_{\Upsilon(1S)}$. As such, the measured values at rather large $P_T$, respectively $30\pm5$\%,  $40\pm10$\% and $4.5\pm1.5$\%, can be used at low(er) $P_T$. However, dedicated measurements will be useful in confirming such extrapolations and we have thus arbitrarily doubled in our tables the uncertainty attached to these extrapolations. We note that these --which
are purely based on data-- seem to be supported by NLO NRQCD computations~\cite{Gong:2013qka,Han:2014kxa,Feng:2015wka} for which the LDMEs are fit on the (large $P_T$) data and which can then be extended to lower $P_T$.

\begin{figure}[!htb]
\begin{center}
\subfloat[Low $P_T$ $\Upsilon(1S)$]{\includegraphics[width=0.33\columnwidth]{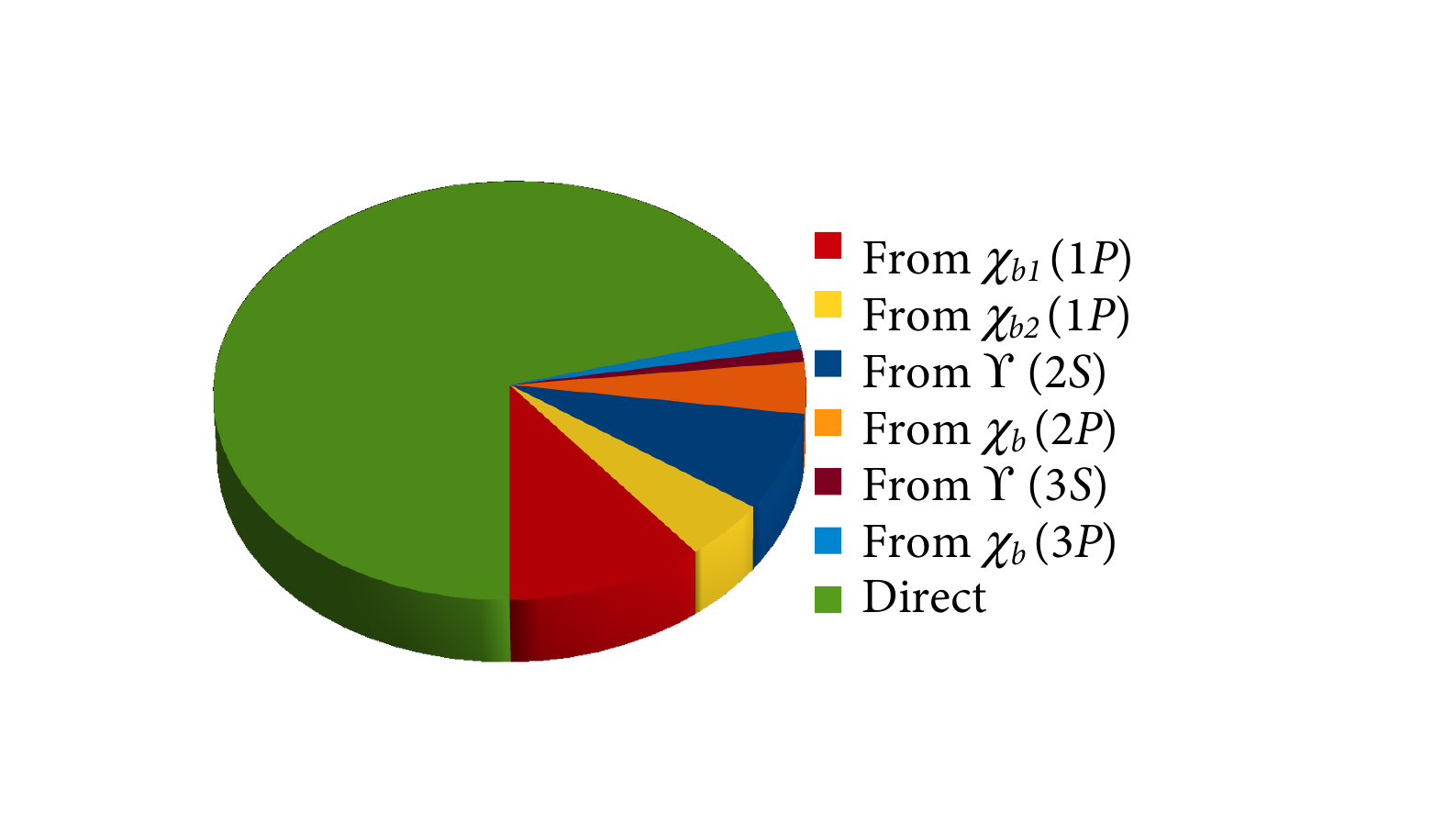}}
\subfloat[Low $P_T$ $\Upsilon(2S)$]{\includegraphics[width=0.33\columnwidth]{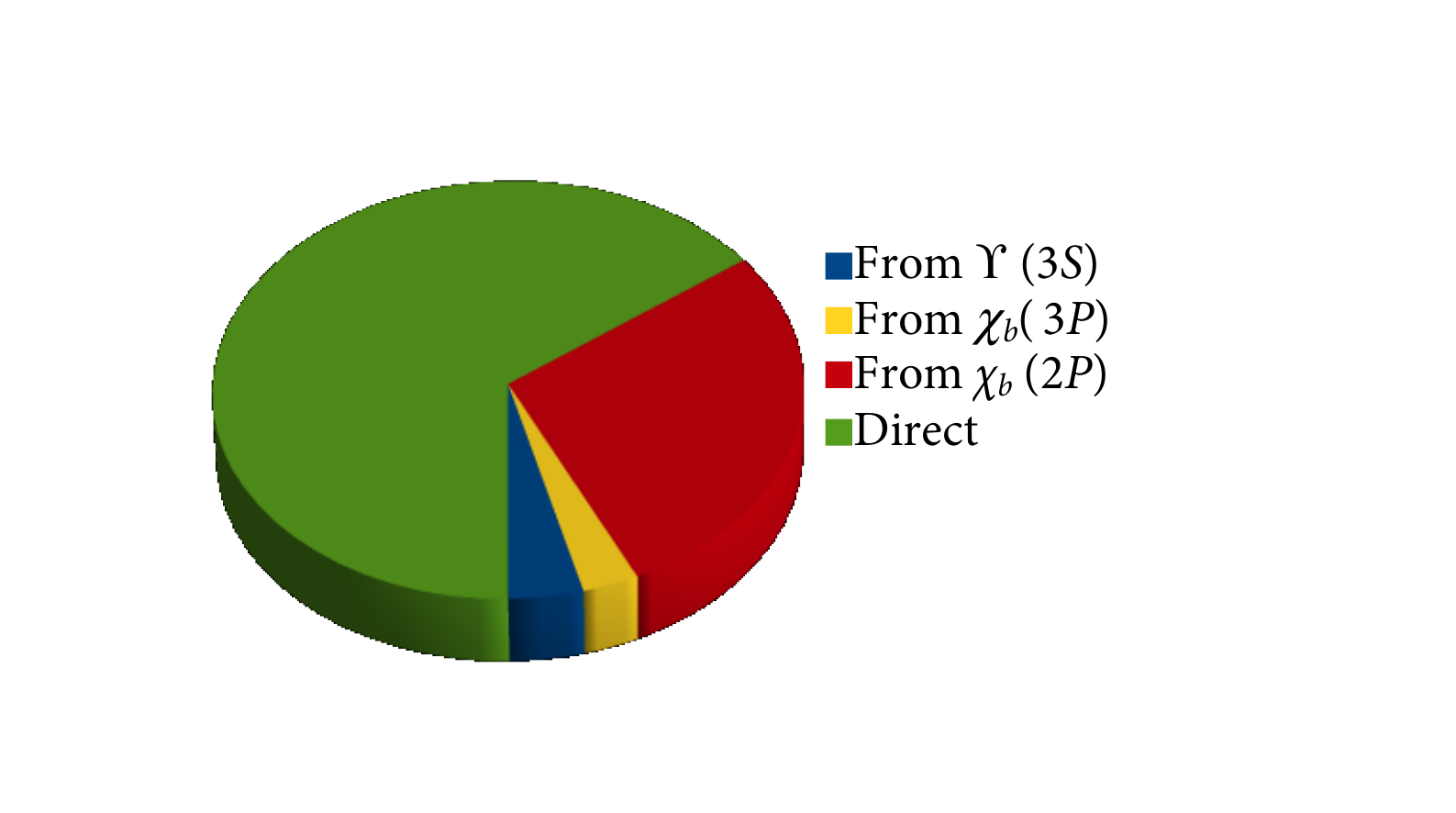}}
\subfloat[Low $P_T$ $\Upsilon(3S)$]{\includegraphics[width=0.33\columnwidth]{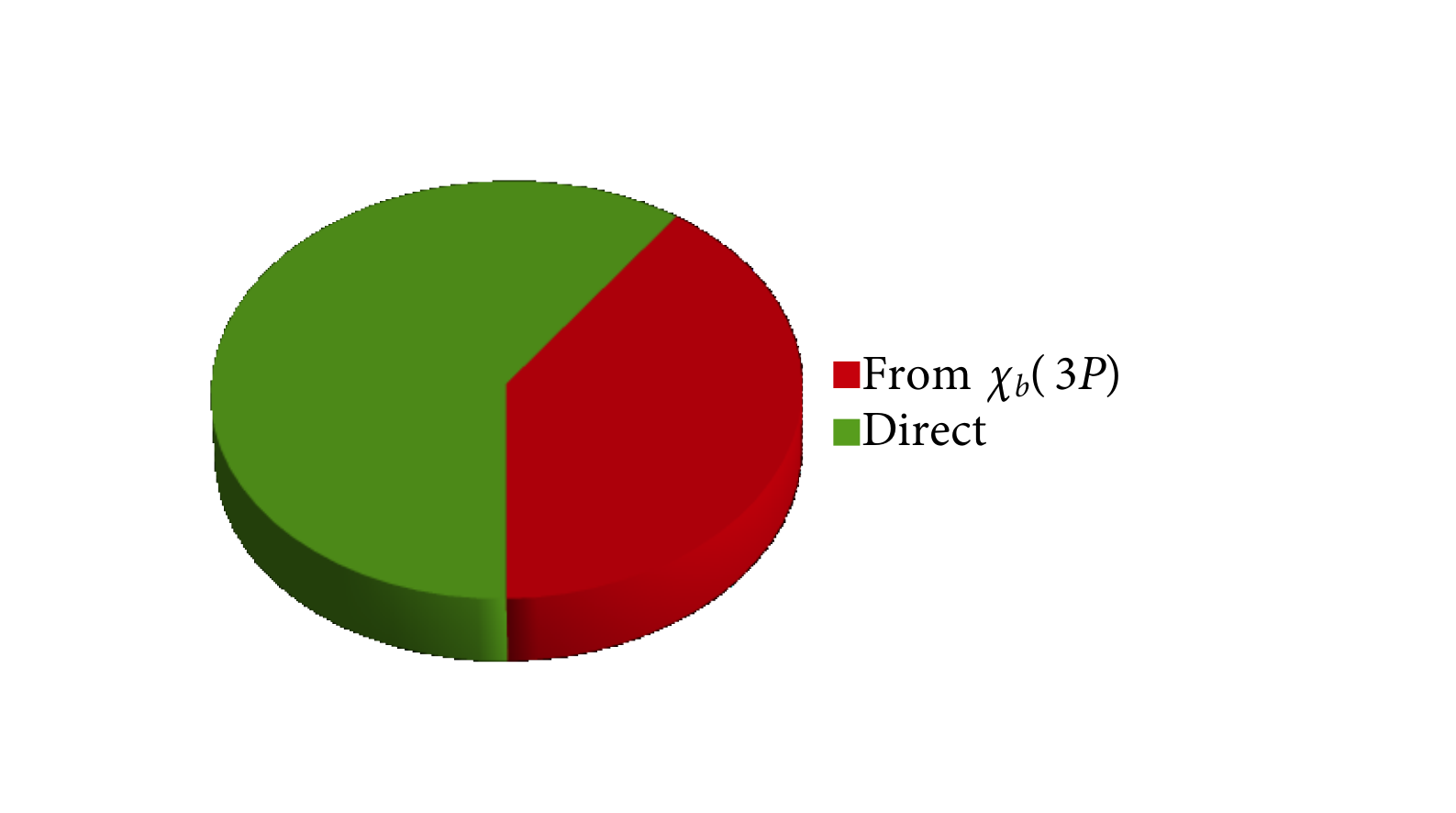}}\\
\subfloat[High $P_T$ $\Upsilon(1S)$]{\includegraphics[width=0.33\columnwidth]{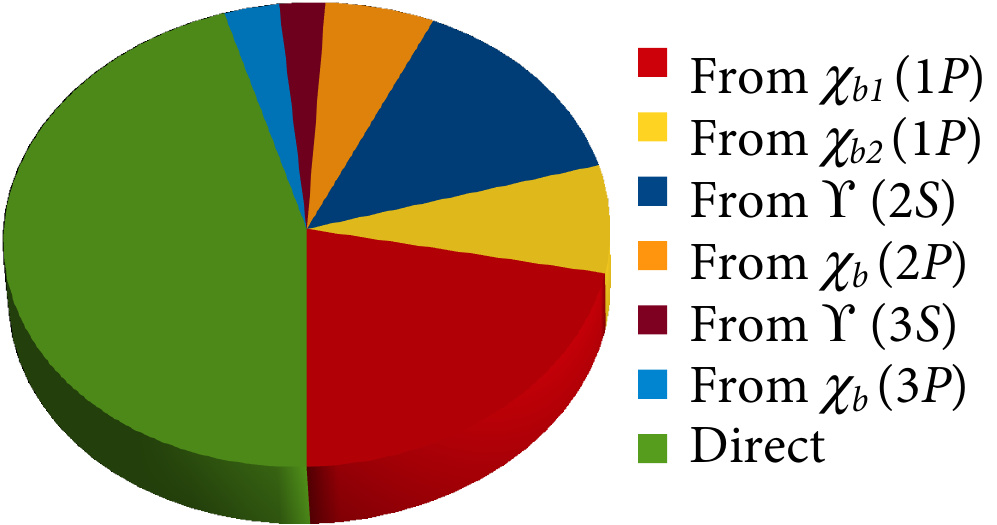}}
\subfloat[High $P_T$ $\Upsilon(2S)$]{\includegraphics[width=0.33\columnwidth]{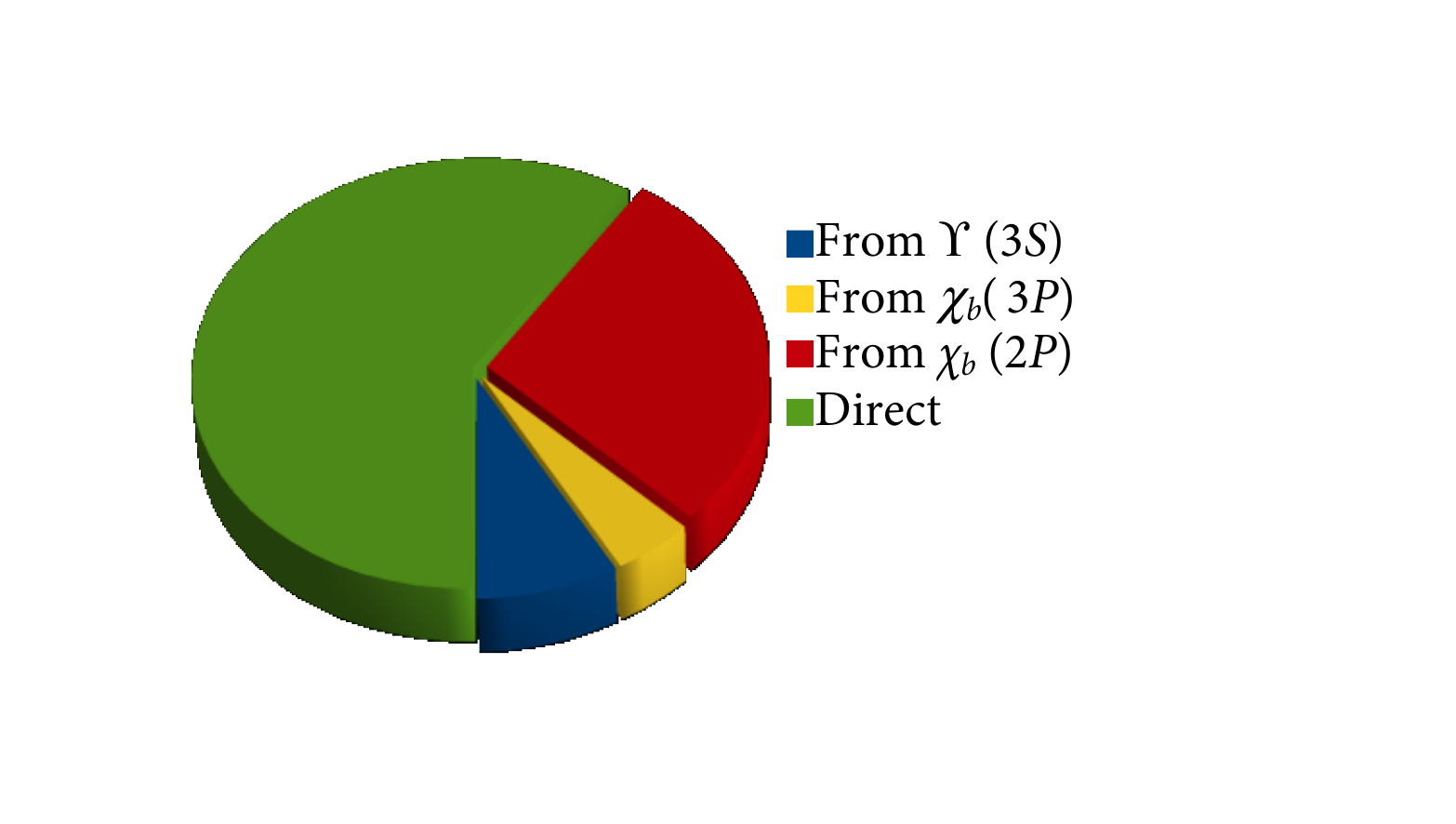}}
\subfloat[High $P_T$ $\Upsilon(3S)$]{\includegraphics[width=0.33\columnwidth]{FD-Upsi3S-AllPT.pdf}}
\caption{
        \label{fig:pp:FeedDownFraction-upsi}
Typical sources of hadroproduced  $\Upsilon(nS)$ at low and high \pt. These numbers are mostly derived from
LHC measurements~\cite{Aaij:2014caa,Aad:2012dlq,Aad:2011xv, Chatrchyan:2013yna,Abelev:2014qha,Khachatryan:2010zg,Aaij:2013yaa,Aaij:2012se,LHCb:2012aa} assuming  an absence of a significant rapidity dependence.
}
\end{center}
\end{figure}

\cf{fig:pp:FeedDownFraction-upsi} outline the FD pattern of the $\Upsilon(nS)$ based on the above considerations. \ct{tab:FD-fractions-Upsi1S}, \ref{tab:FD-fractions-Upsi2S} \& \ref{tab:FD-fractions-Upsi3S} gather similar information as in the plots along with the corresponding estimated uncertainties. 
Owing to the lack of measurements, we do not discuss the other FD among of the bottomonium states.

\begin{table}[!htb]
\begin{center}\renewcommand{\arraystretch}{1.3}
\begin{tabular}{c|cccccccc}
                        & $F^{\rm direct}_{\Upsilon(1S)}$ &  $F^{\chi_{b1}(1P)}_{\Upsilon(1S)}$ &  
$F^{\chi_{b2}(1P)}_{\Upsilon(1S)}$  & $F^{\Upsilon(2S)}_{\Upsilon(1S)}$ & $F^{\chi_{b}(2P)}_{\Upsilon(1S)}$ & $F^{\Upsilon(3S)}_{\Upsilon(1S)}$      & $F^{\chi_{b}(3P)}_{\Upsilon(1S)}$\\\hline\hline
``low''  $P_T$  & $71 \pm 5$     & $10.5 \pm 1.6  $ & $4.5 \pm 0.8  $ & $7.5 \pm 0.5  $ & $4 \pm 1  $&
$1 \pm 0.5  $& $1.5 \pm 0.5  $\\
``high'' $P_T$  & $45.5 \pm 8.5$   & $21.5 \pm 2.7  $ & $7.5 \pm 1.2  $ & $14  \pm 2    $ & $6 \pm 2  $&
$2.5 \pm 0.5  $ & $3 \pm 1  $ 
\end{tabular}
\caption{$\Upsilon(1S)$ FD  fraction [in \%] in hadroproduction at Tevatron and LHC energies. \label{tab:FD-fractions-Upsi1S} }
\end{center}
\end{table}

\begin{table}[!htb]
\begin{center}\renewcommand{\arraystretch}{1.3}
\begin{tabular}{c|cccccc}
                        & $F^{\rm direct}_{\Upsilon(2S)}$  & $F^{\chi_{b}(2P)}_{\Upsilon(2S)}$ & $F^{\Upsilon(3S)}_{\Upsilon(2S)}$      & $F^{\chi_{b}(3P)}_{\Upsilon(2S)}$\\\hline\hline
``low''  $P_T$  & $65 \pm 20$       & $28 \pm 16  $ & $4 \pm 1  $ & $3 \pm 3$\\
``high'' $P_T$  & $59.5 \pm 11.5$   & $28 \pm 8  $ & $8 \pm 2  $ & $4.5 \pm 1.5$
\end{tabular}
\caption{$\Upsilon(2S)$ FD  fraction [in \%] in hadroproduction at Tevatron and LHC energies. We have doubled the uncertainties on $F^{\chi_{b}(2P)}_{\Upsilon(2S)}$ and $F^{\chi_{b}(3P)}_{\Upsilon(2S)}$ at low  $P_T$ since they are extrapolated. \label{tab:FD-fractions-Upsi2S} }
\end{center}
\end{table}

\begin{table}[!htb]
\begin{center}\renewcommand{\arraystretch}{1.3}
\begin{tabular}{c|cccccc}
                        & $F^{\rm direct}_{\Upsilon(3S)}$  & $F^{\chi_{b}(3P)}_{\Upsilon(3S)}$\\\hline\hline
``low''  $P_T$  & $60 \pm 20$     & $40 \pm 20  $ \\
``high'' $P_T$  & $60 \pm 10$     & $40 \pm 10  $ 
\end{tabular}
\caption{$\Upsilon(3S)$ FD  fraction [in \%] in hadroproduction at Tevatron and LHC energies. We have doubled the uncertainties on $F^{\chi_{b}(3P)}_{\Upsilon(3S)}$ at low  $P_T$ since it is extrapolated. \label{tab:FD-fractions-Upsi3S} }
\end{center}
\end{table}

\subsection{Recent developments in the CSM phenomenology}
\label{subsec:CSM_updates}

In the CSM~\cite{Chang:1979nn,Berger:1980ni,Baier:1983va}, the matrix element to create
for instance a $^3S_1$  quarkonium ${\Q}$ with a momentum $P_\Q$ and a polarisation $\lambda$
 possibly accompanied by other partons, noted $j$, 
is the product of the amplitude to create the corresponding heavy-quark pair, ${\cal M}(ab \to Q \bar Q+j)$, a spin
 projector $N(\lambda| s_1,s_2)$ and $R(0)$, the radial wave function at the origin in the configuration
space, obtained from the leptonic width or from potential models, namely 
\eqs{ \label{eq:CSM_generic}
{\cal M}&(ab \to {\Q}^\lambda(P_\Q)+j)=\!\sum_{s_1,s_2,i,i'}\!\!\frac{N(\lambda| s_1,s_2)}{ \sqrt{m_Q}} \frac{\delta^{ii'}}{\sqrt{N_c}} 
\frac{R(0)}{\sqrt{4 \pi}}
{\cal M}(ab \to Q^{s_1}_i \bar Q^{s_2}_{i'}(\mathbf{p}=\mathbf{0})  + j),
}
where $P_\Q=p_Q+p_{\bar Q}$, $p=(p_Q-p_{\bar{Q}})/2$, $s_1$ and $s_2$ are the heavy-quark spins, and $\delta^{ii'}/\sqrt{N_c}$ 
is the projector onto a CS state. 
$N(\lambda| s_i,s_j)$ 
is the spin projector, which has a simple expression in the non-relativistic limit: 
$\frac{1}{2 \sqrt{2} m_{Q} } \bar{v} (\frac{\mathbf{P}}{2},s_j) \Gamma_{S} u (\frac{\mathbf{P}}{2},s_i) \,\, $ 
with $\Gamma_S=\ep^{\lambda}_{\mu}\gamma^{\mu}$.  In the case of a pseudoscalar $^1S_0$ quarkonium (\eg\ $\eta_c$), 
$\Gamma_S$ instead reads $\gamma_5$. Summing over the quark spin yields to traces which can be evaluated in a standard way. 

In the CSM, quantum-number conservation imposes severe constraints on what the possible partonic reaction $ab \to {\cal Q}+j$ can be at LO, and even which topologies are possible via the number of gluons attached to the heavy-quark line. For the charmonia, here are some examples: 
\begin{itemize}
\item for hadroproduction, $gg \to \eta_c$ is allowed but, for $J/\psi$, one more gluon is needed and should be radiated off the charm line, \ie\ $gg \to J/\psi g$; $gg \to \chi_{c,2,0}$ is allowed but, for $\chi_{c1}$, one more gluon is also needed or, one of the initial gluons should be off-shell;
\item for photoproduction, $\gamma g \to J/\psi g$ is allowed but, for $\eta_c$, one more gluon is needed, \ie\ $\gamma g \to \eta_c g g$;
\item for production in $\gamma\gamma$ collisions, $\gamma \gamma \to \eta_c gg$ is allowed  but, for $J/\psi$, one more gluon is needed, \ie\ $\gamma \gamma \to J/\psi ggg$, etc.
\end{itemize}
Along theses lines, it should not be surprising that the QCD corrections to some processes can be very large since they allow for the first time the production of a given state or they correspond to specific topologies which show a different kinematical dependence than the LO ones. 


\subsubsection{$\psi$ and $\Upsilon$ hadroproduction at finite $P_T$}
\label{subsec:CSM_NLO_hadroproduction}
As we just mentioned, pseudoscalar and vector quarkonia are produced according to the CSM by
different partonic processes at LO. We therefore discuss them separately and start with the $\psi$ and $\Upsilon$
vector states.

\paragraph{$P_T$-differential production cross section at NLO.}
\label{subsec:CSM_NLO_PT}

In the case of $\psi$ or $\Upsilon$  hadroproduction, the LO contributions 
are at $\alphaS^3$ with only a single partonic process at work, 
namely $gg\to \Q g$ with 6 Feynman graphs to be evaluated. 
One of them  is drawn on \cf{diagram-CSM-a}. Like for any other $2\to2$ scattering, the differential partonic cross section is readily obtained from the amplitude
squared\footnote{The momenta of the initial gluons, $k_{1,2}$, are, in collinear factorisation~\cite{Brock:1993sz}, related to those of the colliding hadrons ($p_{1,2}$) through
$k_{1,2}=x_{1,2} \, p_{1,2}$. One then defines the Mandelstam variables for the partonic system: $\hat s = s x_1 x_2$, 
$\hat t=(k_1-P_{\Q})^2$ and $\hat u=(k_2-P_{\Q})^2$.}, 
\be \frac{d \hat \sigma}{d \hat t} = \frac{1}{16 \pi \hat s^2} \left| {\cal M}\right|^2,\label{eq:dsdt} \ee
from which one obtains the double differential cross section in $P_T$ ($P_T\equiv P_{\Q,T}$)  and the $\Q$ rapidity, $y$, for $pp \to \Q g$ after convolution
with the gluon PDFs and a change of variable:
\be 
\frac{d\sigma}{dydP_T}=\int_{x_1^{\rm min}}^1 dx_1 \frac{2 \hat s P_T g(x_1,\mu_F) g(x_2(x_1),\mu_F)}
{\sqrt{s}(\sqrt{s} x_1-m_T e^{y_\Q})}
\frac{d\hat \sigma}{d\hat t},
\ee
where $\displaystyle x_1^{\rm min}= \frac{m_T\sqrt{s}e^{y_\Q}-m_{\Q}^2}{\sqrt{s}(\sqrt{s}-m_T e^{-y_\Q})}$, $\displaystyle m_T=\sqrt{m_{_\Q}^2+P_T^2}$. Using the expression of $\left| {\cal M}\right|^2$ given \eg\ in~\cite{Baier:1983va}, one can easily reproduce the blue LO curve in \cf{fig:Psi2S-Tevatron}. Let us note that, on the way, 
that the inability of this LO CSM computation to reproduce the $\psi(2S)$ CDF data~\cite{Aaltonen:2009dm} is
in fact the original\footnote{except for the fact that we plotted RUN-II data} 
Tevatron puzzle known as the $\psi(2S)$ surplus, which boosted
the usage of the NRQCD and of the COM in particular.

\begin{figure}[hbt!]
\centering
\subfloat[]{\includegraphics[scale=.33]{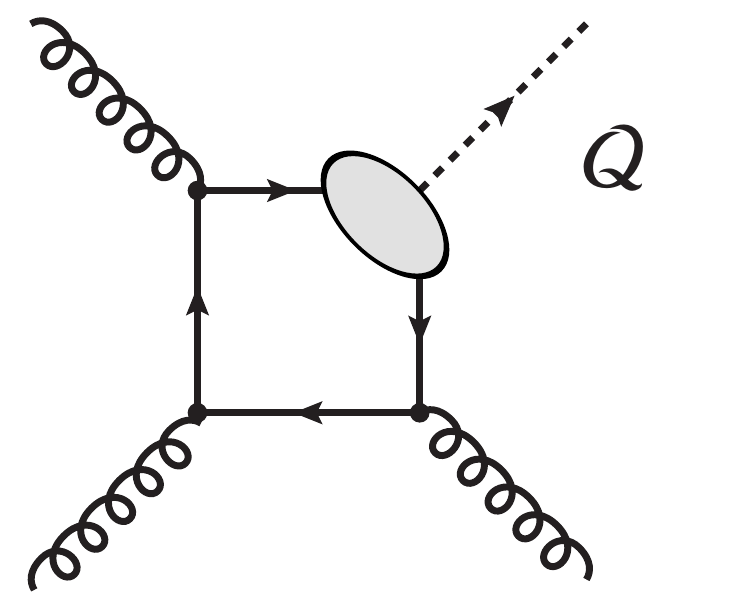}\label{diagram-CSM-a}}
\subfloat[]{\includegraphics[scale=.33]{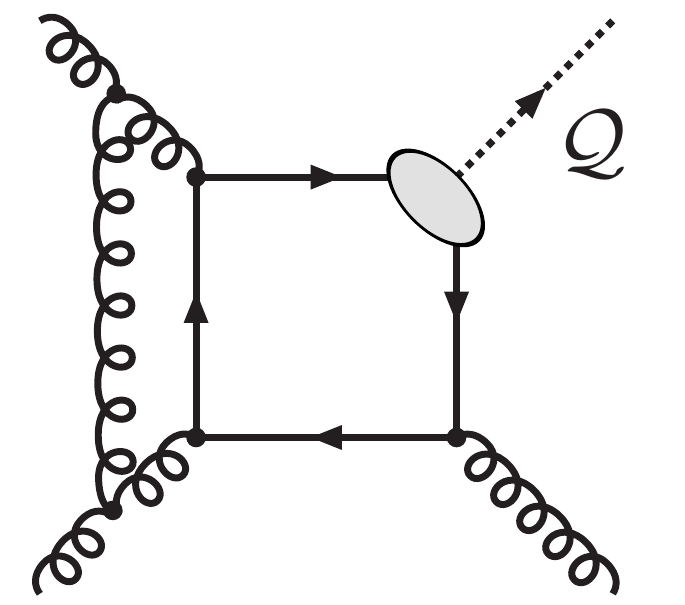}\label{diagram-CSM-b}}
\subfloat[]{\includegraphics[scale=.33]{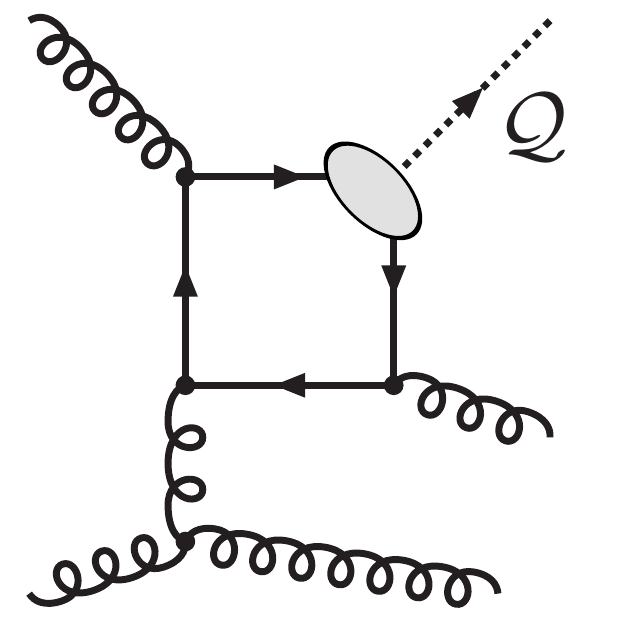}\label{diagram-CSM-c}}
\subfloat[]{\includegraphics[scale=.33]{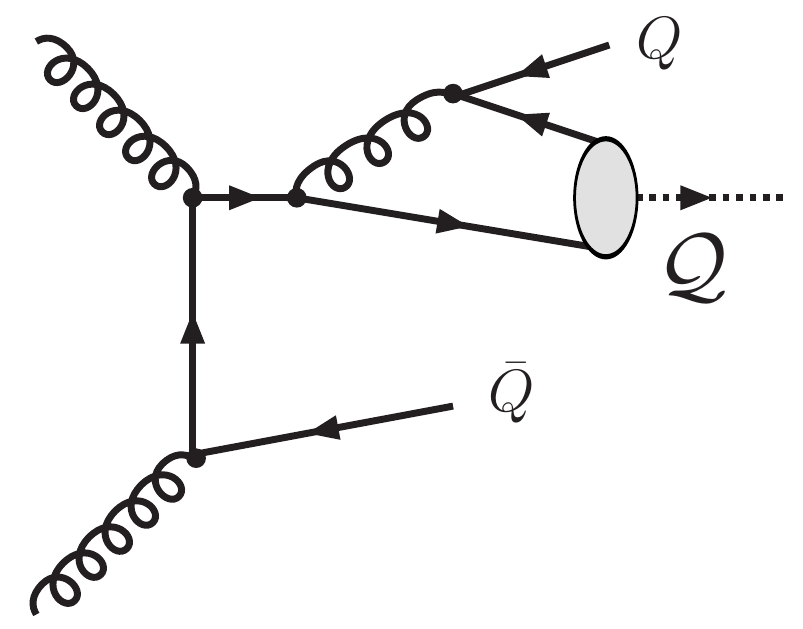}\label{diagram-CSM-d}}
\subfloat[]{\includegraphics[scale=.33]{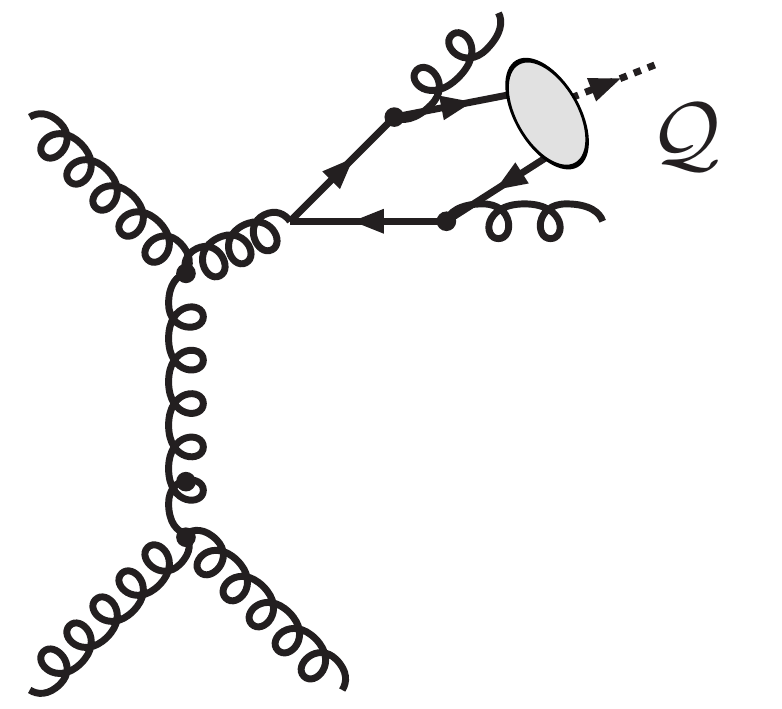}\label{diagram-CSM-e}}
\subfloat[]{\includegraphics[scale=.33]{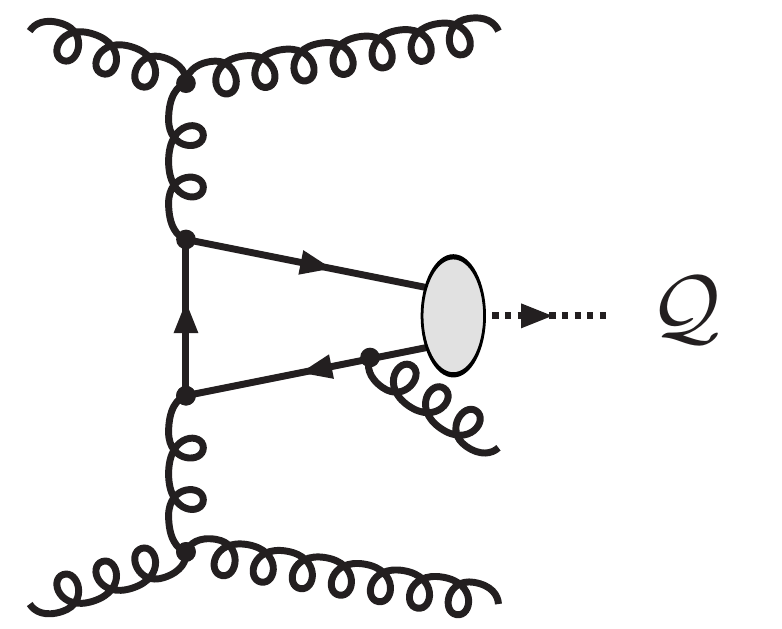}\label{diagram-CSM-f}}
\caption{Representative diagrams contributing to $^3S_1$ hadroproduction via 
CS channels at orders $\alphaS^3$ (a), $\alphaS^4$ (b,c,d), 
$\alphaS^5$ (e,f). 
The quark and antiquark attached to the ellipsis are taken as on-shell
and their relative velocity $v$ is set to zero.}
\label{diagrams}
\end{figure}

Nowadays, automated codes allow one to generate such LO amplitudes, to fold
them with the PDFs and to integrate them in the relevant phase space in order to get differential 
cross sections. Let us cite \MadOnia~\cite{Artoisenet:2007xi} and \HELACOnia~\cite{Shao:2012iz,Shao:2015vga} which have been used in several computations discussed later.

As usual, if one considers ${d\sigma}/{dP^2_T}$, the $P_T$ dependence of 
the differential cross section is simply that of $\left| {\cal M}\right|^2$
up to the decrease of the PDF for increasing $x_i$ when $P_T$ increases. In what follows, we will thus 
indistinctly refer to $\left| {\cal M}\right|^2$ and ${d\sigma}/{dP^2_T}$.

At LO, $\left| {\cal M}\right|^2$ approximately scales as $P_T^{-8}$ with
two far off-shell heavy-quark propagators when the vector quarkonia is produced at large
$P_T$. Compared to the approximate $P_T^{-4}$ scaling of light-meson production
by parton fragmentation, such a scaling is admittedly much softer, yet not trivial
like that of a pseudoscalar quarkonium which is $\delta(P_T)$ at LO (from $gg \to \! ^1S_0$). Most of
the complications of vector-quarkonium production are likely due to this uncommon
scaling.

The situation is absolutely similar for the case of inclusive vector-quarkonium
photoproduction which we discuss in section~\ref{sec:CSM_photoproduction}. At LO, only one partonic process contributes, namely
$\gamma g \to \Q g$ and the cross section also scales as $P_T^{-8}$. 
As first noted by Kr\"amer~\cite{Kramer:1995nb}, QCD corrections are genuinely important 
in such cases. 

\begin{figure*}[ht!]\centering
\subfloat[]{\includegraphics[width=.58\columnwidth]{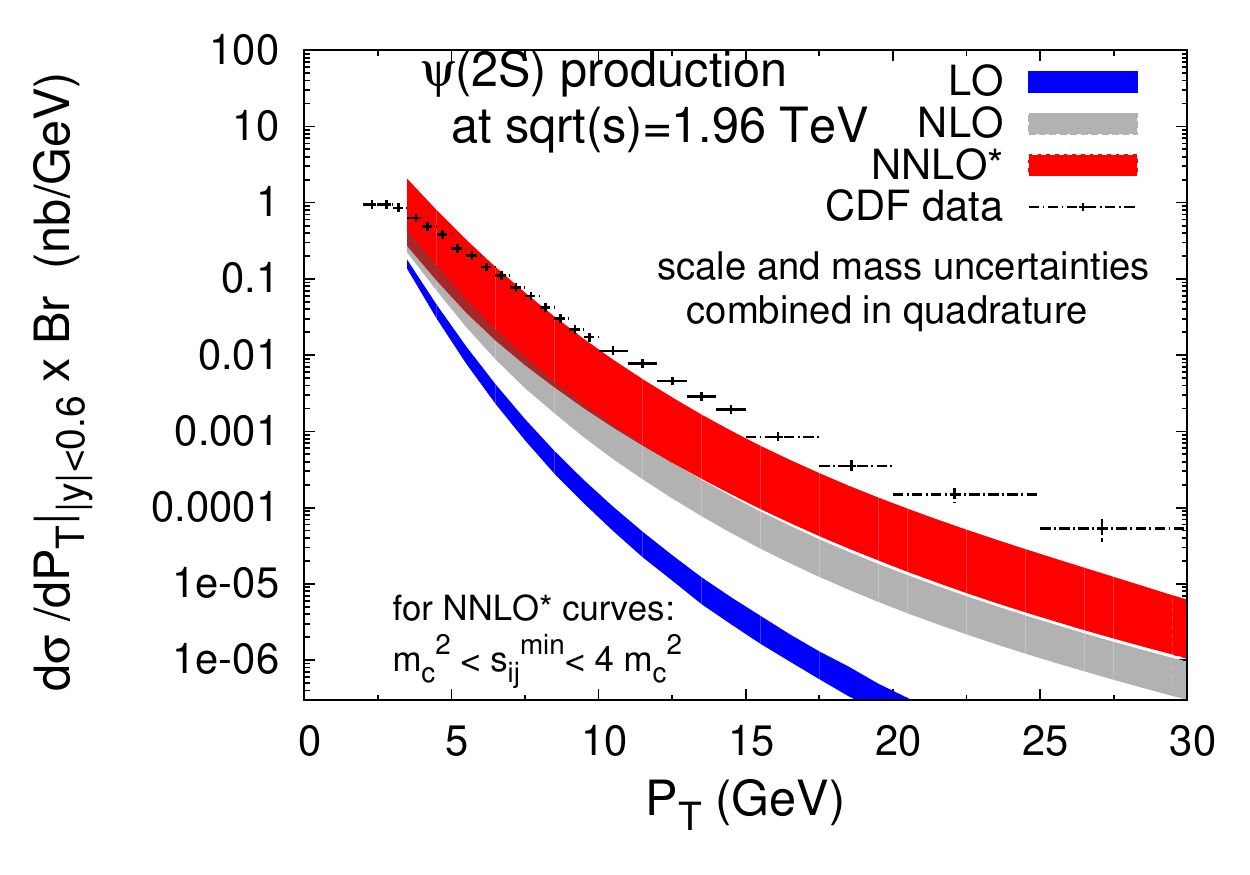}\label{fig:Psi2S-Tevatron}}
\subfloat[]{\includegraphics[width=.42\columnwidth]{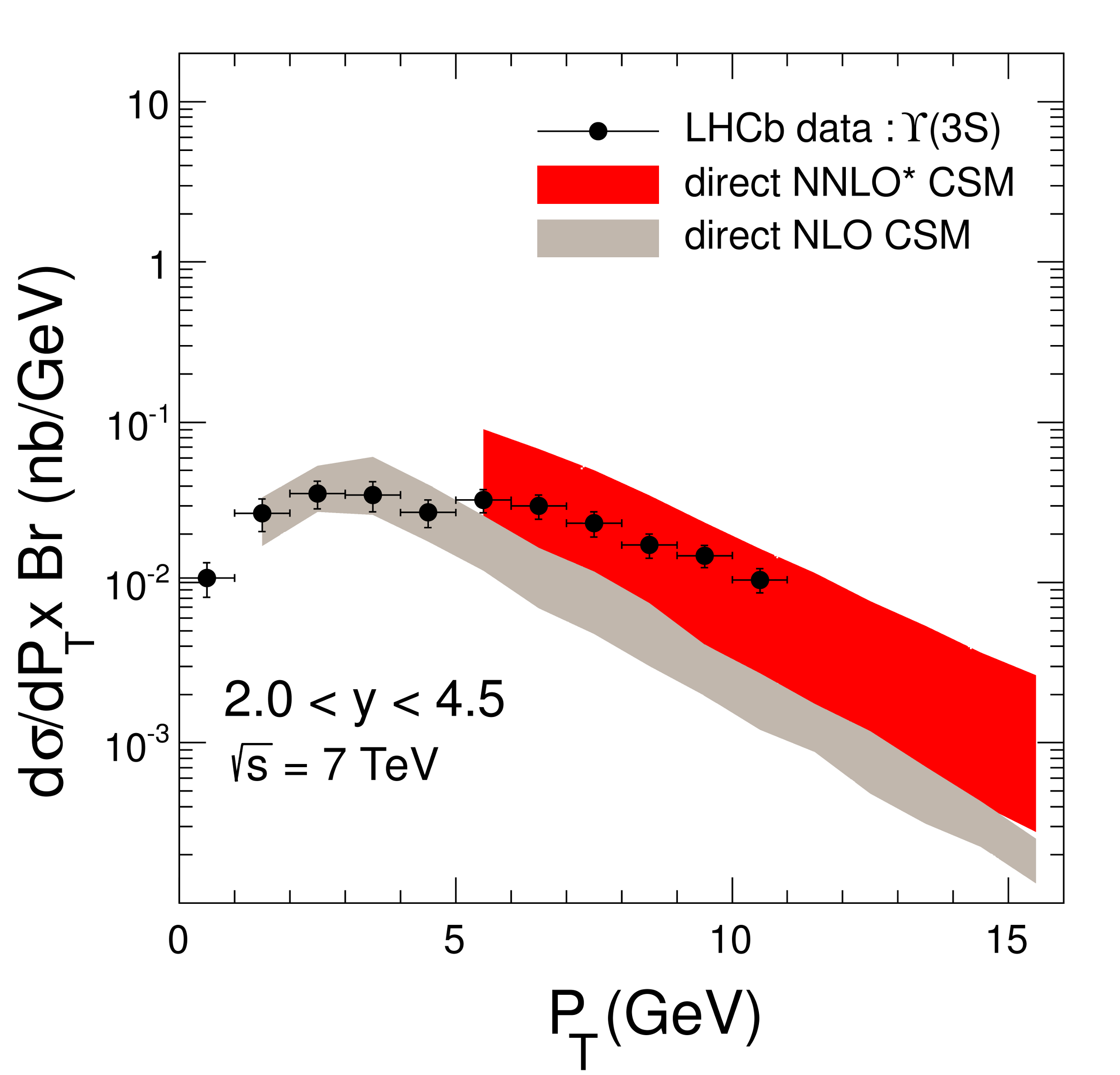}\label{fig:Upsi3S}}
\caption{(a) Comparison between the $\psi(2S)$ CDF data~\cite{Aaltonen:2009dm} and
the CSM prediction at LO (blue band), NLO (grey band) and NNLO$^\star$ (red band) accuracy.
(b) Comparison between the $\Upsilon(3S)$ LHCb data~\cite{LHCb:2012aa} and the NLO (grey band) and NNLO$^\star$ 
  (red band) CSM predictions for the direct yield. See~\cite{Lansberg:2008gk,Artoisenet:2008fc} for details.\label{fig:CSM-dsdpt-1}} 
\end{figure*}

Indeed, at NLO~\cite{Campbell:2007ws,Artoisenet:2007xi}, we can categorise 
the different contributions in different classes with specific $P_T$ scalings. 
The first is that of the virtual-emission contributions 
as shown on \cf{diagram-CSM-b}, which are divergent\footnote{These 
divergences  can be treated as usual using dimensional regularisation, 
see \eg~\cite{Campbell:2007ws}.} but as far their $P_T$ scaling is concerned, 
they would also scale like  $P^{-8}_T$. 
 The second is that of the $t$-channel gluon 
exchange ($t$-CGE) graphs like on  \cf{diagram-CSM-c}, where one initial
parton radiates another parton (or splits). They scale like $P^{-6}_T$. At 
sufficiently large $P_T$, their harder $P_T$ behaviour can easily take over
 their $\alphaS$ suppression compared to the LO ($\alphaS^3$) contributions.
Hence, they are naturally expected to dominate over the whole set of diagrams up to 
$\alphaS^4$. The third interesting class is a subset of the contributions ${\cal Q} + Q \bar Q$ (where $Q$ is of the same flavour 
as the quarks in $\cal Q$) which also appear for the first time at  $\alphaS^4$. 
Indeed, some graphs for 
${\cal Q} + Q \bar Q$ are fragmentation-like  (see \cf{diagrams} (d)) and nearly
scales like $P^{-4}_T$. Such contributions should thus dominate 
at large $P_T$, where the harder behaviour in $P_T$ is enough to compensate 
the suppression in $\alphaS$ and that due to the production of an additional  heavy-quark 
pair. In practice~\cite{Artoisenet:2007xi}, this happens
 at larger $P_T$ than previously thought~\cite{Braaten:1994xb,Cacciari:1994dr}. We shall come back 
to this channel later when discussing the new observables since it can be
studied for itself (see section~\ref{sec:psi-cc}). At $\alphaS^5$, further new  topologies appear, as illustrated 
by \cf{diagram-CSM-e} and \ref{diagram-CSM-f}. We will discuss their impact in section~\ref{subsec:NNLOstar-CSM}.

From the above discussion, one expects some classes of NLO corrections to be enhanced
by factors of $P^{2}_T$ and their consideration is therefore crucial for the
phenomenology at colliders where large values of $P_T$ can be reached. For instance, it would not make
any sense nowadays to compare LO CSM results at mid and large $P_T$  with data, as it was done
in the 1980's and 1990's. This is also true for the polarisation observables (see later) since
the yield happens to be dominated by completely different topologies with a possibly different
quarkonium polarisation~\cite{Gong:2008sn,Artoisenet:2008fc}. A detailed account on how
NLO corrections can be computed can be found in~\cite{Gong:2008hk}. 

It is not the place here to display an exhaustive list of comparisons between NLO CSM computations 
and data. We instead guide the reader to recent reviews where they are compared to 
RHIC and Tevatron data~\cite{Brambilla:2010cs} or LHC data~\cite{Andronic:2015wma}.
To summarise the existing comparisons in a few words: in the $\Upsilon(nS)$ case (see \eg~\cf{fig:Upsi3S}, grey band), the low-$P_T$ data are  well described by the CS NLO yield up to 5 GeV and then a gap opens
for increasing $P_T$. We should however note that the $\Upsilon(3S)$ CS predictions stand for the direct yield which only amounts to 60\% of the inclusive yield (see \ct{tab:FD-fractions-Upsi3S}). For the lighter $\psi(nS)$, the $P_T$ dependence of the predicted CS NLO 
cross section is always softer (see \eg\ \cf{fig:Psi2S-Tevatron}, grey band) and tends to over(under)shoot the data for $P_T$ below(above) a couple of  GeV.

\paragraph{Beyond NLO.}
\label{subsec:NNLOstar-CSM}

As just noted, the discrepancy between the NLO CSM computations 
 and the experimental data, for both $\psi(nS)$ and $\Upsilon(nS)$, gets larger with $P_T$. If 
we parallel this to the existence of these new $P_T^{-4}$ channels at order 
$\alpha^5_S$, it is legitimate to wonder what their sizes effectively are and whether
they can alter the data-theory comparison.

In fact, their approximate contributions can be evaluated in a relatively 
``simple''\footnote{``simple'' compared to a full  --out-of-reach-- NNLO 
computation and thanks to the aforementioned automated tools.} manner by computing the 
$\alpha^5_S$ contributions consisting in the production of a $\cal Q$ with 3 
light partons (noted $j$ thereafter). Among them are the topologies 
of~\cf{diagram-CSM-e} (gluon fragmentation) and~\cf{diagram-CSM-f} 
(``high-energy enhanced''), these close the list of kinematical enhancements 
from higher-order QCD corrections. This $\alpha^5_S$ subset being the Born (LO) contribution for 
a physical process ($pp \to {\cal Q} +jjj$), its contribution is finite except
 for soft and collinear divergences.

\begin{figure*}[ht!]\centering
\subfloat[]{\includegraphics[width=.5\columnwidth]{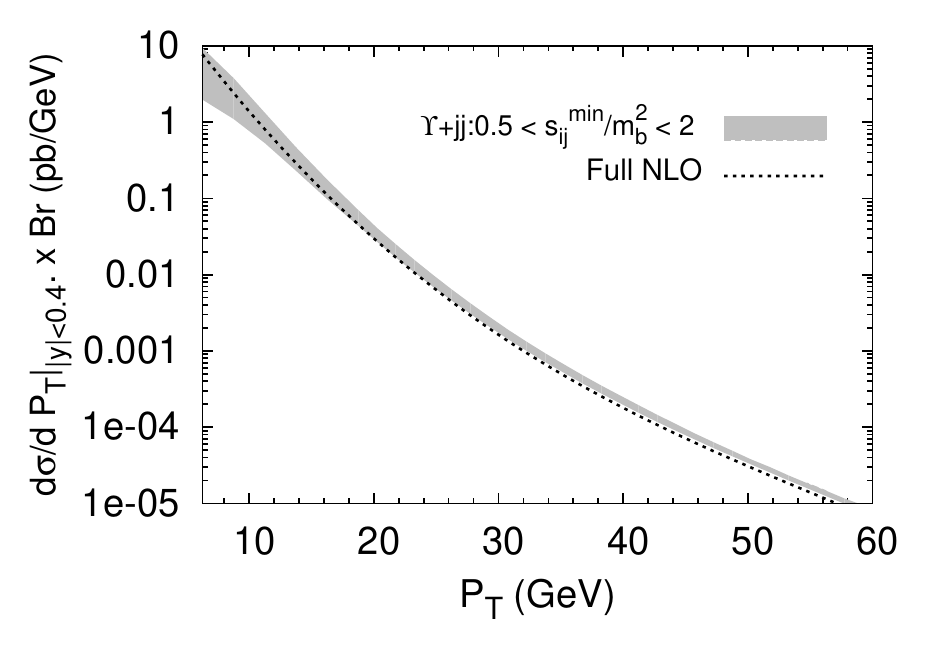}\label{fig:NLO-2jet-upsilon}}
\subfloat[]{\includegraphics[width=.5\columnwidth]{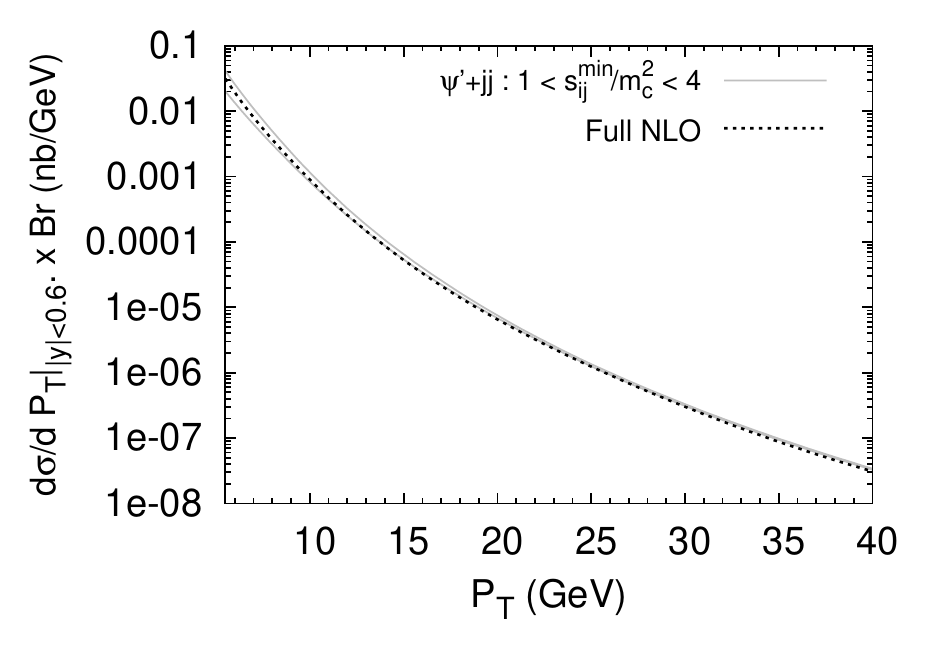}\label{fig:NLO-2jet-psi}}
\caption{Full computation at NLO for (a) $\Upsilon(1S) + X$ (dashed line)  
vs. $\Upsilon(1S)$ + 2 light partons with a cut on $s_{ij}^{\rm min}$ (grey band) and
(b) $\psi(2S) + X$ (dashed line)  vs. $\psi(2S)$ + 2 light partons with a
 cut on $s_{ij}^{\rm min}$ (grey curves) at $\sqrt{s}=14$~TeV. See~\cite{Artoisenet:2008fc,Lansberg:2008gk} for details.
Adapted from (a)~\cite{Artoisenet:2008fc} and (b)~\cite{Lansberg:2008gk}.} \label{fig:NNLOstar-2}
\end{figure*}

\begin{figure*}[ht!]\centering
\subfloat[]{\includegraphics[width=.45\columnwidth]{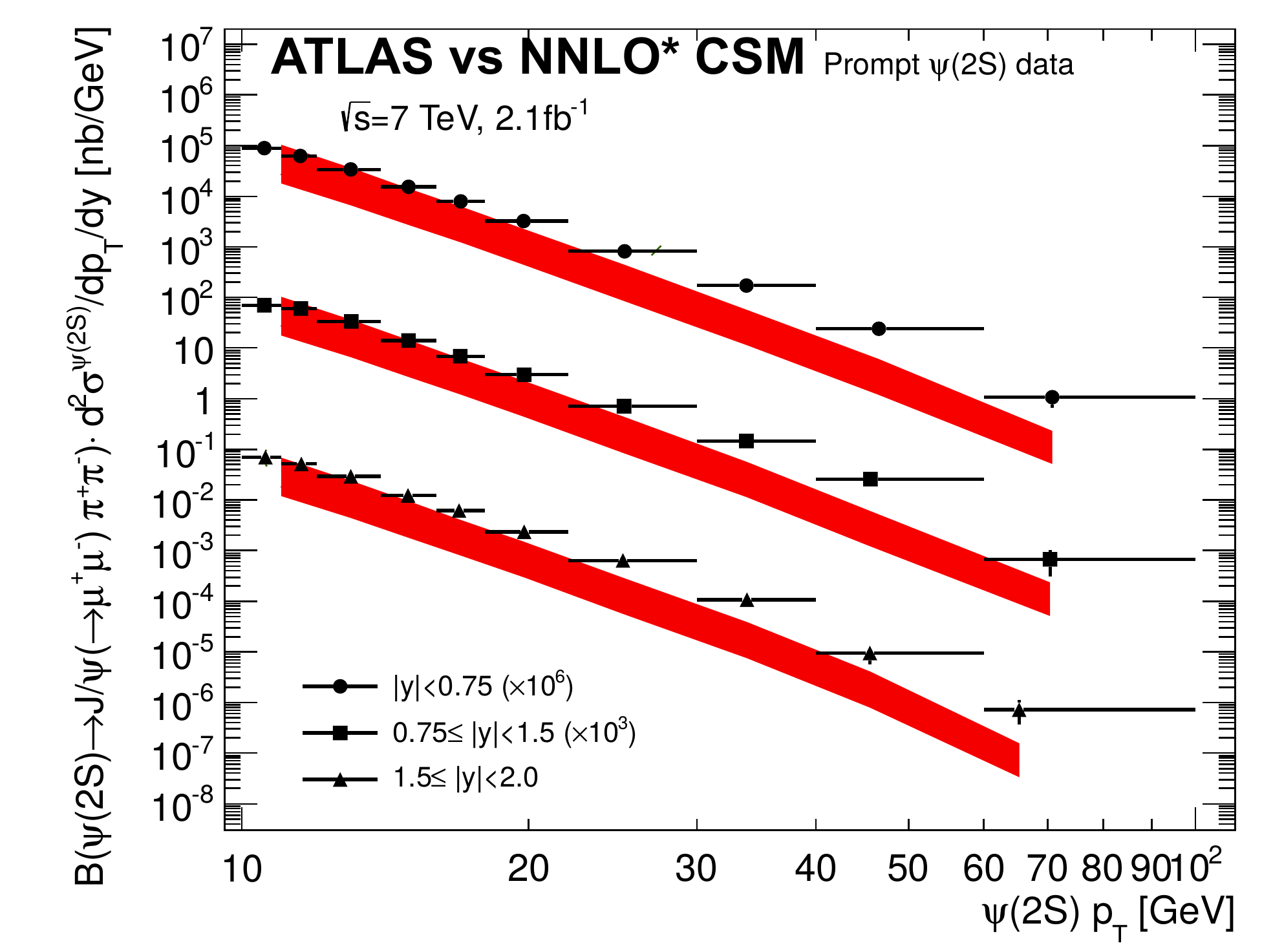}\label{fig:NLO-NNLOstar-ATLAS-psi2S}}
\subfloat[]{\includegraphics[width=.55\columnwidth]{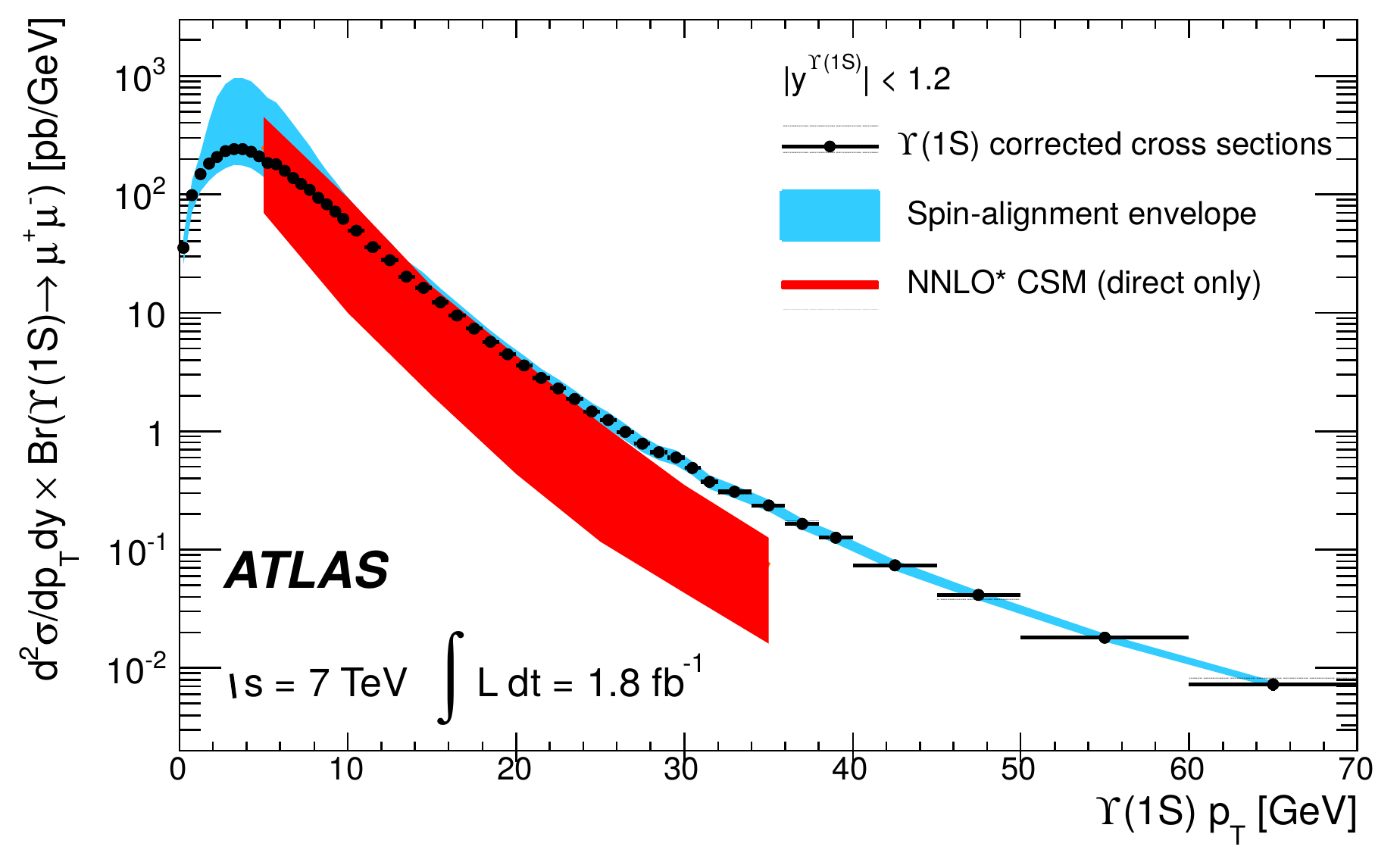}\label{fig:NLO-NNLOstar-ATLAS-upsilon}}
\caption{
(a) Comparison between the $\psi(2S)$ ATLAS data~\cite{Aad:2014fpa} and
the CSM prediction at NNLO$^\star$ accuracy.
(b) Comparison between the $\Upsilon(1S)$ ATLAS data~\cite{Aad:2012dlq} and the NNLO$^\star$ 
CSM predictions for the direct yield. Adapted from ~\cite{Aad:2014fpa} and~\cite{Aad:2012dlq}.\label{fig:NLO-2jet}}
\end{figure*}

To avoid these divergences, one can impose a lower bound on the invariant-mass of 
any light partons ($s_{ij}$). For the new channels opening up at $\alpha^5_S$ with
a $P_T^{-4}$ scaling, 
and whose contributions we are after, the dependence on this cut should diminish 
for increasing $P_T$ since no collinear or soft divergences can appear in this limit
for these topologies.
For other channels, whose LO contribution is at $\alpha^3_S$ or $\alpha^4_S$,
the cut would produce logarithms of $s_{ij}/s_{ij}^{\rm min}$ which can 
be significant. However, one can factorise them over their corresponding LO 
contribution, which scales at most as $P_T^{-6}$. As a result, the sensitivity 
of the yield on the cut $s_{ij}^{\rm min}$ should become negligible at large enough $P_T$.

Thanks to exact NLO computations (\eg~\cite{Campbell:2007ws}), one can partially 
validate  such a procedure for the process $pp \to {\cal Q} +jj$ with one less gluon. 
The differential cross section for the real-emission $\alphaS^4$ corrections, 
$\Upsilon(1S)+jj$ production, is for instance shown in \cf{fig:NLO-2jet-upsilon}. The 
grey band displays the sensitivity of the cross section to the invariant-mass cut 
$s_{ij}^{\rm min}$ between any pairs of light partons for a variation from 
$0.5 m_b^2$ to $2m_b^2$.  It indeed becomes insensitive to the value of 
$s_{ij}^{\rm min}$ as $P_T$ increases, and it very accurately matches  the complete
NLO result. In the charmonium case, the 
corresponding contributions from $pp \to \psi(2S) +jj$ reproduce even better --for lower
 $P_T$ and with a smaller dependence of $s_{ij}^{\rm min}$-- the full NLO 
computation, see \cf{fig:NLO-2jet-psi}.

Anticipating the discussion of associated-production reactions, let us note that the
same procedure was checked at NLO for $J/\psi(\Upsilon)+\gamma$~\cite{Lansberg:2009db}, 
$J/\psi+J/\psi$~\cite{Lansberg:2013qka}
and $J/\psi+Z$~\cite{Gong:2012ah}, after a comparison with the complete corresponding NLO computations~\cite{Li:2008ym,Sun:2014gca,Gong:2012ah}.
Owing to its success and its simplicity, this technique has been used to provide
predictions sometimes with a better numerical accuracy, since it bypasses the complex
computation of the virtual corrections, in more complex 
kinematical phase spaces. We will refer to it as the ``NLO$^\star$'' in the following.

Accordingly, we will refer to as NNLO$^\star$ for the real contribution  
at $\alpha^5_S$ with the same IR cut-off, whose general behaviour is discussed now.
At this stage, it should however be noted that even though such approximate
method allows one to bypass the complication to compute two-loop-virtual 
corrections, it nevertheless amounts to deal with a couple of thousands of 
Feynman graphs. This clearly requires the use of automated tools. Back in 2007, 
we used the tool described in~\cite{Artoisenet:2007qm}, which then became a branch
of \MadGraph~\cite{Alwall:2008pm}. Further results (such as those shown on \cf{fig:NLO-2jet}) 
were later computed with \HELACOnia\ \cite{Shao:2012iz,Shao:2015vga} by Shao.

 In the $\Upsilon$ case, it is fair to say that the contribution 
from $\Upsilon$ with three light partons mostly fills the gap with the LHCb data (\cf{fig:Upsi3S})
and the ATLAS one~\cite{Aad:2014fpa} (see \cf{fig:NLO-NNLOstar-ATLAS-upsilon})  with nevertheless a discrepancy of a factor of $2\sim 3$ remaining at high $P_T$, while for the $\psi(2S)$ there seems to remain a small 
gap opening between the NNLO$^\star$ band and the CDF data~\cite{Aaltonen:2009dm} (\cf{fig:Psi2S-Tevatron})
or the ATLAS data  (\cf{fig:NLO-NNLOstar-ATLAS-psi2S}). These are representative of comparisons which 
could be done with $J/\psi$ once $\chi_c$ and $\psi(2S)$ FDs are subtracted.

In both cases, 
the $\alphaS^5$ contribution is very sensitive
to the choice of the renormalisation scale, $\mu_R$.  This is
easily explained. For moderate values of the $P_T$, the missing virtual part 
can be important and cannot therefore reduce the scale sensitivity as expected in
complete higher-order computations. On the other hand, at large $P_T$, the yield 
is dominated by Born-level $\alphaS^5$-channels from which we expect a 
larger dependence on $\mu_R$ than the lower order contributions.

Clearly, if the NNLO$^\star$ correctly encapsulates the NNLO physics, these
results put the CSM back in the game as the potential leading contribution at mid $P_T$. 
In addition to the comparisons with the polarisation data which we discuss next and
which are not perfect, there are however some drawbacks. As we will discuss in 
section~\ref{sec:psi-gamma}, it is possible that the fragmentation graphs for which
the NNLO$^\star$ method is fully applicable do not dominate our results. Double 
$t$-channel graphs indeed appear at NNLO and may introduce unwanted logarithms
of the cut-off in the results, despite the observed reduction of its dependence
with $P_T$~\cite{Ma:2010jj}. Further kinematical studies are however needed to
draw firmer conclusions in the absence of a full NNLO computation. 

In fact, Shao proposed in 2018 a new method~\cite{Shao:2018adj} to take into account 
such $P_T$-enhanced contributions in a soft- and collinear-safe way, yet with the sole tree-level machinery. His conclusion
is that the corresponding contributions up to $\alphaS^5$, dubbed nnLO\footnote{If $\alphaS^4$ contributions are fully accounted for, one then refers to as nNLO contributions.}, which exhibit the same $P_T$ scaling as the NLO, 
probably hint at a limited impact of NNLO corrections at variance with what the NNLO$^\star$ results lead us to conclude. Shao however observed a significant infrared sensitivity for the $\sa$ channel 
for which a gluon is necessarily radiated from the heavy-quark pair produced by gluon fragmentation. Two such gluons are required
for the production of $\ssnew$ pairs by gluon fragmentation. In any case, it emphasises how important the completion of a full NNLO computation is, not only at low $P_T$ where both the NNLO$^\star$ and nNLO approximations do not hold, but also at large $P_T$.

\paragraph{Polarisation at NLO and beyond.}
\label{subsubsec:polarisation-CSM}

In the case of the vector quarkonia, one can trace back their polarisation 
--also more properly called spin-alignment-- through the angular distribution 
of the dilepton into which they decay. The latter can be parametrised as follows :
\begin{equation}
\frac{d^{2}N}{d\cos\theta d\phi} \propto 1+\lambda_\theta \cos^2\theta + \\ \lambda_\phi \sin^2\theta \cos2\phi + \lambda_{\theta\phi}\sin2\theta \cos\phi \,,
\label{eq:angularDistribution}
\end{equation}
where $\theta$ is the polar angle between the positively charged lepton in the quarkonium rest frame and an 
axis, chosen to be the polarisation (or spin-quantisation) axis, and $\phi$ is the corresponding azimuthal angle defined with respect to the hadron-collision plane. The decay angular  coefficients, $\lambda_\theta$ (sometimes also referred to as $\alpha$), $\lambda_\phi$ and $\lambda_{\theta\phi}$, 
can be related to specific elements of the spin density matrix and are referred to as the polarisation parameters. 
An unpolarised yield would be characterised by ($\lambda_{\theta}, \lambda_{\phi}, \lambda_{\theta \phi}) = (0,0,0)$, thus naturally corresponding to an isotropic decay angular distribution. The cases
 $(1,0,0)$ and $(-1,0,0)$ correspond to fully transverse and fully 
longitudinal polarisation, respectively.  

It is however very important to bear in mind that these coefficients are frame dependent; they depend on the 
polarisation axis like the angular distribution of \ce{eq:angularDistribution} does. In practice, 
the exisiting experimental analyses have been carried in only a few specific reference frames,
 defined by their polarisation axis. Let us name the helicity (HX) frame , the Collins-Soper (CS)~\cite{Collins:1977iv} frame, the Gottfried-Jackson (GJ)~\cite{Gottfried:1964nx} frame as well as the perpendicular helicity (PX)~\cite{Braaten:2008mz} frame. We guide the reader to \cite{Faccioli:2010kd} for the definitions of the corresponding axes, their naturalness in the context of some production mechanisms as well as a detailed discussion about the frame dependence and about
how measurements in different frames can be compared and to~\cite{Andronic:2015wma} for an exhaustive
list of the existing measurements with their main characteristics.

Most of the measurements --and theory predictions-- were nevertheless carried out in the helicity frame where the 
spin-quantisation axis points in the direction of the quarkonium in the {\it hadron} c.m.s.. At
high $P_T$ and in the central rapidity region, it is essentially orthogonal to the beam axis.
Moreover, many of them also only consisted of what one refers to as 1D analyses where the sole polar
anisotropy $\lambda_{\theta}$ is extracted and the $\phi$ dependence is integrated over.

\begin{figure}[hbt!]
\centering
\subfloat[]{\includegraphics[width=0.5\textwidth]{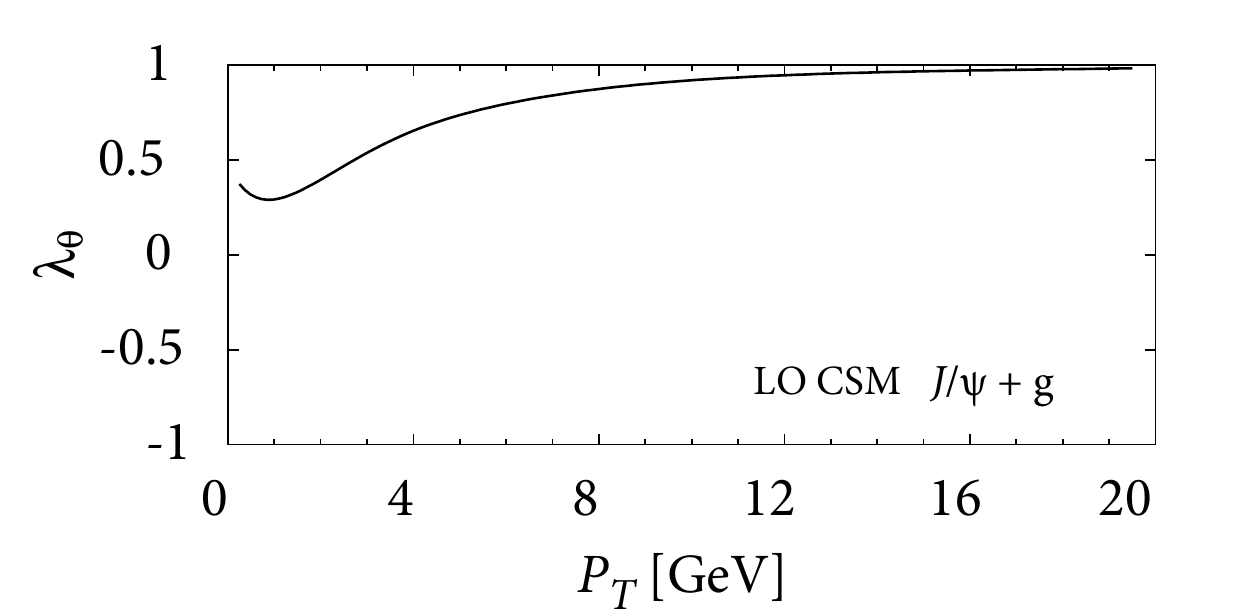}}
\subfloat[]{\includegraphics[width=0.5\textwidth]{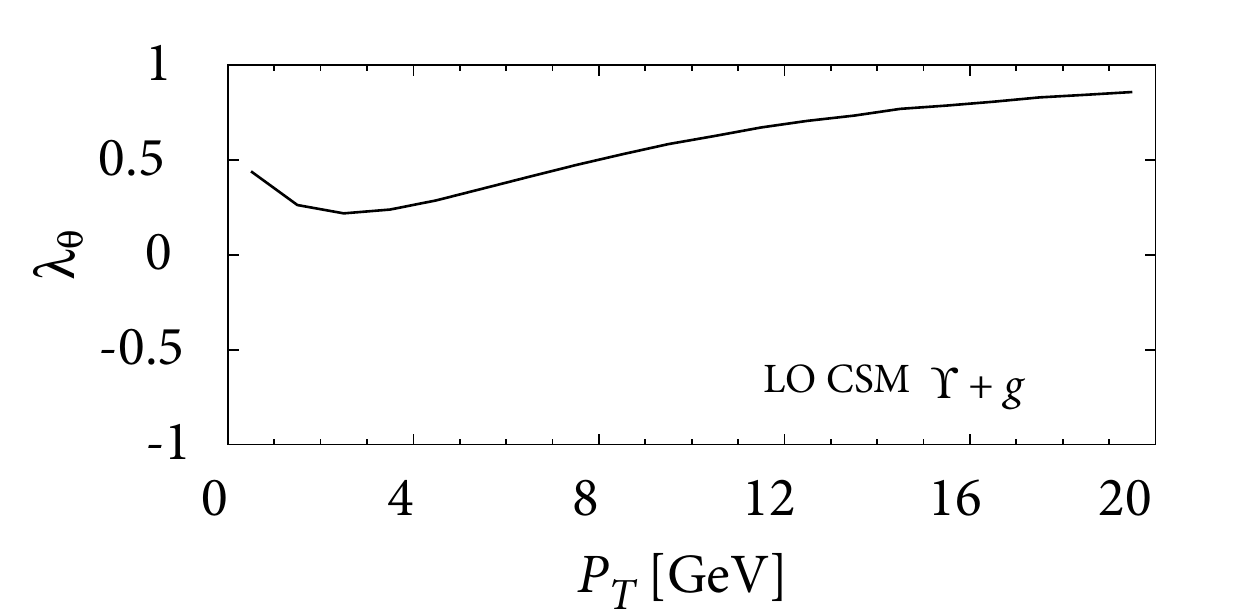}}
\caption{Polar anisotropy $\lambda_\theta$ at LO in the CSM from $p\bar{p} \rightarrow {\cal Q} + X$ for $J/\psi$ (a)
and $\Upsilon$ (b) at the Tevatron, $\sqrt{s}=1.96$~TeV. Plots adapted from~\cite{Artoisenet:2007xi}.}
\label{fig:pol-gg3S1g}
\end{figure} 

Polarisation measurements, despite their complexity, have been the object of much attention since
20 years because of the drastic LO prediction of the COM whereby $J/\psi$ should be transversely
polarised (in the helicity frame) at high $P_T$. We will come back to this in section~\ref{subsubsec:COM_3S1_NLO_PT}.

LO CS computations in the CS and GJ frames can be found in~\cite{Mirkes:1994jr} 
and in the HX frame in~\cite{Leibovich:1996pa}. \cf{fig:pol-gg3S1g} shows the behaviour
of $\lambda_\theta$ (or $\alpha$) at the Tevatron. A very similar 
trend is observed at the LHC. As soon as $P_T$ gets larger than $M_\Q$, 
the yield is nearly purely transverse. However, in this region, the LO
yield is very small and the $P_T^{-6}$ NLO contributions will take over.

\begin{figure}[hbt!]
\centering
\subfloat[]{\includegraphics[width=0.5\textwidth]{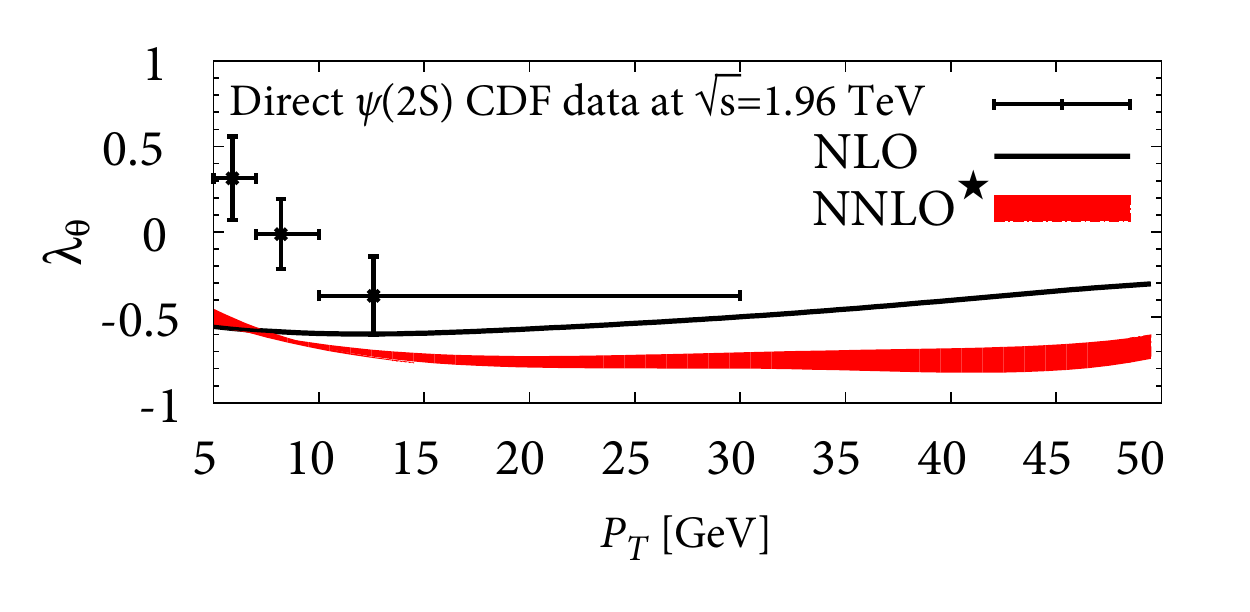}}
\subfloat[]{\includegraphics[width=0.5\textwidth]{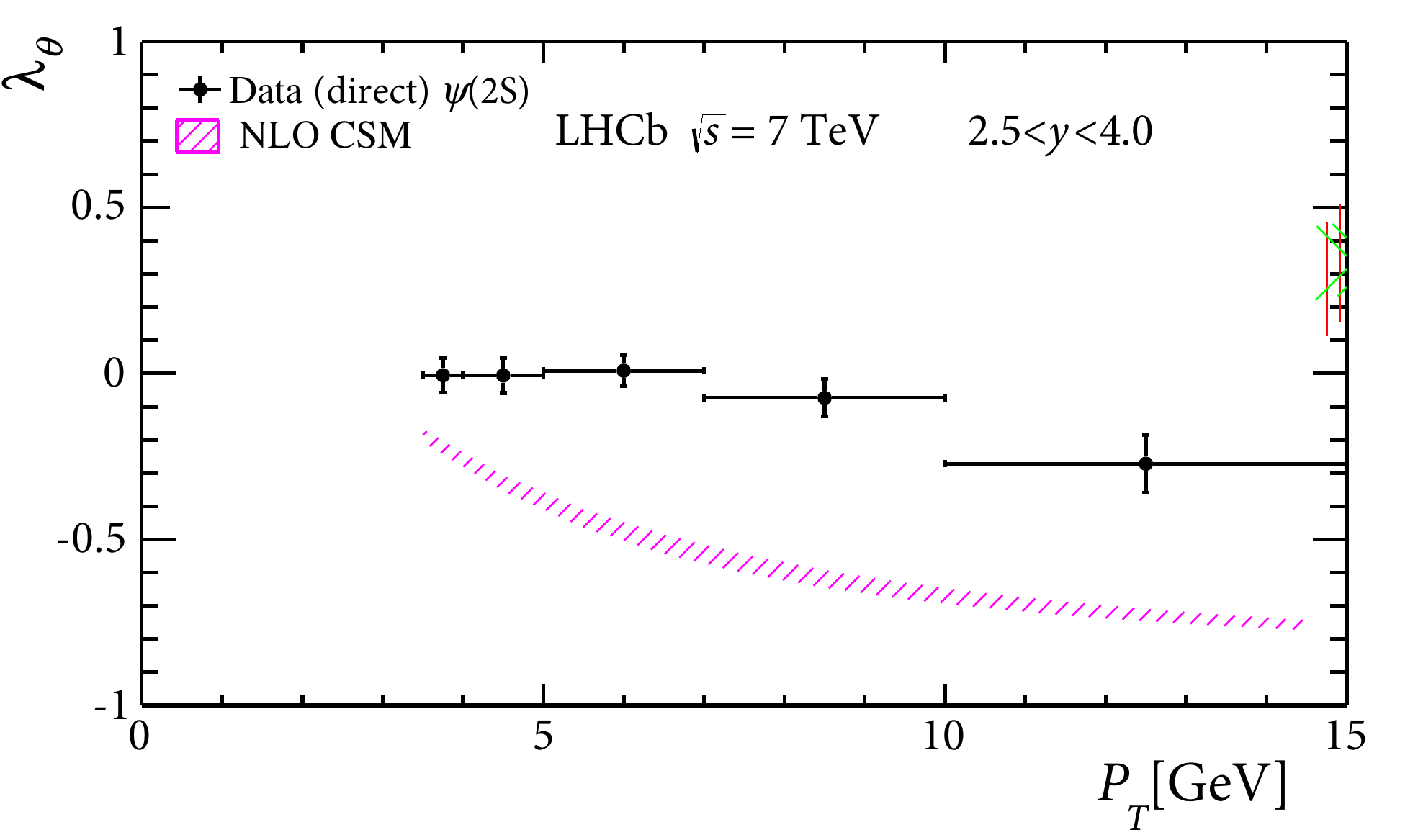}}
\caption{Polar anisotropy $\lambda_\theta$ for $\psi(2S)$  from (a) $p\bar{p} \rightarrow \psi(2S) + X$ 
at the Tevatron at $\sqrt{s}=1.96$~TeV from CDF compared to the CSM at NLO and NNLO$^\star$ 
and  (b) $p{p} \rightarrow \psi(2S) + X$ from LHCb compared to the CSM at NLO.
The  CDF $\psi(2S)$ data are from~\cite{Abulencia:2007us} and the LHCb ones from ~\cite{Aaij:2014qea}. Plots adapted from~\cite{Lansberg:2008gk,Aaij:2014qea}}
\label{fig:pol-NNLOstar-psi2S}
\end{figure}

\begin{figure}[hbt!]
\centering
\subfloat{\includegraphics[width=0.5\textwidth]{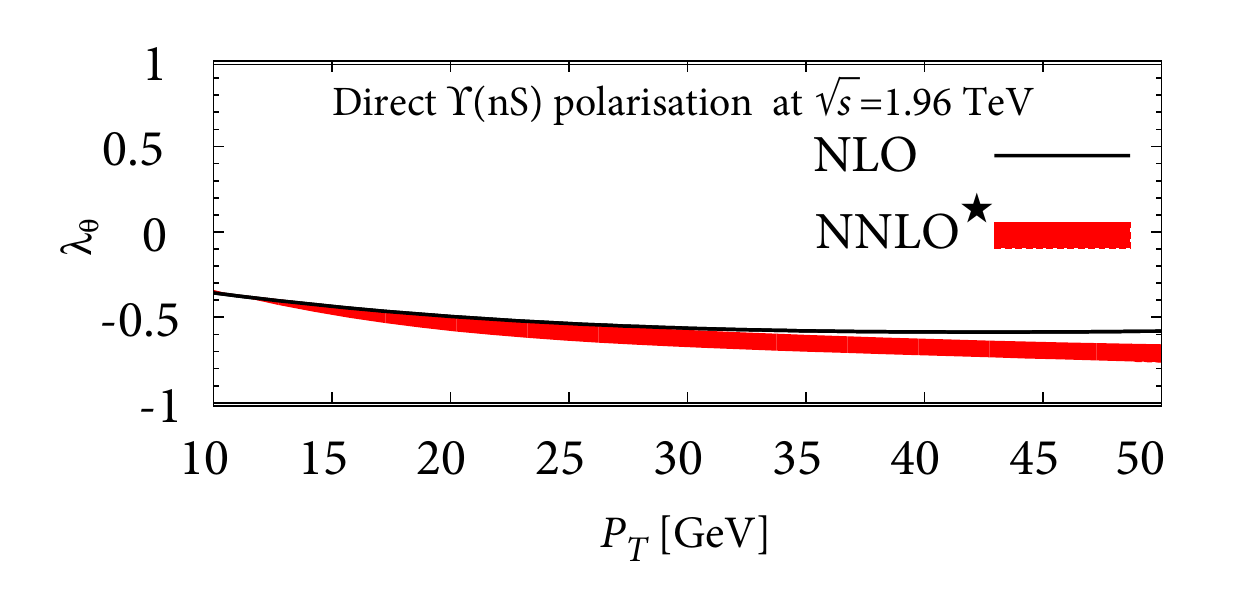}}
\caption{Polar anisotropy $\lambda_\theta$ at NLO and NNLO$^\star$ 
in the CSM  from $p\bar{p} \rightarrow {\cal Q} + X$ for direct $\Upsilon(nS)$  
at the Tevatron at $\sqrt{s}=1.96$~TeV. 
Plot adapted from~\cite{Lansberg:2008gk}}
\label{fig:pol-NNLOstar-Upsilon}
\end{figure}

The first NLO computation of the polarisation was performed by Gong and Wang 
in 2008~\cite{Gong:2008sn} for the $J/\psi$. Still in 2008, we performed the first 
one~\cite{Artoisenet:2008fc} for the $\Upsilon$ along with a study at NNLO$^\star$.
\cf{fig:pol-NNLOstar-psi2S} (a) shows the predicted polar anisotropy at NLO and NNLO$^\star$
for the $\psi(2S)$ --that for direct $J/\psi$ would be the same-- 
and \cf{fig:pol-NNLOstar-Upsilon} (a) shows that for the $\Upsilon(nS)$. 
The trend is similar in both cases: the NLO polarisation is getting 
longitudinal at variance with the LO result. In the case of the $\psi$,
$\lambda_\theta$ goes back to 0 at high $P_T$ since the $\psi+c+\bar c$ contribution 
is becoming the dominant NLO contribution and since it produces unpolarised $\psi$
as we showed in~\cite{Artoisenet:2007xi}. In the $P_T$ region accessed by LHCb, 
this effect is not visible (\cf{fig:pol-NNLOstar-psi2S} (b)). By comparing, the CSM NLO
results in the CDF and LHCb kinematics, one does not notice any significant rapidity nor
energy dependence. Finally, one observes that the NNLO$^\star$ yield is also longitudinal. 

We only show a comparison with data for the $\psi(2S)$ case since
 it is the only one where FD effects do not matter. 
In the $J/\psi$ case, the 30\% of $\chi_c$ FD cannot be neglected (see its impact in~\cite{Lansberg:2010vq} for RHIC).
In the bottomonium sector, all the state yields also contain a non-negligible  FD contribution. Comparisons 
to direct CSM predictions are not possible as long as the polarisation for the $P$-wave 
states are not measured. We refer to section~\ref{sec:FD} for details on FDs. 

In addition, there is surprisingly 
no published NNLO$^\star$ computations for the LHC energies. Yet, these presumingly
would not differ much, at a fixed $P_T$, from those made for the Tevatron energies.

To date, most of the measurements are compatible with a vanishing polarisation 
(see Table 1-5 of~\cite{Andronic:2015wma}). Yet, 
some $J/\psi$ measurements at RHIC, the Tevatron and the LHC seem to indicate a slightly
negative value of $\lambda_\theta$ --much less than the direct CSM computation though. 
In the case of the $\psi(2S)$, one sees on \cf{fig:pol-NNLOstar-psi2S} that both 
the CDF and LHCb data end up to be longitudinal~\cite{Abulencia:2007us,Aaij:2014qea} when $P_T$ increases.

\subsubsection{$\eta_Q$ hadroproduction at finite $P_T$}
\label{subsubsec:etaQ-CSM}

In this section we briefly present the corresponding results for the pseudoscalar
charmonia, $\eta_c$ and $\eta_c'$, at NLO. The $P_T$-differential cross section
 for $\eta_c$ was computed for the 
first time after the first measurement of $\eta_c$ production
($P_T>6.5$~GeV) by LHCb in 2014~\cite{Aaij:2014bga}. Their computation is however 
quite standard and can, 
for instance, be carried out with the semi-automated code FDC~\cite{Wang:2004du}.
As far as the parameters involved are concerned, they are exactly the same as for the 
$J/\psi$, with the same wave function at the origin $|R(0)|^2$. 

\begin{figure}[hbt!]
\centering
\subfloat[]{\includegraphics[scale=.33]{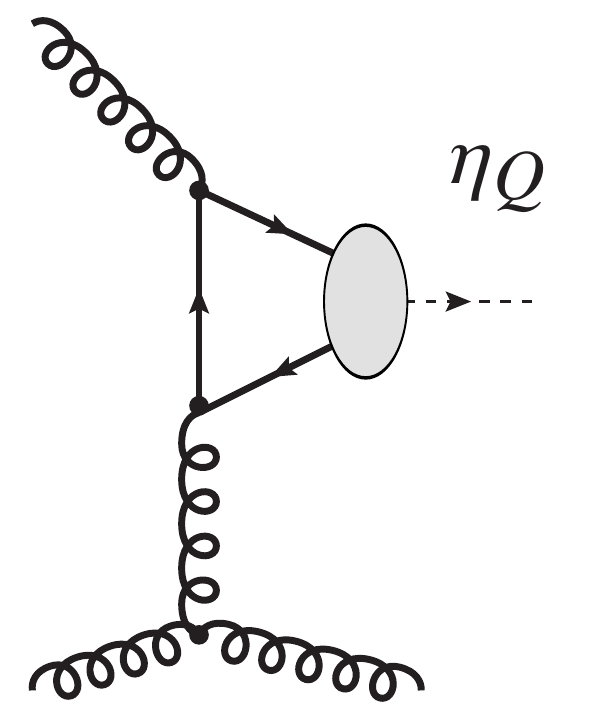}\label{diagram-CSM-etac-a}}
\subfloat[]{\includegraphics[scale=.33]{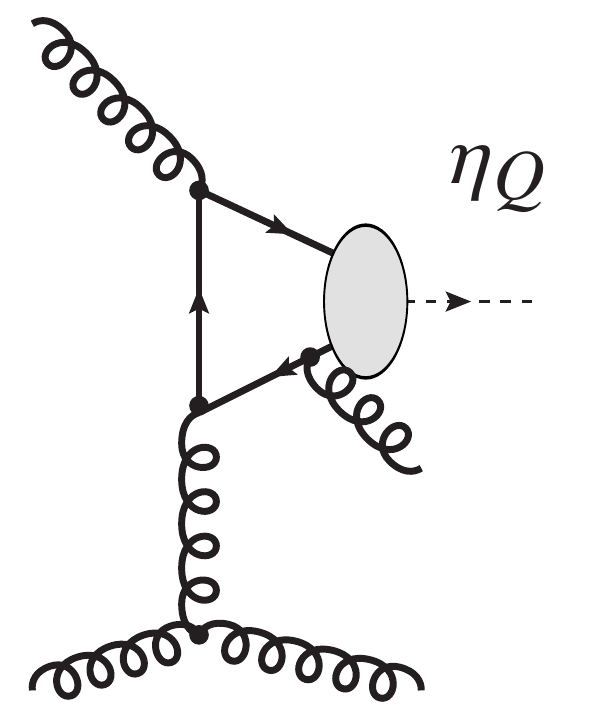}\label{diagram-CSM-etac--b}}
\subfloat[]{\includegraphics[scale=.33]{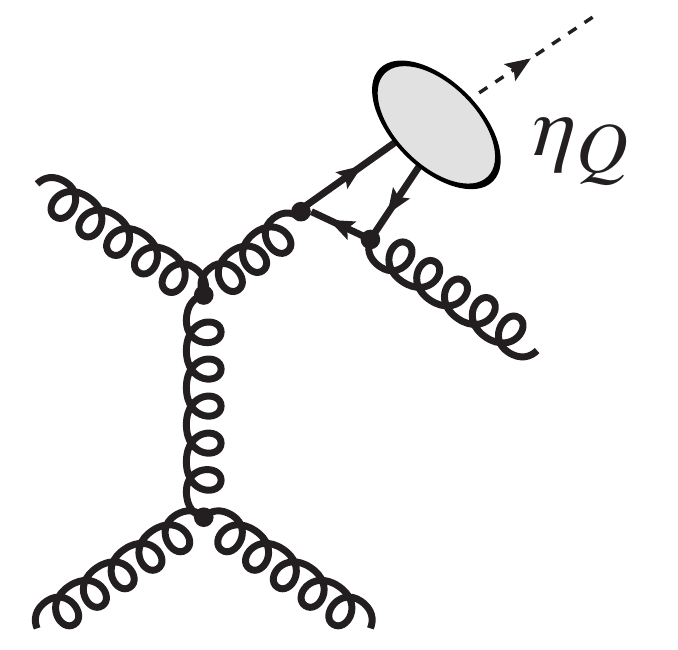}\label{diagram-CSM-etac--c}}
\subfloat[]{\includegraphics[scale=.33]{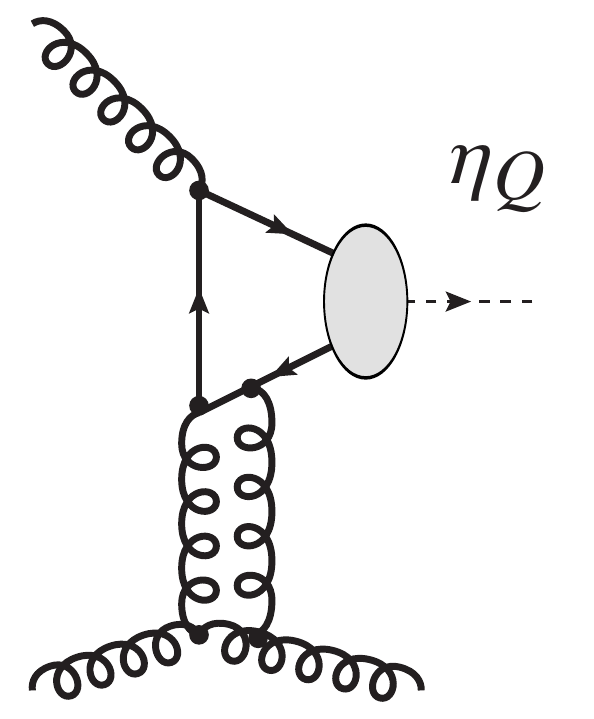}\label{diagram-CSM-etac-d}}
\caption{Representative diagrams contributing to $^1S_0$ $\Q$ hadroproduction via 
CS channels at orders $\alphaS^3$ (a), $\alphaS^4$ (b,c,d).
The quark and antiquark attached to the ellipsis are taken as on-shell
and their relative velocity $v$ is set to zero.}
\label{diagrams-CS-etac}
\end{figure}

\begin{figure}[hbt!]
\centering
\subfloat[]{\includegraphics[width=0.47\textwidth]{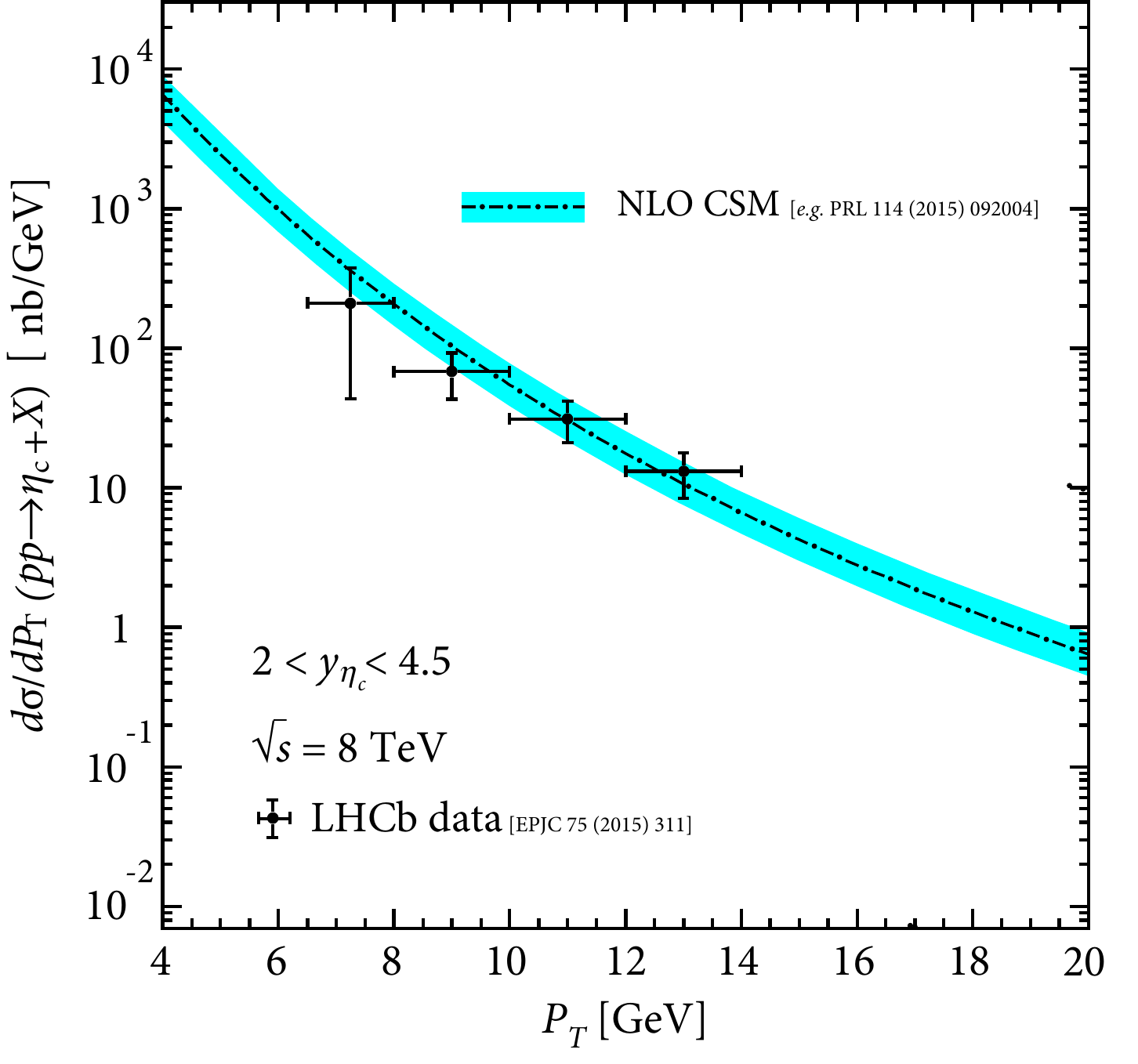}\label{fig:etac-CS-LO}}
\subfloat[]{\includegraphics[width=0.52\textwidth]{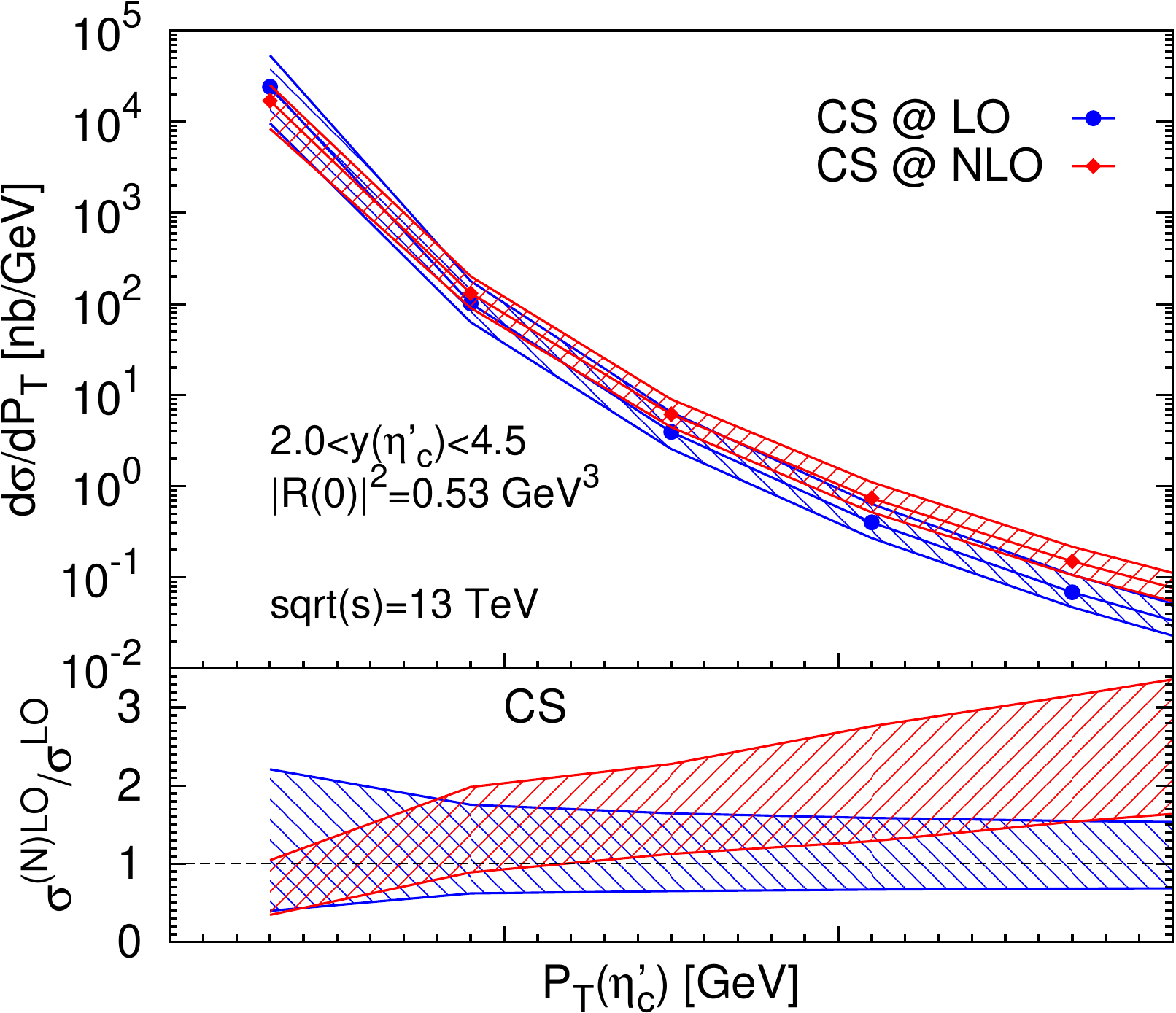}\label{fig:etacp-CS-LO}}
\caption{
(a) NLO CS $P_T$-differential cross section for $\eta_c$~\cite{Han:2014jya,Butenschoen:2014dra,Zhang:2014ybe}
in the LHCb acceptance at $\sqrt{s}=8$~TeV compared to the LHCb measurement~\cite{Aaij:2014bga}.
(b) LO and NLO CS $P_T$-differential cross section for $\eta'_c$ at $\sqrt{s}=13$~TeV.
 (a) Adapted from~\cite{Butenschoen:2014dra}, (b) taken from~\cite{Lansberg:2017ozx}}
\label{fig:etac-dsdpt-CS-NLO}
\end{figure}

The essential difference with the $\psi$ or $\Upsilon$ cases is that 
$P_T^{-6}$ topologies (\cf{diagram-CSM-etac-a}) are already 
contributing  at $\alphaS^3$ (LO) in addition to those contributing to vector states (\cf{diagram-CSM-a}).  
Their $P_T$ spectrum is thus naturally harder. Moreover, at $\alphaS^4$, real-emission $P_T^{-4}$ topologies (\cf{diagram-CSM-etac--c}) appear
and further harden the spectrum. In fact, it is a little softer than for 
$P_T^{-4}$ fragmentation topologies since the gluon attached to the heavy-quark line
cannot be soft, otherwise the amplitude vanishes. The invariant mass of the fragmenting
system is thus a little larger than that of a single heavy-quark pair with 
zero relative momentum and this induces a small suppression.

Strangely enough, at the time the LHCb data came out, nobody had anticipated this and, as a matter of fact,
nobody had studied the corresponding NLO corrections to the $P_T$-differential cross section. Soon after, 
along with that of the CO channels which we discuss later, 3 NLO studies~\cite{Han:2014jya,Butenschoen:2014dra,Zhang:2014ybe} came out with the same conclusion as what regards the NLO CS contribution: it agrees 
with the LHCb data (\cf{fig:etac-CS-LO}) with a $K$ factor slightly
increasing with $P_T$. \cf{fig:etacp-CS-LO} clearly shows the NLO effect
in the case of $\eta_c'$ where the hard scattering is identical. One also
sees that, at low $P_T$, where the virtual corrections can be important, the scale uncertainties (depicted
by the bands) is smaller than at larger $P_T$ where the dominant process is the Born
contribution to $gg \to \eta_c'gg$.
To avoid the same situation, we have indeed studied~\cite{Lansberg:2017ozx}, in 2017, 
the potentialities and the motivations (see also later) to observe the $\eta_c'$ at the LHC, which
is definitely at reach in several decay channels with branchings on the order of $10^{-4}$.
Further predictions for $\eta_c$ production at $\sqrt{s}=13$ TeV have also been provided in~\cite{Feng:2019zmn}.

\subsubsection{$\psi$ photoproduction at finite $P_T$} 
\label{sec:CSM_photoproduction}

Quarkonium inclusive photoproduction, whereby a quasi real photon hits and breaks a proton and then produces the quarkonium, 
is very similar to hadroproduction where two hadrons collide. The fact that one of the hadrons is replaced by
a photon introduces some simplification in the theory computations of such direct photon-proton interactions. However, the cross sections are necessarily smaller which requires large luminosities to produce objects like quarkonia. As such, the statistical samples collected at HERA are smaller than at the Tevatron and the LHC. In addition, the hadronic content of the photon can be ``resolved'' during these collisions and the theory treatment of these resolved-photon interactions with the proton becomes as complicated as that of hadroproduction. Finally, the inclusive contributions compete with the exclusive or diffractive ones whose cross sections for vector mesons like the $J/\psi$ are extremely large. Yet, as we discussed  in section~\ref{sec:FD}, the possible FDs seem to be limited which reduces the complication inherent to their consideration in photoproduction--most of the time, they are ignored or encompassed in a global normalisation factor. This seems reasonable except possibly at the largest $P_T$ accessed at HERA where one cannot exclude that the nonprompt FD  become dominant.

As we already discussed, the first NLO computation of any quarkonium-production process at finite $P_T$ was carried out in the mid 1990's in the case of direct\footnote{In this context ``direct'' applies to the photon and should not be confused with the notion of ``direct'' yield where FD contribution are removed. This is admittedly a unfortunate naming choice which is however very customary.} photoproduction by Kr\"amer~\cite{Kramer:1995nb}. We will outline it here as well as recent discussions which occurred when computations of the yield polarisation at NLO appeared.

\paragraph{$P_T$-differential production cross section at NLO.}

In many respects, the impact of the NLO corrections to the CSM in direct photoproduction and hadroproduction at colliders are similar. In the case of the resolved-photon contributions, it is obvious as the graphs are the same. The only difference comes from the treatment of the gluon flux from a photon instead of a proton. In the direct-photon cases, it appears as obvious when one looks at the different possible contributing graphs of \cf{diagrams-CSM-photoproduction}. The essential difference comes from the smaller number of possible graphs in the photoproduction case. Besides this, the discussion of the kinematical enhancement of some NLO topologies when $P_T$ grows is the same. In fact, the discussion we made in the previous section was already done for photoproduction in the Kr\"amer's studies~\cite{Kramer:1995nb,Kramer:2001hh}. 

\begin{figure}[hbt!]
\centering
\subfloat[]{\includegraphics[scale=.33]{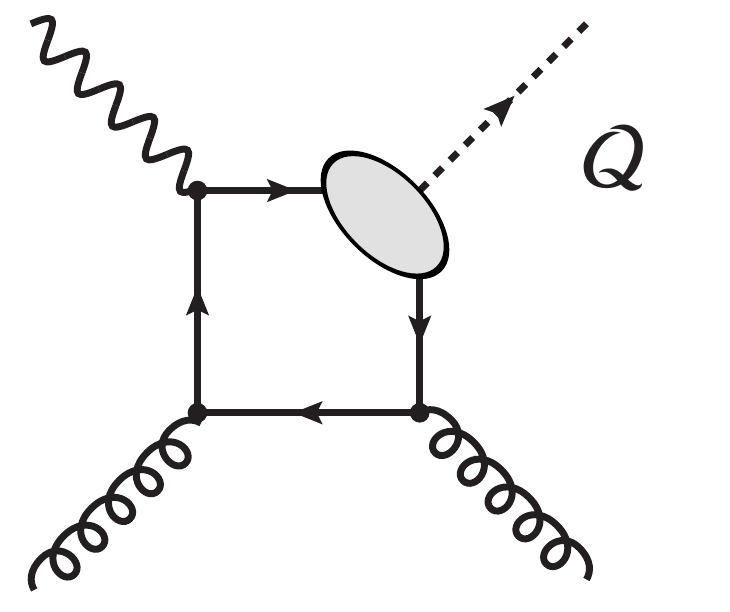}\label{diagram-CSM-photoprod-a}}
\subfloat[]{\includegraphics[scale=.33]{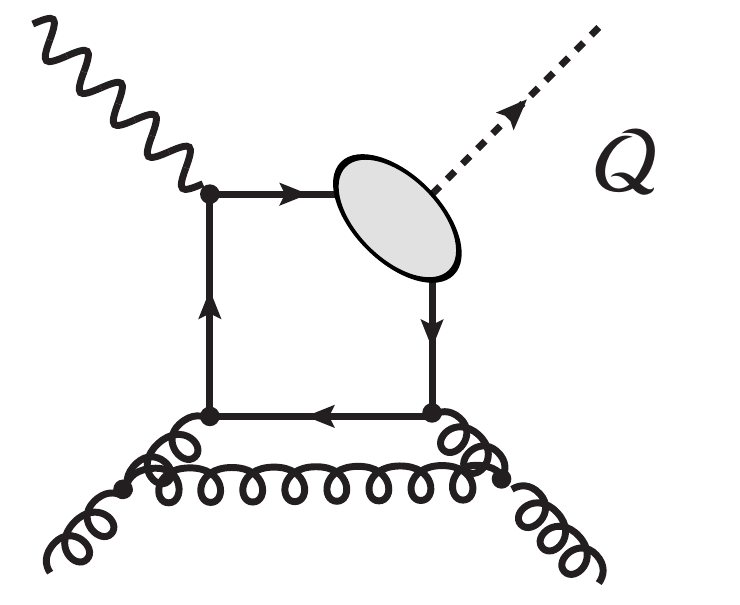}\label{diagram-CSM-photoprod-b}}
\subfloat[]{\includegraphics[scale=.33]{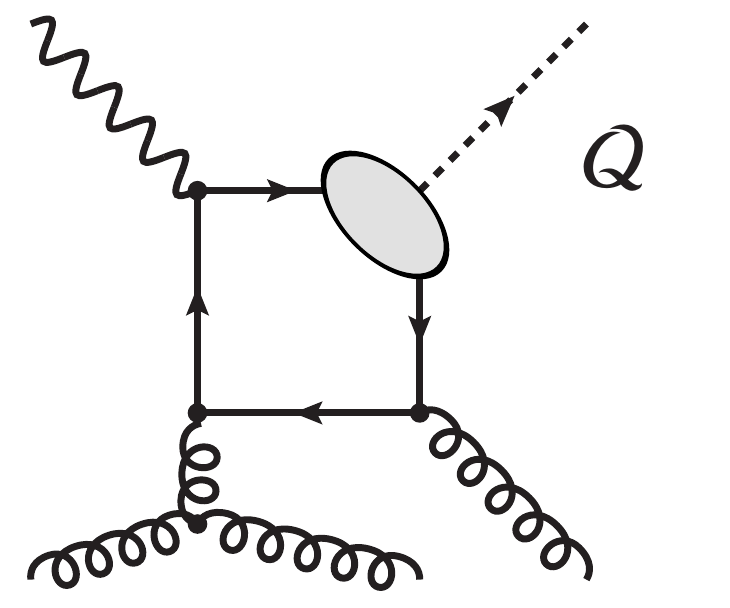}\label{diagram-CSM-photoprod-c}}
\subfloat[]{\includegraphics[scale=.33]{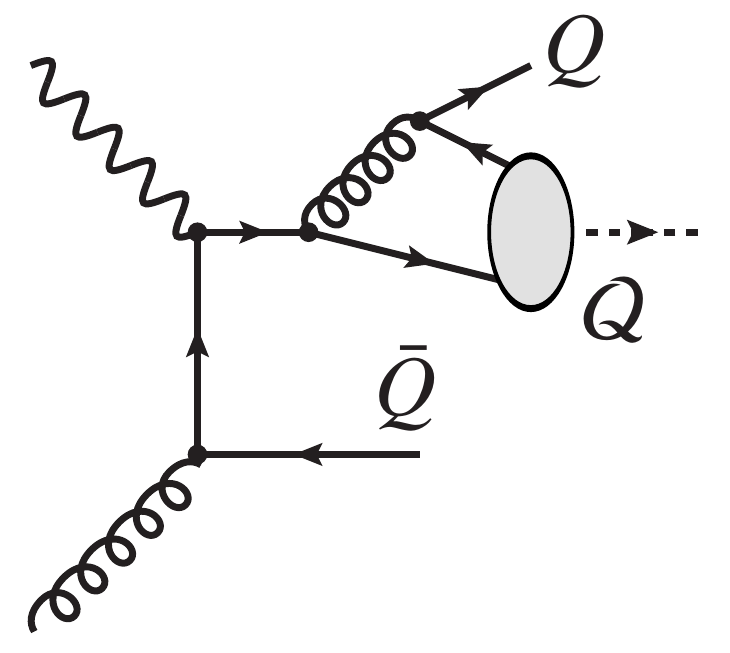}\label{diagram-photoprod-CSM-d}}
\subfloat[]{\includegraphics[scale=.33]{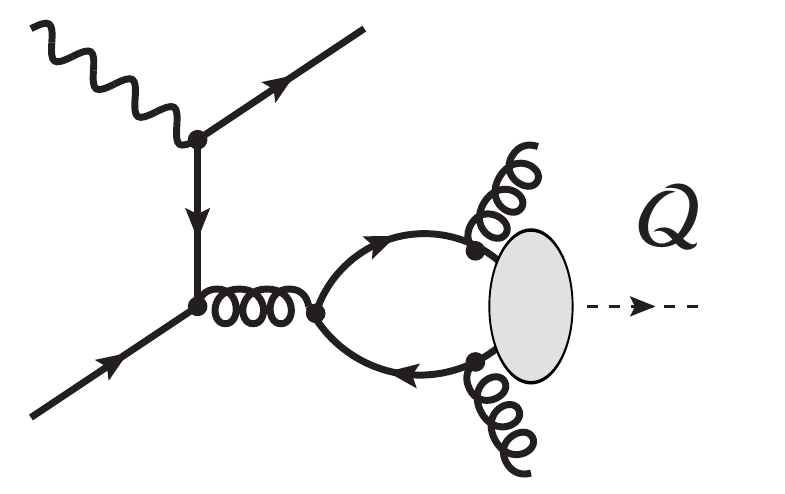}\label{diagram-photoprod-CSM-e}}
\caption{Representative diagrams contributing to $^3S_1$ direct photoproduction via 
CS channels at orders $\alpha\alphaS^2$ (a), $\alpha\alphaS^3$ (b,c,d), 
$\alpha\alphaS^4$ (e). 
The quark and antiquark attached to the ellipsis are taken as on-shell
and their relative velocity $v$ is set to zero.}
\label{diagrams-CSM-photoproduction}
\end{figure}

In practice, like in hadroproduction, the differential partonic cross section is readily obtained from the amplitude
squared\footnote{The momenta of the initial partons, $k_{1,2}$, are related to those of the colliding particles ($P_\gamma$ for the photon and
$P_p$ for the proton) through $k_{\gamma,p}=x_{\gamma,p} \, P_{\gamma,p}$. One then defines $s_{(\gamma p)}=(P_{\gamma}+P_{p})^2=W^2_{(\gamma p)}$ and the Mandelstam variables for the partonic system, as usual: $\hat s = s_{\gamma p} x_\gamma x_p$, 
$\hat t=(k_1-P_{\Q})^2$ and $\hat u=(k_2-P_{\Q})^2$. The proton momentum defines the $z$ axis.}, 
like in \ce{eq:dsdt}.
As it brings in additional information compared to the quarkonium rapidity, $y$, one defines the variable $z=(P_\Q.P_P)/(P_\gamma.P_P$), called inelasticity, which ranges between 0 and 1 and which corresponds to the energy, in the proton rest frame, taken away by the quarkonium $\Q$ off the photon. One also has $z=\frac{2 E_P m_T}{W^2_{\gamma p} e^{y}}$ [$E_P$ and $y$ are obviously defined in the same frame]. Large $z$ corresponds to negative rapidities. At small $z$, resolved-photon contributions take over the direct ones. When $z$ approaches 1, one gets closer to the exclusive limit where the proton stays intact. In between these regions, the direct/unresolved-photon contributions
are expected to be dominant. In such a case, the photon flux is simply assumed to be $\gamma(x_\gamma)=\delta(x_\gamma-1)$. 

Hence, one easily obtains differential cross sections, \eg\ in $P_T$ and $z$ (or $y$), for $\gamma p \to \Q g$ after convolution
with the PDFs and a change of variable from the Mandelstam variables:
\be 
\frac{d\sigma}{dzdP_T}=\int_{x_\gamma^{\rm min}}^1 dx_\gamma \frac{2 x_\gamma x_p P_T \gamma(x_\gamma,\mu_F) g(x_p(x_\gamma),\mu_F)}
{z(x_1-z)}
\frac{1}{16 \pi \hat s^2} \left| {\cal M}\right|^2,
\ee
where $x_p=\frac{x_\gamma P_T^2+m_{_\Q}^2 (x_\gamma-z)}{s_{\gamma p} z (x_\gamma-z)}$, $x_\gamma^{\rm min}= \frac{z(s_{\gamma p}z-m_{_\Q}^2)}{s_{\gamma p}z-m_T^2}$, $m_T=\sqrt{m_{_\Q}^2+P_T^2}$ (see \eg~\cite{Beneke:1998re}). Using the expression of $\left| {\cal M}\right|^2$ given \eg\ in~\cite{Berger:1980ni} for the direct/unresolved-photon case\footnote{In what follows, we will always refer to this case unless explicitly stated otherwise.} (with $x_\gamma=1$), one can easily reproduce the dotted LO curve in \cf{fig:NLO_photoproduction_vs_data}. 

Since data exist essentially only\footnote{The ratio $\psi(2S)/J/\psi$ was measured for 3 points in $P_T$ by ZEUS in 2012~\cite{Abramowicz:2012dh}.} for $J/\psi$, we will restrict the discussion to its production.
Exactly like in hadroproduction, $\left| {\cal M}\right|^2$ approximately scales at LO as $P_T^{-8}$ with
two far off-shell heavy-quark propagators in order to produce a large-$P_T$ $J/\psi$. 
The different NLO contributions can be grouped in different classes with the same $P_T$ dependence. 
The  virtual-emission contributions as shown on \cf{diagram-CSM-photoprod-b} scale like  $P^{-8}_T$. 
 The $t$-CGE topologies (see~\cf{diagram-CSM-photoprod-c}) scale like $P^{-6}_T$ and, at 
sufficiently large $P_T$, they dominate over the LO contributions.
Such topologies exist for $\gamma g$ or $\gamma q$ fusion.
These are expected to dominate over the whole set of diagrams up to 
$\alpha\alphaS^3$. 

\begin{figure}[hbt!]
\centering
\subfloat[]{\includegraphics[scale=.4]{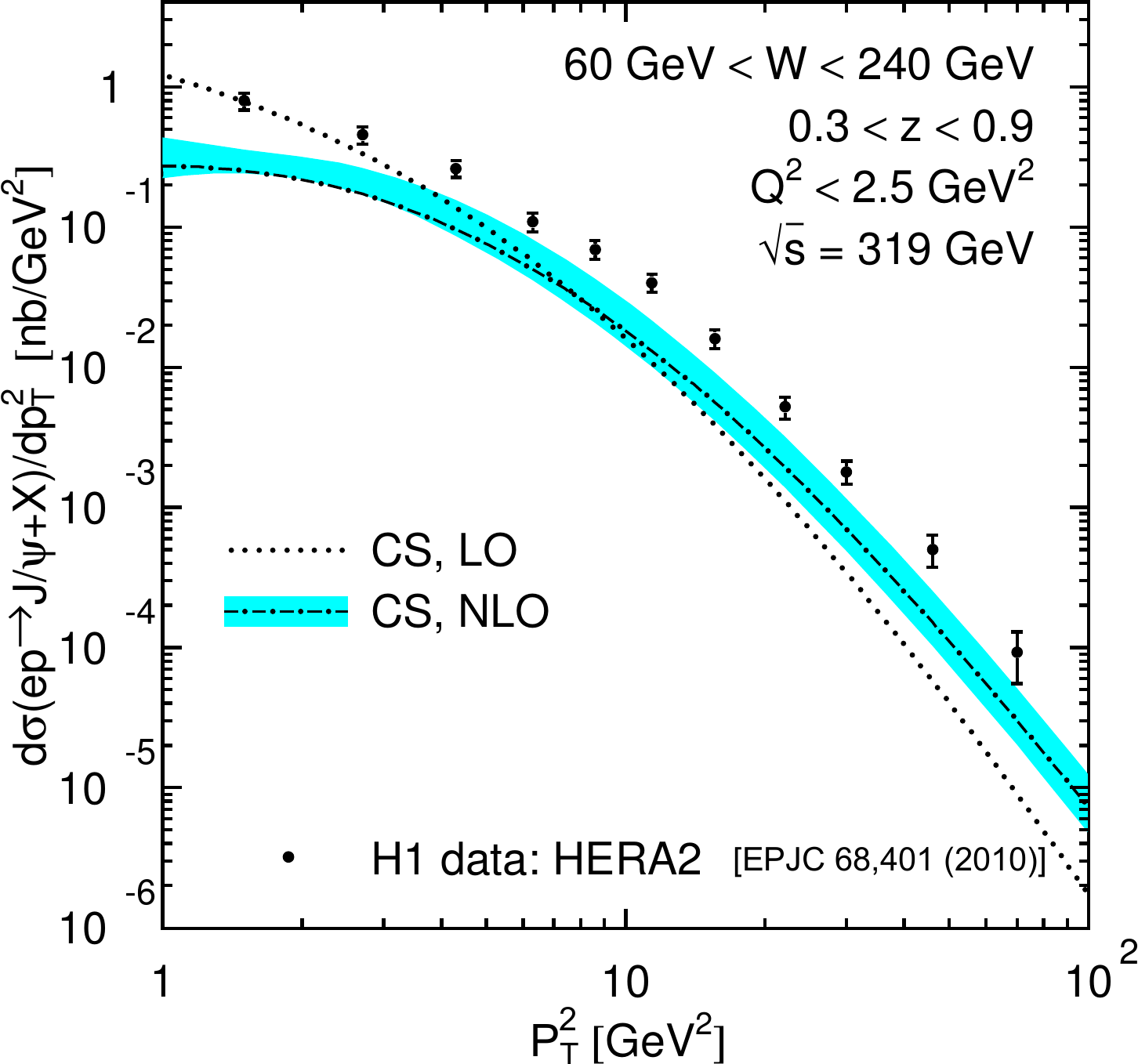}\label{fig:BK_CSM_NLO_h1_hera2_pt}}
\subfloat[]{\includegraphics[scale=.4]{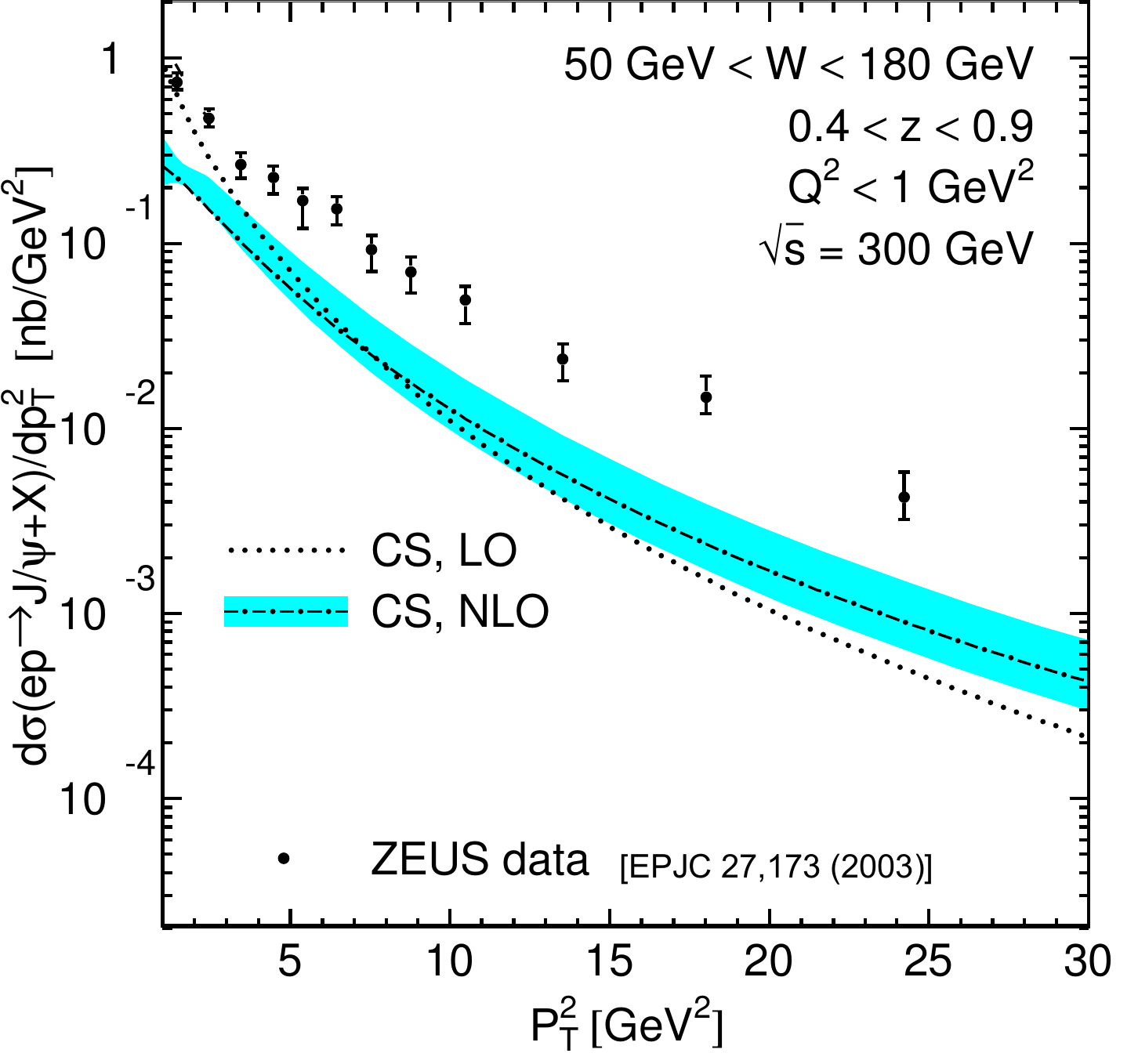}\label{fig:BK_CSM_NLO_zeus_pt}}
\caption{$P_T$-differential $J/\psi$ (unresolved) photoproduction cross sections predicted by the CSM at LO (dotted line) and NLO (dot-dashed line and blue band) computed for the kinematics of the (a) H1 Hera2~\cite{Aaron:2010gz} and (b) Zeus Hera1~\cite{Chekanov:2002at} data and compared to them. 
The theory parameters used for the curves are detailed in~\cite{Butenschoen:2012qh}. Figures adapted from~\cite{Butenschoen:2012qh}.}
\label{fig:NLO_photoproduction_vs_data}
\end{figure}

However, one can also consider a subset of the contributions ${\cal Q} + Q \bar Q$ (where $Q$ is of the same flavour 
as the quarks in $\cal Q$) appearing at  $\alpha\alphaS^3$ and which contains  fragmentation-like topologies (see \cf{diagram-CSM-d}) which nearly scale like $P^{-4}_T$. A similar contribution in a 4-flavour scheme, $\gamma c \to J/\psi c$ was already identified as a potential production source by Berger and Jones as early as on 1982~\cite{Berger:1982fh}. They found that the partonic amplitude was as large as the $\gamma g$ one but that its contribution to the $P_T$-integrated cross section was suppressed by the smallness of the charm PDF. We will re-discuss this channel later as a new observable since it can be studied for itself (see section~\ref{sec:psi-cc}). 

At $\alpha\alphaS^4$, further new  topologies appear as illustrated by \cf{diagram-CSM-e}. Contrary to the hadroproduction case, they have never been the objects of dedicated studies, although they are expected to become dominant at very large $P_T$. We however note that the corresponding fragmentation topologies are not from gluon fusion.
The CS fragmentation  contributions were discussed in 1996 by Godbole \etal~\cite{Godbole:1995ie} and the charm ones were found to be larger than the gluon fragmentation ones at all $P_T$ and larger than the LO gluon fusion ones for $P_T>10$~GeV (see also~\cite{Kniehl:1997gh}). If studies are carried out at large $P_T$ at a future EIC, such fragmentation contributions should be taken into account~\cite{forthcoming-photoproduction}. On the way, we note that the only study of inclusive photoproduction at a future EIC that we are aware of is a LO evaluation by Rajesh \etal\ \cite{Rajesh:2018qks} in the context of gluon Sivers TMD studies (see \cf{fig:LO-CS-dpt_EIC} for some predictions). 

\begin{figure}[hbt!]
\centering
\subfloat{\includegraphics[scale=.6]{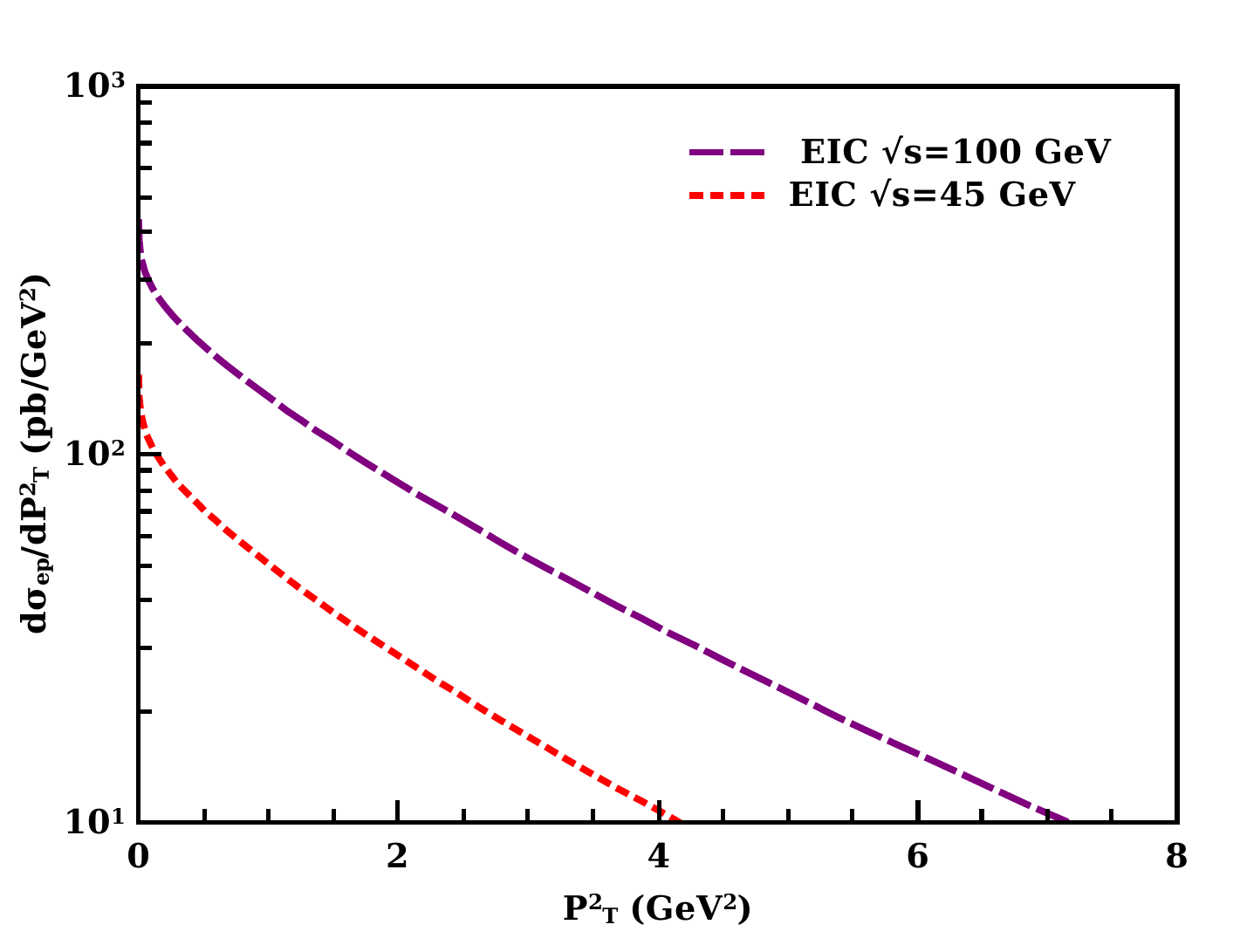}}
\caption{$P^2_T$-differential $J/\psi$ (unresolved) photoproduction cross sections predicted by the CSM at LO for 2 EIC $\sqrt{s}$. 
The theory parameters used for the curves are detailed in~\cite{Rajesh:2018qks}. Plot courtesy of S.~Rajesh.}
\label{fig:LO-CS-dpt_EIC}
\end{figure}

As we already wrote, Kr\"amer performed the first corresponding NLO study in 1995~\cite{Kramer:1995nb} which pointed at a $K$ factor (or equivalently the NLO/LO cross-section ratio) steadily increasing with $P_T$, resulting in a reasonably good agreement with the first H1~\cite{Aid:1996dn} and ZEUS~\cite{Breitweg:1997we} data. This is easily understood from the above arguments. In 2009, this NLO photoproduction study was revisited by 3 groups~\cite{Artoisenet:2009xh,Chang:2009uj,Butenschoen:2009zy} in order to compute  polarisation observables at NLO (see next). In these, they noted that the choice made by Kr\"amer of a constant --and small, $\mu_R=\mu_F=\sqrt{2} m_c$-- scale choice was probably leading to an overestimation of the QCD corrections. With seemingly more standard choices, on the order of $m_T$, the NLO results rather look like those on~\cf{fig:NLO_photoproduction_vs_data} from Butensch\"on and Kniehl~\cite{Butenschoen:2012qh} compared to the HERA2 H1~\cite{Aaron:2010gz} and HERA1 ZEUS data~\cite{Chekanov:2002at}. Whereas the CS NLO predictions are very close to the H1 points (in particular at the largest accessible $P_T$), it lies a factor of 3 below the ZEUS points. This can simply traced back by the fact that the ZEUS cross sections are larger than the H1 ones. So far, this remains unexplained. ZEUS also reported in 2012~\cite{Abramowicz:2012dh} on the cross section as a function of both $z$ and $P_T$ with the same conclusion as above.

We also note that the theory uncertainties are as large as a factor of 2 and that the $K$ factor surprisingly gets much smaller than unity at low $P_T$ which indicates that care should be taken when discussing $P_T$-integrated cross section (see also~\ref{section:CSM-NLO_tot}). In these data, FDs were not subtracted. One expect 15\% FD from $\psi(2S)$ and a $b$ hadron FD increasing with $P_T$ --and then possibly dominant-- following the H1 simulations. None of them could experimentally be measured, especially differentially in $P_T$.

Overall, whereas NLO corrections appear as fundamental in photoproduction than in hadroproduction to get the dominant CS contribution at finite $P_T$, it is difficult to make strong conclusions given the small remaining gap between the theory and the data, the size of the theory uncertainties, an apparent discrepancy between experimental datasets and
our limited knowledge of the $b$ FD at large $P_T$.

\paragraph{Polarisation.} The study of the polarisation of photoproduced $J/\psi$ exactly follows the same lines as for hadroproduction with the measurement of the angular coefficients $\lambda_\theta$ (or $\alpha$, $\lambda$), $\lambda_\phi$ (or $\nu/2$) and $\lambda_{\theta\phi}$ (or $\mu$) of the leptons from the $J/\psi$ decay in a given frame. For LO results and the definition of frequently used frames, we guide the reader to~\cite{Beneke:1998re}. Following the first NLO polarisation hadroproduction studies in 2008, which we previously discussed, $\lambda_\theta$ and $\lambda_\phi$ were evaluated at NLO in 2009~\cite{Artoisenet:2009xh,Chang:2009uj}.

\begin{figure}[hbt!]
\centering
\subfloat[]{\includegraphics[scale=.4]{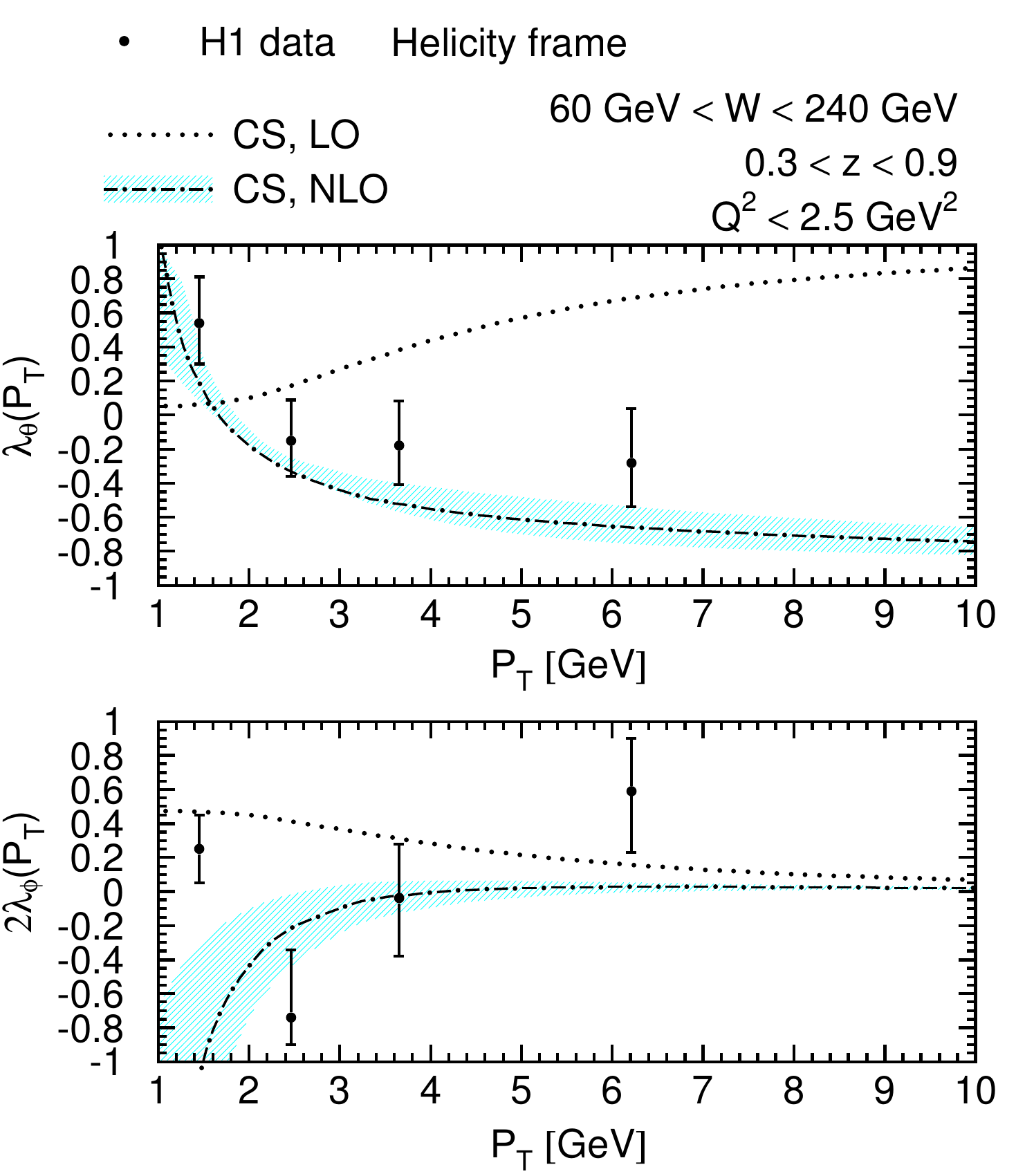}\label{fig:NLO_pol_photoproduction_vs_data-Hera2-pt}}
\subfloat[]{\includegraphics[scale=.4]{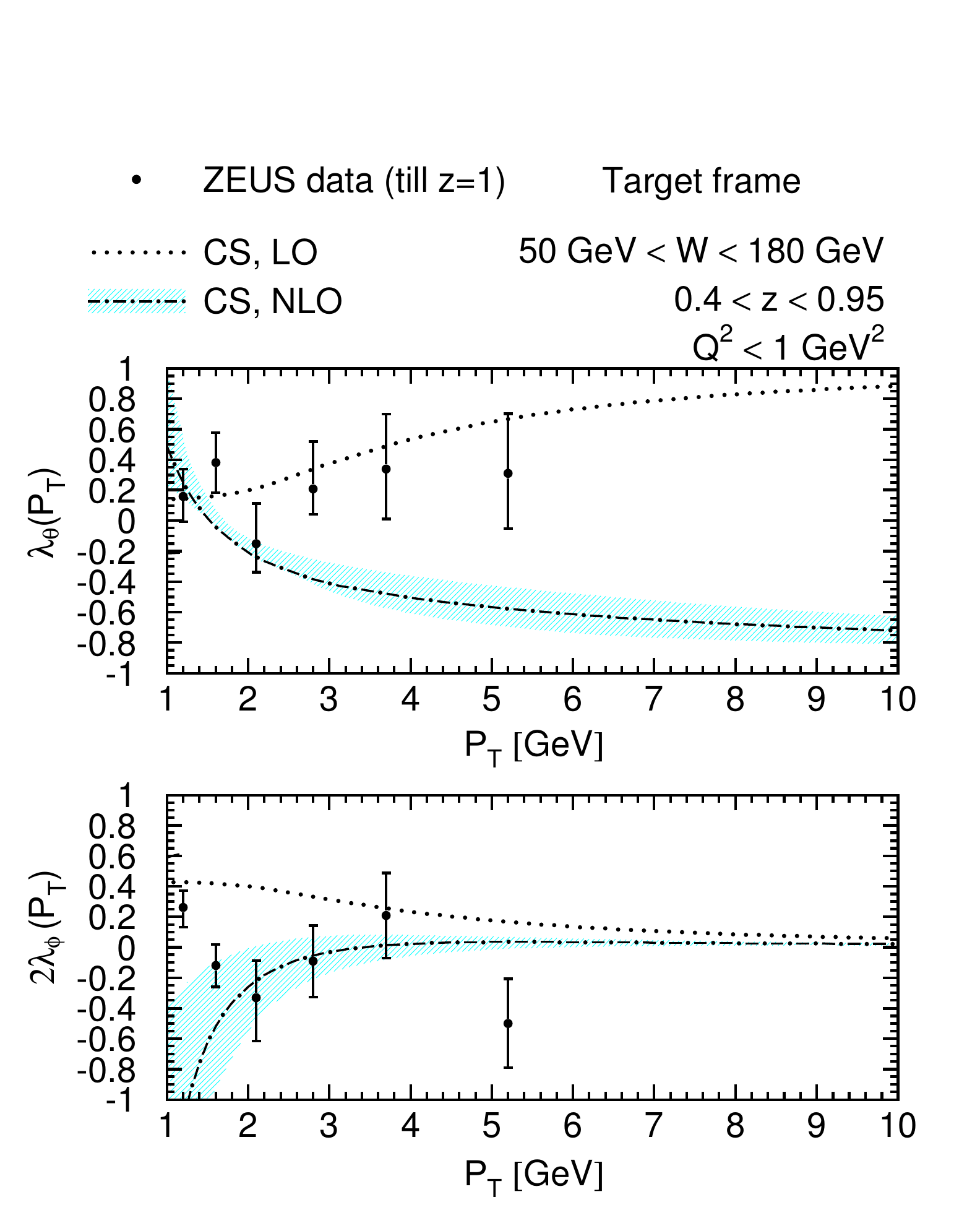}\label{fig:BK_CSM_NLO_zeus_pt_target}}
\caption{$P_T$-differential polarisation parameters for photoproduced $J/\psi$ predicted by the CSM at LO (dotted line) and NLO (dot-dashed line and blue band) computed for the kinematics of the (a) H1 HERA2~\cite{Aaron:2010gz} and (b) Zeus Hera1~\cite{Chekanov:2009ad} data and compared to them. 
The theory parameters used for the curves are detailed in~\cite{Butenschoen:2011ks}. Adapted from~\cite{Butenschoen:2011ks}.}
\label{fig:NLO_pol_photoproduction_vs_data}
\end{figure}

Unsurprisingly, like for hadroproduction, the impact of the dominant NLO real-emission corrections is to render the yield mostly
longitudinally polarised in the helicity frame for increasing $P_T$. 
\cf{fig:NLO_pol_photoproduction_vs_data} shows the LO and NLO CSM computation of $\lambda_\theta$ and $\lambda_\phi$ as a function of $P_T$ by Butensch\"on and Kniehl (similar to that of Artoisenet \etal~\cite{Artoisenet:2009xh} and Chang \etal~\cite{Chang:2009uj}) compared to two HERA datasets: one from H1~\cite{Aaron:2010gz} and another from ZEUS~\cite{Chekanov:2009ad}. The H1 study was carried in both the helicity (or recoil) and Collins-Soper frames and that of ZEUS in the target frames (see appendix A. of \cite{Beneke:1998re}\footnote{At small $P_T$ and for the HERA conditions, the helicity, target and Gottfried-Jackson frames should yield similar results since the polar axis vector would point in similar directions (up to a sign change between the target and Gottfried-Jackson frames).}). The NLO CS band for $\lambda_\theta$ agrees reasonably well with the H1 data in the helicity frame but the agreement is deteriorated in the Collins-Soper frame (not shown). The ZEUS data (in the target frame) tends to positive values of $\lambda_\theta$ at variance with the NLO CS. Like for the cross sections, the H1 and ZEUS datasets do not completely agree. We note that the ZEUS dataset includes data at large $z$ up to unity. As such, it is likely polluted, but only at low $P_T$, by diffractively produced $J/\psi$ whose polarisation is likely  different than inclusive ones. The interpretation of $\lambda_\phi$ is not easy owing to the large theory uncertainties at low $P_T$ and the large experimental uncertainties.

\subsubsection{$\psi$ leptoproduction at finite $P_T^\star$} 
\label{sec:CSM_leptoproduction}

As compared to photoproduction, leptoproduction --also referred to quarkonium production in deep inelastic scatterings (DIS)-- exhibits two important assets. First, the fact of dealing with a virtual photon enables one to disregard the resolved contributions. Second, the photon virtuality likely renders the heavy-quark-pair production more perturbative
than in the real-photon case, with a smaller expected impact of QCD-radiative and higher-twist corrections.

There are however counterparts. First, the cross sections are necessarily smaller --the suppression scaling like $1/Q^4$.
Only 4 experimental studies are available: 3 by H1~\cite{Adloff:1999zs,Adloff:2002ey,Aaron:2010gz} and 1 by ZEUS~\cite{Chekanov:2005cf}.
Second, the theory is getting significantly more complex\footnote{Additionally, the treatment of the photon emission is more complex.  Indeed, it was recently pointed out by Zhang and Sun~\cite{Zhang:2017dia} that the structure functions, $F_1$, $F_2$, and $F_3$ are not sufficient to describe the kinematic distribution of hadron production in DIS in the laboratory frame; azimuthally dependent structure functions (or terms in the hadronic tensor) are also needed. This is of relevance for $J/\psi$ leptoproduction.}
 with the appearance of a new scale in the computation. As a matter of fact the first NLO study~\cite{Sun:2017wxk}, on which we report below, was only completed in 2017.

\paragraph{$P_T^\star$-differential production cross section at NLO.}

The treatment of the kinematics of leptoproduction is usually a little different than that of photoproduction. Obviously, 
one introduces the photon virtuality as $Q^2$; correspondingly, ones refers to the photon momentum as $q$. $k_\ell$ is the lepton momentum. The Bjorken variable $x_B$ is the same as in inclusive DIS, namely  $x_B=Q^2/(2P_P.q)$, but does not coincide with the momentum fraction of the struck parton in the proton, $x_P$. One also introduces $y=P_P.q/P_p.k_\ell$ which helps study the effect of the off-shell photon polarisation and the overall c.m.s. energy squared $s_{\ell p}=(P_P+k_\ell)^2 \simeq2 P_P.k_\ell$. The variable $z$ and $W^2=W^2_{\gamma^\star p}=s_{\gamma^\star p}$ are the same as in photoproduction and can be used to characterise, to some extent, the exclusivity of the event. In addition, one defines "starred" variables, like $P_T^\star$, when they are measured in the $\gamma^\star p$ c.m.s.. In the latter frame, the leptons have a nonzero transverse momentum. Accordingly in photoproduction, $P_T^\star \simeq P_T$.

\begin{figure}[hbt!]
\centering
\subfloat{\includegraphics[width=0.6\columnwidth]{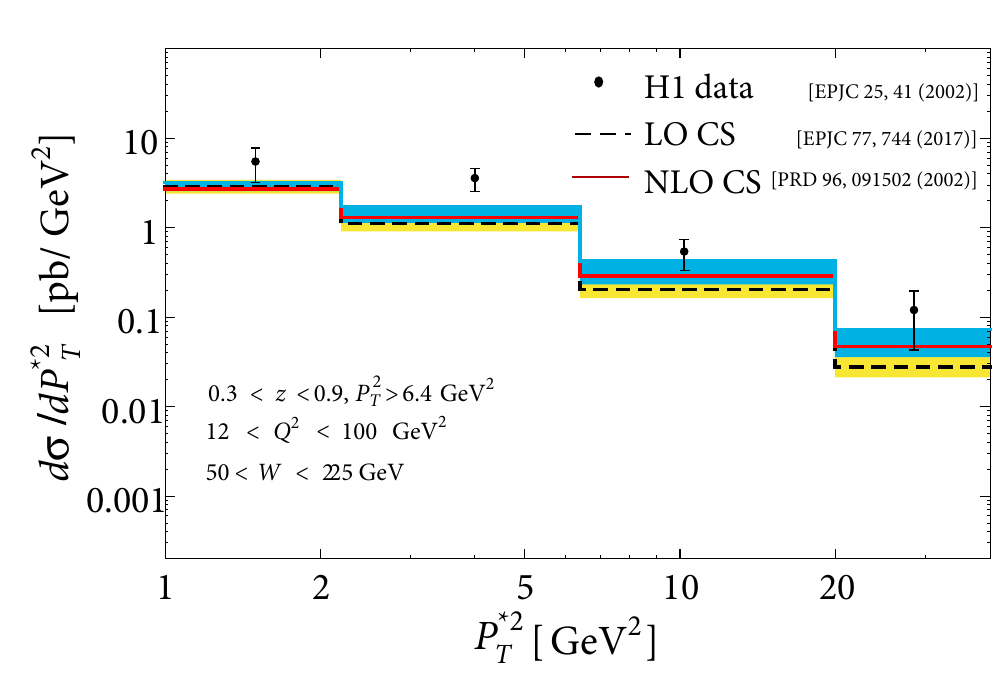}\label{fig:pts2dis-CS-LO-NLO}}
\caption{ $P_T^\star$-differential $J/\psi$ leptoproduction cross sections predicted by the CSM at LO (dashed line and yellow band) and NLO (dot-dashed line and blue band) computed for the kinematics of the H1 HERA1~\cite{Adloff:2002ey} data and compared to them. 
The theory parameters used for the curves are detailed in~\cite{Sun:2017nly,Sun:2017wxk}. Adapted from~\cite{Zhang:2017dia,Sun:2017wxk}.}
\label{fig:NLO_leptoproduction_vs_data}
\end{figure}

As we noted, care should be taken in decomposing the $ep\to e J/\psi X$ cross section in leptonic and hadronic tensors when some cuts are applied --which is nearly always {\it de facto} the case. We refer to~\cite{Sun:2017nly} for analytic expressions of the  CS LO hadronic tensors corresponding to a proper decomposition to deal with cross sections with  cuts. We note that, following the pioneering study of Baier and R\"uckl~\cite{Baier:1981zz}, several CS LO studies~	\cite{Korner:1982fm,Guillet:1987xr,Merabet:1994sm,Yuan:2000cn} were  carried out but they were shown by Kniehl and Zwirner~\cite{Kniehl:2001tk} to disagree between each others. Yet, the latter was shown by Sun and Zhang~\cite{Sun:2017nly}  to be affected by the aforementioned issue related to the hadronic tensor forms.

The same graphs as \cf{diagrams-CSM-photoproduction}, but with a virtual photons, contribute to the cross section. For 
$P_T^{\star 2} \ll Q^2$, the same considerations as for photoproduction with respect to the $P_T^{\star 2}$ scaling of the different contributions can be done. Once again, the essential difference with photoproduction is the absence of resolved-photon contributions. 
The dashed curve and the yellow band of \cf{fig:NLO_leptoproduction_vs_data} show the LO CS yield in the corresponding acceptance of the H1 2002 dataset~\cite{Adloff:2002ey} which is then compared to these data.  One clearly sees that the LO CS cross section has a softer $P_T^{\star2}$ spectrum than the H1 data. 
Comparisons to other data sets can be found in~\cite{Sun:2017nly}.

The first NLO analysis by Sun and Zhang~\cite{Sun:2017wxk} confirmed the expectations for a $K$ factor growing with $P_T^{\star 2}$. A representative comparison between the CSM at NLO and the data is shown \cf{fig:NLO_leptoproduction_vs_data}. Unfortunately the analysed data set in their study ($0.3 < z < 0.9$, $12 < Q^2 < 100$ GeV$^2$, $P_T^2 > 6.4$ GeV$^2$) is
admittedly restricted. As far as the $P_T^{\star 2}$ dependence is concerned, we do not agree  with the conclusion of Sun and Zhang that ``the CS contributions are still much small than the experimental data''. As a final note, let us recall that the FDs have never been measured (and are not included in the theory curves nor subtracted from the data points). They are expected to be 15\% for the $\psi(2S)$ and, 
for the $b$ hadron, 17 \% in $0.3 < z < 0.45$ according to MC simulations tuned to $B$ production cross sections by H1 in 2002~\cite{Adloff:2002ey}. By analogy with photoproduction, they might become significant at large $P_T^{\star 2}$ (see section~\ref{sec:FD}).
\footnote{The very recent preprint by Qiu~\etal~\cite{Qiu:2020xum} reports on a NLO study of the $Q^2$-integrated
yield assuming that the corresponding Born-order contributions are from $2\to 1$ partonic processes, thus at zero $P_T^\star$. As such their virtual corrections only contribute at $P_T^\star=0$ and their real-emission corrections are Born-order contributions to the $P_T^\star$ distributions.}   

\paragraph{Polarisation} To date, there has been only a single study of the polarisation of the $J/\psi$ produced in inclusive DIS. Based on a 1D analysis of the polar distribution on their sample with $P_T^{\star2}> 1 $ GeV$^2$, H1 reported in 2002~\cite{Adloff:2002ey}, $\lambda_\theta=−0.1^{+0.4}_{-0.3}$ in the helicity frame. Such a measurement is compatible with an unpolarised yield but hardly discriminant between the different models. LO CS predictions have been provided by Yuan and Chao~\cite{Yuan:2000cn} and showed an increasingly longitudinal yield for increasing $Q^2$, except at very large $z$. However, QCD corrections may significantly alter such LO-based computations like for photoproduction. Unfortunately, later HERA analyses did not address polarisation measurements.

\subsubsection{$\psi$ production in $e^+e^-$ annihilation} 
\label{sec:CSM_eeproduction}

In addition to quarkonium-production studies in $pp$ and $ep$ collisions, quarkonia were of course also  studied in $e^+e^-$ annihilation. Let us recall that one of the discovery mode of the $J/\psi$ and $\psi(2S)$ was $e^+e^-$ annihilation. Inclusive data have been collected at LEP at $\sqrt{s}\simeq 200$ GeV both via $\gamma \gamma$ collisions and via $Z$ decays as well as at $B$ factories at $\sqrt{s}\simeq 10.6$ GeV. We briefly review below the recent phenomenology of such studies.

\paragraph{$B$ factories}

The first measurements of inclusive $J/\psi$ production cross sections at $B$ factories were released as early as in 2001 by the {\sc BaBar}~\cite{Aubert:2001pd} and Belle~\cite{Abe:2001za} collaborations. They respectively reported $2.52 \pm 0.21 \pm 0.21$ pb and $1.47 \pm 0.10 \pm 0.13$ pb. These values were found to be significantly larger than the LO CS prompt\footnote{CS-based theoretical estimation states that the $\psi(2S)$ FD increases the cross section by 1.3~\cite{Ma:2008gq,Gong:2009kp}.}
 $J/\psi$ predictions~\cite{Driesen:1993us,Yuan:1996ep,Cho:1996cg}, on the order of $0.2\div0.5$ pb, following the lines of earlier works~\cite{Keung:1980ev,Kuhn:1981jn,Kuhn:1981jy} (see the dashed curves of \cf{fig:psigg-Belle-depmu}).

\begin{figure}[hbt!]
\centering
\subfloat[]{\includegraphics[width=0.4\columnwidth]{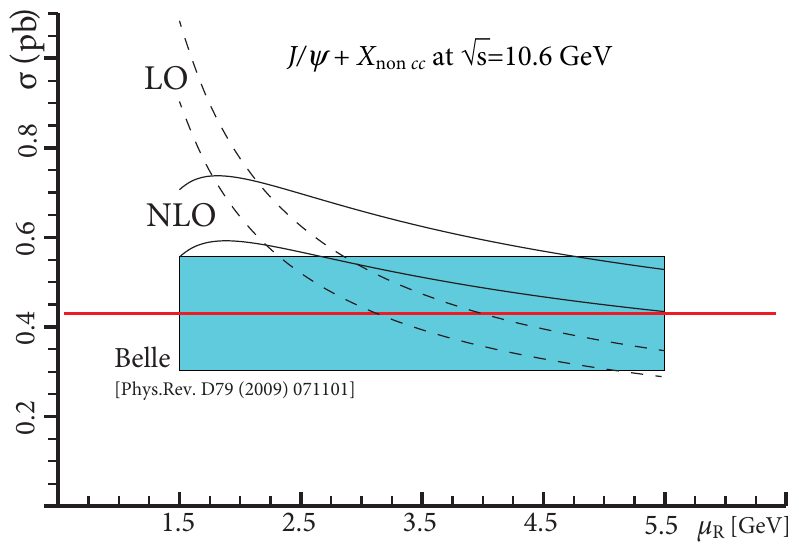}\label{fig:psigg-Belle-depmu}}
\subfloat[]{\includegraphics[width=0.6\columnwidth]{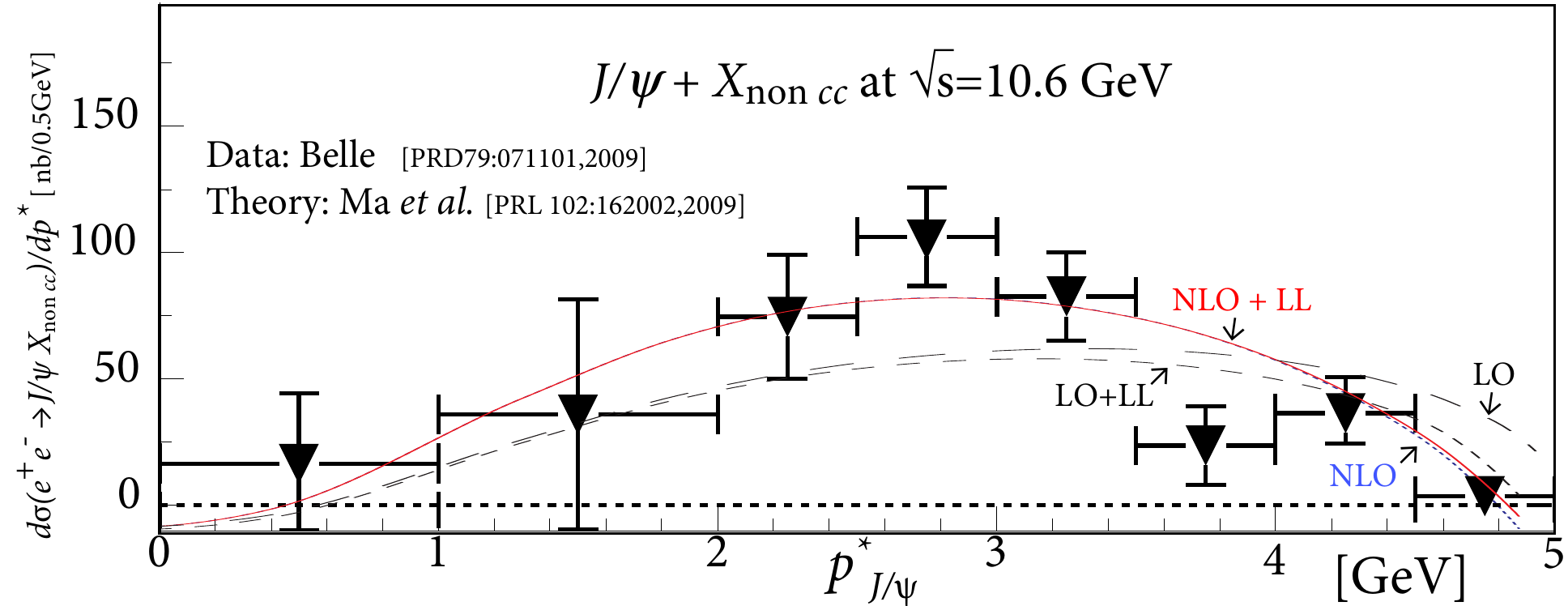}\label{fig:psiX_dsigdp_Belle_NLO-Ma}}
\caption{(a) Prompt $\sigma(J/\psi+X_{\text{non } c \bar c})$ cross section in units of pb from the CS at LO (dashed lines) and NLO (solid lines) compared to the Belle measurement. The region spanned by the curves is indicative of the theoretical uncertainties.
(b) $J/\psi$-momentum differential cross section for $J/\psi+X_{\text{non } c \bar c}$ in units nb/0.5 GeV as measured
by Belle compared to the CS LO (long dash gray), LO+LL (short dash gray),  NLO (blue small dash), NLO+LL (red solid)
predictions by Ma \etal. The theory parameters used for the curves are detailed in~\cite{Ma:2008gq}. The Belle experimental uncertainties~\cite{Pakhlov:2009nj} have been combined in quadrature. Adapted from (a)~\cite{Ma:2008gq} and (b)~\cite{Pakhlov:2009nj}.}
\label{fig:results-CS-psigg}
\end{figure}

In 2002, it was however found out by Belle~\cite{Abe:2002rb} that more than half of the inclusive $J/\psi$ were created with another charm pair. These will specifically be discussed in section~\ref{sec:psi-cc}. When their contribution are subtracted, the discrepancy is drastically reduced. In fact, in 2009, Belle released their last analysis~\cite{Pakhlov:2009nj}, with more than 30 times more data,  where $\sigma(J/\psi+X_{\text{non } c \bar c})=0.43 \pm 0.09 \pm 0.09$~pb which is well compatible with the LO CS predictions, but for the fact that the momentum distribution of the $J/\psi$ tended to be slightly harder than that of the data. This issue has been anticipated
by Lin and Zhu~\cite{Lin:2004eu} as well as Leibovich and Liu \cite{Leibovich:2007vr} who carried out the resummation
of the large $\log(E-E_{\rm max})$ --referred to as leading logarithm (LL) resummation in the following.  Ma \etal\ performed~\cite{Ma:2008gq} a LL resummation and obtained the LO+LL curve shown on \cf{fig:psiX_dsigdp_Belle_NLO-Ma}. Whereas the  dependence on the $J/\psi$ momentum in the $e^+e^-$ c.m.s., $p^\star_{J/\psi}$, is visibly affected, the total cross section is only modified (reduced, to be precise) by about 7\%.

\begin{figure}[hbt!]
\centering
\subfloat[]{\includegraphics[scale=.33,draft=false]{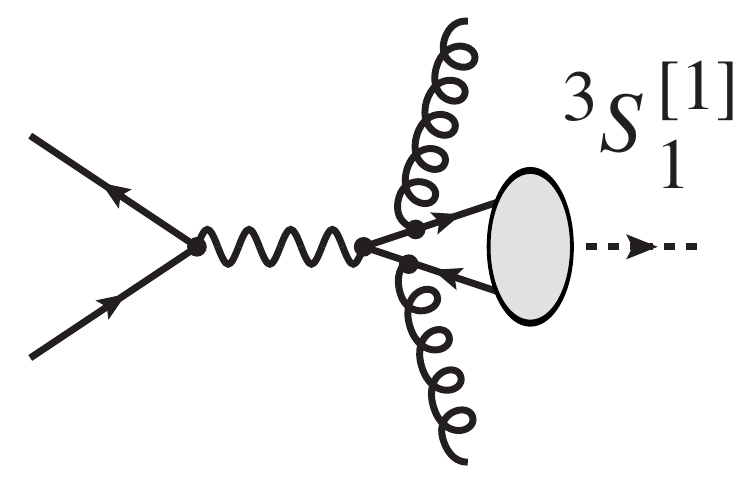}\label{diagram-CSM-eeprod-a}}
\subfloat[]{\includegraphics[scale=.33,draft=false]{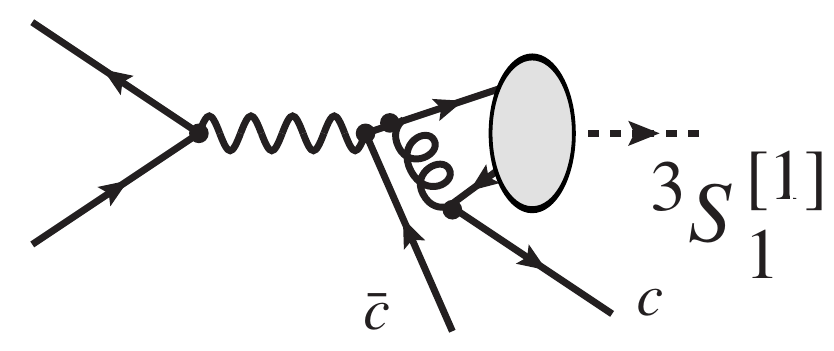}\label{diagram-CSM-eeprod-b}}
\subfloat[]{\includegraphics[scale=.33,draft=false]{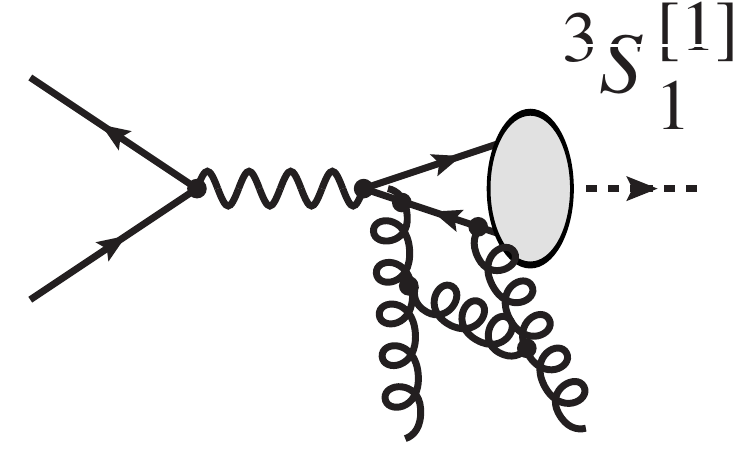}\label{diagram-CSM-eeprod-c}}
\subfloat[]{\includegraphics[scale=.33,draft=false]{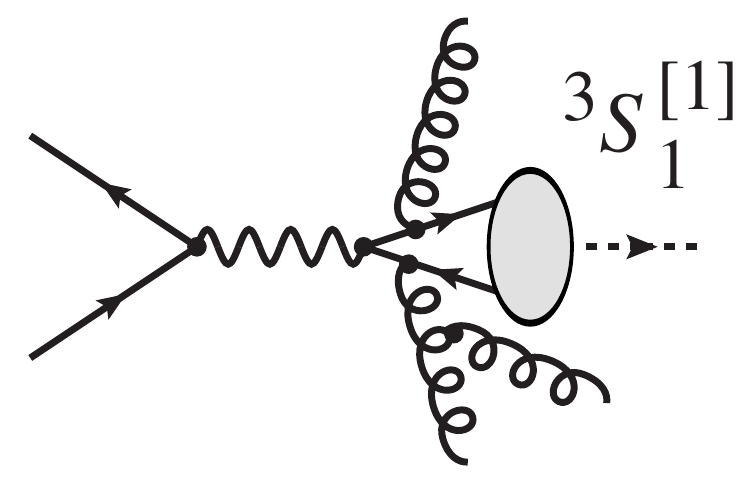}\label{diagram-CSM-eeprod-d}}
\subfloat[]{\includegraphics[scale=.33,draft=false]{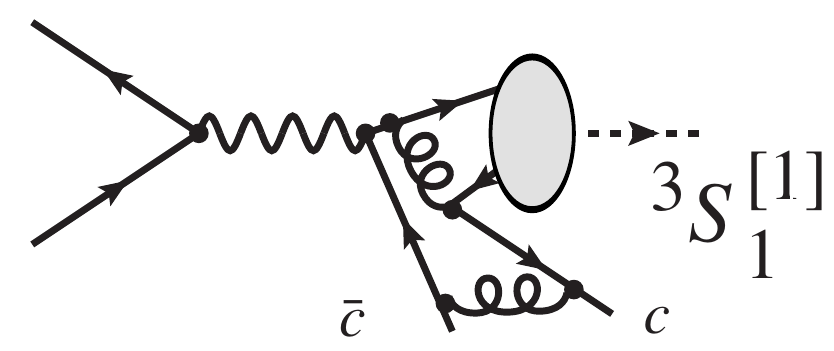}\label{diagram-CSM-eeprod-e}}
\subfloat[]{\includegraphics[scale=.33,draft=false]{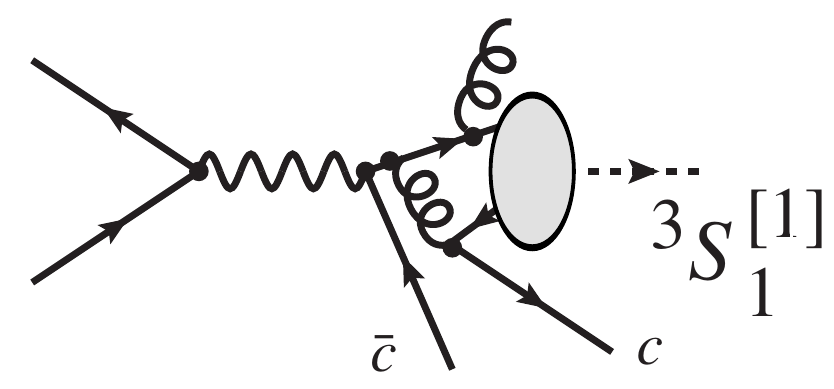}\label{diagram-CSM-eeprod-f}}
\caption{Representative diagrams contributing to $^3S_1$ production in $e^+e^-$ annihilation via 
CS channels at orders $\alpha^2\alphaS^2$ (a,b), $\alpha^2\alphaS^3$ (b,c,d,d). 
The quark and antiquark attached to the ellipsis are taken as on-shell
and their relative velocity $v$ is set to zero.}
\label{diagrams-CSM-eeproduction}
\end{figure}

 In the meantime, the CS cross section could be computed at NLO accuracy~\cite{Ma:2008gq,Gong:2009kp}. A selection of NLO graphs is shown on~\cf{diagrams-CSM-eeproduction} along with a LO graph. A $K$ factor close to 1.2 was found in both studies as well as a good convergence of the perturbative series for $\mu_R$ close to $m_c$ for which the cross section is the largest. A representative data-theory comparison is shown on \cf{fig:psigg-Belle-depmu} which shows that the CS easily saturates the Belle measurement and the effect of the NLO corrections. The NLO+LL curve included the resummation effect which are quasi irrelevant for the total cross section (increase by 0.5\%) and the shape.

Two comments are in order at this stage. First, it will be very important to have a further confirmation of the latest Belle cross-section measurement as it happens to be significantly below the {\sc BaBar} one. In particular, it will be useful to have a study of the impact of requiring more than 4 charged tracks in the final state (in order to cut off initial-state radiations from the $e^\pm$).
Second, whereas resumming LL corrections does not affect the CS NLO spectrum (compare the blue and red curves on \ref{fig:psiX_dsigdp_Belle_NLO-Ma}), $v^2$ relativistic corrections have been shown~\cite{He:2009uf,Jia:2009np} to further enhance the CS cross section by $20\div 30\%$. Taking into account the $\psi(2S)$ FD, one therefore obtains : 
\eqs{
\sigma^{\text{ NLO, }v^2 \text{ corr.}}(J/\psi+X_{\text{non } c \bar c})=0.55\pm 0.15 \text{ pb}.
}
It therefore seems that, after a rather long period of confusion, the CS computations are in good agreement with the experimental data as what concerns the $J/\psi$ inclusively produced without additional charm. In 2014, Shao further
noted~\cite{Shao:2014rwa} that an additional effect had been overlooked, namely that from QED ISR. Since the predicted CS cross section for $J/\psi+X_{\text{non } c \bar c}$ steadily decreases with $\sqrt{s}$ away from the threshold, the expected effect of these ISR is to increase the cross section. Quantitatively, this increase depends on the momentum distribution and he found enhancement values from 15 to 30 \% which should further be combined to the aforementioned QCD and relativistic corrections. Overall, one gets
\eqs{
\sigma^{\text{ NLO, }v^2 \text{ corr., ISR}}(J/\psi+X_{\text{non } c \bar c})=0.65 \pm 0.2 \text{ pb},
}
whose central values now even tends to be above the Belle measurement.

As what concerns the $\psi(2S)$ case, only the inclusive differential cross section (for $p^\star_{\psi(2S)} > 2$ GeV) has been measured by Belle in 2001~\cite{Abe:2001za}. As we have just seen, it is however essential that the $\psi(2S)+X_{c \bar c}$ cross sections could be measured and subtracted. Hopefully, this could be done at Belle-II.

Finally, Belle performed~\cite{Pakhlov:2009nj} a first measurement of the polarisation of the $J/\psi+X_{\text{non } c \bar c}$ yield and reported  $\lambda_{\theta}=0.41^{+0.60}_{-0.45}$ in the helicity frame. This could certainly be improved at Belle-II in spite of the complexity of such a measurement.
On the theory side, Gong and Wang performed~\cite{Gong:2009kp} in 2009 the first NLO CS angular-distribution analysis and found out that $\lambda_{\theta}$ at  NLO and LO~\cite{Baek:1998yf} are very similar and both get negative for increasing momentum. Predictions for $\lambda_{\theta}$ integrated over $p^\star_\psi$ were not provided but we can estimate, since the polarisation does not vary much with $p^\star_\psi$, that it should lie between -0.3 and -0.5, thus about 1.5-$\sigma$ lower than the Belle measurement.

\paragraph{$\gamma\gamma$ collisions at LEP}

Inclusive-$J/\psi$ production has also been studied at higher energies at LEP via $\gamma\gamma$ fusion. However,a single measurement exists by DELPHI~\cite{Abdallah:2003du} for $P_T$ up to barely 3 GeV. For $P_T>1$ GeV, this data set only comprises 16 events. In addition, the $P_T$-differential measurement was only released as an event distribution owing to the uncertainty in the evaluation of the efficiency corrections. As such, conclusions based on this unique data set may need to be taken with a grain of salt. 

Like for photoproduction, resolved photons can contribute to the cross section. With a single resolved photon --a priori the dominant contribution here, the process is akin to photoproduction and, with two resolved photons, the process is like hadroproduction. In the CSM, $C$-parity conservation severely restricts the possible direct-photon channels (leaving aside the FD). As such, $\gamma \gamma \to J/\psi g g$ is forbidden, only $\gamma \gamma \to J/\psi ggg$
at $\alpha^2\alphaS^3$ is allowed. $\gamma \gamma \to J/\psi \gamma$  is however allowed at $\alpha^3$ as well as
$\gamma \gamma \to J/\psi c \bar c$ at $\alpha^2\alphaS^2$. The latter is the dominant contribution to direct $J/\psi$ via direct-photon fusion according to a quick comparison made by Artoisenet in 2009 \cite{Artoisenet:2009zwa}. $\gamma \gamma \to J/\psi ggg$ seemingly had never been studied before. Based on a LO analysis, Klasen \etal\ had found~\cite{Klasen:2001cu} out in 2001 that the CS yield --essentially from single-resolved contributions-- fell short compared to the DELPHI data by a factor about $5\div10$. Yet, we should keep in mind the above comment about the data and that both the data and the theory show uncertainties of a factor of $2\div3$.

QCD corrections to the direct-photon process were then computed by Klasen \etal~\cite{Klasen:2004tz}. In 2009, Li and Chao~\cite{Li:2009fd} performed a complete study of the possible contributions from $J/\psi + c \bar c$ which were found to increase the CS yield a little but not enough to match the DELPHI data. In 2011, bearing on the breakthrough of the NLO computations discussed in section~\ref{subsec:CSM_NLO_hadroproduction}, Kniehl and Butensch\"on provided~\cite{Butenschoen:2011yh} the first complete --\ie\ including the resolved contributions-- NLO CS computation  which they found very close to LO CS result.

In 2016, Chen \etal~\cite{Chen:2016hju} performed the first NLO study of the sub-contribution from $\gamma \gamma \to J/\psi + c \bar c$, thus up to $\alpha^2 \alphaS^3$ . They found out that it could indeed be the dominant source of the CS yield with a $K$ factor close to 1.5. We note on the way that the theory uncertainties on the $J/\psi + c \bar c$ contributions are much larger than that on the other CS contributions.
Overall, the CS yield remains below the DELPHI data whose reliability is under debate.

\subsubsection{The $P_T$-integrated yields and their energy dependence up to NLO}
\label{section:CSM-NLO_tot}

In the previous sections, we have discussed how $\alpha^4_s$ and $\alpha^5_s$ corrections to the CSM
drastically alter the $P_T$-differential $\psi(nS)$ and $\Upsilon(nS)$ cross sections 
in high-energy hadron collisions.
However, if anomalously large contributions
to the total, $P_T$-integrated\footnote{In what follows we shall indeed employ ``total"
for ``$P_T$ integrated", even though $y$ is not ``totally" integrated over. In practice, 
many experimental data sets allow one to recover the $P_T$-integrated yield, whereas
none allows for a complete $y$ integration at colliders.}, 
cross section arise from QCD corrections, this would cast doubt on the convergence
of the expansion in $\alpha_s$. To be more precise, the enhancement of these 
higher-order QCD corrections come along with kinematical factors such as
$P_T/m_{\{Q,\Q\}}$ whose impact is necessarily reduced when 
$P_T$ is integrated over. It is thus important to verify that LO and NLO
predictions to the total cross section are close to each other and agree with
experimental data.

Let us also recall that, unlike other quarkonium states, the $\psi(nS)$ and $\Upsilon(nS)$ 
$P_T$-integrated and -differential cross sections are computed from the same set of 
Feynman graphs in the CSM. Indeed the graphs allowing for their production recoiling against a parton,
which then provides their $P_T$, are also the first non-vanishing ones in the $\alpha_s$ expansion.
Formally, this means that the total cross section of vector quarkonia in the CSM is an
$\alpha_s^3$ process whereas, for pseudoscalar states or $P$ waves, it appears at $\alpha_s^2$.
This lead some to conclude~\cite{Cooper:2004qe} that the direct vector-quarkonium production was sub-leading and they ought to be produced
either via decay from states produced at $\alpha_s^2$ or from other 
mechanisms like the CEM or COM. We will see that it is far from being obvious. 

Indeed, as it was briefly mentioned in the introduction, the LO CSM computations, 
exactly following the lines of the 1980's computations with updated PDFs, 
reproduce  well the $P_T$-integrated yields measured at colliders, RHIC, 
Tevatron and the LHC~\cite{Brodsky:2009cf,Lansberg:2010cn,Feng:2015cba}, without
any parameter adjustement (see \eg\ the LO CSM curves in \cf{fig:energy_dependence_CSM}). The comparison
however gets worse at low energies.

Although the central LO CSM 
predictions agree with the data, the conventional theoretical uncertainties 
--from the arbitrary scales and the heavy-quark mass-- are large, in any case too large
to make strong statements about a possible hierarchy at a precision commensurable with
a suppression by a single power of $\alpha_s$, by the number of graphs, by different branching
ratios or by other nonperturbative parameters. It is thus appealing to wonder whether 
these uncertainties are reduced at one-loop/NLO accuracy relying on the NLO breakthrough
from 2007~\cite{Campbell:2007ws}. This is what we discuss now along with
the case of the pseudoscalar case for reasons which will become clear later.

As just discussed, we have to consider the same graphs as those used to compute
the $P_T$-differential cross section. We stress that there is no 
specific difficulty to integrate the $\alpha_S^3$ and $\alpha_S^4$ contributions 
in $P_T$ since they are indeed finite at $P_T=0$. The only potential issue
is the stability of the result at very low $P_T$ where it is known that some fixed-order computations can become negative. 
We will come back to this later. 
All the required information were known since~\cite{Campbell:2007ws} and can 
even be computed with the semi automated tool FDC~\cite{Wang:2004du} as done in~\cite{Gong:2008sn}.

To the best of our knowledge, the first one-loop CSM study of the $P_T$-integrated cross section was done
in~\cite{Brodsky:2009cf} using the results of~\cite{Campbell:2007ws} for the RHIC 
conditions, \ie~$\sqrt{s}=200$~GeV. 
It was then shown that the $\psi$ and $\Upsilon$ total cross section at LO and NLO
accuracy were indeed compatible and, as expected, the scale uncertainty was reduced at NLO. 

\begin{figure}[hbt!]
  \centering
\subfloat{\includegraphics[trim = 0mm 0mm 0mm 0mm, clip,width=.5\columnwidth,draft=false]{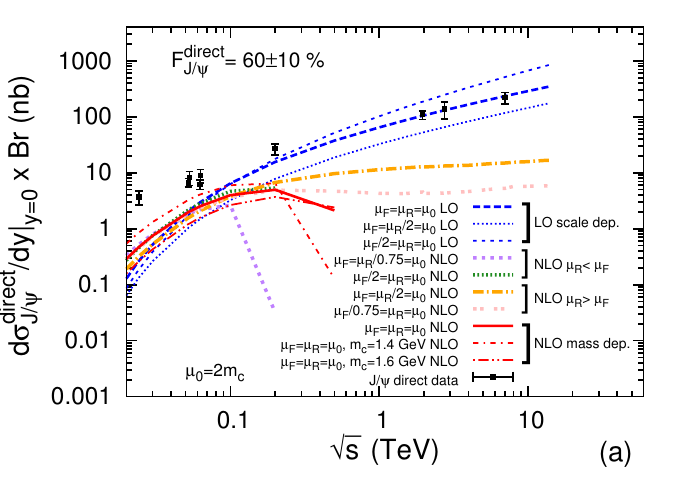}}
\subfloat{\includegraphics[trim = 0mm 0mm 0mm 0mm, clip,width=.5\columnwidth,draft=false]{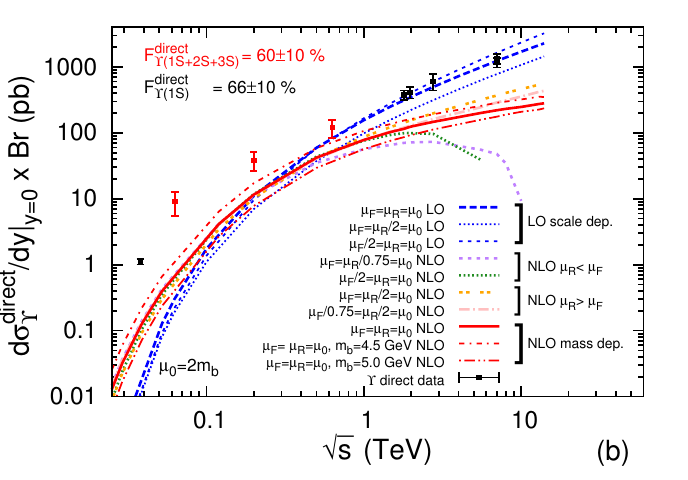}}
  \caption{The cross section for direct (a) $J/\psi$ and (b) $\Upsilon(1S)$ as a function of the 
centre-of-momentum energy in the CSM at LO and NLO for various
choices of the mass and scales compared with the existing experimental measurements 
(see~\cite{Feng:2015cba} for their selection). Taken from~\cite{Feng:2015cba}.}
  \label{fig:energy_dependence_CSM}
\end{figure}

In~\cite{Feng:2015cba}, the corresponding energy dependences were studied from 40 GeV
all the way up to 13 TeV. Let us briefly look at these.
\cfs{fig:energy_dependence_CSM} show the energy dependence of the NLO CSM (7 curves\footnote{If
a curve is not shown until 14 TeV, it means that the total yield got {\it negative}.}).  The 3 red 
curves display the default scale choices ($\mu_R=\mu_F=2 m_Q$) with different heavy-quark-mass values.
From them, one sees that the mass uncertainty is on the order of a factor of 
2 for the $J/\psi$ and 1.6 for the $\Upsilon$. Such uncertainties are admittedly large
but are probably correlated between observables. 

In the $J/\psi$ case, all 3 red curves end up to be negative somewhere 
between 500 GeV and 2 TeV. Note also that the curve for the lowest mass, $m_c=1.4$ GeV, 
is the first to become negative despite being the upper curve at low energies. 
In the $\Upsilon$ case, these 3 curves remain positive
at high energies. 
They nevertheless start to significantly deviate from the LO curves (3 blue curves) 
above 1 TeV, contrary to the aforementioned good LO vs NLO convergence found 
at RHIC energies in~\cite{Brodsky:2009cf}.
One might thus be tempted to identify this weird energy behaviour 
to a {\it low-$x$ effect} or to an effect related to the low-$x$
behaviour of the PDFs at small scales~\cite{Ozcelik:2019qze}.

Such a behaviour was also previously respectively discussed by Sch\"uler~\cite{Schuler:1994hy} and Mangano~\cite{Mangano:1996kg}  in the case of $\eta_c$ (see below) and $C=+1$ CO production.  Similar negative cross sections have been observed at NNLO in the hadroproduction of $c\bar c$ pair (see Fig. 17 of~\cite{Accardi:2016ndt}). It indeed seems to be connected with the behaviour of the PDFs at small scales. If it is the case, one could be entitled to impose during the fitting of the PDFs that such cross sections be positive definite. It remains to be seen if doing so their energy dependence would become more natural and whether the NLO CSM cross section would agree with the data at Tevatron and LHC energies.

In~\cite{Khoze:2004eu}, it was argued\footnote{see however some comments by the same authors in~\cite{Shuvaev:2015fta}.} that some NNLO corrections are kinematically enhanced
at large energies (low $x$). It is not clear if they could provide a solution to this issue. 
Another possible explanation might be the importance of initial-state 
radiations (ISR), which should resummed as done in the CEM~\cite{Berger:2004cc} 
and for some CO channels~\cite{Sun:2012vc}. It is however not clear
if the conventional resummation methods can be applied to the present case
with a non-diverging Born-order amplitude squared at $P_T=0$.

In this context, one can advance the conclusion that the NLO CSM results available in the literature
may be reliable for $\Upsilon$ up to 200~GeV and for $J/\psi$ up to 60~GeV. This corresponds to
$\sqrt{s}$ about 20 times the quarkonium mass. At higher energies, NLO results are
not trustworthy and the Born order results remain the best we have at hand. It also clear that
further investigations are needed to trace back the source of this problem.
As such, analogies with the pseudoscalar are insightful.

As opposed to the spin-triplet vector case, analytical formulae~\cite{Kuhn:1992qw,Petrelli:1997ge}
exist for the partonic-scattering amplitude for the spin-singlet pseudoscalar production 
cross section such as that of $\eta_c$ and $\eta_b$. 
This can be helpful to understand the weird energy behaviour of the CS $^3S_1$ yield 
which we have just discussed (see also~\cite{Schuler:1994hy}). Indeed, the LO production occurs without final-state-gluon 
radiation and is easier to deal with. In fact, this process is the first for which 
inclusive hadroproduction NLO cross sections were obtained as early as in 1992~\cite{Kuhn:1992qw}.

\begin{figure}[hbt!]
  \centering
\subfloat{\includegraphics[trim = 0mm 0mm 0mm 0mm, clip,width=.5\columnwidth,draft=false]{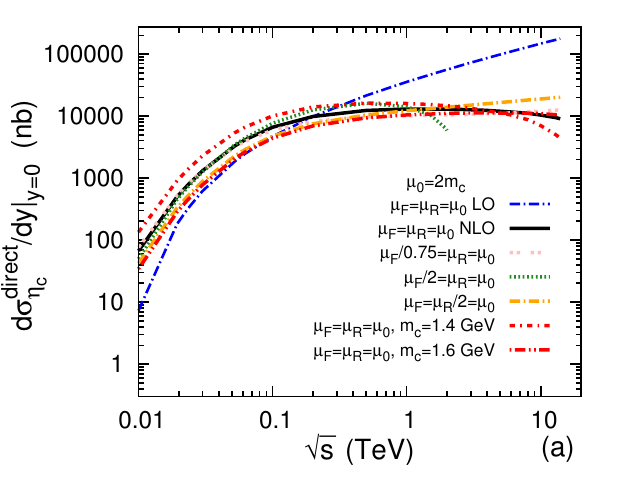}}
\subfloat{\includegraphics[trim = 0mm 0mm 0mm 0mm, clip,width=.5\columnwidth,draft=false]{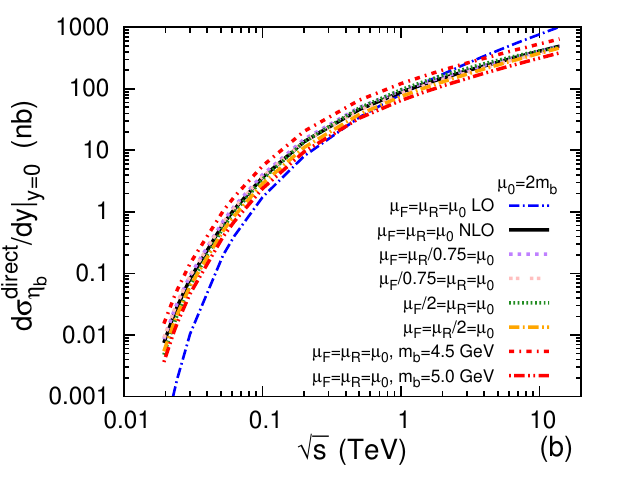}}
  \caption{The cross section for direct (a) $\eta_c$ and (b) $\eta_b$ as a function of the centre-of-momentum energy in the CSM at LO and NLO  for various
choice of the mass and scales.Taken from~\cite{Feng:2015cba}.}
  \label{fig:energy_dependence_CSM-etaQ}
\end{figure}

In fact, as can be seen on \cfs{fig:energy_dependence_CSM-etaQ}, the situation is in many respects
similar.
To go further, one can look at the different NLO contributions
(the real emissions from $gg$ and $qg$ fusion as well as the virtual (loop) contributions) in order
to see which channels generate the largest negative contributions and for which scale/mass values. 
We must stress here that the decomposition between these different (real vs. virtual) contributions depends 
on the regularisation method used. 
As a case in point, the decomposition is drastically different when using FDC~\cite{Wang:2004du,Wan:2014vka} --with sometimes a very large 
cancellation between the {\it positive} real-emission $gg$  contribution and the {\it negative} sum of the 
Born and loop $gg$  contributions -- and the formulae of~\cite{Kuhn:1992qw,Petrelli:1997ge} 
--where all the $gg$ contributions are grouped. Let us note that both methods
give the same results which rules out the possibility for numerical instabilities.

As what concerns the $qg$ contribution, its sign is uniquely determined by $\mu_F/m_Q$.
For $\mu_F$ close to $m_Q$ and below, it will be positive (negative) at small (large) $\sqrt{s}$.
For $\mu_F$ larger than $m_Q$, it remains negative for any $\sqrt{s}$. The value of $\mu_R$ 
is only relevant for the normalisation.

As for $gg$ contributions, supposed to be dominant at high energies, both $\mu_F/m_Q$ and
$\mu_R/\mu_F$ matter for the sign of its contribution. For $\mu_F \simeq m_Q$, 
it monotonously increases as function $\sqrt{s}$ irrespective of $\mu_R/\mu_F$. 
For $\mu_F \simeq  2 m_Q$, the $gg$ contribution gets negative at
large $\sqrt{s}$ for $\mu_R \leq \mu_F$. For $\mu_R > \mu_F$, 
it remains always positive. Yet the sum $gg +gq$ can still become negative since,
in some cases, $-gq$ increases faster with $\sqrt{s}$. However, for $\mu_F \gtrsim m_Q$,  
the $gq$ yield is rather small and, even though it is negative, it
does not affect the increase of the cross section. 

Looking at the analytical results ~\cite{Petrelli:1997ge}, both contributions 
indeed exhibit logarithms of $\mu_F/m_Q$ multiplied by a factor function of $m_Q/\hat s$ which
induces this intricate behaviour. This issue is still an open one and may indeed be related 
to the PDF behaviour at low scales and thus may not be  specific to quarkonium production. Looking
for analogies with TMD factorisation may help us understand its origin~\cite{forthcoming-TMD-etac}.

In the case of double $J/\psi$ production, the energy dependence at one loop
seems well-behaved~\cite{Sun:2014gca}. However there is no existing public code
allowing one for more extensive studies of the scale dependence and the natural scale
is rather on the order of 6 GeV than 3 GeV. As we will discuss 
later (see section~\ref{sec:onium_Z}), 
$\Q+Z$ hadroproduction is also known at NLO and its cross section can be evaluated
down to zero $P_T$. At low $P_T$, the $K$ factor is indeed smaller than unity but the
scale is likely too large to exhibit a similar behaviour.
Another path towards a solution may also be higher-twist contributions~\cite{Alonso:1989pz}
where two gluons come from a single proton as recently re-discussed in~\cite{Motyka:2015kta,Schmidt:2018gep,Levin:2018qxa}. 
Whatever the explanation is, understanding the mechanism of low-$P_T$-quarkonium production 
remains absolutely essential for heavy-ion 
and spin studies.

\subsection{Recent developments in the COM-NRQCD phenomenology}
\label{subsec:COM_updates}

\subsubsection{$\psi$ and $\Upsilon$ hadroproduction at finite $P_T$}
\label{subsubsec:COM_3S1_NLO_PT}

One year after the first NLO CS computation of the \pt-differential cross section, 
Gong and Wang performed~\cite{Gong:2008ft} the first --partial-- NLO COM-based study
focusing only on the $S$-wave-octet states, $\so$ and $\sps$. We know now that
leaving aside the $\pj$ contributions is likely not a good approximation --except under
the complete dominance of the $\sps$ channel which we discuss later. 

Between 2010 and 2012, the progress towards complete NLO hadroproduction studies including the 3 leading CO
transitions, the effect of the FD and the polarisation information were steady. One can even add
the list of computations regarding $\gamma p$ and $e^+e^-$ collisions~\cite{Ma:2008gq,Artoisenet:2009xh,Butenschoen:2009zy,Chang:2009uj,Zhang:2009ym,Gong:2009ng,Gong:2009kp} since they add precise experimental constraints on some CO LDMEs as we shall see later in dedicated sections. Let us recall that these LDMEs cannot be computed from first principle --on the lattice for instance-- and remain to be fixed
from the data. 

To summarise, 3 groups (Hamburg, IHEP and PKU) 
carried out these NLO studies in parallel. Whereas their analyses rely on the same
hard-scattering results\footnote{A public code even exists to perform such computations for the hadroproduction case, FDCHQHP by Wan and Wang~\cite{Wan:2014vka}.}, their studies differ 
in the way the experimental data are considered ($P_T$ cut, whether or not other colliding systems are considered and
which observables are considered: yield, polarisation, FD, other quarkonia related by symmetries) and, most importantly, their conclusions differ both qualitatively and quantitatively.

It would be too tedious to go through all the aspects of such NLO studies. 
We will thus limit to explain what is particular to the NLO corrections 
and how characteristic trends --like a specific high-$P_T$ or
low-$P_T$ behaviour or a specific polarisation-- can be achieved.

\paragraph{LO COM phenomenology.}
Before addressing this, let us perform a brief reminder of what happens at LO.
If one focuses on the discussion of the $P_T$-differential cross section, 
the LO contributions for the 3 leading CO states are at $\alpha_s^3$. Their
topologies are however driven by their quantum number --not any more by that of the observed
quarkonium,  like in the CSM. In particular, the graphs where a single gluon connects to the heavy-quark line
only contribute to $\so$ states. This means that the graph of~\cf{diagram-COM-PT-c},
with an expected $P_T^{-4}$ scaling, does not couple to the $\pj$ and
$\sps$ channels. As such, both channels contribute at LO at best with a 
$P_T^{-6}$ scaling through the topologies of \cf{diagram-COM-PT-b} akin to the 
NLO CS ones of~\cf{diagram-CSM-c}.

\begin{figure}[bt!]
\centering
\subfloat[]{\includegraphics[scale=.35,draft=false]{LO-CSM-new.pdf}\label{diagram-COM-PT-a}}
\subfloat[]{\includegraphics[scale=.35,draft=false]{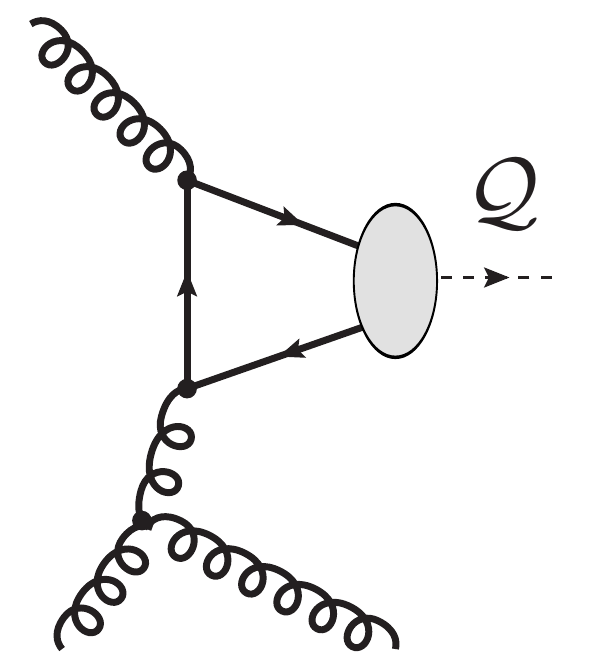}\label{diagram-COM-PT-b}}
\subfloat[]{\includegraphics[scale=.35,draft=false]{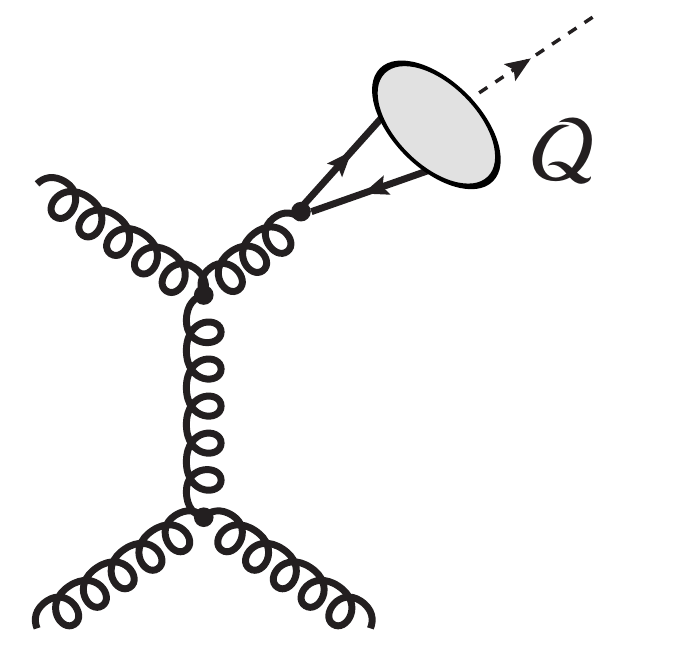}\label{diagram-COM-PT-c}}
\subfloat[]{\includegraphics[scale=.35,draft=false]{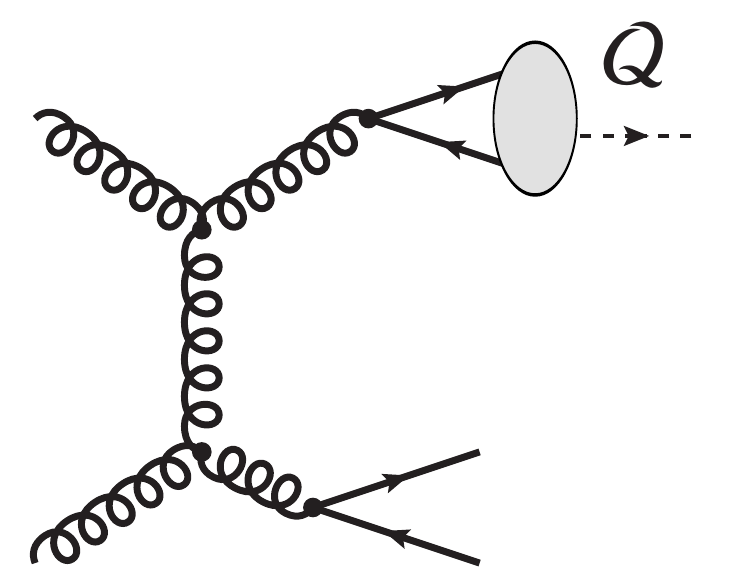}\label{diagram-COM-PT-d}}\\
\subfloat[]{\includegraphics[scale=.35,draft=false]{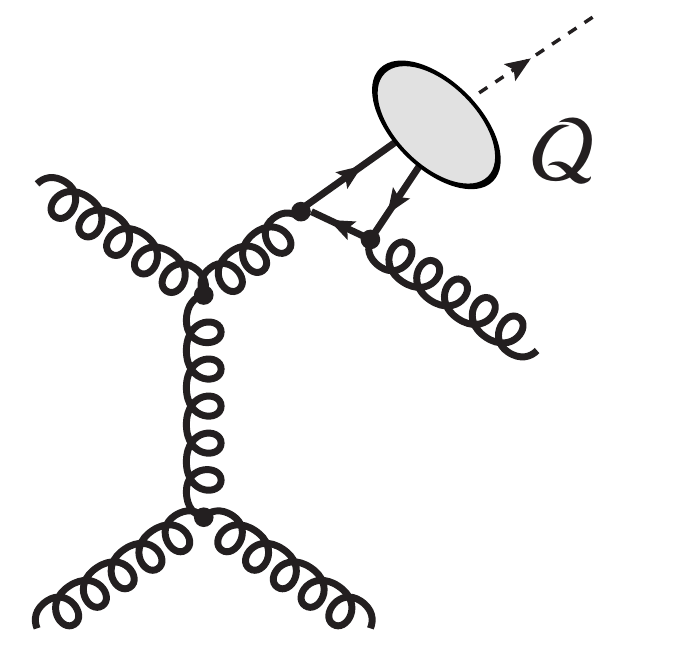}\label{diagram-COM-PT-e}}
\subfloat[]{\includegraphics[scale=.35,draft=false]{NLO-PT6-CSM.pdf}\label{diagram-COM-PT-f}}
\subfloat[]{\includegraphics[scale=.35,draft=false]{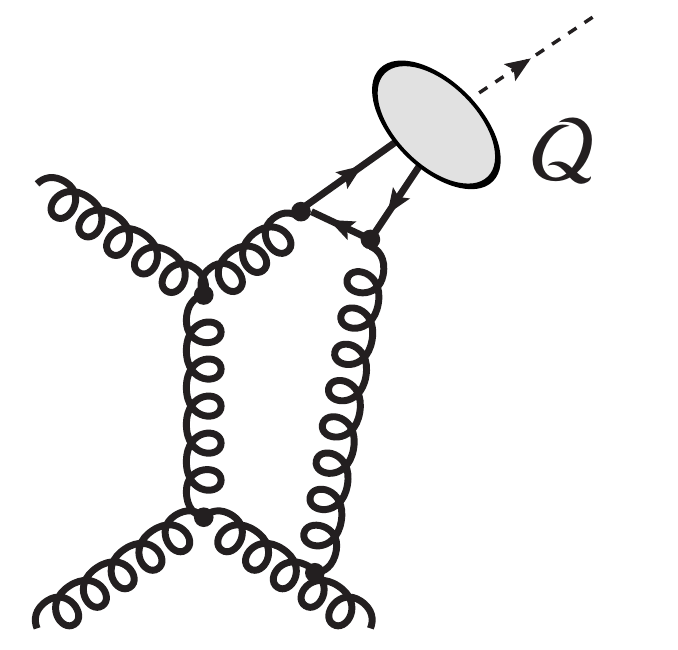}\label{diagram-COM-PT-g}}
\subfloat[]{\includegraphics[scale=.35,draft=false]{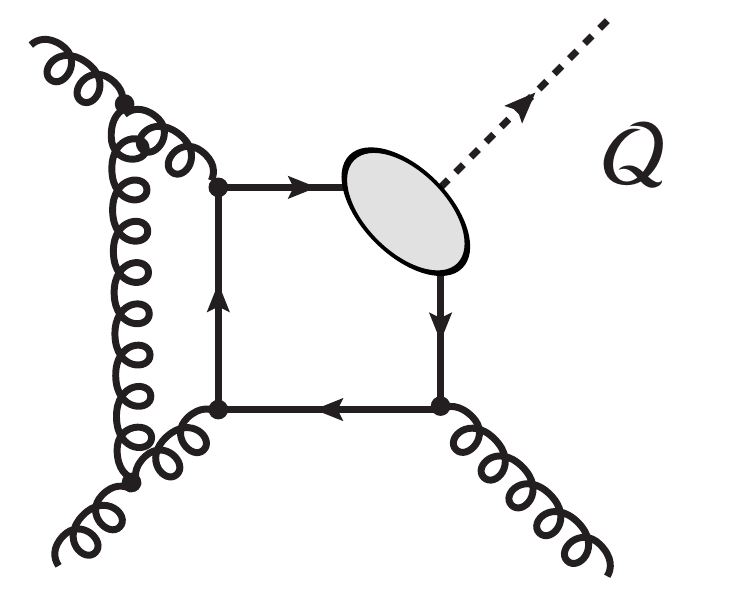}\label{diagram-COM-PT-h}}
  \caption{Representative diagrams contributing in the COM (a-c) at Born order to $i+j\to \Q+$jet  and (d-h) at one loop to $i+j\to \Q+$jet. ``jet" here refers to the unobserved parton which recoils on the $\Q$
in order to provide its finite $P_T$.
} 
\end{figure}

These well known $P_T$-scaling properties are at the basis of all the LO COM analyses, starting from the seminal
studies~\cite{Cho:1995vh,Cho:1995ce} of Cho and Leibovich in 1995 who noted that the COM contributions can be 
split in two classes in order to carry out the fit to the data. In particular, the
experimental data on the $P_T$-differential cross section of the $\psi$ and $\Upsilon$
cannot discriminate between the $\pj$ and $\sps$ contributions. These were then fit
through the combination $M_{0,r_0}=\mops + (r_0/m^2_c) \mop0$,\footnote{Because of 
the similar topologies contributing to the different CO states, it happens that
if the $\pj$ cross-section SDC are decomposed as a linear combination of the $\sps$ and $\so$ SDCs, \ie\
$\hat\sigma[\pj]=(r_0/m_Q^2)\hat\sigma[\sps]+(r_1/m_Q^2)\hat\sigma[\so]$, 
the dimensionless quantities, $r_0$ and $r_1$, depends very weakly on the kinematics. As such, fitting LDMEs using $\hat\sigma= \hat\sigma[\p0] \mop0 + \hat\sigma[\sps] \mops +\hat\sigma[\so]\mopb$ is equivalent to fitting $\hat\sigma=\hat\sigma[\sps] M_{0,r_0} +\hat\sigma[\so] M_{1,r_1}$ with $ M_{0,r_0} \equiv \mops + (r_0/m^2_c) \mop0$ and $M_{1,r_1}\equiv\mopb + (r_1/m^2_c) \mop0$. At LO, $r_1=0$ (or equivalently $\hat\sigma[\p0]$ does not contain the leading $P_T^{-4}$ topologies
of $\hat\sigma[\so]$) and inclusive cross-section fits can only fix $M_{0,r_0}$ and $\mopb$.
In what follows, we will discuss another such linear combination for the transverse cross section, thus related to the polarisation predictions, with primed $r'$. To sum up, $r$'s are essentially ratios of cross-section partial SDCs; they would  differ for different reactions (photoproduction, associated production, etc.) and in fact are not strictly constant if one probes sufficiently large phase space. However, in practice, when their variations are smaller than the theoretical and experimental uncertainties, considering them constant and fitting the $M$'s are probably more reasonable than trying to extract specific information on the LDMEs themselves.
} which happens here to be quasi independent of $P_T$.

Another well-known consequence of these different scalings is that the $\so$ 
transition is expected to be dominant as soon as $P_T$ is getting significantly 
larger than $M_\Q$. By virtue of Heavy-Quark-Spin Symmetry (HQSS) --a fundamental
symmetry of NRQCD--, Cho and Wise anticipated~\cite{Cho:1994ih} in 1994 that, when 
the gluon fragmenting in the $\so$
pair is so energetic that it essentially becomes on-shell, it can only fragment
into a transversally polarised $\Q$ (in the helicity frame). For a long time, 
the observation (or not) of a transversally polarised $J/\psi$ (and $\psi(2S)$) 
yield was considered to  be a smoking gun signal of the COM, which motivated
studies at larger and larger $P_T$.
Just before the LHC start-up, the first tangible observations against such LO COM predictions 
came from the latest CDF $\psi(2S)$ data~\cite{Aaltonen:2009dm} which were later confirmed by
LHCb ones~\cite{Aaij:2014qea}. As discussed before along the CS polarisation results, 
in fact, no measurement indicates a marked transversally polarised yield of $J/\psi$
and $\Upsilon(nS)$. The $\psi(2S)$ data are however easier to interpret since they are FD free.

To end the discussion of the LO results, let us add that the COM fits~\cite{Cho:1995ce,Beneke:1996yw,Braaten:1999qk} yield
to a $P_T^{-4}$ scaling at high $P_T$ and to a $P_T^{-6}$ scaling at lower $P_T$ and
this seems compatible with the Tevatron data up to 20 GeV. We guide the reader to~\cite{Sharma:2012dy} for one of the few LO fits using LHC data.

\paragraph{NLO COM phenomenology.}
The $\sps$  case is in fact nearly identical to that of the NLO CS corrections to
the $\eta_c$ $P_T$-differential cross section previously discussed. The gluon fragmentation in this case
indeed goes along with a gluon emission from the heavy-quark line (\cf{diagram-COM-PT-e})
which softens its $P_T$ spectrum, but remains harder than $P_T^{-6}$.
It is also important to note that $\sps$ transitions produce unpolarised $\Q$ by virtue of HQSS.

At NLO, the $\pj$ pairs can also
be produced by gluon fragmentation as shown on \cf{diagram-COM-PT-e}. The $P$-wave case is however particular
since it comes along with a differentiation of the amplitude. As such, the IR
divergences of these real emissions are no more systematically cancelled by 
that of the loop contributions (\eg\ \cf{diagram-COM-PT-g}). The remaining 
IR divergences are in fact cancelled by soft divergences associated
with the running of the $\so$ LDME.

Indeed, just as the PDF evolution satisfies an evolution equation involving
the (factorisation) scale, $\mu_F$, the LDMEs
also evolve as a function of the NRQCD scale $\mu_\Lambda$ and this evolution
comes along with IR divergences which are, for the $\so$ states, those  
cancelling those of the $\pj$ transitions. We refer 
to the seminal work of Petrelli \etal\ \cite{Petrelli:1997ge} in 1997 for details. 

In this context, it is therefore natural to find out that the normalisation of the
$P_T^{-4}$ behaviour is no more only driven by the $\so$ but also by the $\pj$ transition
whose $K$ factor is thus getting large at larger $P_T$. The first studies which took this effect
into account appeared in 2010~\cite{Ma:2010yw,Butenschoen:2010rq,Ma:2010jj}. It was 
noted by Ma \etal\ (PKU) that, for the $J/\psi$  and $\psi'$ cases, the $P_T$ dependence of the yield was sensitive to another
linear combination of LDMEs, $M_{1,r_1}=\mopb + (r_1/m^2_c) \mop0$, in addition to $M_{0,r_0}$.
  They also noted that the $J/\psi$ data could be fit with $M_{0,r_0} \gg M_{1,r_1}$ which can be realised
either by a cancellation between the $\pj$ and $\so$ LDMEs within $M_{1,r_1}$  or
by setting both these LDMEs to small values. The latter option would in turn mean that the yield would 
essentially be dominated by the $\sps$ transition, thus also possibly in agreement with the
polarisation data. We stress that since $r_1$ is found to be negative in the kinematic domain 
where the data were fit, the cancellation can be realised with both LDMEs $\pj$ and $\so$ being positive.

This is in fact in line with the fact that the $P_T$-differential data tend 
to be compatible with a scaling slightly 
softer than $P_T^{-4}$ but harder than $P_T^{-6}$ as we noted with the comparison
to the NLO CS computations. Such a behaviour is indeed similar to what one would 
expect from the $\sps$ fragmentation contributions.

Such an idea of a large $\sps$ LDMEs was pushed further by  Bodwin \etal~\cite{Bodwin:2014gia} and 
Faccioli \etal~\cite{Faccioli:2014cqa}. In the $J/\psi$ case, the global 
analysis of Butensch\"on and Kniehl (Hamburg) however
showed~\cite{Butenschoen:2010rq,Butenschoen:2011yh} that this would drastically contradict
the $B$-factories $e^+e^-$ and HERA $\gamma p$ data, whose NRQCD studies were also then promoted up to NLO 
accuracy~\cite{Ma:2008gq,Artoisenet:2009xh,Butenschoen:2009zy,Chang:2009uj,Zhang:2009ym,Gong:2009ng,Gong:2009kp}
with stringent constraints on $\sps$ and $\pj$~\cite{He:2009uf}. It would also yield to predictions badly
overshooting  the low- and mid-$P_T$ hadroproduction data. We will come back these aspects in the next sections. 

Anticipating the discussion of the $\chi_Q$, the PKU group also performed in 2010 the first
NLO study~\cite{Ma:2010vd} of $\chi_c$ production which allowed them to fully account for the FD 
effects~\cite{Ma:2010jj} in the $J/\psi$ case --the $\psi(2S)$ case was of course also accounted for. They obtained 
\eqs{
M^{J/\psi}_{0,r_0}= 7.4 \pm 1.9 \text{ (theo.} ) \pm 0.4 \text{ (fit) } 10^{-2} \text{ GeV}^2 \text{ and }
M^{J/\psi}_{1,r_1}= 0.05 \pm 0.02 \text{ (theo.} ) \pm 0.02 \text{ (fit) } 10^{-2} \text{ GeV}^2.
}
whereas the global Hamburg fit (after an approximate FD subtraction) was yielding values on the order the  
\eqs{
M^{J/\psi}_{0,r_0} \simeq  1.5  \times 10^{-2} \text{ GeV}^2 \text{ and }
M^{J/\psi}_{1,r_1} \simeq 0.4 \times 10^{-2} \text{ GeV}^2
}
notably with a negative value for $\pj$ (see \ct{tab:NLO-LDME-psi})  which solves the excess of the NRQCD predictions in 
the $z\to 1$ limit in $\gamma p$ production (see section \ref{subsubsec:COM_Psi_NLO_PT-photoprod}). Note however that the Hamburg fit 
results in a significant impact of the $P_T^{-4}$ component, via $M_{1,r_1}$, which tends to overshoot 
the LHC data at large $P_T$ --much larger $P_T$ than the Tevatron ones. A possible
solution for this excess may be the need to resum the logarithms of $P_T/m_Q$ but it is not shared
by the other groups.

NLO studies of the $\psi$ polarisation appeared in 2012\footnote{A first partial polarisation study had been done in 2008 by the IHEP group with the sole consideration of the CO $S$-wave states~\cite{Gong:2008ft}.} by the Hamburg~\cite{Butenschoen:2012px}, PKU~\cite{Chao:2012iv,Shao:2012fs} and IHEP~\cite{Gong:2012ug} groups with different interpretations, though. Whereas the Hamburg and IHEP fit results exhibit increasingly 
transverse $\psi$ yields which departs from the polarisation data for increasing $P_T$, 
the PKU group performed a fit of the Tevatron polarisation data and 
obtained a yield quasi unpolarised, still with $M_{0,r_0} \gg M_{1,r_1}$. Let us also note
that the transverse component of the yield is almost proportional to another linear combination
$M_{1,r'_1}=\mopb + (r'_1/m^2_c) \mopj$ with $r'_1$ very close to $r_1$ (both negative, \ie\ $-0.56$ vs. $-0.52$ in the central rapidity region) --which explains
that the polarisation is less discriminant than initially thought. 
If $\pj$ is negative, one thus necessarily 
generates transverse polarised $J/\psi$, hence the Hamburg fit conclusion. 
When $\mops$ is chosen to be its maximal value, the yield is unpolarised;
when $\mops$ vanishes, $\lambda_\theta$ increases from -0.25 at
$P_T \simeq 5$ GeV to 0 at $P_T > 15$ GeV at the Tevatron.

The LHC measurements which came afterwards (see~\cite{Andronic:2015wma}) just 
reinforced these observations as what
regards the trend of the $P_T$-differential cross section and of the polarisation. The cross sections 
can mostly be described by a global fit of the world $J/\psi$ data (Hamburg fit)
which is however in dramatic contradiction with the unpolarised  character of the hadroproduced yield --if not slightly longitudinal for the $J/\psi$. If one wishes to reproduce the polarisation data, one is required to drop
some cross-section measurements from the fits. Moreover, such a global approach is further challenged by $\eta_c$ data which we discuss in section~\ref{subsec:COM_1S0_NLO_PT}. A new fit by the IHEP people by Zhang~\etal~\cite{Zhang:2014ybe} was carried out using these constraints and illustrates how the picture changes with very different LDMEs, in particular a much smaller  $\mops$ --still compatible~\cite{Han:2014jya} with the allowed PKU-fit range. We will discuss later why.

\begin{table}[hbt!]
  \centering\renewcommand{\arraystretch}{1.5}
\subfloat[${J/\psi}$]{\begin{tabularx}{\textwidth}{p{0.9cm}|p{0.9cm}|p{1.6cm}|XXX|XX}
 	   & $P_{T,\rm min.}^{J/\psi}$ [GeV] & $\mspb$ [GeV$^3$] & $\mops$ [$10^{-2}$~GeV$^3$]& $\mop0/m_c^2$ [$10^{-3}$~GeV$^3$]& $\mopb$ [$10^{-3}$~GeV$^3$] & $M_{0,r_0}$   [$10^{-2}$~GeV$^3$]   & $M_{1,r_1}$ [$10^{-2}$~GeV$^3$]\\ \hline \hline
BK \cite{Butenschoen:2011yh} & \multirow{2}{*}{3} & \multirow{2}{*}{1.32} &  \multirow{2}{*}{$3.04\pm 0.35$} & \multirow{2}{*}{$-4.04 \pm 0.72$} & \multirow{2}{*}{$1.68\pm 0.46$} & \multirow{2}{*}{{\it 1.46}}& \multirow{2}{*}{{\it 0.39}} \\ \hline
IHEP \cite{Gong:2012ug} & \multirow{2}{*}{7} & \multirow{2}{*}{1.16} &  \multirow{2}{*}{$9.7\pm 0.9$} & \multirow{2}{*}{$-9.5 \pm 2.5$} & \multirow{2}{*}{$-4.6\pm 1.3$} & \multirow{2}{*}{{\it 13}}& \multirow{2}{*}{{\it 0.07}} \\\hline
IHEP \cite{Zhang:2014ybe} & \multirow{2}{*}{7} & \multirow{2}{*}{$0.65 \pm 0.41$} &  \multirow{2}{*}{$0.78\pm 0.34$} & \multirow{2}{*}{$17 \pm 5$} & \multirow{2}{*}{$10\pm 3$} & \multirow{2}{*}{{\it 7.4}}& \multirow{2}{*}{{\it 0.05}} \\\hline
IHEP \cite{Feng:2018ukp} & \multirow{2}{*}{7} & \multirow{2}{*}{1.16} &  \multirow{2}{*}{$5.66\pm 0.47$} & \multirow{2}{*}{$3.42 \pm 1.02$} & \multirow{2}{*}{$1.77\pm 0.58$} & \multirow{2}{*}{{\it 7}}& \multirow{2}{*}{{\it -0.015}} \\\hline
\multirow{2}{0.9cm}{PKU \cite{Shao:2014yta}} & \multirow{2}{*}{7} & \multirow{2}{*}{1.16}     &   {\it 7.4}  &  {\it 0} & {\it 0.5}    & \multirow{2}{*}{$7.4 \pm 1.9$} & \multirow{2}{*}{ $0.05 \pm 0.02$} \\
                   &  &     &   {\it  0 }   &  {\it 18.9}  & {\it 11.1 } & &  \\\hline
\end{tabularx}\label{tab:NLO-LDME-psi}}
\\
\subfloat[$\psi(2S)$]{\begin{tabularx}{\textwidth}{p{0.9cm}|p{1.0cm}|p{1.5cm}|XXX|XX}
 	   & $P_{T,\rm min.}^{\psi(2S)}$ [GeV] & $\mspb$ [GeV$^3$] & $\mops$ [$10^{-2}$~GeV$^3$]& $\mop0/m_b^2$ [$10^{-3}$~GeV$^3$]& $\mopb$ [$10^{-3}$~GeV$^3$] & $M_{0,r_0}$   [$10^{-2}$~GeV$^3$]   & $M_{1,r_1}$ [$10^{-2}$~GeV$^3$] \\ \hline \hline
\multirow{2}{0.9cm}{PKU \cite{Shao:2014yta}} & \multirow{2}{*}{7} & \multirow{2}{*}{0.76}     &   {\it 2}  &  {\it 0} & {\it 1.2}    & \multirow{2}{*}{$2.0 \pm 0.6$} & \multirow{2}{*}{ $0.12 \pm 0.03$} \\
                   &  &     &   {\it  0 }   &  {\it 5.1}  & {\it 4.1} & &  \\
\multirow{2}{0.9cm}{PKU \cite{Shao:2014yta}} & \multirow{2}{*}{11} & \multirow{2}{*}{0.76}     &   {\it 3.8}  &  {\it 0} & {\it 0.6}    & \multirow{2}{*}{$3.82 \pm 0.78$} & \multirow{2}{*}{ $0.059 \pm 0.029$} \\
                   &  &     &   {\it  0 }   &  {\it 9.8}  & {\it 6.1} & &  \\
\hline
IHEP \cite{Gong:2012ug} &\multirow{2}{*}{7} & \multirow{2}{*}{0.76} &  \multirow{2}{*}{$-0.01\pm 0.87$} &\multirow{2}{*}{$4.2\pm 2.4$} & \multirow{2}{*}{$3.4\pm 1.2$} & \multirow{2}{*}{{\it 1.6}}& \multirow{2}{*}{{\it 1.1}} \\
\end{tabularx}\label{tab:LDME-psi2S-NLO}}
\caption{Selection of $J/\psi$ and $\psi(2S)$ LDMEs from NLO fits using Tevatron (and LHC) data. Numbers in italic are derived from $M_{i,r_i}$ (PKU) or from $\langle {\cal O} \rangle$ (IHEP) with $r_0=3.9$ and $r_1=-0.56$. It is important to note that the $\langle {\cal O} \rangle$ values derived from $M_{i,r}$ are boundary values. Any triplet of $\langle {\cal O} \rangle$ resulting in the quoted values of $M_{i,r_i}$ for the IHEP fit are equally good.  $P_{T,\rm min.}^\Q$ indicates the minimum $P_T$ of the fit data. The CS LDMEs are not fit.}
\end{table}

So far, we only discussed the aspects of the polarisation related to the polar
anisotropy and $\lambda_\theta$. In 2018, the IHEP group performed~\cite{Feng:2018ukp} the first
NLO computation of $\lambda_{\theta\phi}$. They restated on the way that by virtue
of $P$-party invariance $\lambda_{\theta\phi}$ should vanish in symmetric rapidity intervals -- a feature
which could be used to improve experimental analyses. They also updated their previous LDME determination with a joint fit of the yield and the 3 polarisation parameters. The resulting LDMEs are drastically different than those of their first fit~\cite{Gong:2012ug}. For instance they are now all positive (this has implication \eg\ for associated production with a photon, see section~\ref{sec:psi-gamma}) but are not compatible with $\eta_c$ data. \ct{tab:NLO-LDME-psi} displays the LDME values of the all aforementioned fits.

\begin{table}[hbt!]
  \centering\renewcommand{\arraystretch}{1.5}
\subfloat[$\Upsilon(1S)$]{\begin{tabularx}{\textwidth}{p{0.9cm}|p{1.0cm}|p{1.5cm}|XXX|XX}
 	   & $P_{T,\rm min.}^\Upsilon$ [GeV] & $\mspb$ [GeV$^3$] & $\mops$ [$10^{-2}$~GeV$^3$]& $\mop0/m_b^2$ [$10^{-2}$~GeV$^3$]& $\mopb$ [$10^{-2}$~GeV$^3$] & $M_{0,r_0}$   [$10^{-2}$~GeV$^3$]   & $M_{1,r_1}$ [$10^{-2}$~GeV$^3$]\\ \hline
\multirow{2}{0.9cm}{PKU \cite{Han:2014kxa}} & \multirow{2}{*}{15} & \multirow{2}{*}{9.28}     &   {\it 13.7}  &  {\it 0} & {\it 1.17}    & \multirow{2}{*}{$13.70 \pm 1.11$} & \multirow{2}{*}{ $1.17 \pm 0.02$} \\
                   &  &     &   {\it  0 }   &  {\it 3.61}  & {\it 3.04} & &  \\
\hline
IHEP \cite{Feng:2015wka} & \multirow{2}{*}{8} & \multirow{2}{*}{9.28} &  \multirow{2}{*}{$11.6\pm 2.61$} & \multirow{2}{*}{$-0.49 \pm 0.59$} & \multirow{2}{*}{$0.47\pm 0.41$} & \multirow{2}{*}{{\it 9.74}}& \multirow{2}{*}{{\it 0.72}} \\
\end{tabularx}}
\\
\subfloat[$\Upsilon(2S)$]{\begin{tabularx}{\textwidth}{p{0.9cm}|p{1.0cm}|p{1.5cm}|XXX|XX}
 	   & $P_{T,\rm min.}^\Upsilon$ [GeV] & $\mspb$ [GeV$^3$] & $\mops$ [GeV$^3$]& $\mop0/m_b^2$ [$10^{-2}$~GeV$^3$]& $\mopb$ [$10^{-2}$~GeV$^3$] & $M_{0,r_0}$   [$10^{-2}$~GeV$^3$]   & $M_{1,r_1}$ [$10^{-2}$~GeV$^3$] \\ \hline
\multirow{2}{0.9cm}{PKU \cite{Han:2014kxa}} & \multirow{2}{*}{15} & \multirow{2}{*}{4.63}     &   {\it 6.07}  &  {\it 0} & {\it 1.08}    & \multirow{2}{*}{$6.07 \pm 1.08$} & \multirow{2}{*}{ $1.08 \pm 0.20$} \\
                   &  &     &   {\it  0 }   &  {\it 1.6}  & {\it 1.91} & &  \\
\hline
IHEP \cite{Feng:2015wka} &\multirow{2}{*}{8} & \multirow{2}{*}{4.63} &  \multirow{2}{*}{$-0.53\pm 2.31$} &\multirow{2}{*}{$0.28\pm0.52$} & \multirow{2}{*}{$2.94\pm 0.4$} & \multirow{2}{*}{{\it 0.47}}& \multirow{2}{*}{{\it 2.79}} \\
\end{tabularx}}
\\
\subfloat[$\Upsilon(3S)$]{\begin{tabularx}{\textwidth}{p{0.9cm}|p{1.0cm}|p{1.5cm}|XXX|XX}
	   & $P_{T,\rm min.}^\Upsilon$ [GeV] & $\mspb$ [GeV$^3$] & $\mops$ [$10^{-2}$~GeV$^3$]& $\mop0/m_b^2$ [$10^{-2}$~GeV$^3$]& $\mopb$ [$10^{-2}$~GeV$^3$] & $M_{0,r_0}$   [$10^{-2}$~GeV$^3$]   & $M_{1,r_1}$ [$10^{-2}$~GeV$^3$]\\ \hline
\multirow{2}{0.9cm}{PKU \cite{Han:2014kxa}} & \multirow{2}{*}{15} & \multirow{2}{*}{3.54}     &   {\it 2.83}  &  {\it 0} & {\it 0.83}    & \multirow{2}{*}{$2.83 \pm 0.07$} & \multirow{2}{*}{ $0.83 \pm 0.02$} \\
                   &  &     &   {\it  0 }   &  {\it 0.74}  & {\it 1.21} & &  \\
\hline
IHEP \cite{Feng:2015wka} &8 & 3.54 &  $-0.18\pm1.4$ &$-0.01\pm 0.30$ & $1.52\pm 0.33$ & {\it -0.22}& {\it 1.53} \\
\end{tabularx}}
\caption{Selection of $\Upsilon(nS)$ LDMEs from 2 NLO fits using LHC and Tevatron data. Numbers in italic are derived from $M_{i,r_i}$ (PKU) or from $\langle {\cal O} \rangle$ (IHEP) with $r_0=3.8$ and $r_1=-0.52$. It is important to note that the $\langle {\cal O} \rangle$ values derived from $M_{i,r}$ are boundary values. Any triplet of $\langle {\cal O} \rangle$ resulting in the quoted values of $M_{i,r_i}$ for the IHEP fit are equally good. For the IHEP fit, we have chosen to quote only the values for which the FD treatment is similar to that of the PKU fit. $P_{T,\rm min.}^\Upsilon$ indicates the minimum $P_T$ of the fit data. The CS LDMEs are not fit~\cite{Han:2014kxa,Feng:2015wka}.}\label{tab:LDME-Upsilon-NLO}
\end{table}

As what concerns the $\psi(2S)$, 2 NLO studies exists by the IHEP~\cite{Gong:2012ug} and PKU~\cite{Shao:2014yta} groups. In fact, they were also carried out to feed in the prompt $J/\psi$ fits. \ct{tab:LDME-psi2S-NLO} gathers the corresponding LDME values. Two sets have been released by the PKU group with different $P_T$ cuts. Whereas $M_0$ are similar, $M_1$ are very different. The PKU fit with the larger  $P_{T,\rm min.}$  cut, with a very small $M_0$, 
shows the best account of the polarisation data, which however is far from being perfect in particular for the CDF~\cite{Abulencia:2007us} and LHCb~\cite{Aaij:2014qea} data.  Let us also mention that Sun and Zhang~\cite{Sun:2015pia} proposed a new {\it modus operandi} to fit the 
$J/\psi$ and $\psi(2S)$ yield and polarisation. They indeed proposed to first determine, from the polarisation data, the LDME ratio $\mops/\mop0$ 
and the fix $\mopb$ from the yields. Doing so, they could achieve a reasonable description of both the yields and polarisations at central rapidities. The CDF polarisation data however cannot be described.

As for the $\Upsilon$, the situation is less clear-cut because of the more 
intricate FD pattern and less precise data. 3 complete fits exist, those of the IHEP group~\cite{Gong:2013qka,Feng:2015wka} and that of the PKU group~\cite{Han:2014kxa}.
These fits result in slightly transversely polarised $\Upsilon(nS)$, which tends to agree with the CMS data~\cite{Chatrchyan:2012woa}. The fits are made with a $P_{T,\rm min.}^\Upsilon$ cut of respectively 
8 and 15 GeV. The corresponding LDMEs are collected in \ct{tab:LDME-Upsilon-NLO} .

Overall, the trend of the LHC data (in the region which is fit) seems to be reproduced, but not that of 
the Tevatron CDF polarisation data~\cite{CDF:2011ag}. Besides, 
a quick look at the corresponding $P_T$-differential cross sections displayed in the
papers reveals that the predicted cross sections overshoot the data at low $P_T$ by factors as large as 5 at \eg\ $P_T \simeq 5$ GeV. NRQCD factorisation may indeed not applicable in this region which would explain why fixed-order COM computations cannot reproduce such data. However, no solid argument to explain such a large necessary reduction  of these predicted differential yields at low $P_T$ has so far been proposed.  Yet, as we have seen in section \ref{subsec:CSM_NLO_PT}, the sole NLO CS contributions do reproduce them down to $P_T=0$ --and thus the total amount of produced $\Upsilon$. The possible inapplicability of collinear factorisation in this region is thus maybe a little too expediently invoked. In addition, let us stress that all the data which generate tensions for the $J/\psi$ case, in particular the $e^+e^-$ and $\gamma p$ ones, are not available for the $\Upsilon$. $P_T$-integrated cross sections will be discussed in section \ref{subsec:COM_NLO_tot}.

\subsubsection{$\psi$ photoproduction}
\label{subsubsec:COM_Psi_NLO_PT-photoprod} 

The phenomenology of the COM in $J/\psi$ photoproduction started as early as in 1996. Cacciari and Kr\"amer~\cite{Cacciari:1996dg}, Ko \etal~\cite{Ko:1996xw} and then Amundson \etal~\cite{Amundson:1996ik} performed the first studies of the impact of the COM in these reactions and compared their evaluations to preliminary HERA data\footnote{See~\cite{Aid:1996dn} for the first published data.}. At the time, Cho and Leibovich had made~\cite{Cho:1995vh,Cho:1995ce} a first determination of 
the 3 leading CO LDMEs (see above) at LO in $\alphaS$ using the Tevatron data. These were used to perform predictions for photoproduction in 2 regimes. 

The first is that of forward $J/\psi$ production where the inelasticity $z$ is close to unity and the $J/\psi$ $P_T$ is very small. In this kinematical region, the inclusive production is expected to be dominated by the LO $\alpha\alphaS$ COM contributions --a LO graph is shown on \cf{gammag-oniumg-LO-COM}. The corresponding cross section is then found to be proportional to $M^{J/\psi}_{0,7}$. In the 3 aforementioned studies, it was noted that the existing HERA measurements were significantly below the CO predictions using the CO LDMEs from the hadroproduction fit. Some important caveats were however noted. In this region, large diffractive contributions can indeed be present. Yet, subtracting them would increase the discrepancy. In addition, because of nonperturbative emissions during the hadronisation of the CO pairs, the average $z$, $\langle z \rangle$, for these processes may rather be $1 - {\cal O}(v^2)$ than 1 and thus may affect the $z$ dependence away from 1.  

\begin{figure}[hbt!]
\centering
\subfloat[]{\includegraphics[scale=.33,draft=false]{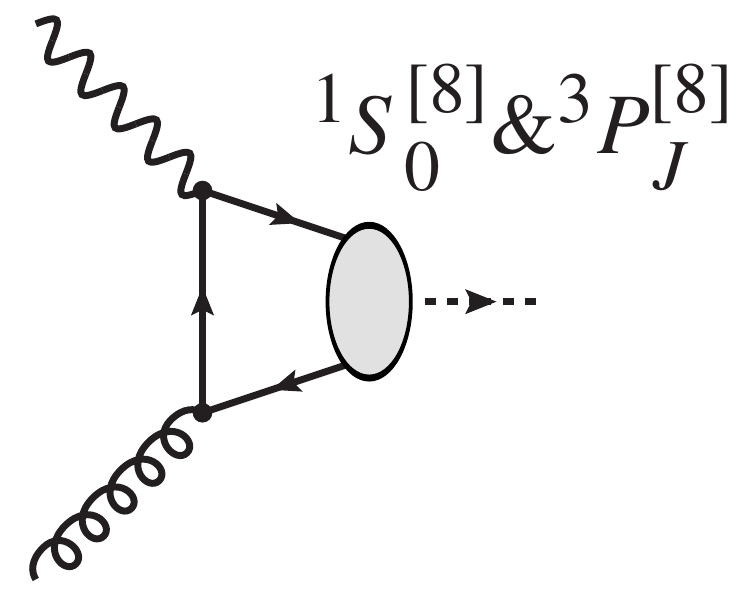}\label{gammag-oniumg-LO-COM}}
\subfloat[]{\includegraphics[scale=.33,draft=false]{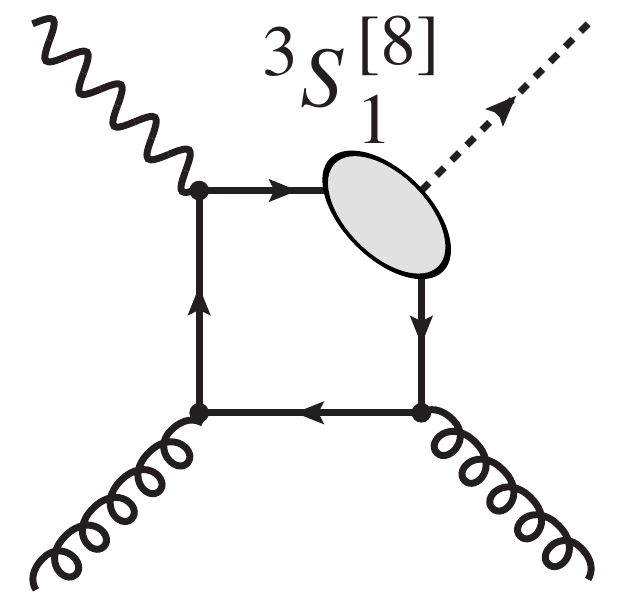}\label{gammag-oniumg-NLO-PT8-COM}}
\subfloat[]{\includegraphics[scale=.33,draft=false]{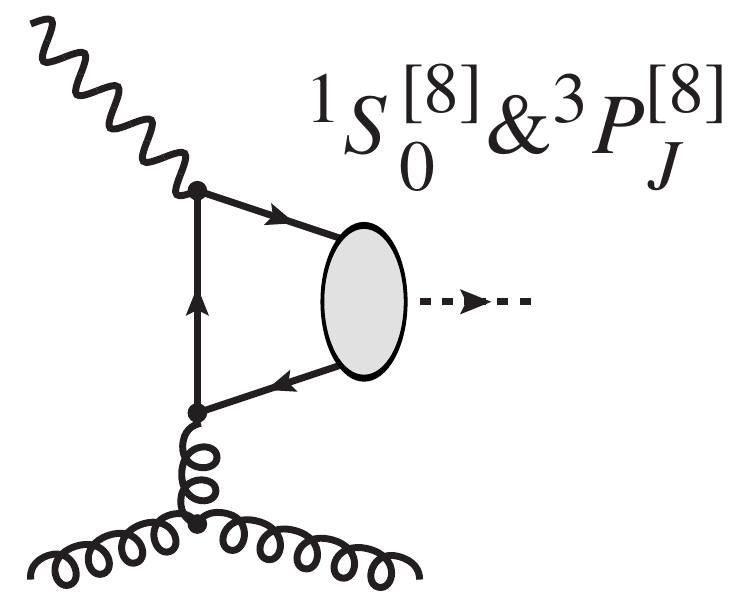}\label{gammag-oniumg-NLO-PT6-COM}}
\subfloat[]{\includegraphics[scale=.33,draft=false]{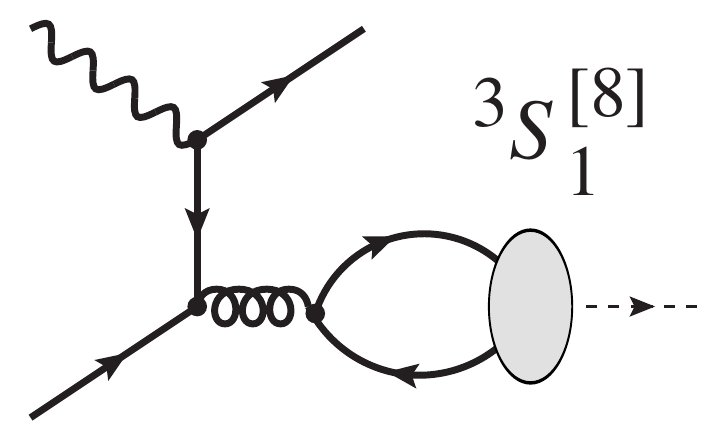}\label{gammaq-oniumq-frag-COM}}
\caption{Representative diagrams contributing to $^3S_1$ direct photoproduction via 
CO channels at orders $\alpha\alphaS$ (a) and  $\alpha\alphaS^2$ (b,c,d). 
The quark and antiquark attached to the ellipsis are taken as on-shell
and their relative velocity $v$ is set to zero.}
\label{diagrams-COM-photoproduction}
\end{figure}

The second regime is that of the production of $J/\psi$ at finite $P_T$ --in practice $P_T> 1$ GeV-- and for $z>0.3$ while excluding the high-$z$ region. In this region, the direct-photon contributions should be dominant. Typical LO $\alpha\alphaS^2$ COM  graphs contributing to this region are shown on \cf{gammag-oniumg-NLO-PT8-COM}-\ref{gammaq-oniumq-frag-COM}. Like in the first regime, the CO predictions using the CO LDMEs from the hadroproduction fit were found~\cite{Cacciari:1996dg,Ko:1996xw} to be significantly above the data. 

The objective of this section is to report on the evolution of our understanding of this apparent breakdown of the NRQCD LDME universality between photoproduction and hadroproduction, and in particular the impact the NLO corrections computed in the 2010's.

\paragraph{$P_T$-differential cross section.} Let us start with the discussion of what we referred above to as the second regime since it is the one for which pQCD and NRQCD should apply better. In addition, let us focus on the description of the $P_T$-differential cross section for which NLO CS computations are available since 1995. As we discussed in section \ref{sec:CSM_photoproduction}, when these are accounted for, the HERA data and the CS predictions are in the same ballpark, despite some observed deviations. In this context, a significant CO contribution could easily generate the excess which was observed as early as in 1996. 
Following the aforementioned analyses, other LO ones appeared and we guide the reader to the following reviews~\cite{Kramer:2001hh,Brambilla:2004wf} where these are discussed. Polarisation observables were subsequently predicted~\cite{Beneke:1998re}. By analogy with hadroproduction, CO fragmentation channels were studied~\cite{Godbole:1995ie,Kniehl:1997fv,Kniehl:1997gh} but were shown to be relevant only for $P_T > 10$~GeV, thus outside the region covered by the HERA data. Until the advent of NLO analyses including the CO contributions, the situation could well be summarised by~\cf{fig:H1_2002-dsigdpt-photoprod_LO_COM} where both LO CS and CO contributions are added together in the blue band. The agreement seems to be quite good\footnote{bearing in mind that the CS NLO is also close to the data and is not shown here.}. However, the CO LDMEs used --more precisely $M_{0,3.5}^{J/\psi}$ following the notation introduced in the previous section-- for this plot are 5 to 15 times smaller~\cite{Adloff:2002ex} than those obtained from the Tevatron hadroproduction fits\footnote{except if parton shower effects in hadroproduction are accounted for, which results in LDMEs 10 times smaller than those obtained via collinear factorisation. See also \cite{Kniehl:1998qy}.}~\cite{Kramer:2001hh}. One further notes two additional trends: the blue band tends to overshoot the data at low $P_T$ and to move further up at larger $z$, which is not surprising since the LO $\alpha \alphaS$ COM contributions lie at $\langle z \rangle \simeq 1$. We will come back to this when discussing the $z$ dependence.

\begin{figure}[hbt!]
\centering
\subfloat[NRQCD @ LO]{\includegraphics[width=0.66\textwidth]{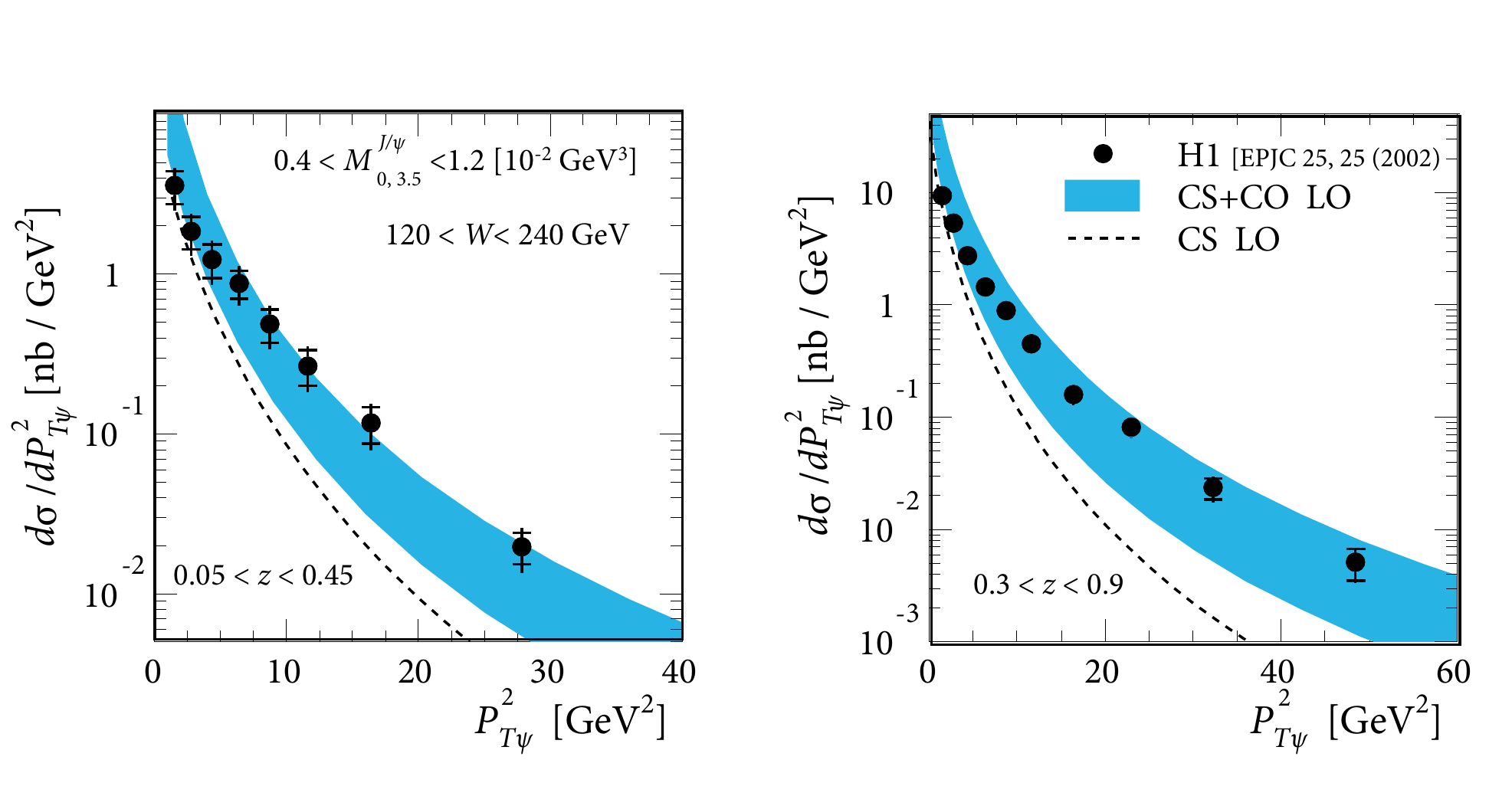}\label{fig:H1_2002-dsigdpt-photoprod_LO_COM}}
\subfloat[NRQCD @ NLO]{\includegraphics[width=0.33\textwidth]{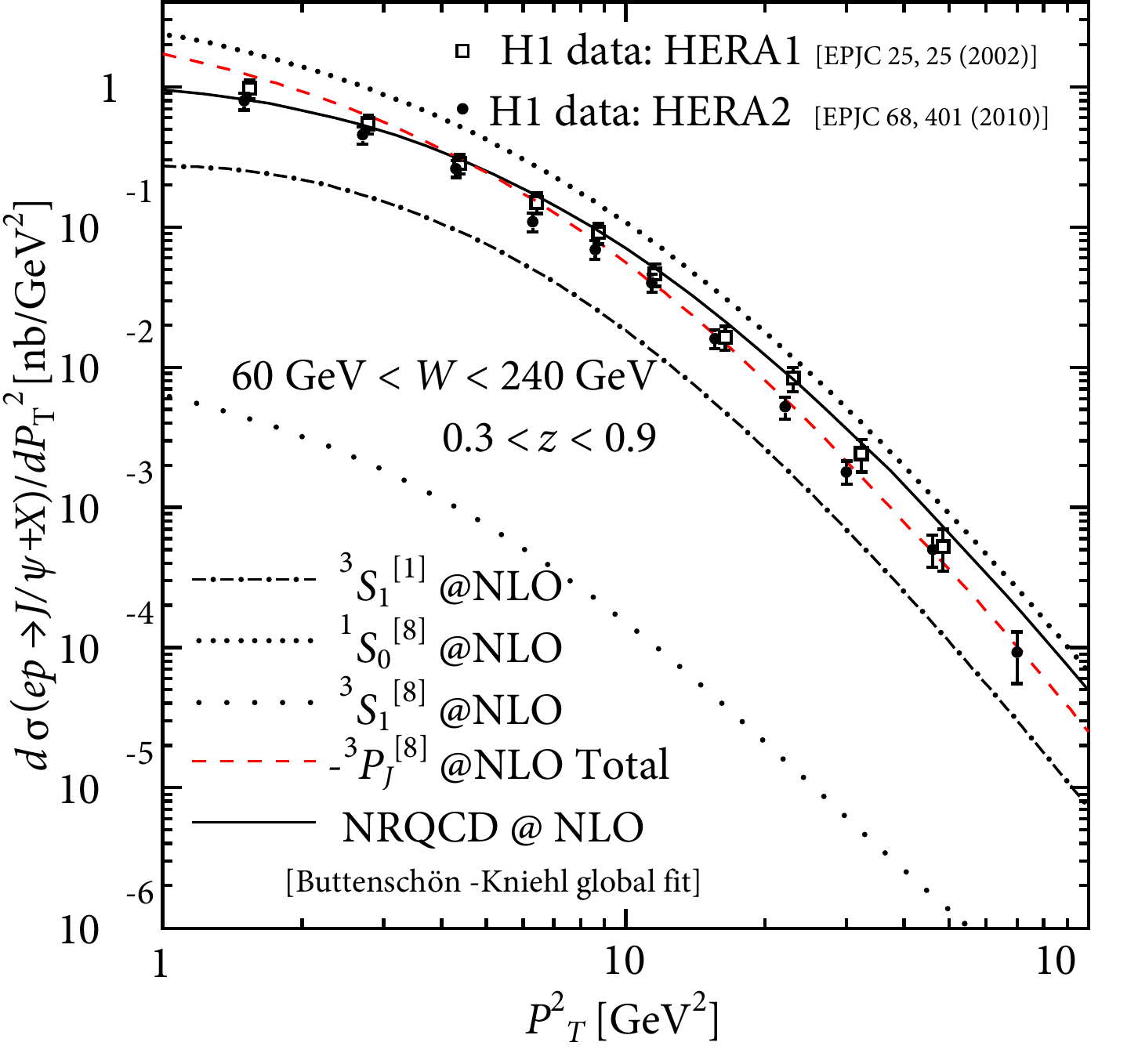}\label{fig:h1_pt_CO_states-NLO-BK}}
\caption{ $P_T^2$-differential cross section for $\gamma p \to J/\psi X$ as measured (a) by H1 at HERA1 for $120 < W_{\gamma p} < 240 $~GeV for 2 $z$ ranges (left \& right) compared to LO CS and CS+CO computations along the lines of \cite{Kramer:2001hh} with $0.4\times 10^{-2} < M_{0,3.5}^{J/\psi} < 1.2 \times 10^{-2}$; (b) by H1 HERA1 \& 2 compared to NLO CS \& CO channel contributions using the central value of NLO globally fit LDMEs~\cite{Butenschoen:2010rq} (no theory uncertainties are shown). Adapted (a) from~\cite{Adloff:2002ex} and (b) from~\cite{Butenschoen:2010rq}.}
\label{fig:COM-photoproduction-dpt}
\end{figure}

In 2009, Butensch\"on and Kniehl performed the first complete photoproduction analysis of the COM contributions at NLO, including for the first time the contribution from the $\pj$ states to any NLO $2\to 2$ NRQCD computations~\cite{Butenschoen:2009zy}. This allowed them to perform in 2010 a first global fit~\cite{Butenschoen:2010rq} of photo- and hadroproduction data. \cf{fig:h1_pt_CO_states-NLO-BK} shows the various contributions to photoproduction of the 3 leading CO states fit to reproduce both HERA and Tevatron data. What appeared to be impossible at LO, namely to account for both hadro- and photoproduction data with the same LDMEs, became possible at NLO. 

The reason why this is so lies in the subtle interplay between the $\pj$ and $\sb$ contributions. As we discussed in the previous section, these interfere and a NLO fit benefits a much wider parameter space to describe the hadroproduction data. In particular, the data allow for negative $\mop0$ yielding a rather small value for $M_{0,r_0}$  and a fair description of the low $P_T$ $pp$ and $\gamma p$ data. 

There is however a triple cost for this. First, at large $P_T$, both $\sb$ and $\pj$ dominate hadroproduction.  The resulting large-$P_T$ spectrum is thus harder ($\propto P_T^{-4}$) than the data and one needs to invoke the resummation of $\log(P_T)$ to justify such a discrepancy. Second, the hadroproduction yield increasingly becomes transversely polarised at large $P_T$ at variance with the Tevatron and LHC data. The $\sb$ yield is transversely polarised and the $\pj$ one is longitudinal but comes with a negative weight at large $P_T$.  
Third, the $\eta_c$ data cannot be reproduced~\cite{Butenschoen:2014dra}
since $\sa$ is too large (see section~\ref{subsec:COM_1S0_NLO_PT}). 
In other words, a ``global'' fit can only be realised
at the cost of dropping some hadroproduction data, namely those on the polarisation and the $\eta_c$ and, to a lesser extent, the large $P_T$ data. Yet, one can then obtain a good description of low $P_T$ and $P_T$-integrated hadroproduction and photoproduction data as well as to some extent the $e^+e^-$ data (see section~\ref{subsec:COM_Psi_NLO_ee}). 
The PKU and IHEP fits simply cannot account for these data.

We note that none of these studies explicitly took into account the complexity of the FD effects. FD from excited states are not expected to qualitatively change these conclusions. However, there remains a big question mark with regards a possible dominant non-prompt contribution at the largest $P_T$ hinted by the H1 simulations (see section~\ref{sec:FD}). Even though corresponding data will probably never be available, a fresh phenomenological look at this issue may be expedient~\cite{forthcoming-photoproduction}.

\paragraph{$z$-differential cross section.} As we wrote in the introduction of this section, 
forward $J/\psi$ photoproduction ($z\to1$) has been the object of several theoretical works in the mid 1990's. 
At the time, this was believed to be a clean probe of the COM as its LO contribution should lie at $z=1$.
As we wrote above, data were found to be smaller than the NRQCD expectations based on early Tevatron LDME fits~\cite{Cho:1995vh,Cho:1995ce}. The 
community started to wonder about the reliability of NRQCD computations in this region since its velocity expansion would eventually break down there~\cite{Beneke:1997qw}. Yet, NLO $\alpha \alphaS^2$ corrections at the end point were computed by Petrelli \etal\ in 1997~\cite{Maltoni:1997pt}. The study of the resummation of large logarithm of $1-z$ was carried out by Fleming \etal\ \cite{Fleming:2006cd} in 2006 which relies on the introduction of a shape function which is supposed to be the same than in $e^+ e^-$ collisions where similar end-point divergences occur (see section~\ref{subsec:COM_Psi_NLO_ee}). They showed that such a resummation can significantly affect the $z$ distribution at large $z$ but that the (LO) LDMEs needed to describe the data remained one order of magnitude smaller that those fit (at LO) to the Tevatron data.

Nevertheless, the study of the $z$-dependence of the yield of photoproduced $J/\psi$ remained an object of attention
in particular when a $P_T$ cut was imposed on the data to make sure that the diffractive contribution could safely
be neglected. Based on LO considerations, it was generally admitted that $d\sigma/d z$ should significantly increase with $z$ if the COM was to be dominant. \cf{fig:H1_1996-dsigdz-photoprod_LO_COM_NLO_CSM} shows a comparison between the first published H1 data~\cite{Aid:1996dn} and 
predictions by Cacciari and Kr\"amer with $\mops=\mop0/m_c^2=10^{-2}$~GeV$^2$ (thus $ M_{3.5,r_0}=4.5 \times 10^{-2}$ GeV$^2$)~\cite{Cacciari:1996dg} along with the NLO CS computation also by Kr\"amer~\cite{Kramer:1995nb}.

\begin{figure}[hbt!]
\centering
\subfloat[CO @ LO \& CSM @ NLO]{\includegraphics[width=0.42\textwidth]{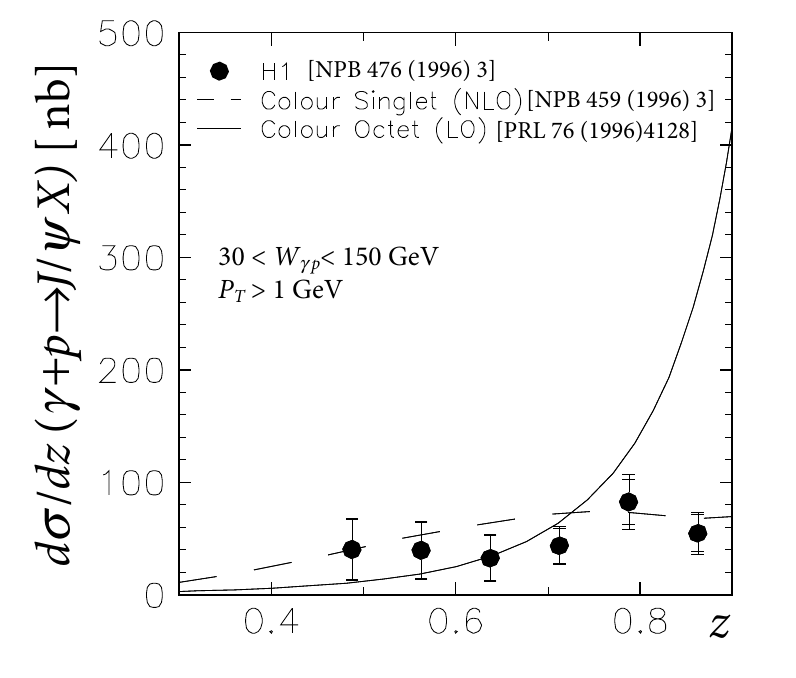}\label{fig:H1_1996-dsigdz-photoprod_LO_COM_NLO_CSM}}
\subfloat[NRQCD @ NLO]{\includegraphics[width=0.37\textwidth]{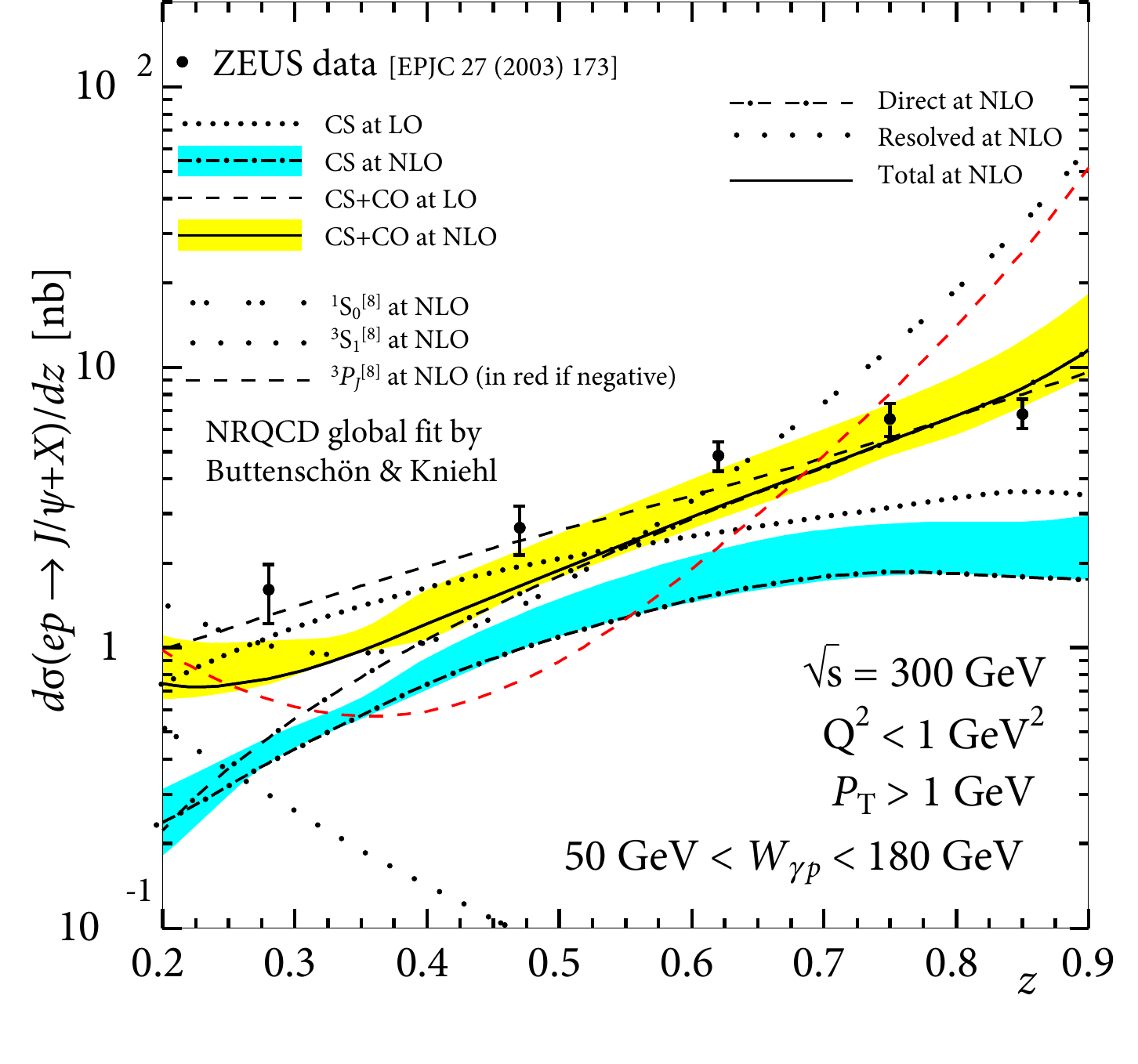}\label{fig:ZEUS_2003-dsigdz-photoprod_NLO_COM_NLO_CSM}}
\caption{$z$-differential cross section for $\gamma p \to J/\psi X$ as measured by (a) H1 at HERA1~\cite{Aid:1996dn} for $30 < W_{\gamma p} < 150 $~GeV compared to NLO CS and LO CO computations with $M_{0,3.5}^{J/\psi} = 4.5 \times 10^{-2}$~GeV$^2$ [See~\cite{Kramer:1995nb,Cacciari:1996dg} for details on the theory curves], by (b) ZEUS at HERA1 compared to NLO CS \& CO channel contributions using NLO globally fit LDMEs~\cite{Butenschoen:2010rq}. The plots also show the contributions of the resolved-photon contributions. Adapted (a) from~\cite{Aid:1996dn} and (b) from~\cite{Butenschoen:2012qh}.}
\label{fig:COM-photoproduction-dz}
\end{figure}

Butensch\"on and Kniehl also considered the $z$-dependence in their global NLO analysis~\cite{Butenschoen:2010rq}.
Like for the $P_T$-differential cross section, a good agreement is obtained as compared to the LO situation with the total NLO NRQCD cross section shown by solid line and yellow band of \cf{fig:ZEUS_2003-dsigdz-photoprod_NLO_COM_NLO_CSM} lying very near the data. The reason for the absence of a peak above 0.7 is due to the cancellation between
the $\sa$ and $\pj$ contributions owing to the negative value of the $\p0$ LDME. This good agreement also {\it de facto} comes along with severe issues in hadroproduction, \ie\ wrong polarisation predictions, the impossibility to account for the $\eta_c$ data and some difficulties to account for data at the large $P_T$ accessed at the LHC, and again issues in $e^+e^-$ annihilation. We will come back to these later. As above, the PKU and IHEP fits simply cannot account for these data.

\paragraph{Polarisation.} In 2012, Butensch\"on and Kniehl advanced further the idea of a global NRQCD NLO analysis by computing the QCD corrections to the yield polarisation of photoproduced $J/\psi$. Two of their plots are shown on~\cf{fig:NLO_NRQCD_pol_photoproduction_vs_data}. It is not clear owing to the uncertainties on both in the theory and the data whether the inclusion of CO improves or not the agreement with the HERA polarisation data. More precise data from an EIC would be highly desirable.

\begin{figure}[hbt!]
\centering
\subfloat[]{\includegraphics[scale=.4]{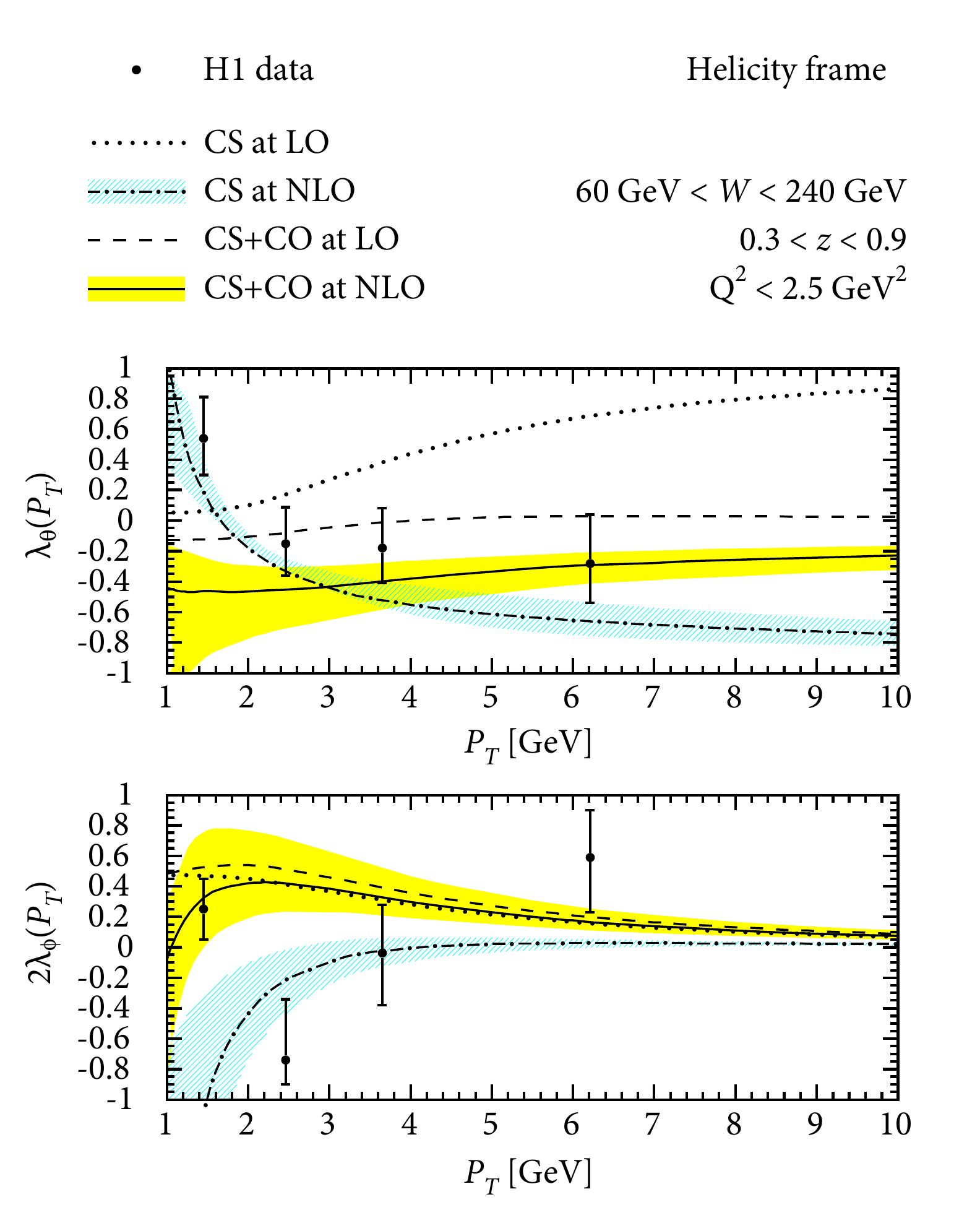}\label{fig:butenschoen_photopol_h1_pt_heli}}
\subfloat[]{\includegraphics[scale=.4]{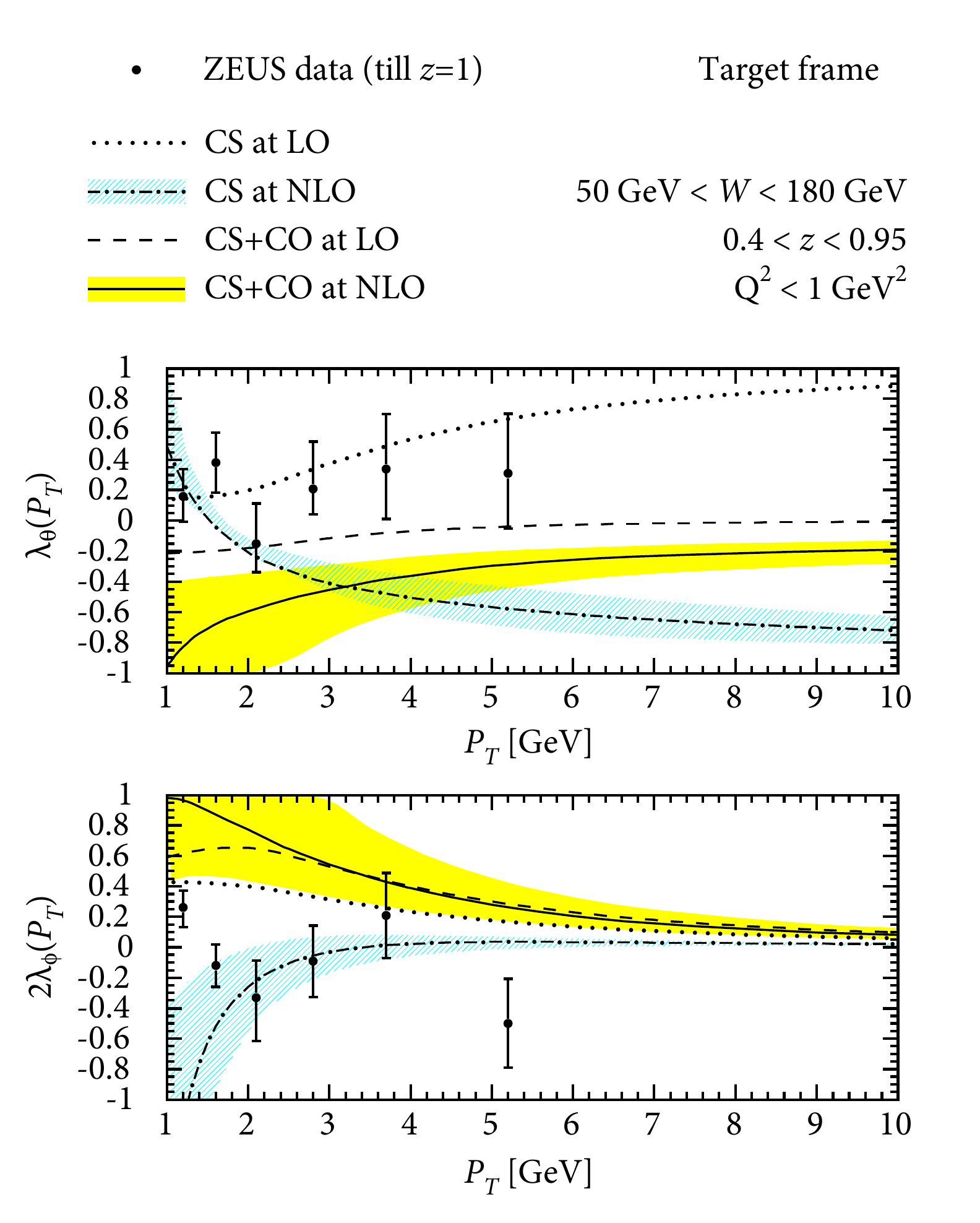}\label{fig:butenschoen_photopol_zeus_pt95_target}}
\caption{$P_T$-differential polarisation parameters for photoproduced $J/\psi$ predicted by the CSM at LO (dotted line) and NLO (dot-dashed line and blue band)  and by NRQCD at LO (dashed line) and NLO (solid line and yellow band) computed for the kinematics of the (a) H1 HERA2~\cite{Aaron:2010gz} and (b) ZEUS HERA1~\cite{Chekanov:2009ad} data and compared to them. Note that the ZEUS data extend up to $z=1$ where diffractive events can be dominant.
See ~\cite{Butenschoen:2011ks} for the theory parameters used for the curves. Figures adapted from~\cite{Butenschoen:2011ks}.}
\label{fig:NLO_NRQCD_pol_photoproduction_vs_data}
\end{figure}

\subsubsection{$\psi$ production in $e^+e^-$ annihilation}
\label{subsec:COM_Psi_NLO_ee} 

\paragraph{$B$ factories.}

As early as 1995, Braaten and Chen~\cite{Braaten:1995ez} proposed to look for a CO signal via an excess of $J/\psi$ production at large
momenta, near the end point, in $e^+e^-$ annihilation. In fact, this expected excess is akin to that at $z\to 1$ in photoproduction
and similar doubts about the applicability of NRQCD in this limit have also been raised. Yet, a crucial advantage is the absence of a possible diffractive backgrounds contrary to the photoproduction case. 

A first LO NRQCD study was performed by Yuan~\etal\ in 1996~\cite{Yuan:1996ep} in which the relative contribution of the CS and CO contributions were compared using the LO CO LDMEs available at the time. Then, a first attempt to compare with experimental data was done~\cite{Yuan:1997sn}. As we emphasised in section \ref{sec:CSM_eeproduction}, for a proper data-theory comparison, it is crucial that the yield associated with additional charm be properly substracted. This could only be done with good accuracy by Belle in 2009 which found $\sigma(J/\psi+X_{\text{non } c \bar c})=0.43 \pm 0.09 \pm 0.09$~pb. Earlier conclusions based on more inclusive cross sections are likely misleading because of the large measured difference between $\sigma(J/\psi+X)$ and $\sigma(J/\psi+X_{\text{non } c \bar c})$. The evaluation which we discussed here apply to $\sigma(J/\psi+X_{\text{non } c \bar c})$ and $\sigma(J/\psi+ c \bar c+X)$ will be addressed in section~\ref{sec:psi-cc} . 

As far as the LO CO cross section are concerned at $\sqrt{s}=10.6$~GeV, it is helpful to quote the computed cross section: 
\eqs{
\sigma(J/\psi+X_{\text{non } c \bar c})=\left [ 11 \frac{\mops}{\text{GeV}^3} + 18  \frac{\mop0}{\text{GeV}^5} \right] \text{pb},
}
for $M_\psi=2 m_c$, $m=1.55$ GeV and $\alphaS(M_\psi)=0.245$. To fix the idea if one takes $M_{0,3.5}^{J/\psi} = 4.5 \times 10^{-2}$~GeV$^2$~\cite{Kramer:2001hh} as we did for the LO photoproduction discussion and which is among the lowest found values from LO hadroproduction studies, one gets~\cite{Zhang:2009ym}
\eqs{
\sigma(J/\psi+X_{\text{non } c \bar c}) \simeq 0.5 \text{ pb},
}
which is already above the Belle experimental measurement. Since the CS yield is on the same order, the addition of the CO generate a clear excess over the data. In addition, the predicted enhancement at the end-point is clearly absent in the data. However, it was shown in 2003 by Fleming~\etal\  that the resummation of large $\log(E-E_{\rm max})$ with SCET~\cite{Fleming:2003gt} could explain the momentum distribution of the data with the introduction of a so-called shape function which should however be universal and thus similar than in photoproduction.

In this context, the first NLO COM study was performed in 2009 by Zhang \etal~\cite{Zhang:2009ym} who found, for the same parameters as above : 
\eqs{
\sigma(J/\psi+X_{\text{non } c \bar c})=\left [ 21 \frac{\mops}{\text{GeV}^3} + 35  \frac{\mop0}{\text{GeV}^5} \right] \text{pb}.
}
For a detailed discussion of the scale dependence, see~\cite{Zhang:2009ym}. The $K$ factors are thus close to 2 and the resulting cross section using similar LDMEs would increase to reach about 1 pb, thus more than twice larger than the Belle measurement even without including the CS contributions.  This lead Zhang \etal\ to derive an upper limit on the CO LDMEs as : 
\eqs{
M_{0,4.0}^{J/\psi} < (2.0 \pm 0.6) \times 10^{-2}\text{ GeV}^2.
\label{eq:LMDE_bound_ee_Zhang}
}
It should however be clear that this bound was derived by assuming a vanishing contribution of the CS yield, whereas the latter already saturates the Belle measurement.  Instead, we suggest the following upper limit 
\eqs{
M_{0,4.0}^{J/\psi} < 5  \times 10^{-3}\text{ GeV}^2
\label{eq:LMDE_bound_ee_JPL}
}
assuming the CS NLO cross section with $v^2$ relativistic corrections and QED ISR effects\footnote{So far, such ISR effects have not been studied for the CO contributions. Here, we made the hypothesis that they were small. Should the associated correction factor be larger than unity, the upper bound on $M_{0,4.0}^{J/\psi}$ would be tighter.} to be $0.65\pm 0.2$~pb~\cite{Ma:2008gq,Gong:2009kp,He:2009uf,Jia:2009np,Shao:2014rwa} and taking the 1-$\sigma$ range for the CSM value and the data. It is 3.7 times lower\footnote{2.8 if one considers the uncertainty in \ce{eq:LMDE_bound_ee_Zhang}.} than the bound quoted by Zhang~\etal. We find it more fair since we are not aware of any justification to set the CS contribution to zero.

Clearly, the Hamburg fit (with the FD subtracted) with $\mops=(3.04\pm 0.35)\times 10^{-2}$ GeV$^3$ and
$\mop0=(-9.08 \pm 1.61)\times 10^{-3}$ GeV$^3$, thus $M^{J/\psi}_{0,4.0}=(1.43\pm 0.64)\times 10^{-2}$ GeV$^3$, despite of being 5 times smaller than that of the PKU fit, does not comply with the latter bound. It barely complies with that of Zhang \etal. Indeed, in their study, Butensch\"on and Kniehl ignored~\cite{Butenschoen:2011yh} both the effect of the NLO and relativistic corrections and therefore considered a CS yield whose central value was barely 0.25 pb.

Although inclusive $\psi(2S)$ production in $e^+e^-$ annhiliation was studied by Belle in 2002~\cite{Abe:2002rb}, the $\psi(2S)+X_{\text{non } c \bar c}$ yield could not be measured. Since $P_T$-differential $\psi(2S)$ photoproduction measurements also do not exist, global fits are currently impossible. In this context, a possible future measurement of  $\psi(2S)+X_{\text{non } c \bar c}$ production by Belle-II~\cite{Kou:2018nap} would be extremely instructive and would certainly allow one to derive tight constraints on the $\psi(2S)$ LDMEs.

\paragraph{$\gamma\gamma$ scatterings at LEP.} As discussed in section \ref{sec:CSM_eeproduction}, inclusive $J/\psi$ production at LEP in $\gamma \gamma$ scatterings was only measured by DELPHI~\cite{Abdallah:2003du} at low $P_T$, with a very small number of events and the $P_T$-differential measurement was not normalised. Whereas, based on this data sample confronted to a LO NRQCD analysis, Klasen \etal\ claimed~\cite{Klasen:2001cu} in 2002 evidence for COM, the corresponding NLO analysis using the LDME globally fit by Butensch\"on and Kniehl predicts~\cite{Butenschoen:2011yh} cross sections several times below the DELPHI data. This comes from a cancellation between the $\sa$ and $\pj$ contributions owing to the negative value of $\mop0$ obtained from the global fit. We recall that such a negative value is necessary to obtain a good description of the low-$P_T$ hadroproduction data and to damp down the peak at large $z$ in photoproduction. However, it yields~\cite{Li:2014ava} negative cross sections for $J/\psi+\gamma$ at large $P_T$ (see section \ref{sec:psi-gamma})

\subsubsection{$\psi$ leptoproduction}
\label{subsec:COM_Psi_NLO_leptoprod} 

 Leptoproduction is probably the last system for which inclusive quarkonium production has not  yet been evaluated at NLO within NRQCD. A couple of LO NRQCD studies have been carried out in the past~\cite{Fleming:1997fq,Yuan:2000cn,Kniehl:2001tk}. 
Sun and Zhang performed an up-to-date LO analysis~\cite{Sun:2017nly} in 2017 emphasising potential issues in the definition of the correct hadronic tensor when cuts are imposed. They performed their LO analysis using NLO fit LDMEs from the PKU~ \cite{Chao:2012iv}, Hamburg~\cite{Butenschoen:2011yh}
 and IHEP~\cite{Zhang:2014ybe} groups, which may be arguable in terms of the coherence of the analysis. Yet, one can outline some trends which should be preserved in a full NLO analysis. 

The NRQCD results generally describe better the $P_T^\star$ dependence than the NLO CSM~\cite{Sun:2017wxk} but systematically overshoot the data at small $P_T^\star$ and increasing $z$. As expected, the issue is less marked for the Hamburg fit with the same cancellation between the
$\sa$ and $\pj$ channels than for photoproduction. As for the $Q^2$  and $W$ dependences, the Hamburg fit clearly provides the best description and the PKU one, the worse. The rapidity dependences ($y_\psi$ and $y_\psi^\star$) display similar features as the $z$ distribution with marked discrepancies for the PKU and IHEP fits at backward (negative) $y_\psi$ and at the largest (positive) $y_\psi^\star$. This is easily understood if one notes that $z=e^{y_\psi^\star} {\sqrt{P_T^{\star 2}+M_\psi^2}}/{W}$ and that the largest $y_\psi^\star$ correspond to the largest $z$. To go further in the interpretation, a first NLO NRQCD study is clearly awaited for as well as potential EIC data.

\subsubsection{$\eta_c$ hadroproduction at finite $P_T$ and its unexpected impact on the $J/\psi$ phenomenology}
\label{subsec:COM_1S0_NLO_PT}

Another important piece came in the puzzle in 2014: the LHCb $\eta_c$-production data
for $P_T>6.5$~GeV~\cite{Aaij:2014bga}. Not only, this data set was shown to be nearly perfectly described,
at both 7 and 8 TeV, by the CS NLO computations but it was quickly realised that these
$\eta_c$ data were posing stringent constraints on the size of the 
$J/\psi$ LDME $^1S_0^{[8]}$.

Strangely enough, the relevance of the connection between the $J/\psi$ and the $\eta_c$ LDMEs --beyond
the obvious CS one with the same value of $|R(0)|^2$ for both spin-triplet and spin-singlet $S$ states-- had
indeed been overlooked since the mid 2000's. The interest of $\eta_c$ hadroproduction studies was for instance clearly stated in~\cite{Brambilla:2004wf} written in 2004. If one restricts to the leading CO LDMEs, the HQSS of NRQCD indeed implies that
\eqs{
 \langle\mathcal{O}^{\etac}(\bigl.^3\! S_1^{[8]})\rangle=&{\langle\mathcal{O}^{\jpsi}(\bigl.^1\! S_0^{[8]})\rangle},\\
3\langle\mathcal{O}^{\etac}(\bigl.^1\! S_0^{[8]})\rangle=&{\langle\mathcal{O}^{\jpsi}(\bigl.^3\! S_1^{[8]})\rangle},\\
\langle\mathcal{O}^{\etac}(\bigl.^1\! P_1^{[8]})\rangle=& \frac{3}{2J+1} {\langle\mathcal{O}^{\jpsi}(\bigl.^3\! P_J^{[8]})\rangle},\\
}
up to ${\cal O}(v^2)$ corrections.

Along the same lines as discussed above, explicit NLO $(\alpha_s^4)$ computations~\cite{Han:2014jya,Butenschoen:2014dra,Zhang:2014ybe} of the 
short-distance coefficients (SDC) in the LHC kinematics 
show that the sole $\pseudos$ and $\so$ states 
are relevant to predict the $\eta_c$ $P_T$-differential cross section -- 
unless unrealistically large LDMEs are allowed 
for the other transitions. The discussion of the results can thus be simplified by only 
considering these transitions. At NLO, we have seen that the CS $\pseudos$ contributions
saturate the data. The $\so$ contribution
will scale like $P_T^{-4}$ and it happens that its natural normalisation is
not reduced by any unexpected factor. As such, the data are easily
overshot if $\langle\mathcal{O}^{\etac}(\bigl.^3\! S_1^{[8]})\rangle$ is not small.
In turn, this constrains $\langle\mathcal{O}^{\jpsi}(\bigl.^1\! S_0^{[8]})\rangle$.
In fact, it severely constrains it. By fitting the LHCb data --and choosing
a CS LDME based on the $\eta_c\to \gamma \gamma$ branching, \ie\ a little lower 
than what is expected by HQSS -- the Hamburg group reported 
\eqs{
\langle\mathcal{O}^{\etac}(\bigl.^3\! S_1^{[8]})\rangle = 3.3 \pm 2.3 \times 10^{-3} \text{ GeV}^3
}
and thus a value of $\langle\mathcal{O}^{\jpsi}(\bigl.^1\! S_0^{[8]})\rangle$ that is one order of magnitude
lower than their extraction $3.0  \pm 0.35 \times 10^{-2}$~GeV$^3$, and even more when compared to 
those of the IHEP group $9.7 \pm 0.9 \times 10^{-2}$~GeV$^3$
and of the Bodwin~\etal\ fit, driven by the $\sps$ dominance, $9.9 \pm 2.2 \times 10^{-2}$~GeV$^3$.
The PKU group instead derived \cite{Han:2014jya} a conservative upper limit by neglecting
the --however dominant-- CS yield and found out 
\eqs{
0< \langle\mathcal{O}^{\etac}(\bigl.^3\! S_1^{[8]})\rangle \lesssim 1.5 \times 10^{-2} \text{ GeV}^3
}
This constraint further shrunk the uncertainties of their polarisation range without 
affecting the differential cross section since $M_{0,r_0}$ and $M_{1,r_1}$ can be left unchanged
in most of the relevant kinematical ranges owing to the small variations of $r_0$ and $r_1$. The effect
on the polarisation for central rapidities where $r_1 \simeq r'_1$ is clear: $M_{1,r'_1}\simeq M_{1,r_1}$ remains constant but, since $\sps$ decreases, the yield becomes slightly more transverse, still
within the original uncertainty band of their pre-$\eta_c$ fit. At forward rapidities, it has 
the opposite effect and the polarisation gets slightly more longitudinal. We note that 
a very small value of $\langle\mathcal{O}^{\jpsi}(\bigl.^1\! S_0^{[8]})\rangle$ below $10^{-3}$ still remains
within their band and is not excluded. 

It goes without saying that neglecting the CS yield is 
a drastic and likely unjustified way to assess the theoretical uncertainties.
Indeed, if one believes that these computations are so unreliable as to disregard 
their NLO uncertainties or if one even drops HQSS, the whole NRQCD edifice falls apart, 
at least as long as its  phenomenology is concerned
since HQSS enters not only here, but in polarisation computations as well as in the fits
involving the $\pj$ states.
As such, we believe that a more fair upper value is 
\eqs{\label{eq:LMDE_bound_etac_JPL}
\langle\mathcal{O}^{\etac}(\bigl.^3\! S_1^{[8]})\rangle \lesssim 5 \times 10^{-3} \text{ GeV}^3.
}
We note on the way that, for $\langle\mathcal{O}^{\jpsi}(\bigl.^3\! P_0^{[8]})\rangle<0$, the resulting constraint on 
$\langle\mathcal{O}^{\jpsi}(\bigl.^1\! S_0^{[8]})$ is less strong
than that from $e^+e^-$ data (see \ce{eq:LMDE_bound_ee_JPL}) but becomes stronger otherwise. If $\langle\mathcal{O}^{\jpsi}(\bigl.^3\! P_0^{[8]})\rangle$ is negligible, both constraints quasi coincide.

As a conclusion, a global description of the world $J/\psi$ data is not possible as it cannot
describe the polarisation data\footnote{and to some extent the $e^+e^-$ B-factory and $\gamma\gamma$ LEP data (see above)}, it is even more dramatic when $\eta_c$ data are considered. Once
believed to be a solution to describe the {\it sole} hadroproduction data --yield and polarisation
altogether--, the elegant and --maybe too-- simple solution of the dominance of the $\sps$ contributions
is completely annihilated by the constraints set in by the $\eta_c$ data with
$\langle\mathcal{O}^{\jpsi}(\bigl.^1\! S_0^{[8]})\rangle \lesssim 5 \times 10^{-3}$ GeV$^3$.

As for now, similar conclusions cannot be drawn for the $\psi(2S)$ case even in the absence of
FD since constraints from $\gamma p$ and $e^+e^-$ data are lacking. In this context, 
we proposed~\cite{Lansberg:2017ozx} in 2017 to study $\eta_c(2S)$ production 
which we demonstrated to be feasible very soon at the  LHC. Indeed, \cfs{fig:3-CO} show the 
projected statistical uncertainties expected using the 3 existing $\psip$ NLO 
fits~\cite{Shao:2014yta,Gong:2012ug,Bodwin:2015iua} similar to the $J/\psi$ fits discussed above.
Clearly, the $\etacp$ cross section is measurable up to $P_T=20$ GeV with a competitive precision, and significantly further in $P_T$ if other decay channels can be used as already done~\cite{Aaij:2017tzn} for the nonprompt sample.

\begin{figure*}[hbt!]
\centering
\subfloat{\includegraphics[width=0.33\textwidth,draft=false]{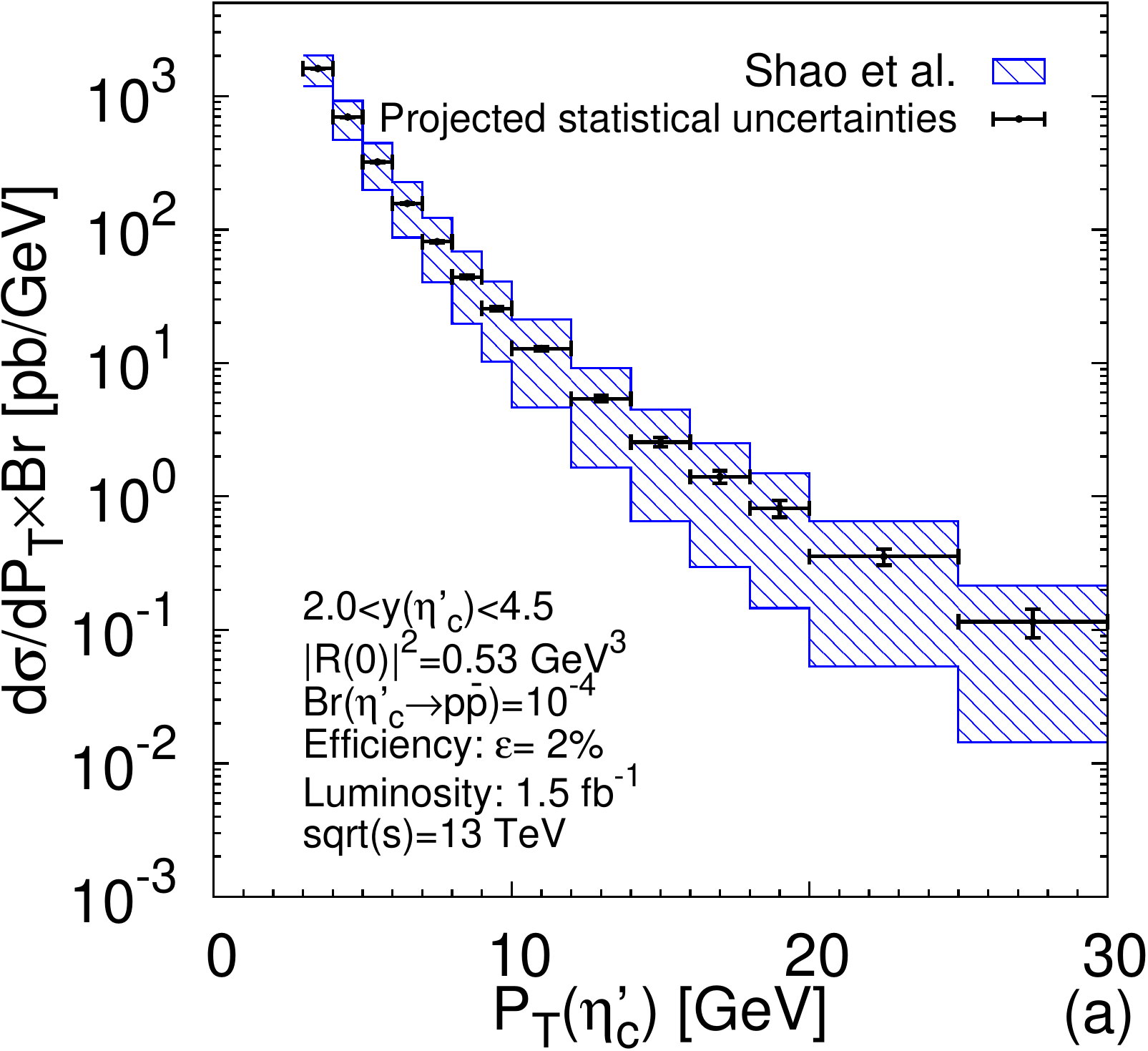}\label{fig:Shao}}
\subfloat{\includegraphics[width=0.33\textwidth,draft=false]{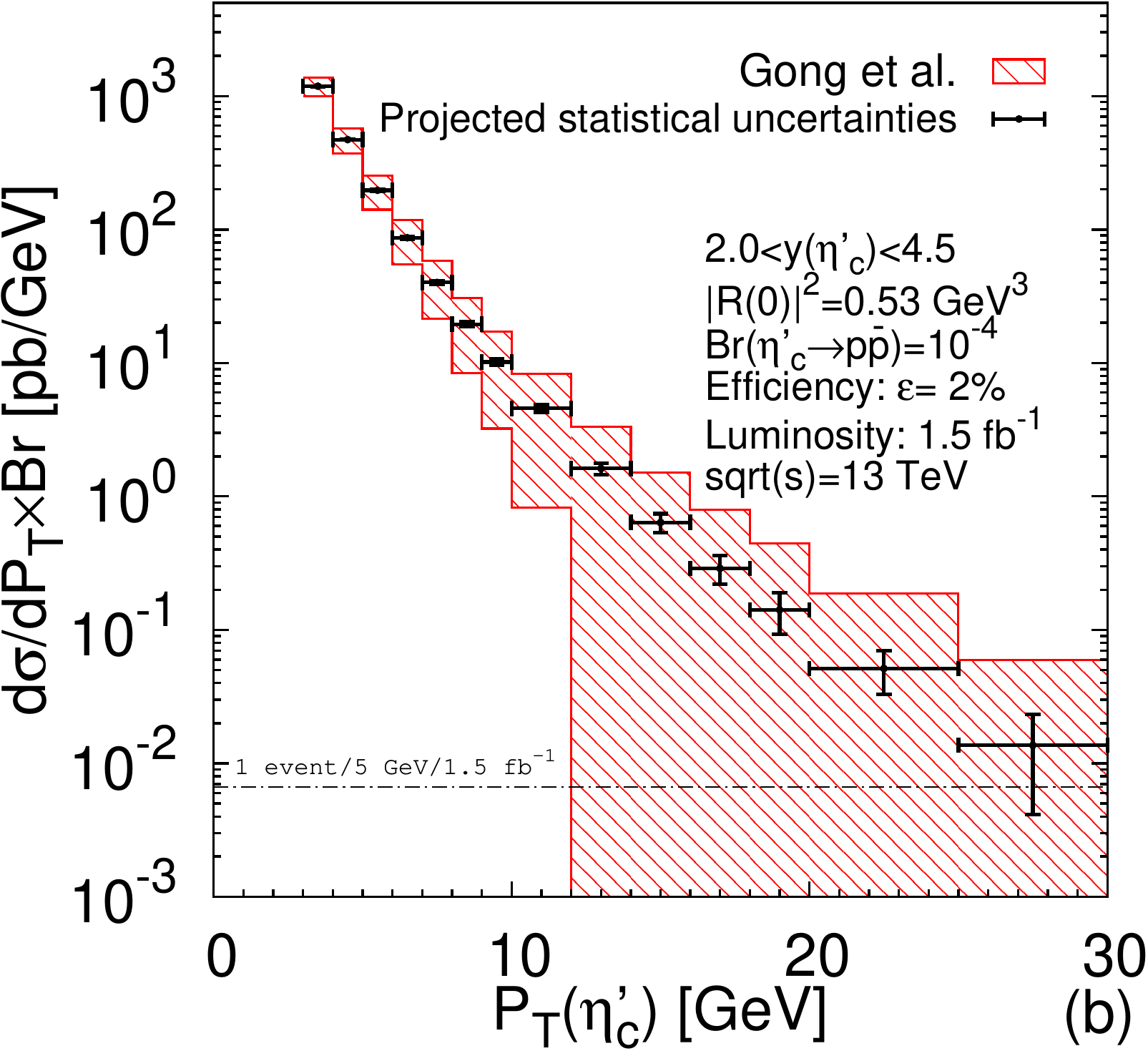}\label{fig:Gong}}
\subfloat{\includegraphics[width=0.33\textwidth,draft=false]{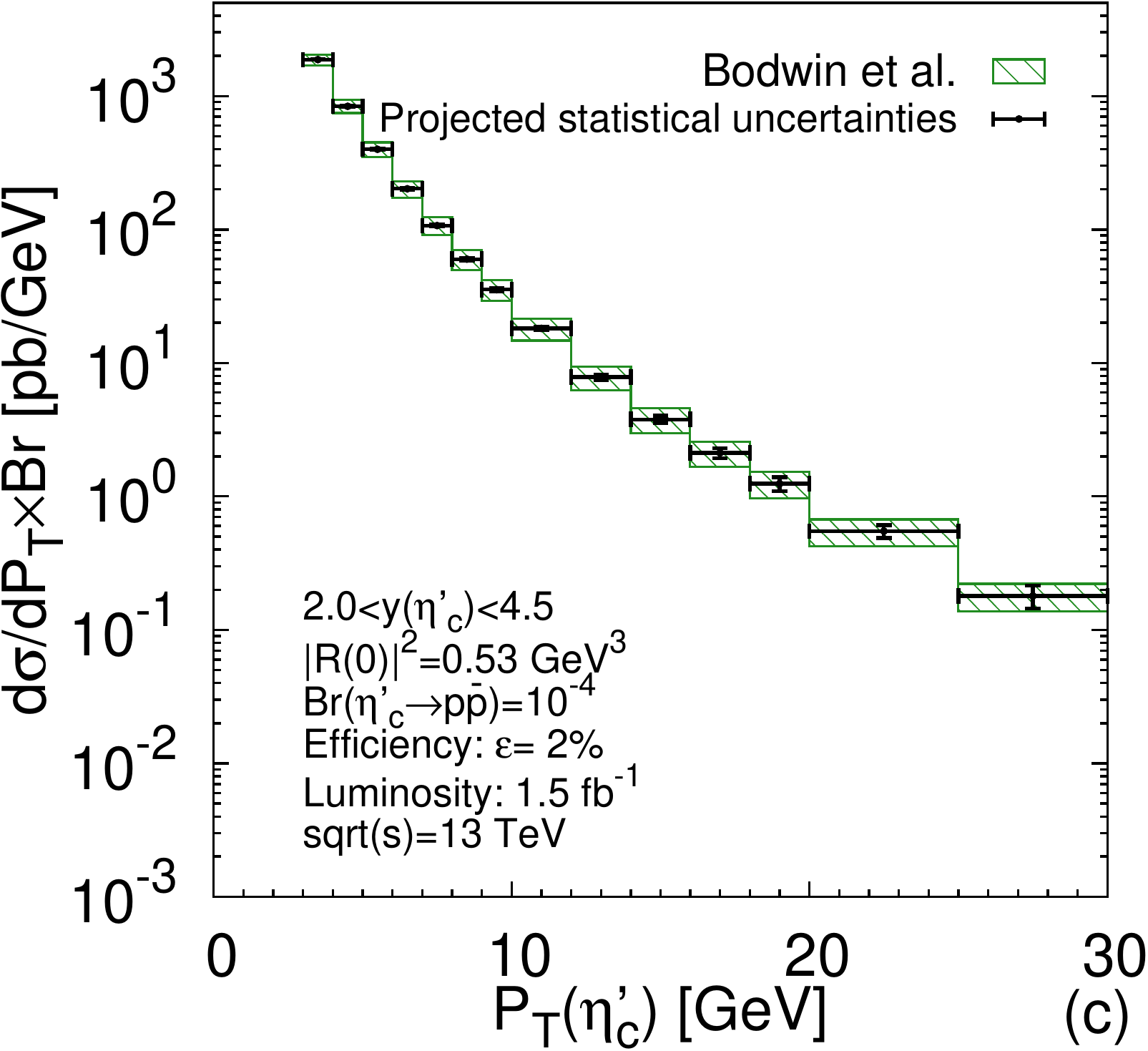}\label{fig:Bodwin}}
\caption{Differential-$P_T$ cross section for $\etacp$ production times ${\cal B}(\etacp \to p \bar p)$ 
for the 3 ranges of $\coetacp$ (a,b,c) along the 
projected statistical uncertainties using the central theoretical values in each cases, with
an assumed efficiency of 2\% and the luminosity collected so far by LHCb at $\sqrt{s}=13$~TeV, 1.5~fb$^{-1}$.
Taken from \cite{Lansberg:2017ozx}.
\label{fig:3-CO}}
\end{figure*}

\subsubsection{$\chi_Q$ production at NLO at finite $P_T$}
\label{subsec:COM_3PJ_NLO_PT}

We have not previously discussed the NLO corrections to $\chi_Q$ in the CSM by itself since it is a well known
fact that IR divergences from NLO soft-gluon emissions arise in the CS contributions 
because the differentiation of the amplitude owing to the vanishing of the wave function 
at the origin. NRQCD naturally cures this issue via the $\so$ transitions
whose LDME evolution also goes  with IR divergences with an opposite sign which cancel 
the CS $P$-wave ones. It is in fact the same mechanism as the one involved in the cancellation 
of the IR divergences associated with $\pj$ production at NLO.

However, in the present case, it happens between  CO and CS contributions. These 
mix and the trade off is essentially driven by the NRQCD scale $\mu_\Lambda$ on 
which the final result should not depend. Yet, the theoretical ratio of the CO vs the CS yield
does depend on it. If $\mu_\Lambda$ was to be artificially taken extremely small, the CO contributions
would become extremely small.  

Similarly to the case of the $\pj$ transition, HQSS also enters the play as 
it relates the CS LDMEs of the different $\chi_{QJ}$ such as
\eqs{
{\langle\mathcal{O}^{\chi_{QJ}}(\bigl.^3\hspace{-0.5mm}P_J^{[1]})\rangle}=
\frac{(2J+1)3 N_C}{2\pi} |R'_P(0)|^2.
}
As what concerns the CO states, the situation is a little simpler as 
one expects a single leading LDME, namely $\so$. At LO, the CS contributions arise 
via graphs scaling like $P_T^{-6}$ (\cf{diagram-COM-PT-b}) and the CO ones 
via gluon fragmentation (\cf{diagram-COM-PT-c}).

At NLO, the gluon fragmentation followed by CS transitions naturally sets in. As we mentioned 
above, Ma \etal\  performed the first NLO study~\cite{Ma:2010vd} of $\chi_c$.
They essentially tackled the description of $R_{\chi_c}=\sigma_{\chi_{c2}}/\sigma_{\chi_{c1}}$
whose experimental value does not agree with $5/3$ which is the expected value
from the $\so$ contribution owing to the HQSS prediction dating back
to Cho and Leibovich. 

At variance with the CO channels, the CS $\pones$ contributions show a harder spectrum than 
the $\ptwos$ ones which gives a handle in describing $R_{\chi_c}$ via the ratio $r_{nP}$ of 
the CO and CS LDMEs (defined in the $\bar{MS}$ scheme and for $\mu_\Lambda=m_c$~\cite{Ma:2010vd})
\eqs{
r_{1P}=\frac{\mopbchiQ}{\mspzerochiQ/m^2_q}.
} They obtained, for the $\chi_c$, $r_{1P}=0.27 \pm 0.06$ using the 
CDF data~\cite{Abulencia:2007bra}. Similar values are obtained~\cite{Shao:2014fca} at mid and high $P_T$
with the LHCb~\cite{LHCb:2012ac} and CMS~\cite{Chatrchyan:2012ub} data. 
In 2016, Zhang \etal~\cite{Jia:2014jfa} performed an exhaustive comparison with the existing data from the Tevatron and the LHC. We refer to this nice survey for more details on the $\chi_c$ NRQCD studies. We however note that 
the $\chi_{c0}$ case has never been addressed although its yield has been measured by LHCb~\cite{Aaij:2013dja}.

Overall, the comparison with the LHC data following such fits are good, including the 
absolute $P_T$-differential cross sections (see~\cite{Jia:2014jfa,Andronic:2015wma}). Until now, 
the polarisation of the $\chi_c$ has been predicted~\cite{Shao:2014fca} but not yet measured. Suggestions 
to measure it indirectly via the $J/\psi$ angular distribution are discussed in~\cite{Faccioli:2011be}.
First constraints relating the $\chi_{c1}$ and $\chi_{c2}$ polarisations have been obtained in 2019 by CMS~\cite{Sirunyan:2019apc}.

\begin{table}[hbt!]\renewcommand{\arraystretch}{1.4}
  \centering
\subfloat[$\chi_b(1P)$]{\begin{tabularx}{0.7\textwidth}{X|p{1.2cm}|p{2.2cm}|p{1.9cm}|c}
 	   & $P_{T,\rm min.}^\Upsilon$ [GeV] & $\msp0/m_b^2$ [GeV$^3$] & $\mopb$ [$10^{-2}$~GeV$^3$] & $r_{1P}$     \\ \hline\hline
PKU \cite{Han:2014kxa} & 15 & $0.39/m_b^2$     &   {\it 0.73}  &  $0.42 \pm 0.05$\\
\hline
IHEP \cite{Feng:2015wka} & 8 & $0.33/m_b^2$    &  $1.16 \pm 0.07$  & {\it 0.79} \\
\end{tabularx}}
\\
\subfloat[$\chi_b(2P)$]{\begin{tabularx}{0.7\textwidth}{X|p{1.2cm}|p{2.2cm}|p{1.9cm}|c}
 	   & $P_{T,\rm min.}^\Upsilon$ [GeV] & $\msp0/m_b^2$ [GeV$^3$] & $\mopb$ [$10^{-2}$~GeV$^3$] & $r_{2P}$     \\ \hline\hline
PKU \cite{Han:2014kxa} & 15 & $0.39/m_b^2$     &   {\it 1.07}  &  $0.62 \pm 0.08$\\
\hline
IHEP \cite{Feng:2015wka} & 8 & $0.39/m_b^2$    &  $1.50 \pm 0.21$  & {\it 0.87} \\
\end{tabularx}}
\\
\subfloat[$\chi_b(3P)$]{\begin{tabularx}{0.7\textwidth}{X|p{1.2cm}|p{2.2cm}|p{1.9cm}|c}
 	   & $P_{T,\rm min.}^\Upsilon$ [GeV] & $\msp0/m_b^2$ [GeV$^3$] & $\mopb$ [$10^{-2}$~GeV$^3$] & $r_{3P}$     \\ \hline\hline
PKU \cite{Han:2014kxa} & 15 & $0.39/m_b^2$     &   {\it 1.43}  &  $0.83 \pm 0.22$\\
\hline
IHEP \cite{Feng:2015wka} & 8 & $0.42/m_b^2$    &  $1.92 \pm 0.34$   & {\it 1.03} \\
\end{tabularx}}
\caption{Selection of $\chi_{b0}(nP)$ LDMEs --related to those of $\chi_{bJ}(nP)$ via a $2J+1$ factor by virtue of HQSS-- from 2 NLO fits using LHC and Tevatron data. Numbers in italic are derived from $r_{nP}=(\mopb)/(\msp0/m_b^2)$ (PKU) or from $\langle {\cal O} \rangle$ (IHEP). In both fits, $\mu_\Lambda=m_b$. For the IHEP fit, we have chosen to quote only the values for which the FD treatment is similar to that of the PKU fit. $P_{T,\rm min.}^\Upsilon$ indicates the minimum $P_T$ of the fit data. The CS LDMEs are not fit~\cite{Han:2014kxa,Feng:2015wka}.
}\label{tab:LDME-chib-NLO}
\end{table}
 
Along these lines, we note that the latest low-$P_T$ $\chi_c$ LHCb data show~\cite{LHCb:2012af} a $R_{\chi_c}$
increasing for $P_T$ getting closer to zero as expected from the Landau-Yang suppression
of the $\chi_{c1}$ production via gluon fusion. This indicates that, in complex
environments such as $pp$ collisions at the LHC, the quantum numbers of the final state remain
relevant even at low $P_T$. Along these lines, the prospects for a lower reach in $P_T$ offered by the use of the newly observed~\cite{Aaij:2017vck} $\chi_c$ decay into $J/\psi+\mu^+\mu^-$ are truly interesting.

In the bottomonium sector, like for the $\Upsilon$,  3 studies exist from the IHEP group~\cite{Gong:2013qka,Feng:2015wka} and from the PKU group~\cite{Han:2014kxa}. The corresponding values of the LDMEs
and of $r_{nP}$ are given in \ct{tab:LDME-chib-NLO}. The measured FD fractions at the Tevatron and the LHC --thus the $\chi_b(nP)$ yields-- are relatively well described, except maybe for $F^{\chi_b(2P)}_{\Upsilon(2S)}$.

\subsubsection{The $P_T$-integrated $J/\psi$ and $\Upsilon$ production and their energy dependence up to NLO}
\label{subsec:COM_NLO_tot}

As we wrote in the introduction, understanding  how the low-$P_T$ quarkonia  are
produced in nucleon-nucleon collisions is extremely important
to efficiently use them  as probes of deconfinement or collectivity in nucleus-nucleus collisions. 
Most of the analyses of quarkonium production
in nucleus-nucleus collisions are indeed carried out~\cite{Andronic:2015wma} 
on the bulk of the cross section where the yields are the highest, namely at low $P_T$. 

It is thus not satisfactory to leave out these data from the NRQCD realm just because one
observes that fixed-order COM computations do not properly describe the $P_T$-differential 
cross section in this region where admittedly one may need to account for initial-state radiation (ISR)
for instance. We recall that the NLO CS predictions accurately describe the low-$P_T$ spectrum of
the $\Upsilon(nS)$ at the LHC.

Yet, an intermediate test of the theory is to compare the $P_T$-integrated yield predicted
by NRQCD. It would not be acceptable that the introduction of the CO transitions in order to solve 
high-$P_T$ puzzles would
generate a significant excess in the quarkonium total yields. In this context, despite 
the possibility that the NRQCD factorisation would not be valid at low $P_T$, 
several analyses have been carried in the past to address this question. For the $J/\psi$ case,  
let us cite those of Beneke and Rothstein~\cite{Beneke:1996tk}, Cooper \etal\ \cite{Cooper:2004qe}, and Maltoni \etal\ \cite{Maltoni:2006yp}. Going further,  the impact of ISR at very low $P_T$
was studied by Sun \etal\ \cite{Sun:2012vc} in 2012 along with the introduction 
of additional non-perturbative parameters which however complicate the interpretation 
of these results.

Cooper~\etal\ in fact argued~\cite{Cooper:2004qe}, using only early RHIC data,
that the universality of NRQCD was safe and that the CS contributions to the 
$P_T$-integrated $J/\psi$ yields were negligible. On the contrary, the NLO global analysis of Maltoni~\etal~at 
NLO highlighted~\cite{Maltoni:2006yp} that the  CO LDMEs compatible with
the total prompt $J/\psi$ yield from fixed-target 
energies to RHIC were {\it one tenth} of that we discussed above from the LO fits of
$P_T$-differential cross sections at Tevatron energies -- which in any case cannot 
reproduce the polarisation data. 

As we have just seen, the inclusion of the NLO corrections to the 
$P_T$-differential  cross sections drastically changes the picture with 
subtle interplays between the different CO contributions. Despite the remaining puzzles, 
one can say now that we have LDMEs determined at NLO.

Using these, we could then significantly extend~\cite{Feng:2015cba} in 2015 
the aforementioned existing NRQCD studies of the $P_T$-integrated 
cross section by coherently combining the hard-scattering coefficients, known
up to $\alpha_S^3$ since 1997~\cite{Petrelli:1997ge} and 
checked with FDC~\cite{Wang:2004du}, with the aforementioned NLO
NRQCD LDMEs extracted from the $P_T$ dependence of the yields. The logic
behind is that the {\it $P_T$-integrated} and the {\it $P_T$-differential} cross sections
can be considered as two different observables with notably 
different Born contributions.

\paragraph{A full one-loop cross-section study.}

\begin{figure*}
\centering
\subfloat[]{\includegraphics[scale=.33,draft=false]{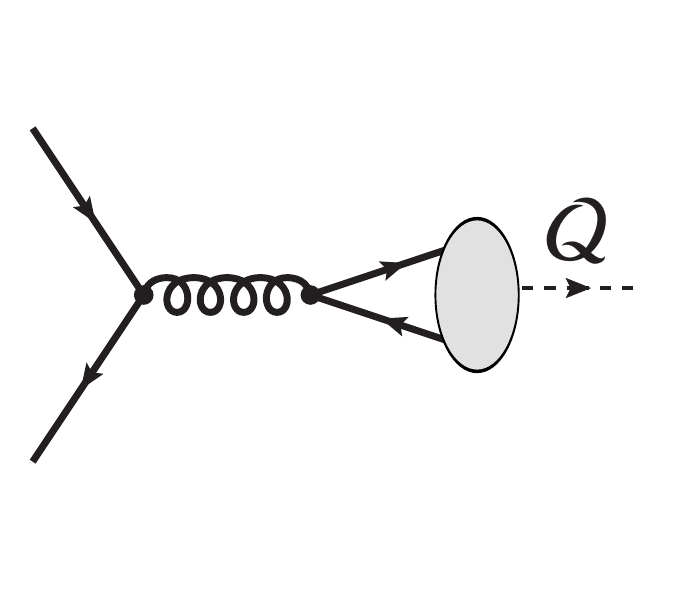}\label{diagram-tot-COM-a}}\hspace*{-.3cm}
\subfloat[]{\includegraphics[scale=.33,draft=false]{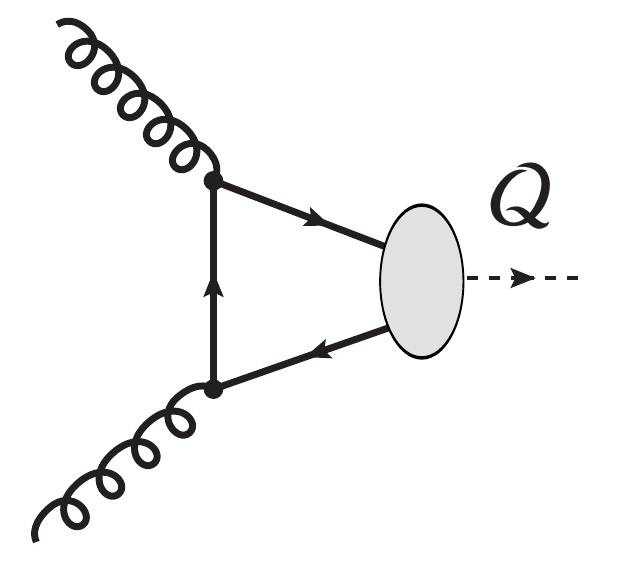}\label{diagram-tot-COM-b}}\hspace*{-.3cm}
\subfloat[]{\includegraphics[scale=.33,draft=false]{LO-CSM-new.pdf}\label{diagram-tot-COM-c}}\hspace*{-.3cm}
\subfloat[]{\includegraphics[scale=.33,draft=false]{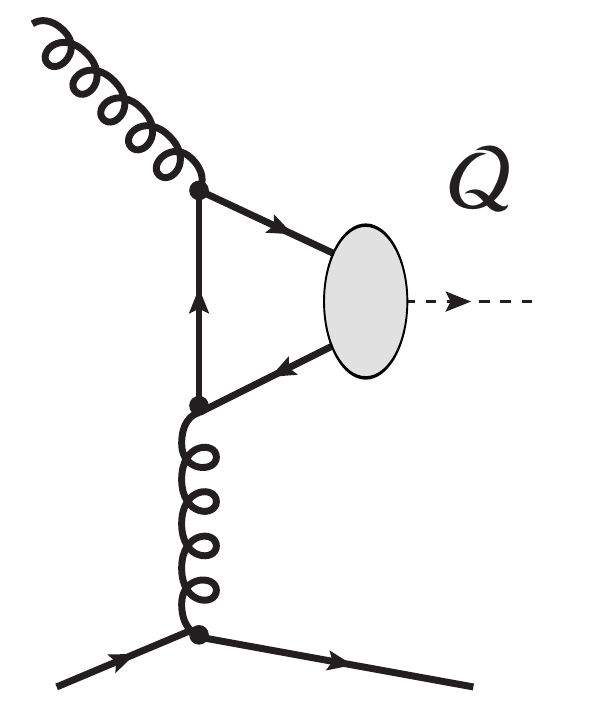}\label{diagram-tot-COM-d}}\hspace*{-.3cm}
\subfloat[]{\includegraphics[scale=.33,draft=false]{ij-oniuml-LO-COM.pdf}\label{diagram-tot-COM-e}}\hspace*{-.3cm}
\subfloat[]{\includegraphics[scale=.33,draft=false]{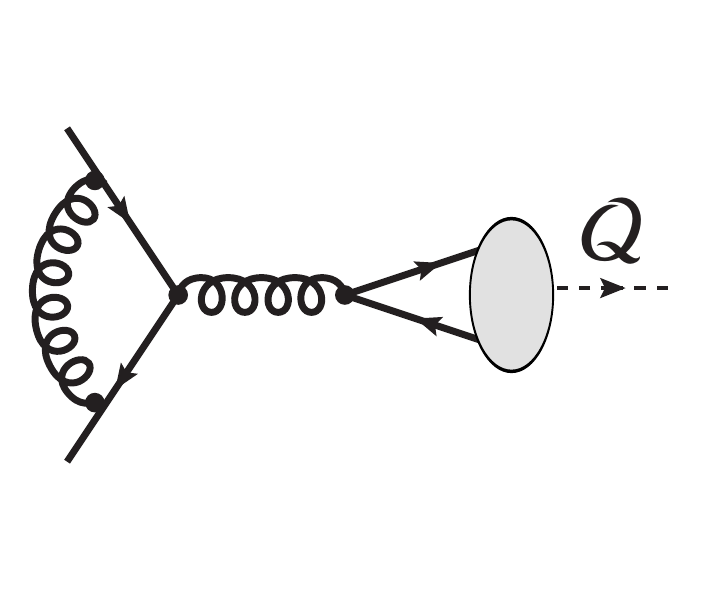}\label{diagram-tot-COM-f}}\hspace*{-.3cm}\\
\subfloat[]{\includegraphics[scale=.33,draft=false]{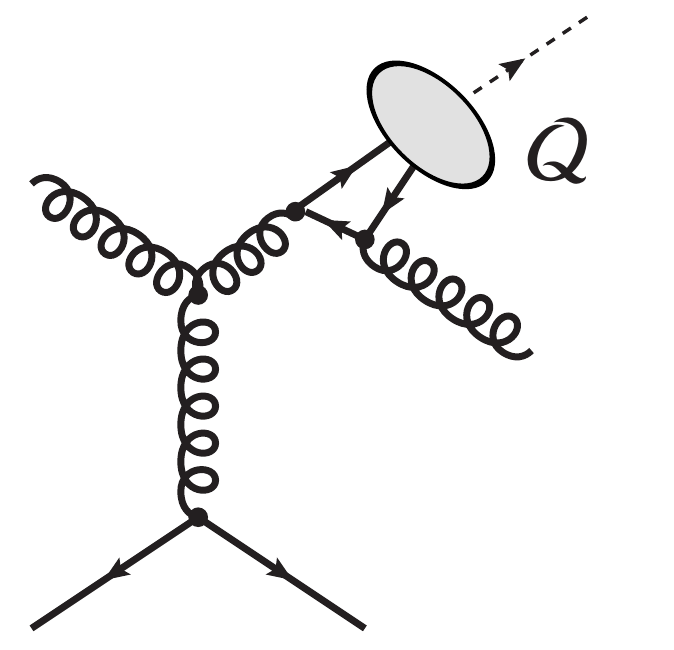}\label{diagram-tot-COM-g}}\hspace*{-.3cm}
\subfloat[]{\includegraphics[scale=.33,draft=false]{ij-oniuml-NLO-COM.pdf}\label{diagram-tot-COM-h}}\hspace*{-.3cm}
\subfloat[]{\includegraphics[scale=.33,draft=false]{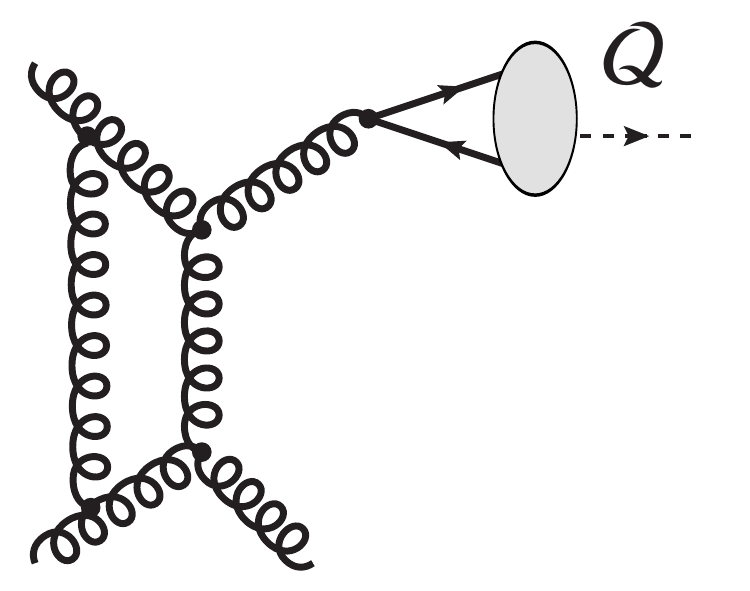}\label{diagram-tot-COM-i}}\hspace*{-.3cm}
\subfloat[]{\includegraphics[scale=.33,draft=false]{NLO-PT8-CSM-new.pdf}\label{diagram-tot-COM-j}}\hspace*{-.3cm}
\subfloat[]{\includegraphics[scale=.33,draft=false]{NLO-PT6-CSM.pdf}\label{diagram-tot-COM-k}}\hspace*{-.3cm}
  \caption{Representative diagrams contributing (a-b) at Born order to $i+j\to \Q$, (c-e) both
at Born order to $i+j\to \Q+$jet and at one loop to $i+j\to \Q$, (f) at one loop to $i+j\to \Q$, (g-k) at one loop to $i+j\to \Q+$jet. \label{fig:diagram-tot-COM}
} 
\end{figure*}

For the $^3S_1$ quarkonium states, we have seen that the first CO states which appear 
in the $v$ expansion are the  $^{1}\!S^{[8]}_{0},^{3}\!S^{[8]}_{1}$ and $^{3}\!P^{[8]}_{J}$ 
states, in addition to the leading $v$ contribution $^{3}S^{[1]}_{1}$ from a CS transition. 
One however has to note that, for hadroproduction, whereas the CO contributions 
already appear at $\alpha_S^2$ (\cf{diagram-tot-COM-a} \& \ref{diagram-tot-COM-b}), the CS ones 
only appear at $\alpha_S^3$ (\cf{diagram-tot-COM-c}). These $\alpha_S^2$ CO graphs 
nevertheless do not contribute to the production of quarkonia with a {\it nonzero} 
$P_T$ since they would be produced alone without any other hard particle to recoil on.

The Born contributions from CS and CO transitions are indeed different in nature:
the former is the production of a quarkonium in association with a recoiling gluon, which 
could form a jet, while the latter\footnote{except for the $gg$ induced $^{3}S^{[8]}_{1}$ production which also proceeds at LO via a recoiling gluon}  is the production of a quarkonium essentially alone
at low $P_T$. 

Let us now have a look at the $\alpha_S^3$ CO contributions  
(\cf{diagram-tot-COM-d}-\ref{diagram-tot-COM-f}) which are then NLO --or one loop-- 
corrections to quarkonium production and which are potentially plagued 
by the typical divergences of radiative corrections. Yet, the real-emission 
$\alpha_S^3$ corrections to CO contributions (\cf{diagram-tot-COM-d} \& \ref{diagram-tot-COM-e}) 
can also be seen as Born-order contributions to the production of a 
quarkonium + a jet --or, equally speaking, of a quarkonium with $P_T \gg \Lambda_{\rm QCD}$.
As such, they do not show any soft divergences for  $P_T \neq 0$. These 
are supposed to be the leading contribution to the $P_T$-differential cross section
in most of the datasets taken at hadron collider (Tevatron, RHIC and LHC). 
As we have seen in section~\ref{subsubsec:COM_3S1_NLO_PT}, these are now known up to one-loop accuracy, namely up to $\alpha_S^4$ 
(\cf{diagram-tot-COM-g} \& \ref{diagram-tot-COM-h}). 

It is important to note that one cannot avoid dealing with the divergences 
appearing at $\alpha_S^3$ if one studies the $P_T$-integrated cross section.

\paragraph{Different contributions up to $\alpha^3_S$.}

In order to discuss the energy dependence down to low energies, one may need to consider quark-induced channels. At $\alpha_S^2$, the CO partonic processes are: 
\eqs{&q+\bar q \rightarrow Q \bar Q[^{3}S^{[8]}_{1}]  \text{ (\cf{diagram-tot-COM-a})}, \quad
g+g\rightarrow Q\bar Q[^{1}\!S^{[8]}_{0}, ^{3}\!P^{[8]}_{J=0,2}] \text{ (\cf{diagram-tot-COM-b})} }
where $q$ denotes $u,d,s$. 

At $\alpha_S^3$, the QCD corrections to the aforementioned channels include 
real (\cf{diagram-tot-COM-d} \& \ref{diagram-tot-COM-e}) and virtual (\cf{diagram-tot-COM-f})
corrections. One encounters UV, IR and Coulomb singularities in
the calculation of the virtual corrections. The UV-divergences from the self-energy 
and triangle diagrams are removed by the renormalisation procedure (see~\cite{Gong:2010bk,Gong:2012ug}).
As regards the real-emission corrections, they arise from 3 kinds of processes (not all drawn): 
\eqs{&g+g\rightarrow Q \bar Q[^{1}\!S^{[8]}_{0},^{3}\!S^{[8]}_{1},^{3}\!P^{[8]}_{J=0,2}]+g, \quad
g+q(\overline{q})\rightarrow Q \bar Q[^{1}\!S^{[0]}_{8},^{3}\!S^{[8]}_{1},^{3}\!P^{[8]}_{J=0,2}]+q(\overline{q}), \\
&q+\overline{q}\rightarrow Q \bar Q[^{1}\!S^{[8]}_{0},^{3}\!S^{[8]}_{1},^{3}\!P^{[8]}_{J=0,1,2}]+g.}
As usual, the phase-space integrations generate IR singularities, either soft or collinear.
We refer to the analysis Petrelli \etal~\cite{Petrelli:1997ge} for a discussion of their treatment which
is standard for quarkonium production.

As we previously alluded to, the $\alpha_S^3$ CS contribution is particular since
it would be strictly speaking Born order for both the production 
of a quarkonium and of a quarkonium + a jet. It arises from the well-known process: 
\eqs{&g+g\rightarrow Q\bar Q[^{3}\!S^{[1]}_{1}]+g  \text{ (\cf{diagram-tot-COM-c})}}

Overall, we proposed to use the LDMEs fitted at the same order (\ie~one loop or NLO) to the $P_T$-differential cross sections in order to perform a global NLO analysis of hadroproduction, and hence go beyond the phenomenological study made by Maltoni~\etal\ \cite{Maltoni:2006yp} in 2006.

At rather low energies, CSM contributions via $\gamma^\star$ exchange may also be relevant. Indeed, as we noted in a different 
context in~\cite{Lansberg:2013wva},  the QED CS contributions 
via $\gamma^\star$ are naturally as large as the CO $^3S_1^{[8]}$ transition 
via $g^\star$ -- the $\alpha_{\rm em}$ suppression being compensated by the small 
relative size of the $^3S_1^{[8]}$ CO LDME (${\cal O}(10^{-3})$) as compared to 
the $^3S_1^{[1]}$ CS LDME (${\cal O}(1)$). The real-emission contributions arise from
\eqs{&q+\overline{q}\rightarrow Q \bar Q[^{3}\!S^{[1]}_{1}]+g,  \quad
g+q(\overline{q})\rightarrow Q \bar Q[^{3}\!S^{[1]}_{1}] +q(\overline{q}),}
whereas the loop contributions are only from 
\eqs{&q+\overline{q}\rightarrow Q \bar Q[^{3}\!S^{[1]}_{1}].} \cf{diagram-tot-COM-a} (\cf{diagram-tot-COM-f}) 
with the $s$-channel gluon replaced by a $\gamma^\star$ would depict the Born (a one-loop) 
contribution. As noted in~\cite{Feng:2015cba}, they however do not matter for $pp$ collisions at $\sqrt{s}  > 40$~GeV.

\paragraph{Complete NLO results within NRQCD.}

Direct $J/\psi$, $\psi(2S)$ and $\Upsilon(1S)$ cross section are shown on respectively 
\cf{fig:energy_dependence} (a), (b) and (c) for five different NLO LDME sets. 
We refer to~\cite{Feng:2015cba} for details on the LDMEs and on other theory parameters as well as for a  description of the experimental data used for 
the comparison (and the assumed FD fractions).

\begin{figure*}[hbt!]
  \centering
\subfloat[$J/\psi$]{\includegraphics[trim = 0mm 0mm 0mm 0mm, clip,scale=1.1,draft=false]
{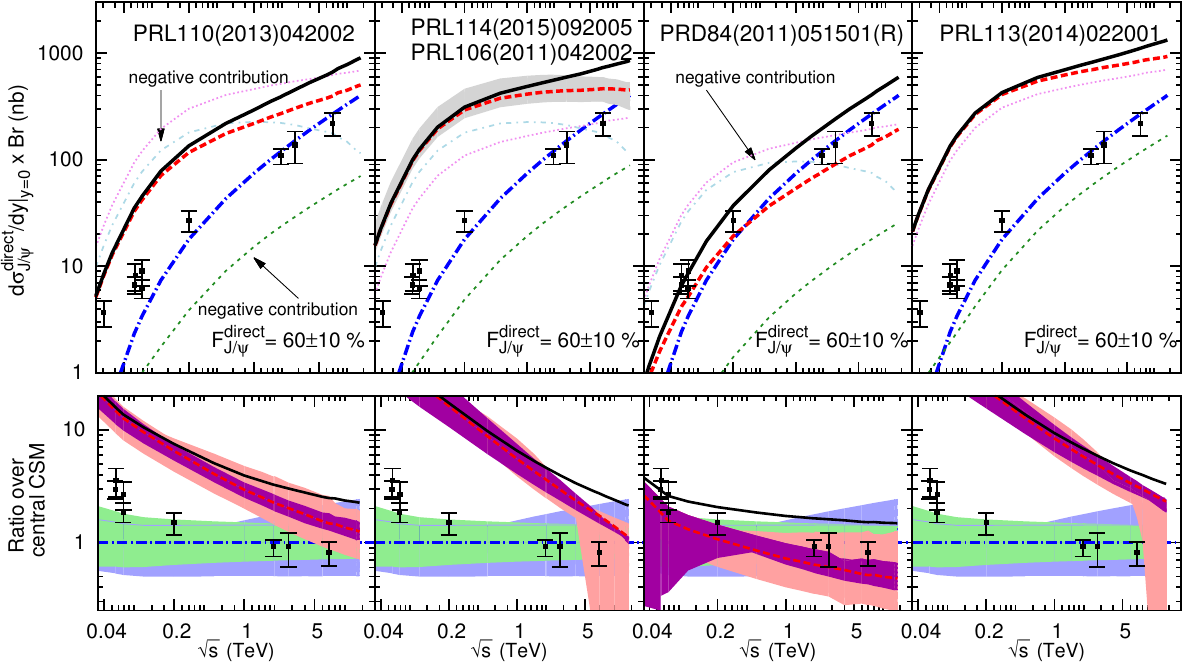}} \\
\subfloat[$\psi(2S)$]{\includegraphics[trim = 0mm 0mm 0mm 0mm, clip,scale=1.1,draft=false]
{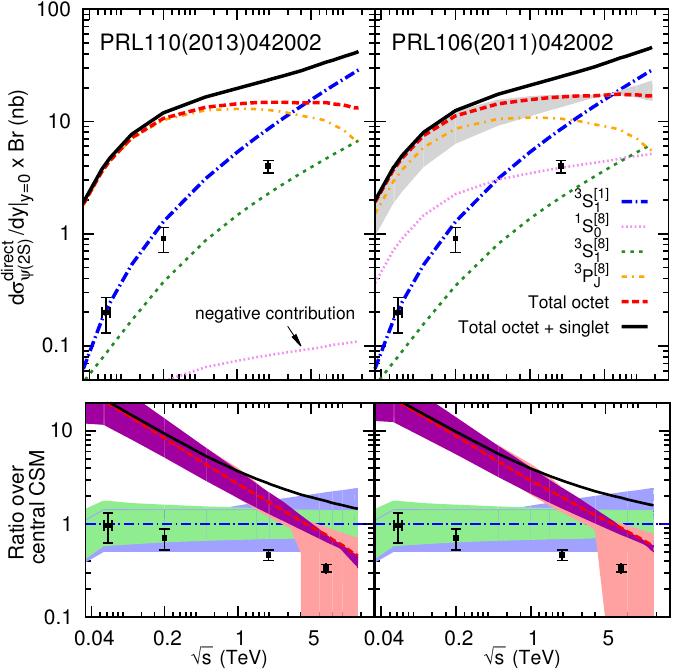}} \hspace{0.4cm}
\subfloat[$\Upsilon(1S)$]{\includegraphics[trim = 00mm 0mm 0mm 0mm, clip,scale=1.1,draft=false]
{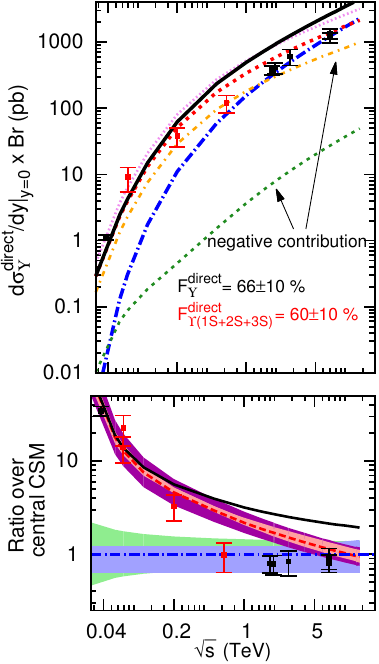}}
  \caption{The cross section for direct (a) $J/\psi$, (b) $\psi(2S)$ and (c) $\Upsilon(1S)$ as a function of $\sqrt{s}$.
The blue dot-dashed curve is the central CS curve. Its relative uncertainty is shown in the lower panels; the light green (light blue) band shows the scale (mass) uncertainty.
The dashed red curve is the total CO contribution from 3 channels: $^{3}\!P^{[8]}_{0}$ 
(thin dot-dashed orange), $^{1}\!S^{[8]}_{0}$ (thin dotted magenta) and  
$^{3}\!S^{[8]}_{1}$ (thin dashed green). The total CO uncertainty relative 
to the CS central curve is shown in the lower panels; the light pink 
(purple) band shows the scale (mass) uncertainty. The black is the total 
contribution (CS+CO) at one loop. These are compared to experimental data 
(see~\cite{Feng:2015cba}) multiplied by a direct fraction factor (when applicable) and 
normalised to the central CS curve in the lower panels.
[Negative CO contributions are indicated by arrows]. Taken from~\cite{Feng:2015cba}.}
  \label{fig:energy_dependence}
\end{figure*}

Let us first discuss the comparison between the five fits and the 
$J/\psi$ data (\cf{fig:energy_dependence} (a)). Without any surprise, the Hamburg global fit~\cite{Butenschoen:2011yh},
 including rather low $P_T/m_\Q$ data, provides the only acceptable description of the 
total cross section. We however recall that the latter fit is unable to describe
polarisation data and $\eta_c$ data~\cite{Butenschoen:2014dra}. 
The IHEP~\cite{Gong:2012ug} and PKU~\cite{Ma:2010yw,Han:2014jya} fits greatly overshoot 
the data in the energy range between RHIC and the Tevatron, whereas these fits a priori
provide a good description of the $P_T$-differential cross section at these energies.

The fit of Bodwin \etal~\cite{Bodwin:2014gia} gives the worse account of the $P_T$-integrated $J/\psi$ data
in the whole energy range.
Indeed, the new ingredient of~\cite{Bodwin:2014gia} allows one to describe high-$P_T$ spectrum with
a large $^1S_0^{[8]}$ LDME ($9.7 \times 10^{-2}$ GeV$^3$) --like for~\cite{Gong:2012ug} but without negative LDMEs for the other octet LDMEs-- and a large $M_{0,r_0}^{J/\psi}$ which results in too large a yield at low $P_T$. This fit also
largely overshoots the LHCb $\eta_c$ cross section (see section~\ref{subsec:COM_1S0_NLO_PT}).

In addition, we also note the strange energy dependence of at least the $P$-wave octet channel
which is reminiscent of the discussion we had about the CS channels. We will not detail its study
here and rather guide the reader to~\cite{Feng:2015cba} and~\cite{Mangano:1996kg}.

As what regards the $\psi(2S)$ (\cf{fig:energy_dependence} (b)), the NLO NRQCD 
results do not reproduce the data at all at RHIC energies and, since both fits as 
dominated by the $P$-wave octet channel, show a nearly unphysical behaviour
at LHC energies.

The comparison for the $\Upsilon(1S)$ (\cf{fig:energy_dependence} (c)), is more encouraging. 
At RHIC energies and below, the agreement is even quite good, while at 
Tevatron and LHC energies, the NLO NRQCD curves {\it only} overshoot
the data by a factor of 2.

We finally note that from RHIC to LHC energies, the 
LO CSM contributions (the blue in all the plots) accounts well for the data. This
is in line with the observations made in section \ref{section:CSM-NLO_tot}. The 
agreement is a little less nice for $\psi(2S)$ if we stick only to the default/central
value. This is not at all a surprise and agrees with the previous conclusions 
made in~\cite{Brodsky:2009cf,Lansberg:2010cn,Lansberg:2012ta,Lansberg:2013iya}. 
In fact, strangely enough, it seems that it is only at low energies (below $\sqrt{s}=100$~GeV) 
that the CO contributions would be needed to describe the data. The more recent data from the LHC
and the Tevatron tend to agree more with the LO CSM.

Overall, this confirms --unless the resummation of ISR or CGC effects (see below) modify these results by a factor of ten-- that 
it would be difficult to achieve a global description of the total and $P_T$-differential yield 
and its polarisation at least for the charmonia.

As we discussed above, a first resummation study performed in 2012
within NRQCD allowed Sun \etal\ \cite{Sun:2012vc} to perform a dedicated LDME fit. They reported 
$M^{J/\psi}_{0,r_0=7}=(1.97 \pm 0.09) \times 10^{-2}$ GeV$^3$ and $M^{\Upsilon}_{0,r_0=7}= (3.21 \pm  
 0.14) \times 10^{-2}$ GeV$^3$. Let us note that this fit was done by setting the CS contribution to zero which {\it 	a priori} not justified. Since $r_0$ is quite different than that for the $P_T$ 
differential cross section, these constraints can be combined to extract individual LDMEs. 
For the $J/\psi$, $\mop0$ and $\mopb$ are found to be negative while $\mops \simeq 0.14$ GeV$^3$, 
which nearly 30 times above the upper limit set by the $\eta_c$ data.
It should also be stressed that this study introduces 
3 new parameters $g_{1,2,3}$ to parametrise the
so-called $W^{\rm NP}$ function used the Collins-Soper-Sterman-resummation procedure. Moreover, 
we stress that such negative values of the CO LDMEs would result in a negative NLO $P_T$-differential cross section 
for $J/\psi+\gamma$ at large $P_T$ where NRQCD factorisation should normally hold. 
We will come back to this in section~\ref{sec:psi-gamma}.

Besides, in 2014, Ma and Venugopalan obtained~\cite{Ma:2014mri} a good description of the low-$P_T$ $J/\psi$ data 
over a wide range of energy by, on the one hand, using the LDMEs from~\cite{Ma:2010yw} (PKU) and, on the other, 
a CGC-based computation of the low-$P_T$ dependence. In reproducing the data, they found that
the CS contribution is only 10\% of the total yield\footnote{Recently they complemented their study in providing computations of the polarisation~\cite{Ma:2018qvc}.}. This 10\% is reminiscent of the
factor 10 between the CS and CO in the ``collinear'' study which we just discussed. This could be interpreted as if the 
specific ingredient of this  CGC-based computation corresponded to an effective reduction 
of the two-gluon flux by a factor of 10. However it may be a little more complex: in this CGC-based 
evaluation~\cite{Ma:2014mri}, the CSM yield is from contributions which are higher twist, thus suppressed, in the collinear limit. 
The leading-twist contributions corresponding to the LO $\alphaS^3$ contributions in the collinear limit would only appear in a complete NLO CGC analysis . We also note that this reduction factor seems to be the same
at 200 GeV than at 7 TeV which is surprising if it arises from low-$x$ effects.
It is therefore very interesting to find out new processes which
would be sensitive to this physics in order to test their proposed solution.  

In any case, whatever the explanation for this situation --excess and energy dependence-- may be, past claims that the COM  is dominantly responsible for low-$P_T$ quarkonium production 
were premature in  light of the results presented in this section.

\subsection{Recent developments in the CEM phenomenology}
\label{subsec:CEM_updates}

\subsubsection{$P_T$-integrated hadroproduction cross section at NLO}
\label{subsec:CEM_NLO_total} 

Contrary to the CSM and COM cases, let us start with the 
discussion of  the total cross section which has been the object
of several NLO studies for heavy-ion physics applications~\cite{Vogt:2015uba,Vogt:2010aa,Bedjidian:2004gd,Vogt:2004dh}.

\begin{figure}[ht!]
\centering
\subfloat[]{\includegraphics[scale=.33]{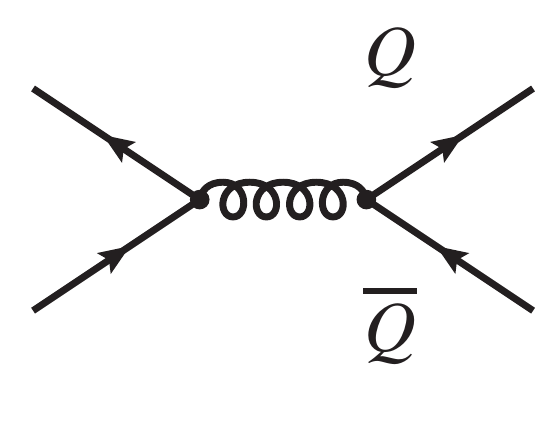}\label{diagram-CEM-a}}
\subfloat[]{\includegraphics[scale=.33]{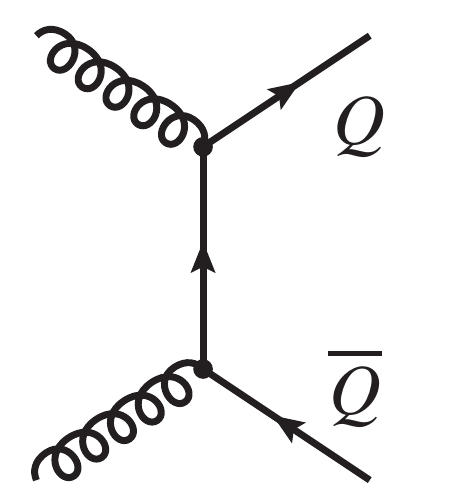}\label{diagram-CEM-b}}
\subfloat[]{\includegraphics[scale=.33]{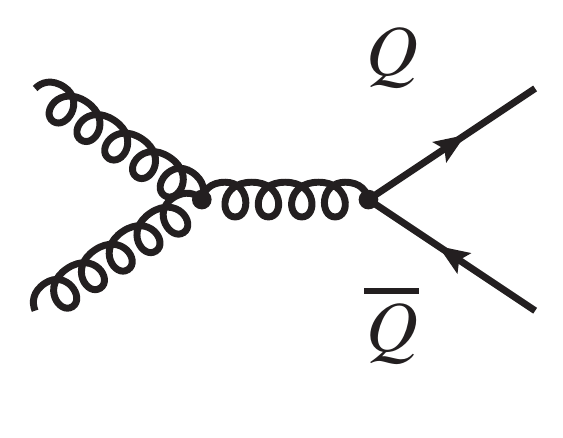}\label{diagram-CEM-c}}
\subfloat[]{\includegraphics[scale=.33]{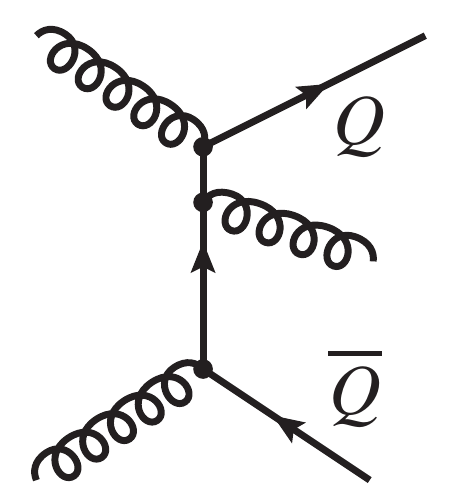}\label{diagram-CEM-d}}
\subfloat[]{\includegraphics[scale=.33]{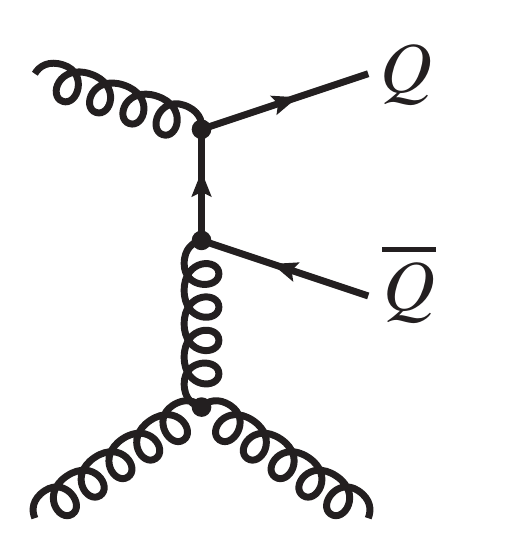}\label{diagram-CEM-e}}
\subfloat[]{\includegraphics[scale=.33]{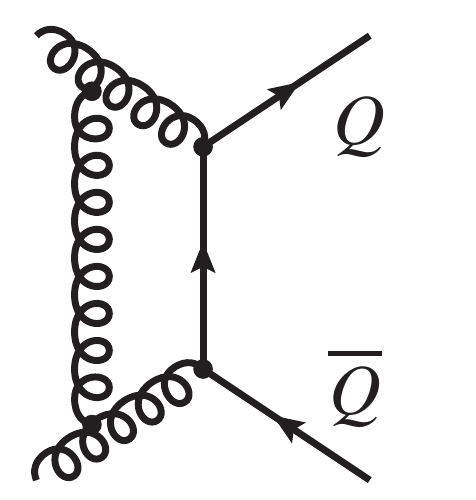}\label{diagram-CEM-f}}
\subfloat[]{\includegraphics[scale=.33]{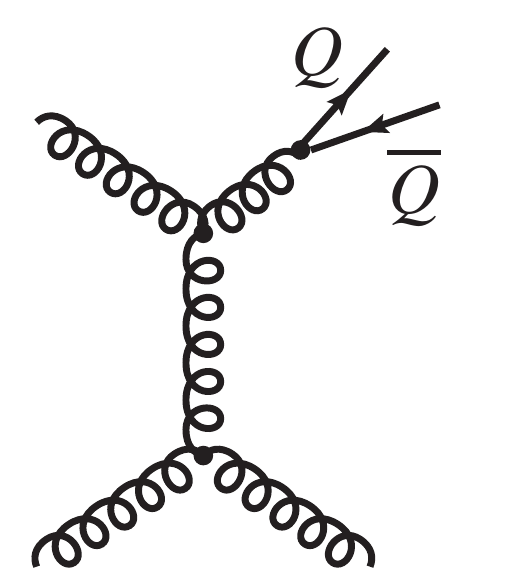}\label{diagram-CEM-g}}
\subfloat[]{\includegraphics[scale=.33,draft=treu]{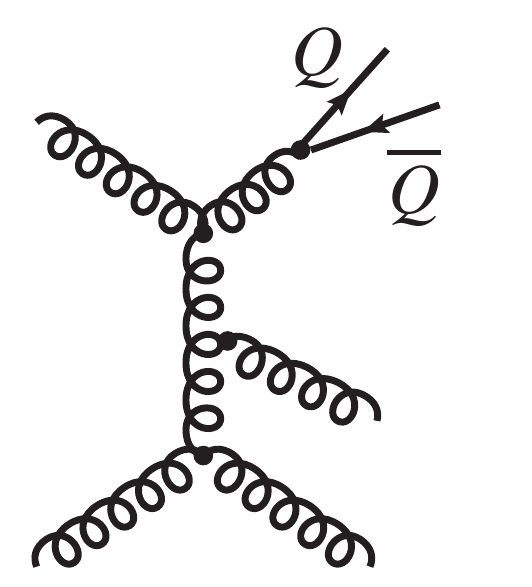}\label{diagram-CEM-h}}
\subfloat[]{\includegraphics[scale=.33,draft=treu]{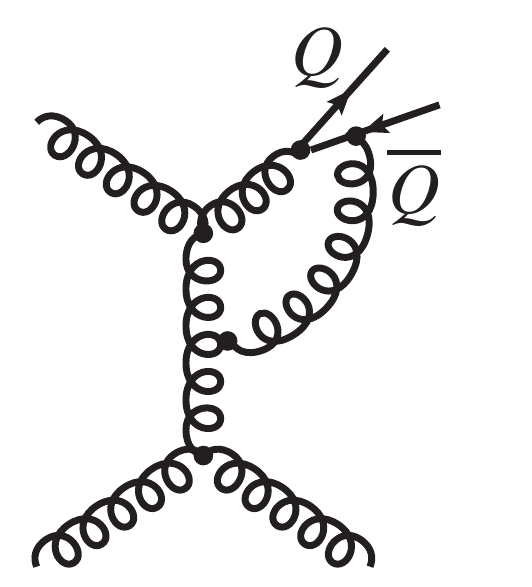}\label{diagram-CEM-i}}
\caption{Representative diagrams contributing to quarkonium hadroproduction via 
CEM at orders $\alpha_S^2$ (a-c), $\alpha_S^3$ (d-f), 
$\alpha_S^4$ (h,i). 
In the CEM, the invariant mass of the heavy-quark pair will
be integrated over within the $2m_Q$ and $2m_{H(Q)}$ where $H(Q)$ is the lightest hadron
with the flavour $Q$.}
\label{diagrams-CEM}
\end{figure}

\paragraph{Conventional approach.}
In the CEM, the total, $P_T$-integrated, cross section simply follows
from that of a pair of heavy quarks at small invariant masses. Unlike the CSM and COM computations, 
no symmetry can prevent simple reactions like $gg \to Q \bar{Q}$ or
$q \bar{q} \to Q \bar{Q}$ to contribute to a quarkonium production at $\alpha_s^2$ (\cf{diagram-CEM-a} -- \ref{diagram-CEM-c}), 
whatever its quantum numbers. Analytical results
for such amplitudes squared are known since the late 1970's (see \eg~\cite{Combridge:1978kx}) and
the differential partonic cross section as a function of the pair invariant mass 
can be integrated under the CEM conditions~\cite{Fritzsch:1977ay,Gluck:1977zm,Babcock:1977fi}.

NLO ($\alpha_s^3$) corrections (\cf{diagram-CEM-d} -- \ref{diagram-CEM-g}) can in addition be taken into account by using for instance
the MNR computation~\cite{Mangano:1991jk} or any automated NLO tool exclusive
enough to set cuts on the pair invariant mass. Two examples from which we will show results
 are the library MCFM~\cite{Campbell:1999ah} and \MG5aMC~\cite{Alwall:2014hca}.
Fits of the non-pertubative parameters $\P^{\rm direct/prompt}_{J/\psi}$ 
have been performed at NLO 
by Vogt in \cite{Bedjidian:2004gd} using $P_T$-integrated data up to 
$\sqrt{s}=62$ GeV. Values between 1.5~\% and 2.5~\% were obtained. It is worth
noting that the uncertainty follows from that of the heavy-quark production rather
than from the data which are fit.

As discussed in the introdution, a simple statistical counting rule based on the number
of quarkonium states below the open-heavy-flavour threshold 
works remarkably well for the $J/\psi$, whereas it does not work for 
$P$-waves as discussed in~\cite{Lansberg:2006dh,Brambilla:2010cs} 
as well as for $\psi(2S)$. This exhibits the limit of the model in incorportating  
differences in the hadronisation of different quarkonium states. 
 
For the $\Upsilon$, Vogt's fit yields a similar magnitude, 
 around 2~\%. Following the state-counting argument, one would however expect 
a smaller number than for $J/\psi$. At this stage, it is important to reiterate that 
such consideration ignore phase-space constraints in the hadronisation process. What 
the CEM really predicts is that $\P^{\rm direct}_{\cal Q}$ should be process independent.
Going further, the fact that a given yield ratio of 2 quarkonia would not follow the state counting is not necessarily problematic unlike a possible observation that such a yield ratio would depend on some kinematical variables or
would vary from one collision system to another.

\begin{figure}[hbt!]
\begin{center}
\includegraphics[width=11cm]{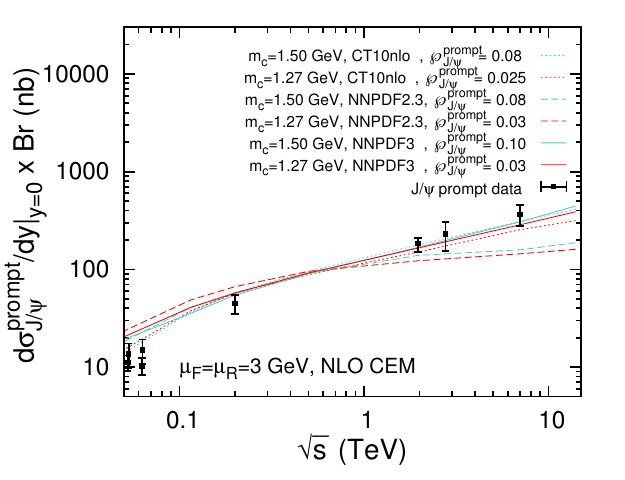}
\caption{Comparison between a selection~\cite{Lansberg:2016rcx} of $P_T$-integrated prompt $J/\psi$ cross 
sections measured  at $y=0$ and the NLO CEM results computed with {\small MCFM}~\cite{Campbell:1999ah}  
with two choices of the 
charm-quark mass ($m_c$) and with the PDFs~\cite{Ball:2014uwa,Ball:2012cx,Lai:2010vv} as a function 
of $\sqrt{s}$. In each case, $\P^{\rm prompt}_{J/\psi}$ has been fit to the data. Taken from~\cite{Lansberg:2016rcx}.}\label{fig:sigma_CEM_vs_s}
\end{center}
\end{figure}

A NLO comparison with data from fixed-target experiments as well as from colliders at central rapidities is shown 
on~\cf{fig:sigma_CEM_vs_s} for the $J/\psi$ case for different choices of the 
charm-quark mass and of the PDFs. 
First, one notes that the mass $m_Q$ dependence is nearly completely absorbed in the fit value of 
$\P^{\rm prompt}_{J/\psi}$. Let us stress that using different values of $m_Q$ 
mainly affect the cross-section via the integration range over the pair invariant 
mass, not much via the partonic cross section itself. 
In any case, the mass dependence does not show up in the energy dependence which is rather similar
for 1.27 and 1.5~GeV. 

A thorough study of the connection between heavy-flavour and quarkonium 
production in the CEM can be found in~\cite{Nelson:2012bc}. Along these lines, 
taking a rather small value for the charm quark mass, $m_c=1.27$~GeV, seems to be 
indicated. Let us however note that results with $m_c=1.5$ GeV are sometimes slightly different but 
never such as to modify the physics conclusion of a study.

On the contrary, one observes a significant dependence on the PDF set. 
This can be traced back to the rather low scale 
$\mu_R$ of the process and the very low probed $x$ values ($x \simeq M_\psi / \sqrt{s}$).
This reminds us that quarkonia, as well as heavy 
quarks~\cite{Garzelli:2016xmx,Gauld:2016kpd}, remain potential excellent probes 
of gluons inside hadrons.

\paragraph{CEM-NRQCD vs CEM vs ... CSM.}\label{subsec:CEM_vs_NRQCD}
As we have just seen and in section \ref{subsec:COM_NLO_tot},
the energy dependence of the CEM and COM/NRQCD total cross sections at NLO are rather different.
That of the CEM nicely follows the data and could even be used to bring in new constraints
on the gluon PDFs. That of the COM channels is ill-behaved, does not follow the data and
certainly calls for more theoretical efforts. In principle, the comparison 
should end here. Indeed, there are other observables where the CEM predictions disagree
with data whereas the COM ones do not. 

Along the lines of \cite{Bodwin:2005hm}, the CEM predictions should however match with a specific
realisation of NRQCD. It is therefore tempting to compare how the energy 
dependence of this ``CEM-NRQCD" compares with the genuine CEM. The connection is in fact
rather simple. 
Up to $v^2$ corrections, only 
4 intermediate $Q\Bar{Q}$ states contribute to $^3\!S_1$ quarkonium production in
a CEM-like implementation of NRQCD, namely 
$^{3}\!S^{[1]}_{1}$, $^{1}\!S^{[8]}_{0}$, $^{3}\!S^{[8]}_{1}$ and $^{1}\!S^{[1]}_{0}$.
As compared 
to usual NRQCD computations for vector quarkonia, the state $^{1}\!S^{[1]}_{0}$, 
which normally needs not to be considered for $^3\!S_1$ production 
at this level of accuracy in $v$, contributes. 
One then needs to impose the following
set of constraints between the LDMEs: 
\eqs{\label{eq:CEM-LDME}
\langle{\cal O}_{^3\!S_1}(^{3}\!S^{[1]}_{1})\rangle =3  \langle{\cal O}_{^3\!S_1}(^{1}\!S^{[1]}_{0})\rangle, \quad
\langle{\cal O}_{^3\!S_1}(^{1}\!S^{[8]}_{0})\rangle =\frac{4}{3}  \langle{\cal O}_{^3\!S_1}(^{1}\!S^{[1]}_{0})\rangle, \quad
\langle{\cal O}_{^3\!S_1}(^{3}\!S^{[8]}_{1})\rangle =4 \langle{\cal O}_{^3\!S_1}(^{1}\!S^{[1]}_{0})\rangle.
}

Just like a CEM computation only depends on $\P^{\rm direct/prompt}_{J/\psi}$, 
all these nonvanishing LDMEs are fixed once one of the LDMEs is fixed since they are related to each others.
One can even go further since, in principle, $\langle{\cal O}_{^3\!S_1}(^{3}\!S^{[1]}_{1})\rangle$ should be 
the usual CS LDME fixed by potential models or the leptonic decay width of the quarkonium, 
\ie~$\frac{2 N_C}{4 \pi}~(2J+1)~|R(0)|^2$. 
Finally, the contribution from $^{1}\!S^{[1]}_{0}$ can for instance be computed with FDC~\cite{Wang:2004du}
without any specific difficulty as in~\cite{Feng:2015cba}.

\begin{figure}[hbt!]
  \centering
\subfloat{\includegraphics[trim = 0mm 0mm 0mm 0mm, clip,width=0.5\columnwidth,,draft=false]{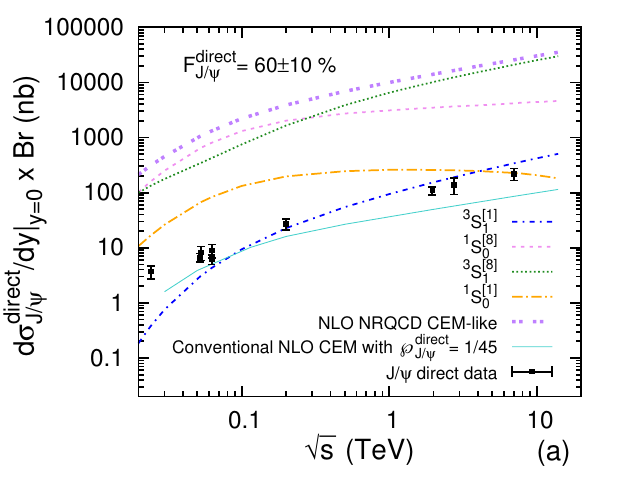}}
\subfloat{\includegraphics[trim = 0mm 0mm 0mm 0mm, clip,width=0.5\columnwidth,,draft=false]{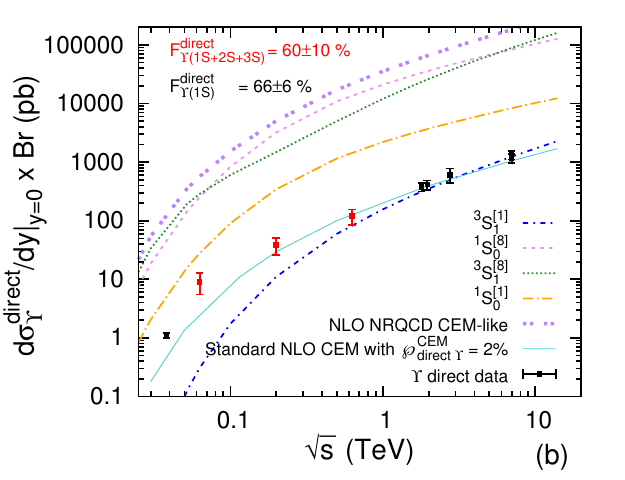}}
  \caption{The cross section for direct (a) $J/\psi$ and (b) $\Upsilon(1S)$ as a function of $\sqrt{s}$ from NLO NRQCD using CEM-NRQCD to be compared with the conventional NLO CEM (see section \ref{subsec:CEM_NLO_total}) and the existing experimental measurements. Taken from~\cite{Feng:2015cba}.}
  \label{fig:energy_dependence_CEM}
\end{figure}

\cf{fig:energy_dependence_CEM} (a) \& (b) show the resulting cross sections 
for the $J/\psi$ and $\Upsilon(1S)$ cases (the relevant channels and their 
sum), to be compared to the world data (see~\cite{Feng:2015cba}) and the conventional
CEM (solid light blue curve) evaluated with $\P^{\rm direct}_{J/\psi}=  \frac{1}{45}$ and 
$\P^{\rm direct}_{\Upsilon(1S)}=  2\%$ both in lines with Vogt's fits.

One directly observes that the NRQCD-CEM total cross section greatly 
overshoots the data, by a factor as large as 100, both at low and high energies.
Yet, such a behaviour could have been anticipated. Indeed, following \ce{eq:CEM-LDME},  
the 4 LDMEs are on the same order. We know that the $^{3}\!S^{[1]}_{1}$ 
yield (the NLO CSM one) is roughly compatible with the data and that  
the other hard coefficients (i) appear at $\alpha_S^2$, (ii)
are not suppressed as $P_T \to 0$ and (iii) are thus expected to give a larger contribution than 
the $^{3}\!S^{[1]}_{1}$ transitions. Taken together, the NRQCD-CEM yield should be much larger
than the data.

Of course, one could drop the assumption  that 
$\langle{\cal O}_{^3\!S_1}(^{3}\!S^{[1]}_{1})\rangle =\frac{2 N_C}{4 \pi}~(2J+1)~|R(0)|^2$ 
and rather fit
$\langle{\cal O}_{^3\!S_1}(^{1}\!S^{[1]}_{0})\rangle$ as one does for 
$\P^{\rm prompt}$. The corresponding LDMEs would then roughly be
100 times smaller and the CS transition would be 100 
times less probable that what one needs to get the measured leptonic decay. This 
would be a clear violation of NRQCD factorisation. 
We are therefore tempted to doubt the practical applicability of  the relations
 derived in~\cite{Bodwin:2005hm}. 

Yet, the constraints of~\ce{eq:CEM-LDME} encapsulate a fundamental difference
between the CEM and the CSM which normally should be visible by analysing different sets
of data, with otherwise extreme colour-flow configurations driven by the initial state.
In the CEM, all the colour configurations are summed over and then
considered on the same footage. It is of course not the case within the CSM, nor
within NRQCD if one follows the velocity scaling rules. It was for instance noted
in~\cite{Bodwin:2005hm} that these scaling rules were dramatically violated
in NRQCD-CEM.

In a reaction where the heavy-quark pairs are systematically produced colourless
with the correct spin to produce a vector state, such as $e^+e^- \to 
\gamma^\star \to Q \bar Q$, CSM and CEM predictions necessarily differ.
Whereas one is entitled to see NRQCD as a natural extension of the CSM --it even coincides 
with it at LO in $v$-- the CEM does not encompass the CSM.
If one approach agrees with some specific data, the other cannot. What matters then is how precise the predictions
and the data to rule out one approach or the other can be.

\subsubsection{Hadroproduction at NLO at finite $P_T$}
\label{subsec:CEM_NLO_PT}

In contrast to the $P_T$-integrated cross section, the $P_T$-differential cross section
has only been studied at NLO (or one-loop) accuracy in 2017 for $J/\psi$~\cite{Lansberg:2016rcx} and 
in 2020 for $\Upsilon$~\cite{Lansberg:2020rft}.

As we have seen, the comparisons of the CEM $P_T$-integrated yields with the data
is overall good. Most of 
the Tevatron, RHIC and LHC data for $P_T$-differential cross sections have also been 
confronted to (LO) CEM predictions (see~\cite{Andronic:2015wma} for a representative 
selection). Leaving aside the spectrum below a few GeV where a phenomenological 
parton-$k_T$-smearing procedure is usually applied~\cite{Bedjidian:2004gd} to reproduce the cross section
behaviour, the CEM predictions usually tend to overshoot the data for increasing $P_T$. 
Its spectrum always ends up to be harder than the data. 
This issue was already identified
as early as in the late 1990's and different mechanisms~\cite{Edin:1997zb,Damet:2001gu,BrennerMariotto:2001sv} 
have then been considered (see~\cite{Lansberg:2006dh} for a brief overview) to address this issue but none 
was the object of a consensus. More recently, an ``improved" version of the CEM~\cite{Ma:2016exq} was proposed by Ma and Vogt. We will briefly report on it in section~\ref{sec:CEM_additions}.

The origin of the problem is easy to find and arises from the existence at $\alpha_s^3$ 
of leading-$P_T$ topologies (\cf{diagram-CEM-g}) scaling 
like $P_T^{-4}$, just like those associated with the $^3S_1^{[8]}$ octet states in 
NRQCD (see section~\ref{subsec:COM_updates}). Whereas these were originally thought to solve the $\psi(2S)$ surplus 
at the Tevatron, we have seen that the recent NLO NRQCD fits to the LHC and Tevatron data point at
a need to reduce their impact, because they generate too hard a spectrum and because
they produce transversely polarised vector quarkonia. 

In NRQCD, the reduction of their impact is realised through a partial 
cancellation between both channels which show leading-$P_T$ contributions
 at NLO, namely the $^3S_1^{[8]}$ and $^3P_J^{[8]}$ states, opening the possibility 
for a dominant $^1S_0^{[8]}$ contribution in agreement with a softer $P_T$ 
spectrum.
In the CEM, owing to the simplicity of 
the model, no such cancellation can happen. The fragmentation contributions are 
obviously there, at any non-trivial order where the computation is carried out.

Until recently, 
in all Tevatron and LHC CEM computations~\cite{Bedjidian:2004gd,Brambilla:2004wf}, 
the employed hard-scattering matrix element was at NLO for 
inclusive heavy-quark production, namely $\alpha_s^3$ from the 
well-known MNR computation~\cite{Mangano:1991jk} using the specific invariant-mass 
cut of \ce{eq:sigma_CEM}, along the same lines as for some computations
of the total cross section at NLO discussed above.

However, at $\alpha_s^3$, the only graphs which contribute to the 
production of the heavy-quark pair  with a finite $P_T$ (with or without invariant-mass cut) 
are those from $2\to 3$ processes (\cf{diagram-CEM-d}, \ref{diagram-CEM-e} \& \ref{diagram-CEM-g}). 
The $\alpha_s^3$ loop 
contributions (\cf{diagram-CEM-f}) are therefore excluded as soon as finite $P_T$ is requested. 
As such, all these existing computations of the $P_T$ differential cross section of the pair 
are effectively Born-order/tree-level computation from $gg [q \bar q] \to (Q \bar Q) g$
 or $gq \to (Q \bar Q) q$, these are not truly speaking NLO computations in this context.

It is therefore legitimate to wonder whether the 
resulting LO $P_T$ spectrum, which disagrees with data,
 could be affected by large QCD $\alpha_s^4$ corrections (\cf{diagram-CEM-h} \& \ref{diagram-CEM-i}), 
effectively NLO and not NNLO for this quantity. In particular, one could wonder 
whether the data can be better described at NLO and whether 
$\P^{\rm NLO}_{J/\psi}$ is different than  $\P^{\rm LO}_{J/\psi}$ ? 

Given the direct connection between the CEM and heavy-quark production, 
it is not too demanding to perform such computations with modern tools of automated 
NLO frameworks, at the minimum cost of some slight tunings.  In particular, in~\cite{Lansberg:2016rcx}, we 
have employed \MG5aMC~\cite{Alwall:2014hca} for these 
(N)LO CEM calculations for $J/\psi $ + a recoiling parton with a finite $P_T$.

\begin{figure}[hbt!]
\begin{center}
\subfloat[LO]{\includegraphics[width=7.5cm]{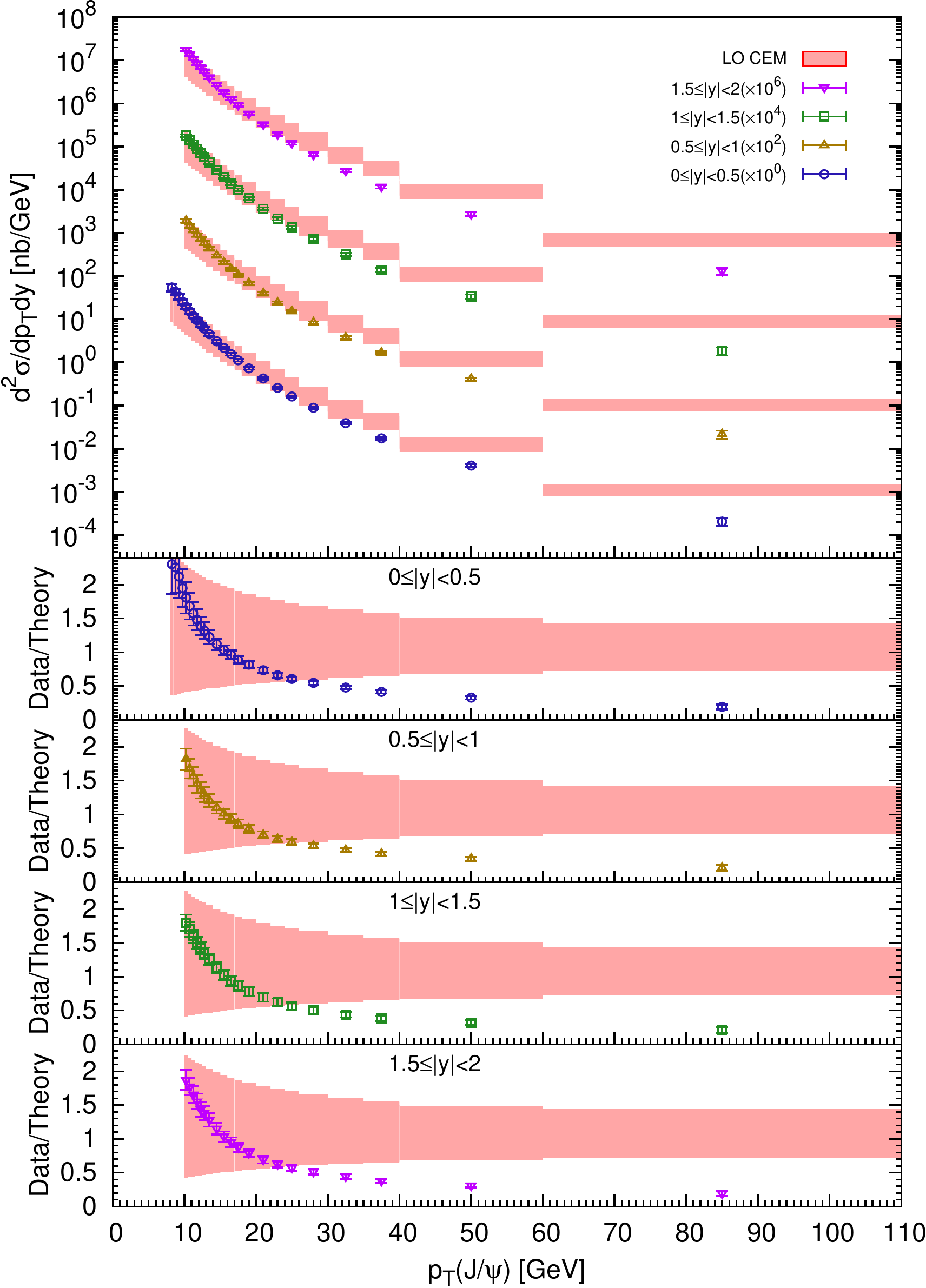}}
\subfloat[NLO]{\includegraphics[width=7.5cm]{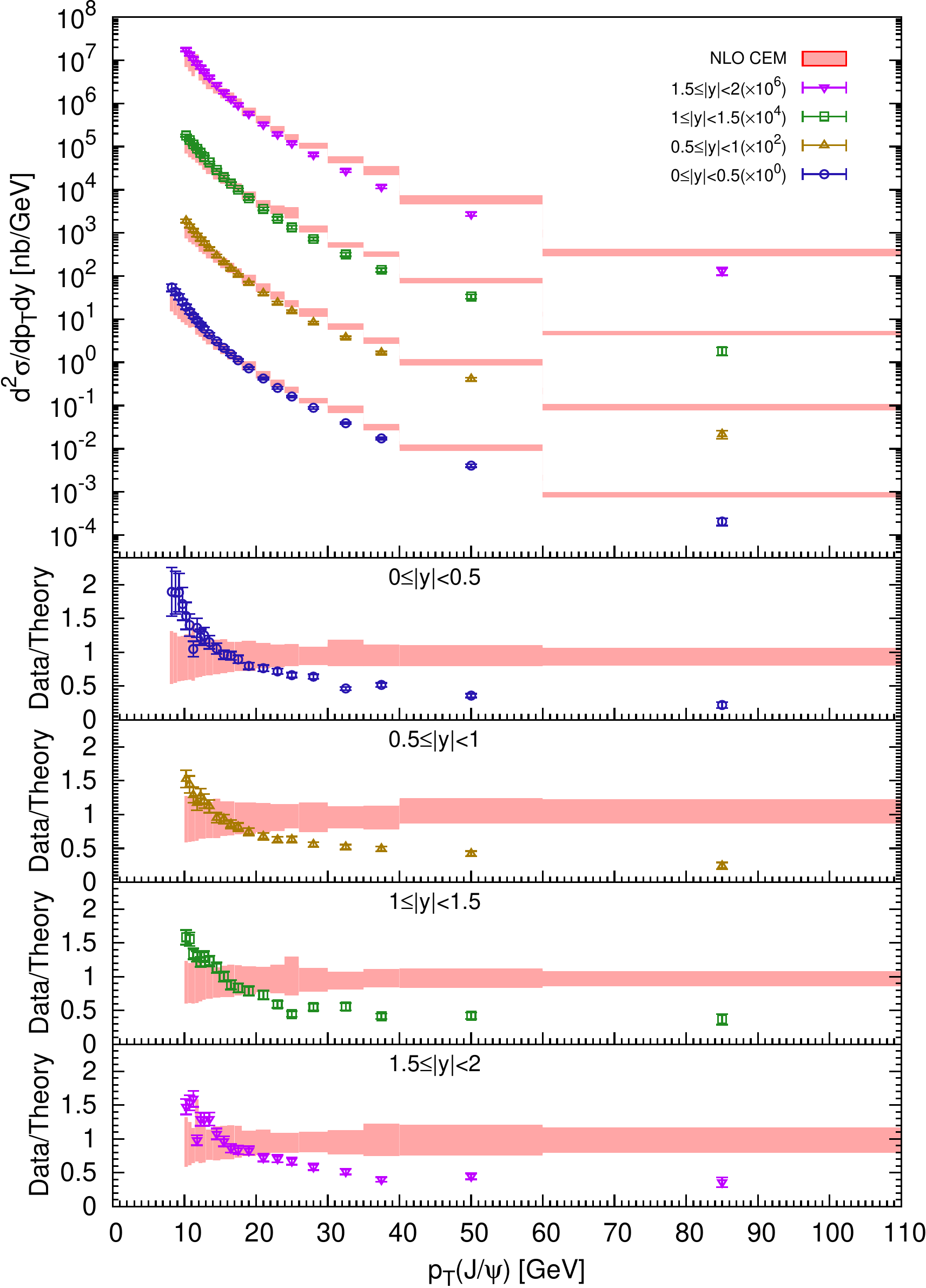}}
\caption{Comparison between 
the ATLAS data~\cite{Aad:2015duc} and the CEM results for $\displaystyle \frac{d^2\sigma}{dy dP_T}$ of 
$J/\psi$ + a recoiling parton at (a) LO and (b) NLO at $\sqrt{s}=8$ TeV. [The 
theoretical uncertainty band is from the scale variation (see the text) with $\P$ fit
using the central curve.] Taken from~\cite{Lansberg:2016rcx}.
\label{fig:single-Jpsi-pT-spetrum}}
\end{center}
\end{figure}

This allowed us to perform the first NLO fit of $\P^{\rm prompt}_{J/\psi}$ 
using $P_T$ differential ATLAS single-$J/\psi$ data at $\sqrt{s}=8$ TeV~\cite{Aad:2015duc} 
corresponding to 11.4 fb$^{-1}$. As a comparison, we have also carried out a LO fit
of the $P_T$-differential cross section. We stress that such a study can easily 
 be extended to other quarkonia in other kinematical regions as recently done~\cite{Lansberg:2020rft}.

A comparison with the data is shown on \cf{fig:single-Jpsi-pT-spetrum}.
As suggested in~\cite{Nelson:2012bc}, we have taken 
$m_c=1.27$~GeV. Another fit with a different value, \eg\  $m_c=1.5$ GeV, 
could be carried out. The $P_T$ dependence is however very similar and only 
the normalisation would differ. The difference would then be absorbed in 
the fit of $\P^{\rm prompt}_{J/\psi}$.
As what regards the parton distribution function (PDF), the NLO 
NNPDF 2.3 PDF set~\cite{Ball:2012cx} with $\alpha_s(M_Z)=0.118$ provided by 
LHAPDF~\cite{Buckley:2014ana} was used. 
Let us add that, since the heavy-quark-mass dependence 
is {\it de facto} absorbed in the CEM parameter, the main theoretical 
uncertainties arise from the renormalisation $\mu_R$ and 
factorisation $\mu_F$ scale variations expected to account for the unknown higher-order 
corrections. In practice,  they were varied within 
$\frac{1}{2}\mu_0\le \mu_R,\mu_F \le 2\mu_0$ where the central scale $\mu_0$ 
is the transverse mass of the $J/\psi$ in $J/\psi$ + parton.

Without much surprise, 
the CEM yields start to depart from the data when $P_T$ increases 
(see \cf{fig:single-Jpsi-pT-spetrum}). This is indeed happening in the region where 
fragmentation contributions are dominant. 
In addition, we note that the LO and NLO behaviours are similar and that the NLO uncertainties 
are smaller than the LO ones. Let us recall that the LO results (\ie\ for prompt 
$J/\psi$ + a recoiling parton) would coincide with the $P_T$ spectrum obtained 
from a NLO code for inclusive prompt $J/\psi$ results~\cite{Bedjidian:2004gd,Brambilla:2004wf}.
This reduction of uncertainties is the main added value of this first NLO CEM study of the $P_T$
spectrum of single $J/\psi$ using a one-loop evaluation of $J/\psi$ + a recoiling 
parton. The same conclusions hold for the other states~\cite{Lansberg:2020rft}.

Fitting this ATLAS data with $m_c=1.27$ GeV, we have obtained~\cite{Lansberg:2016rcx} $\P^{\rm LO, prompt}_{J/\psi}=0.014\pm 0.001$ and
 $\P^{\rm NLO, prompt}_{J/\psi} =0.009 \pm 0.0004$. These values depend on the used $P_T$ range~\cite{Lansberg:2020rft}. From them, one can deduce 
that the $K$ factor affecting the $d \sigma/dP_T$ is close to 1.6. It is also interesting
to compare these values with those obtained from the $P_T$-integrated total yields. 
The value of $\P^{\rm (N)LO, prompt}_{J/\psi}$ obtained  from the
$P_T$-differential yields are about a factor of 2--3 smaller than that obtained
from the $P_T$-integrated yields. This is not surprising since the CEM tends 
to overshoot the high-$P_T$ data.  This trend is in fact opposed 
to that of NRQCD where the CO LDME 
fit values from the $P_T$-differential yield overshoot those fit 
from the $P_T$-integrated yields. We conclude that the 
universality of $\P^{\rm prompt}_{J/\psi}$ seems to be challenged in the CEM as well. 

Yet, this shortcoming of the CEM, namely to generate too many quarkonia by
gluon fragmentation, can be an useful tool to set upper limits of cross sections
which cannot reliably be computed within NRQCD (see sections \ref{sec:onium_Z} \& \ref{sec:onium_W}).

\subsubsection{CEM phenomenology in $\gamma p$, $e^+e-$ and $\gamma \gamma$ collisions}
\label{sec:CEM_NLO_other}

In addition to hadroproduction, the CEM phenomenology has been extended to production in $\gamma p$, $e^+e^-$ and $\gamma \gamma$ collisions. To date, none of these studies for differential distributions have been performed at one loop in QCD. Only the total photoproduction study  by Amundson \etal\ in 1996~\cite{Amundson:1996qr} was carried out using a NLO heavy-quark photoproduction code~\cite{Ellis:1988sb}  in the same fashion as the $P_T$-integrated hadroproduction cross section discussed in section~\ref{subsec:CEM_NLO_total}.

\begin{figure}[hbt!]
\centering
\subfloat[]{\includegraphics[width=.33\textwidth]{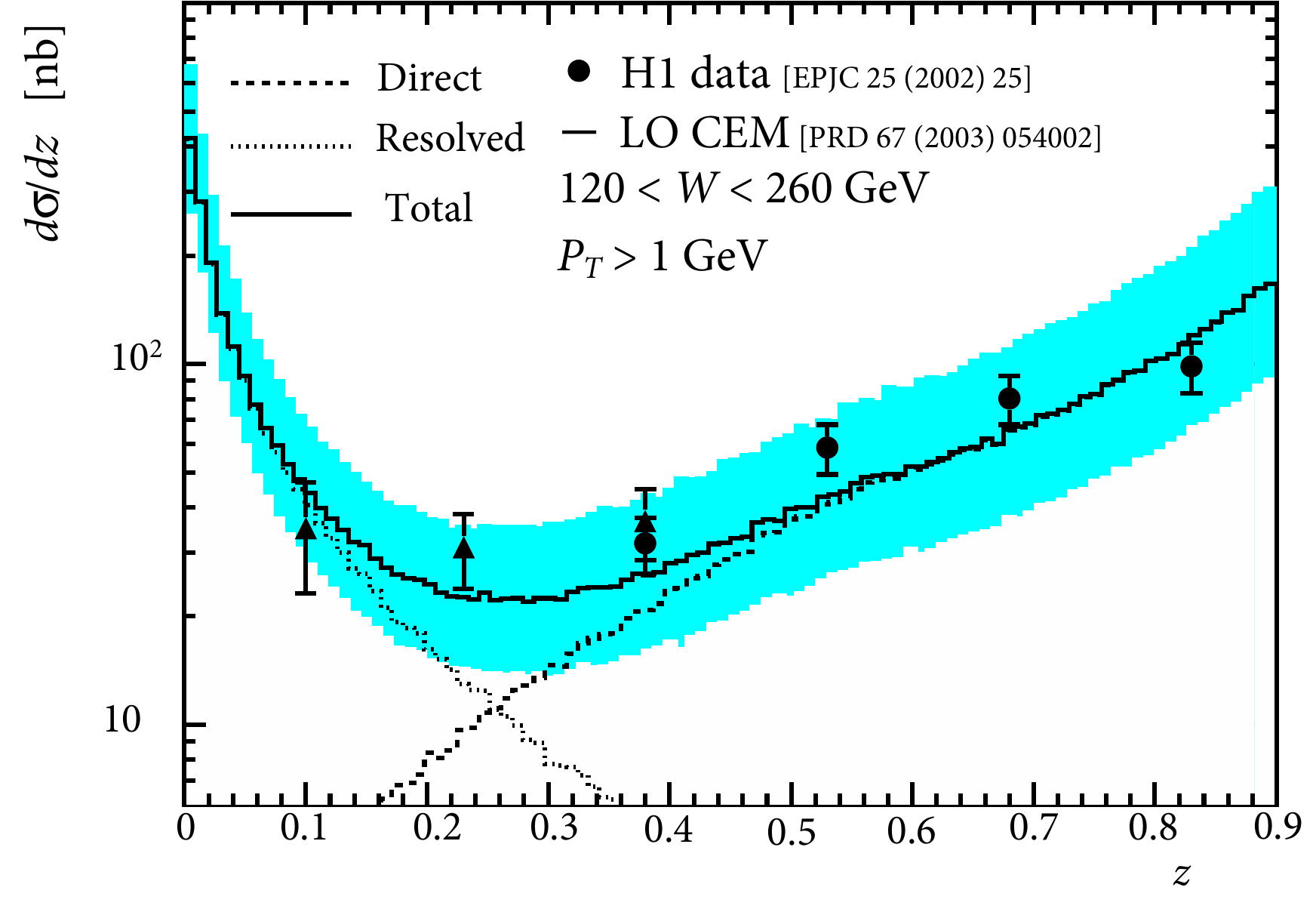}\label{fig:Eboli-0211161-zall}}
\subfloat[]{\includegraphics[width=.33\textwidth]{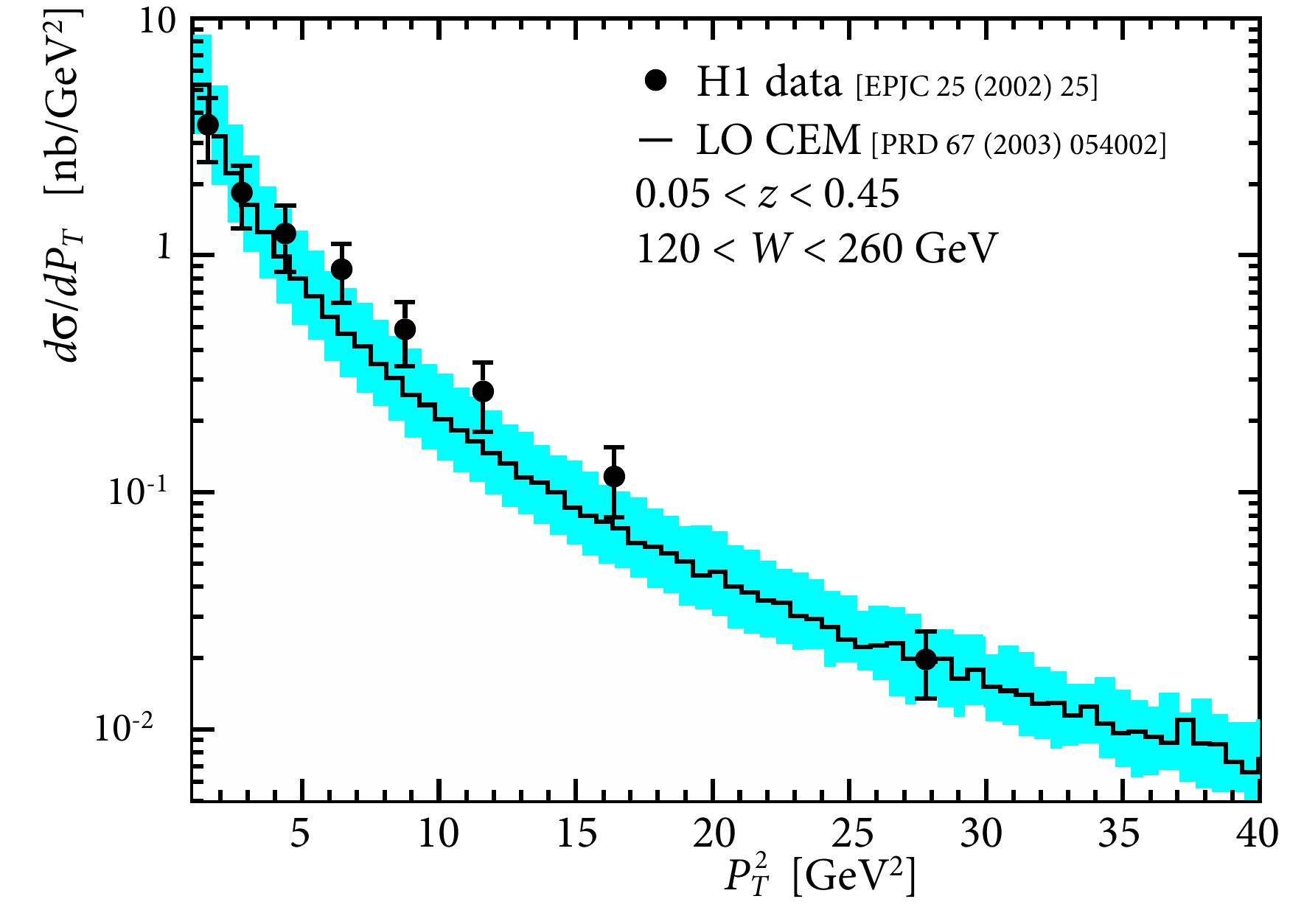}\label{fig:Eboli-0211161-ptlow}}
\subfloat[]{\includegraphics[width=.33\textwidth]{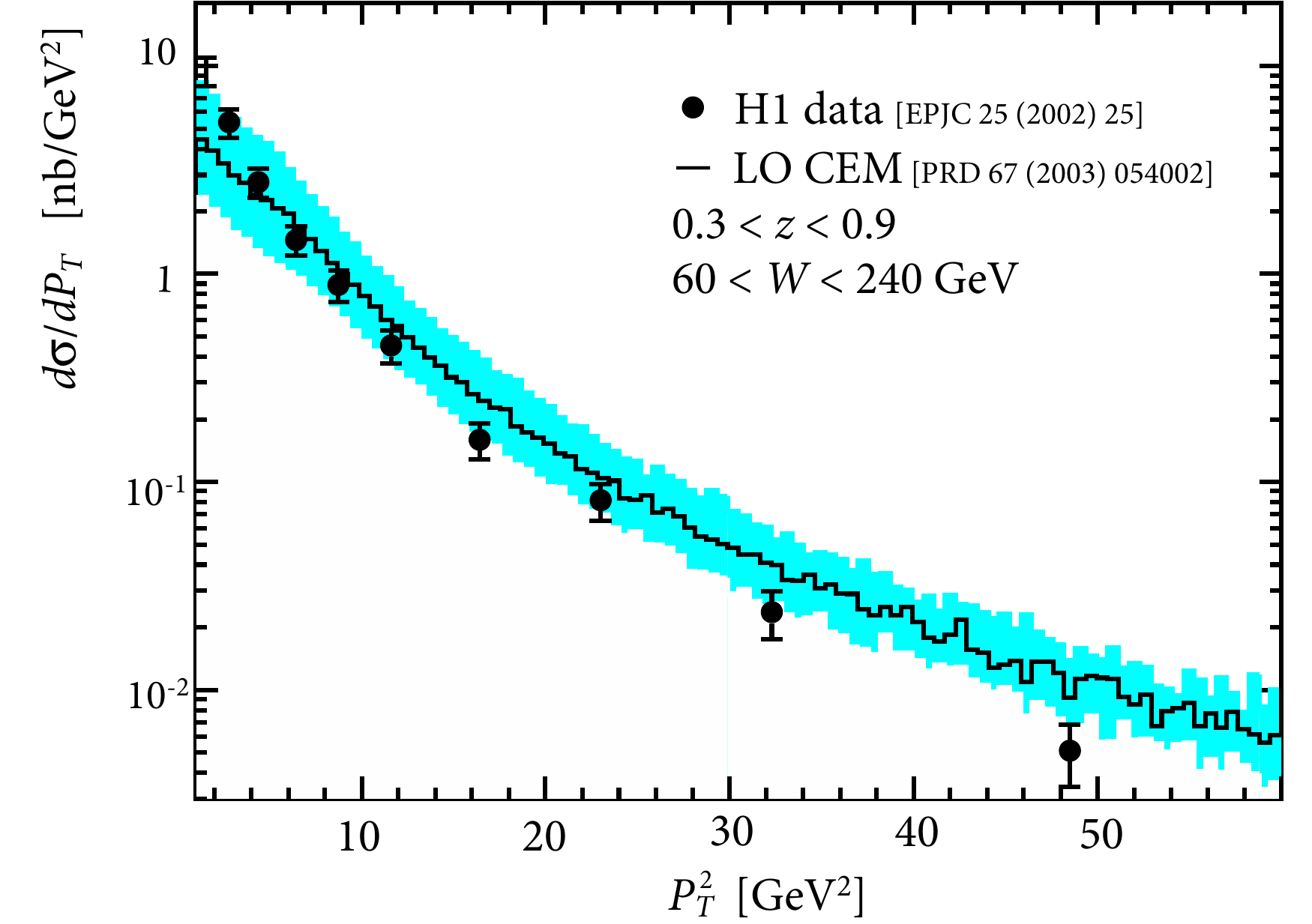}\label{fig:Eboli-0211161-ptmid}}
\caption{
$z$ and $P^2_T$-differential cross section for photoproduced $J/\psi$ predicted by the CEM at LO compared to H1 data~\cite{Adloff:2002ex}. The blue bands result from the variation of $m_c$ between 1.3 and 1.5 GeV. See~\cite{Eboli:2003fr} for the theory parameters used for the curves. Figures taken from~\cite{Eboli:2003fr}.}
\label{fig:CEM-photoprod-2003}
\end{figure}

Let us briefly review the main features of these studies. First, the total $J/\psi$ photoproduction data are well described by the CEM at NLO using $\P_{J/\psi} =0.048 \div 0.056$ with $m_c \simeq 1.45$~GeV. We refer to~\cite{Amundson:1996qr} for more details, in particular about the experimental data used for the comparison. In 1998, Eboli \etal~\cite{Eboli:1998xx} studied the $z$-differential cross section and tackled a similar peak at large $z$ as that found in NRQCD. Their procedure was inspired by approaches to reproduce the Drell-Yan $P_T$-differential cross section at low $P_T$ with a free parameter setting the relative size of the perturbative and non-perturbative
contributions at low $P_T$. Doing so, with $\P_{J/\psi}=0.048 \div 0.056$ (but with $m_c=1.3$~GeV) and  after fixing the corresponding free parameter, they were able to reproduce the H1~\cite{Aid:1996dn} and ZEUS~\cite{Breitweg:1997we}
 data differential in $z$ down to $z=0.4$. In their study, they also included the resolved-photon contributions which result in a prominent peak at low $z$.  In 2003, Eboli \etal~\cite{Eboli:2003fr} updated their study and compared their CEM cross section with new H1 data~\cite{Adloff:2002ex}. \cf{fig:Eboli-0211161-zall} shows a comparison of the inelasticity distribution. The CEM accounts relatively well for the H1 data --keeping in mind that the shape mainly results from a tuned parameter. \cf{fig:Eboli-0211161-ptlow} and \cf{fig:Eboli-0211161-ptmid} show comparisons for the $P_T^2$-differential cross section. At low $z$, where the yield is dominated by the resolved-photon contribution (see the dashed curve on \cf{fig:Eboli-0211161-zall}), the CEM is particularly successful. At larger $z$, the CEM exhibits the same issue as for hadroproduction with too a hard
$P_T^{(2)}$ spectrum.

Besides photoproduction, Eboli \etal~\cite{Eboli:2003qg} studied in 2003 the case of $\gamma \gamma$ scattering at LEP. Still with $\P_{J/\psi} =0.048 \div 0.056$ and $m_c=1.3$ GeV, they obtained for the CEM LO a very good description of the DELPHI data, very similar to the LO COM one of Klasen \etal\ \cite{Klasen:2001cu}. They also showed that the resolved-photon contributions were contributing about 10\% of the yield.

Finally, the CEM phenomenology was explored in $e^+e^-$ annihilation at $B$ factories by Kang \etal~\cite{Kang:2004zj} who found 0.095 pb for $\sigma(J/\psi+X_{\text{non } c \bar c})$ with $\P_{J/\psi}=0.025$ and $m_c=1.4$ GeV. This is more  4.5 times below the latest Belle measurement --and contrary to the COM case, the CSM cross section should not be added to this CEM value.

\subsubsection{Improving the CEM}
\label{sec:CEM_additions}

Our discussion has so far focused on the $J/\psi$ case with some comments on the $\Upsilon$. One of the main failure of the CEM phenomenology is however to be found on the treatment of the excited states. It is often reported that the CEM cannot explain of $\chi_Q$ yields which do not simply follow the $1:3:5$ ratio rule from the spin counting. In fact, whereas this is a little suspicious, it is not critical. What is however more problematic is that these ratios have been seen to depend on kinematical variables. We guide the reader to different reviews~\cite{Andronic:2015wma,Brambilla:2010cs,Brambilla:2004wf} for some specific cases, in particular that of the $\psi(2S)/J/\psi$ yield ratio. 

Motivated by this shortcoming of the conventional implementation of the CEM, Ma and Vogt suggested~\cite{Ma:2016exq} in 2016 to improve it by explicitly taking into account, in the kinematics, the fact that the invariant mass of the pair produced at short distances differs from that of the physical bound state. Such an effect can be paralleled to the mass effect in the decay of an excited state to a lower-lying state. This shifts momenta, modifies their spectrum and thus generates kinematical dependences in yield ratios between different states.

In order to take into account the kinematical shift between the pair before and after hadronisation, the improved CEM equation relating the cross section to produce a heavy quark pair $Q \bar Q$ and that to produce the quarkonium reads: 
\eqs{\label{eq:iCEM}
\frac{d \sigma_\Q(P_\Q)}{d^3 \vect P_\Q}= \P_\Q \int^{2M_{H(Q)}}_{2 M_Q} d^3 \vect P_{Q \bar Q} dM_{Q \bar Q}
\frac{d\sigma(M_{Q \bar Q},P_{Q \bar Q})}{dM_{Q \bar Q}d^3\vect P_{Q \bar Q}} 
\delta^{(3)}\left(\vect P_{\Q} - \frac{M_\Q}{M_{Q \bar Q}} \vect P_{Q \bar Q} \right),
}
where $P_{Q \bar Q}=P_Q + P_{\bar Q}$. This approaches has been dubbed improved CEM (ICEM). One can see on \cf{fig:ratio_1S2S_7TeV_ICEM}, the effect on the $\psi(2S)/J/\psi$ yield ratio as a function of $P_T$ in the LHCb acceptance at $\sqrt{s}=7$~TeV -- the conventional CEM predicts a constant. The effect is manifest at low $P_T$ when $P_T$ dependence quickly varies and then vanishes at higher $P_T$. Yet, CMS and ATLAS measurements at higher $P_T$ seem to indicate that the ratio further increases for $P_T$ between 10 and 20~GeV (see \eg\ \cite{Woehri:QWG2014}). So far, the ICEM has only been applied to hadroproduction.

\begin{figure}[ht!]
\centering
\subfloat{\includegraphics[width=0.65\textwidth]{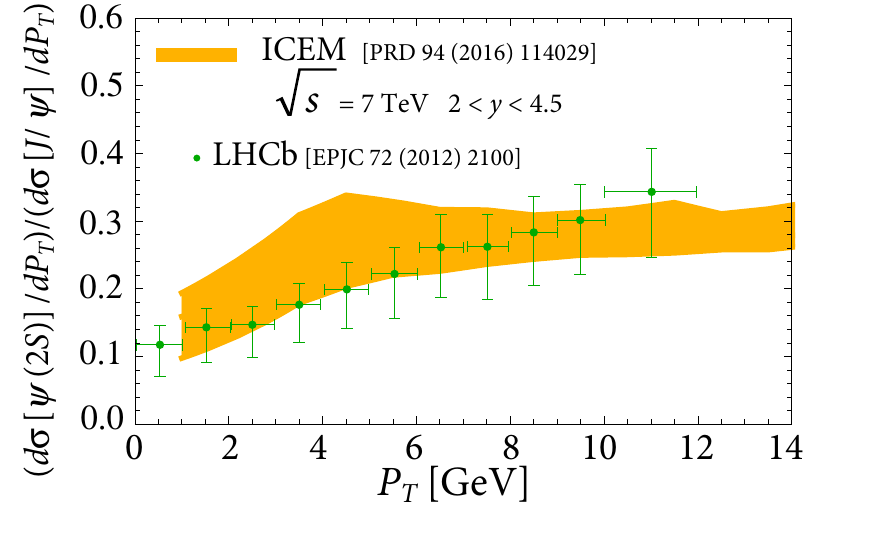}\label{fig:ratio_1S2S_7TeV_ICEM}}
\caption{$\psi(2S)/J/\psi$ yield ratio as a function of $P_T$ predicted by the ICEM at $\sqrt{s}=7$~TeV in the LHCb acceptance compared with LHCb data~\cite{Aaij:2012ag}. Adapted from~\cite{Ma:2016exq}.}
\label{ratio_1S2S_7TeV_ICEM}
\end{figure}

As what regards the individual cross sections, the ICEM dependence is slightly softer but we have checked that the effect is not strong enough --on the order of 20\% at $P_T\simeq 15$ GeV-- to solve the excess of the CEM over the data discussed in section~\ref{subsec:CEM_NLO_PT}. Modifying the kinematics of the initial gluons with a $k_T$-smearing (see \eg~\cite{Bedjidian:2004gd}) or $k_T$-factorisation (see \cite{Maciula:2018bex} and references therein) can also obviously affect the $P_T$ distribution of the produced quarkonium. In some implementations of the unintegrated gluon densities used in $k_T$-factorisation (compare the KMR- and JH-like curves in \cite{Maciula:2018bex}), the effect --which can consist in altering the yield by as much as one order of magnitude-- can propagate up to 20 GeV and above. However, such an effect would not impact the problematic double ratios.

Another improvement in the CEM phenomenology was recently brought in by Cheung and Vogt who pioneered (hadroproduction) polarisation studies~\cite{Cheung:2018upe,Cheung:2018tvq,Cheung:2017osx,Cheung:2017loo}. With this regard, two somewhat opposed views exist. On the one hand, it is often suggested (see \eg\ \cite{Brambilla:2004wf}) that, to be coherent with the randomisation effects induced the numerous gluon emissions during hadronisation, the CEM physics should correspond to the production of unpolarised quarkonia. Strictly speaking, this cannot be exact because of FD effects. For instance, a sample of unpolarised $P$ waves will feed into  $S$ waves with specific polarised yields. On the other hand, one can consider that only the colour is randomised by the numerous incoherent gluon emissions associated with the heavy-quark pair hadronisation. As such, one can keep track of the heavy-quark helicities and, under specific assumptions, derive predictions for the quarkonium polarisation. Cheung and Vogt assumed that the quark spins are ``frozen'' during hadronisation. We refer to~\cite{Vogt:2019zmr} for a neat summary of their results and we note that their computations at finite $P_T$ were done under the $k_T$ factorisation~\cite{Cheung:2018upe,Cheung:2018tvq}.

 In the $J/\psi$ case, the CEM yield is found to be unpolarised at mid $P_T$ and slightly
transverse at large $P_T$ in the helicity frame. The agreement with the CMS data is good but bad with those from ALICE, LHCb at the LHC and CDF at the Tevatron.
They also found the prompt $\Upsilon(nS)$ production to be unpolarised at $P_T \gtrsim M_\Upsilon$ irrespective of the chosen frame.  The $\Upsilon(1S)$ results agree well with the CMS values.  It is less true for the $\Upsilon(2S)$ and $\Upsilon(3S)$.  At low $P_T$, the yield is found to be markedly transversely polarised, a feature which is not observed in the LHCb and CDF data. In both cases, the computed polarisation was also found to be independent of the rapidity or the  collision energy at RHIC, the Tevatron and the LHC.

\section{Associated production of quarkonia}
\label{ch:associated}

As we have seen in the previous sections, quarkonium physics has reached a precision era
since a long time, with momentum spectra predicted and measured over as much as 7 orders of magnitude. Yet, a number of puzzles still challenge our understanding of their production mechanism,
and {\it de facto} of QCD at the interplay between its short- and long-distance domains.

With the advent of the LHC, data at higher energies, at higher transverse 
momenta, with higher precision and with more exclusivity towards direct 
production are now flowing in. Unfortunately, all this may not be sufficient 
to pin down the complexity of the quarkonium-production mechanism. In this context, 
much hope is put into the study of associated-quarkonium 
production, which is the object of this section.

\subsection{Digression about double parton scatterings}

Most of the discussion of single-quarkonium production made in the previous sections
relied on the picture that it arises from a single scattering of two partons. 
This is considered to be the leading-twist contribution as other contributions
like scatterings involving three partons~\cite{Alonso:1989pz,Motyka:2015kta,Schmidt:2018gep,Levin:2018qxa} 
should be suppressed by inverse powers of the hard scale of the process, $Q$. 
In the present case, $Q$ is at least the quarkonium mass, which is significantly
larger than the  hadronic scale, $\Lambda$, on the order of $m_p$ or $\Lambda_{\rm QCD}$. 

Along these lines, the cross section for a Single Parton Scattering (SPS) is expected to
scale as $\sigma^{\rm SPS} \sim 1/Q^2$ and, for a Double Parton Scattering (DPS), it is expected
to scale as $\sigma^{\rm DPS} \sim \Lambda^2/Q^4$, thus to be suppressed. From such dimensional
arguments, one would thus expect that, if one observes two perturbatively produced particles, 
they are more likely to come from a SPS than a DPS. Yet, one should keep in mind that
$\sigma_{\rm DPS}$ is proportional to the the fourth power of the PDFs which increase
with the collision energy. Second, for configurations where the momenta of both observed particles
are equal and smaller than $Q$, the DPS and SPS differential cross sections can be of similar magnitudes with: 
\eqs{
\frac{d\sigma^{\rm SPS}}{d^2q_1 d^2q_2} \sim \frac{d\sigma^{\rm DPS}}{d^2q_1 d^2q_2} 
\sim \frac{1}{Q^4\Lambda^2}.
}

In the context of our discussion of quarkonia hadroproduced in association with another particle, 
it therefore makes perfect sense to include the DPS in our considerations. As such 
let us elaborate a little more on their expected contributions.
The description of such a mechanism is usually done by assuming that DPSs can be 
factorised into two SPSs for the production of {\it each} of both observed particles. 
This factorisation massively used at the Tevatron and the LHC 
--and from which results distinctive kinematic distributions-- is
certainly a first rough approximation. It is however sometimes justified by the fact 
that possible unfactorisable corrections due to parton correlations are larger
in the valence region than at low $x$ -- the realm of the LHC. 
In the case of the associated-quarkonium hadroproduction, 
the master formula from which one starts under the factorisation 
assumption is~(see e.g.~\cite{dEnterria:2013mrp}) [$\P_1$ and $\P_2$ label both
observed particles]
\bqa
\sigma^{\rm DPS}_{\P_1\P_2}&=&\frac{1}{1+\delta_{\P_1\P_2}}\sum_{i,j,k,l}{\int{dx_1dx_2dx_1^{\prime}dx_2^{\prime}}d^2{\bold b_1}d^2{\bold b_2}d^2{\bold b}}
\nonumber\\
&&\times \, \Gamma_{ij}(x_1,x_2,{\bold b_1},{\bold b_2}) \, \hat{\sigma}^{\P_1}_{ik}(x_1,x_1^{\prime})\, \hat{\sigma}^{\P_2}_{jl}(x_2,x_2^{\prime})\, 
\Gamma_{kl}(x_1^{\prime},x_2^{\prime},{\bold b_1}-{\bold b},{\bold b_2}-{\bold b}) ,
\eqa
where   $\Gamma_{ij}(x_1,x_2,{\bold b_1},{\bold b_2})$ is the generalised double distributions with the longitudinal fractions $x_1$,$x_2$ and the transverse impact parameters ${\bold b_1}$ and ${\bold b_2}$, $\hat{\sigma}^{\P_i}_{jk}(x_l,x_l^{\prime})$ are the usual {\it partonic} cross sections for the inclusive production 
of the particle $\P_i$ and $\delta_{\P_1\P_2}$ is the Kronecker symbol. To go further, 
one decomposes $\Gamma_{ij}(x_1,x_2,{\bold b_1},{\bold b_2})$ into independent longitudinal and transverse parts
\bq
\Gamma_{ij}(x_1,x_2,{\bold b_1},{\bold b_2})=D_{ij}(x_1,x_2)T_{ij}({\bold b_1},{\bold b_2}),
\eq
where  $D_{ij}(x_1,x_2)$ is the double-parton distribution functions (dPDF)~\cite{Gaunt:2009re}. 

If one then ignores the correlations between the partons emerging from each hadrons and which trigger
both scatterings, one can also assume that
\bqa
D_{ij}(x_1,x_2)=f_i(x_1)f_j(x_2),~~~
T_{ij}({\bold b_1},{\bold b_2})=T_i({\bold b_1})T_j({\bold b_2}),
\eqa
where $f_i(x_1)$ and $f_j(x_2)$ are the normal single PDFs. This yields to [$\sigma_{ik\to\P_j}$
is the partial contribution of the {\it hadronic} cross section to produce  $\P_j$ from $ik$ fusion]
\bqa
\sigma^{\rm DPS}_{\P_1\P_2}=\frac{1}{1+\delta_{\P_1\P_2}}\sum_{i,j,k,l}{\sigma_{ik\to\P_1}\sigma_{jl\to\P_2}}\int{d^2{\bold b}}\!\!
\int{\!T_i({\bold b_1})T_k({\bold b_1}-{\bold b})d^2{\bold b_1}}\!
\int{\!T_j({\bold b_2})T_l({\bold b_2}-{\bold b})d^2{\bold b_2}}.
\eqa
If the parton-flavour dependence in $T_{i,j,k,l}({\bold b})$ is also ignored, 
one can define a single overlapping function, 
\bq
F({\bold b})=\int{T({\bold b_i})T({\bold b_i}-{\bold b})d^2{\bold b_i}},
\eq
whose inverse of the integral, 
\bq
\sigma_{\rm eff}=\left[\int{d^2{\bold b}F({\bold b})^2}\right]^{-1}.
\eq
is a parameter characterising the effective spatial area of the parton-parton interactions,
from which one derives the so-called ``pocket formula"
\bq
\sigma^{\rm DPS}_{\P_1\P_2}=\frac{1}{1+\delta_{\P_1\P_2}}\frac{\sigma_{\P_1}\sigma_{\P_2}}{\sigma_{\rm eff}},\label{eq:dpseq}
\eq
where $\sigma_{\P_1}$ and $\sigma_{\P_2}$ are the {\it hadronic} 
cross sections for respectively single $\P_1$ and $\P_2$ production. This means that knowing
$\sigma_{\P_1}$ and $\sigma_{\P_2}$ and $\sigma_{\rm eff}$, one can predict $\sigma^{\rm DPS}_{\P_1\P_2}$.

Under these assumptions, $\sigma_{\rm eff}$ is only related to the initial state and should be independent of the final state. However, the validation of its universality (process independence as well as energy independence) and the factorisation in \ce{eq:dpseq} should be verified case by case. In fact, some factorisation-breaking effects have recently been identified (see \eg\ \cite{Gaunt:2011xd,Diehl:2011tt,Blok:2013bpa,Kasemets:2012pr,Diehl:2014vaa}).

\subsection{Quarkonium-pair production}
\label{sec:onium-pair}
Inclusive quarkonium-pair production
has been the object of a number of recent studies at the LHC and the 
Tevatron~\cite{Aaij:2011yc,Abazov:2014qba,Khachatryan:2014iia,Aaboud:2016fzt,Aaij:2016bqq,Sirunyan:2020txn}.
We should however stress that the first observation of $J/\psi$-pair events dates back to that of the 
CERN-NA3 Collaboration~\cite{Badier:1982ae,Badier:1985ri} in the early 1980's. Strictly speaking, this is not a ``new'' observable,
but it is probably the richest of all quarkonium-associated-production channels.

In the case of di-$J/\psi$ production,
all the experimental results for small rapidity separations ($\Delta y$) agree with 
the SPS CSM contributions, known up to NLO accuracy~\cite{Lansberg:2013qka,Sun:2014gca,Likhoded:2016zmk}.
However, for increasing $\Delta y$, they point at a significant DPS -- 
in accordance with previously studied jet observables~\cite{Akesson:1986iv,Alitti:1991rd,Abe:1993rv,Abe:1997xk,Abazov:2009gc,Aad:2013bjm,Chatrchyan:2013xxa}, 
except for a likely smaller $\sigma_{\rm eff}$.
In addition, the measured yields at large pair transverse momenta, which agree with the
NLO CSM contributions at $\alpha_s^5$, tell us that the pair is produced with a hard recoiling parton.
At the same time, the debate remains open about the relevance
of COM contributions in some part of the phase space~\cite{He:2015qya}. 
In the following, we will review all these aspects.
Recently, we also showed~\cite{Lansberg:2017dzg,Scarpa:2019fol} 
that di-$J/\psi$ production can also tell us much information about the gluon TMDs.

Besides the specific case of $J/\psi$ pairs, $\Upsilon+J/\psi$ hadroproduction --expected
to be a golden probe of the COM~\cite{Ko:2010xy}-- has also been 
seen by the D0 collaboration~\cite{Abazov:2015fbl} with a claim that the 
yield is highly dominated by DPS contributions which was further
confirmed by Shao and Zhang~\cite{Shao:2016wor}. 
Finally, $\Upsilon$-pair hadroproduction was measured for the first time in 2016 by CMS with 
a limited sample~\cite{Khachatryan:2016ydm} of 40 events though. A second CMS study was released in 2020 based
on 150 events~\cite{Sirunyan:2020txn}. With the current
admittedly large experimental and theoretical uncertainties, the measured
cross sections agree with the CS based computations~\cite{Li:2009ug,Ko:2010xy,Berezhnoy:2012tu,Shao:2016knn}.

\subsubsection{DPS in $J/\psi$-pair hadroproduction}

Even before the final LHCb data~\cite{Aaij:2011yc} were released, Kom \etal\ suggested~\cite{Kom:2011bd} in 2011
that a significant fraction of the di-$J/\psi$ events seen by LHCb might be from DPS origin. 
This thus drew the attention 
on  the possibility that quarkonium-pair hadroproduction in general could also
come from two independent quarkonium-production scatterings. Kom \etal\ also proposed
the yield as a function of the rapidity difference, $\Delta y$, 
between both $J/\psi$'s as a good observable to distinguish DPS and SPS events.
The DPS  events  compared to the SPS ones indeed have a broader $\Delta y$ distribution. 
In a single scattering, large values of 
$\Delta y$ imply large momentum transfers, thus highly off-shell particles\footnote{Except in the case
of ``pseudo-diffractive" reactions~\cite{Baranov:2012re} appearing at NNLO.}, and are strongly suppressed. It is not 
the case for DPSs where the rapidity of both particles is independent. 
Large rapidity differences are only suppressed because the individual yields are suppressed for increasing rapidities. In turn, 
Kom \etal\ also warned about the usage of the azimuthal angular separation $\Delta \phi$, which
might lead to wrong conclusions about the DPS/SPS ratio. This was confirmed later on by further 
computations~\cite{Lansberg:2013qka,Baranov:2015cle,Borschensky:2016nkv}. 
Using $\sigma_{\rm eff} = 14.5$ mb from the CDF 
$\gamma+3 \text{ jets}$ analysis \cite{Abe:1997xk}, they obtained a yield close to that of LHCb
which however shows large uncertainties. Let us note that such a value also
agrees with other jet-related observables~\cite{Akesson:1986iv,Alitti:1991rd,Abe:1993rv,Abe:1997xk,Abazov:2009gc,Aad:2013bjm,Chatrchyan:2013xxa}.

In 2014, D0~\cite{Abazov:2014qba} --thanks to the wide (pseudo)rapidity coverage 
of about 4 units ($|\eta|\leq 2$) of their detector-- reported the first extraction 
of the DPS contributions to  $J/\psi$-pair production.
By fitting the  $\Delta \eta$  
distribution of their data, D0 managed to separate out the DPS and the SPS yields\footnote{We however note
that this separation relied on Monte Carlo SPS templates derived from LO CS computations. As we will see later
NLO corrections to the SPS yield can also alter the $\Delta y$ distribution.}, \ie\
 $\sigma^{\rm DPS}_{\psi\psi}$ and $\sigma^{\rm SPS}_{\psi\psi}$ and reported 
that a little less than half of the events was from DPSs.

Injecting their measured $\sigma_\psi$ and $\sigma^{\rm DPS}_{\psi\psi}$ in the DPS ``pocket formula", \ce{eq:dpseq},
 D0 reported $\sigma_{\rm eff} \simeq 5.0 \pm 2.75~{\rm mb}$. Such a value is 3 times {\it lower}
than those of the jet analyses which we just mentioned. In other words, the DPS yield 
in the D0 acceptance is roughly 3 times {\it higher} than what could have been naively expected. 
Yet, as we will see later, such a low value of $\sigma_{\rm eff}$ is in line with other
quarkonium-associated-production  measurements, including other quarkonium-pair data.

\subsubsection{LO and NLO SPS computations of di-$J/\psi$ hadroproduction}

\paragraph{CSM LO and NLO evaluations.}

Historically, the first CSM computations for direct di-$J/\psi$ hadroproduction 
were carried out in the early 1980's 
by Kartvelishvili \& Esakiya~\cite{Kartvelishvili:1984ur} and Humpert~\cite{Humpert:1983yj}.
Analytical results for the partonic cross section were published by Qiao in 2002~\cite{Qiao:2002rh}.
The $P_T$-integrated rates obtained by recent theoretical expectations 
at the LHC~\cite{Li:2009ug,Ko:2010xy,Berezhnoy:2011xy} follow the same lines and, in fact, 
happen to agree with the LHCb data --the only ones extending down 
to $P_T=0$. There is clearly no deficit compared to the data, especially
since DPS are also expected to populate the yield.

In practice, the amplitude to produce of a pair of $S$-wave 
quarkonia denoted ${\Q_1}$ and  ${\Q_2}$, of given momenta $P_{1,2}$ and of polarisation $\lambda_{1,2}$ 
accompanied by other partons --inclusive production--, noted $k$, is then given by the product 
of (i) the amplitude to create the corresponding double heavy-quark pair, in the specific kinematic 
configuration where the relative momenta of these heavy quarks ($p_{1,2}$) in each pairs is zero, (ii) 
two spin projectors $N(\lambda_{1,2}| s_{1,3},s_{2,4})$ and (iii) $R_{1,2}(0)$, 
the radial wave functions at the origin in the configuration space for both quarkonia. 
At the partonic level, the SPS amplitude thus reads: 
\eqs{\label{eq:amplitude_di-onium-CS}
{\cal M}(ab \to {\Q_1}^{\lambda_1}(P_1)+{\Q_2}^{\lambda_2}(P_2)+k)=
&\sum_{s_1,s_2,c_1,c_2}\sum_{s_3,s_4,c_3,c_4}\!\!\!\!\frac{N(\lambda_1| s_1,s_2)N(\lambda_2| s_3,s_4)}{ \sqrt{m_{\Q_1}m_{\Q_2}}} \frac{\delta^{c_1c_2}\delta^{c_3c_4}}{3}\frac{R_1(0)R_2(0)}{4 \pi}\\
&\times {\cal M}(ab \to Q^{s_1}_{c_1} \bar Q^{s_2}_{c_2}(\mathbf{p_1}=\mathbf{0}) + Q^{s_3}_{c_3} \bar Q^{s_4}_{c_4}(\mathbf{p_2}=\mathbf{0}) + k),}
where one defines from the heavy-quark momenta, $q_{1,2,3,4}$, 
$P_{1,2}=q_{1,3}+q_{2,4}$, $p_{1,2}=(q_{1,3}-q_{2,4})/2$, and where $s_{1,3}$,$s_{2,4}$ are the 
heavy-quark spin components and $\delta^{c_ic_j}/\sqrt{3}$  is the colour projector. $N(\lambda| s_i,s_j)$ 
is the same as for single quarkonium production (see section~\ref{subsec:CSM_NLO_PT}). 
Such a partonic amplitude is then squared, summed over the
colour and spin of external partons and, under the collinear factorisation, convoluted with the PDFs in the allowed
kinematical phase space. Analytical results for the uncontracted amplitude for $gg \to ^3\!\!S_1 + ^3\!\!S_1$ 
are available in~\cite{Qiao:2009kg} and those for the amplitude squared in~\cite{Qiao:2002rh,Qiao:2009kg,Li:2009ug}.
Let us also note the existence, since 2010, of the event generator {\small \sc DJpsiFDC}~\cite{Qiao:2010kn} 
specific to this process. Nowadays, \HELACOnia~\cite{Shao:2012iz,Shao:2015vga} also allows for the generation
of di-$J/\psi$ events but it is much more general.

\HELACOnia~\cite{Shao:2012iz,Shao:2015vga} 
 allows anyone to numerically compute cross sections for any tree-level process 
involving pairs of quarkonia. In particular, this allowed us to 
perform~\cite{Lansberg:2013qka} the first evaluation of the impact of the QCD 
corrections 
and to demonstrate that, like for single vector quarkonium production, QCD corrections
drastically affect the $P_T$-differential cross section\footnote{Unless specified otherwise,  $P^\psi_T$
is the $P_T$ of one $J/\psi$ randomly chosen among both and the $P_T$ cuts discussed here apply to both.}. It is in fact expected from the number of off-shell quark propagators of typical LO and $P_T$-enhanced real-emission NLO graphs, depicted on \cf{diagram-psi-psi-LO} and \cf{diagram-psi-psi-NLO-PT6}.

\begin{figure*}[hbt!]
\centering
\subfloat[]{\includegraphics[width=.15\textwidth,draft=false]{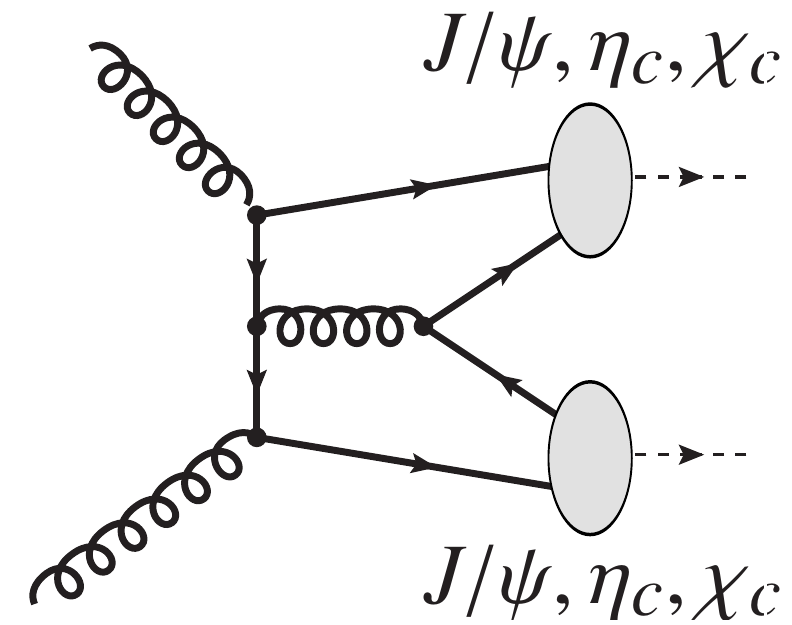}\label{diagram-psi-psi-LO}}
\subfloat[]{\includegraphics[width=.15\textwidth,draft=false]{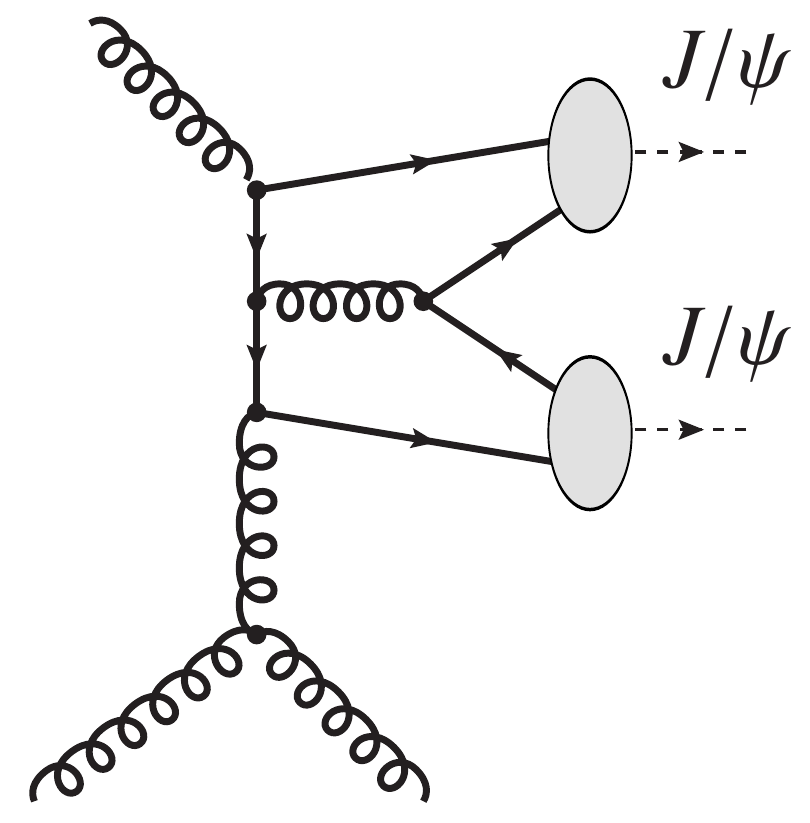}\label{diagram-psi-psi-NLO-PT6}}
\subfloat[]{\includegraphics[width=.15\textwidth,draft=false]{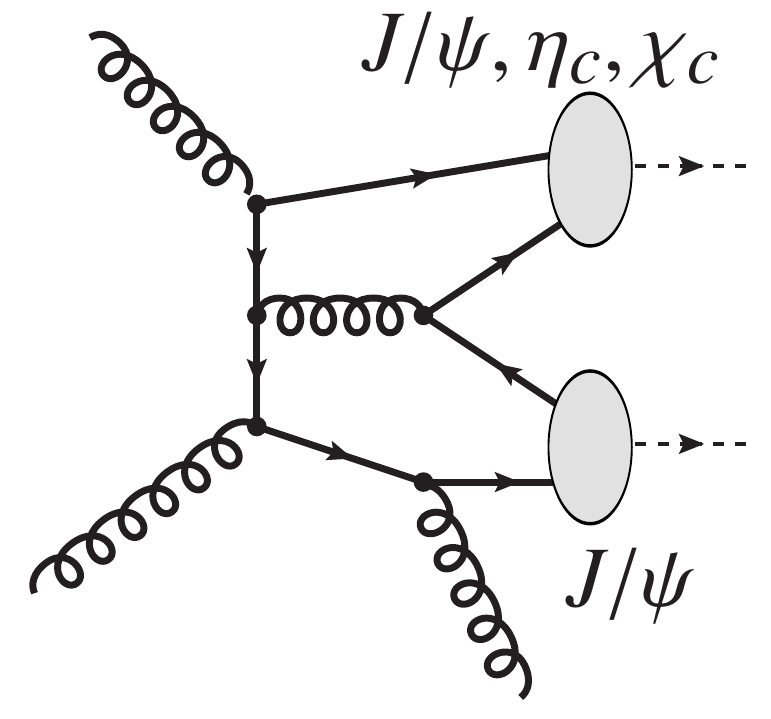}\label{diagram-psi-etac}}
\subfloat[]{\includegraphics[width=.15\textwidth,draft=false]{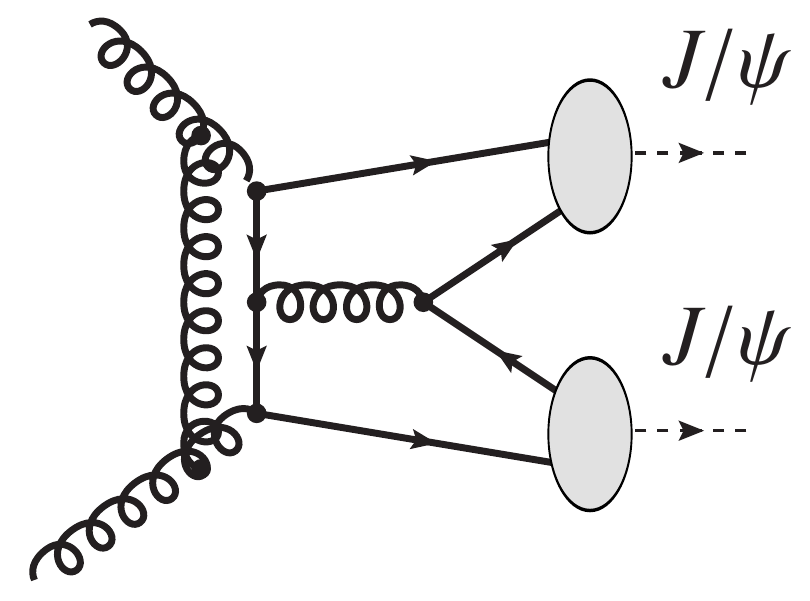}\label{diagram-psi-psi-NLO-loop}}\\
\subfloat[]{\includegraphics[width=.15\textwidth,draft=false]{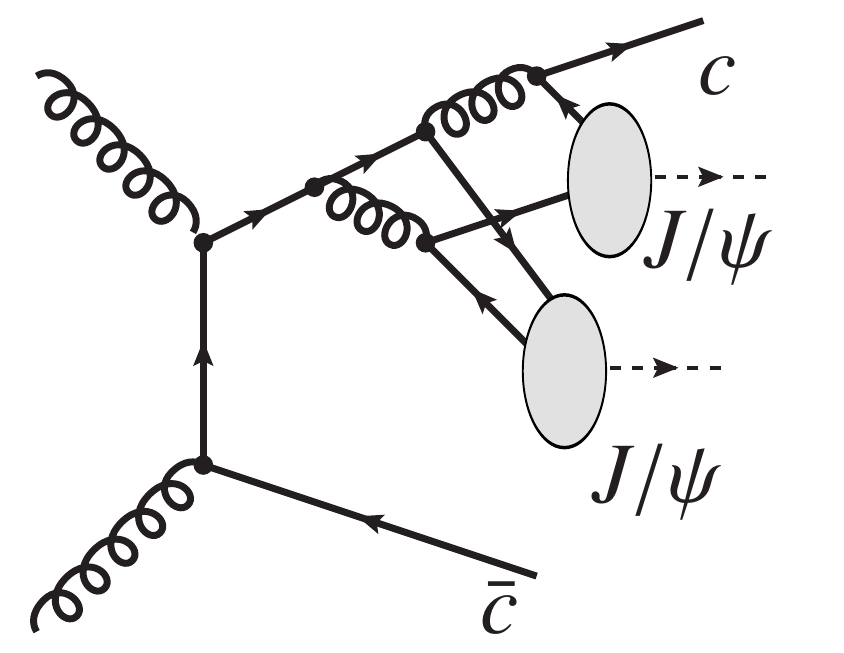}\label{diagram-psi-psi-e}}
\subfloat[]{\includegraphics[width=.15\textwidth,draft=false]{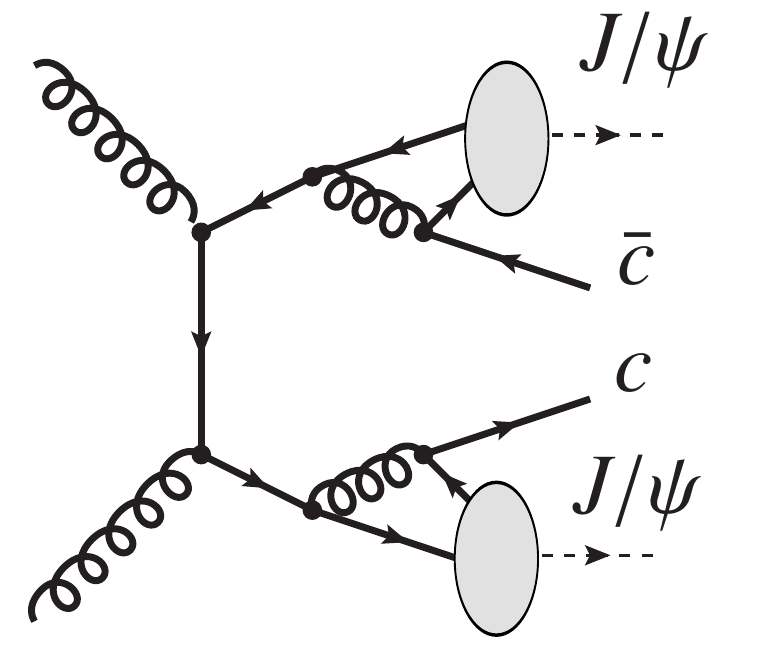}\label{diagram-psi-psi-f}}
\subfloat[]{\includegraphics[width=.15\textwidth,draft=false]{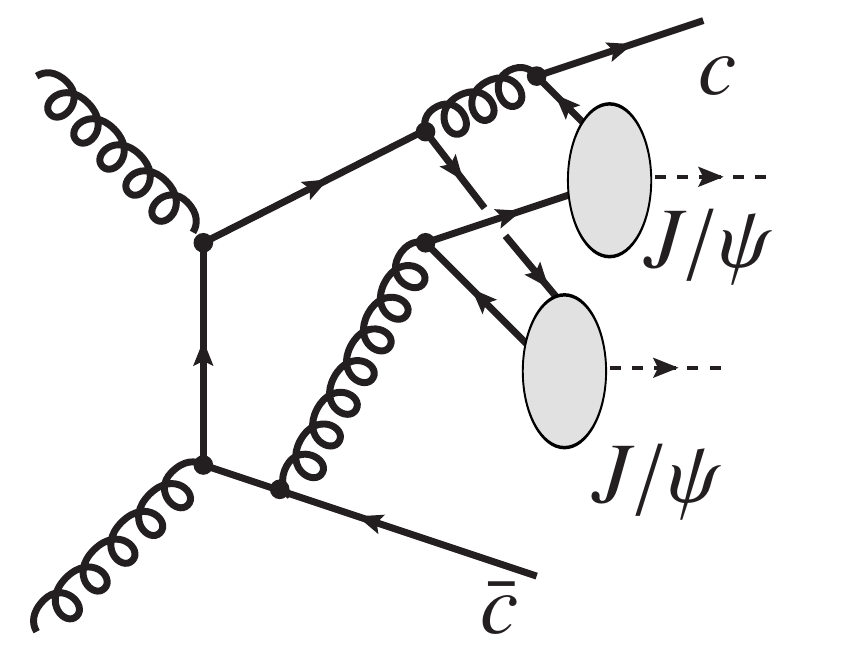}\label{diagram-psi-psi-f2}}
\subfloat[]{\includegraphics[width=.15\textwidth,draft=false]{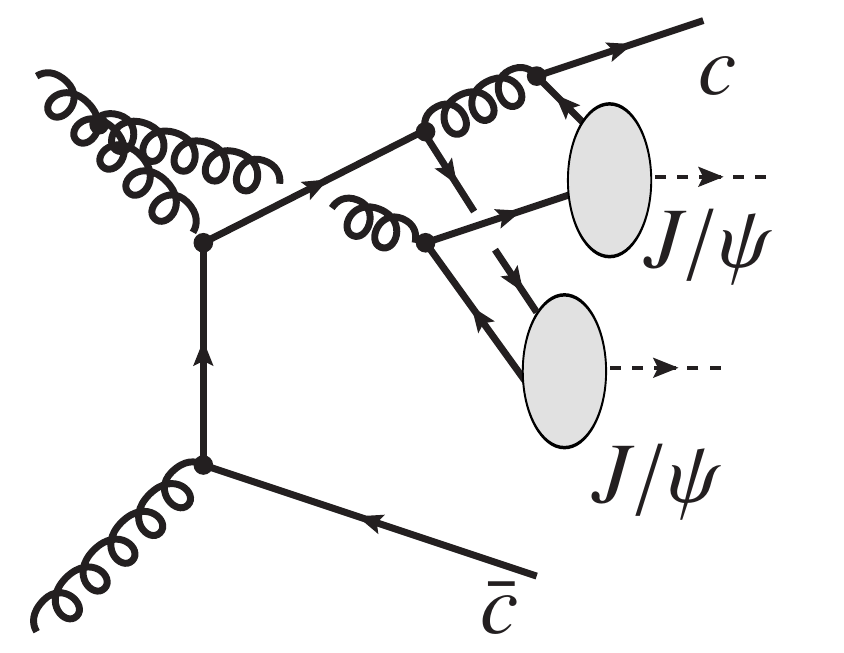}\label{diagram-psi-psi-f3}}
\caption{Representative diagrams for the hadroproduction of charmonium pairs via SPSs at $\mathcal{O}(\alpha_s^4)$~(a), at $\mathcal{O}(\alpha_s^5)$ (b-d) and 
 $\mathcal{O}(\alpha_s^6)$ (e-h).}
\label{diagrams-psipsi}
\end{figure*}

Thanks to this observation, a NLO$^\star$ computation as in~\cite{Artoisenet:2008fc} (see section \ref{subsec:NNLOstar-CSM}) is applicable. The IR divergences  of the real emissions are cut
with the constraint that the invariant mass of any light-parton pair, $s_{ij}$, 
be larger than the IR cut-off $s_{ij}^{\rm{min}}$. By inspecting all the propagators in \cf{diagram-psi-psi-NLO-PT6} or
\cf{diagram-psi-etac} , 
one can indeed see that the condition $s_{ij}>s_{ij}^{\rm{min}}$ regulates 
all the collinear and soft divergences in the real-emission corrections 
to $J/\psi$-pair production. In fact, 
for a cross section differential in $P^{\psi\psi}_T$, the IR divergences are simply absent
and the loop corrections do not contribute for finite $P^{\psi\psi}_T$.

Just like single $J/\psi$ production, 
such an IR treatment is expected to give a reliable estimation of the NLO result at 
least at large $P_T$ --and probably at mid $P_T$. Indeed, for the new $P_T$-enhanced 
topologies appearing at NLO, from \eg~the 
$t$-channel-exchange diagram shown \cf{diagram-psi-psi-NLO-PT6}, $s_{ij}$ will necessarily be large for any 
light-parton pair at large $P_T$ and such topologies would not be affected by the cut-off. For the remaining real-emission topologies, one may encounter large logarithms of
$s_{ij}/s_{ij}^{\rm{min}}$. These are however factors of the amplitudes of these $P_T$-suppressed topologies. As a consequence,
the dependence on $s_{ij}^{\rm{min}}$ should vanish as inverse powers of $P_T$. 
Lastly, the virtual corrections, with the same $P_T$-scaling as the LO contributions, 
are also $P_T$ suppressed compared to the $P_T$-enhanced real-emission contributions. 
One can simply neglect them since they are no longer necessary to regulate the IR divergences
of the real-emission contributions.

 \cf{fig:psi-psi-dsigdPT-LHCb} clearly shows\footnote{As regards the parameters entering the computations, $|R_{J/\psi,\eta_c}(0)|^2=0.81$~GeV$^3$ and
$M_{J/\psi,\eta_c}=2m_c$. The uncertainty bands are obtained from the {\it combined} 
variations of $m_c=1.5\pm 0.1$ GeV, with the factorisation $\mu_F$ and the renormalisation $\mu_R$ 
scales chosen among the couples $(0.5 \mu_0, 2 \mu_0)$, where $\mu_0=m_T=\sqrt{(4m_c)^2+p_T^2}$.}
 how the harder scaling of NLO$^\star$ yield (gray band)
compares to that of the LO yield (blue band). We also noted~\cite{Lansberg:2013qka} that the IR cut-off 
sensitivity vanishes extremely quickly and is in practice negligible for any purpose
for $P_T > 10$~GeV.

\begin{figure}[hbt!]\centering
\subfloat[]{\includegraphics[width=0.425\textwidth]{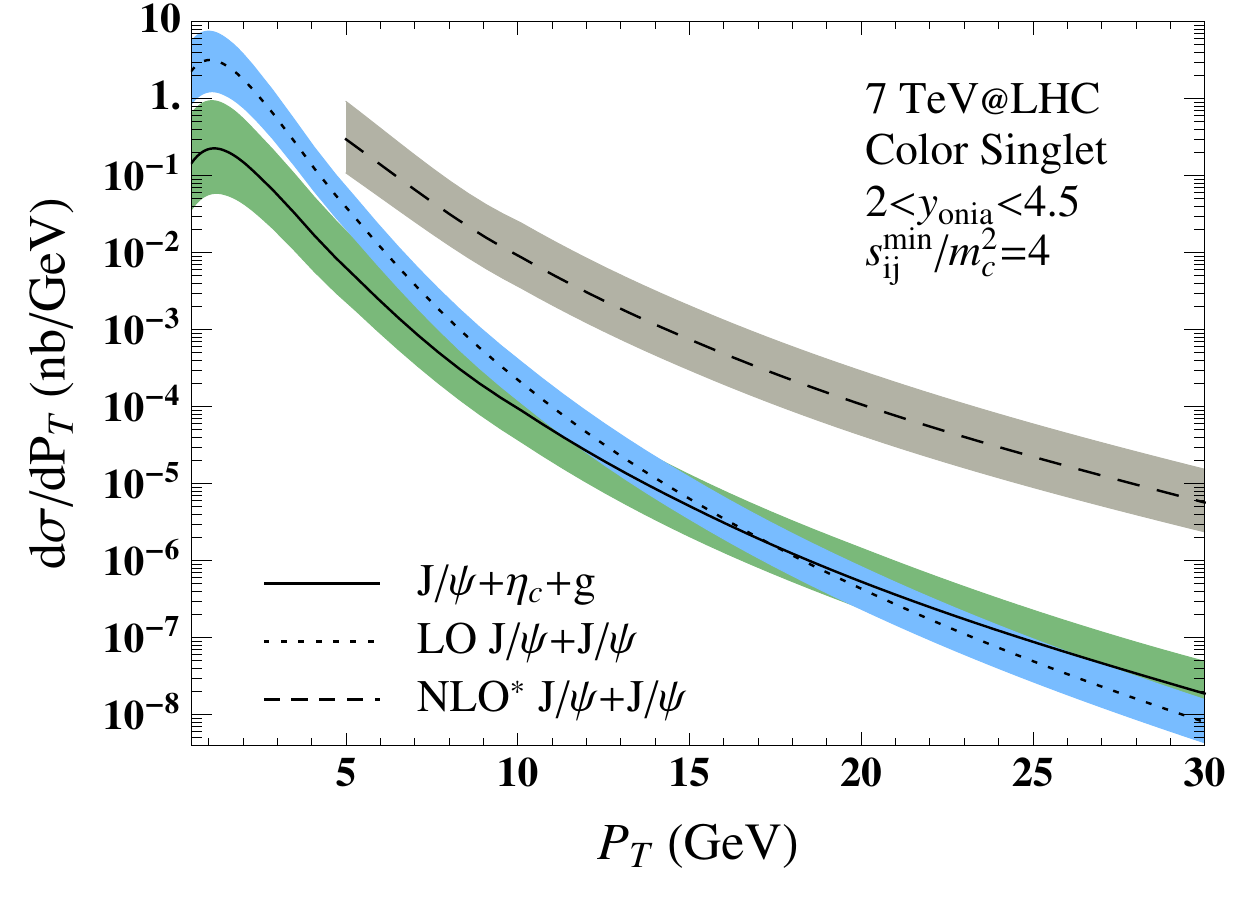}\label{fig:psi-psi-dsigdPT-LHCb}}
\subfloat[]{\includegraphics[width=.4\textwidth]{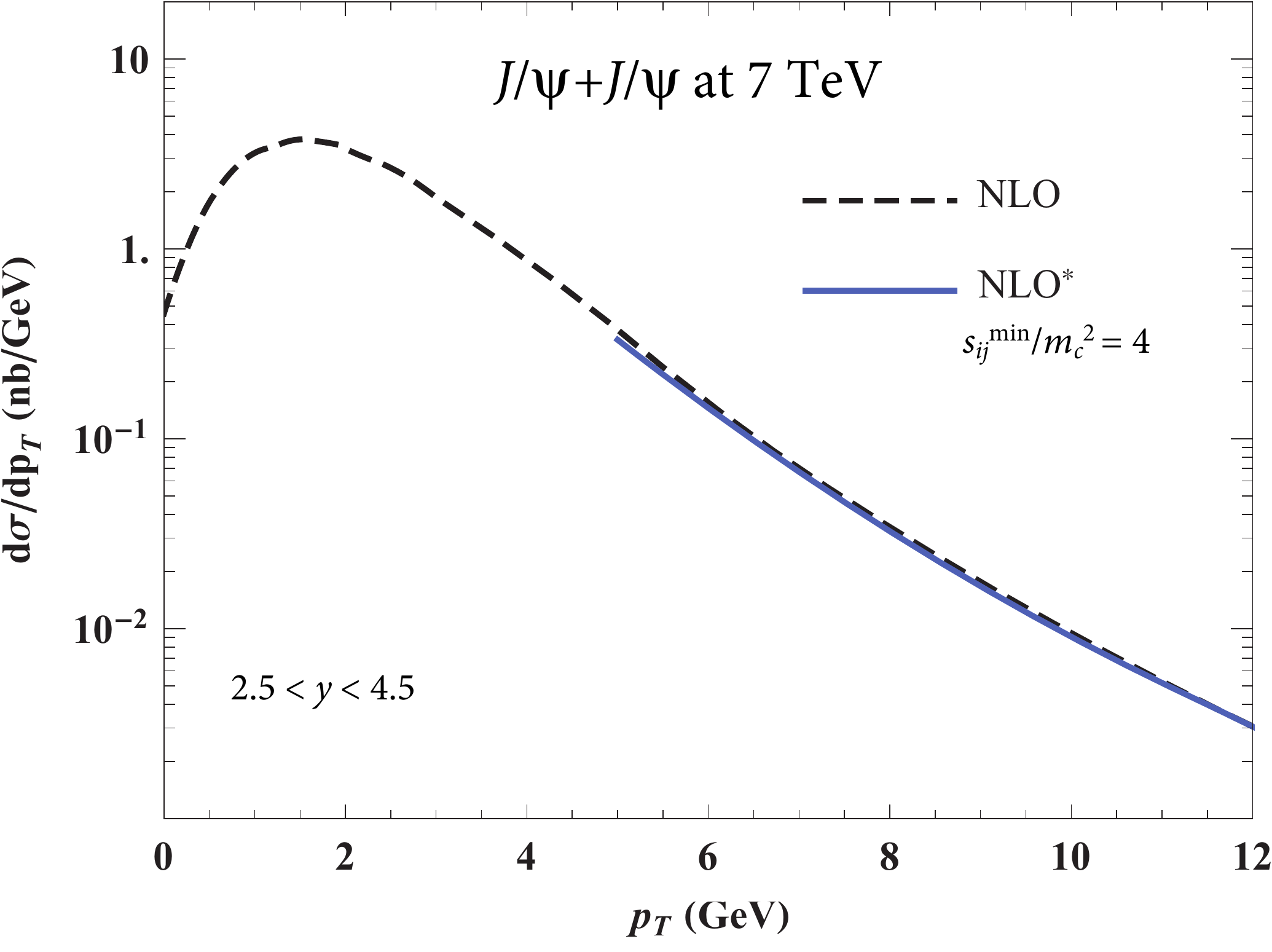}\label{fig:psi-psi-NLO-vs-NLOstar}}
\caption{Di-$J/\psi$ production at $\sqrt{s}=7$~TeV in the LHCb rapidity region (a) comparison between the LO and NLO$^\star$
predictions along with the first computation of the $J/\psi+\eta_c$ cross section; (b) comparison between the $P_T$-differential cross sections 
from NLO and NLO$^\star$ computations. Taken from (a) \cite{Lansberg:2013qka} and (b) \cite{Sun:2014gca} v1.}
\end{figure}

In 2014, Sun, Hao and Chao~\cite{Sun:2014gca} completed the full NLO computation, \ie\ the 
first complete one-loop computation for the hadroproduction of a pair of $J/\psi$. Without any surprise 
their computations confirmed our NLO$^\star$ evaluation (see~\cf{fig:psi-psi-NLO-vs-NLOstar}) which was then subsequently used for many other phenomenological studies, which we discuss later.

In the context of the discussion of the CMS data~\cite{Khachatryan:2014iia} extending to higher $P_T$, \HELACOnia~allows one to advance the investigations further and study the possible relevance
of $P_T^{-4}$ contributions in the region of large values of $P_{T\rm min}=\min(P_{T1},P_{T2})$. 
In fact, we could perform~\cite{Lansberg:2014swa} the first exact --gauge-invariant and IR-safe-- evaluation
of a class of NNLO contributions at $\alpha_s^6$, namely those from $gg\to J/\psi J/\psi+c\bar c$.
In this case, there exist a couple of thousands of graphs (see \cf{diagrams-psipsi} (e-h)) 
which remain non-trivially zero even after the removal of the topologies with a single gluon connected to an individual heavy-quark lines. For these $\psi c \bar c \psi$ contribution, there is no need to apply any IR cut-off. We found out that these can indeed become the dominant CSM contributions but likely not in
the phase space accessible at the LHC. 

In 2019, we have examined~\cite{Lansberg:2019fgm} another CS SPS contributions that may be kinematically enhanced 
where DPS are thought to be dominant. These also appear at NNLO and consists in a gauge-invariant and
infrared-safe subset of the loop-induced topologies via CS transitions. We have found it
to become the leading CS SPS contributions at $\Delta y$, yet too small to account for the
data without invoking the presence of DPS yields (see later).

\paragraph{Expectations for FD fractions.}
Like we have defined in section~\ref{sec:FD}, the fractions, $F^{\rm direct}_\psi$, $F^{\chi_c}_\psi$ and $F^{\psi'}_\psi$,  
of single-$J/\psi$ directly produced, from $\chi_c$ decay or from $\psi'$ decay, 
 one can  define various FD fractions for quarkonium-pair production. However, 
it would  probably be experimentally very challenging
to measure (and subtract) the $\chi_c+\chi_c$ yield as well as that for $\chi_c+\psi'$. 
We therefore define 
$F^{\chi_c}_{\psi\psi}$ (resp. $F^{\psi'}_{\psi\psi}$)
as the fraction of  $J/\psi+J/\psi$ events from the FD of {\it at least} a $\chi_c$ (or resp.  a $\psi'$) decay. $F^{\chi_c}_{\psi\psi}$ is thus the fraction of events containing one prompt $J/\psi$ (direct or from 
$\chi_c$ and $\psi'$ FD) and one $J/\psi$ identified as from a $\chi_c$.  We 
also define $F^{\rm direct}_{\psi\psi}$ as being the pure  direct 
component, excluding all the possible FDs, which can be easier to theoretically predict 
but which  is probably very difficult to measure.

$C$-parity conservation in the CSM imposes that $J/\psi+\eta_c$ and 
$J/\psi+\chi_c$ production necessarily goes along with an additional gluon --in 
complete analogy with single-$J/\psi$ production. As such, the LO contributions 
to $J/\psi+\eta_c$  only appear at $\alpha_s^5$ --one graph is 
drawn on \cf{diagram-psi-etac}. In addition, the production of $J/\psi+\eta_c$
usually comes with another factor of $1/3$ due to the spin counting.

In the case
of $J/\psi+\chi_c$, another suppression comes from the vanishing of the $\chi_c$ wave function
at the origin. However, the latter process could in principle be a source of
FD. Although we have explicitly checked~\cite{Lansberg:2014swa} that 
its contribution is negligible in the phase space accessed 
by the LHC experiments, one cannot exclude that it might be enhanced in some of
its corners.  On the contrary, one expects a significant FD 
from the $\psi'$ within the SPS CSM contributions.
Indeed, the CSM hard part for $\psi'+J/\psi$ and $J/\psi+J/\psi$ 
are identical and only $|R(0)|^2$ differ in the calculation at LO in $v$.

With these definitions, taking $|R_{\psi'}(0)|^2=0.53~{\rm GeV}^3$~\cite{Eichten:1995ch}, 
whereas $|R_{J/\psi}(0)|^2=0.81~{\rm GeV}^3$,
and ${\cal B}(\psi'\to J/\psi)=61.4 \%$~\cite{Tanabashi:2018oca} as well as accounting for 
a factor $2$ from the final-state symmetry, the ratio of $F^{\psi'}_{\psi\psi}/F^{\rm direct}_{\psi\psi}$  is expected
to be as large as $0.53/0.81 \times 0.61 \times 2 + (0.53/0.81 \times 0.61)^2\simeq 1$. 
In principle, it is a little larger since $\sigma(\chi_c+\psi')$ was neglected 
in the evaluation of $F^{\psi'}_{\psi\psi}$. 
The latter approximation is justified in view of the above discussed and we 
explicitly checked it with \HELACOnia: neither $\sigma(\chi_c+J/\psi)$ nor 
$\sigma(\chi_c+\psi')$ are relevant. 

If one thus considers that CO contributions are negligible (see next), one can state 
that $\sigma_{\rm SPS}^{\rm prompt}= 2 \times \sigma_{\rm SPS}^{\rm direct}$ and 
$F^{\psi'}_{\psi\psi} \simeq 50 \%$ at any order in $\alpha_s$ --but at LO in $v$.

If one sticks to the simplistic --although widely used-- view of the DPS production mechanisms 
from 2 independent scatterings, one can derive general relations between the 
FD fractions of the DPS yields for double- and single-$J/\psi$ production. 
Such relations can be used 
to evaluate the FD impact, but also, by returning the argument, to test a possible 
DPS-dominance hypothesis by directly measuring pair productions involving the excited states, 
instead of looking for a possibly flat $\Delta \phi$ distribution --keeping in mind
the aforementioned caveats-- or analysing
that as a function of $\Delta y$.

Assuming \ce{eq:dpseq} holds for all charmonia, one gets
\eqs{\label{eq:FD_fraction_DPS}
F^{\chi_c}_{\psi\psi}&= F^{\chi_c}_\psi \times \big(F^{\chi_c}_\psi + 2 F^{\rm direct}_\psi + 2 F^{\psi'}_\psi\big),\\
F^{\psi'}_{\psi\psi}&= F^{\psi'}_\psi \times \big(F^{\psi'}_\psi + 2 F^{\rm direct}_\psi + 2 F^{\chi_c}_\psi\big),\\
F^{\rm direct}_{\psi\psi}&= (F^{\rm direct}_\psi)^2.}
The derivation of \ce{eq:FD_fraction_DPS} for $\chi_c$ follows from the decomposition of the different sources of a prompt $J/\psi$ + a $J/\psi$ from a $\chi_c$. These are the direct $J/\psi+\chi_c$, the $\chi_c+\chi_c$ and the $\psi'+\chi_c$ FD. Their cross section with the relevant branchings can then be written in terms of single-quarkonium cross sections using \ce{eq:dpseq} taking care of not double counting $\chi_c+\chi_c$. Their sum divided by the cross section for a pair of prompt $J/\psi$ decomposed in this way then reads as \ce{eq:FD_fraction_DPS} following the standard definitions of $F^{\chi_c}_\psi$, $F^{\psi'}_\psi$ and $F^{\rm direct}_\psi$.

In order to obtain numbers,  let us recall (see section~\ref{sec:FD}) that
$F^{\rm direct}_\psi$, $F^{\chi_c}_\psi$ and $F^{\psi'}_\psi$ are respectively close to 80\%, 13\% and 7\%
at low $P_T$ and  65\%, 28\% and 7\% at high $P_T$. 
Accordingly, we thus have, at low $P_T$, $F^{\chi_c}_{\psi\psi} \simeq 25 \%$, $F^{\psi'}_{\psi\psi} \simeq 15\%$, $F^{\rm direct}_{\psi\psi} \simeq 65 \%$ and, at low $P_T$, $F^{\chi_c}_{\psi\psi} \simeq 50 \%$, $F^{\psi'}_{\psi\psi} \simeq 15\%$ 
and $F^{\rm direct}_{\psi\psi} \simeq 40 \%$. 
Let us however note that, although $F^{\chi_c}_{\psi\psi}$ and $F^{\psi'}_{\psi\psi}$ 
are experimentally accessible via $\sigma((\chi_c \to J/\psi)+J/\psi)$ and $\sigma((\psi' \to J/\psi)+J/\psi)$, 
they are not sufficient to determine the pure direct yield since 
$F^{\rm direct}_{\psi\psi} \neq 1 - F^{\chi_c}_{\psi\psi} - F^{\psi'}_{\psi\psi}$. Its extraction also requires
the measurement of $\sigma(\chi_c+\psi')$.

The figures from the above discussion are gathered in \ct{tab:FD_table}.

\begin{table}[hbt!]\centering\renewcommand{\arraystretch}{1.5}
\begin{tabular}{c|c|c|c}
  & (CSM) SPS  & ``low''-$P_T$ DPS & ``high''-$P_T$ DPS  \\
\hline \hline
$F^{\psi'}_{\psi\psi}$  & 50\% & 15\%  & 15 \%\\
$F^{\chi_c}_{\psi\psi}$ & small & 25\% & 50\%
\end{tabular}

\caption{Expectations for two FD factors in di-$J/\psi$ production.}
\label{tab:FD_table}
\end{table}

\paragraph{COM/NRQCD studies.}

COM contributions to di-$J/\psi$ hadroproduction were first computed at LO 
by Barger~\etal~\cite{Barger:1995vx} in 1996 for the Tevatron, and then by Qiao~\etal~\cite{Qiao:2009kg} and Li \etal~\cite{Li:2009ug}
in 2009 for the LHC. However, their computations were based on the fragmentation approximation
which did not allow them to compute $P_T$-integrated cross section which could be compared
to the LHCb data for instance. In fact, the comparison done by Li \etal~\cite{Li:2009ug}
at low $P_T$ between the CS and CO rates was misleading as shown in 2010 by Ko \etal~\cite{Ko:2010xy}
who performed the first complete calculation of the $\so+\so$ channel.
The formula for the $S$-wave CO amplitude is similar to that for CS state production with the following formal 
replacements for CO in \ce{eq:amplitude_di-onium-CS}
\bq
\frac{\delta_{c_i,c_j}}{\sqrt{N_c}}\to \sqrt{2}T^a_{c_ic_j},
\frac{R_i(0)}{\sqrt{4\pi}}\to  \frac{\sqrt{\langle\mathcal{O}^i(\soge) \rangle}}{\sqrt{(2J+1)(N_c^2-1)}},
\eq
where $T^a_{c_ic_j}$ denotes the usual Gell-Mann matrix elements. 
Because they only considered the $\so+\so$ channel, their computation is also probably reliable {\it only} at large $P_T$. 
In addition,
one could argue~\cite{He:2015qya} that NLO corrections to sub-leading $P_T$ channels at LO
could be large like in the inclusive $J/\psi$ case. 

That being stated, Ko \etal~\cite{Ko:2010xy} found out that the $\so+\so$ channel
was contributing less than 0.5~\textperthousand\ of the CS yield for the $P_T$ integrated cross section
at $\sqrt{s}=7$~TeV for $|y_\psi| < 2.4$ using $\mopb = 3.9  \times 10^{-3}$~GeV$^3$ 
from~\cite{Braaten:1999qk}\footnote{The other relevant parameters used were $m_c=1.5$~GeV, $\mu_F=\mu_R=\sqrt{4m_c^2+(P_T^{J/\psi})^2}$ and $|R(0)|^2=0.94$~GeV$^3$. As for the PDFs, CTEQL was used.}.  Of course, 
such a statement heavily depends on the chosen LDME values as the cross section is proportional
to its square.

We have also investigated~\cite{Lansberg:2014swa} the possible impact of CO channels 
in the context of the comparison with CMS data which we discuss later. Having at hand
a NLO$^\star$ computation we could, for the first time, make reliable CO vs CS comparisons at the level of $P_T$-differential 
cross section. In particular, we have evaluated the contribution from  $^3S_1^{[8]}+\ ^3S_1^{[8]}$ 
and $^1S_0^{[8]}+\ ^1S_0^{[8]}$ --the latter was considered for the first time by He and Kniehl in~\cite{He:2015qya} which constituted the first complete LO NRQCD study.

In order to make quantitative statements, we used the 1-$\sigma$ upper value of the  
$^3S_1^{[8]}$ (resp. $^1S_0^{[8]}$) LDMEs
of the NLO prompt fit of \cite{Butenschoen:2011yh}, \ie\ $2.83\times 10^{-3}$ GeV$^3$ (resp. $5.41 \times 10^{-2}$ GeV$^3$).
These are compatible with the LO direct fit of~\cite{Sharma:2012dy} 
and are the only ones not  dramatically overshooting
the low-$P_T$ single $J/\psi$ data~\cite{Feng:2015cba} (see section \ref{section:CSM-NLO_tot}).
We however stress that such a value of $\mops$ is 10 times larger than the upper limit set by the $\eta_c$ 
hadroproduction data (see section ~\ref{subsec:COM_1S0_NLO_PT}) and should thus be considered as a very conservative upper limit.

As we were looking for an upper value, we disregarded the  $^3P_J^{[8]}+\ ^3P_J^{[8]}$ 
contribution which is negative with this LDME choice. We found that
CO+CO  channels are nowhere important when $P_T^{\psi}<50~{\rm GeV}$ (see~\cf{fig:CompareCMS}). 
We will come back to the description of these results later.

This upper limit of the COM yield  always remains smaller than the CS one except for $|\Delta y| > 2.5$ (last bin in \cf{fig:dsigCMSb}) and $M_{\psi\psi} > 40$~GeV 
(last two bins in \cf{fig:dsigCMSc}).  
These regions are however prone to DPS contributions as we have seen and will see again.
In addition, the CS SPS can receive significant $\alpha_s^6$ contributions there. 
The only kinematic distribution where the CO contributions might show up is that of $P_{T \rm min}$ (\cf{fig:dsigCMSd}) as it has 
a similar size and the same dependence as our partial NNLO evaluation. 
The cross section is however probably not measurable there. 
About possible mixed CO+CS channels, let us add that, if one analyses their graphs, one can see that they do not benefit from any 
$P_T^{\psi}$ enhancement at LO and these are thus simply suppressed by the LDME. 
We expect them to contribute at the per cent level, not more.

As aforementioned, the first complete LO NRQCD study was performed in 
2015~\cite{He:2015qya} --using NLO LDMEs. Their results are qualitatively similar
to ours which are described above but their conclusion differ from ours as regards the possible relevance
of the COM in di-$J/\psi$ production at colliders. 

In 2019, we have made a survey~\cite{Lansberg:2019fgm} the possible CO contributions 
using both old and up-to-date LDMEs and found that the pure 
CO yields crucially depend on the LDMEs. Among all the LDMEs
we used, only two result into a visible modification of the LO NRQCD (CS+CO) yield, but only in two
kinematical distributions measured by ATLAS which we discuss below, those of the rapidity separation and of the pair invariant mass. We stress that these modifications do not impact the control region used for their DPS study.
In any case, we
think that only a full NLO NRQCD analysis could close the debate. 

In addition, it was shown in 2018 by He \etal\ \cite{He:2018hwb} that NRQCD breaks down in presence of two $P$ waves and that new 
types of operators should be introduced to extend NRQCD. Whereas possible similar issues 
are anticipated for di-$\psi$ production, it is not clear yet how much the phenomenology will be affected and whether
it would be for the entire phase space or for specific configurations where the quarkonia are for instance close to each other and whether it will also affect the CS contribution which dominates in most of the phase space.

\paragraph{CEM expectations.}

In 2020, we performed the first computation~\cite{Lansberg:2020rft} of the 
production of di-$\psi$ in hadroproduction within the CEM both at LO and NLO. In general, it amounts to compute
the cross section for $ij \to c\bar{c} c\bar{c}$ at $\alphaS^3$ with the usual CEM invariant-mass
cut. Considering $ij \to c\bar{c} c\bar{c}$ is sufficient to compute $\Delta y$ or $m_{\psi\psi}$ distribution. To obtain a non-trivial  $P_T^{\psi\psi}$ spectrum, it is however necessary 
to consider partonic sub-processes such as $ij \to c\bar{c} c\bar{c} k$ ($i,j$ and $k$ label any possible parton) at $\alphaS^4$.

\begin{figure}[hbt!]
\begin{center}
\subfloat[]{\includegraphics[width=0.5\textwidth,draft=false]{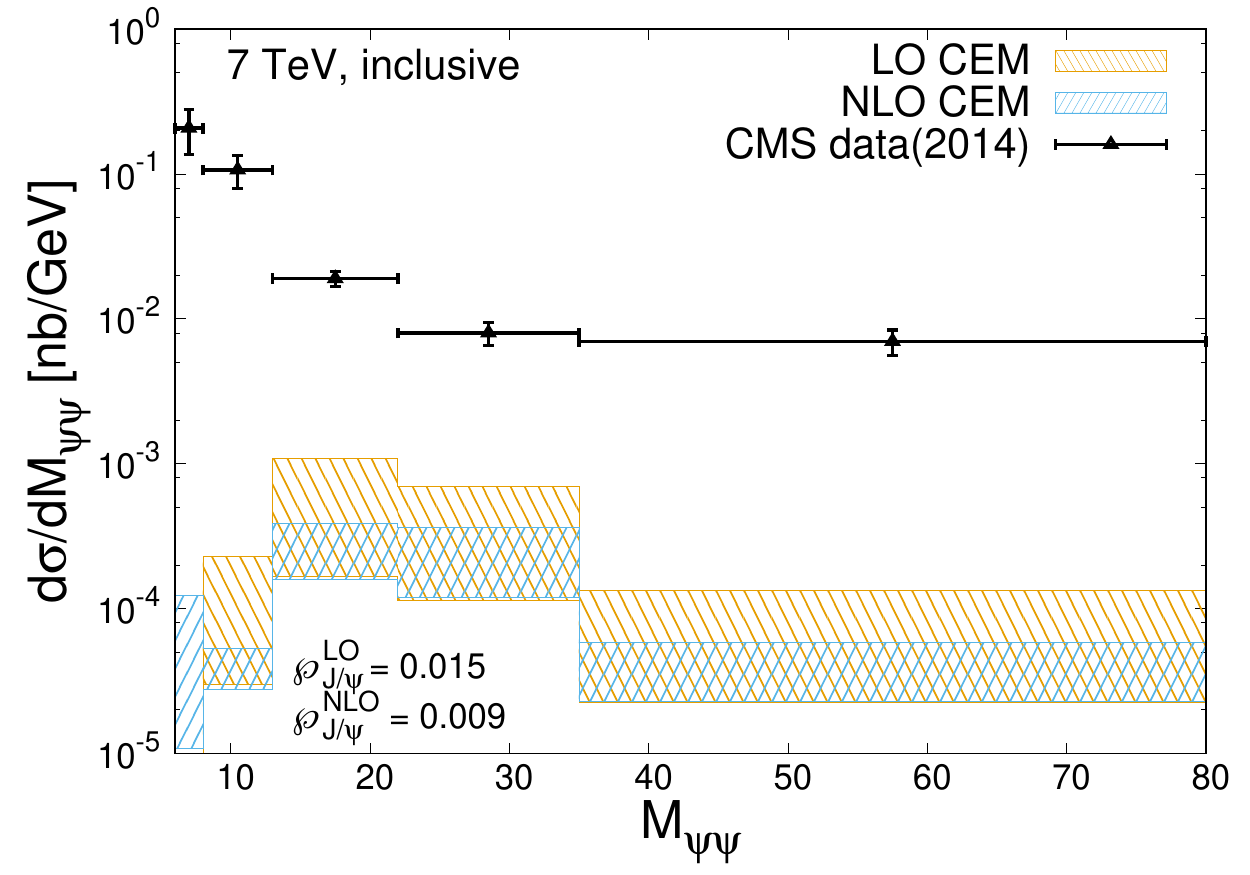}}
\subfloat[]{\includegraphics[width=0.5\textwidth,draft=false]{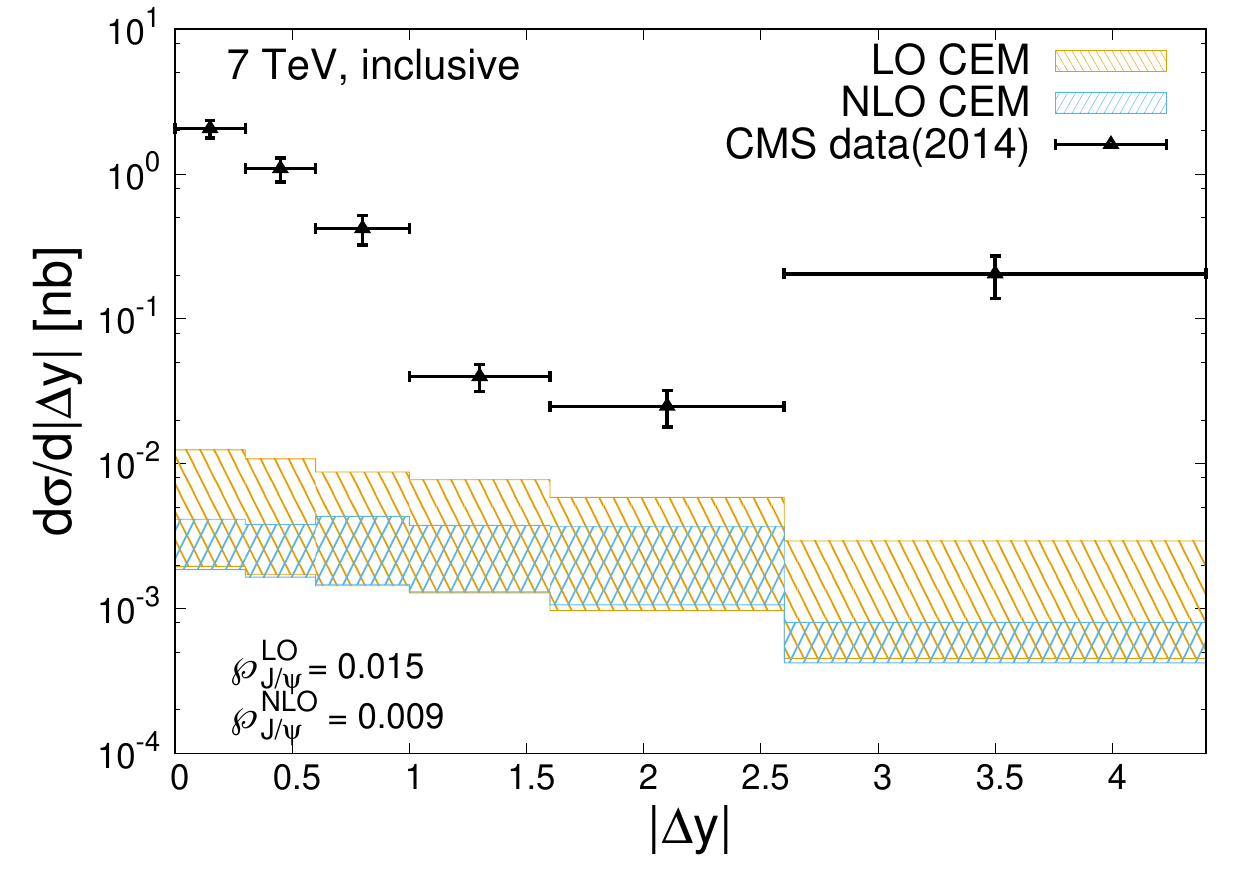}}
\caption{(a) $M_{\psi\psi}$ and (b) $\Delta y$ differential cross section for di-$J/\psi$ production
compared to the CMS data (see later) at $\sqrt{s}=7$~TeV in the CEM at LO (orange) and NLO (light blue).
Taken from~\cite{Lansberg:2020rft}.}
\label{fig:dsigdM-psipsi-CEM-CMS}
\end{center}
\end{figure}

Using \MG5aMC~\cite{Alwall:2014hca} slightly tuned to apply the
CEM invariant-mass cuts (see section~\ref{subsec:CEM_NLO_PT}), one can carry out LO and NLO studies. 
\cf{fig:dsigdM-psipsi-CEM-CMS} show the invariant-mass  and rapidity-separation dependences of the
di-$J/\psi$ in the CMS acceptance (namely with their $y_{\jpsi}$ and $P_T^{\jpsi}$ cuts). We discuss them later along
with the other model results.

\paragraph{Comparisons with the experimental data.}

Having set the theory scene, we can now discuss the comparisons with the LHC and Tevatron data, 
which we have already anticipated for the case of the first LHCb data at 7 TeV. 
We have concluded that both CSM SPS and DPS contributions were in the ballpark 
of the data showing large uncertainties. They are summarised on the first line of \ct{xsections-psipsi}.
The theory is clearly above the data. However, the DPS yield is quoted 
for $\sigma_{\rm eff}=8.2$~mb for a reason which we explained later. Yet, even with 
$\sigma_{\rm eff}=14.5$~mb, as used by Kom \etal~\cite{Kom:2011bd},
the sum of the SPS and DPS would be above the data.

\begin{table}[hbt!] 
  \begin{center} \renewcommand{\arraystretch}{1.4}\setlength\tabcolsep{5pt}\small
\begin{tabularx}{\textwidth}{cp{4.5cm}|p{3cm}cccp{0cm}} 
\hline  \hline      & Energy and quarkonium cuts   &$\sigma_{\rm exp.}$ &$\sigma^{\rm SPS, prompt}_{\rm LO}$ &
$\sigma^{\rm SPS, prompt}_{\rm NLO^{(\star)}}$ & $\sigma^{\rm DPS, prompt}$  {\tiny [$\sigma_{\rm}=8.2$~mb]} \\
\hline\hline 
LHCb
 &  $\sqrt{s}=7$~TeV, $P^{\psi_{1,2}}_T < 10$~GeV, \hspace*{0.82cm} $2<y_\psi < 4.5$~\cite{Aaij:2011yc} &$18 \pm 5.3$ pb  & $41^{+51}_{-24}$  pb  & $46^{+58}_{-27}$ pb &   $31^{+24}_{-15}$ pb \\\hline 
\multirow{2}{*}{D0} & $\sqrt{s}=1.96$~TeV, $P^{\psi_{1,2}}_T > 4$~GeV, & SPS: $70 \pm 23$ fb  & $53^{+57}_{-27}$ fb 
& $170^{+340}_{-110}$ fb &  -- \\ 
                    & $|\eta_{\psi}|<2.0$ ~\cite{Abazov:2014qba} (+ $\mu$ cuts in caption)& DPS: $59 \pm 23$ fb           &        --           & --  & 
$44^{+7.5}_{-5.1}$ fb&                               \\ \hline 
CMS  & $\sqrt{s}=7$~TeV, $P_T^{\psi_{1,2}}> 6.5 \rightarrow 4.5$~GeV depending on $|y_{\psi_{1,2}}| \in [0,2.2]$ (see the caption)~\cite{Khachatryan:2014iia}&$5.25\pm 0.52$  pb & $0.35^{+0.26}_{-0.17}$  pb  & $1.5^{+2.2}_{-0.87}$  pb & $0.69^{+0.039}_{-0.027}$ pb  \\\hline 
ATLAS & $\sqrt{s}=8$ TeV,  $P_T^{\psi_{1,2}}>8.5$ GeV and $|y_{\psi_{1,2}}|<2.1$~\cite{Aaboud:2016fzt} 
& $570 \pm 70$ fb &  -- & $200^{+350}_{-120}$ fb  & 
$40 \pm 0.3$ fb \\ 
\hline
LHCb & $\sqrt{s}=13$ TeV,   $P^{\psi_{1,2}}_T < 10$~GeV, \hspace*{0.82cm} $2<y_\psi < 4.5$~\cite{Aaij:2016bqq}&  $53.6 \pm 3.5 \text{ (stat.) } \pm 3.2 \text{ (syst.) }$ pb & -- & $77^{+100}_{-45}$ pb 
&  $51 \pm 5.7$ pb 
\\
\hline 
\hline 
\end{tabularx}
\caption{Comparison for $\sigma(pp(\bar{p})\to J/\psi+J/\psi+X) \times {\cal B}^2(J/\psi \to \mu\mu)$ between the LHCb, CMS, D$0$ and ATLAS data and theory in the relevant kinematic regions.
The theory predictions are: the SPS prompt yields at LO and NLO$^{\star}$ [For LHCb @ 7 and 13~TeV, the evaluation is a based on LO value rescaled by the NLO $K$ factor from~\cite{Sun:2014gca}], the DPS prompt yields with $\sigma_{\rm eff}$ fit to the CMS differential distributions (\ie\ $8.2$ mb). For the DPS yields for LHCb @ 7 TeV, CMS, D$0$, 
the uncertainty is a systematical uncertainty from the 3 $\sigma_\psi$ fits used in the pocket formula (see~\cite{Lansberg:2014swa}), and for LHCb @ 13 TeV and ATLAS these are from the experimental uncertainty on $\sigma_\psi$.
{ [For D$0$, the additional uncorrected $\mu$ cuts are $P_T^{\mu}> 2$~GeV when $|\eta_{\mu}|<1.35$ and total momenta $|p^{\mu}|>4$~GeV when $1.35<|\eta_{\mu}|<2.0$. For CMS, the detailed cuts are $P_T^{\psi}> 6.5$~GeV if $|y_{\psi}|<1.2$; $P_T^{\psi}> 6.5 \rightarrow 4.5$~GeV 
if $1.2<|y_{\psi}|<1.43$; $P_T^{\psi}> 4.5$~GeV if $1.43<|y_{\psi}|<2.2$ where in $1.2<|y_{\psi}|<1.43$, 
the $P_T^{\psi}$ cutoff scales linearly with $|y_{\psi}|$]}.
 }
\label{xsections-psipsi}
\end{center}
\end{table}

The CMS sample~\cite{Khachatryan:2014iia} significantly differs to that of LHCb
because of $P_T$ cuts ranging from 6.5 to 4.5~GeV. Not only this significantly reduces
the cross section but this imposes to rely on an NLO$^{(\star)}$ analysis. The values quoted 
in~\ct{xsections-psipsi} clearly shows the importance of the QCD corrections with an integrated
cross section multiplied by 5 from LO to NLO$^\star$. Such an enhancement is in fact necessary
to get a near agreement with the data.

CMS released in addition 17 data points of differential cross sections as a function of 
$P_T^{\psi \psi}$, $|\Delta y|$ and $M_{\psi \psi}$ which we compared with our NLO$^{(\star)}$ CS
evaluation~\cite{Lansberg:2014swa}\footnote{As regards the PDFs, we used the set CTEQ6L1 for  LO (${\cal O}(\alpha_s^4)$) calculations and 
CTEQ6M~\cite{Pumplin:2002vw} for NLO$^{\star}$ (${\cal O}(\alpha_s^5)$)  and the ${\cal O}(\alpha_s^6)$ 
$\psi c \bar c \psi$ calculations. The SPS uncertainties were obtained by a common variation of $m_c$
and $\mu_F=\mu_R$ as $((1.4\hbox{ GeV},0.5 \times m_T^{\psi\psi});(1.5\hbox{ GeV}, m_T^{\psi\psi});(1.6\hbox{ GeV},2 \times m_T^{\psi\psi}))$
with $m_T^{\psi\psi}=\big((\sum m_i)^2+(P_T^{\psi})^2\big)^{1/2}$ 
with $\sum m_i =4 m_c$, but for $\psi c \bar c \psi$ where  $\sum m_i =6 m_c$. }. 
To predict the corresponding DPS spectra, we have used~\cite{Lansberg:2014swa} 
single $J/\psi$  experimental data to fit $\sigma_\psi$ as done by Kom \etal~\cite{Kom:2011bd}
which we injected in the DPS pocket formula. The normalisation was then fixed by
$\sigma_{\rm eff}$ which was obtained by fitting the di-$J/\psi$ data subtracted from our theoretical 
evaluations of the SPS NLO$^\star$ yield (green band). We refer to~\cite{Lansberg:2014swa} 
for the detail of this phenomenological study. Overall, 
we obtained $\sigma_{\rm eff}=8.2 \pm 2.0 \pm 2.9$~mb, the first uncertainty being
from the CMS-data and the SPS-theory uncertainties, the second being 
from the single $J/\psi$ $\sigma_\psi$ templates. In \ct{xsections-psipsi}, the DPS
cross sections are quoted with this value.

\begin{figure}[hbt!]
\begin{center}
\subfloat{\includegraphics[width=0.48\columnwidth,draft=false]{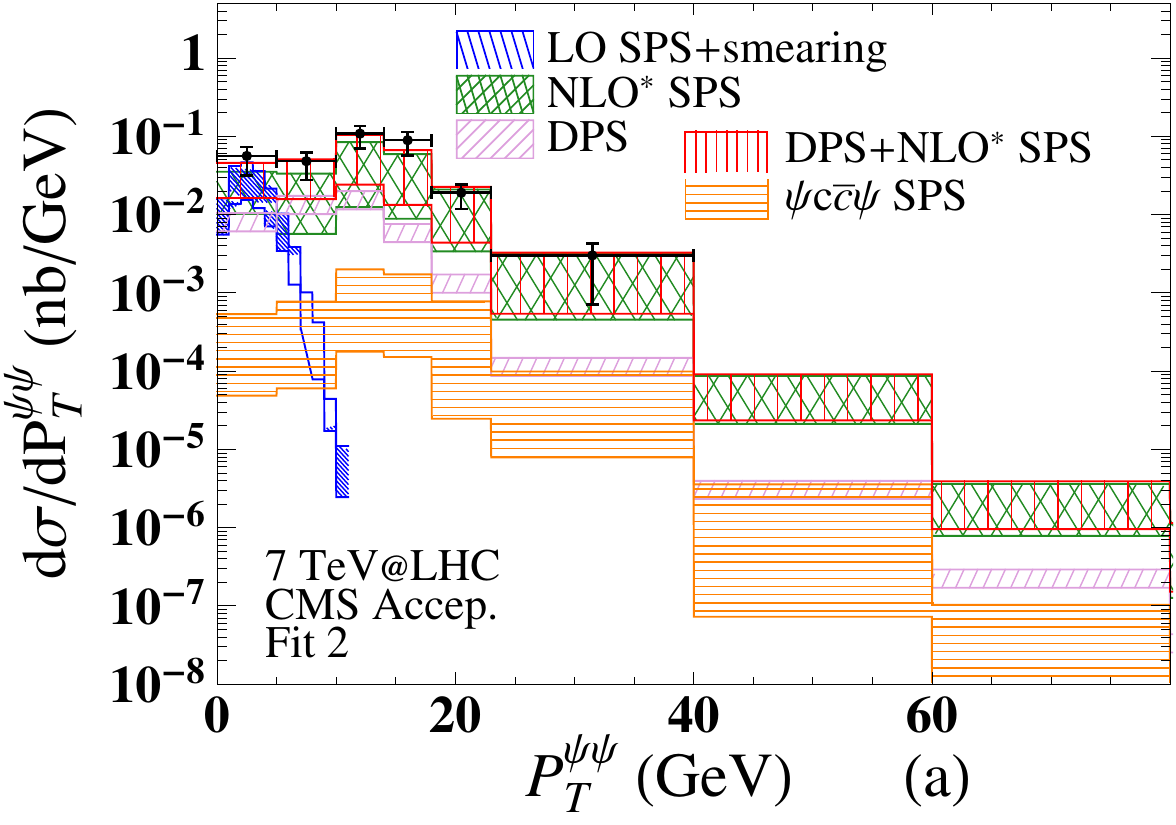}\label{fig:dsigCMSa}}\quad
\subfloat{\includegraphics[width=0.48\columnwidth,draft=false]{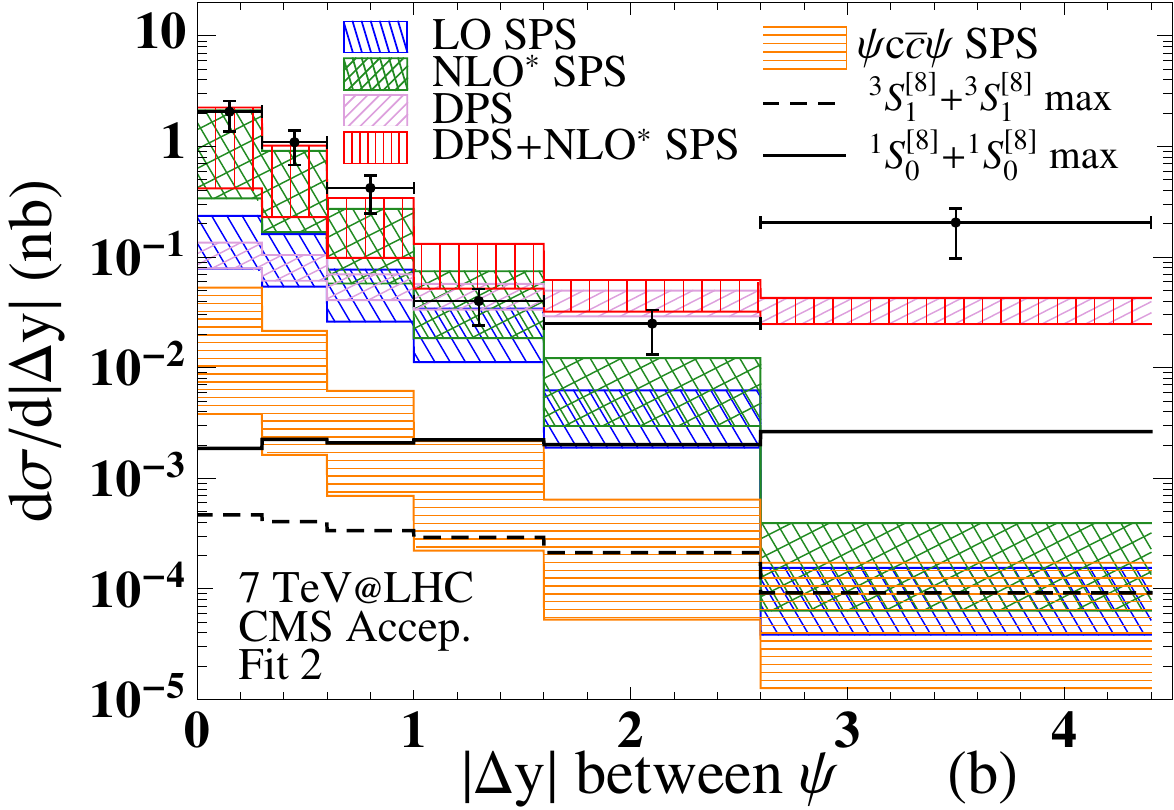}\label{fig:dsigCMSb}}\\
\subfloat{\includegraphics[width=0.48\columnwidth,draft=false]{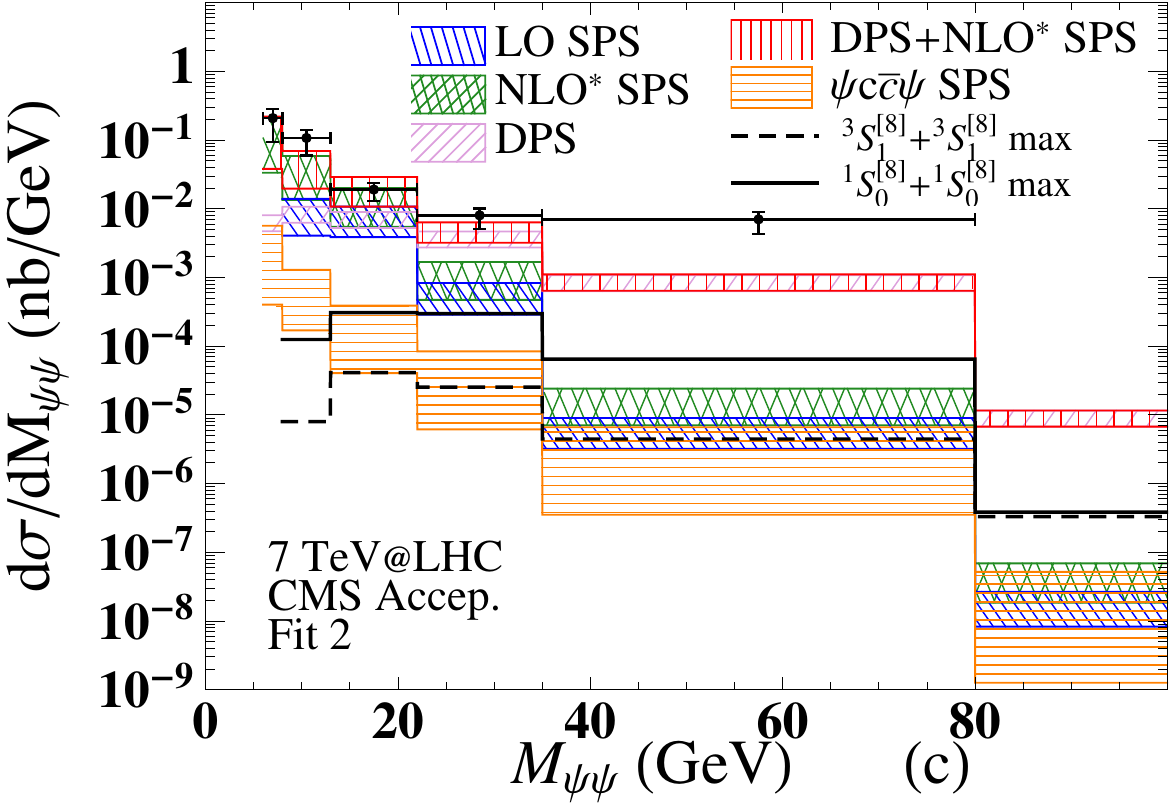}\label{fig:dsigCMSc}}\quad
\subfloat{\includegraphics[width=0.48\columnwidth,draft=false]{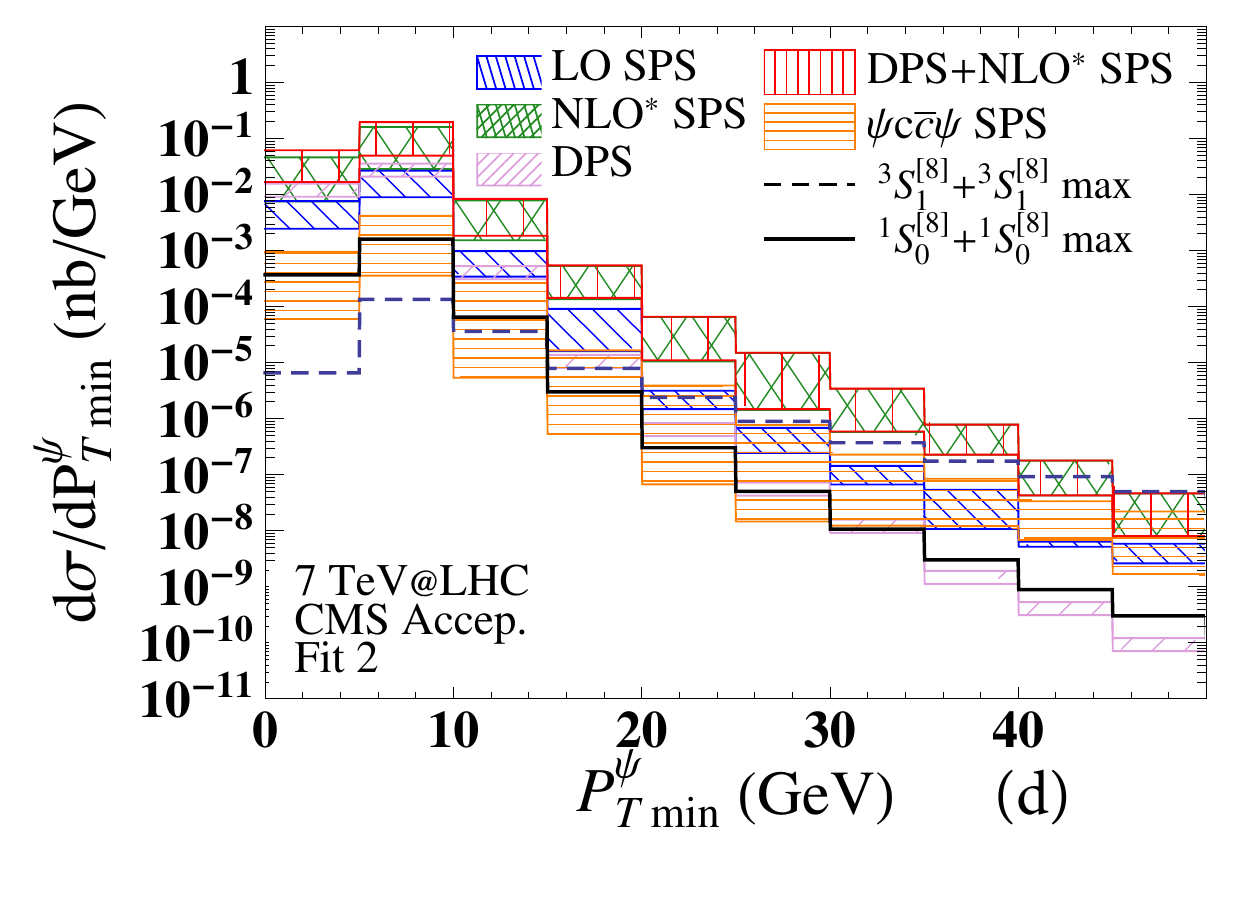}\label{fig:dsigCMSd}}
\caption{Comparison of different theoretical contributions with the CMS measurement as a function of the (a)  pair transverse momentum; (b) absolute-rapidity difference ; (c) pair invariant mass, (d) $P_T$ of the sub-leading $J/\psi$.
Taken from~\cite{Lansberg:2014swa}.}
\label{fig:CompareCMS}
\end{center}
\end{figure}

 As regards the differential cross section as a function of the $J/\psi$-pair $P_T$, $P_T^{\psi \psi}$, \cf{fig:dsigCMSa}
clearly shows how important the $\alpha_s^5$ QCD corrections to the SPS yield are. 
Phenomenologically accounting for the initial parton $k_T$ does not reproduce the data
as the smeared LO curve shows\footnote{The LO kinematics is a $2\to 2$ one. As such it generates a trivial 
$P_T^{\psi \psi}$ dependence ($\delta(P_T^{\psi \psi})$) in the collinear factorisation. 
We have therefore accounted for the $k_T$'s of the initial partons with a Gaussian 
distribution with $\langle k_T \rangle=2$~GeV (see~\cite{Shao:2015vga}) to obtain
fairer comparisons of the $P_T^{\psi \psi}$ spectra.}. 
At NLO, hard real-emissions recoiling on the $J/\psi$ pair in
the real-emission topologies (\cf{diagram-psi-psi-NLO-PT6}) do produce two 
{\it near} $J/\psi$ --as opposed to back-to-back-- with a large $P_T^{\psi \psi}$. 
When one $J/\psi$ has a large $P_T$, there should be another one near it.
On the contrary for the DPSs, such correlations are absent and low $P_T^{\psi \psi}$ configurations are favoured. There is no mechanism to enhance the  large $P_T^{\psi \psi}$ configurations and
the DPS band exhibits a softer spectrum than the  NLO$^\star$ SPS one at large $P_T^{\psi \psi}$. 
The ``bump'' around $P_T^{\psi \psi}\simeq12~{\rm GeV}$ simply reflects the kinematic cuts
in the CMS acceptance.  Overall, we obtained a good agreement ($\chi^2_{\rm d.o.f.}\simeq 1.1$) with the $P_T^{\psi \psi}$ distribution when DPS and (NLO$^\star$) SPS are included. 

CMS also analysed the relative-rapidity spectrum, $d\sigma/d|\Delta y|$. As aforementioned, 
the SPS contributions dominate when $|\Delta y| \to 0 $, 
while the DPS ones are several orders of magnitude above the SPS ones at large $|\Delta y|$. 
\cf{fig:dsigCMSb} displays the comparison with the CMS data.
Most of the data are consistent with the sum of SPS and DPS  results, except for the last bin. 
For this distribution, $\chi^2_{\rm d.o.f.}\simeq 2.1$ with DPS for the 3 $\sigma_\psi$ fits and about 
2.6 without  DPS. This clearly emphasises the relevance of the DPS to account for large $|\Delta y|$
events. 
The CMS acceptance with a rapidity-dependent $P_T$ cut renders $d\sigma/d|\Delta y|$ 
flatter but this effect is apparently not marked enough in our theory curves. 
More data are however needed to confirm the discrepancy in the last $|\Delta y|$ bin.

At $P_T^{\psi \psi}=0$ --where the bulk of the yield lies--, the $J/\psi$-pair invariant mass, $M_{\psi \psi}$, 
is closely related to $|\Delta y|$ and its distribution provides similar information than above 
(see~\cf{fig:dsigCMSc}). One indeed has
$M_{\psi \psi}=2 m_T^{\psi}\cosh{\frac{\Delta y}{2}}$. 
Large $\Delta y$ --\ie~large relative {\it longitudinal} momenta-- correspond to 
large $M_{\psi \psi}$. For instance, at $\Delta y=3.5$ and $P_T=6$ GeV, $M_{\psi \psi}\simeq 40$ GeV. 
The  $M_{\psi \psi}$  and $d\sigma/d|\Delta y|$ spectra of the CMS indeed reveal the same conclusion: 
the DPS contributions is dominant in the region of large momentum differences. At small $M_{\psi \psi}$, 
SPS contributions dominate and NLO corrections are large 
--essentially because the CMS data do not cover low $P^\psi_T$.
For this data set, $\chi^2_{\rm d.o.f.}\simeq 3.0$ with DPS for the 3 fits 
and about 5.3 without (only NLO$^\star$), which again highlights the importance
of the DPS yield. 

On~\cf{fig:CompareCMS} (b-d), the black dashed and solid lines represent what we consider to be
the (conservative) maximum CO yield. These were computed with $\mopb= 2.8 \times 10^{-3}$~GeV$^3$ and $\mops= 5.4\times 10^{-2}$~GeV$^3$.
We recall that the latter value is 10 times higher than the (NLO) bound set by the $\eta_c$ data (see section~\ref{subsec:COM_1S0_NLO_PT}).
\cf{fig:dsigdM-psipsi-CEM-CMS} clearly shows that, for both the $M_{\psi \psi}$ and $|\Delta y|$ distributions,
the CEM is also unable to account for the data. 
We note that the LO differential cross section is unsurprisingly similar that of the COM channels and 
that the QCD corrections do not change the observed LO trend. The addition of (a significant) 
DPS would not solve the issue with severe deficit at low $M_{\psi \psi}$ and $|\Delta y|$.

Based on this discussion, we confirm our conclusion~\cite{Lansberg:2014swa} that, in the CMS kinematics, the excess between the data and
these SPS evaluations was due to DPS and that $\sigma_{\rm eff}$ was smaller than the value extracted with jet-related analyses, namely $\sigma_{\rm eff}=8.2 \pm 2.0 \pm 2.9$~mb.

As we explained in the introduction of this section, D0 reported~\cite{Abazov:2014qba} on a value of
$\sigma_{\rm eff} \simeq 5.0 \pm 2.75~{\rm mb}$ using their separation of the
DPS and SPS yields. Such a value is in line with our extraction from the CMS data. In fact, 
in~\ct{xsections-psipsi}, one sees that our ``predicted" DPS value agrees with theirs.

Without entering the details of the SPS/DPS ratio D0 extraction,
let us however note that it required~\cite{Abazov:2014qba} them some information on the $\Delta y$ distribution
of the SPS yield which was taken from LO SPS computations. Given the 
impact of the QCD corrections in the D0 acceptance (compare the LO and NLO$^\star$ 
figures in~\ct{xsections-psipsi}), 
such a choice may not be an optimal one.

\begin{figure}[hbt!]
\begin{center}
\subfloat[]{\includegraphics[width=0.4\columnwidth,draft=false]{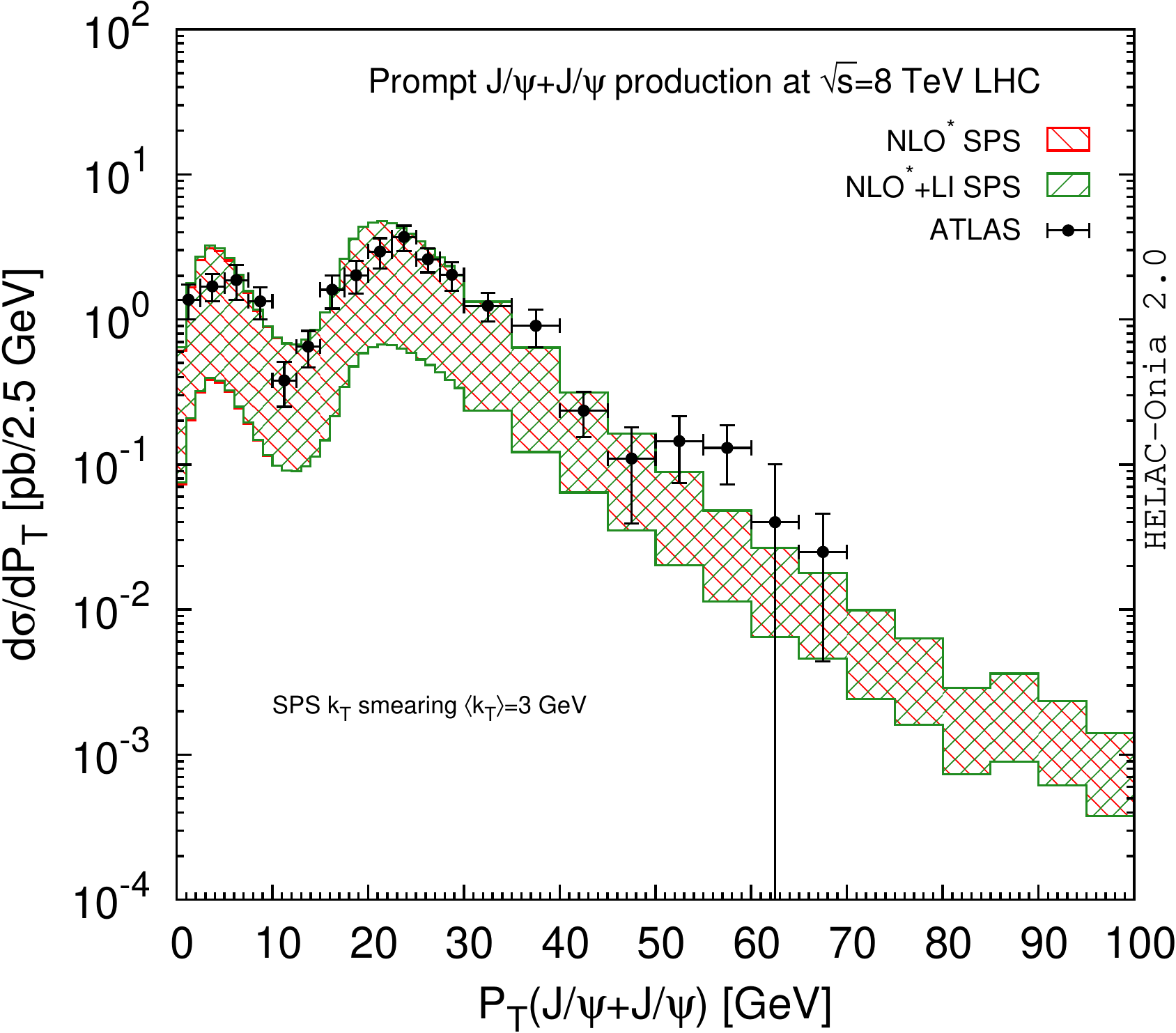}\label{fig:dsigATLAS-Pt}}\quad
\subfloat[]{\includegraphics[width=0.4\columnwidth,draft=false]{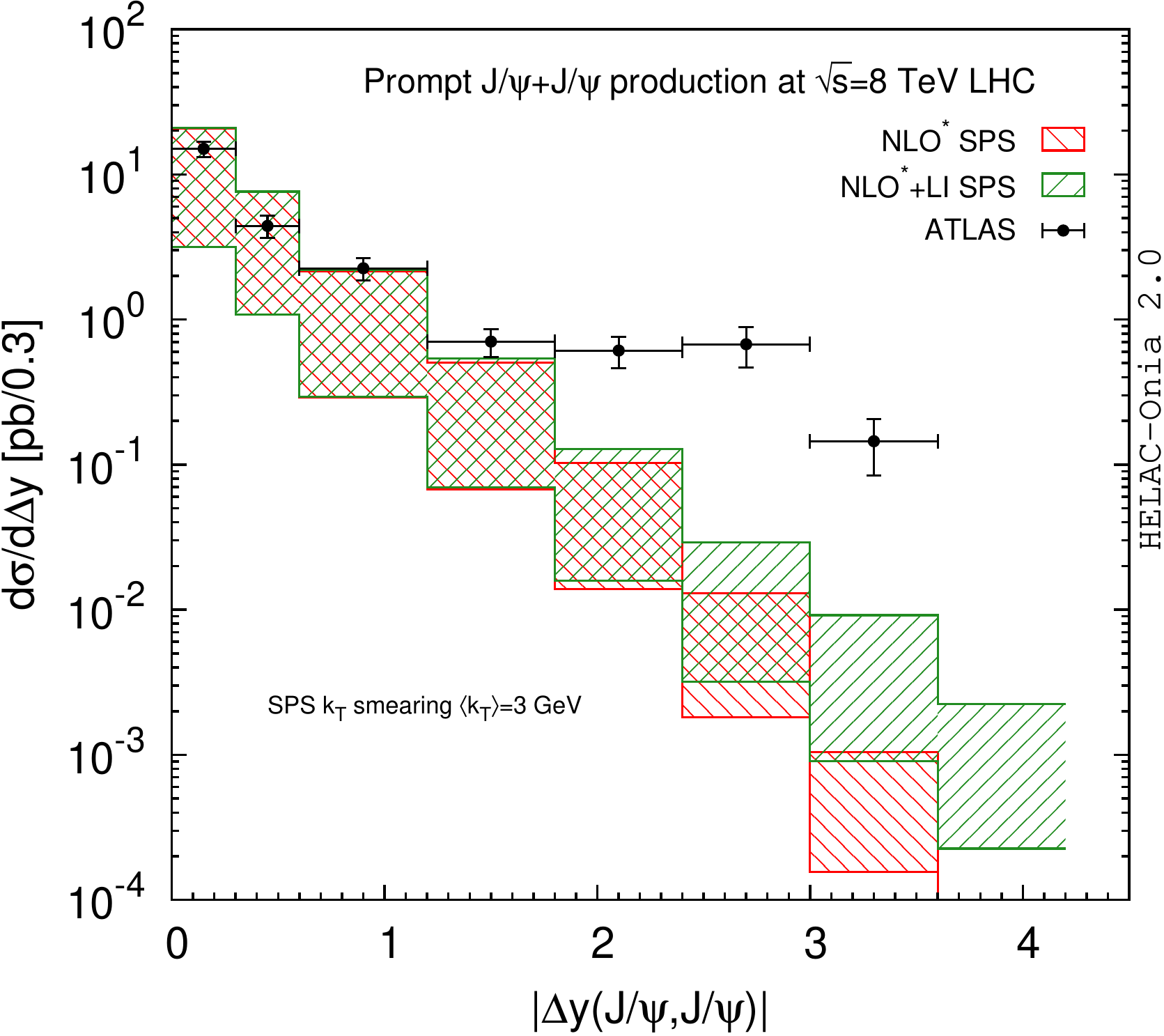}\label{fig:dsigATLAS-dy}}\\
\subfloat[]{\includegraphics[width=0.4\columnwidth,draft=false]{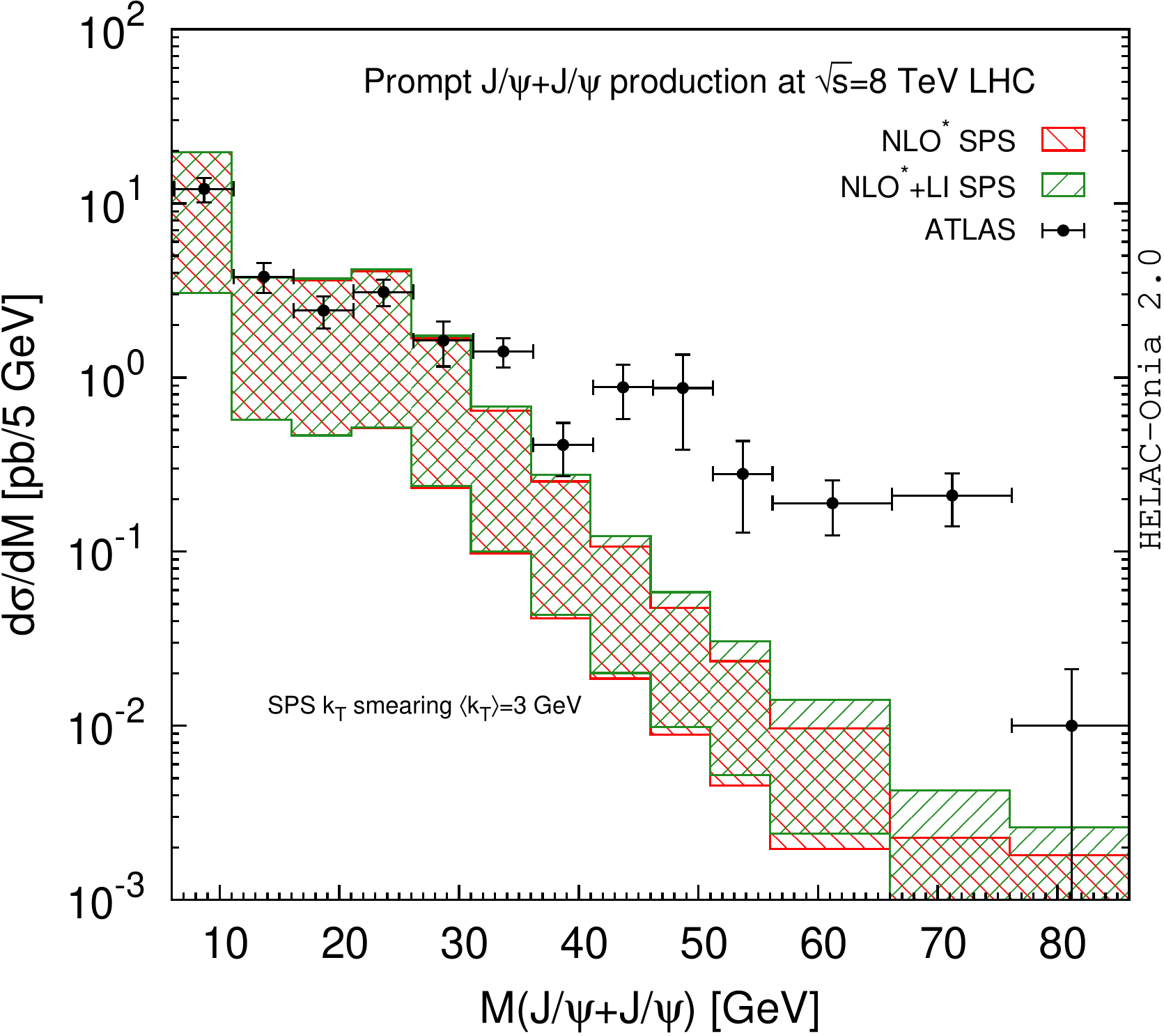}\label{fig:dsigATLAS-M}}\quad
\subfloat[]{\includegraphics[width=0.4\columnwidth,draft=false]{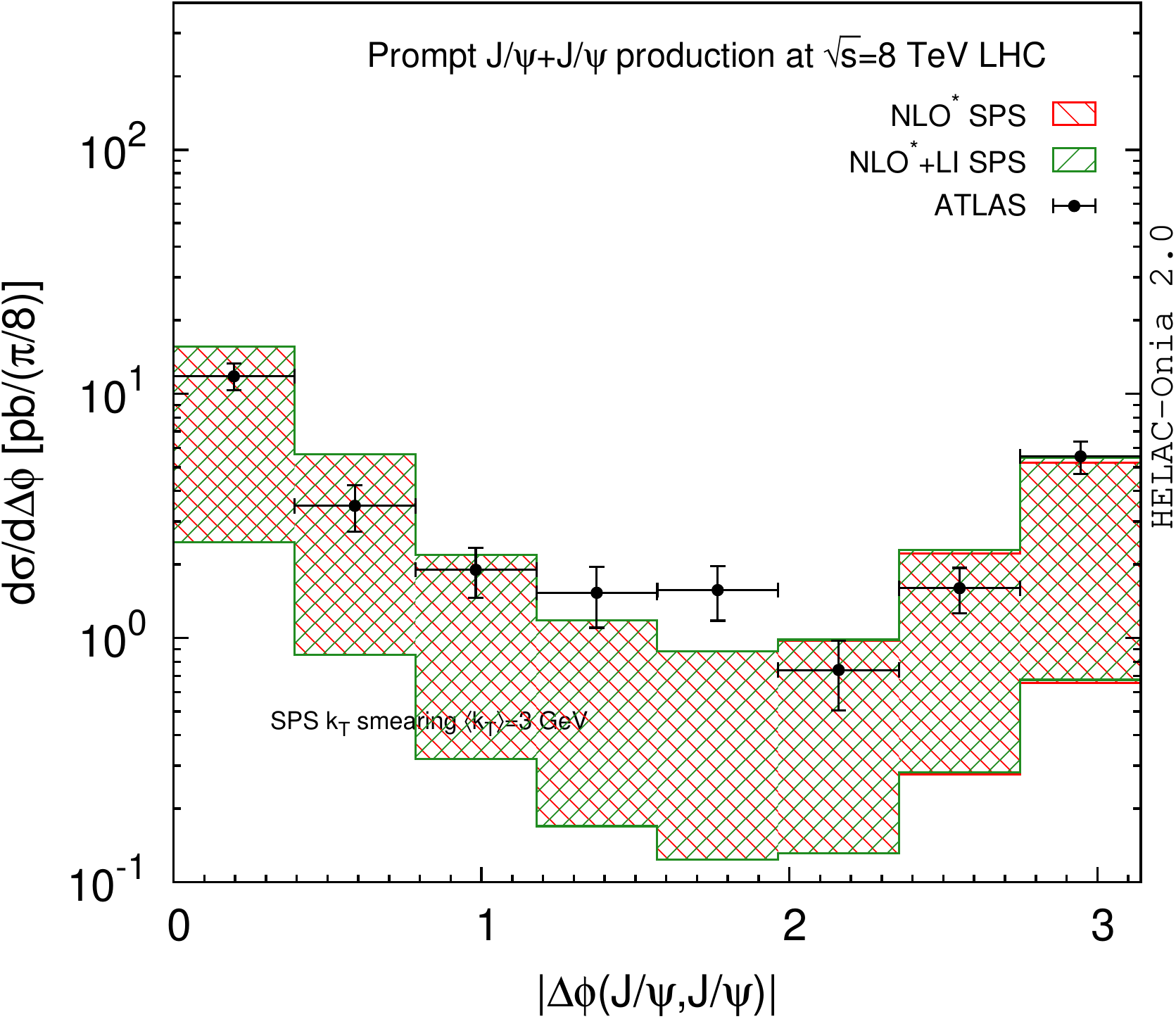}\label{fig:dsigATLAS-dphi}}
\caption{Comparison of the NLO$^\star$ (red) and NLO$^\star$+NNLO LI (green) CSM theoretical predictions with the ATLAS measurement as a function of the (a)  pair transverse momentum; (b) absolute-rapidity difference ; (c) pair invariant mass; (d) azimuthal separation. From~\cite{Lansberg:2019fgm}.}
\label{fig:CompareATLAS}
\end{center}
\end{figure}

In 2016~\cite{Aaboud:2016fzt}, ATLAS reported on an analysis at 8 TeV including
the separation of the DPS and SPS yields relying on a DPS template which they applied
in a region where the SPS is expected to be negligible. Based on this 
extraction they stated that the DPS accounted for 9.2 \%
of the yield in their acceptance. Using their measured value of $\sigma_{J/\psi}$
in the same kinematical conditions, they could perform their own extraction
of  $\sigma_{\rm eff} = 6.3 \pm 1.6 \text{ (stat.) } \pm 1.0 \text{ (syst.)}$~mb.

The $P_T$-integrated cross section is reported in~\ct{xsections-psipsi}
along with a DPS cross section with $\sigma_{\rm eff} = 8.2$~mb. Theirs is just 
$8.2/6.3$ times larger. Comparisons between their cross section and
our NLO$^\star$ predictions were shown in~\cite{Aaboud:2016fzt}. Let us however note that we  discovered
that there had been a confusion during the plotting procedure. The NLO$^\star$
cross sections shown in the figures of~\cite{Aaboud:2016fzt} hold for the inclusive yield, not the fiducial one. As such, all
the red theory histograms in the figures of~\cite{Aaboud:2016fzt} should a priori be scaled down by a factor on the order of 6.
\cf{fig:CompareATLAS} shows the proper comparison with NLO$^\star$ and NLO$^\star$+NNLO LI CSM fiducial yields.
The CSM predictions are not any more systematically above the data which looked a bit surprising
since only the lower limit of the CSM yield was in agreement with the ATLAS data.
Like for the CMS case, the $P_T^{\psi \psi}$ distribution is  well 
described by the NLO$^\star$ all the way up to 80 GeV. This is absolutely remarkable. The NNLO LI 
contributions~\cite{Lansberg:2019fgm} can be neglected as what regards the $P_T^{\psi \psi}$ distribution.

At low $P_T^{\psi \psi}$, we
 note that the $P_T^{\psi}$ cut at 8.5 GeV  induces a very strong cross-section 
modulation --as large as a factor of 10--
from the regime where both $J/\psi$ are back to back to the regime where
they are near each other. In this region ISR contributions may be relevant. 
This is why we complemented our NLO$^\star$ prediction by a smearing which indeed brings the unsmeared
NLO$^\star$ much closer to the data. We note that it has no effect as we go away 
from the ``bum''.

The $\Delta y$ spectrum  is also
well accounted for except for the 3 bins at the largest $\Delta y$ where the DPS contributions
are expected to fill the gap with the data. We note that, in the last 2 bins, the NNLO LI contributions
are~\cite{Lansberg:2019fgm} the leading SPS CSM ones, which shows that kinematically enhanced topologies
can matter in these extreme configurations but remain insufficient to account for the data.

From the ATLAS DPS extraction, 
it however seems~\cite{Aaboud:2016fzt} that the DPS is not completly filling the gap between the SPS yield 
and the data. Yet one may wonder
whether the ATLAS technique with a single $(\Delta y,\Delta \phi)$ template
is accurate enough in such corners of the phase space and whether the counts 
in these bins are not too small. 

One observes a similar issue for the large $M_{\psi\psi}$ values 
of the corresponding spectrum. However, in this case, the counts are really low 
with large fluctuations -- some bins only gather a handful of events. Similar 
fluctuations are also observed by ATLAS between their
extracted DPS and their predicted DPS yields (see~\cite{Aaboud:2016fzt}). As such, 
it would be delicate to draw any further conclusions.
Finally, the $\Delta \phi$ distribution (dominated
by the lowest $P_T$ data) is very well accounted for by the NLO$^\star$ CSM complemented 
by a smearing. In particular let us note that both the near and away peaks are 
reproduced.

Finally, let us also note the significant dependence (up to 70\%) of the ATLAS acceptance
on the $J/\psi$ polarisation, especially near the acceptance edges. 
If the DPS are 
dominant in these regions, one is then entitled to use the polarisation 
measurements of single $J/\psi$. If the SPS contributions are significant, one does
not know their polarisation as no measurement exist. Polarisation predictions for the NLO$^\star$ CSM 
have been provided in~\cite{Lansberg:2013qka}.
  
In 2016 as well, LHCb released~\cite{Aaij:2016bqq} the first study of 
the RUN-II LHC at 13~TeV. This study is particularly instructive
as it suggests that their 7 TeV cross section was probably a lower fluctuation. 
Indeed, the 13 TeV cross section is 3 times larger than the 7 TeV one; one would 
have expected a factor close to 2.

Correspondingly, one sees that the NLO CSM cross section
now agrees well with the data leaving some room for the DPS contribution.
Yet the sum of the SPS and the expected DPS contributions with $\sigma_{\rm eff}=8.2$~mb
(see~\ct{xsections-psipsi}) remains a little higher than the data. 
With $\sigma_{\rm eff}=14.5$~mb and the 1-$\sigma$ lower CSM value, 
data and theory nevertheless agree. This would correspond to a DPS fraction of 45~\%.
We stress that this value is purely indicative. 

In addition, it is very important 
to note that the scale-induced uncertainties, which indicate the impact of the missing
higher-order QCD corrections, do not give enough information on how the kinematical
{\it distributions} may be altered by these QCD corrections. The theory uncertainty band 
should really be considered as a range
within which the distributions may fluctuate at higher orders. To phrase it differently,
the scale theory correlations are not reliable at all. Given
that these are as large as a factor 2 (if not more), we would be a little less assertive
of the LHCb interpretation~\cite{Aaij:2016bqq} of the kinematic distributions as what regards the agreements
or disagreements with the existing theory predictions.

\paragraph{Some words of conclusion.}

Taken together, these 5 data sets indicate 
the presence of different production mechanisms in different kinematic regions. 
First, the NLO CS contributions at $\alpha_s^5$ dominate the yield at large transverse momenta 
of the pair, up to 80 GeV. Second, the DPS contributions are clearly important 
at large $J/\psi$-rapidity differences -- which also correspond to large invariant masses of 
the pair. In the CMS and ATLAS acceptances, with 
a significant $P_T$ cut on the $J/\psi$, their integrated contribution 
is however limited  to 10 \% at most. 

In addition, we have seen that the CO contributions to the integrated yield is 
negligible. It may compete with specific NNLO CS and DPS contributions
in some corners of the phase space. However, given the lack of NLO CO
computations and the current precision of the data, it would be adventurous
to draw final conclusions for these very specific regions. Let us also note that the CMS, ATLAS and D0 data point
at a rather low value of $\sigma_{\rm eff}$ between 5 and 10 mb, whereas
the LHCb data would better agree  with 15 mb than with lower values.

Based on a first exact --gauge-invariant and infrared-safe-- evaluation
of a class of leading-$P_T$ ($P_T^{-4}$) NNLO contributions at $\alpha_s^6$, one does not expect
NNLO CSM corrections to qualitatively change these conclusions. These happen to be only 
relevant in the region of large values of $P_{T\rm min}=\min(P_{T1},P_{T2})$.

We have also shown how sensitive the FD fractions from the production of an
excited charmonium state with a $J/\psi$ are to the 
relative DPS/SPS abundances. This follows from the fact that the DPS and SPS mechanisms are characterised
by very different sizes of the $J/\psi+\chi_c$ and $J/\psi+\psi'$ yields. These should be measured in regions where the DPS is thought to be dominant. $J/\psi+\psi'$ yields are  probably the easiest to measured.

In this context, we have finally studied~\cite{Lansberg:2015lva} the energy dependence
of the  DPS \& SPS cross sections  for prompt $J/\psi$  production. \cf{fig:energydep} shows both SPS and DPS results (for $5 < \sigma_{\rm eff} < 15$~mb).
As expected from the increase of the gluon PDFs, the DPS yield takes over the SPS one 
at high energies. However, we note that the crossing point heavily depends on 
$\sigma_{\rm eff}$. Studies at lower energies (RHIC and LHC in the fixed-target mode, AFTER@LHC~\cite{Brodsky:2012vg,Hadjidakis:2018ifr})
may crucially complement those at Tevatron and LHC energies.

\begin{figure}[hbt!]
\begin{center}
\subfloat{\includegraphics[width=0.6\textwidth,draft=false]{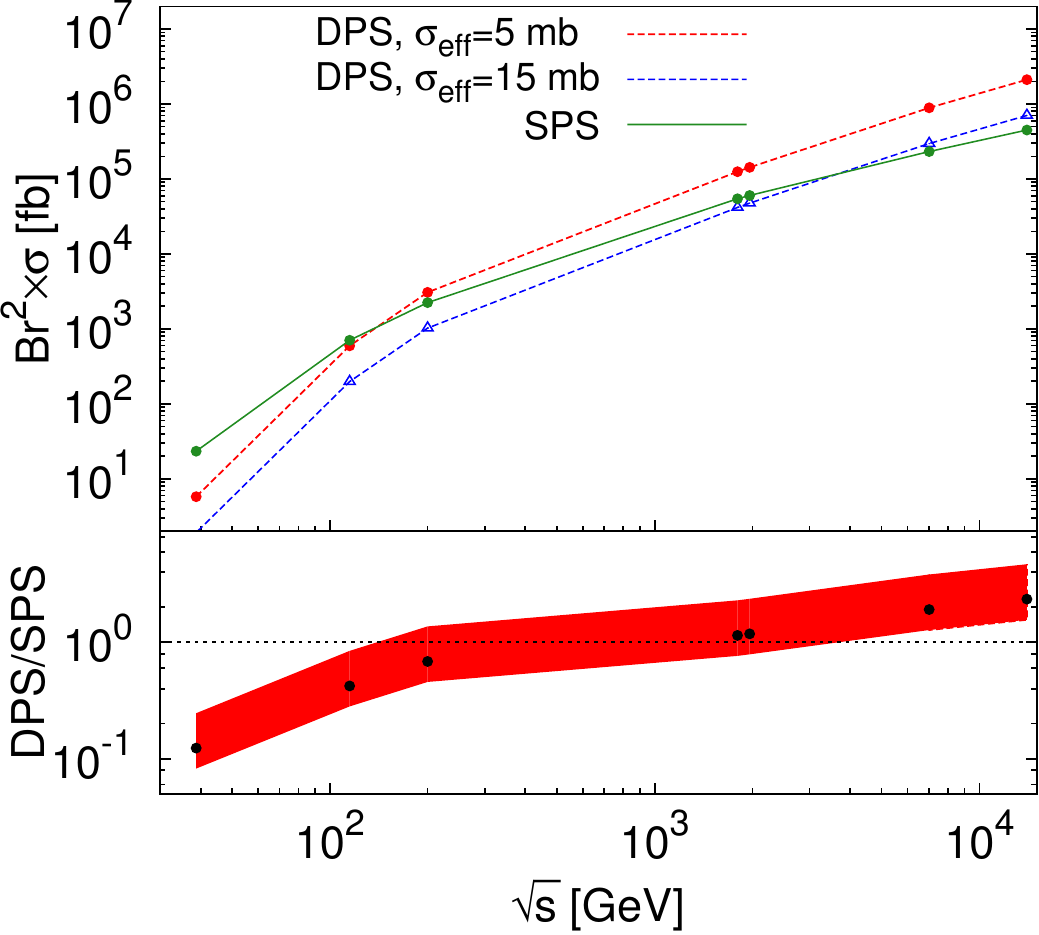}}
\caption{(Upper panel) Prompt $J/\psi$-pair production cross sections via SPS and DPS
 for two values of $\sigma_{\rm eff}$ as a function of $\sqrt{s}$. (Lower panel)
DPS over SPS yield ratio for $5 < \sigma_{\rm eff} < 15$~mb. The black circles correspond to 10 mb. [No theoretical uncertainties are shown]. Taken from~\cite{Lansberg:2015lva}.
}
\label{fig:energydep}
\end{center}
\end{figure}

\subsubsection{Bottomonium-pair hadroproduction}

The first theoretical analysis of di-$\Upsilon$ hadroproduction in the CSM was carried out by Li 
\etal~\cite{Li:2009ug} back in 2009 in which they evaluated the cross section 
for the Tevatron and LHC kinematics. Ko \etal~\cite{Ko:2010xy} confirmed their results
one year later. These computations exactly follow  the same lines as 
of the di-$J/\psi$ case and we do not repeat their description here. 

Unfortunately, these studies did not address the theoretical uncertainties
which are expected to be large for such processes with a strong sensitivity
on the $b$-quark mass and the renormalisation scale, nor did they consider the FD effects.
A third study by Berezhnoy~\etal~\cite{Berezhnoy:2012tu} went into more details 
with an estimation of the FD from di-$\Upsilon(nS)$ on the order of 30\% --
that from $\chi_b(nP)+\Upsilon(nS)$ or $\chi_b(nP)+\chi_b(nP)$ is expected to be
smaller. They also evaluated the theoretical uncertainties on the order of 50\%
upwards and downwards -- not including those from the $b$ quark mass, though.
The choices of the parameters were essentially standard in all 3 computations
and the possible differences would in any case be irrelevant in view of the expected
uncertainties.

In 2015~\cite{Lansberg:2015lva} (see also~\cite{Shao:2016knn}), we performed a study (with 
a complete uncertainty analysis) for 
the conditions reached in the fixed-target mode at the LHC (AFTER@LHC),
namely with $\sqrt{s}=115$~GeV. The obtained cross section are however 
too small to expect to measure this reaction with the projected yearly 
luminosities~\cite{Brodsky:2012vg,Kikola:2017hnp,Trzeciak:2017csa,Hadjidakis:2018ifr} of 10~fb$^{-1}$.

As far as the CO computations are concerned, that of Li~\etal~\cite{Li:2009ug}, 
being based on the fragmentation approximation, does not provide any informative results
for the regions currently accessible at the LHC. On the other hand, Ko \etal~\cite{Ko:2010xy},
who however performed a fixed-order computation, only considered the $\so$ channels. 
We believe this to be only justified at large $P_T$ as well. 

In any case, the CO contributions to the $P_T$-integrated cross section are down 
by a factor as large as 100 compared to the CS ones. The inclusion of the $\sps$ and $\pj$ channels or 
the mixed CS+CO contributions should also have a very minor impact. 
Ko \etal~also analysed the $P_T$-differential cross section and claimed
that the CO $\so$ contributions could become dominant for $P_T$ around 20~GeV.
We nevertheless do not believe that it will be the case since, in this region,
the NLO CS yield will significantly be larger due to the $P_T$-enhanced $\alpha_s^5$ corrections as 
observed for the di-$J/\psi$ case~\cite{Lansberg:2013qka,Lansberg:2014swa}.
There is currently no evaluation of the FDs when CO are considered. We 
however do not see specific reasons why it would deviate from that evaluated using
the CS channels only.

Let us now briefly compare these expectations with both CMS analyses.
At $\sqrt{s}=8$~TeV for $|y_\Upsilon|<2.0$, they collected a limited sample~\cite{Khachatryan:2016ydm} 
of 40 events which however allowed them to perform the following determination of the cross section (without any branching):
\eqs{
\sigma_{\Upsilon\Upsilon} = 68.8 \pm 12.7 \text{ (stat) } \pm 7.4 \text{ (syst) } \pm 2.8 (\Br) \text{ pb}. 
}
In addition to the quoted uncertainties, let us note that a significant 
uncertainty arises from the unknown $\Upsilon$ polarisation in this reaction. 
It nearly amounts to an additional 40 \% uncertainty upwards and downwards.
At $\sqrt{s}=13$~TeV, still for $|y_\Upsilon|<2.0$, they collected~\cite{Sirunyan:2020txn} a slightly larger 
sample of 150 events corresponding to
\eqs{
\sigma_{\Upsilon\Upsilon} = 79 \pm 11 \text{ (stat) } \pm 6 \text{ (syst) } \pm 3 (\Br) \text{ pb},
} assuming the $\Upsilon$ to be unpolarised.

According to the aforementioned theoretical 
computations~\cite{Li:2009ug,Ko:2010xy,Berezhnoy:2012tu}, slightly corrected to match the
rapidity range of the CMS measurement, the expected CS prompt yield at 8 TeV would be $26\pm 13$~pb 
(not accounting for the unknown mass uncertainty which should be large). This 
number is slightly below the CMS data. Given the uncertainties, this was 
only a hint of a possible underestimation. Indeed, the 13~TeV data seems to be comparatively lower.
A NLO$^\star$ computation using \HELACOnia\ following the lines of~\cite{Lansberg:2015lva} with $m_b=4.75$~GeV and NNPDF3.0 indeed yields at 13~TeV
\eqs{
\sigma^{\rm NLO^\star ~ CSM}_{\Upsilon\Upsilon} = 56 _{-33}^{+82} \text{ (scale) } \pm 1.3  \text{ (PDF) }  \text{ pb }
}
with an admittedly larger scale uncertainties than the former expectations.
LO and NLO CEM results were published in 2020~\cite{Lansberg:2020rft}. The expected CEM yields are on the order
of 1 pb thus much smaller than the CMS results.	

A natural question then concerns the expected size of the DPS contributions. In fact, at 8~TeV,
it can easily be estimated thanks to the quoted single $\Upsilon$ cross section
in the same kinematical conditions, namely $\sigma_{\Upsilon}=7.5\pm0.6$~mb. 
Let us add that we do not agree with the way $\sigma_{\rm eff}$
is estimated by CMS in their first study~\cite{Khachatryan:2016ydm}. We prefer to use $7.5$~mb inspired from the 
di-$J/\psi$ case discussed above. Doing so, the DPS cross section is readily evaluated to be 
$4 \pm 2$ pb --our quoted uncertainty is meant to account for the possibility 
of $\sigma_{\rm eff}$ ranging from $4$ to $15$~mb. It thus seemed that the impact
of the DPS in di-$\Upsilon$ production at the LHC in the central rapidity and low $P_T$ 
region would be below $5\%$. 

In their second study, CMS quoted a DPS fraction ($f_{DPS}$) of $39 \pm 14 \%$ obtained from fitting
the normalisations of both SPS NLO$^\star$ CSM and DPS $\Delta y$ distributions. This is reasonable given the
large scale uncertainty of the SPS but it assumes that the SPS scale uncertainty is correlated bin by bin, which
is not strictly true. Fitting the $M_{\Upsilon\Upsilon}$ yields $f_{DPS}=27\pm22\%$. More
statistics are therefore needed for a better determination of the DPS fraction within the range 5 to 50\%. Unfortunately, the single
$\Upsilon$ cross section, $\sigma_{\Upsilon}$, was not measured at 13~TeV in the same fiducial region which would allow one to derive $\sigma_{\rm eff}$.
Extrapolating from LHCb measurements at the same energy but at forward rapidity would 
however yield $\sigma_{\rm eff}$ on the order of 3mb~\cite{private:Caillol}. 

\subsubsection{Charmonium-bottomonium-pair hadroproduction}

Before the measurement by D0~\cite{Abazov:2015fbl}, theory predictions for the simultaneous production 
of a charmonium  and a bottomonium were reported in~\cite{Ko:2010xy,Likhoded:2015zna}. 
This reaction is remarkable in the sense that its CS contributions are expected 
to be suppressed since the direct CS LO contributions only appear at $\mathcal{O}(\alpha_s^6)$.
This is an additional $\alpha_s^2$ suppression compared to double-charmonium 
and double-bottomonium production, which we have just discussed. 

As such, it was proposed to be a golden channel to probe the COM at the 
LHC~\cite{Ko:2010xy}. So far, the fact that the (SPS) yield should be dominated
by COM contributions has never been contradicted
--unlike the case of $J/\psi+W$ which we discuss later. 
However, whether or not this makes this reaction a golden-plated probe of the COM
 crucially depends on the observability of the SPS yield 
in the presence~\cite{Likhoded:2015zna} of substantial DPS contributions .

\begin{figure}[hbt!]
\begin{center}
\subfloat[]{\includegraphics[width=0.35\columnwidth,draft=false]{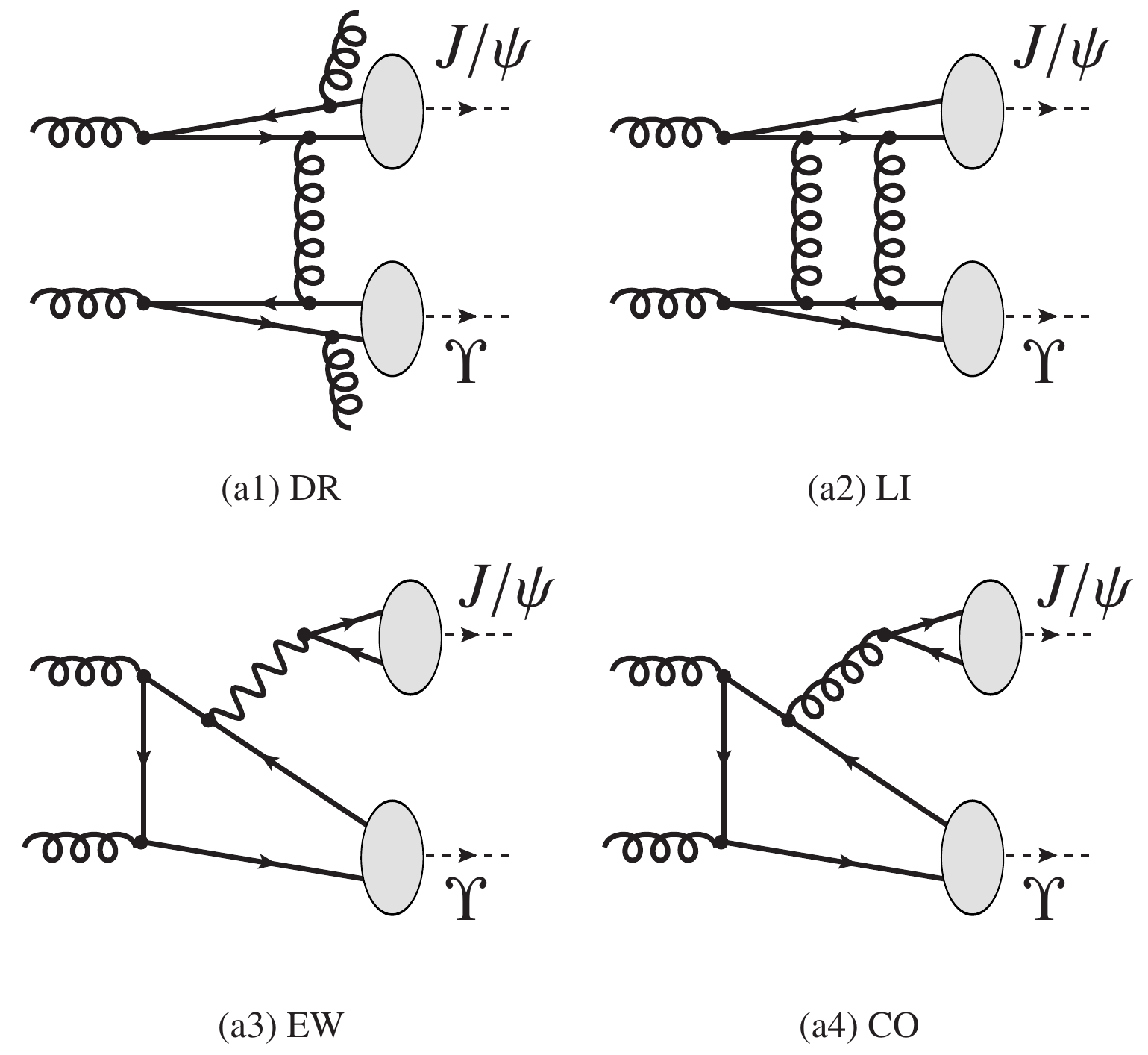}\label{fig:Upsi-Jpsi-diag}}\quad \quad
\subfloat[]{\includegraphics[width=0.48\columnwidth,draft=false]{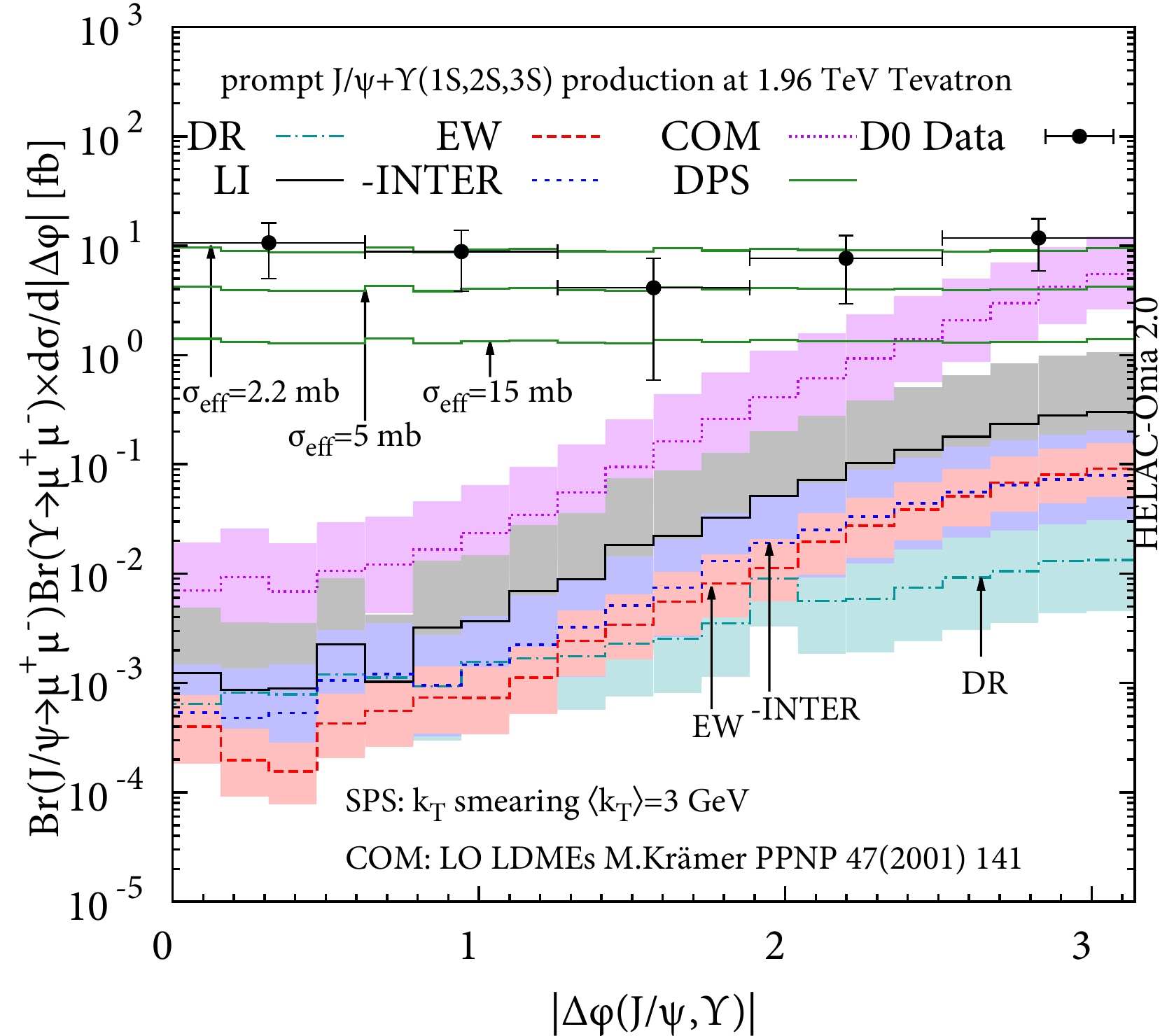}\label{fig:dphi_JpsiY_D0}}
\caption{(a) Representative graphs specific to $J/\psi+\Upsilon$ production: (a1) CS Double Real emissions (DR); (a2) CS Loop Induced (LI); (a3) CS Electroweak (EW); (a4) Colour Octet (CO); (b) $\Delta \phi_{\Upsilon J/\psi}$-differential cross section in the D$0$ acceptance
compared to various SPS contributions and that of the DPS for different $\sigma_{\rm eff}$. Adapted from~\cite{Shao:2016wor}.}
\label{fig:Upsi-Psi}
\end{center}
\end{figure}

Colour octet (CO) contributions can appear at $\mathcal{O}(\alpha_s^4)$ but 
remain suppressed by the small size of the CO LDMEs. In~\cite{Ko:2010xy}, 
Ko \etal\ only considered the  $c\bar{c}(\so)+b\bar{b}(\so)$, $c\bar{c}(\ssnew)+b\bar{b}(\so)$ and $c\bar{c}(\so)+b\bar{b}(\ssnew)$ channels and neglected the other CO channels which are not $P_T$
enhanced. Such an 
approximation may only make sense at large enough $P_T$ provided
that the CO LMDEs are of similar sizes. Until recently, 
the complete computation accounting for all the possible channels up to $v^7$ in 
NRQCD was lacking in the literature: there are indeed more than 50 channels at LO in $\alpha_s$ 
contributing to $\psi+\Upsilon$ production. Thanks to the automation 
of \HELACOnia~\cite{Shao:2012iz,Shao:2015vga}, we carried out~\cite{Lansberg:2015lva}
the first computation for AFTER@LHC at $\sqrt{s}=115$~GeV. $J/\psi+\Upsilon$ events
are at the limit of detectability with two years of data taking (20 fb$^{-1}$).

Following the D0 measurements~\cite{Abazov:2015fbl}, Shao and Zhang~\cite{Shao:2016wor}
performed the same computation for the Tevatron energies, with the notable addition
of the $\alpha_s^6$ CS contribution (DR and LI) as well as EW contributions.  These are depicted
on~\cf{fig:dphi_JpsiY_D0} and the corresponding $\Delta \phi_{\Upsilon J/\psi}$-differential cross section is shown on \cf{fig:Upsi-Psi} along with the D0 data. This computation remains the NRQCD state-of-the-art for this process.  First LO and NLO CEM computations were released in 2020~\cite{Lansberg:2020rft}.
At $|\Delta \phi_{\Upsilon J/\psi}| =\pi$, the NLO CEM is at best 5 times smaller than the data while, at $|\Delta \phi_{\Upsilon J/\psi}| =0$, it is more than 100 times smaller. This tends to confirm tha the D0 data are dominated by DPS contributions.

D0 reported an extraction of $\sigma_{\rm eff}=2.2 \pm 0.7 \text{ (stat.) }
\pm 0.9 \text{ (stat.)  nb}$ assuming their yield to be dominated by DPSs.
Thanks to their complete SPS analysis, Shao and Zhang~\cite{Shao:2016wor} instead
derived an upper limit, $\sigma_{\rm eff} \leq 8.2$~mb at 68\% CL without the need
for such an assumption.

\subsection{Quarkonia and vector bosons}

The associated hadroproduction of a $J/\psi$ or an $\Upsilon$ along
with a $W$ boson or a $Z$ boson has been motivated by many theoretical 
studies. The measurement of $J/\psi+W$ was for instance proposed to probe the COM via
gluon fragmentation. $J/\psi+\gamma$ hadroproduction and photoproduction 
should allow one to probe many novel aspects
of the gluon content in the proton or to test our understanding of the quarkonium-production mechanism. 
$J/\psi+\gamma$ production in $e^+e^-$ annihiliation and $\gamma\gamma$ fusion was also proposed as a test of NRQCD.

On the experimental side, with the advent of the LHC, the observation of the associated production of a quarkonium 
and a $W/Z$ became possible. Pioneering studies of CDF at 
the Fermilab-Tevatron~\cite{Acosta:2003mu,Aaltonen:2014rda} were motivated by 
the search for a charged Higgs, $H^\pm$, decaying in a pair of $\Upsilon+W^\pm$ for 
instance. In 2014, ATLAS succesfully observed for the first time the simultaneous 
production of $J/\psi+W$~\cite{Aad:2014rua} and $J/\psi+Z$~\cite{Aad:2014kba}. 
Similar processes involving bottomonia have never been 
observed.

As often in quarkonium physics, these new measurements seemed to introduce
new puzzles when compared the existing computations. However, just like for
quarkonium-pair production, it is very likely that the DPS in fact play the major
role, at least at the LHC. As a consequence, this forces us to reconsider
the earlier motivations brought in for their studies. 

In the following, we will briefly review these theoretical predictions and
the impact of DPS in the light of both ATLAS measurements

\subsubsection{Associated hadroproduction with a $Z$}
\label{sec:onium_Z}

Whereas the corresponding SPS process may give us complementary information on quarkonium 
production if it happens to be experimentally accessible at the LHC, it
also offers an interesting theoretical playground for the understanding of the QCD corrections in 
quarkonium-production processes. Indeed, as can be seen on \cf{diagrams-psiZ-CSM}, the structure
of the QCD corrections is very similar to that of single $J/\psi$ or $\Upsilon$ with the notable
difference --apart from the emission of a colourless particle in the final state-- that the natural scale of the process would be much higher.

\begin{figure}[htb!]
\centering
\subfloat[Born]{\includegraphics[width=0.17\textwidth]{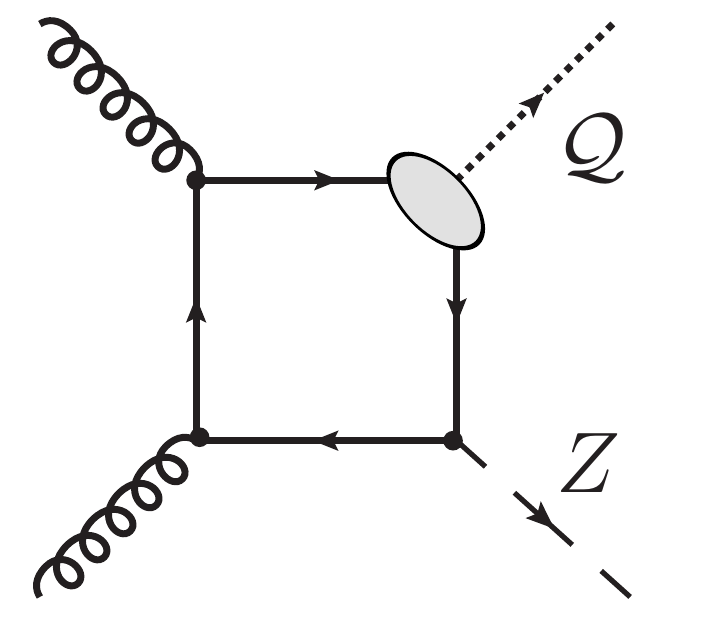}\label{fig:LO-CSM-psiZ}}\hspace*{-.2cm}
\subfloat[NLO loop ]{\includegraphics[width=0.17\textwidth]{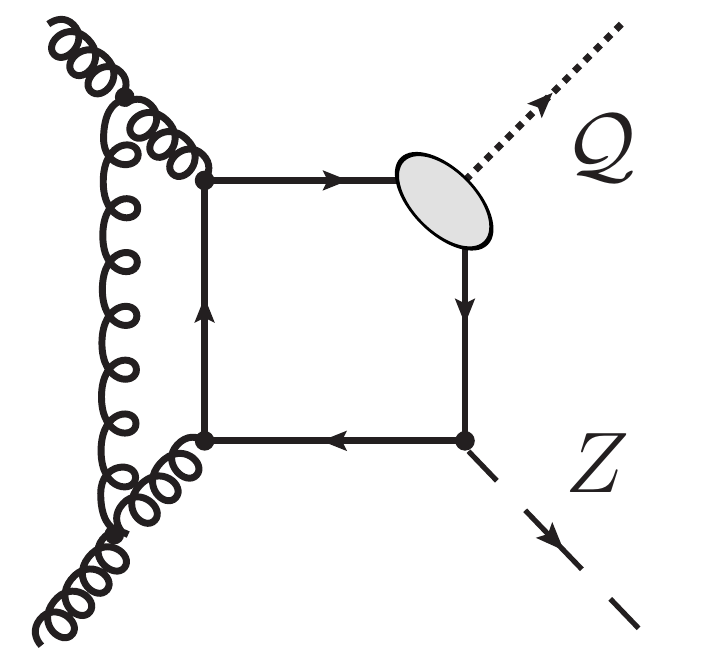}\label{NLO-PT8-CSMpsiZ}}\hspace*{-.2cm}
\subfloat[NLO real emission from the heavy quark]{\includegraphics[width=0.2\textwidth]{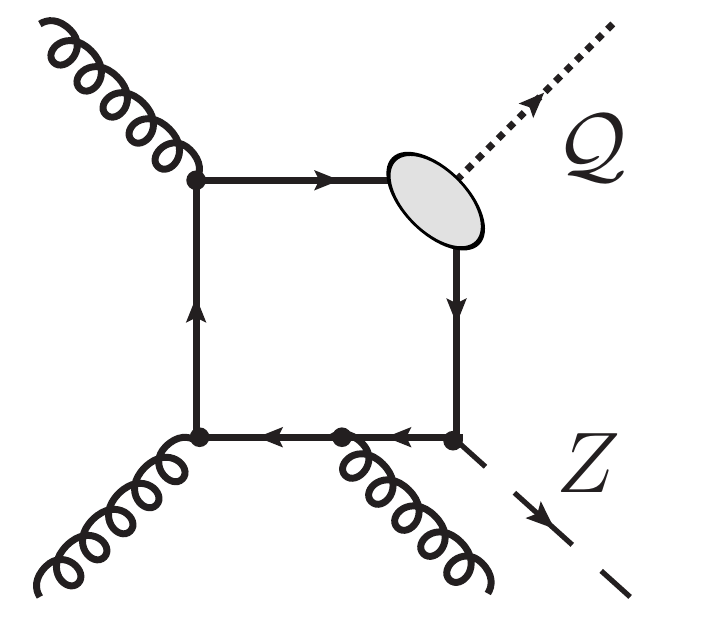}\label{NLO-PT8-real-CSMpsiZ}}\hspace*{-.2cm}
\subfloat[$t-$CGE]{\includegraphics[width=0.17\textwidth]{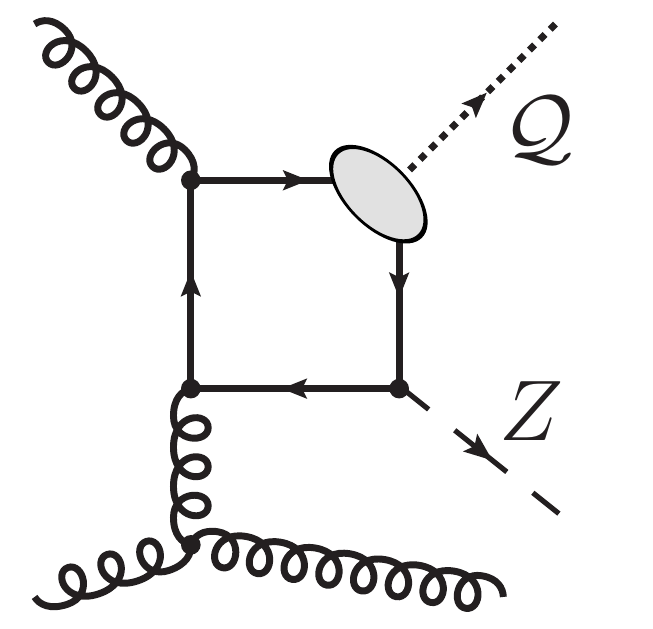}\label{NLOg-PT6-CSMpsiZ}}\hspace*{-.2cm}
\subfloat[$t-$CGE]{\includegraphics[width=0.17\textwidth]{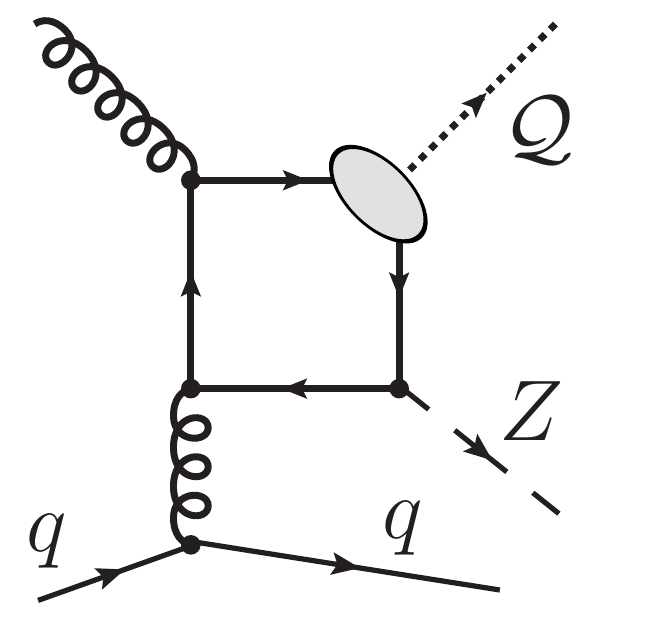}\label{NLOq-PT6-CSMpsiZ}}\hspace*{-.2cm}
\caption{Representative diagrams contributing to $J/\psi$ and $\Upsilon$ {(denoted $\Q$)} hadroproduction with a $Z$ boson in the CSM
 by gluon fusion at orders $\alpha \alpha_s^2$ (a), $\alpha \alpha_s^3$ (b,c,d) and initiated
by a light-quark gluon fusion at order $\alpha \alpha_s^3$ (e).
The quark and anti-quark attached to the ellipsis are taken as on-shell
and their relative velocity $v$ is set to zero.}
\label{diagrams-psiZ-CSM}
\end{figure}

In particular, it offers a nice case to study the alteration of the quarkonium polarisation by the QCD corrections at large $P_T$. It also allows for a check, with yet another process, of the
NLO$^\star$ procedure and for the confirmation of our understanding 
of the effect of new topologies opening at NLO and of the scale dependence in such a multi-scale process.

Pioneer studies of its SPS cross section were carried at LO in 
 NRQCD (CS+CO channels) around 2000's~\cite{Braaten:1998th,Kniehl:2002wd}
with the conclusion that ``{\it The dominant production mechanism involves the binding  of a 
color-octet $b \bar b$ pair into a P-wave bottomonium state 
which subsequently decays into $\Upsilon$}"~\cite{Braaten:1998th}.
To date, only two NLO analyses were reported, in fact bearing on the direct yield.
The first in fact only focused on $J/\psi+Z$~\cite{Mao:2011kf} 
 and it is not strictly speaking a complete
 NLO NRQCD analysis despite having considered both CS and CO contributions. 
Indeed the $^3P_J^{[8]}$ transitions --notably more complicated to evaluate-- 
were not computed and, as we know now, a small value of the corresponding 
LDME would contradict single-$J/\psi$ data (see section~\ref{subsec:COM_updates}).
 One year later, we reconsidered~\cite{Gong:2012ah} the sole CS contributions
both for $J/\psi+Z$ and $\Upsilon+Z$, 
 corrected a mistake in~\cite{Mao:2011kf}, presented first polarisation 
results and argued, owing to the scale dependence of the process, that a 
reasonable choice for the scales is rather $M_Z$ than $\sqrt{M_{J/\psi}^2+P_T^2}$ 
as used in~\cite{Mao:2011kf}.

When the ATLAS collaboration released its study~\cite{Aad:2014kba}, they obviously 
attempted to compare their data to these NLO predictions. They first realised 
that DPS contributions were probably significant, then noted that the aforementionned 
predictions were well below their analysis. We will review all these aspects in the following
and argue that in fact the ATLAS data agree with theory, but not quite how
predicted.

In such electroweak processes, the identification  of the leading contributions
is quite intricate. The gluon fusion does not necessarily dominate under all production
mechanisms. Owing to the presence of a quark line, from where the electroweak
boson is emitted, the off-shell-photon production can also contribute. For all these
reasons, we find it better to start the discussion with the simpler case
of direct production by CS transitions, leaving aside first these 
QED contributions. 

This is what we did in 2012~\cite{Gong:2012ah} since we were intrigued by the
unusual behaviour of the CS $P_T$ differential cross section --with a LO yield
larger than the NLO at large $P_T$ found by
Mao \etal~\cite{Mao:2011kf} --which was in fact later corrected. Indeed, 
as aforementioned, the normal expectation was a very similar behaviour
as that of single-vector-quarkonium production, with a large --positive-- impact
of the NLO corrections, at least at large $P_T$. It was then natural to check
that nothing unusual occurred for this process and that we could rely
on the order of magnitude of the LO evaluation as the expectation of
possible experimental measurements.

\paragraph{CSM contributions.}
 
We will outline its computation with some basic reminders. The matrix element to create
a $^3S_1$  quarkonium ${\Q}$ with a momentum $P_\Q$ and a polarisation $\lambda$
 accompanied by other partons, noted $j$, and a $Z$ boson of momentum $P_Z$
reads
\eqs{ \label{eq:CSM-PsiZ}
{\cal M}&(ab \to {\Q}^\lambda(P_\Q)+Z(P_Z)+j)=\!\sum_{s_1,s_2,i,i'}\!\!\frac{N(\lambda| s_1,s_2)}{ \sqrt{m_Q}} \frac{\delta^{ii'}}{\sqrt{N_c}} 
\frac{R(0)}{\sqrt{4 \pi}} 
{\cal M}(ab \to Q^{s_1}_i \bar Q^{s_2}_{i'}(\mathbf{p}=\mathbf{0}) +Z(P_Z) + j),
}
with the same definitions as in \ce{eq:CSM_generic}. LO cross sections can be evaluated
by automated code such as \MadOnia~\cite{Artoisenet:2007qm}, \HELACOnia~\cite{Shao:2012iz,Shao:2015vga}. The semi-automated framework FDC~\cite{Wang:2004du} further
allows one to advance its computation to NLO in $\alpha_s$ along with that of the polarisation
coefficient $\lambda_\theta$ (or $\alpha$).

At LO, there is only a single partonic process at work, namely $gg\to \Q Z$ --completely analogous to
$gg\to \Q \gamma$ for $\Q$-prompt photon associated production-- with 4 Feynman graphs to be evaluated. 
One of them  is drawn on \cf{fig:LO-CSM-psiZ}.  
The NLO contributions can be divided in two sets: that of the virtual corrections which arise from loop
diagrams (\cf{NLO-PT8-CSMpsiZ}), the other gathers the real (emission) corrections where one more particle appears in the final state (\cf{NLO-PT8-real-CSMpsiZ}-\ref{NLOq-PT6-CSMpsiZ}).  
To be complete, the real corrections (\cf{NLO-PT8-real-CSMpsiZ}-\ref{NLOq-PT6-CSMpsiZ}) come from three parton level sub-processes:
\bea
g+g&\rightarrow & \ssnew + Z + g, \label{eq:ggpsiZg} \\
g+q(\bar{q})&\rightarrow &\ssnew + Z + q(\bar{q}),\label{eq:gqpsiZq} \\
q+\bar{q}&\rightarrow & \ssnew + Z + g,\label{eq:qqpsiZg}
\eea
where $q$ denotes light quarks with different flavours ($u,d,s$). Since the scale 
of the process is large because of the presence of the $Z$ boson, one could consider the
heavy-quark-gluon fusion contribution, $g Q \to \ssnew + Z +Q$, in a 4(5) flavour scheme (FS), resumming logarithms of $\mu_F$ and $m_Q$. 
In a 3(4) FS, it corresponds
to $gg \to \ssnew + Z + Q \bar{Q}$ analogous to $gg \to \ssnew + g + Q \bar{Q}$ which
we considered in~\cite{Artoisenet:2007xi}. The latter however appears
at $\alpha_s^3 \alpha$ and was left for future studies. Such channels might become
relevant when high $P_T^{\Q}$ associated events  can be collected.

In the following, we show our results  for $|y^{J/\psi}|< 2.4$ --the usual $J/\psi$ acceptance for the CMS and ATLAS detectors--
at 8 TeV \footnote{The parameters entering the evaluation  of the cross section as well as additional plots for 14 TeV  can be found in~\cite{Gong:2012ah}. The cross section at 13 TeV is 12 \% smaller than at 14 TeV.} for the 
renormalisation and factorisation scales set at $m_Z$. We have 
kept\footnote{Note that we could have evaluated the cross section for lower $P_T$ where the cross section is well behaved. 
However, we do not expect --at least in the central region-- any experimental measurement 
to be carried out in this region owing to the momentum cut on the muons because of the strong magnetic fields in the ATLAS and CMS detectors.} 
the cut $P_T^{J/\psi} > 3$ GeV.
Our result at $\sqrt{s}=8$ TeV are depicted
on \cf{fig:dsigpdt-NLO-psiZ-8TeV}. The dotted blue line is our LO result and the solid grey line is our prediction at NLO. It is 
obvious, contrary to what was obtained in~\cite{Mao:2011kf}, that the yield at NLO is getting larger than at LO for increasing
$P_T$. This is similar to what happens in the inclusive case. This is also indicative that new leading-$P_T$ topologies, 
in particular $t$-CGE, start dominating rather early in $P_T$ despite of the presence of a $Z$ boson in the process. 
At $P_T =150$~GeV, the NLO yield is already ten times that of LO. 

\begin{figure}[htb!]
\centering
\subfloat[$J/\psi+Z \text{ @ } \sqrt{s}=8$ TeV]{\includegraphics[width=0.47\textwidth]{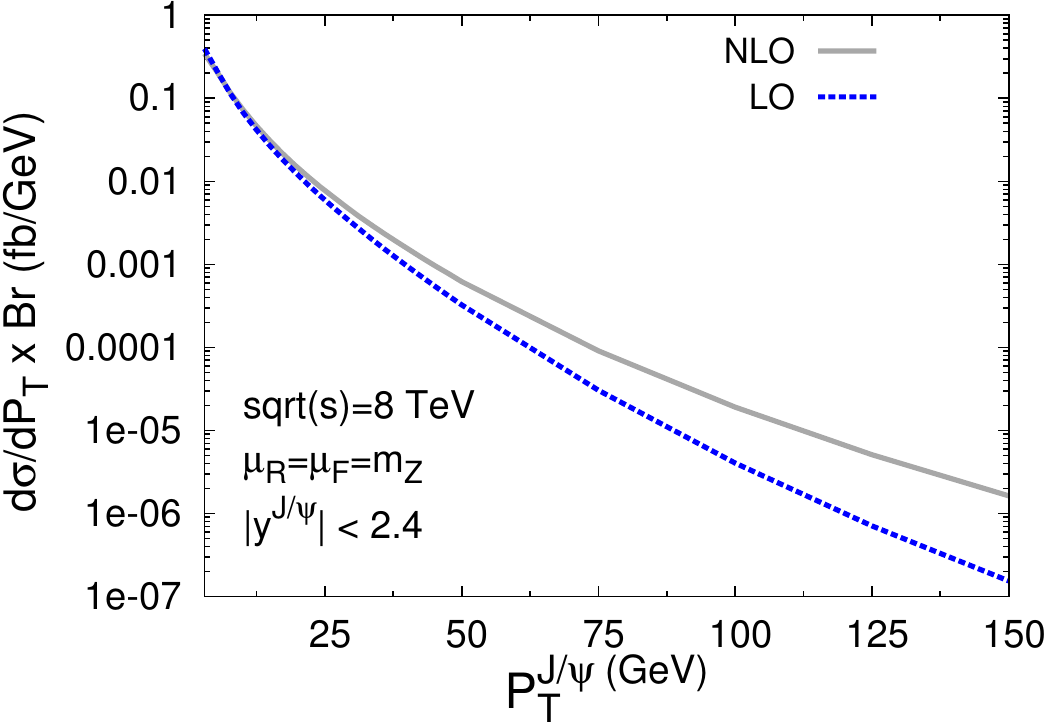}\label{fig:dsigpdt-NLO-psiZ-8TeV}}
\subfloat[$\Upsilon+Z \text{ @ } \sqrt{s}=8$ TeV]{\includegraphics[width=0.5\textwidth]{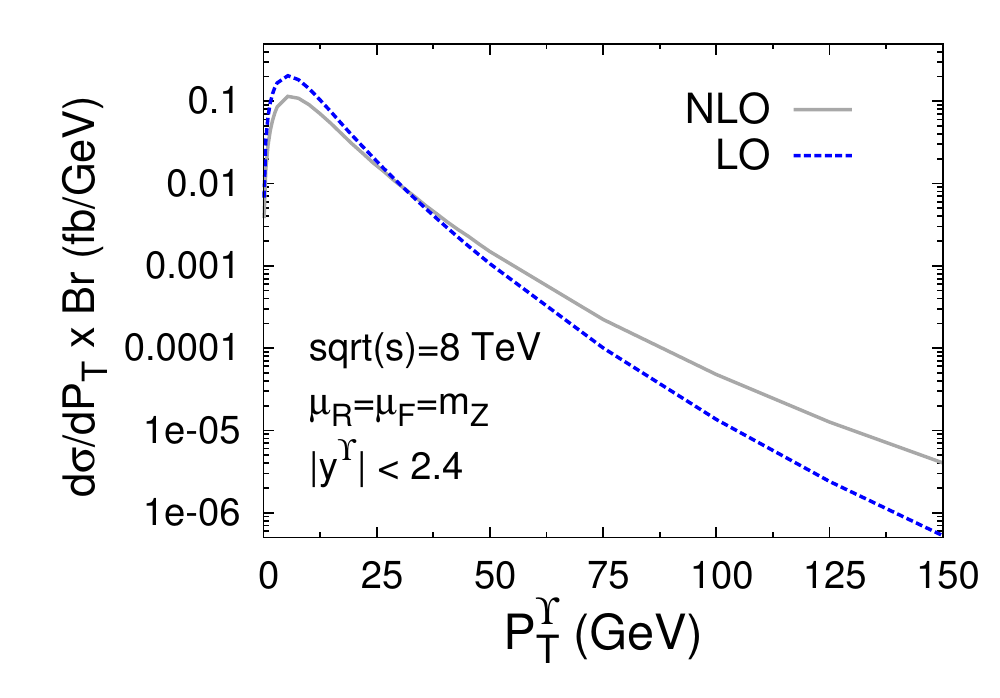}\label{fig:dsigpdt-NLO-upsiZ8TeV}}
\caption{Differential cross section for direct $\Q+Z$ in the CSM vs. $P_T$ at LO (blue-dashed) and NLO (grey-solid)
with  $\mu_F=\mu_R=m_Z$. Taken from~\cite{Gong:2012ah}.}
\label{dsdpt-OniumZ-CSM}
\end{figure}

Along the same lines as for $J/\psi$, we have also evaluated the cross section 
for direct-$\Upsilon+Z$  (see~\cf{fig:dsigpdt-NLO-upsiZ8TeV}). We have set the factorisation and renormalisation scales at the same value, namely $\mu_F=\mu_R=m_Z$, and we used the same PDF sets as for the $J/\psi+Z$ case. 
Let us note that the CS cross sections for $J/\psi+Z$ and $\Upsilon+Z$ production are of similar size
for this specific process where the $Z$ mass is the most relevant scale . 

As we announced, the current process is very interesting theory-wise to
study the relevance of the NLO$^\star$ method, which we evaluated with \MadOnia\ \cite{Artoisenet:2007qm},
 slightly tuned to implement an IR cut-off
on all light parton-pair invariant mass. On the way, let us note the LO cross section has also been checked with \MadOnia.
The procedure used here to evaluate the leading-$P_T$ NLO contributions is exactly the same as 
in~\cite{Artoisenet:2008fc} (see section~\ref{subsec:NNLOstar-CSM}) but for the process, 
$pp\to J/\psi+Z+\hbox{jet}$. Namely, the 
real-emission contributions at $\alpha \alpha_s^3$ are evaluated using \MadOnia\ by 
imposing a lower bound on the invariant mass of 
any light parton pair ($s^{\rm min}_{ij}$). The underlying idea in the single quarkonium case 
was that for the new channels opening up at NLO which have a leading-$P_T$ behaviour \emph{w.r.t.} to LO ones (for instance the $t$-CGE),  the cut-off dependence should decrease for increasing $P_T$ 
since no collinear or soft divergences can appear with this topologies at large $P_T$.
For other NLO channels, whose Born contribution is at LO, the cut would 
produce logarithms of $s_{ij}/s_{ij}^{\rm min}$, which are not necessarily negligible. Nevertheless, 
they can be factorised over their corresponding Born contribution, which scales as $P_T^{-8}$, and hence they are
suppressed by at least two powers of $P_T$ with respect to the leading-$P_T$ contributions ($P_T^{-6}$) at this order. 
The sensitivity on $s_{ij}^{\rm min}$ should vanish at large $P_T$. As we saw, this argument 
has been checked in the inclusive case for $\Upsilon$~\cite{Artoisenet:2008fc} and $\psi$~\cite{Lansberg:2008gk}
  as well as in association with a photon~\cite{Lansberg:2009db} (see section~\ref{sec:psi-gamma}).
Because of the presence of the $Z$ boson mass, it is not a priori obvious that $t-$CGE topologies  dominate over
the LO ones. It was thus not clear at all how such a procedure to evaluate
the NLO$^\star$ yield can provide a reliable evaluation of the full NLO of $J/\psi +Z$. In fact, at mid $P_T$, significantly below
the $Z$ boson mass, the difference of the $P_T$ dependence of the NLO and LO cross 
sections is maybe not large enough for the dependence on 
$s_{ij}^{\rm min}$ to decrease fast. Having at hand a full NLO computation, we could carry out such a
comparison and better investigate the effect of QCD corrections in quarkonium production. 

\begin{figure}[htb!]
\centering
{\includegraphics[width=0.6\textwidth]{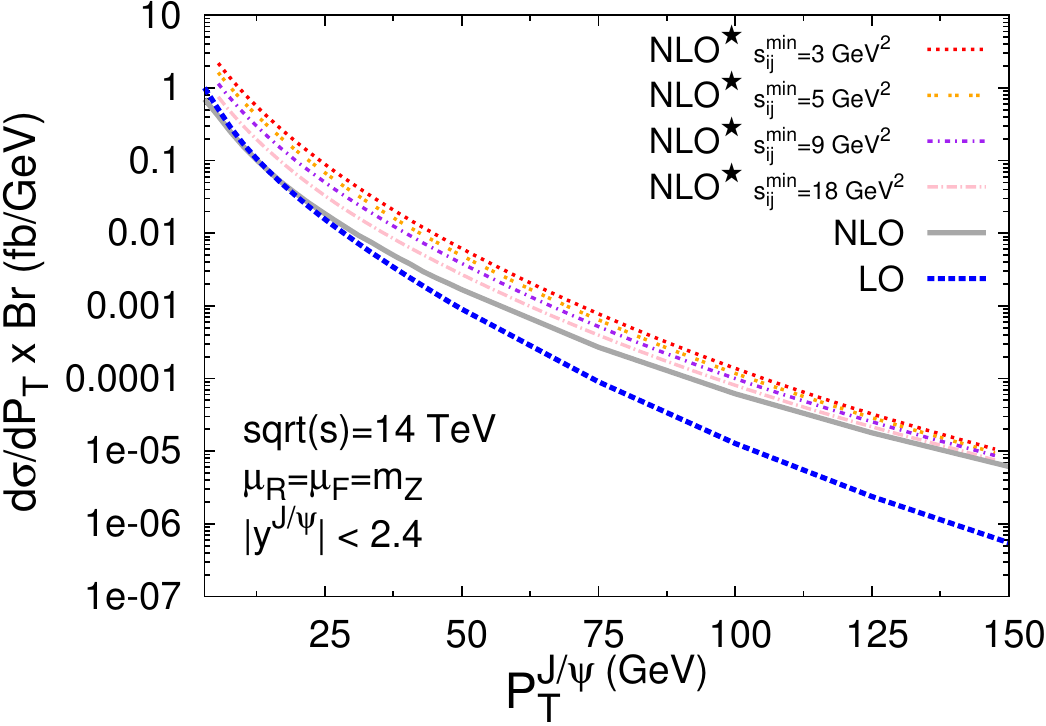}}
\caption{Differential cross section for $J/\psi+Z$ vs. $P_T$ at $\sqrt{s}=14$ TeV at LO (blue dashed) and NLO (grey solid)
with  $\mu_F=\mu_R=m_Z$ along with the NLO$^\star$ for different values of $s_{ij}^{\rm min}$ (red dotted, yellow double dotted, 
purple dash dotted and pink long-dash dotted).  Taken from~\cite{Gong:2012ah}.
}
\label{fig:dsigpdt-NLO-14TeV}
\end{figure}

The dominance of $t$-CGE topologies hinted at by the LO vs NLO comparison can be quantified by another comparison to the results from the NLO$^\star$ evaluation. 
The various dotted lines on \cf{fig:dsigpdt-NLO-14TeV} show the NLO$^\star$ evaluation for different 
cut-off values. Two observations can be made: 1) they converge to the NLO steadily for increasing 
$P_T$, 2) for $P_T> m_Z$, the NLO$^\star$ evaluations are within a factor of 2 compatible with the complete NLO
yield. This confirms that loop corrections are sub-leading in $P_T$ and can be safely neglected 
for $P_T$ larger than all the masses relevant for the process under consideration and that new 
topologies appearing at NLO, the $t$-CGE ones, dominate at large $P_T$. At low $P_T$, where the NLO and LO yields 
are similar, the NLO$^\star$ overestimate the NLO. It would be very interesting to apply the method recently proposed
 by Shao~\cite{Shao:2018adj} for a ``nLO'' evaluation.

As announced, we believe that a scale close to $m_Z$, rather than the 
transverse mass of the $J/\psi$ taken in~\cite{Mao:2011kf},
is more appropriate.  A similar choice was for instance made 
for $Z+b-$jet~\cite{Campbell:2003dd} which is a very similar process.
Whereas the final numbers do not vary too much from one choice to the other, 
the default scale choice significantly affects the scale sensitivity.

\begin{figure}[htb!]
  \centering 
\subfloat[Scale dependence with $\mu_0=m_T^{J/\psi}$]{\includegraphics[width=0.5\textwidth]{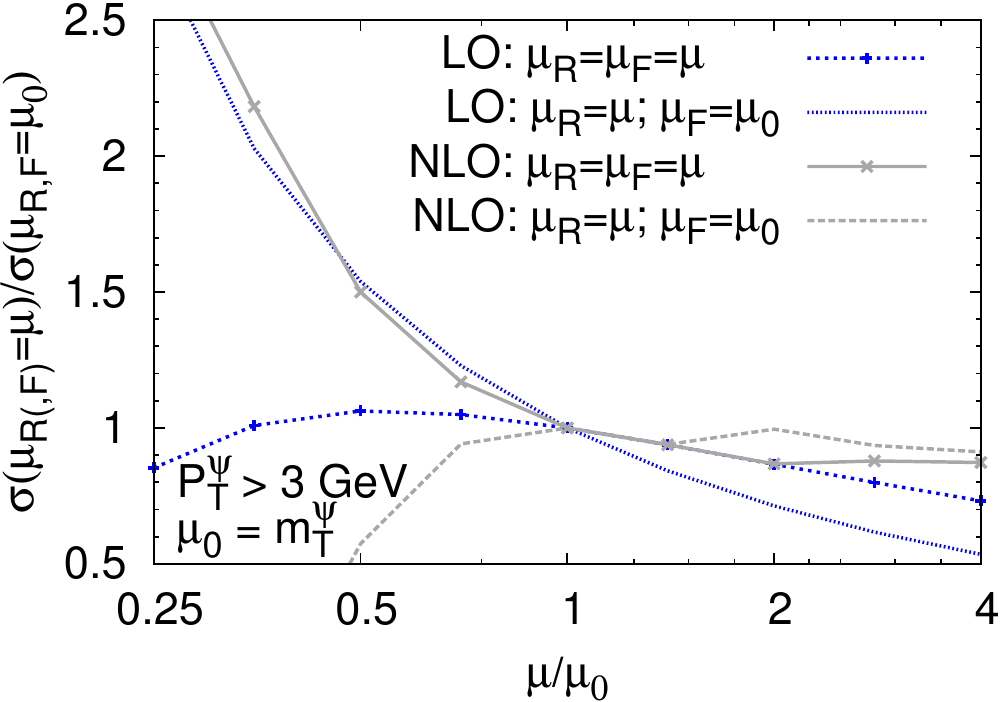}}
\subfloat[Scale dependence with $\mu_0=m_Z$]{\includegraphics[width=0.5\textwidth]{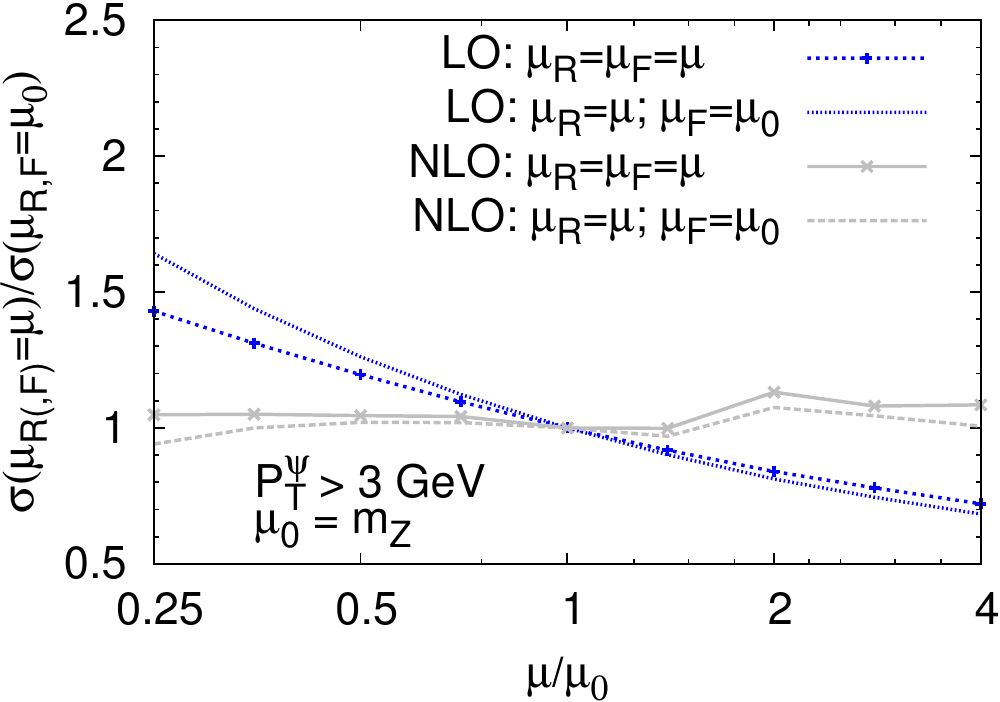}}
\caption{(a) Renormalisation and factorisation scale dependence of 
the LO and NLO yield for $P_T>3$ GeV with $\mu_0=m_T^{J/\psi}$. (b) Same plot as (a) for $\mu_0=m_Z$. Taken from~\cite{Gong:2012ah}.}
\label{fig:comp-mT-mZ}
\end{figure}

\cf{fig:comp-mT-mZ} displays the scale sensitivity at low $P_T$ around both these choices
of a ``default'' scale, $\mu_0$, (a) $m_T$ and (b)  $m_Z$. 
With $\mu_0=m_Z$, the cross section at NLO is more stable, 
except for a bump at $m_t$ [to be corrected by properly setting the value of $\Lambda^{[6]}$ in the running 
of coupling constant (currently 0.151 MeV with $m_t=180$~GeV), which matters for $\mu_R > m_t$]. Taking
$\mu_0=m_T$, the NLO results are clearly unstable and they even may then artificially be enhanced. 

Since the real-emission contributions
 at $\alpha \alpha^3_s$ dominate for $P_T\, \gtrsim\, m_Z/2$,
the yield scale dependence should correspondingly be affected. At low $P_T$ ($\ll m_Z$), 
we expect a reduced scale dependence since we indeed benefit from the NLO precision. 
At large $P_T$, the leading process becomes  $pp \to J/\psi +Z+\hbox{parton}$. 
The loop contributions are small and are not expected to reduce the scale sensitivity. 
In fact, we expect a larger sensitivity
 on the renormalisation scale, $\mu_R$, as the leading process shows an additional power of $\alpha_s(\mu_R)$.
Such an evolution of the scale sensitivity can explicitly be analysed by 
varying $\mu_F$ and $\mu_R$ together and then $\mu_R$ 
alone by a factor 2 about the ``default'' scale $m_Z$ for 3 $P_T$ regions, 
\ie\ $P_T \geq 3, 50 \text{ and } 150$~GeV.

\begin{figure}[htb!]
\centering 
\subfloat[Scale dependence at LO]{\includegraphics[width=0.5\textwidth]{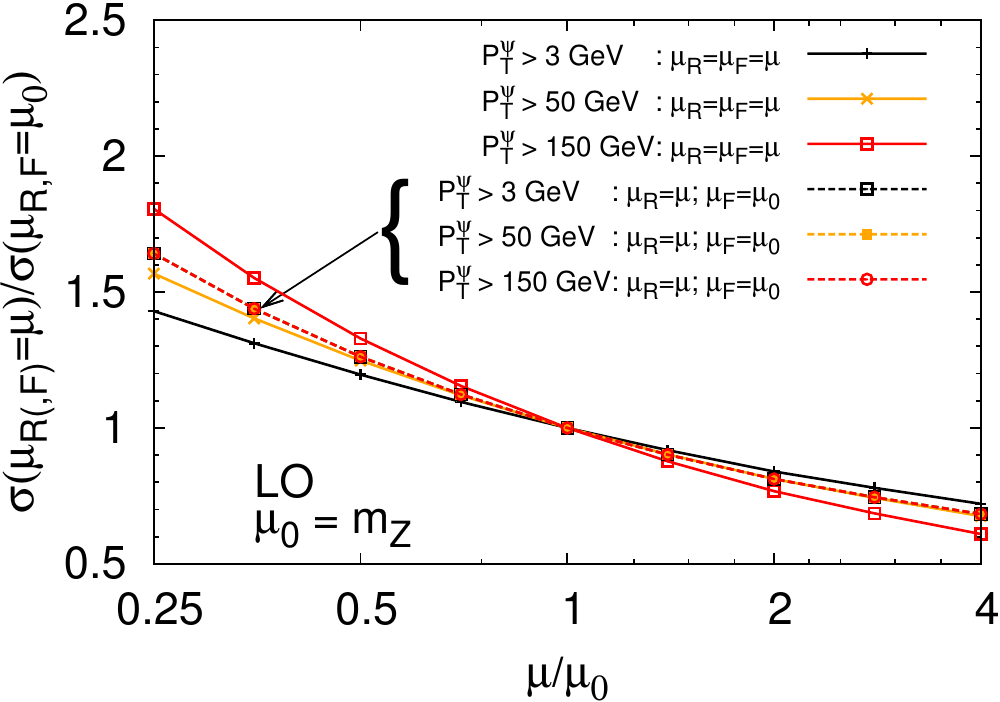}}
\subfloat[Scale dependence at NLO]{\includegraphics[width=0.5\textwidth]{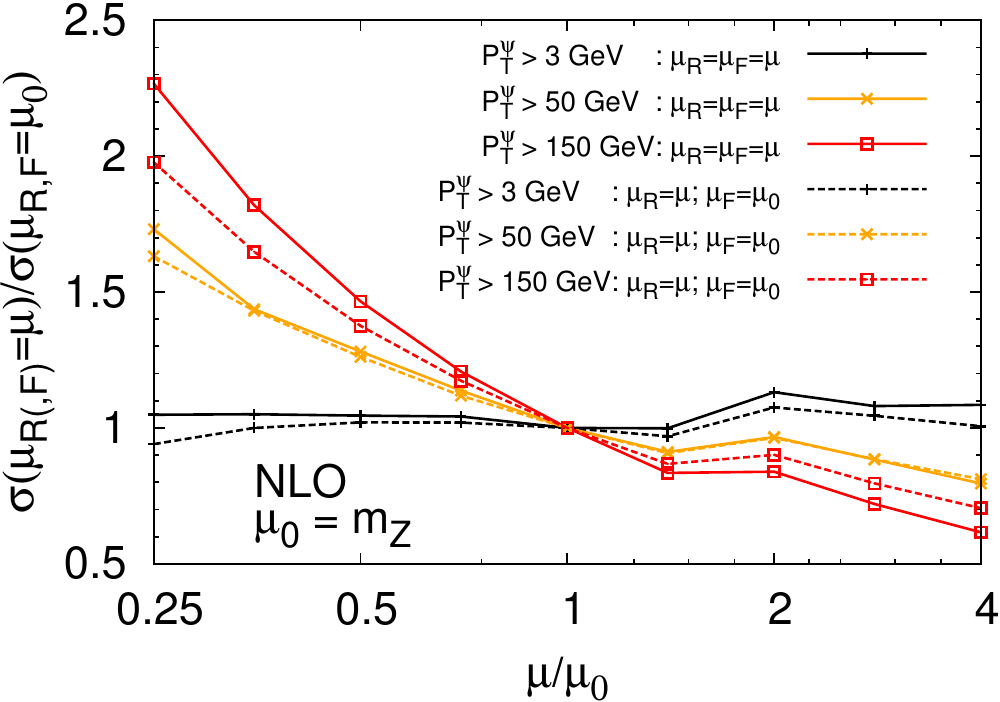}}
\caption{ Scale dependence of 
the yield at LO (a) and NLO (b) for $P_T>3$~GeV, $P_T>50$~GeV, $P_T>150$~GeV 
where both $\mu_R$ and $\mu_F$ are varied together ($\mu_F=\mu_R$, solid lines)
about  $\mu_0=m_Z$ and only $\mu_R$ is varied ($\mu_F$ fixed, dashed lines). For
simplicity,  $\alpha$ 
has been kept fixed. Taken from~\cite{Gong:2012ah}.}
\label{fig:scale-sensitivity}
\end{figure}

As anticipated for $P_T\,\gsim\, m_Z$ (red curves), we observed on~\cf{fig:scale-sensitivity},  
a stronger scale sensitivity of the NLO yield (b) --at $\alpha \alpha_s^3$-- 
than of the LO yields (a)-- at $\alpha \alpha_s^2$. The NLO curve with $\mu_F$
fixed clearly shows that it essentially comes from $\mu_R$.
At mid $P_T$ (orange curves), it is similar at LO and NLO, while at low $P_T$ (black curves), the NLO yield is less scale dependent than the LO. 

Although it is likely not observable until the high lumonisity phase of the LHC,
we have also analysed the effect of the QCD correction to the polar anisotropy 
of the dilepton decay of the $\Q$, $\lambda_\theta$ or $\alpha$. The latter  
can be evaluated from
the polarised hadronic cross sections:
\be
\lambda_\theta(P_T)=\frac{\frac{d\sigma_T}{d P_T}-2 \frac{d\sigma_L}{d P_T}}
                 {\frac{d\sigma_T}{d P_T}+2 \frac{d\sigma_L}{d P_T}}.
\ee

\begin{figure}[htb!]
\centering
\subfloat[$J/\psi+Z$]{\includegraphics[width=0.5\textwidth]{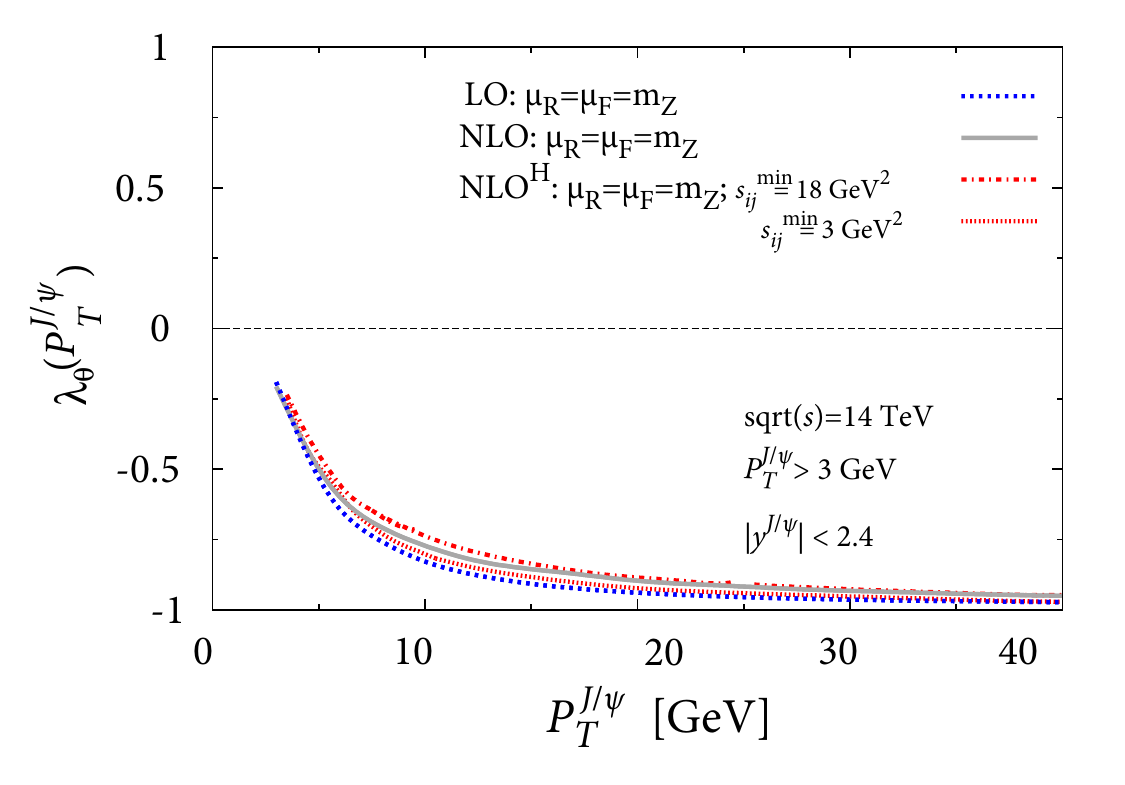}\label{fig:alpha-psiZ-LHC-14TeV}}
\subfloat[$\Upsilon+Z$]{\includegraphics[width=0.5\textwidth]{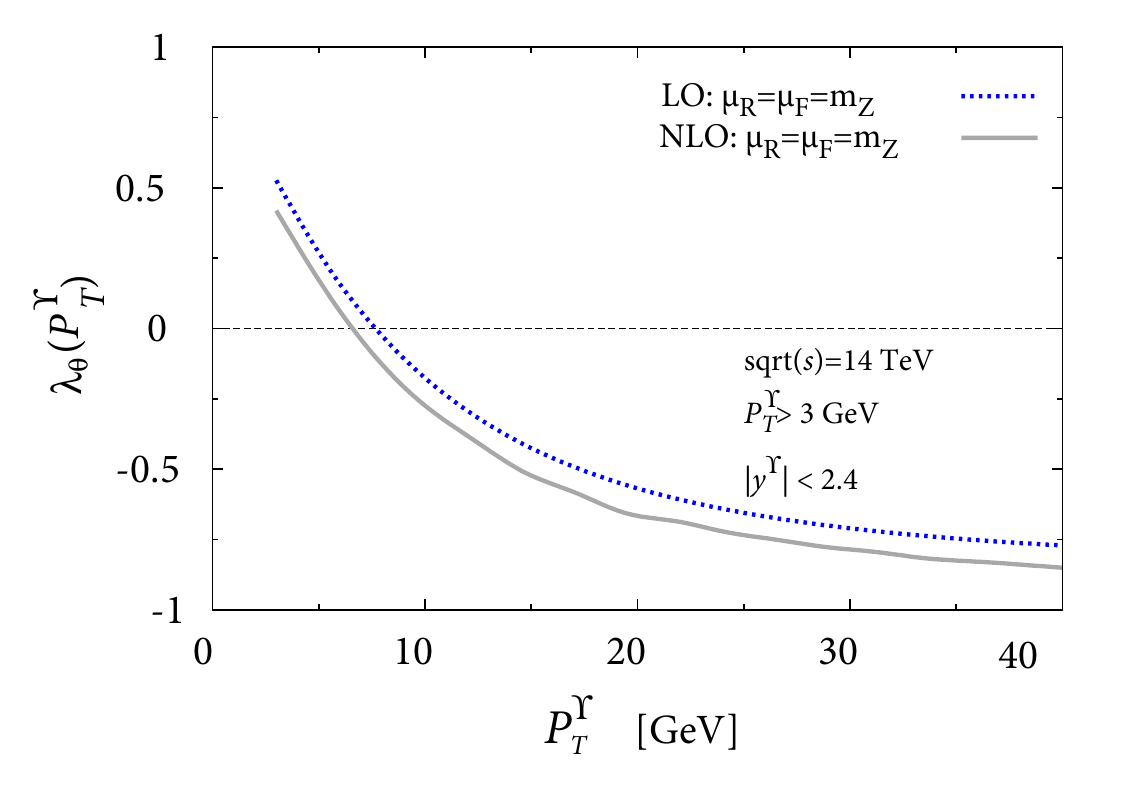}\label{fig:alpha-upsiZ-LHC-14TeV}}
\caption{$P_T$ dependence of the polarisation (or azimuthal anisotropy) in the helicity frame of the direct (a) $J/\psi$ and (b) $\Upsilon$ 
produced with a $Z$ boson at LO and NLO  at $\sqrt{s}=14$ TeV. Taken from~\cite{Gong:2012ah}.}
\label{fig:pol-psiZ}
\end{figure}

Its evaluation follows the same line as in the inclusive case (see \eg\ \cite{Gong:2008hk}) 
and we do not  repeat the procedure here.  Several choices of the polarisation 
frames are commonly used (see~\cite{Beneke:1998re,Faccioli:2010kd,Faccioli:2012nv}) 
and we have chosen to work in the helicity frame.
\cf{fig:pol-psiZ}  shows our results at 14 TeV for both direct $J/\psi+Z$ and $\Upsilon+Z$.
The direct-$J/\psi$ yield with a $Z$ boson gets strongly longitudinal for $P_T^{J/\psi}$ larger than 5 GeV. The NLO and  NLO$^\star$ results coincide and the latter is nearly insensitive to the IR cutoff. 
Interestingly, the NLO and the LO results are also very similar. This is the 
first time that such an observation is made for the CSM. 
For the $J/\psi+X$  or $J/\psi+\gamma$, the LO and NLO yields have 
a completely different polarisation. It therefore seems that 
when a $Z$ boson is emitted by one of the charm quarks forming the $J/\psi$, the latter is longitudinally 
polarised, irrespective of the off-shellness and of the transverse momentum of the gluons producing the 
charm-quark pair. This is not so when a photon or a gluon is emitted in the final
state. A very similar behaviour is observed for the $\Upsilon$ where one however
notes (\cf{fig:alpha-upsiZ-LHC-14TeV}) a slight gap between the LO and NLO polarisation.

\paragraph{The predictions including COM contributions.}

\begin{figure}[hbt!]
\centering
\subfloat[]{\includegraphics[scale=.375,draft=false]{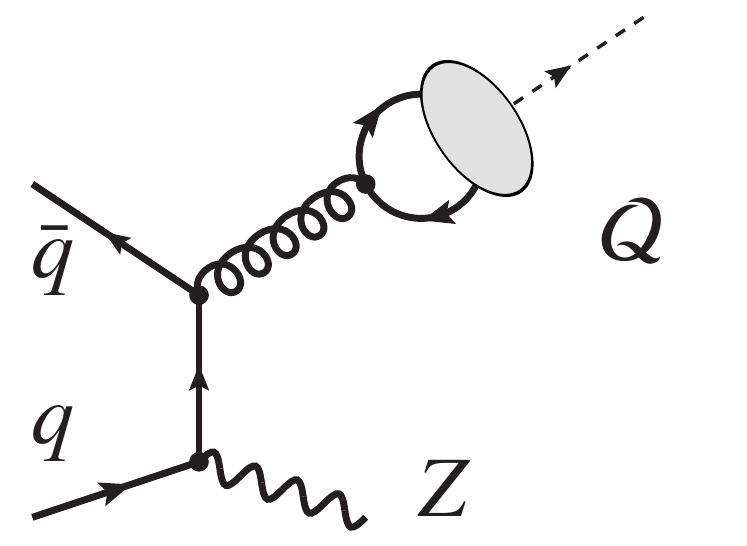}\label{diagram-a-psiZ-COM}}
\subfloat[]{\includegraphics[scale=.375,draft=false]{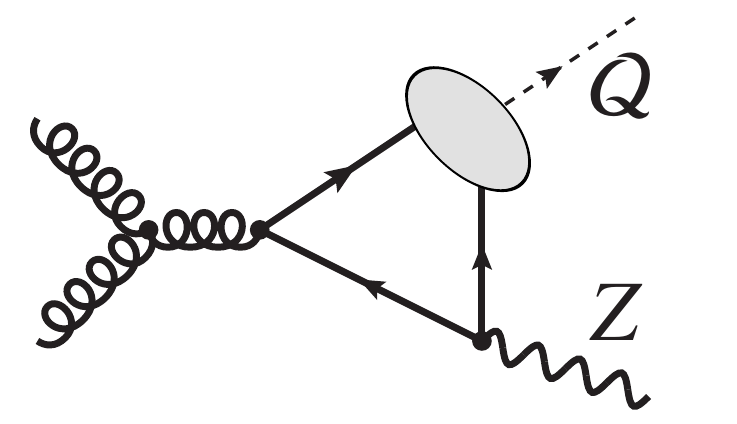}\label{diagram-b-psiZ-COM}}
\subfloat[]{\includegraphics[scale=.375,draft=false]{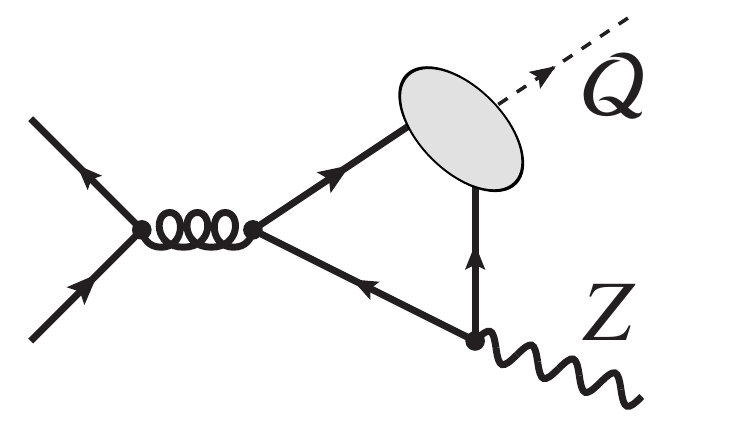}\label{diagram-c-psiZ-COM}}
\subfloat[]{\includegraphics[scale=.375,draft=false]{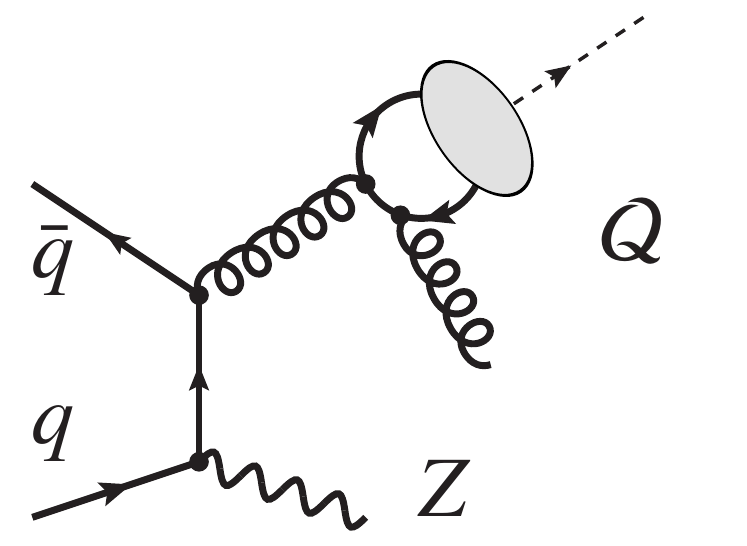}\label{diagram-d-psiZ-COM}}
\subfloat[]{\includegraphics[scale=.375,draft=false]{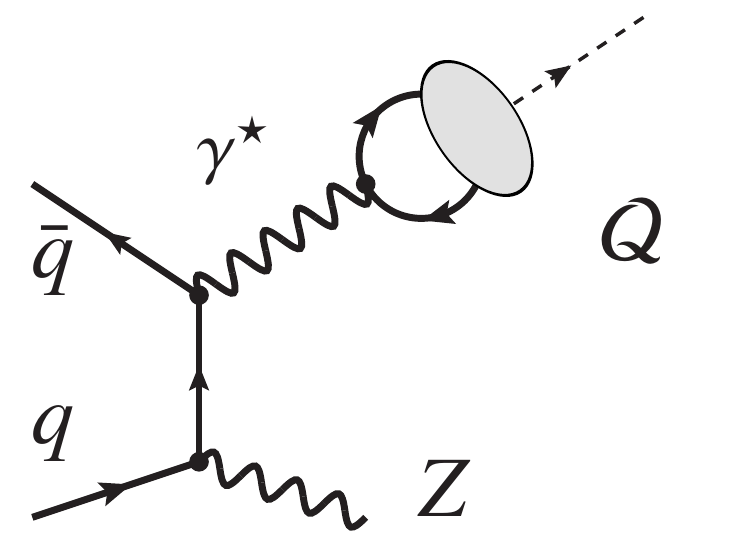}
\label{diagram-qq-psiZ-CSM}}
\caption{Representative diagrams contributing to $\Q+ Z$ hadroproduction in the COM at orders (a-c) $\alpha_s^2\alpha$, (d) $\alpha_s^3\alpha$  and in the CSM at orders (e) $\alpha^3$, which was not considered previously. The quark and anti-quark attached to the ellipsis are taken as on-shell and their relative velocity $v$ is set to zero.
[$\Q$ stands for $\Upsilon$ or $\psi$].}
\end{figure}

Let us now move on to the discussion of the computations taking into the 
COM contributions. 
Even though an NLO evaluation exist, it is instructive to look back
at LO computations which in fact envisioned more channels. 
To the best of our knowledge, the first one was reported in 1998 by Braaten \etal~\cite{Braaten:1998th} and
focused on the $\Upsilon$. It covered both the CS and CO channels with a complete
consideration of the FD, namely for CO channels (all at $\alpha_s^2 \alpha$, see~\cf{diagram-a-psiZ-COM}-\ref{diagram-c-psiZ-COM}) 
\eqs{u \bar u, d \bar d, s \bar s, c \bar c \to& Z + \Upsilon \text{ via }(\so,\sps,\pj) \\
g g \to& Z +\Upsilon \text{ via }(\so,\sps,\pj) \\
}
and, for the CS channels, 
\eqs{
u \bar u, d \bar d, s \bar s, c \bar c \to& Z + \gamma^\star \to Z + \Upsilon \text{ via } \ssnew ~~[\text{ at } \alpha^3] \\
g g \to& Z +\Upsilon \text{ via } \ssnew,\pseudos, \pjs ~~[\text{ at } \alpha_s^2 \alpha] \\
}
It is understood that the CS channels other than $\ssnew$ can only contribute to the $\Upsilon$ yield 
with a FD. These were however found to be small. At the Tevatron, thus for
$p\bar{p}$ collisions, the photon fragmentation process in $q\bar{q}$ annihilation (see~\cf{diagram-qq-psiZ-CSM}) contributes 15\% of the 
CS $P_T$ integrated yields. Its contribution is obviously much smaller
 in $pp$ collisions  and at higher energies. 

For the CO, the dominant contribution is from $\so$ via $q\bar{q}$ annihilation (see~\cf{diagram-a-psiZ-COM}) if one
uses the LO LDMEs of Cho and Leibovich~\cite{Cho:1995ce}, yielding
a CO cross section about 20 times larger than CS one. 
Since the CO yield is dominated by $q\bar{q}$ annihilation, 
this ratio drastically decreases to a mere
factor 2-3 at the LHC at 14 TeV. Given the admittedly large uncertainties
on the CO LDMEs (easily one order of magnitude), this first study
could not support the conclusion of the dominance of the COM at the LHC whereas,
at the Tevatron, the COM was probably the most important. 

Another study by Kniehl \etal\ later considered~\cite{Kniehl:2002wd} the $J/\psi+Z$ and $\chi_{cJ}+Z$
 cases whose cross-section ratio via CO is approximately 0.2 which, in turn, means that
the FD from $\chi_c$ is expected to be close to 50 \% after the consideration
of the branchings.  
Predictions for the $P_T$- and $y$-differential cross sections 
at the Tevatron and at the LHC were also reported. Like for the $\Upsilon$ case,
the CO yield largely dominates at the Tevatron, less than at the LHC, with the same caveat 
concerning the precision at which the CO LDMEs are known. 

We finally note that the EW CS contribution was not considered in the latter study --
like in our NLO CS one-- although it could be relevant at the Tevatron. As noted above, 
it contributes 15\% of the $\Upsilon+Z$ yield. With the smaller mass of the $J/\psi$
and thus the reduced $1/m_\Q^4$ penalty for the off-shell photon, this fraction 
should be larger for the $J/\psi$ case. In fact, if as stated in~\cite{Mao:2011kf} for the LHC, 
the $q\bar q \to \so + Z$ channel dominates the CO $J/\psi+Z$ yield, both CS and CO contributions
via photon (resp. gluon) fragmentation should be of similar magnitude if $\mopb \simeq 10^{-3}$ GeV$^3$ 
as we noted for the $J/\psi+W$ case~\cite{Lansberg:2013wva}. We thus believe that a detailed study
of this CS contributions is still needed.

Let us now address the NLO COM computations. 
As announced, Mao \etal\ in 2011 reported~\cite{Mao:2011kf} on the first NRQCD one-loop 
computations for $J/\psi+Z$ --so far, there is no such computations for $\Upsilon+Z$.
However, because of the small impact at LO of the $\pj$ and $\sps$ transitions\footnote{Such an observation
however does not seem to be obvious along the lines of \cite{Alexopoulos:2013hxa}, where different LDMEs were used.}, they did not consider them in their computation. 
Whereas this drastically simplifies the NLO computation, it ignores the appearance
of gluon-fragmentation contribution via $\pj$ and $\sps$ (via the graph depicted on~\cf{diagram-d-psiZ-COM}) which have a crucial importance
in single $J/\psi$ contribution. In particular, it softens the $P_T$
dependence of all gluon-fragmentation contribution, including those from $\so$. 
We do not see any argument why it would not be so for $J/\psi+Z$.

We will not detail here the methodology which they use, which is similar to that we
briefly summarised above for the CS channel. Instead, we summarise their main results.
First, using $\mopb \simeq 2.7 \times 10^{-3}$~GeV$^3$, their LO CS and CO yields tend to be similar for 
$P_T^\psi$ below 5 GeV. Second, with an admittedly very small scale choice, namely $m_T$ and not $m_Z$, 
they observed a large $K$ factor close to 5 at low $P_T^\psi$,  reaching 3 at high $P_T^\psi$. Since
the $K$ factor for the CS yield at $P_T^\psi$ is significantly below unity (see also~\cite{Gong:2012ah}) 
at such low scales, the ratio CS/CO at NLO is significantly reduced. 

\paragraph{CEM as the SPS upper limit.}

The procedure to compute the $J/\psi+Z$ CEM cross section exactly follows from the 
same lines as for $J/\psi$ + a recoiling parton discussed in section~\ref{subsec:CEM_NLO_PT} --here
it recoils on the $Z$-- with the Born hard scattering $ij \to c \bar c + Z $  where $i$, $j$  stand for $g$, $q$ or 
$\bar q$ (see \cf{diagrams-CEM-Z}). The procedure 
(invariant-mass cut, CEM non-perturbative parameter, PDFs, scale variation, etc.) follow the same lines, although the central scale we take now is 
$\mu_0=M_Z$~\cite{Gong:2012ah} instead of the transverse mass of 
$J/\psi$ as done in~\cite{Mao:2011kf} for NRQCD. To carry out the computation, we have also used 
{\small \sc MadGraph5\_aMC@NLO} slightly tuned to account for the CEM invariant-mass cut.

\begin{figure}[ht!]
\centering
\subfloat[]{\includegraphics[scale=.42]{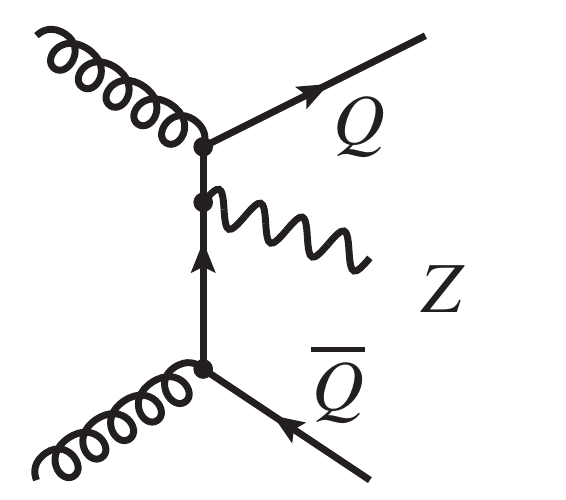}\label{diagram-CEM-Z-a}}
\subfloat[]{\includegraphics[scale=.42]{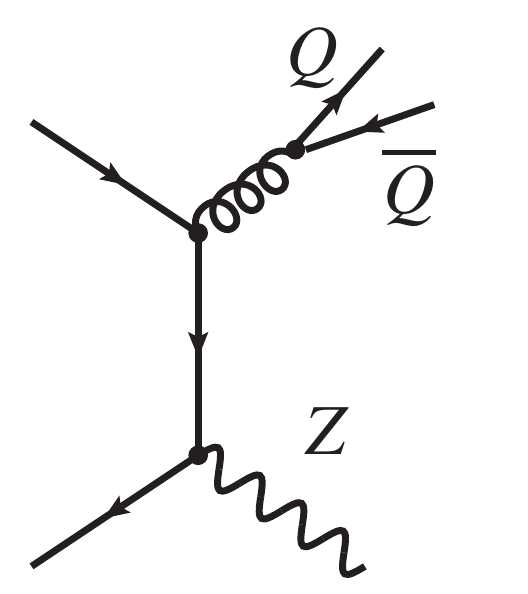}\label{diagram-CEM-Z-b}}
\subfloat[]{\includegraphics[scale=.42]{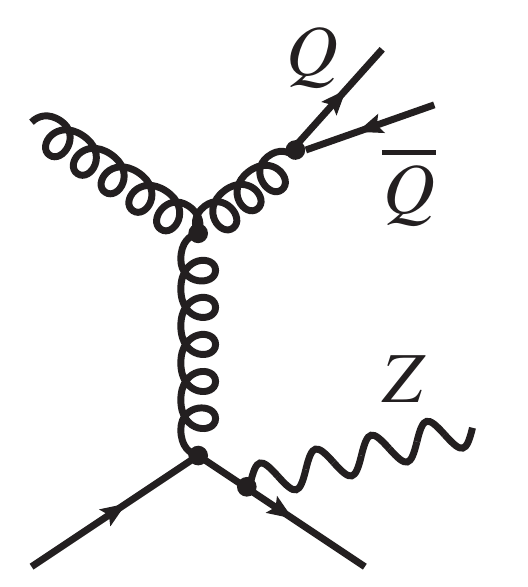}\label{diagram-CEM-Z-c}}
\subfloat[]{\includegraphics[scale=.42]{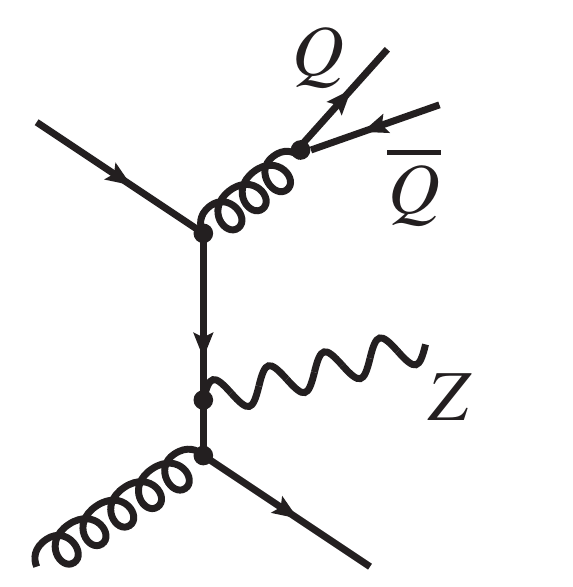}\label{diagram-CEM-Z-d}}
\caption{Representative diagrams contributing to $\Q+Z$ hadroproduction in the 
CEM at order $\alpha_s^2\alpha$ (LO). 
In the CEM, the invariant mass of the heavy-quark pair will
be integrated over within the $2m_Q$ and $2m_{H(Q)}$ where $H(Q)$ is the lightest hadron
with the flavour $Q$.}
\label{diagrams-CEM-Z}
\end{figure}

The NLO prompt cross section in the ATLAS inclusive acceptance~\cite{Aad:2014kba} is $0.19^{+0.05}_{-0.04}$~pb~\cite{Lansberg:2016rcx}. This is twice our estimation of the upper limit of the COM cross section.  
The $K$ factor for the hard  part is 2.8 
and the LO CEM yield (with $\mathcal{P}^{\rm LO,prompt}_{J/\psi}$) is
1.9 times smaller than the NLO yield (with $\mathcal{P}^{\rm NLO,prompt}_{J/\psi}$). 
The quark-gluon fusion channel at NLO is responsible for this large $K$ factor 
(see~\cf{diagram-CEM-Z-c} \& \ref{diagram-CEM-Z-c}).

As we discussed in section~\ref{subsec:CEM_NLO_PT}, the CEM $P_T$ spectrum of single $\Q$ production is too hard because 
of the presence of gluon-fragmentation channels scaling like $P_T^{-4}$. For the same reason for which
we believe that the computation of Mao \etal~\cite{Mao:2011kf} is an upper limit
of the COM SPS yield, a CEM computation of the $J/\psi+Z$ $P_T$-differential 
cross section is bound to overshoot the data --and any realistic computation-- because of the presence
of the very same topologies (see~\cf{diagram-CEM-Z-b}). The virtues of using the CEM in the present case are 
its simplicity, the fact it includes, in a transparent and coherent manner, the FDs
and, finally, the fact that we can carry out complete NLO computations.

\paragraph{Comparison with the ATLAS measurement\protect\footnote{We noticed that the theory numbers quoted in 
\cite{Aad:2014kba} have probably been misconverted into the ratio $R$  which is 
compared to the data (see below).}.}

In their publication~\cite{Aad:2013bjm}, ATLAS compared
their data to these NLO predictions. However they did not directly compare them to
their (prompt) yield since they suspected that a non-negligible part of the yield was probably from DPS contributions. Indeed, their distribution 
of the events as a function of the azimuthal angle between both detected particles,
$\Delta \phi$, was showing a plateau close to 0, whereas a dominant 
SPS yield --especially from $2\to 2$ topologies-- would show a peak 
at $\pi$, \ie\ for back-to-back events. We indeed have to note here
that the ATLAS acceptance imposes a rather large $P_T^\psi$ cut of 8.5 GeV, 
which makes it unlikely that initial-state radiations smear this peak
like it was noted~\cite{Kom:2011bd,Lansberg:2013qka,Baranov:2015cle} for instance 
for di-$J/\psi$ at low $P_T^\psi$.

Hence, they evaluated the DPS contribution with the  pocket formula~\ce{eq:dpseq} applied to the 
present case (in each point of the phase space):
\begin{eqnarray}
\sigma^{\rm DPS}(J/\psi+Z)=\frac{\sigma(J/\psi)\sigma(Z)}{\sigma_{\rm eff}}.
\end{eqnarray}
by using a value of $\sigma_{\rm eff}$ extracted from their $W+$ 2-jet analysis~\cite{Aad:2013bjm}, 
that is  $15 \pm 3 \hbox{(stat.)} ^{+5}_{-3} \hbox{(sys.)}$ mb.
$\sigma(J/\psi)$ and $\sigma(Z)$ were then taken from experimental data
with the same cuts as for $\sigma(J/\psi+Z)$.

Based on this DPS cross-section evaluation, they could quote a corresponding DPS-subtracted 
cross section to be compared with the SPS predictions discussed above.
In~\cite{Aad:2013bjm} and in some plots which will be shown later, the data-theory comparison was
done with the ratio of the $J/\psi+Z$ yield over that for $Z$ in order to cancel
some of the experimental uncertainty related to the $Z$ observation, \ie\
\eqs{
^{\rm prompt}R^{\rm DPS\ sub}_{J/\psi+Z}&=\mathcal{B}(J/\psi\to\mu^+\mu^-)\,\frac{\sigma(pp\to Z+J/\psi)}{\sigma(pp\to Z)}.} 
This requires one
to evaluate $\sigma(Z)$ under their kinematical conditions\footnote{In~\cite{Lansberg:2016rcx}, we employed 
{\small \sc MadGraph5\_aMC@NLO}~\cite{Alwall:2014hca} to calculate 
it up to NLO. The spin-correlated decay $Z\rightarrow e^+e^-$ was done 
thanks to {\small\sc MadSpin}~\cite{Artoisenet:2012st} and the 
NLO calculation of $pp\rightarrow Z \rightarrow e^+e^-$ was matched to the 
parton showers provided by {\small\sc Pythia8.1}~\cite{Sjostrand:2007gs} via the 
MC@NLO method~\cite{Frixione:2002ik}. The corresponding cross section 
was found to be 427 pb with a $20\%$ increase when the spin correlation and
 the parton-shower effects are taken into account, i.e. $505-520$ pb.
This value is to compared to 533.4 pb used by ATLAS for their data-theory comparison.}, which is on the
order of 500 pb when multiplied by the $Z\to e^+e^-$ branching fraction.

However, to ease the present discussion, we have converted back all the numbers into 
the simple $\sigma(pp\to Z+J/\psi)$ cross section without any branching but in the ATLAS 
acceptance\footnote{For the $Z$ selection : $P_T$(trigger lepton)$>25$~{\rm GeV}, $P_T$(sub-leading 
lepton)$>15$~{\rm GeV}, $|\eta(\mathrm{lepton~from}~Z)|<2.5$. For the $J/\psi$ selection: 
 $8.5<P_T^{J/\psi}<100 \,{\rm GeV}$ and $|y_{J/\psi}|<2.1$.}. These are gathered in~\ct{tab:psiZ}.
In view of the above discussion of the computation uncertainties (CO LDMEs, scale ambiguity, 
incomplete computations and intrinsic uncertainties, \dots), we prefer to quote a range
for the CS yield and an upper limit for the COM owing to the missing cancellation
in the current NLO computation between the leading-$P_T$ $S$-wave contributions and the 
uncalculated $P$-wave ones. Another reason why it would not make much sense to quote 
 precise values anyway is the large experimental uncertainties the ATLAS data -- 
much larger in absolute value than the theory ones.

\begin{table}[hbt!]
\centering
\def\arraystretch{1.5}
\begin{tabular}{c|cccc}
ATLAS            & DPS ($\sigma_{\rm eff}$ = 15 mb) & CSM                    & COM   & CEM \\\hline\hline
$1.6 \pm 0.4$ pb & $0.46$ pb                 & $0.025 \div  0.125$ pb & $< 0.1$ pb & $0.19^{+0.05}_{-0.04}$ pb
\end{tabular}
\caption{Comparison of the measured $P_T$-integrated cross section of prompt $J/\psi+Z$ production at 8 TeV 
by ATLAS to the CSM, COM and CEM evaluations discussed in the text.\label{tab:psiZ}. We have not assigned
any uncertainty to the DPS yield which is commensurate to that of the data and to that on $\sigma_{\rm eff}$.}
\end{table}

From the numbers in~\ct{tab:psiZ}, one could conclude that the SPS evaluations are misleading because some channels
have maybe been overlooked or because the quoted COM cross section is not reliable. Such a conclusion
would even be supported by the fact that ATLAS ``see'' SPS events. Indeed, beside exhibiting
a plateau close to 0, the $\Delta \phi$ distribution of their events exhibits a significant
peak at $\pi$. The fact that the CEM cross section is also way below the ATLAS cross section 
is a clear indication against this. Apart from questioning the reliability of the 
ATLAS measurement --bearing on an admittedly restricted
number of events and sizeable uncertainties-- or from 
invoking new physics contributions, the only other possibility left to solve 
the puzzle is thus to question the size of the DPS yield quoted by ATLAS 
with $\sigma_{\rm eff}=15$ mb. This is tempting since we have seen that recent quarkonium-related 
analyses~\cite{Lansberg:2014swa,Abazov:2014qba,Aaboud:2016fzt} have pointed at 
values smaller than 10 mb. 

One could  indeed argue that the DPS cross section could be 3 times higher, 
\ie\ close to 1.5 pb with a smaller $\sigma_{\rm eff}$, on the order of 5 mb close to the values observed
for some quarkonium-pair samples but this would normally not agree with a peak in the $\Delta \phi$ distribution.
However, in~\cite{Lansberg:2016rcx}, we in fact showed that the measured $\Delta \phi$ distribution
could --contrary to the appearances just discussed above-- accommodate 
a DPS cross section of 1.5 pb, thus compatible with the ATLAS total cross-section. Let us explain now how it is possible.

If one fits $\sigma_{\rm eff}$ to the total ``inclusive" 
ATLAS $J/\psi+Z$ yield, from which one has subtracted  the NLO SPS CEM one, one obtains
 $\sigma_{\rm eff} = 4.7$ mb.  To do so, one completely relies 
on the ATLAS evaluation of the DPS. With $\sigma_{\rm eff} = 4.7$ mb, the 
estimated DPS total cross section gets close to 1.5 pb, more than 5 times larger 
than the most optimistic estimation of the SPS one.

Going further, one can also derive an upper limit on $\sigma_{\rm eff}$ (corresponding
to the smallest acceptable DPS yield) by subtracting the 1-$\sigma$ higher value
of the NLO CEM yield from the 1-$\sigma$ lower value of the ATLAS measurements\footnote{
Under the assumption of such a large DPS fraction, the polarisation of the $J/\psi$ in association
with $Z$ should be the same as that produced alone. In such a case, we know that the yield cannot be strongly polarised~\cite{Chatrchyan:2013cla}, which narrows the spin-alignment uncertainties closer to the central value quoted by ATLAS. This justifies to leave aside such uncertainties from our discussion. If the SPS was dominant, this would not be so.}. 
This results in 7.1 mb for the upper limit. If, instead, one assumes that the SPS is negligibly small, one can extract a lower value
for $\sigma_{\rm eff}$ as low as 3.2 mb. Overall, we obtained the following range : $\sigma_{\rm eff}=4.7^{+2.4}_{-1.5}$~mb.

Clearly, with a DPS yield (with $\sigma_{\rm eff} =4.7$ mb) up to five times larger 
than SPS yield, the azimuthal distribution should not exhibit a structure at $\Delta \phi$ close
to $\pi$. Before addressing the solution to this issue,  it is however needed to discuss the yield as a 
function of $P^{J/\psi}_T$.

\begin{figure}[hbt!]
\begin{center}
\subfloat[ATLAS inclusive]{\includegraphics[width=0.45\textwidth]{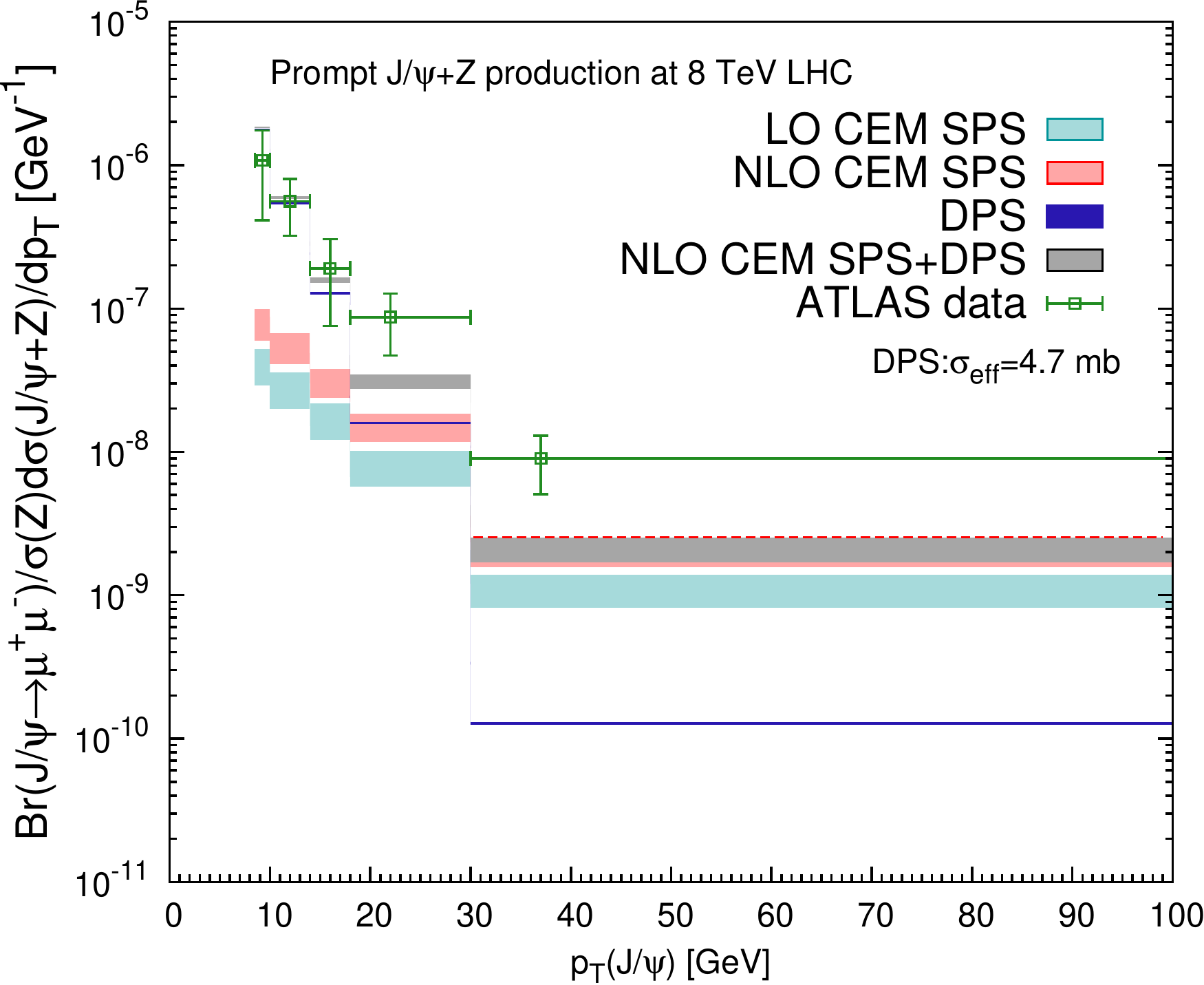}\label{fig:dRdpsipt-psiZ}}
\subfloat[NLO]{\includegraphics[width=0.45\textwidth]{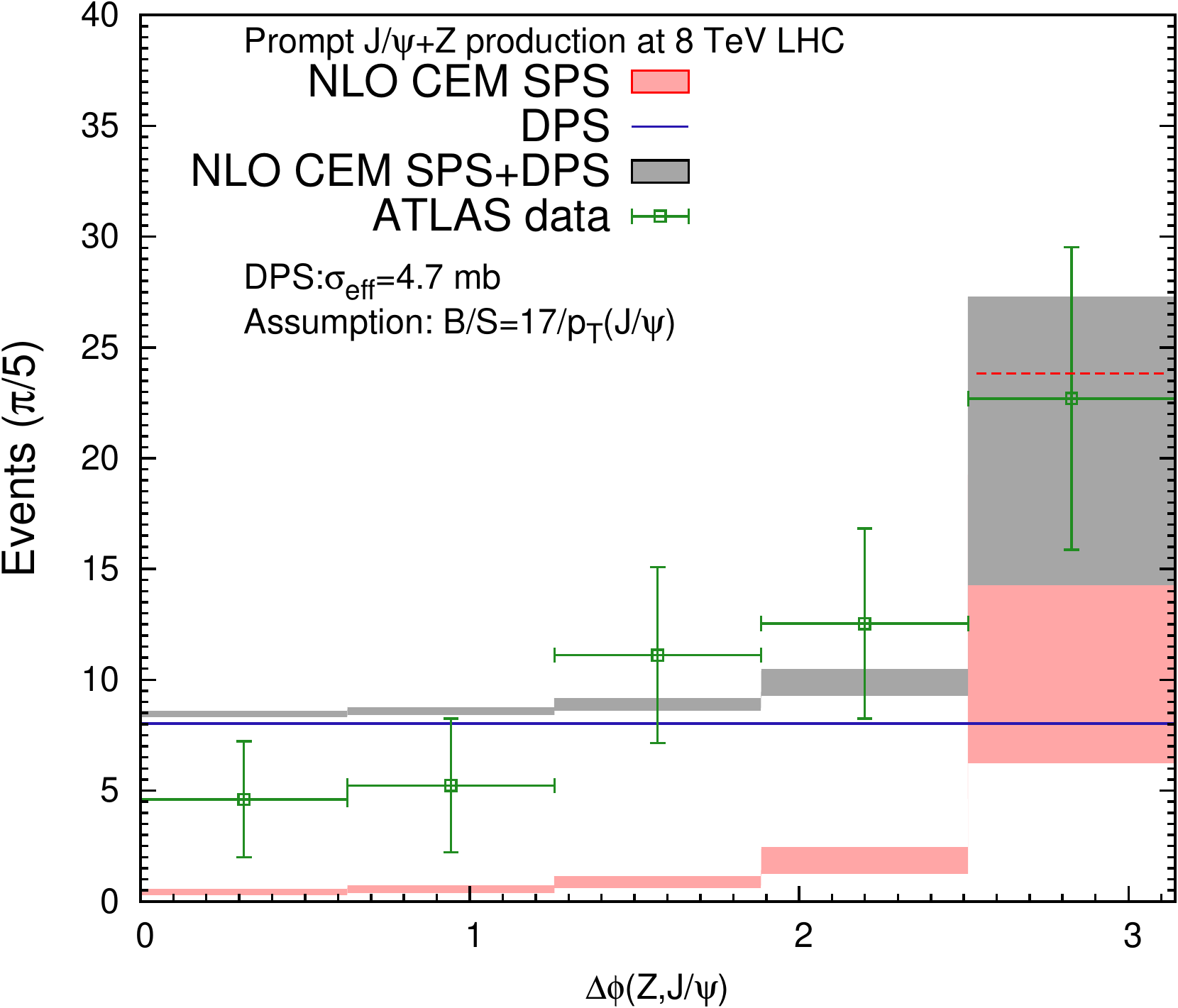}\label{fig:dphi-psiZ}}
\caption{(a) The $J/\psi$ $P_T$ dependence of $R$:  comparison between the ATLAS 
data~\cite{Aad:2014kba}, the CEM results for $J/\psi+Z$  at NLO (and LO) and our 
fit DPS yield in the ATLAS acceptance  (b) Comparison between the (uncorrected) ATLAS azimuthal event distribution
 and our theoretical results for $J/\psi+Z$  at NLO CEM SPS + DPS
effectively folded with an assumed ATLAS efficiency. Taken from~\cite{Lansberg:2016rcx}.
}
\end{center}
\end{figure}

\cf{fig:dRdpsipt-psiZ} shows the $P^{J/\psi}_T$ spectrum of $R$, 
where the SPS CEM cross section was also computed with {\small \sc MadGraph5\_aMC@NLO} and
the $P_T$ dependence of the DPS yield follows from the one of ATLAS (Table 5 of~\cite{Aad:2014kba})
 with a simple rescaling due to the change in $\sigma_{\rm eff}$.
One notes that the sum of the SPS and DPS cross sections (in gray) nearly agrees with the ATLAS data, 
with a slight gap opening at large $P^{J/\psi}_T$. 
The good agreement at low $P^{J/\psi}_T$  just reflects the fit of $\sigma_{\rm eff}$. 
As such, the comparison is rather a consistency check than a test and the discrepancy 
at high-$P^{J/\psi}_T$ should be further checked both from experimental and 
theoretical sides. On the experimental side, let us note that the systematical uncertainties 
may be a little underestimated; there is indeed a slight discrepancy in the inclusive 
single $J/\psi$ production measured by CMS~\cite{Khachatryan:2015rra} and ATLAS~\cite{Aad:2015duc}. 
On the theoretical side, large logarithms at high $P_T$ may need to be resummed.

That being said, \cf{fig:dRdpsipt-psiZ} clearly shows that the low-$P^{J/\psi}_T$ yield is 
completely dominated by the DPS contributions --and thus the total yield-- and that 
the high-$P^{J/\psi}_T$ yield is exclusively from SPS contributions. 
Based on this observation, we can now turn to the discussion of the problematic $\Delta \phi$
distribution. This difference of the SPS and DPS $P^{J/\psi}_T$ spectra indeed has an unexpected 
consequence. First, one should note that the ATLAS azimuthal distribution shown in~\cite{Aad:2014kba} was 
done with event counts, without efficiency correction. Second, the ATLAS efficiency is 
much higher for the last bin in $P^{J/\psi}_T$ than for 
the first bin, up to 3 times in fact. 
This is visible from the statistical uncertainties in \cf{fig:dRdpsipt-psiZ} 
which remain more or less constant with a much smaller cross section in the last bin.
This means that the events used for the  $\Delta \phi$ distribution are taken from a  biased
sample --that collected by ATLAS--, strongly enriched in high-$P^{J/\psi}_T$ events. 

These high-$P^{J/\psi}_T$ events are essentially of SPS origin --thus 
mostly populating the  $\Delta \phi \sim \pi$ region. Our claim is that the peak 
is only visible because of the ATLAS acceptance, not because of a large SPS yield in general. 
To phrase it differently, such $P^{J/\psi}_T$-integrated raw-yield distributions 
{\it cannot} be used to disentangle DPS from SPS contributions, unless they 
have the same $P^{J/\psi}_T$ distribution or the detector has a flat $P^{J/\psi}_T$
acceptance. The same of course applies for any other kinematical variable which would be
integrated over. 

To check our hypothesis,  we folded our DPS and SPS $P^{J/\psi}_T$ spectra with 
an estimation of the ATLAS efficiency~\cite{Lansberg:2016rcx}, plotted and 
summed the DPS and SPS $\Delta \phi$ distributions.
Thanks to an estimation of the yield in each bin and the computed SPS/DPS fraction
 in each $P^{J/\psi}_T$ bin, our theoretical SPS 
and DPS events can be added, bin by bin in $P^{J/\psi}_T$, in the $\Delta \phi$ plot with 
their specific $\Delta \phi$ distributions --flat for DPS, peaked at 
$\Delta \phi \simeq \pi$ for the SPS following our NLO CEM computation.

The resulting distribution is shown on \cf{fig:dphi-psiZ} and demonstrates 
that increasing the DPS yield by a factor of 3 does not create any tension 
with the observed ATLAS {\it event} 
$\Delta \phi$  distribution if the efficiency corrections are approximately 
accounted for in the theory evaluations. Future ATLAS data are eagerly awaited
to confirm this proposed solution to the puzzle.

\subsubsection{Associated hadroproduction with a $W^\pm$}
\label{sec:onium_W}

Compared to the associated production with a $Z$ which we just discussed, the production along with a $W^\pm$
exhibits some interesting differences and similarities as far as the SPS reactions 
are concerned. First, like for the $Z$ boson case, the production of a $W$ likely sets a much larger
scale than in single quarkonium production --even at low $P_T$. This is useful
in investigating the behaviour of the QCD corrections in this region. 

Second, the emission of the boson strongly enhances the likelihood for a light quark line
in the reaction, from which a gluon fragmentation in a CO pair can be emitted (see \cf{diagram-a-psiW}). In addition, in the case
of the $W$, the flavour change however also prevents it to be emitted from the heavy-quark line
forming the quarkonia. This drastically reduces the number of LO CO topologies compared to the
$Z$ and $\gamma$ cases and facilitates the completion of NLO NRQCD computations~\cite{Li:2010hc,Gang:2012ww}.

\begin{figure}[h!]
\centering
\subfloat[]{\includegraphics[scale=.375,draft=false]{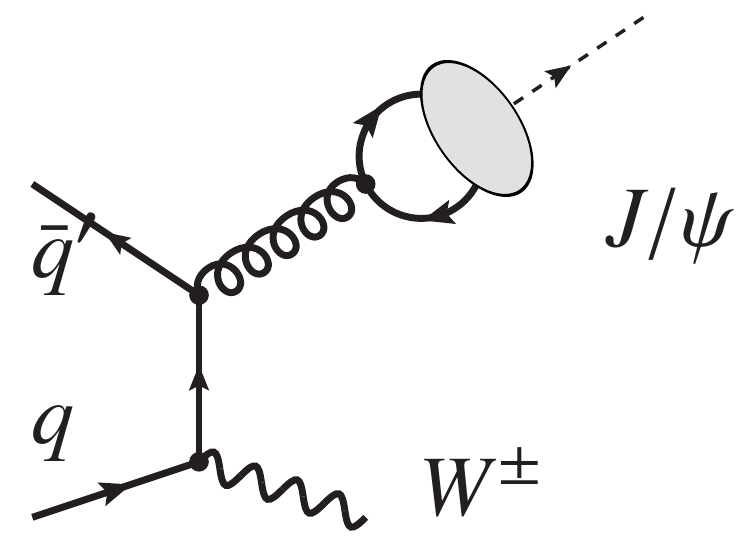}\label{diagram-a-psiW}}
\subfloat[]{\includegraphics[scale=.375,draft=false]{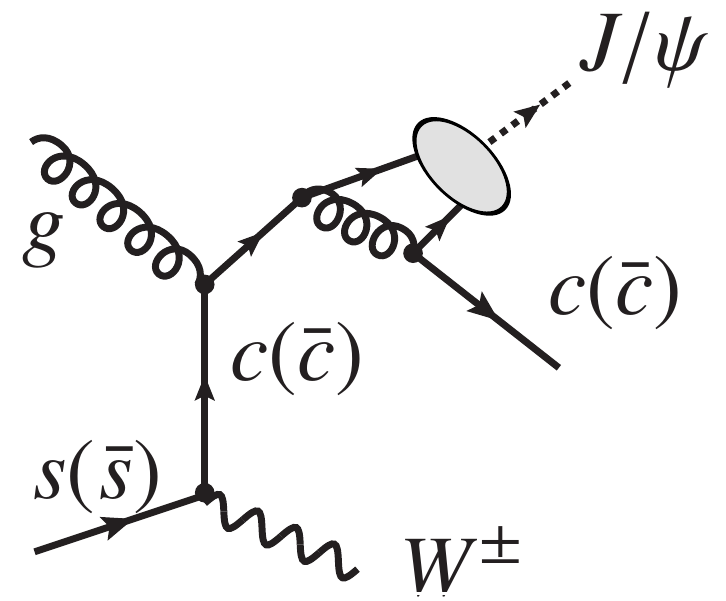}\label{diagram-b-psiW}}
\subfloat[]{\includegraphics[scale=.375,draft=false]{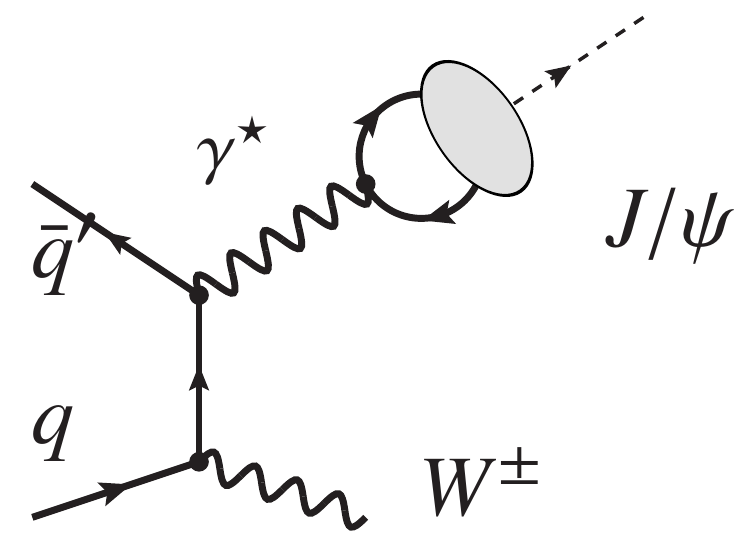}\label{diagram-c-psiW}}\hspace*{-.2cm}

\caption{Representative diagrams contributing to $J/\psi+ W^\pm$ hadroproduction (a) in the COM at orders $\alpha_S^2\alpha$, in the CSM at orders (b) $\alpha_S^3\alpha$ and (c) $\alpha^3$. The quark and anti-quark attached to the ellipsis are taken as on-shell and their relative velocity $v$ is set to zero.
Similar graphs can be drawn for $\Upsilon$ production.}
\label{diagrams-psiW}
\end{figure}

Third, this difference can also impact the CS channels which, for a long time, were thought to be strongly
suppressed as they would only appear through real QCD corrections (see \cf{diagram-b-psiW}) or pure electroweak reactions (see \cf{diagram-c-psiW}).
As such, one can find in the litterature many claims such as {\it ``$\psi+W$ offers a clean test of the colour-octet 
contributions"} from~\cite{Barger:1995vx},  {\it ``If the $J/\psi+W$ production is really 
detected, it would be a solid basis for testing the color-octet mechanism of the NRQCD"} 
from~\cite{Li:2010hc} or the less assertive statements made in \cite{Kniehl:2002wd} where
the CS contributions were however considered to be {\it ``forbidden or exceedingly suppressed"}.
In 2013, we tackled the CS case in detail in~\cite{Lansberg:2013wva} and showed that these claims
were unjustified.

Fourth, $\Upsilon+W^+$ was once proposed to be a decay channel of a charged 
Higgs~\cite{Grifols:1987iq} which triggered first searches at the Fermilab-Tevatron~\cite{Acosta:2003mu,Aaltonen:2014rda}. These were however $J/\psi+W$ events which were 
observed first by ATLAS at 7 TeV~\cite{Aad:2014rua} in 2014.

Of course, all the above discussion applies to the SPS yield, which may not be dominant,
as we saw for the $J/\psi+Z$ case. In fact, like for the discussion of their $J/\psi+Z$
sample, ATLAS uncovered a puzzle.  We will review all these aspects in the following
and parallel them to the previously discussed reactions.

\paragraph{Early theory expectations}

Since, historically, they were the first to be discussed, let us commence 
the discussion with the computations taking into the 
COM contributions. 
As early as in 1995, that is in the early days of NRQCD, Barger~\etal\
analysed~\cite{Barger:1995vx} the relevance of COM gluon fragmentation in a number
of associated production channels, including $J/\psi+W$, in a simplified set-up
only considering the $\so$ contributions in the high-$P^{J/\psi}_T$ limit.

A little later, Braaten \etal~\cite{Braaten:1998th} focused on $\Upsilon+W^\pm$, which could have been 
a background for charged Higgs searches. They considered CS and CO channels with a discussion of the feed-down, namely [for $W^+$] for CO channels  
\eqs{u \bar d, c \bar s \to& W^+ + \Upsilon \text{ via } \so ~~[\text{ at } \alpha_s^2 \alpha]}
and, for the CS channels, 
\eqs{
u \bar d, c \bar s \to& W^+ \gamma^\star \to W^+ + \Upsilon \text{ via } \ssnew ~~[\text{ at } \alpha^3] \\
c \bar b \to& W^+\Upsilon \text{ via } \ssnew,\pseudos,\pjs ~~[\text{ at } \alpha_s^2 \alpha] 
}
  All other CO contributions 
are suppressed by an additional factor of $\alpha_s$.

The CS channels other than $\ssnew$ can obviously also contribute to the $\Upsilon$ yield 
via a feed-down, but these were found to be small. 
For the $p\bar{p}$ collisions at the Tevatron, the photon fragmentation process 
in $q\bar{q}$ annihilation is largely dominant within
the CS channels. Its contribution is obviously much smaller
 in $pp$ collisions  and at higher energies compared to the possible gluon or gluon-light quark 
fusion channels. These are however further suppressed (CKM, $\alpha_s^2$ or PDF), 
unlike the $J/\psi+W$ case. As discussed in~\cite{Braaten:1998th}, $c \bar b \to W^+ +\Upsilon$ --which is also
strongly suppressed by the parton fluxes, is supposed to give a representative account of
the QCD-induced CS channels, including $g g \to W^+ + \Upsilon + b \bar c$. However, this was
not backed up by any computation of the latter.

Using the LO LDMEs fit  on the Tevatron 
data by Cho and Leibovich~\cite{Cho:1995ce}, they found out that the CO contribution was as much as 300 times larger 
than the CS ones. 
This seems a little large to us since both CS and CO contributions
via photon (resp. gluon) fragmentation are dominant. In such a case~\cite{Lansberg:2013wva}, they should not be so different with  $\mopb$ 
on the order of $10^{-2}$~GeV$^3$~\cite{Braaten:2000cm}.
Overall, even though it is likely that the $\so$ CO contribution via gluon fragmentation
dominates, the actual size of the CS/CO ratio remains uncertain in view of 
the number of channels which were neglected and the poorly known bottomonium LDMEs.

In 2002, Kniehl \etal\ considered~\cite{Kniehl:2002wd} the $J/\psi+W$ and $\chi_{cJ}+W$
 cases, but only via CO channels. As found for the $Z$ case, their cross-section 
ratio via CO is directly connected
to the ratio of the corresponding $\so$ LDMEs, which happens to be close to 0.2. 
The feed-down from $\chi_c$ should thus be expected to be close to 50 \% and may play
an important role in describing possible data.
 They also reported on predictions for the $P_T$- and $y$-differential cross sections 
at the Tevatron and at the LHC .

\paragraph{NLO COM computations.}

In this context, two NLO studies were carried out, the first on $J/\psi+W$~\cite{Li:2010hc} and
the second on $\Upsilon+W$~\cite{Gang:2012ww}. As discussed above, the change of the quark-line flavour induced by the $W$ emission drastically simplifies the computation. As such, 
the contribution from $\sps$ and $\pj$ transitions only appear as real-emission corrections, 
where one gluon splits into a heavy-quark pair which then radiates a gluon before forming the
quarkonium. These are precisely the topologies via which $\pj$ transitions damp down
the impact of the gluon fragmentation $\so$ channels in single $\Q$ production. 
All the details of the computations can be found in~\cite{Li:2010hc,Gang:2012ww} and follow
the lines of the NLO COM computations we have discussed in \eg\ section~\ref{subsec:COM_updates}.

Before discussing the results of these computations, two comments are however in 
order since they may question their overall interpretations. First, we think that
the default scale choice employed in~\cite{Li:2010hc,Gang:2012ww} is not realistic. 
Instead of using $m_T=\sqrt{m_\Q^2+(P_T^\Q)^2}$, which completely ignores the emission
of the $W$, we believe that a natural choice for the scale should be close to $m_W$, 
possibly accounting for $P_T^\Q$ variations. Obviously, as we discussed
for the $\Q+Z$ cases, such a low value for the scale overemphasises the impact of the 
QCD corrections with a probably overestimated value of $\alpha_s$. One could however object
that the presence of different scales in the process may generate large logarithms appearing
in further higher-order QCD corrections and whose importance may be signaled by the large size
of QCD corrections observed with a small scale choice. 

Second, the choice of CO LDMEs made in~\cite{Li:2010hc,Gang:2012ww} is arguable 
as LO LDMEs were used. As discussed above, large cancellations occur at NLO between
$\so$ and $\pj$ transitions when fragmentation contributions are involved. 
This is expected to occur both in single $\Q$ and $\Q+W$ production and strongly
motivates for a coherent choice of LDMEs, that is using  LDMEs fit 
with NLO $\Q$-production computations to perform NLO $\Q+W$ production predictions. 
There is no reason not to do so since NLO fits on Tevatron and LHC data exist.

With these caveats in mind, we note that the $K$ factor for $J/\psi+W$~\cite{Li:2010hc} 
is a little above 4 for $P^{J/\psi}_T$ below 5 GeV and, for $P^{J/\psi}_T$ 
above 20 GeV, constant and close to 3. The $\sps$ contributions were found to be  suppressed compared to those from $\so$ and $\pj$ for 
increasing $P^{J/\psi}_T$. Indeed, the gluon radiated by the charm-quark line
forming the $\sps$ pair cannot be soft and the corresponding topology does not
scale as $P_T^{-4}$. The suppression however seems to be stronger than in the 
single $J/\psi$ case. As expected, the $\pj$ contribution is negative and scales
like the LO $\so$. Its overall impact however strongly depends on the choice of the LDMEs.
 
In the $\Upsilon+W$ case~\cite{Gang:2012ww}, the $K$ factor was found to 
be close to 4 for $P^{\Upsilon}_T$ below 5 GeV --the value is larger at 14 TeV than at 8 TeV-- and 
reaching a constant value,  for $P^{\Upsilon}_T$ above 30 GeV, close to 3 
--also larger at 14 TeV than at 8 TeV. The $\pj$ contribution was found to 
be negative and not scaling like LO $\so$, which might appear as surprising. 

\paragraph{The overlooked CSM contributions.}
In 2013, we revisited~\cite{Lansberg:2013wva} the importance of the CSM
to $J/\psi+W$ production, in particular of two classes of 
 CS contributions which had so far been overlooked, at least in this case.
The first arises from the strange-quark--gluon fusion resulting in a $W+c$  pair where the charm quark fragments into a $J/\psi$ (see \cf{diagram-b-psiW})
. This process is to be paralleled to the leading-$P_T$ contribution to $J/\psi + c\bar c$~\cite{Artoisenet:2007xi}. 
In fact, $W+c$ has been in the past  identified as a probe of the strange quark PDF~\cite{Baur:1993zd}. The second class is one of the EW contribution discussed by Braaten \etal\ \cite{Braaten:1998th} for
$\Upsilon+W$. For unclear reasons, it had been overlooked for $J/\psi+W$ for which it should actually matter more.
The $J/\psi$ is simply produced by an off-shell photon emitted by the quark line which also radiates the $W$ boson (see \cf{diagram-c-psiW}); it is like a vector-meson-dominance contribution.

The latter contribution is clearly enhanced in $p \bar p$ collisions at the Tevatron 
thanks to the anti-proton valence anti-quarks. The former contribution 
should be enhanced at LHC energies in $pp$ collisions with the gluon PDF getting large at lower $x$. In any case, as we will see, these CSM processes are not at all negligible compared to the leading CO contributions (see \cf{diagram-a-psiW}) if reasonable values of CO LDMEs are chosen. In addition, both these Born contributions possess a leading-$P_T$ contribution ($P_T^{-4}$). This means that they could be significant at low $P_T^\psi$ and remain important at large $P_T^\psi$. This is at variance with the single-$J/\psi$-production case where the Born contributions are not leading power in $P_T^\psi$ (see section~\ref{subsec:CSM_NLO_PT}).

We will not repeat here the methodology to compute such contributions. It closely follows  the other CS computations which we already discussed, using the analog of \ce{eq:CSM-PsiZ} where the $Z$ boson is
replaced by a $W$ and the parton flavours are set accordingly. For the record, the processes under discussion are
(for $W^+$)
\eqs{
u \bar d \to& W^+ \gamma^\star \to W^+ + J/\psi \text{ via } \ssnew ~~[\text{ at } \alpha^3] \\
g \bar s \to&  W^+ + J/\psi+ \bar{c} \text{ via } \text{ via } \ssnew ~~[\text{ at } \alpha_s^2 \alpha]. 
}
In order to assess their relevance, we also evaluated the LO CO contributions. As we have seen, 
at LO, the sole $\so$ channel contributes via
\eqs{
u \bar d \to& W^+ g^\star \to W^+ + J/\psi \text{ via } \so ~~[\text{ at }\alpha_s^2 \alpha]
}
All these are Born order processes, free of any divergences and we have computed them with \MadOnia~\cite{Artoisenet:2007qm}. They could also be computed with \HELACOnia~\cite{Shao:2012iz,Shao:2015vga}\footnote{
For the cross-section evaluation, apart for the usually CS and SM parameters, we have generated uncertainty bands for the resulting predictions from the {\it combined} variations of the heavy-quark mass within the range $m_c=1.5\pm 0.1$ GeV, with the factorisation $\mu_F$ and the renormalisation $\mu_R$ scales chosen among the couples $((0.75,0.75);(0.75,2);(1,2);(1,1);(2,1);(2,0.75);(2,2))\times m_W$.}. For the CO cross section, we have set $\mopb$  to $2.2 \times 10^{-3}$ GeV$^{3}$, \ie~a value inspired by the recent global NLO analysis of Butensch\"on and Kniehl~\cite{Butenschoen:2011yh}, yet compatible with the central value of the sole LO fit on LHC data by Sharma and Vitev~\cite{Sharma:2012dy} (with extremely large uncertainties though). This value is also on the order of what was obtained by the  NLO PKU fit~\cite{Chao:2012iv}. 
Using a NLO LDME fit for LO computations is however not ideal; it is nevertheless
the best one can do, the reverse not being true. Obviously, one should then avoid negative
NLO LDME values like those of the early IHEP fit~\cite{Gong:2012ug}. It is also important to recall that our choice is also close to the LO analysis of~\cite{Braaten:1999qk} and from analyses which partially took into account QCD corrections \cite{Kniehl:1998qy,Gong:2008ft}

Our LO results for the differential cross sections in $P_T^\psi$ are shown in \cf{fig:dsigdPT-psiW} for the Tevatron $(a)$, and for the LHC $14$ TeV $(b)$. We note that the corresponding COM results are compatible with those discussed above of~\cite{Kniehl:2002wd} and the LO of~\cite{Li:2010hc} provided that 
the differences in the choices of the scales, of the LDME and of the kinematical cuts are accounted for.

\begin{figure}[hbt!]
\begin{center}
\subfloat[1.96 TeV]{\includegraphics[width=0.5\columnwidth,draft=false]{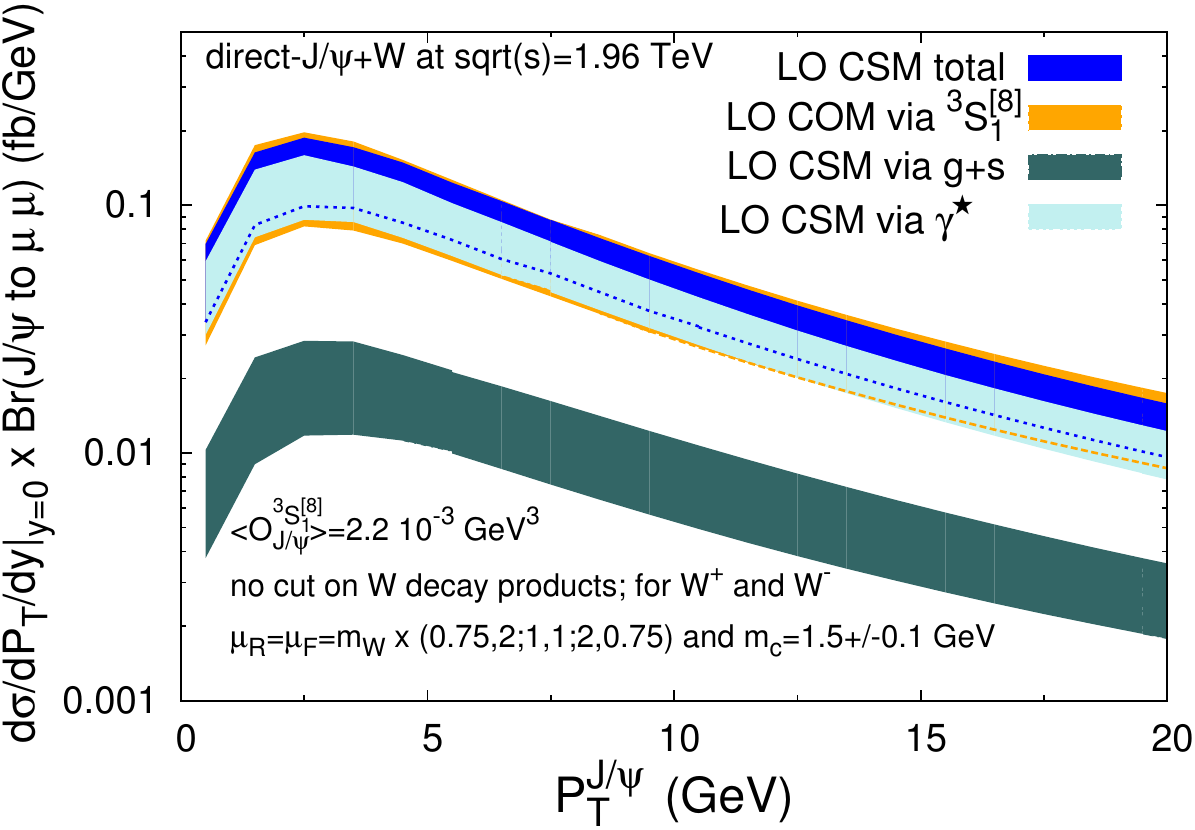}\label{fig:dsigdPTa-psiW}}
\subfloat[14 TeV]{\includegraphics[width=0.48\columnwidth,draft=false]{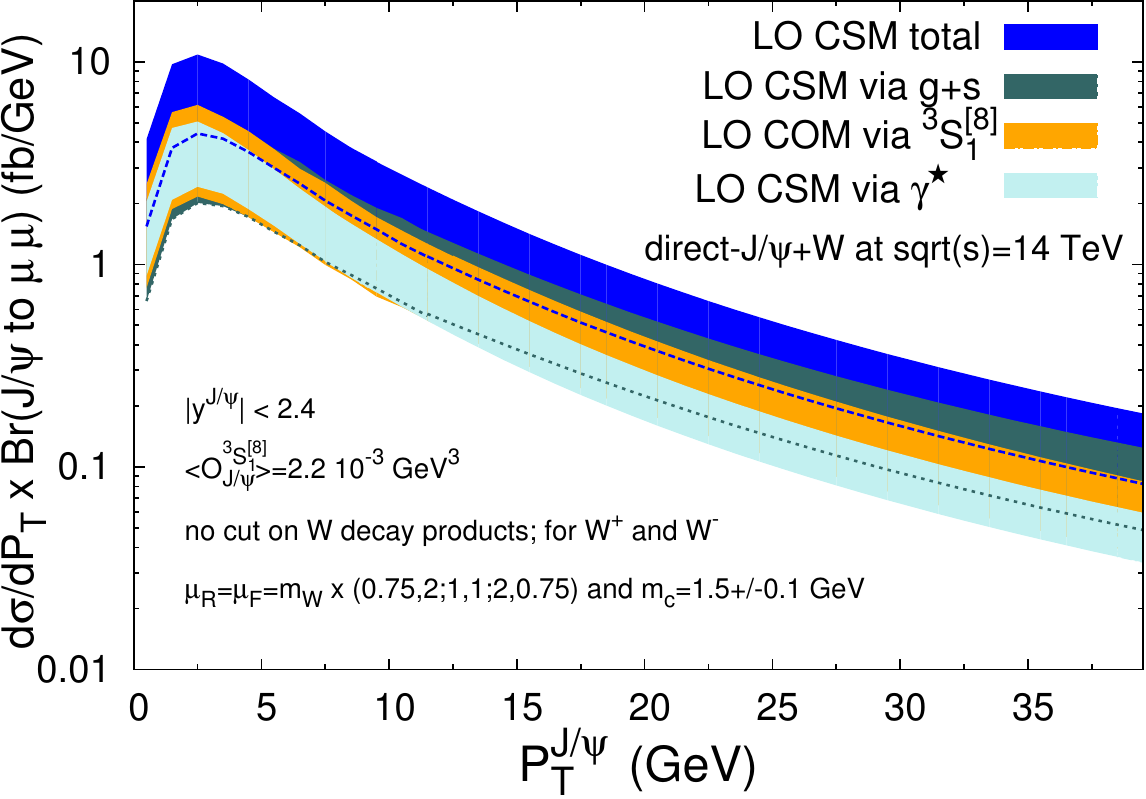}\label{fig:dsigdPTc-psiW}}
\caption{Differential cross section at LO for $J/\psi+W$ vs. $P_T^\psi$ for the Tevatron (a) and the LHC (b). The orange band is for the COM while the light blue, dark green and blue bands are for the CSM via $\gamma^\star$, via $sg$ fusion and total contributions, respectively. Taken from~\cite{Lansberg:2013wva}.}
\label{fig:dsigdPT-psiW}
\end{center}
\end{figure}

At the Tevatron, the COM contribution (orange band) is indeed much larger than that of the CSM via $sg$ fusion (dark green band). Yet, the CSM contribution\footnote{To be precise, let us note that the light-blue band in fact contains also other electroweak contributions appearing at the same order, like  \emph{i.e.} via $Z^\star$. This contributions  is however strongly dominated by processes via $\gamma^\star$.} via $\gamma^\star$ (light blue band) is of similar size.  At LHC energies, the three contributions equally contribute and the CSM cross section is thus about twice as large as the COM one at 14 TeV and at large $P_T$ (see \cf{fig:dsigdPTc-psiW}). 
It is thus clear, contrary to earlier claims in the literature~\cite{Barger:1995vx,Li:2010hc}, that
$J/\psi$ in association with a $W$ boson is not a clean probe of the COM, whatever the  $P_T^\psi$. The Born CSM contributions are indeed already leading $P_T$.

\begin{figure}[htb!]
\centering
\subfloat[8 TeV]{\includegraphics[width=.45\columnwidth,draft=false]{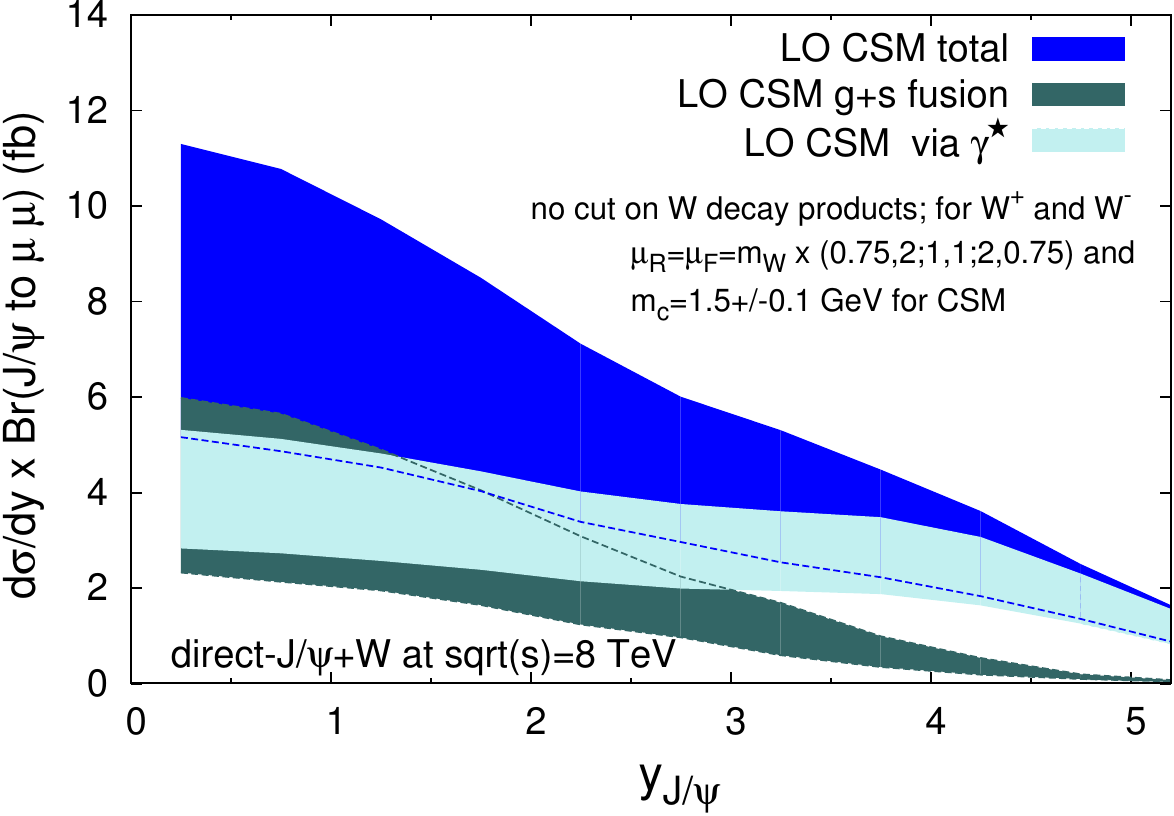}}\quad
\subfloat[14 TeV]{\includegraphics[width=.45\columnwidth,draft=false]{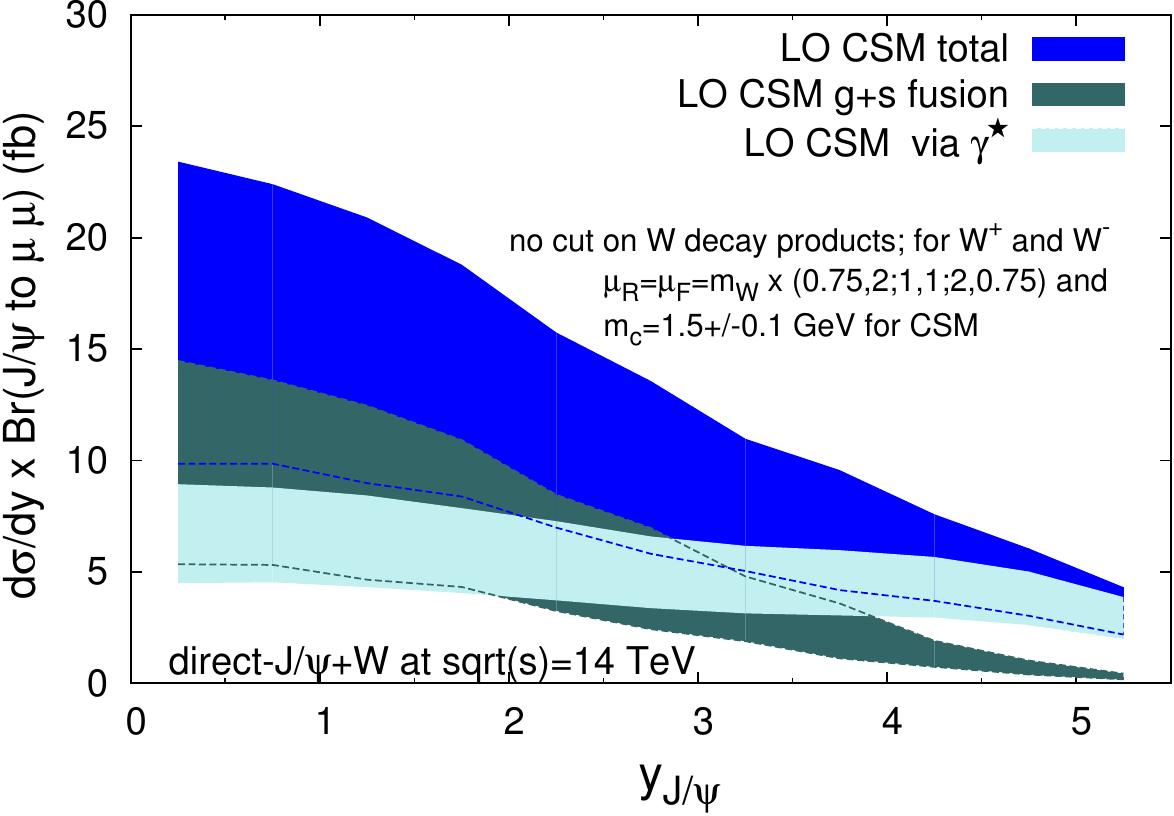}}
\caption{$y_{J/\psi}$ differential cross section at LO for $J/\psi+W$ for the LHC at $8$ TeV $(a)$ and $14$ TeV $(b)$. The colour code is the same as in figure \ref{fig:dsigdPT-psiW}. Note that these results are obtained without cut on the $J/\psi$ $P_T$. Taken from~\cite{Lansberg:2013wva}.}
\label{fig:dsigdy-psiW}
\end{figure}

In addition to the $P_T$ dependence, we present in \cf{fig:dsigdy-psiW}  the CSM results for the differential cross sections in $y_{J/\psi}$ for the LHC at $8$ TeV $(a)$ and $14$ TeV $(b)$. One observes that the CSM yields via $\gamma^\star$ and via $sg$ fusion are of the same order at the LHC energies, with an increasing proportion of $sg$ fusion as the energy increases and the effect of the valence quarks enhancing the $q\bar{q'}$ contribution when one reaches large rapidities.

It is well known that experimental analyses of $W$ production are usually 
performed via their leptonic decay, in particular $\mu+\nu_\mu$. 
Such events are tagged by the missing transverse energy carried by the 
undetected neutrino. As such, one cannot strictly enforce that the 
invariant mass of the $\mu+\nu_\mu$ pair equals that of the $W$. 

This has an unexpected consequence in the case of $J/\psi+W$ since
the rare 3-body decay$ W\to J/\psi+\mu+\nu_\mu$ cannot be disentangled 
from genuine $J/\psi+W \to J/\psi+\mu+\nu_\mu$ events. Let us note that a 
similar $W$ decay channel, $W\to \Upsilon+\mu+\nu_\mu$ 
has previously been considered in~\cite{Qiao:2011yk}.  Its contribution is actually 
not negligible with the ATLAS cut ($E^\text{miss}_T >20$ GeV, $P^\mu_T>25$ GeV, $|\eta^\mu|<2.4$, $m^W_T=\sqrt{2P^\mu_TE^\text{miss}_T[1-\cos(\phi^\mu-\phi^\nu)]}>40$~GeV). We have indeed found~\cite{Lansberg:2013wva} 
that the process $q\bar q' \to W \to J/\psi+\mu+\nu_\mu$ contributes 
nearly equally to that of $q\bar q' \to J/\psi+W\to J/\psi+\mu+\nu_\mu$, 
where ${\cal B}(W \to \mu+\nu_\mu)\simeq 11\%$.

\paragraph{CEM as the SPS upper limit.}

The procedure to compute the $J/\psi+W$ CEM cross section exactly follows from the 
same lines as for $J/\psi+W$ or $J/\psi$ + a recoiling parton (see sections~\ref{sec:onium_Z} and~\ref{subsec:CEM_NLO_PT}). The Born hard-scattering is  $ij \to c \bar c + W $  where $i$, $j$  stand for $g$, $q$ or 
$\bar q$ (see \cf{diagrams-CEM-W}). The procedure 
(invariant-mass cut, CEM non-perturbative parameter, PDFs, scale variation, etc.) are identical with the central scale
$\mu_0=M_W$. The computation can be done with {\small \sc MadGraph5\_aMC@NLO} slightly tuned to account for the CEM invariant-mass cut.

\begin{figure}[ht!]
\centering
\subfloat[]{\includegraphics[scale=.42,draft=false]{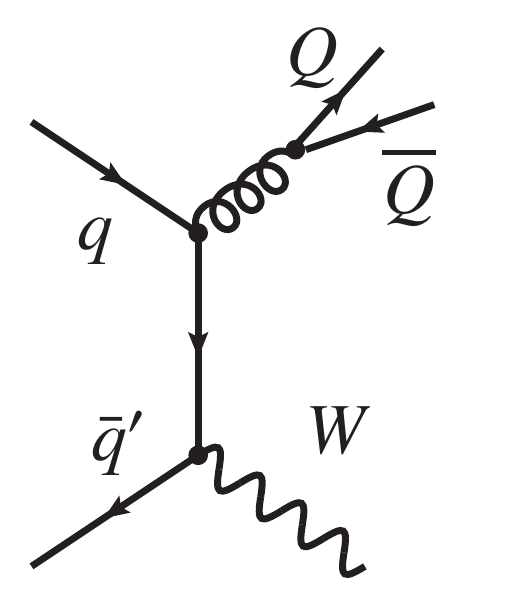}\label{diagram-CEM-W-a}}
\subfloat[]{\includegraphics[scale=.42,draft=false]{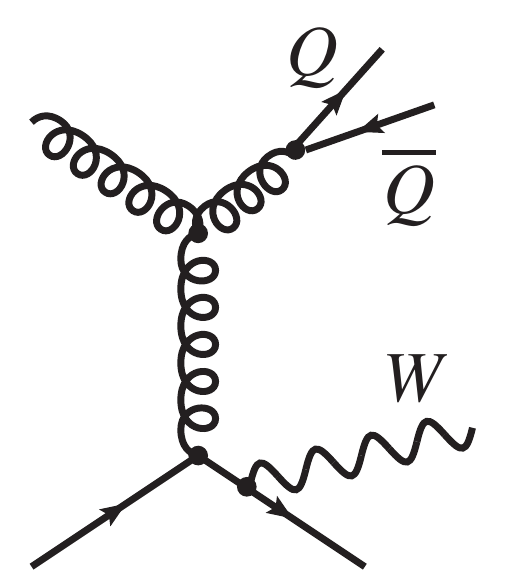}\label{diagram-CEM-W-b}}
\subfloat[]{\includegraphics[scale=.42,draft=false]{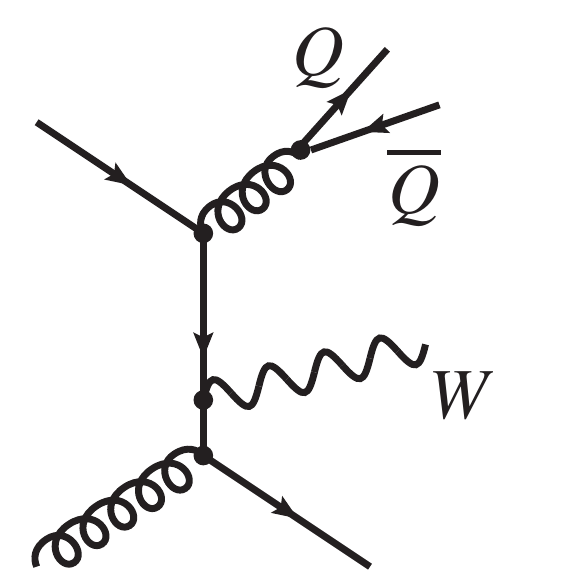}\label{diagram-CEM-W-c}}
\caption{Representative diagrams contributing to $\Q+W$ hadroproduction in the 
CEM at order $\alpha_s^2\alpha$ (LO). 
}
\label{diagrams-CEM-W}
\end{figure}

At 7 TeV, the NLO prompt cross section in the ATLAS inclusive acceptance~\cite{Aad:2014rua} is $0.28\pm0.07$~pb~\cite{Lansberg:2017chq}. This is nearly twice our estimation of the corresponding CSM and COM cross sections (\ct{tab:psiW}).  
The $K$ factor for the hard  part is 2.76 and the LO CEM yield (with $\mathcal{P}^{\rm LO,prompt}_{J/\psi}$) is
1.77 times smaller than the NLO yield (with $\mathcal{P}^{\rm NLO,prompt}_{J/\psi}$). 
This value should be considered as an upper bound for the SPS cross section.

\paragraph{Comparison with the ATLAS measurement\protect\footnote{In 2020, ATLAS published another study~\cite{Aaboud:2019wfr} at 8 TeV which tends to support the following DPS discussion. However at large $P_T$, it seems that SPS contributions are not fully under control yet.} at 7 TeV.}

Just like for their $J/\psi+Z$ study, ATLAS compared~\cite{Aad:2014rua}
their data to the existing (SPS) predictions, assuming a DPS contributions
compatible with their $W+$ 2-jet analysis~\cite{Aad:2013bjm}, thus computed with
$15 \pm 3 \hbox{(stat.)} ^{+5}_{-3} \hbox{(sys.)}$~mb. 
Their distribution  of the events as a function of the azimuthal angle between both detected particles,
$\Delta \phi$, is also showing some events near 0 -- the statistics is however limited. 
In addition the same distribution exhibits a peak at $\Delta \phi \simeq \pi$ typical
of SPS events from $2\to 2$ (or even $2 \to 3$) topologies. Like for the
$J/\psi+Z$ analysis, the ATLAS acceptance imposes a rather large $P_T^\psi$ cut of 8.5 GeV, 
which makes unlikely that initial-state radiations smear this peak.
Their DPS evaluation is absolutely standard and  exactly follows the same line 
as $J/\psi+Z$.

They could then quote a corresponding DPS-subtracted 
cross section to be compared to the SPS predictions discussed above.
In~\cite{Aad:2014rua} and in some plots which will be shown later, the data-theory comparison was
done with a normalised cross section\footnote{Note however that this one is differential in $y$.},
\begin{equation}
\frac{d\sigma (pp \to W^\pm + J /\psi)}{dy_{J/\psi}} \frac{{\cal B}(J/\psi \to \mu^+ \mu^- )}{ \sigma (pp\to W^\pm ) }
,
\end{equation} 
in order to cancel some of the experimental uncertainty related to the $W$ observation.
This requires one
to evaluate $\sigma(W)$ under their kinematical conditions
which is on the
order of 5 nb.

However, like for $J/\psi+Z$, in order to make the discussion easier to follow, we have converted back all the numbers into 
the simple $\sigma(pp\to W+J/\psi)$ cross section without any branching but in the ATLAS 
acceptance\footnote{For the $W$ selection : $P_T(\mu^\pm)>25$~ GeV, $E_T \!\!\!\!\!\!\slash ~> 20$ GeV,
$M_T^W > 40$ GeV, $|\eta(\mu^\pm)|<2.4$. For the $J/\psi$ selection: 
 $8.5<P_T^{J/\psi}<30 \,{\rm GeV}$ and $|y_{J/\psi}|<2.1$. $M_T^W \equiv 2 P_T(\mu^\pm) E_T \!\!\!\!\!\!\slash\ \ (1- \hbox{cos} (\phi^\mu -\phi^{\nu_\mu}))$ should not be confused with $m_T^2=m^2+p_T^2$.}. 
These are gathered in~\ct{tab:psiW}.
In view of the above discussion of the computation uncertainties (LO CO LDMEs, scale ambiguity, \dots), 
we prefer to quote a range for the CS and COM yields. We also make the same remark as for
$J/\psi+Z$ : quoting too precise values would not bring much to discussion given 
 the large polarisation-induced uncertainty affecting 
the ATLAS data; it is not quoted here.

\begin{table}[hbt!]
\centering
\def\arraystretch{1.5}
\begin{tabular}{c|cccc}
ATLAS                   & DPS ($\sigma_{\rm eff}$ = 15 mb) & CSM                    & COM          & CEM \\\hline\hline
$4.5 ^{+1.9}_{-1.5}$ pb & $1.7$ pb                  & $0.11 \div  0.04$ pb & $0.16 \div 0.22$ pb & $0.28 \pm 0.07$ pb
\end{tabular}
\caption{Comparison of the measured $P_T$-integrated cross section of prompt $J/\psi+W$ production at 7 TeV 
by ATLAS to the CSM, COM and CEM evaluations discussed in the text. We have not assigned
any uncertainty to the DPS yield which is commensurate to that of the data and to that on $\sigma_{\rm eff}$.\label{tab:psiW}}
\end{table}

The numbers for the CSM and COM  are extremely small compared to the ATLAS measurement even when 
 the expected DPS yield is subtracted. The CEM value~\cite{Lansberg:2017chq} is compatible with the sum of the CSM and COM cross section, \ie\ more than one order of magnitude below the data. A smaller $\sigma_{\rm eff}$, on the order of 5 mb, with DPS cross section multiplied by 3, thus seems to be the only solution to the problem. 

A similar observation can be made by looking at the distribution in $P_T^{J/\psi}$, shown in \cf{fig:ptspsdps-psiW} 
with the black (green) hatched histograms, resulting in a 3.1~$\sigma$ 
discrepancy between the SPS theory and the data. We can thus really claim for an evidence of the DPS yield in this process provided that
a large DPS cross section is supported by the analysis of the azimuthal dependences
which we now discuss. 

\begin{figure}[hbt!]
\begin{center}
\subfloat[]{\includegraphics[width=0.5\columnwidth]{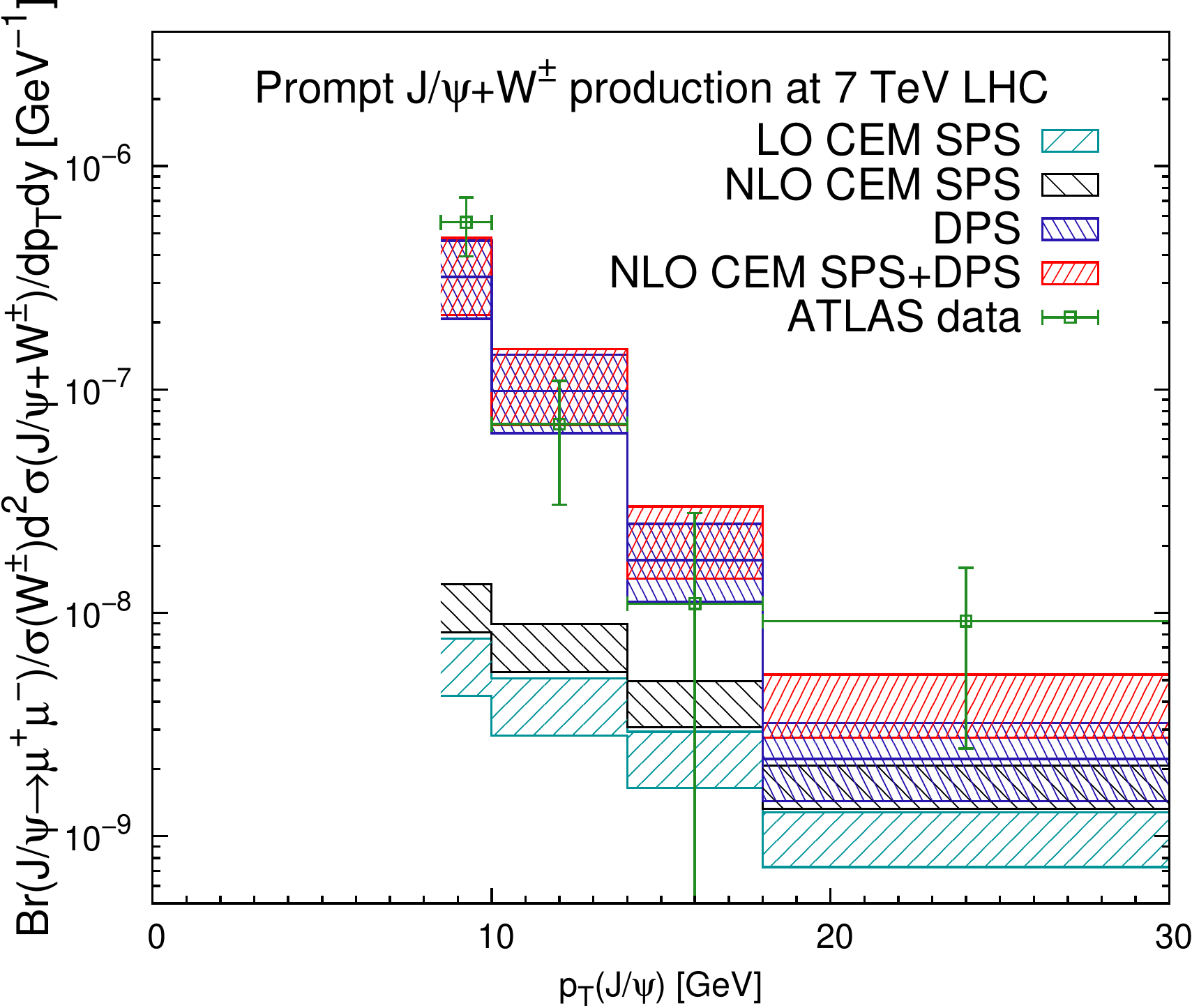}\label{fig:ptspsdps-psiW}}\quad
\subfloat[]{\includegraphics[width=0.47\columnwidth]{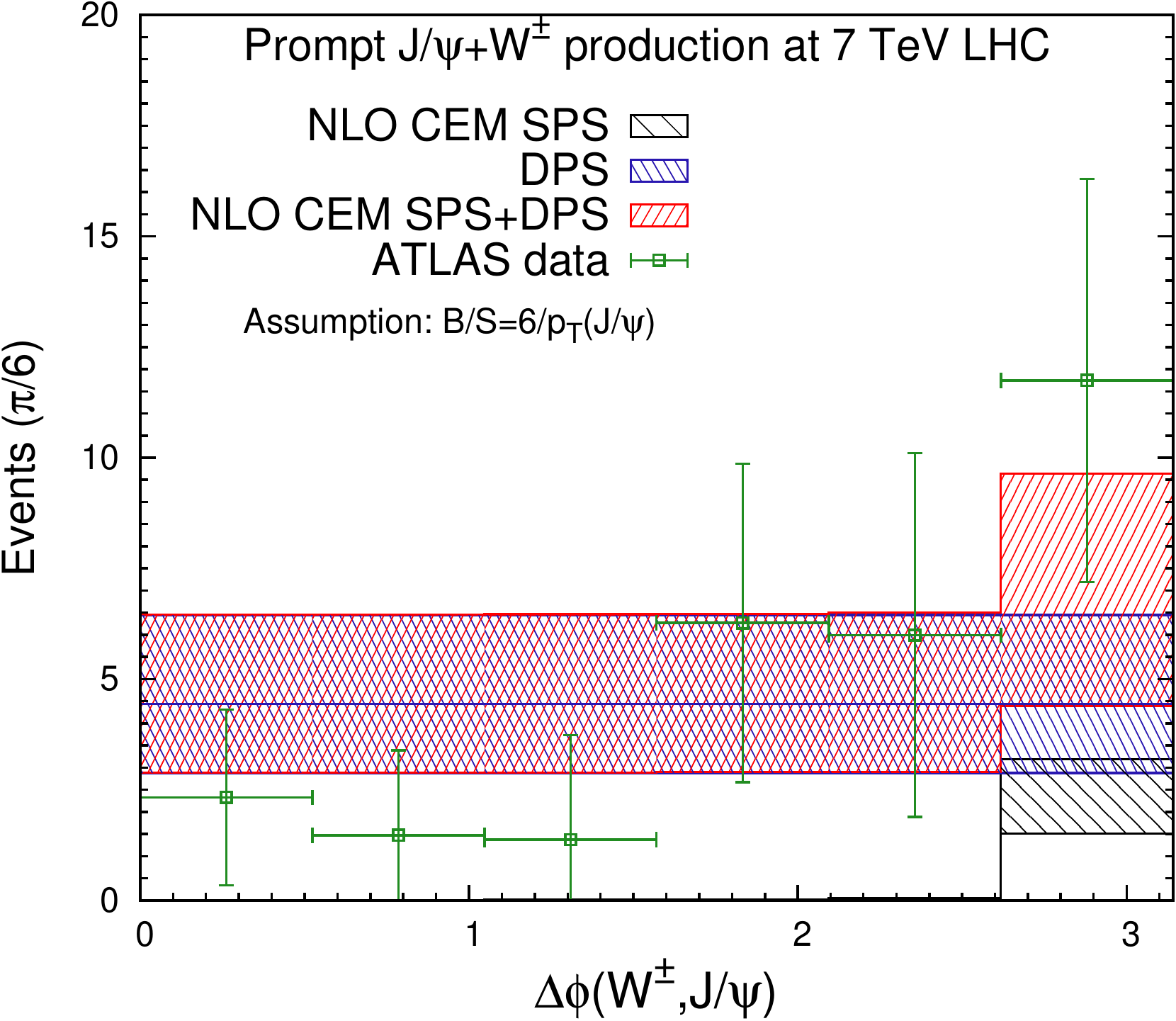}\label{fig:dphi-psiW}}
\caption{ (a) $\frac{d\sigma (pp \to W^\pm + J /\psi)}{dydp_T} \frac{{\rm BR}(J/\psi \to \mu^+ \mu^- )}{ \sigma (pp\to W^\pm ) }$ as a function of $P_T^\psi$: comparison between the ATLAS, the SPS CEM yield and our fit DPS yield (b) Comparison 
between the (uncorrected) ATLAS azimuthal event distribution and our
 NLO theoretical results for $J/\psi+W$ in the CEM (SPS + DPS) effectively
 folded with an assumed ATLAS acceptance. Taken from~\cite{Lansberg:2017chq}.
}

\end{center}
\end{figure}

First, let us evaluate $\sigma_{\rm eff}$ and its uncertainty.
The DPS contributions would now be the (inclusive) cross section of ATLAS
minus the  SPS contributions. As such, the DPS uncertainty, as well as 
that of $\sigma_{\rm eff}$, follows from the --statistical and systematical--
uncertainties of the data~\cite{Aad:2014rua} and from the range spanned by the SPS 
evaluations. We consider the NLO CEM to be the upper limit and the direct 
LO CSM~\cite{Lansberg:2013wva} to be the lower one. Let us 
note that the NLO NRQCD evaluation~\cite{Li:2010hc} lies within this 
range. However, the SPS values are anyway much smaller than the data, these theoretical 
uncertainties are in fact quasi irrelevant in the determination of 
$\sigma_{\rm eff}$. Our combined result for $\sigma_{\rm eff}$ is then
\begin{equation}\label{eq:sigma_eff}
\sigma_{\rm eff}
=
(6.1^{+3.3}_{-1.9\, {\rm exp}} \, ^{+0.1}_{-0.3{\rm theo~} } ) {\rm mb}
.
\end{equation}
which is consistent with our extraction from prompt $J/\psi +Z$ 
production (see section \ref{sec:onium_Z}).

Looking back at \cf{fig:ptspsdps-psiW}, we see that a  
larger DPS yield perfectly fills the gap where needed without creating any 
surplus at large $P_T^\psi$ where the SPS was already close to the data. 
Similar to the $J/\psi+Z$ case, the low $P_T^\psi$ yield is uniquely from 
DPS contributions (as is the total yield) and DPS and 
SPS contribute equally  at high $P_T^\psi$. One can thus anticipate the same effect as 
for $J/\psi+Z$ on the azimuthal distribution which was also generated by ATLAS without efficiency correction.
Following the same procedure as for $J/\psi+Z$, we obtain the resulting 
theoretical distribution shown on \cf{fig:dphi-psiW} which
agrees within uncertainties with the uncorrected ATLAS distribution.
We further note that the number of expected SPS events is as low as $2\pm 1$ 
to be compared to $29^{+8}_{-7}$ events observed by ATLAS~\cite{Aad:2014rua}. 
These $2\pm 1$ SPS events are expected to lie at $\Delta \phi_{W\psi} \sim \pi$. 
An updated ATLAS $\Delta \phi_{W\psi}$ analysis in 2 $P_T^\psi$ bins 
should show a flat behaviour in the lower $P_T^\psi$ bin and a slightly peaked one
for the higher $P_T^\psi$ bin. 

In the ATLAS analysis, both $J/\psi+W^+$ and $J/\psi+W^-$ yields were summed. A future analysis
with separate signals may be useful in the unlikely situation in which the observed excess over the computed SPS cross section would not come from the DPS, but for a new resonance for instance.

\subsubsection{Associated hadroproduction with a photon}
\label{sec:psi-gamma}

Among the associated-production processes, the production with a photon is certainly
that which has been the object of the largest amount of theoretical studies~\cite{Drees:1991ig,Kim:1992at,Sridhar:1992np,Berger:1993sj,Doncheski:1993dj,Kim:1994bm,Mirkes:1994jr,Roy:1994vb,Cacciari:1996zu,Kim:1996bb,Mathews:1999ye,Kniehl:2002wd,Kniehl:2006qq,Li:2008ym,Lansberg:2009db,Baranov:2010zze,Li:2014ava}. The reason 
for this interest are multiple. However, it is clear that the emission of
a photon in a hard reaction where a quarkonium is produced alters
the relative importance of the different expected contributions. 

In hadroproduction, the situation however remains complex and should carefully be analysed
along the lines of the progress of our understanding of single $\Q$ production, in
particular the impact of the QCD corrections on both CS and CO contributions. We will review
this in this section. 

To date, no experimental analysis of photon-quarkonium production exist, apart from those from
decays. A well known example is the decay of $\chi_c$ and $\chi_b$ states into $J/\psi$
and $\Upsilon$ states with a photon emission. Another one is that of a $H^0$ boson into
$J/\psi+\gamma$ and $\Upsilon+\gamma$, whose measurement\footnote{ATLAS performed two searches
one in 2015~\cite{Aad:2015sda} and another in 2018~\cite{Aaboud:2018txb} and CMS one in 2018~\cite{Sirunyan:2018fmm}.} can give a novel access to its coupling 
to the charm and beauty quarks~\cite{Doroshenko:1987nj,Bodwin:2013gca}. These 
happen via exclusive decays whose treatment somewhat differs from
the non-resonant process we are interested here where the invariant mass of the 
quarkonium--photon system is not fixed. 

Finally, let us note that for now there does not exist any --theoretical-- studies of the impact
of the DPS on $\Q+\gamma$.

\paragraph{Early motivations and computations.}

In 1991, Drees and Kim suggested~\cite{Drees:1991ig} that $J/\psi+\gamma$ hadroproduction
--away from the $\chi_c$-mass region--  would be a good probe of the gluon content in the proton.
In 1992, Sridhar extended~\cite{Sridhar:1992np} this idea with a proposal to study doubly
longitudinally polarised proton-proton collisions as a novel means to measure
the gluon helicity distribution $\Delta g$ in the proton. His study was
driven by the possibility of performing such studies with a polarised
target with possible future polarised beams at the Tevatron or HERA.
In 1993, Doncheski and Kim extended~\cite{Doncheski:1993dj} the scope of such ideas 
to the colliders cases of  RHIC and the SSC.

These studies were followed by several proposals~\cite{Kim:1994bm,Mirkes:1994jr,Roy:1994vb} 
to study $J/\psi+\gamma$ hadroproduction in order to probe the production mechanisms in a new manner.
Let us recall that at the time the first puzzling Tevatron observations were appearing
along with the advent of NRQCD and its COM.

The first study of the COM contributions to the associated $J/\psi+\gamma$ 
hadroproduction was carried out by Kim \etal~\cite{Kim:1996bb} in 1996. They concluded
that the COM contributions ought to be small and that the process remained a clean probe of gluon 
content of the proton via the well normalised CS rates. In 1999, Mathews \etal\ pushed~\cite{Mathews:1999ye} the 
studies to large $P_T$ with some considerations on possibly novel topologies --mostly for the COM contributions though. We know now from the previous discussions 
that one should be very cautious in this regime. In 2002, Kniehl \etal\ performed another 
LO study~\cite{Kniehl:2002wd} of the $P_T$-differential cross section. 
Overall, the discussion of CO contributions is complex because of the interplay
between the suppression of (anti)quark PDFs w.r.t. to the gluons ones, of
the scaling of the different LO -- and NLO or even NNLO-- topologies and of the uncertain size of the LDMEs.

One thing nevertheless remains  the object of a consensus: the CS contributions are dominant
in the $P_T$ region below 15 GeV where, in fact, first measurements are possible. 
\cf{fig:psi-gamma-LO} offers a realistic sketch of the situation at LO.

\begin{figure}[hbt!]
\centering
\includegraphics[width=0.6\columnwidth]{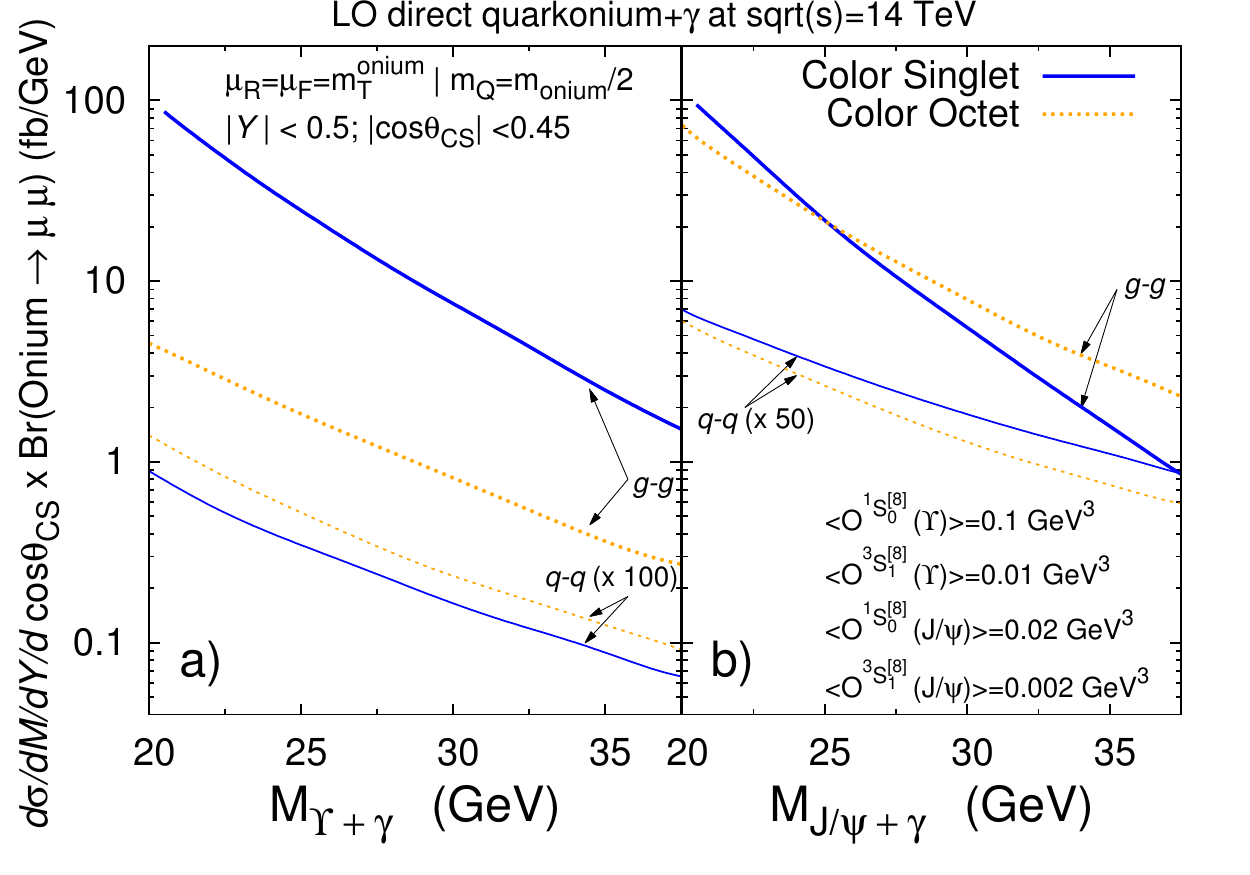}
\caption{Different contributions to the production of $\Q+\gamma$ at LO with a) an $\Upsilon(1S)$ (resp. b) a $J/\psi$)  from $g-g$ and $q-\bar q$ fusion  from the CS and CO channels as function the invariant mass of the pair, $M$. The curves for the $q-\bar q$ fusion are rescaled by a factor $100$ (resp. $50$). The CO matrix elements we used  are very close to those obtained in a recent LO fit of LHC data~\cite{Sharma:2012dy}. $Y$ stands for the rapidity of the pair and the range in $\cos \theta_{CS}$ roughly corresponds to $|\Delta y|<1$.Taken from \cite{Dunnen:2014eta}.}
\label{fig:psi-gamma-LO}
\end{figure}

\paragraph{NLO CS computations}

In 2008, Li and Wang reported~\cite{Li:2008ym} on the first computations of 
the NLO corrections of the hadroproduction of direct $J/\psi+\gamma$ and
$\Upsilon+\gamma$. Like in the inclusive case, they found sizeable 
NLO corrections at large $P_T$ from the new topologies which
appear with a $P_T^{-6}$ scaling  (see \cf{fig:NLO-PT6-CSMpsigamma}), 
in comparison to LO topologies \cf{fig:LO-CSM-psigamma}
and other NLO topologies such as the loop corrections \cf{fig:NLO-PT8-CSMpsigamma}.

\begin{figure*}[ht!]
\centering
\subfloat[]{\includegraphics[scale=.4]{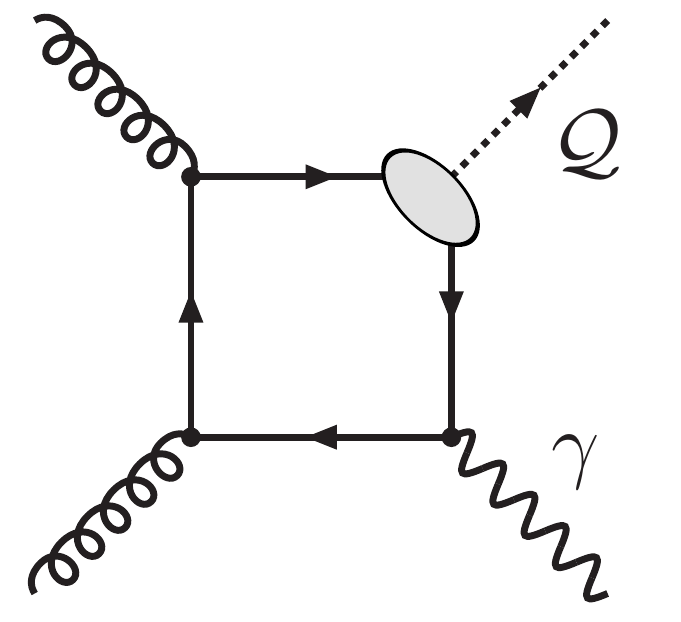}\label{fig:LO-CSM-psigamma}}
\subfloat[]{\includegraphics[scale=.4]{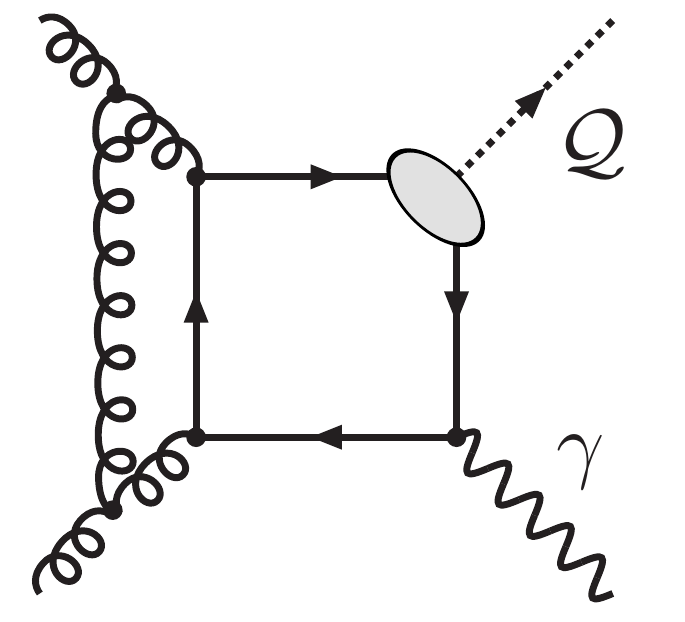}\label{fig:NLO-PT8-CSMpsigamma}}
\subfloat[]{\includegraphics[scale=.4]{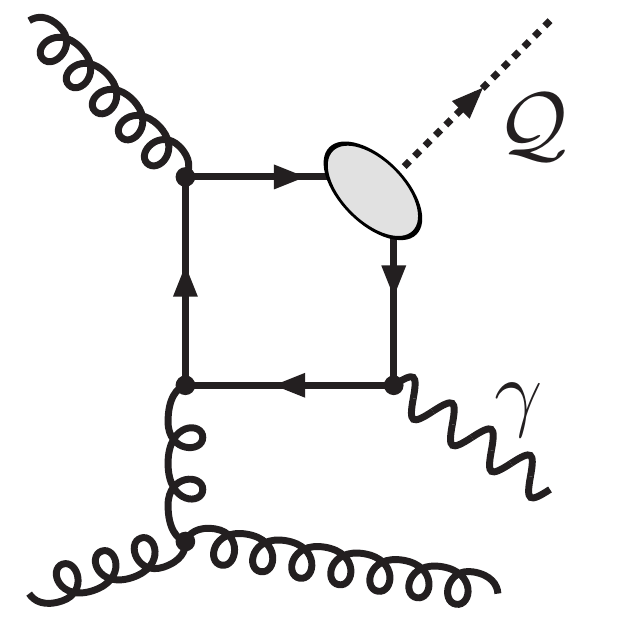}\label{fig:NLO-PT6-CSMpsigamma}}
\subfloat[]{\includegraphics[scale=.4]{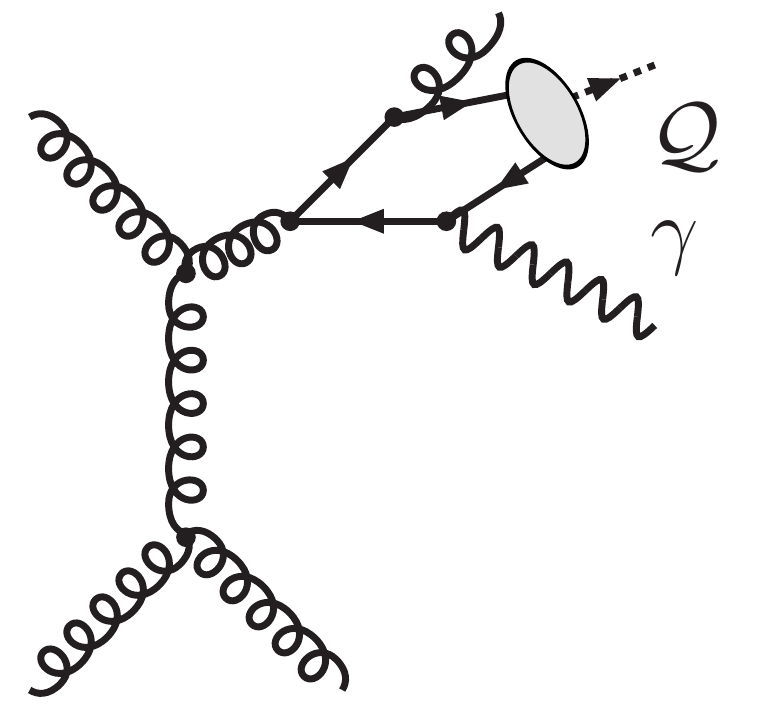}\label{fig:NNLO-PT4g-CSMpsigamma}}
\subfloat[]{\includegraphics[scale=.4]{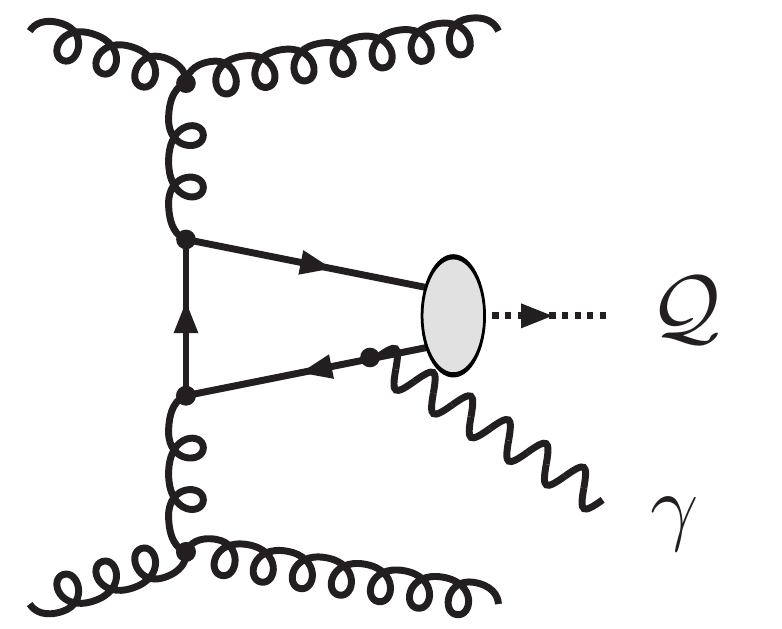}\label{fig:NNLO-logs-gluon-CSMpsigamma}}\\
\subfloat[]{\includegraphics[scale=.4]{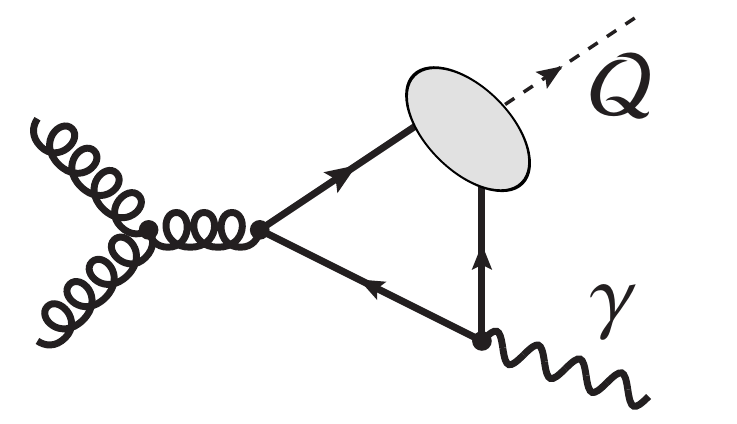}\label{fig:gg-COMpsigamma-viatriangle}}
\subfloat[]{\includegraphics[scale=.4]{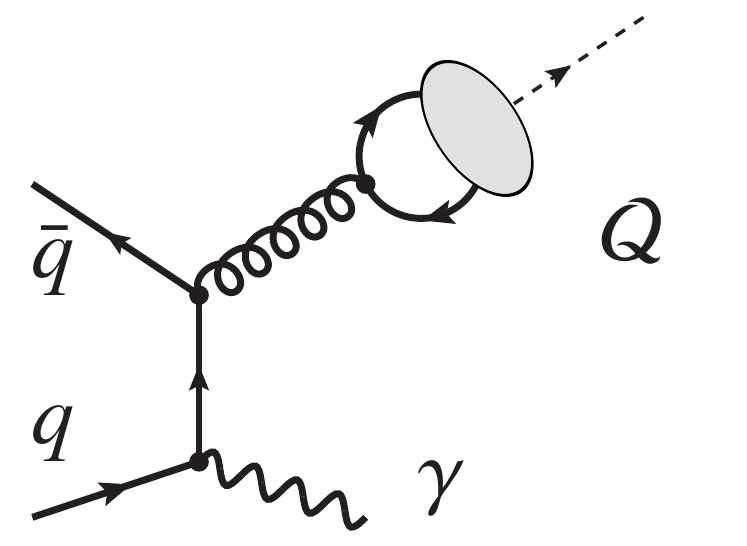}\label{fig:LOqqbar-PT4-COMpsigamma}}
\subfloat[]{\includegraphics[scale=.4]{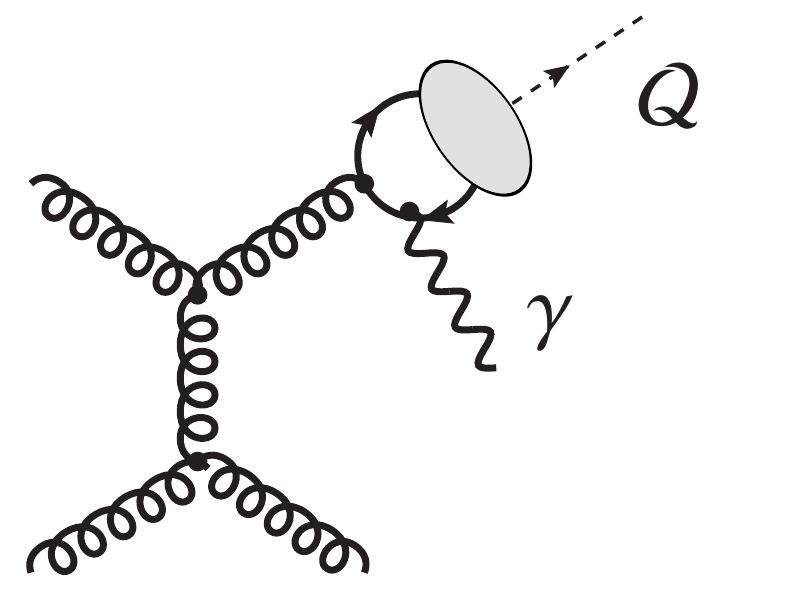}\label{fig:NLOgg-PT4-COMpsigamma}}
\subfloat[]{\includegraphics[scale=.4]{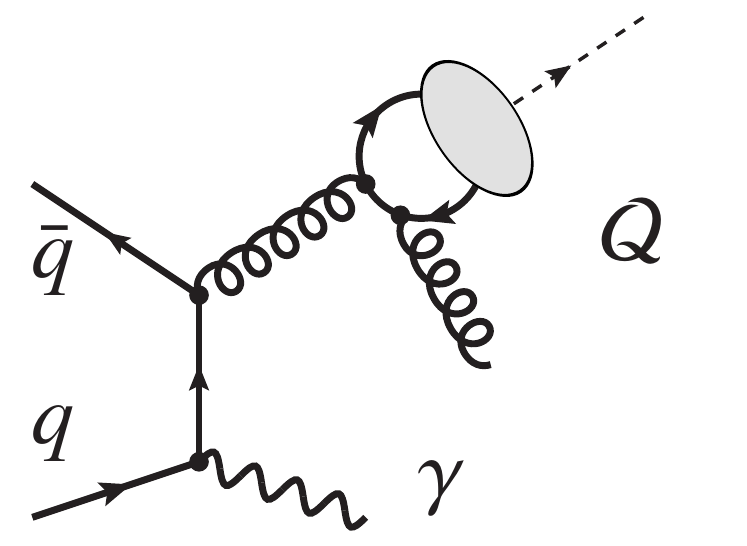}\label{fig:NLOqqbar-PT4-COMpsigamma}}
\caption{Representative diagrams contributing to the hadroproduction of a $J/\psi$ (or $\Upsilon$) in association
with a photon in the CSM at orders $\alpha_s^2 \alpha$ (a), $\alpha_s^3 \alpha$ (b,c), $\alpha_s^4  \alpha$ (d,e), and in the COM at $\alpha_s^2 \alpha$ (a,f,g), $\alpha_s^3 \alpha$ (b,c,h,i), \dots}
\label{diagrams-psigamma}
\end{figure*}

This computation is very similar to that outlined for $\Q+Z$ in section \ref{sec:onium_Z}. We will only mention
here the most relevant differences. First, unlike the $Z$, the emission of the photon can generate
IR divergences. To avoid them, one imposes specific cuts which anyway go along with the detectability 
of the photon which we discuss later. At higher orders, this can generate some subtle issues. 
Second, the natural choice scale for the current process
is likely similar to that of single $\Q$ production, that is $m_T=\sqrt{m_T^2+(P_T^\Q)^2}$, 
unless the cuts on the photon generate new significantly different momentum scales. Last, the resulting colour structure
is a little simpler because of the sole vector nature of the photon coupling.
Their results are shown on \cf{fig:NLO_vs_NLO_star} (dashed line) for both $J/\psi+\gamma$ and $\Upsilon+\gamma$ at the LHC for $p_T^\gamma > 1.5$~GeV\footnote{The corresponding theory parameters are : $\alpha=1/137$ with, for the $J/\psi$, $m_c=1.5$ GeV, $|R(0)|^2=0.810$ GeV$^3$, $\mu_R=\mu_F=\mu_0=m_T$  and with, for the $\Upsilon(1S)$,
 $m_b=4.75$ GeV, $|R(0)|^2=6.48$ GeV$^3$, $\mu_R=\mu_F=\mu_0=m_T$. The PDF used was CTEQ6M~\cite{Pumplin:2002vw}. The theoretical uncertainties
from the mass and scale variations are not shown.}
At the leading order in the heavy-quark velocity ($v$), the direct cross section 
for the $\psi(2S)$, $\Upsilon(2S)$ and 
$\Upsilon(3S)$ are readily obtained by changing $|R_{\Q}(0)|^2$ and the branching
ratio into dileptons. 

\begin{figure*}[hbt!]\centering
\subfloat[$J/\psi + \gamma+ X$]{\includegraphics[width=.45\columnwidth]{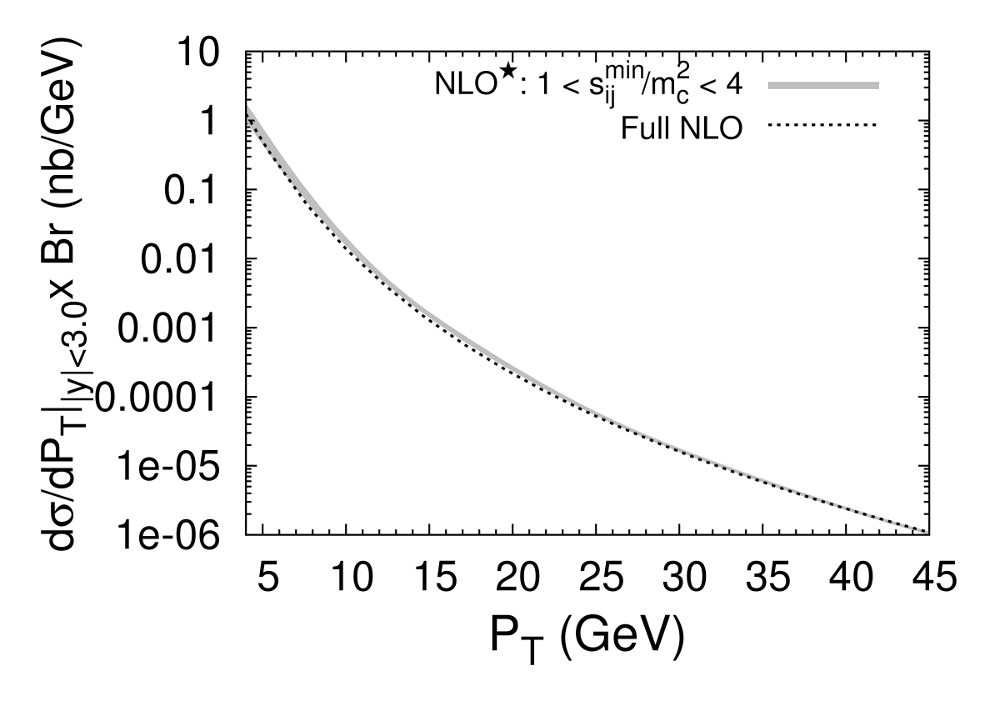}}
\subfloat[$\Upsilon + \gamma+ X$]{\includegraphics[width=.45\columnwidth]{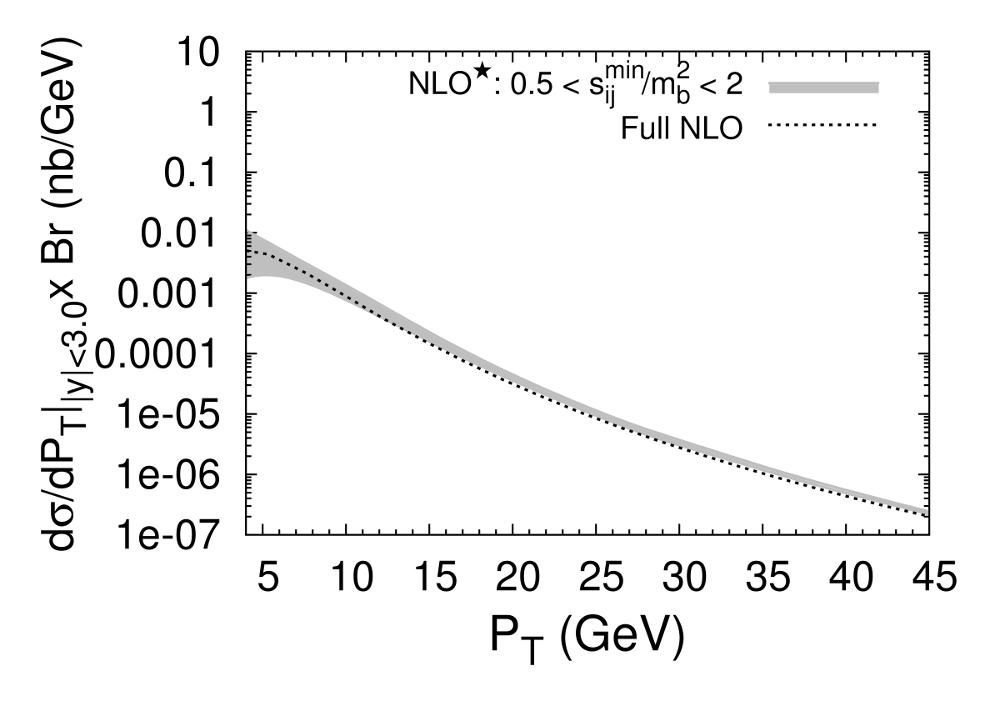}}
\caption{Full NLO computation of  $\Q +\gamma+X$ (dashed line)~\cite{Li:2008ym} 
 vs. $\Q +\gamma$ + 1 light parton~\cite{Lansberg:2009db} with a $s_{ij}$ cut (grey band) at $\sqrt{s}=14$ TeV: (a) for $J/\psi$ 
and (b) for $\Upsilon$. Taken from~\cite{Lansberg:2009db}.
} \label{fig:NLO_vs_NLO_star}
\end{figure*} 
 
Along the same lines as the single $^3S_1$ production case,  one also expects~\cite{Lansberg:2009db} to accurately reproduce  the 
cross section at NLO accuracy ($\alpha^3_S\alpha$) by 
computing the yield from ${\cal Q}+\gamma$ plus
one light parton with the invariant-mass cut $s_{ij} > s_{ij}^{\rm min}$ between any pairs of light partons,
\ie\ with a NLO$^\star$ computation. 

\cf{fig:NLO_vs_NLO_star} confirms these expectations  with a very good agreement
between the dashed line (full NLO) and the NLO$^\star$ (grey band) computed for different 
values of $s_{ij}^{\rm min}$. 
In addition, one notes that the sensitivity on $s_{ij}^{\rm min}$ is quasi irrelevant for the $J/\psi$ case
and quickly vanishes for increasing $P_T$ for the $\Upsilon$. 

This independently confirms --in addition to the aforementioned di-$J/\psi$ and $J/\psi+Z$ cases-- 
the validity of the arguments
initially given in~\cite{Artoisenet:2008fc} according to which the 
NLO QCD corrections to quarkonium-production processes whose LO shows 
a non-leading $P_T$ behaviour can reliably be  computed at mid and large 
$P_T$ with the NLO$^\star$ method. Considering only the real-emission contributions
accompanied with a kinematical cut is sufficient. 
This thus gives us confidence that improved Monte-Carlo simulations  at
mid and large $P_T$ can be achieved likewise. This also lead us 
to evaluate the impact of NNLO contributions by computing the 
NNLO$^\star$ contributions, as we review next.
 
Like for the full NLO computation~\cite{Li:2008ym}, the NLO$^\star$ yield was found to be  dominantly longitudinal, 
at variance with the LO yield which is strongly transverse at nonzero $P_T$. This is to be paralleled with the inclusive 
case. It thus seems that the replacement of the photon by a gluon is indeed irrelevant as
what concerns the produced quarkonium polarisation.

\paragraph{Beyond NLO.}

Like for single $\psi$ and $\Upsilon$,  new contributions appearing at $\alpha^4_S\alpha$, like the topologies 
of~\cf{fig:NNLO-PT4g-CSMpsigamma} (gluon fragmentation) and~\cf{fig:NNLO-logs-gluon-CSMpsigamma} 
(``high-energy enhanced'' or double $t$-channel gluon exchange), benefit from further kinematical enhancements. 
These in fact provide us with novel
mechanisms to produce a high-$P_T$ ${\cal Q}$ with a $\gamma$ with a lower kinematical
suppression, still via CS transitions. As such, they should dominate the
differential cross section at NNLO accuracy at large $P_T^\Q$.
They can be computed via $pp \to {\cal Q} +\gamma+ jj$ ($j$ being any light parton) like 
in section~\ref{subsec:NNLOstar-CSM}.

\begin{figure}[hbt!]\centering
\subfloat[]{\includegraphics[width=.5\columnwidth]{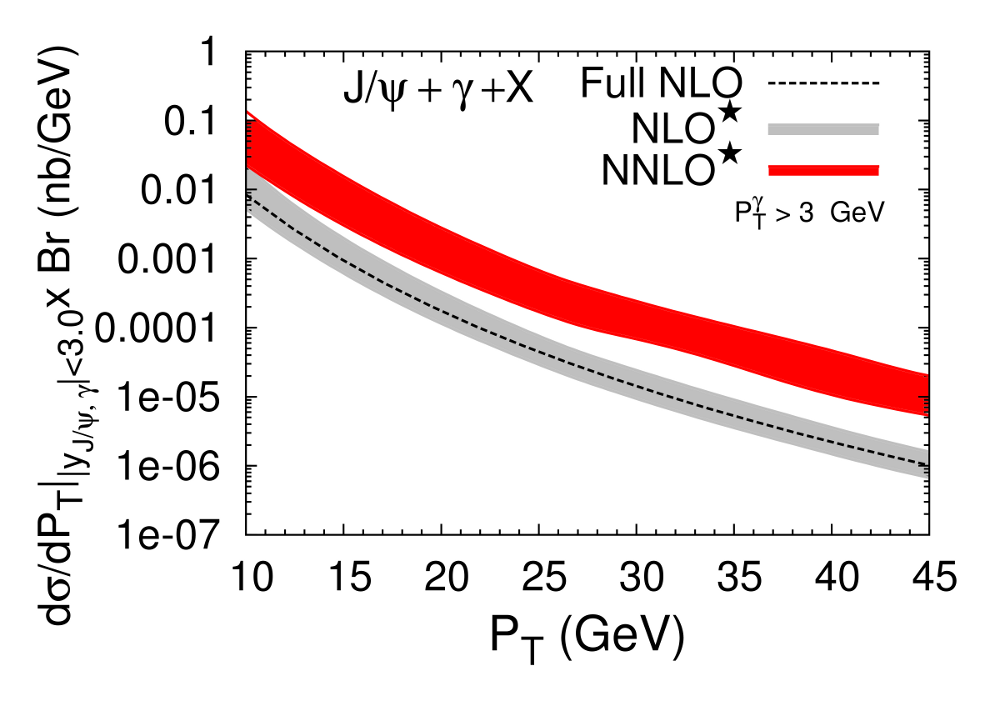}\label{fig:dsdpt_NNLO_star-psigamma}} 
\subfloat[]{\includegraphics[width=.5\columnwidth]{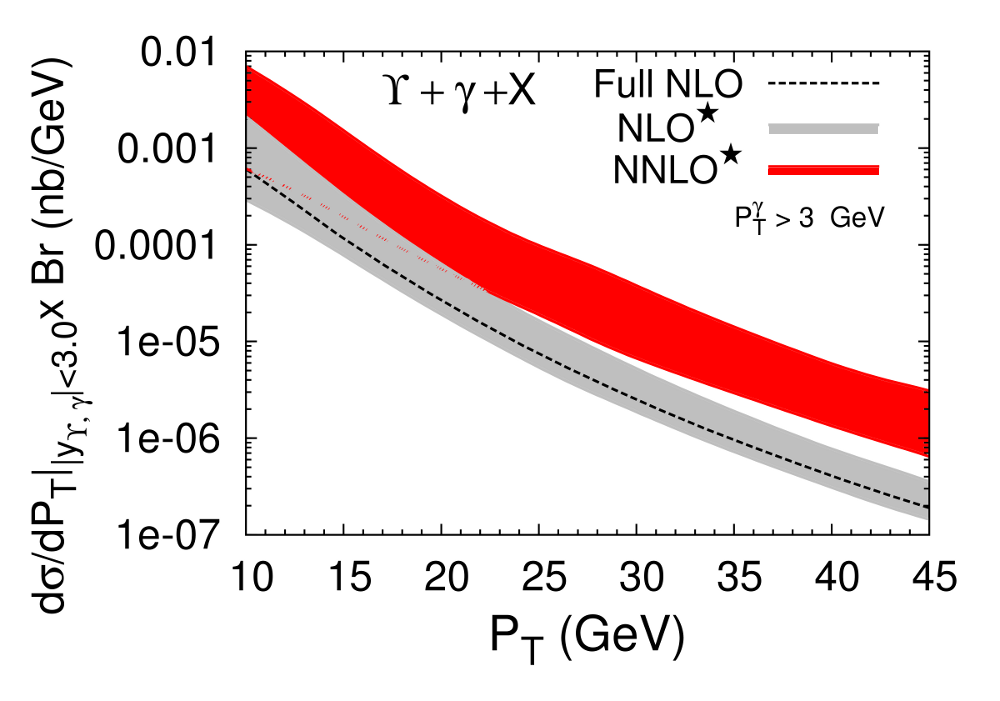}\label{fig:dsdpt_NNLO_star-upsigamma}}
\caption{Comparison between the full NLO, the NLO$^\star$ and the NNLO$^\star$ 
contributions at $\sqrt{s}=14$~TeV for $P_T^\gamma > 3$ GeV: (a) for $J/\psi+\gamma$ and (b) $\Upsilon(1S)+\gamma$.
  A photon isolation cut ($\Delta R>0.1$) was applied on the NLO$^\star$ and NNLO$^\star$ yields (see text). Taken from~\cite{Lansberg:2009db}.} \label{fig:dsdpt_NNLO_star}
\end{figure}

The resulting differential cross-sections for $J/\psi +\gamma$ and $\Upsilon(1S) +\gamma$
are shown in~\cf{fig:dsdpt_NNLO_star}. The grey band  (referred to as
NLO$^\star$) corresponds to the sum of the LO and the real $\alpha_S^3 \alpha$ contributions. 
The red (or dark) band (referred to as NNLO$^\star$) corresponds to the sum of the LO, the real $\alpha_S^3 \alpha$ and 
the real $\alpha_S^4 \alpha$ contributions.  The  $\alpha_S^4 \alpha$  contributions 
in both cases dominate over the yield at large $P_T$. The uncertainty bands result from the combined variations
 $0.5 \mu_0 \leq \mu_{R,F}\leq 2 \mu_0$ with for the $J/\psi$, $m_c=1.5\pm0.1$ GeV  and 
$1 \leq s_{ij}^{\rm min}/(1.5~{\rm GeV})^2 \leq 2$ and,
for the $\Upsilon(1S)$,   $m_b=4.75\pm0.25$ GeV and $0.5 \leq s_{ij}^{\rm min}/(4.5~{\rm GeV})^2 \leq 2$.

Although the uncertainty associated with the choice of the cut 
$s_{ij}^{\rm min}$ is larger than at NLO$^\star$,  it 
is remains smaller than those attached to the mass, the renormalisation scale and the factorisation 
scale --the latter being the smallest. A large dependence on $\mu_R$ is
expected. First, the virtual contributions are missing at low $P_T$  where they are sizeable and where they reduce the $\mu_R$ dependence. Second, the dominant contributions at large $P_T$ are proportional to $\alpha_S^4$.

We also found~\cite{Lansberg:2009db} that the sub-process 
$gg \to \Q+\gamma+gg$ dominates, providing with more than two thirds of the whole
yield in the $J/\psi$ case and that this fraction slightly increases with $P_T$ 
and only weakly depends on the value of 
the invariant-mass cut-off of light partons $s_{ij}^{\rm min}$. This tends to indicate 
that the IR divergences are not --after being cut--  artificially responsible 
for a large part of the NNLO$^\star$ yield. 
It is also possible that the largest part of this contribution is not from
gluon fragmentation topologies, but rather from double $t$-channel gluon exchange ones, keeping in
mind that such a decomposition in terms of the corresponding Feynman graphs is not gauge invariant. 
A couple of indications support this viewpoint: (i) the fragmentation contributions are more likely to provide transverse quarkonia, leaving aside corrections
from the off-shellness of the fragmenting gluons  or possible spin-flip contributions from the radiated gluon. (ii)
processes such as $q q' \to \Q +\gamma+q q'$, proceeding uniquely via  double $t$-channel gluon exchange, have
the same  $P_T$ dependence as the process $gg \to \Q+\gamma+gg$ and the difference in normalisation
seems to follow from the colour factors and
the smaller quark PDFs at low $x$.  
Another similarity with $gg \to \Q+\gamma+gg$ is the  strongly longitudinal
polarisation of the  yield from $qq' \to \Q +\gamma+qq' $, 
for $P_T$ larger than 5 GeV, as observed for the full NNLO$^\star$ yield dominated by $gg \to \Q+\gamma+gg$. 

The previous discussion can in fact be extended to the inclusive case studied in~\cite{Artoisenet:2008fc}
and discussed in the section~\ref{subsec:NNLOstar-CSM}. Indeed, 
the results obtained at NNLO$^\star$ for single $\Q$ production are significantly higher than those using the 
fragmentation approximation --without QCD corrections, though. In addition, 
the polarisation of the yield from $gg \to \Q+ggg$ processes is also getting strongly 
longitudinal at large $P_T$ seemingly contradicting the expectations
for a fragmentation channel. For both processes $\Q+\gamma$ and $\Q+X$ at NNLO$^\star$, 
the yield may partly come from double $t$-gluon channel exchanges appearing 
for the first time at this order. 

Whether or not this questions the validity of the 
NNLO$^\star$ computation would only be answered by a full NNLO computation or maybe by 
a careful kinematical analysis of the yield from $gg \to \Q+\gamma+gg$ or by performing 
a nnLO evaluation~\cite{Shao:2018adj} of the $\Q+\gamma$ yield. 
Such a kinematical analysis would be highly computer-time demanding, especially 
to obtain an invariant-mass distribution of the quarkonia and its 
closest gluon with a sufficient precision to unequivocally attribute the yield to one or the other
topologies. Such a detailed study is likely easier to carry for $\Q+\gamma$ than for
single $\Q$ with a large number of graphs.

\paragraph{Photon detectability.}

In order to detect the photon,  one has to impose a $P_T^\gamma$ cut
for it not to go in the beam pipe and isolation criteria to avoid misidentifications from Bremsstrahlung radiations or hadron decays. In other words, the photons considered here are what is called isolated or direct photons (se \eg\ \cite{Frixione:1998jh}).
Yet, such isolation criteria highly depend on the detector potentialities. 
The determination of
optimum values for $P^\gamma_{T,min}$ and for an isolation criterion for the $\gamma$ improving
the signal over background ratio is beyond the scope of this discussion.

Yet, at LO, the $P_T$ requirement is trivially met for a $\Q$ with a finite $P^\Q_T$ since $P^\gamma_T$ is 
balancing $P^\Q_T$. As discussed by Li and Wang~\cite{Li:2008ym}, this is not automatically the case at NLO 
and a minimum $P^\gamma_T$ cut has to be applied. It was found that it affects the yield vs $P^\Q_T$ up to roughly three  times the cut value. 

As for the isolation cut, configurations with partons ($q$,$g$) within
a cone from the photon $\Delta R=\sqrt{\Delta \eta^2+\Delta \phi^2}$ of a given size have to be excluded
($\Delta R =0.1$ was used in~\cite{Lansberg:2009db}).
Let us note here that such a cut also avoids the QED singularities that may appear 
in $qg\to qg\gamma \Q$ for the $\gamma$ emission by the external quark. Those are anyhow sub-dominant topologies
of suppressed quark-gluon initiated contributions and can safely be  neglected. We will 
come back to this later with the CO computations, where they are not necessarily sub-dominant.

The $\chi_Q$ feed-down should not be significant since it is suppressed by $v^2$ and by the branching while 
having the same $\alpha_s$ suppression than for the $J/\psi$
 for similar topologies. As regards the feed-down of the radially excited states $^3S_1$, they
are readily accounted for by constant multiplicative factors to the cross section of the state fed in : 
$\sim 1.4$ for the $\psi(2S)$ into $J/\psi$,  $\sim 1.1$ for both the $\Upsilon(2S)$ into $\Upsilon(1S)$ and 
$\Upsilon(3S)$ into $\Upsilon(2S)$, provided that one neglects
the mass difference in the decay kinematics.

\paragraph{The impact of the COM.}

As noted in~\cite{Li:2008ym}, the NLO CS and the LO CO yields 
obtained in~\cite{Kniehl:2002wd} are of the same order at the LHC, 
with approximately the same $P_T$ dependence. Yet, contrary to what one could expect~\cite{Li:2008ym} by analogy with the apparent limited impact of 
NLO corrections to CO channels in single $\Q$ production, we anticipated~\cite{Lansberg:2009db} that NLO CO corrections may be important, bringing the NLO CO yields above the NLO CS ones, yet 
below the NNLO$^\star$ ones.

Indeed, for $\Q +\gamma$  production, the $P_T^{-4}$ topologies (\cf{fig:NLOgg-PT4-COMpsigamma}) are not opened at LO, namely at $\alpha\alpha_S^2$, 
except for $q\bar q\to \gamma + \so $ (\cf{fig:LOqqbar-PT4-COMpsigamma}). Compared to the $J/\psi+W$ case where CO contributions were once thought to be dominant, these CO are from gluon fusion and mediated by either $\sps$ or $\pj$ $C=+1$ CO, but they have to compete here with significant CS contributions from gluon fusion as well. However, irrespective of their actual impact, the appearance of $C=+1$ CO transitions as the most important leading-$P_T$ CO topologies, instead of the $C=-1$ $\so$ transition, is a unique feature of this process which has notable consequences which we discuss now.

In 2014, still Li and Wang reported~\cite{Li:2014ava} on the first computation of 
the NLO corrections to the CO contributions to the hadroproduction of direct $J/\psi+\gamma$ and
$\Upsilon+\gamma$. Like the CS case, the computations closely follows from the $\Q+Z$
case. In fact, it was also carried out thanks to the semi-automated framework FDC~\cite{Wang:2004du}.
We will thus only discuss some specificities before analysing their results.

As anticipated, collinear IR divergences due to the radiation of the photon by light quarks appear. 
They thus imposed an isolation cut to the photon according to~\cite{Frixione:1998jh} and a minimum $P_T^\gamma$ cut. Care should also be taken as what regards the cancellation of the (QCD) IR divergences
associated to the fragmentation topologies of  $\pj+\gamma +g$ (\cf{fig:NLOqqbar-PT4-COMpsigamma}) which is specific to 
the $\Q+\gamma$ case.

Since their computations considered the 3 leading CO transitions, namely $\so$, $\sps$, $\pj$, and appeared
after the first complete NLO CO LDME fits~\cite{Ma:2010yw,Chao:2012iv}, they managed to perform a completely coherent 
study, fully accounting for the experimental constraints set by the LHC data. 

To be more precise, they employed the sets of NLO LDMEs 
from \cite{Butenschoen:2011yh} (Hamburg fit), \cite{Gong:2012ug} (IHEP fit)
and \cite{Ma:2010yw,Chao:2012iv} (PKU fit). The later provides a range of values 
from which they extracted two extreme cases which are realistic. We nevertheless note that
the additional constraint from $\eta_c$ data (see section \ref{subsec:COM_1S0_NLO_PT}) disfavours the case where $\pj$ is
zero and $\sps$ is the largest. 

As what concerns their results, they first noted that
the short-distance coefficient (SDC) for the $\pj$ transition is positive for any $P_T^{J/\psi}$.
In the case of the single $J/\psi$ case, the corresponding SDC is negative at large 
$P_T^{J/\psi}$. This feature is fundamental as it allows potential $J/\psi+\gamma$ 
data to break the degeneracy of the data constraints on the linear combination 
of $\sps$ and $\pj$ LDMEs, $M_{0,r_0}$. It is even more relevant than one could think because 
possible LHC data would likely be taken in similar phase spaces than single $J/\psi$ data, 
thus with similar PDFs and where one could not argue that NRQCD is less applicable. 
In the absence of any experimental data, 
this remains a wish, though. 

Yet, their results are so striking that constraints even appear without real data ! Indeed, 
they found out that the NLO LDME sets of the Hamburg and IHEP --\ie\ theirs-- groups result in
negative $P_T$-differential $J/\psi+\gamma$ cross sections for $P_T^{J/\psi}> 13$~GeV, which is of course 
not acceptable in a region where no solid argument could be proposed to invalidate 
the applicability of NRQCD. These unphysical cross-section predictions can in fact be traced back
to the negative values of the $\pj$ LDMEs obtained in these fits. Whereas negative LDMEs are in general not forbidden, as
they encapsulate the hadronisation process where interferences can take place, it
 seems that they can yield to unphysical results for some specific observables, such as
$J/\psi+\gamma$.  

In practice, this would also mean that both Hamburg and IHEP fits have 
to be disregarded. On the way, let us also note that they cannot reproduce the
$J/\psi$ polarisation measurement at the LHC~\cite{Abelev:2011md,Chatrchyan:2013cla,Aaij:2013nlm}, whereas the Hamburg
fit, which uses by far the largest data sample, is the only one which does not
completely disagree with the $P_T$-integrated cross section (see section~\ref{subsec:COM_updates}).

The only caveat which should be solved before drawing such a drastic conclusion is about
the observed sensitivity of the result on the photon $P_T$ cuts. Contrary to
the $J/\psi+\gamma$ CS case at NNLO where the IR QED divergences only arise from sub-leading $P_T$
quark-induced reactions, they arise in the 
CO case from leading-$P_T$ topologies. 
In particular, the sensitivity of the $\pj$ and $\sps$ cross sections on 
$P_T^\gamma$ remains large even for configurations where $P_T^\gamma$ 
and $P_T^{J/\psi}$ are very different. This may cast some doubts on the treatment
of these IR QED divergences and further investigations are probably needed.

However, it should be stressed that the sole consideration of the positivity
of the cross section of this quarkonium-associated process can in principle
provide theoretical constraints --at a given order in $\alpha_s$-- on the 
CO LDMEs. In fact, it would be expedient to attempt to generalise the argument
to other channels, which may or may not be observed one day. In 2019, 
Li and Wang performed another NLO study~\cite{Li:2019anc} of $\Upsilon+\gamma$ hadroproduction 
with similar conclusions regarding the negative character of the cross section at large $P_T$ when $\mopj$ is negative.

\subsection{Quarkonia and heavy quarks}
\label{sec:onium_HQ}

In early 2000's, another puzzle showed up in quarkonium production: the 
rates for inclusive and exclusive $J/\psi$ production 
in association with a $c\bar{c}$ quark pair in $e^+e^-$ annihilation at $B$ factories~\cite{Abe:2002rb,Pakhlov:2009nj}
were found to be much larger than the LO NRQCD
theoretical expectations~\cite{Kiselev:1994pu,Cho:1996cg,Yuan:1996ep,Berezhnoy:2003hz,Hagiwara:2004pf}. 
In this case, contrary to the hadroproduction of a single $J/\psi$,
the COM could not be invoked to bring theoretical
predictions in agreement with the experimental
measurement~\cite{Liu:2003jj}. However, the inclusion of
$\alpha_s^3$ (NLO) CS corrections were a few years later found to largely reduce the
discrepancy between the theory and the measurements~\cite{Zhang:2006ay,Gong:2009ng}.
This triggered many other theoretical studies which we will address in this section.

In fact, the associated production of a $J/\psi$ with charm had also already been discussed as early
in the 1980's in $e^+e^-$ annihilation~\cite{Clavelli:1982hp}, in photoproduction~\cite{Berger:1982fh} and in hadroprodution~\cite{Tuite:1984zz}.
However, as opposed to $\Q+\gamma$, the associated hadroproduction of quarkonium with
another heavy quark has been the object of a limited number of theoretical studies
in spite of the variety of the possible measurable final states at hadron colliders. 

One could directly think of hadroproduction of $J/\psi+c$,
$\Upsilon+b$,  $J/\psi+b$ and $\Upsilon+c$, for which analysed or on-tape LHC data  do exist.
The small amount of theoretical studies is probably due to (i) the  complexity of tackling the  
production of another (massive) strongly-interacting particle produced with the quarkonium, (ii)
the additional complexity involved in the 
production of a quarkonium along with a heavy quark of the same flavour and (iii) the suppresion of the CS channels for the production of a quarkonium along with a heavy quark of a different
 flavour. Nonetheless, as opposed to the production along with $W$ or $Z$ boson whose rates are admittedly
small, those of such channels should not be small. In addition, dedicated techniques to tag
heavy-flavour production are now routinely used with high efficiencies. As such, it is not surprising that
LHCb pioneered such studies with that of the final states sensitive to $J/\psi+c$~\cite{Aaij:2012dz}
and $\Upsilon+c$~\cite{Aaij:2015wpa}. In addition, D0 certainly also observed $\Upsilon+b$,
like LHCb, CMS and ATLAS did for $J/\psi+b$, in reactions where the $b$ decayed into 
a non-prompt $J/\psi$, but so far they did not perform a dedicated analysis to extract 
the corresponding cross sections.

The theory motivations to study these associated channels are however not of minor importance. Beside 
the $e^+e^-$ case, we showed~\cite{Artoisenet:2007xi} that $J/\psi+c\bar{c}$ hadroproduction 
is the dominant $\alpha_s^4$ CS channel 
for single $J/\psi$ production at large $P_T^\psi$. As such, it is worth specific attention. Second, 
the production of 3 heavy quarks surrounding each other may result in an enhanced production rate of the quarkonium with a novel phenomenon called Colour Transfer (CT)~\cite{Nayak:2007mb,Nayak:2007zb}.
Third, since for process like $J/\psi+b$ and $\Upsilon+c$ hadroproduction, the expected 
SPS contributions do not seem to be large, they may indeed be an ideal playground to 
study DPS. Fourth, $J/\psi+c$, and maybe $\Upsilon+b$ hadroproduction and photoproduction, could help determine the momentum
distribution of heavy quarks in the proton, including the possibility 
for the non-perturbative contribution from intrinsic charm~\cite{Brodsky:2009cf}. 
Fifth, in the latter channels, there is no reason for the CO contributions to be enhanced
and it could therefore be a possible channel to test the CSM on its own in different production modes.

Yet, there does not exist any LO CS computation for $J/\psi+b$ hadroproduction, nor 
any NLO computations for any of these channels --except $e^+e^- \to J/\psi +c \bar c$. In the following, we will review the main points of the existent computations and outline the conclusions of both
LHCb analyses in this context.

\subsubsection{Charmonium + charm} 
\label{sec:psi-cc}

\paragraph{Hadroproduction.}

In view of the unexpectedly large measurements for the production of
$J/\psi$ associated with a $c\bar{c}$ quark pair in $e^+e^-$
annihilation, we found it natural~\cite{Artoisenet:2007xi} to see whether the corresponding
production pattern could also be relevant in hadroproduction.  Beside
offering a new interesting experimental signature to be studied, this process is one
of the $\alpha_s^4$ (NLO) corrections to the inclusive CS
hadroproduction of $J/\psi$. Historically,
these higher-order contributions to the cross section at the Tevatron 
had been first considered in the fragmentation approximation as a first attempt 
to solve the so-called $\psi'$ anomaly~\cite{Braaten:1994xb,Cacciari:1994dr} and,
by Berezhnoy~\etal~\cite{Berezhnoy:1998aa}, in the context of double-charm-baryon
production where, on the way, the relevance of such a fragmentation approximation
was questioned at mid $P_T$.
Finally, one year before our study~\cite{Artoisenet:2007xi}, $\Q+Q\bar{Q}$ production 
had been considered by Baranov
in the framework of $k_T$ factorisation~\cite{Baranov:2006dh,Baranov:2006rz}.

\begin{figure}[hbt]
\centering
\subfloat[]{\includegraphics[height=2cm]{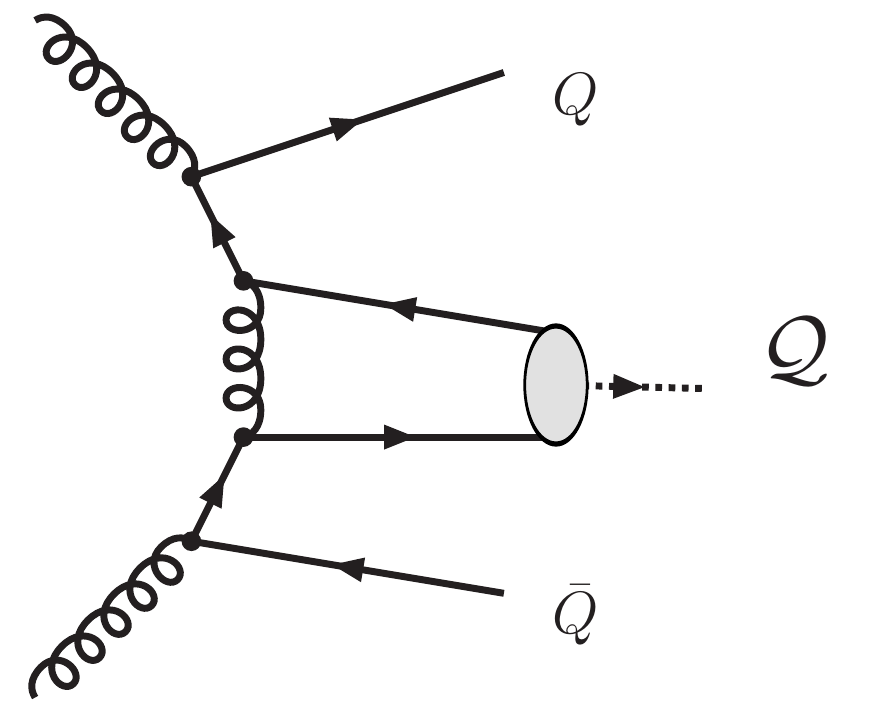}}
\subfloat[]{\includegraphics[height=2cm]{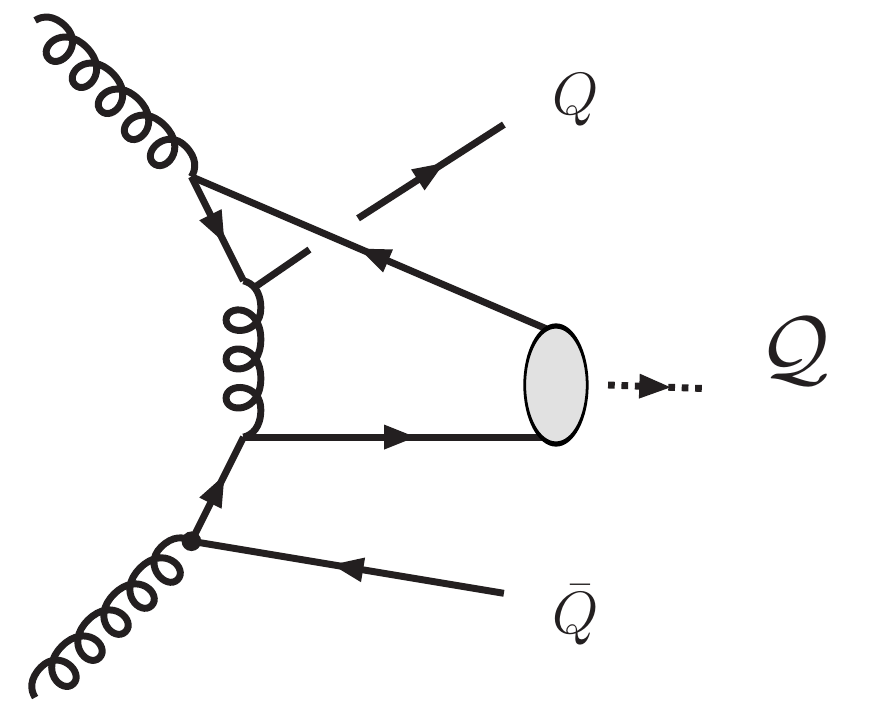}}
\subfloat[]{\includegraphics[height=2cm]{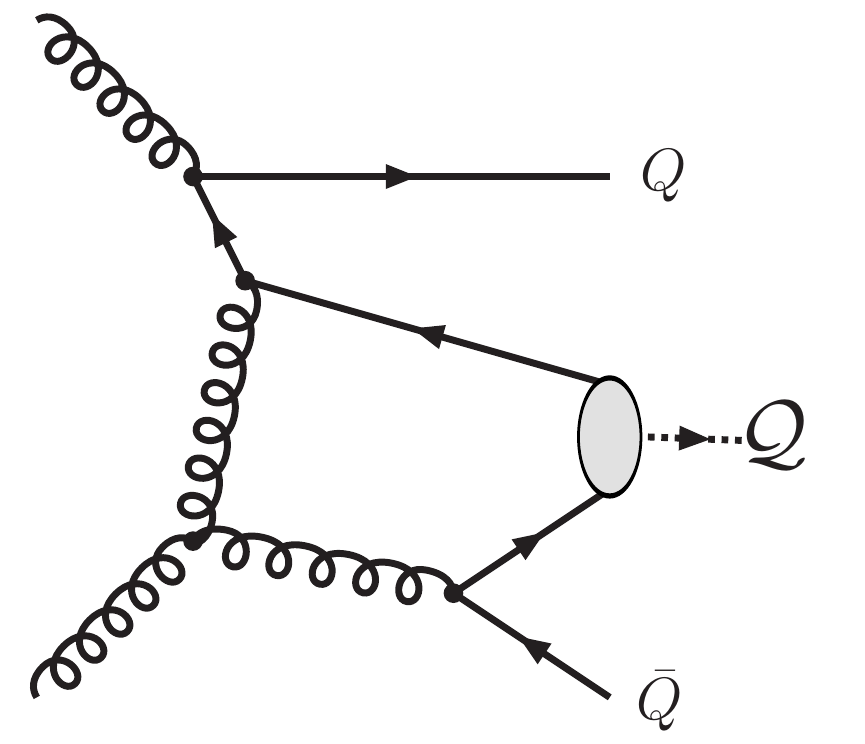}}
\subfloat[]{\includegraphics[height=2cm]{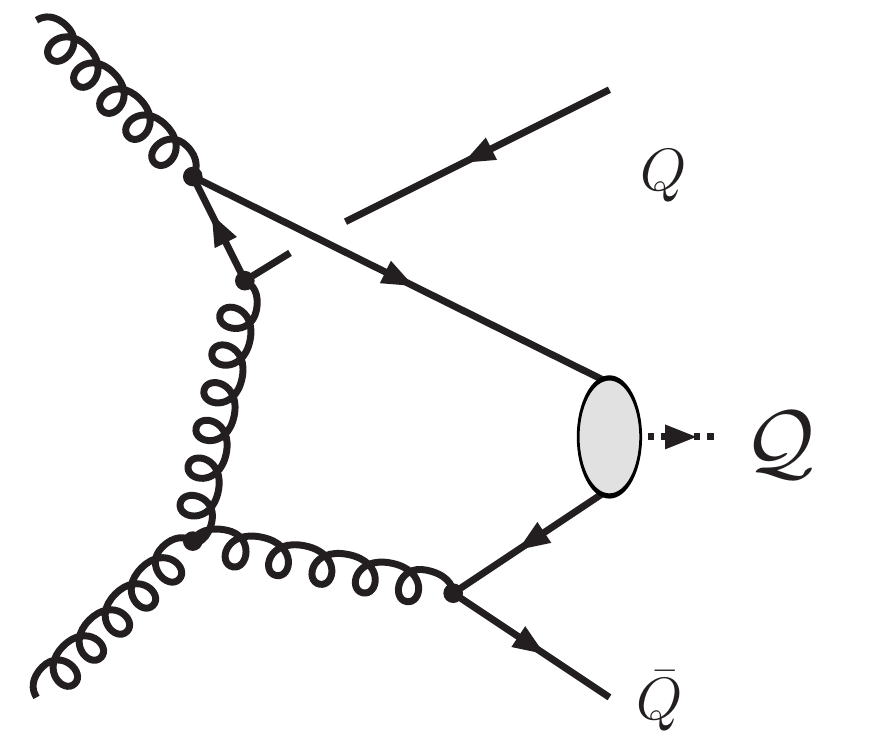}}
\subfloat[]{\includegraphics[height=2cm]{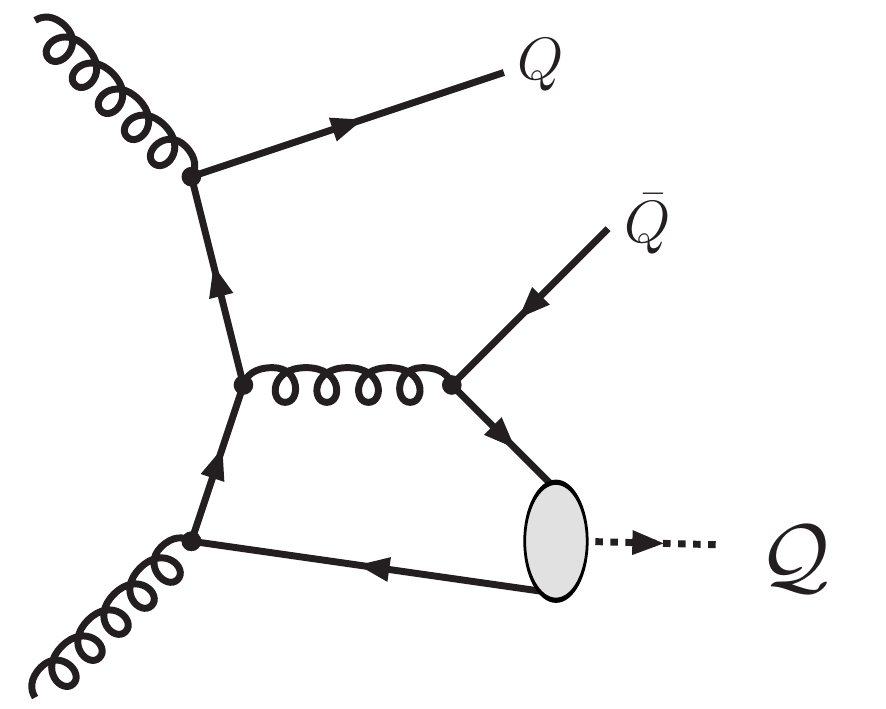}}
\subfloat[]{\includegraphics[height=2cm]{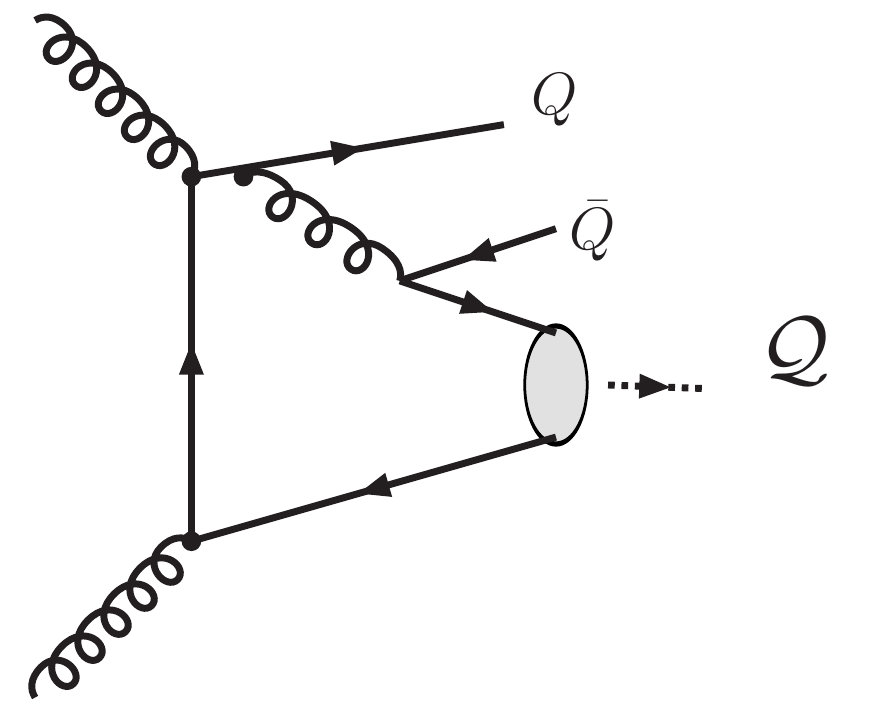}}
\subfloat[]{\includegraphics[height=2cm]{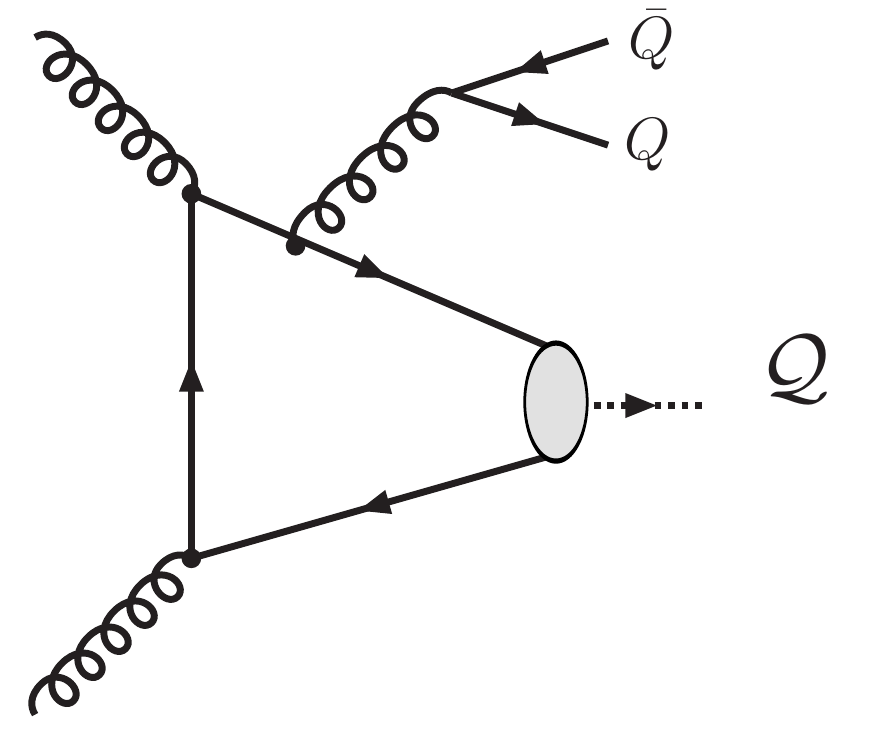}\label{fig:OniumQQ-graph7b}}\\
\subfloat[]{\includegraphics[height=2.cm]{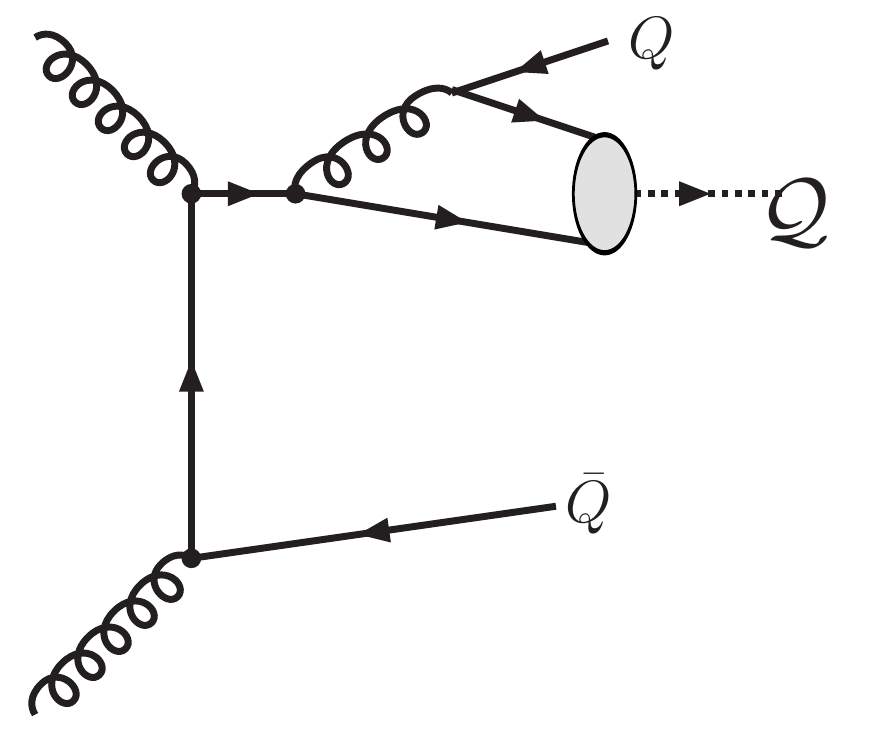}}
\subfloat[]{\includegraphics[height=2cm]{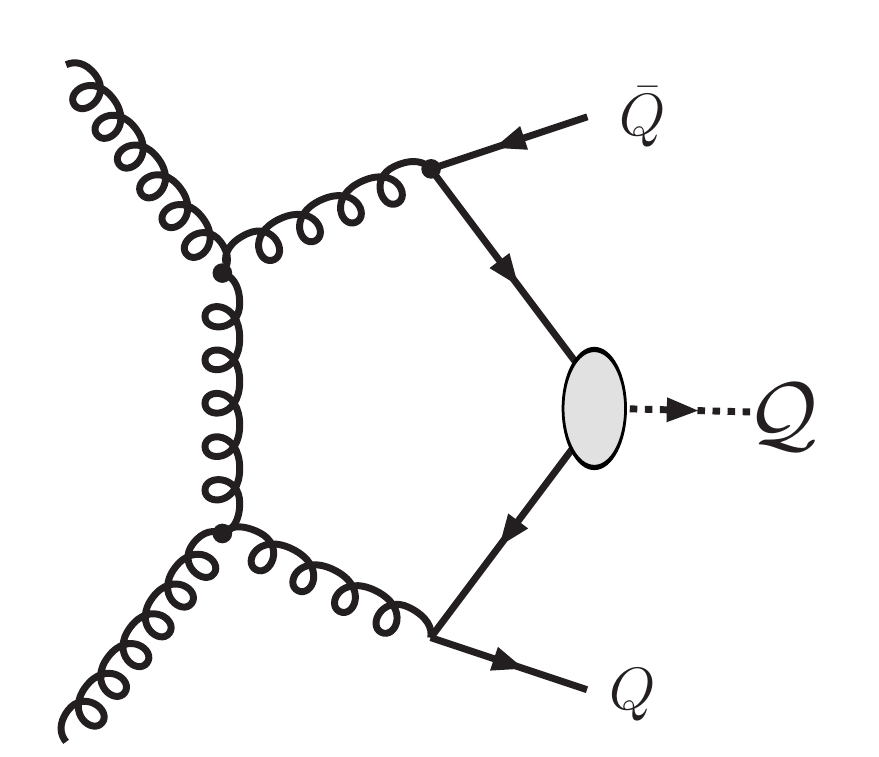}}
\subfloat[]{\includegraphics[height=2cm]{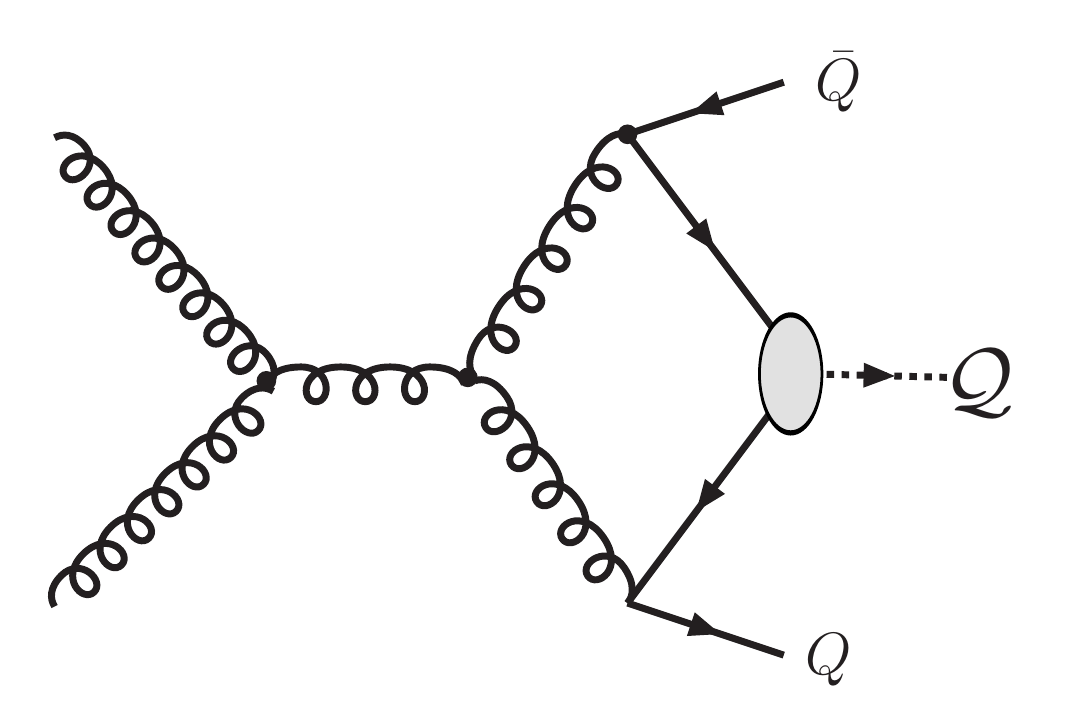}}
\subfloat[]{\includegraphics[height=2cm]{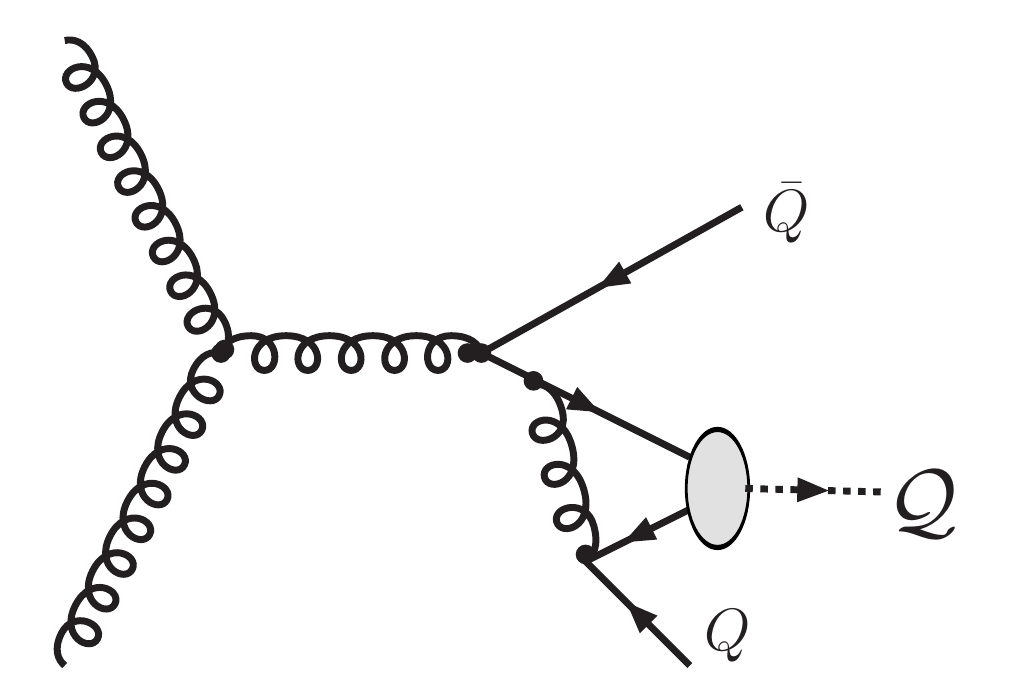}}
\subfloat[]{\includegraphics[height=2cm]{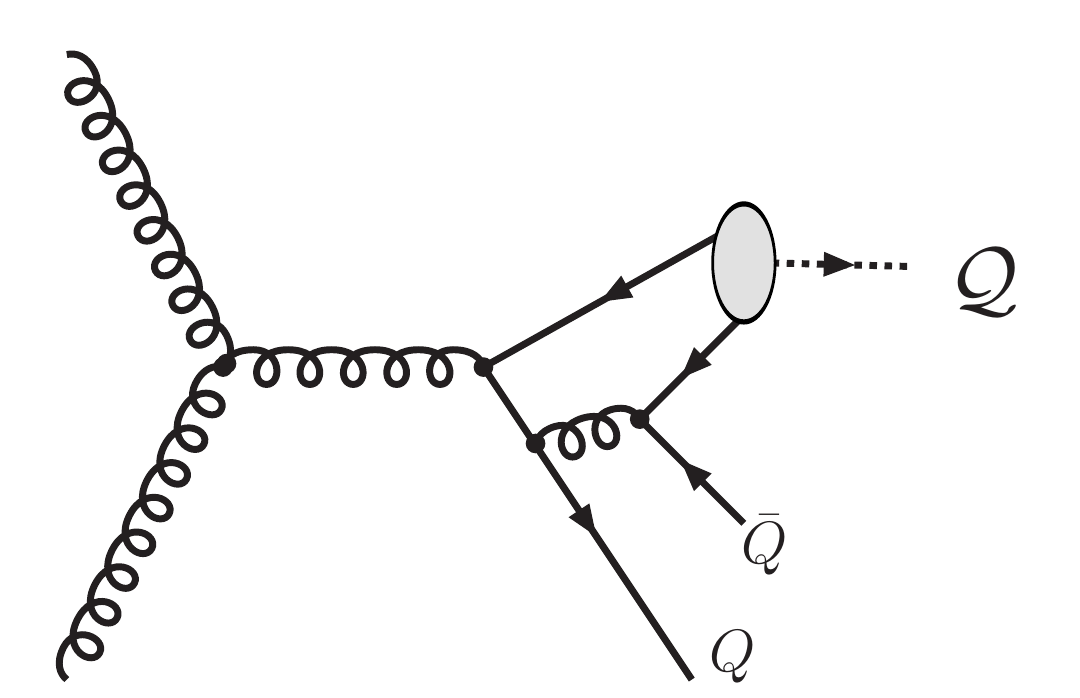}}
\subfloat[]{\includegraphics[height=2cm]{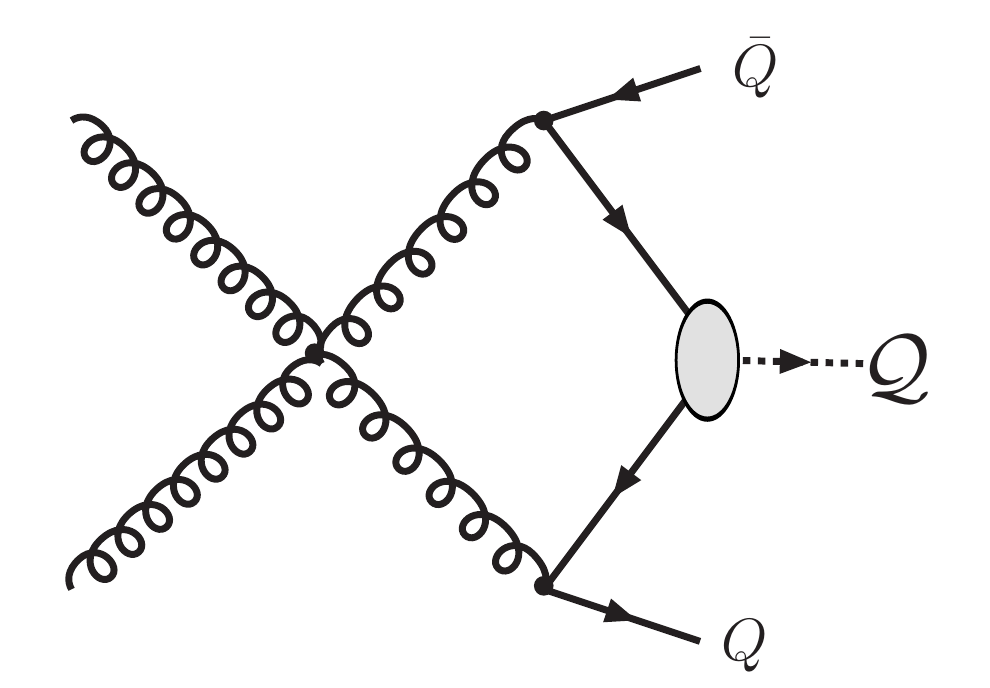}}
\caption{Representative diagrams contributing at LO in $\alpha_s$
to $gg \to {\cal Q} + Q \bar Q $ in CSM ($\Q$ standing for $\psi$).}
\label{fig:gg-Qqq}
\end{figure}

Like for open charm/bottom and quarkonium production, the current reaction is expected
to be dominated by gluon fusion.  We indeed checked that the light-quark initiated
process was suppressed by three orders of magnitude, and thus this contribution was then neglected.
The amplitude for $gg \to {\cal Q}(P) + Q \bar Q$ ($\Q$ standing for $\psi$) involves 42
Feynman diagrams (a representative selection is shown on \cf{fig:gg-Qqq}) which we computed with an early version of \MadOnia\footnote{Let us note that, as a side computation, we computed the cross section and the $P_T$ distribution for $B_c^*$ production at the Tevatron and
found it to agree with the results of Berezhnoy \etal~\cite{Berezhnoy:1996ks}.}.
The computation is quite standard and we will not repeat its description here\footnote{We simply note
that the default scale choice was taken to be $\mu_0=\sqrt{(4 m_Q)^2+P_T^2}$ and
we used NLO PDF (CTEQ6M~\cite{Pumplin:2002vw}) as this process was considered 
to be a NLO correction to the inclusive case. One could argue that using a LO PDF is also licit.
As for the other parameters, their choice is also standard: $|R_{J/\psi}(0)|^2=0.81$~GeV$^3$  and $m_c=1.5$ GeV.}.

\begin{figure}[hbt]
\centering
\subfloat[]{\includegraphics[width=0.47\textwidth]{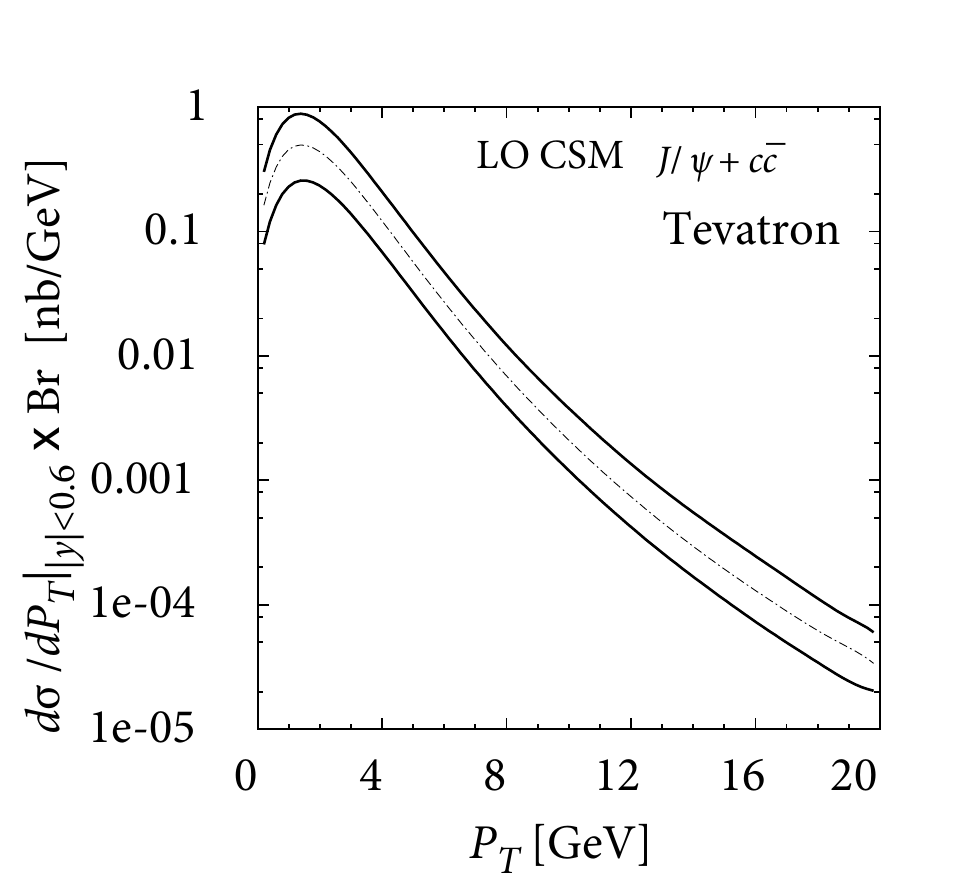}}
\quad
\subfloat[]{\includegraphics[width=0.47\textwidth]{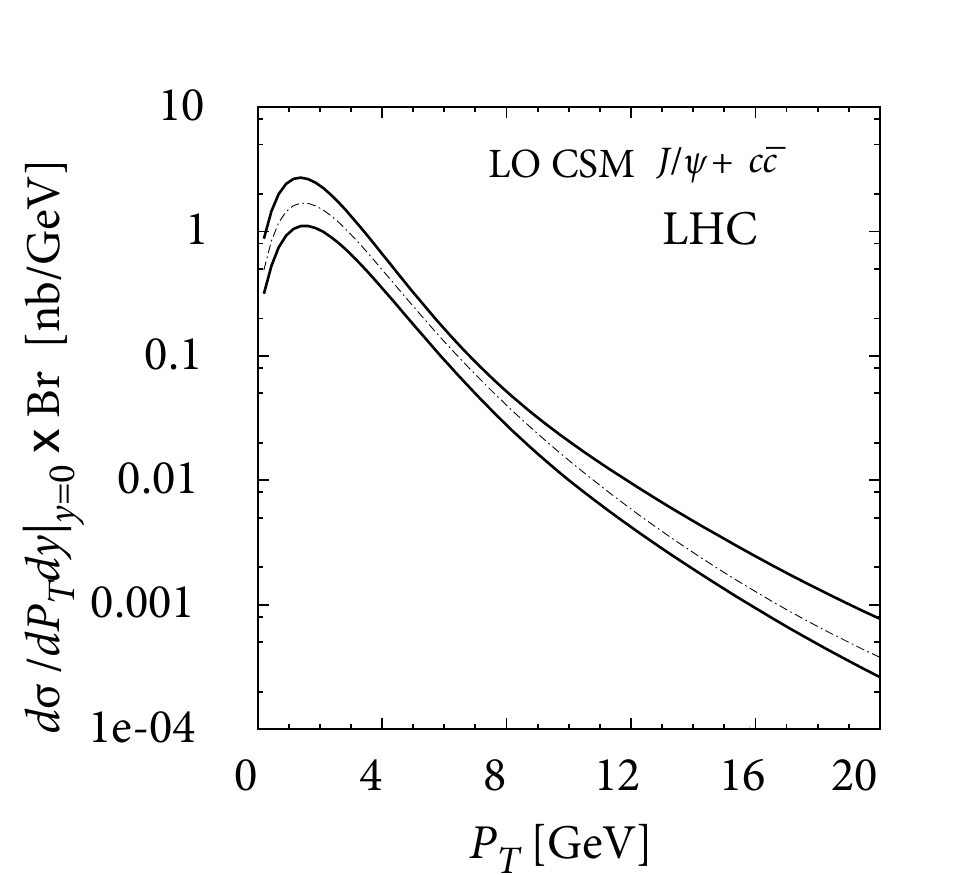}}
\caption{$P_T$-differential cross section for  $pp\to J/\psi +  c\bar c $  
at (a) the Tevatron at $\sqrt{s}=1.96$ TeV and (b)  at the LHC at $\sqrt{s}=14$ TeV.
Adapted from~\cite{Artoisenet:2007xi}.}
\label{fig:3S1QQ}
\end{figure} 

\cf{fig:3S1QQ} displays the $P_T$-differential cross section at central rapidities
at the Tevatron at $\sqrt{s}=1.96$~TeV and the LHC at $\sqrt{s}=14$~TeV. 
Considering the $P_T$-integrated cross sections, we found that analyses at the Tevatron 
(CDF and D0) were possible
for the RUN2 at the Tevatron at
$\sqrt{s}=1.96$ TeV, with a significant integrated cross-section (times the relevant branching) :
\eqs{\sigma(J/\psi +c \bar c) \times {\cal B} (\ell^+\ell^-) \simeq 0.5 \div 2~ 
\hbox{nb}.}
At the LHC, at 14 TeV, it increases to reach (for $|y_\Q|\leq 0.5$): 
\eqs{\sigma(J/\psi +c \bar c) \times {\cal B} (\ell^+\ell^-) \simeq 5\div 15~\hbox{nb}.}

To illustrate the potentialities at RHIC, we also computed~\cite{Lansberg:2008gk}
(see~\cf{fig:associated-sig-STAR})  the differential cross section for 
$pp \to J/\psi +c \bar c$  computed in the STAR kinematics for
$\sqrt{s}=200$~GeV. In this case, the $q\bar{q}$ channel may start to be relevant 
at large $P_T$  --a region which would be difficult to access, though. First studies 
could for instance be carried out by STAR with an integrated 
luminosities of around 50 pb$^{-1}$ if dedicated  triggers are 
available. 

\begin{figure}[hbt!]\centering
\subfloat[]{\includegraphics[height=6cm]{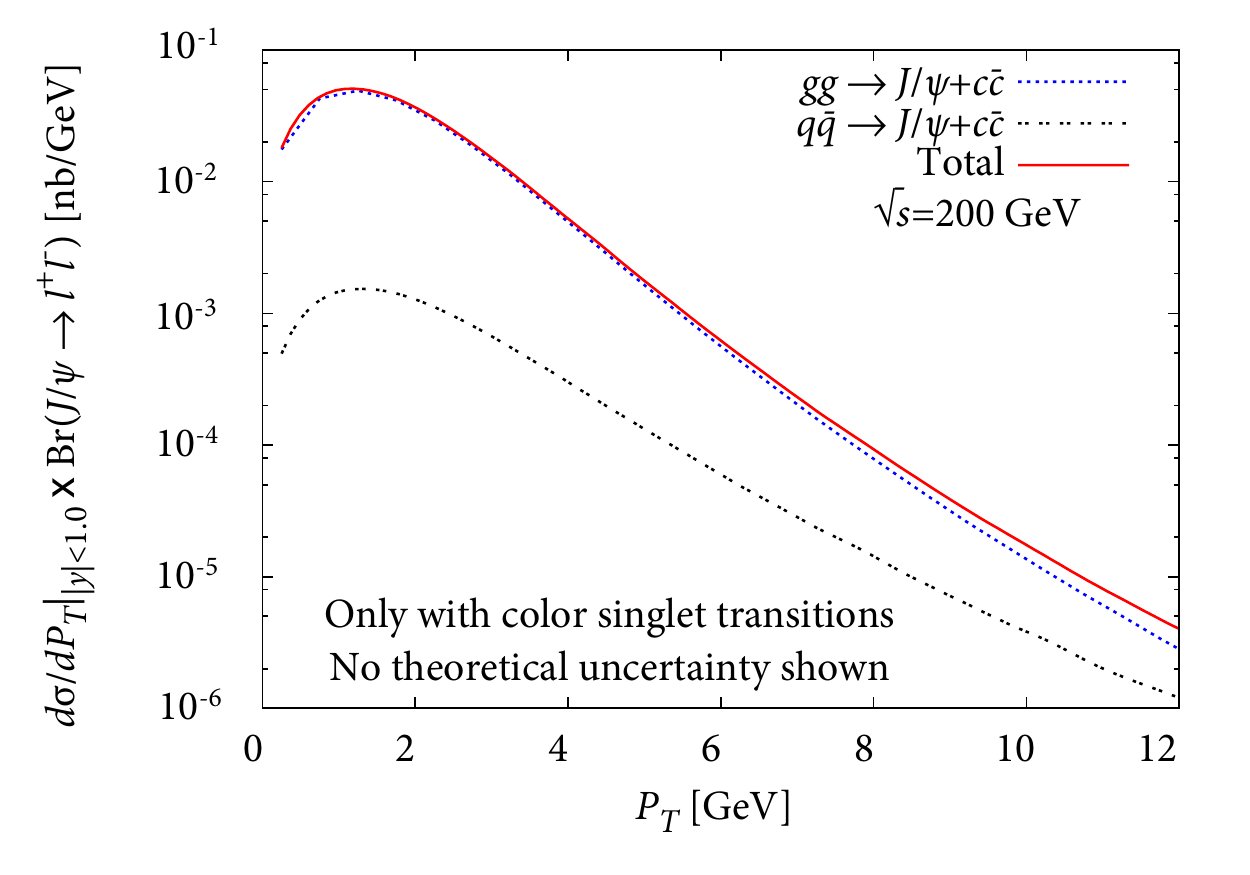}\label{fig:associated-sig-STAR}}
\subfloat[]{\includegraphics[height=6cm]{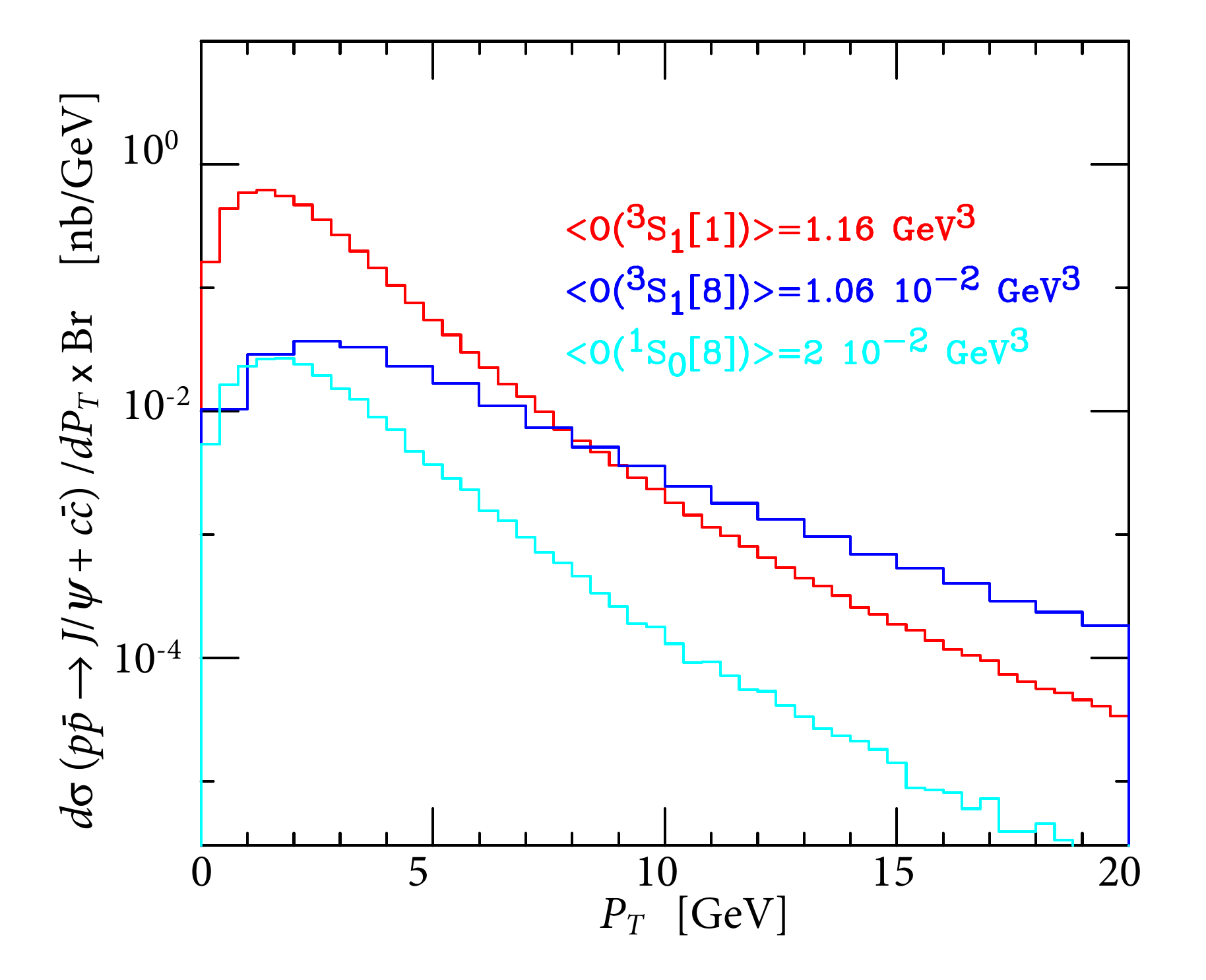}\label{fig:psi-cc-Tevatron}}
\caption{(a) CS $P_T$-differential cross section for $pp \to J/\psi +c \bar c$  for the STAR kinematics 
($\sqrt{s}=200$~GeV, $|y|\leq 1.0$); (b) $P_T$-differential cross section for $pp \to J/\psi +c \bar c$
at the Tevatron: comparison between the CS and 2 CO contributions (see text). In both cases, no theoretical uncertainties are shown. Adapted from (a)~\cite{Lansberg:2008gk} and (b) from~\cite{Artoisenet:1900zz}.} 
\end{figure}

We also compared these LO results with those calculated in the 
fragmentation approximation. The latter were found to be less than half  the 
full computation up to $P_T^{J/\psi}=20$ GeV at the Tevatron. 
Only around $P_T^{J/\psi}=80$ GeV, both computations differ by a 
little less than $10$\%. For the $\Upsilon$, we found that 
much larger values of $P_T$ are required for the approximation to be
close to the full results. To sum up, $p\bar{p}\to J/\psi + c \bar c$, which is
an NLO subset of $p\bar{p}\to J/\psi + X$, is not dominated by the
fragmentation contributions for the $P_T^{J/\psi}$-range accessible at the Tevatron
and mostly accessed at the LHC.

A tentative explanation for such a failure of the fragmentation approximation is 
the large number of Feynman diagrams among which only a few give 
rise to fragmentation topologies.
A similar situation was also found for 
$\gamma \gamma \to J/\psi c \bar{c}$~\cite{Qiao:2003ba} and
and for the $B_c^*$ hadroproduction, where
the fragmentation approximation is not accurate in the $P_T$ range
explored at the Tevatron~\cite{Chang:1994aw,Berezhnoy:1996ks}.

We also analysed the polarisation of the $J/\psi$ and $\Upsilon$ produced
along with a heavy quark pair of the same flavour and found that their azimuthal
anisotropy  in the helicity frame was null at any $P_T^{\Q}$. 
In other words, they are unpolarised.

The existence of a non-perturbative charm content of the proton~\cite{Brodsky:1980pb}, the so-called
intrinsic charm (IC),  is object
of continuous debates~\cite{Pumplin:2007wg,Kniehl:2009ar,Jimenez-Delgado:2014zga,Ball:2016neh,Brodsky:2015uwa,Dulat:2013hea} since the 1980's. In 2009, we made a first 
study~\cite{Brodsky:2009cf} to quantity its possible impact via the charm gluon fusion
in $J/\psi+D$ production, namely $cg\to J/\psi+c$ (see \cf{fig:LO-cg-CSM}),\footnote{In the presence of a genuine
IC charm, that is not from gluon splitting, the consideration of such contributions, which
normally appear in the 4-flavour scheme --meant to be used in high scale processes-- 
is justified.} in the CSM in the RHIC kinematics towards large rapidities. 
In this region, the parton momentum fraction may reach the valence region and
this reaction may be used to test different models of the charm quark PDF $c(x)$. 

\begin{figure}[hbt!]
\vspace*{-1cm}
\centering
\subfloat[]{\includegraphics[width=2.5cm]{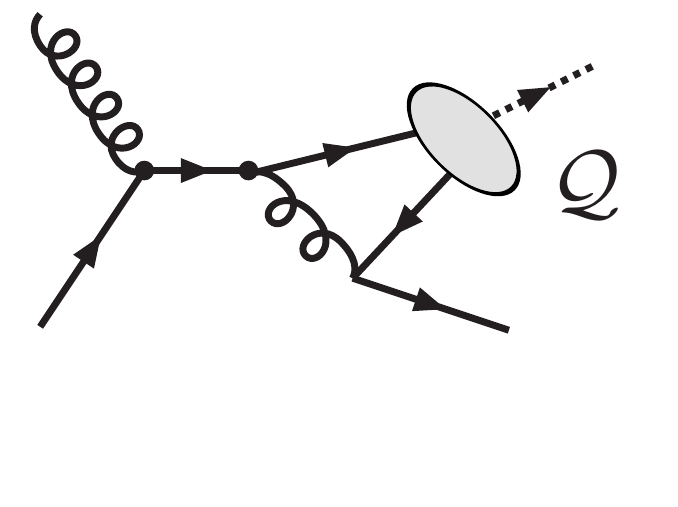}\label{fig:LO-cg-CSM}}
\subfloat[]{\includegraphics[width=2.5cm]{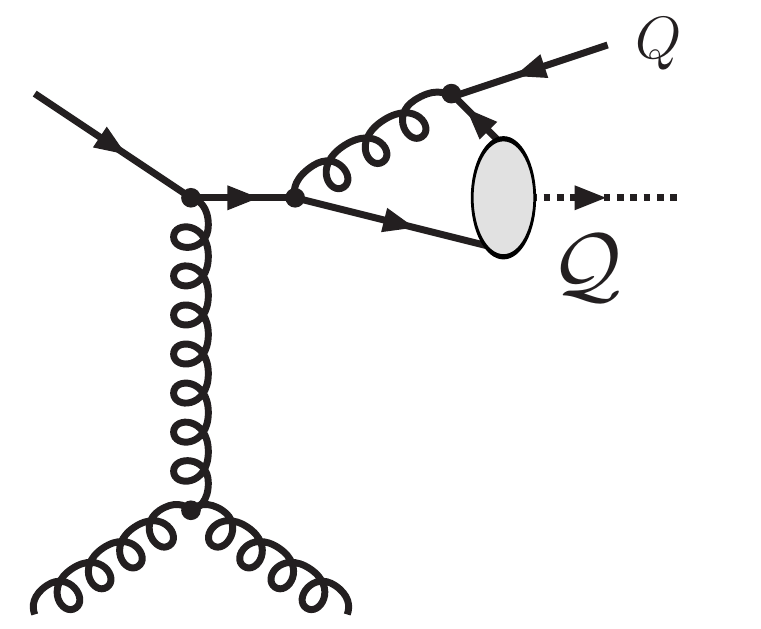}\label{fig:cg-frag}}
\subfloat[]{\includegraphics[width=2.5cm]{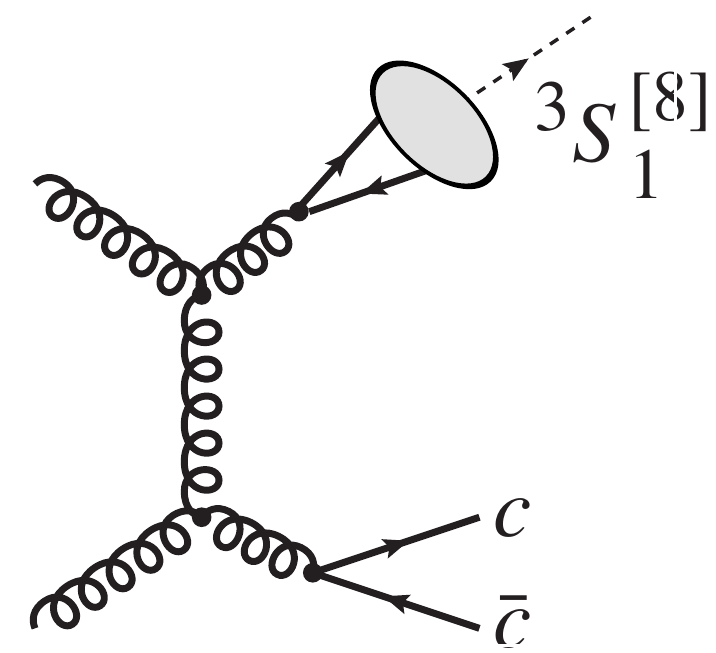}\label{fig:onium-cc-3S18}}
\subfloat[]{\includegraphics[width=2.5cm]{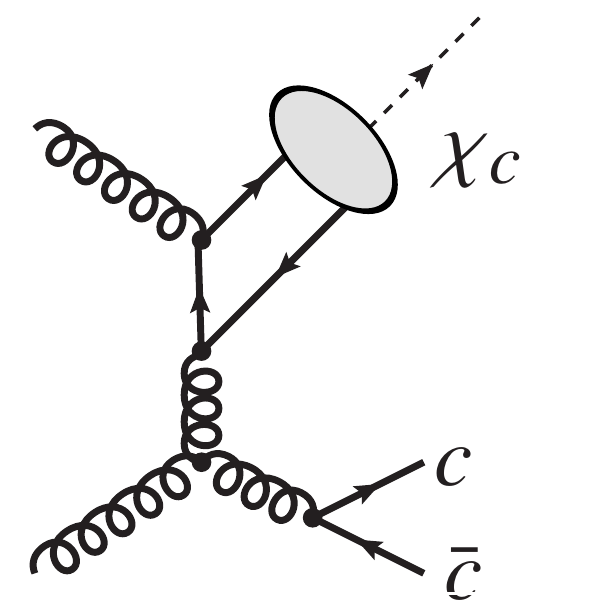}\label{fig:jj-chic-QQ-PT6}}
\caption{Representative diagrams contributing to $^3S_1$ charmonium {(denoted $\Q$)} hadroproduction in the CSM
initiated by a charm quark at orders $\alpha_S^3$ (a) and $\alpha_S^4$ (b).
(c) Gluon fragmentation diagram involved in $J/\psi+c\bar c$ and $\chi_c+c\bar c$ via a $\so$ CO channel. 
(d) One of the rare diagrams of $\chi_c+c\bar c$ which is vanishing for
$J/\psi+c\bar c$, in both cases in the CSM.
The quark and antiquark attached to the ellipsis are taken as on-shell
and their relative velocity $v$ is set to zero.}
\label{diagrams-psi-cc-2}
\end{figure}

\begin{figure}[hbt!]\centering
\subfloat[]{\includegraphics[width=.5\textwidth,angle=0]{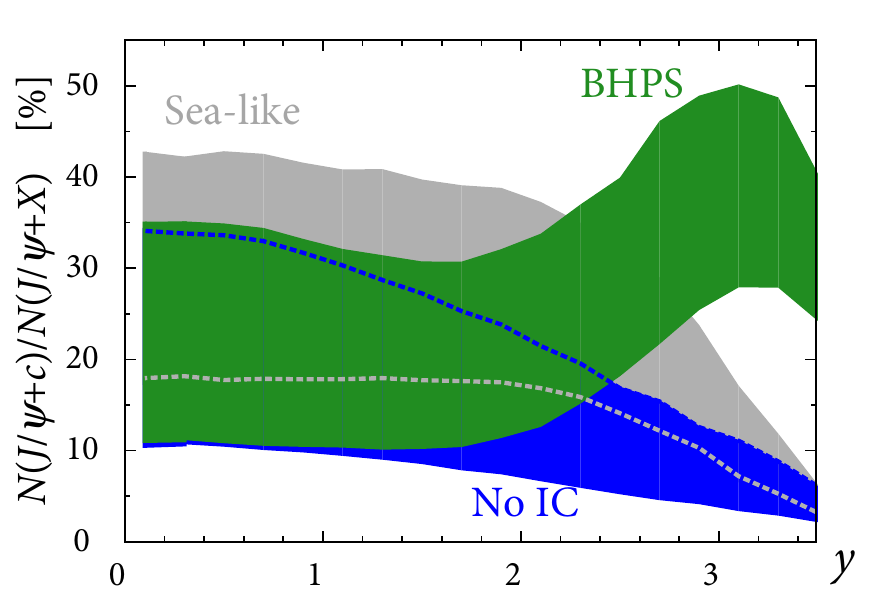}\label{fig:fractionpsic}}
\subfloat[]{{\includegraphics[width=.5\textwidth]{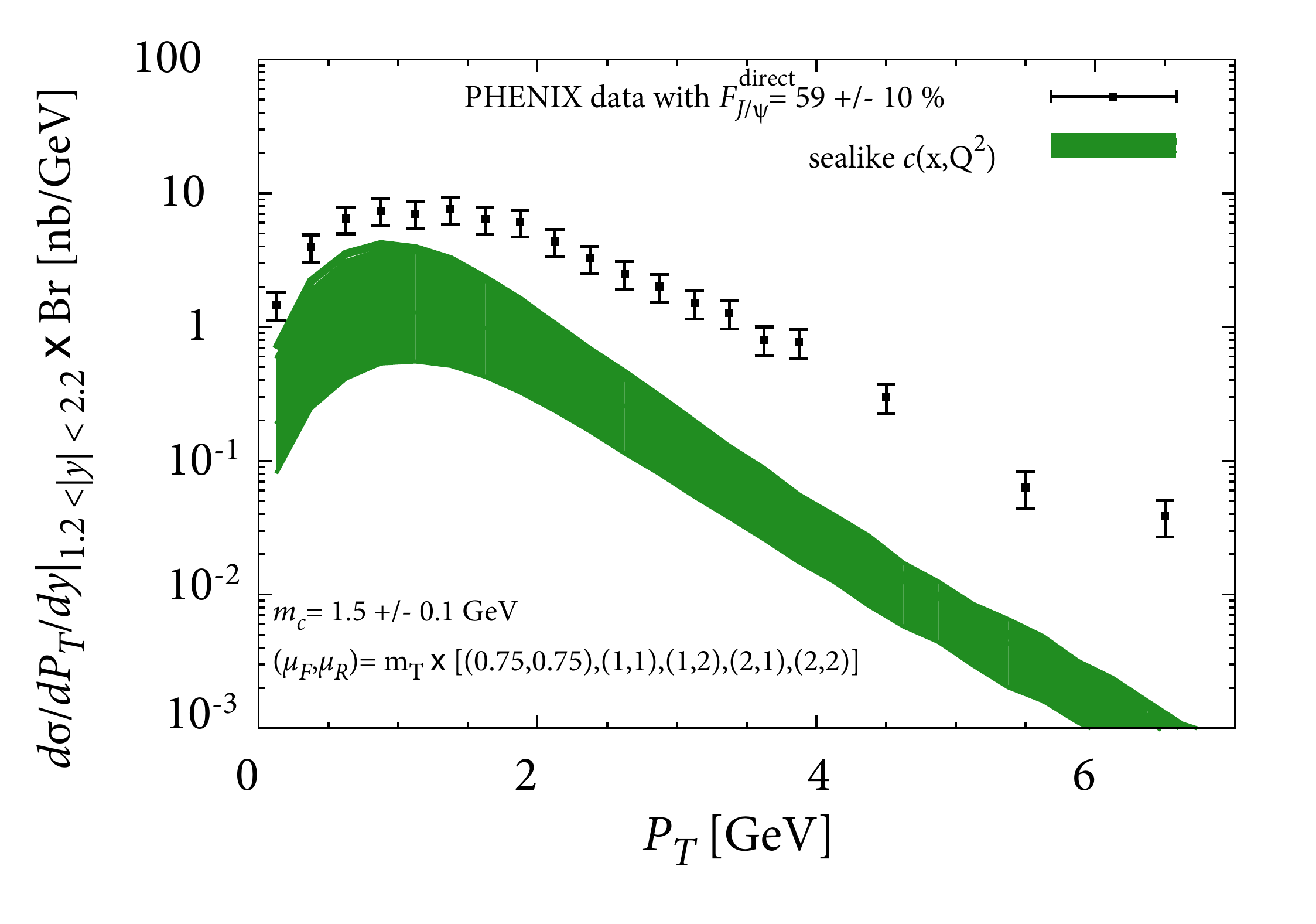}}\label{fig:plot-dsigdpt-cg}}
\caption{(a) Fraction of $J/\psi$ produced in association with a single $c$-quark {(via $gc\to J/\psi c$)} relative to the expected direct inclusive yield  as
a function of  $y_\psi$ in \pp\ at $\sqrt{s}=200\mathrm{~GeV}$ and for three models for $c(x)$: DGLAP-like, without IC (No IC), sea-like and BHPS (see details in~\cite{Brodsky:2009cf}). (b) $d\sigma/dP_T/dy \times {\rm Br}$ from $cg$ fusion in \pp\ at $\sqrt{s}=200\mathrm{~GeV}$ using a sealike charm distribution at forward rapidites compared to the PHENIX~\cite{Adare:2006kf}
data. The theoretical-error bands 
come from combining the uncertainties resulting from the choice of $\mu_F$, $\mu_R$, $m_c$ 
(see~\cite{Lansberg:2010vq}). Adapted from~\cite{Lansberg:2010vq}.}
\end{figure}

Since the $P_T$-integrated NLO (and LO) CSM computations were in agreement with 
the PHENIX data~\cite{Adare:2006kf}, we managed to evaluate the fraction of $J/\psi$ produced 
in association with a single $c$-quark relative to the expected direct yield  as
a function of  $y_\psi$ and for three models of $c(x)$ as encoded 
in {\small CTEQ6.5c}~\cite{Pumplin:2007wg}: (i) without IC {($c(x,\mu_0)=0$ at $\mu_0=$1.2 GeV)}, (ii) with  BHPS IC~\cite{Brodsky:1980pb} ($\langle x \rangle_{c+\bar c}{\equiv\int^1_0  x [c(x)+\bar c(x)] dx=}2\%$) and (iii) with sea-like IC ($\langle x \rangle_{c+\bar c}=2.4\%$).

Our computation clearly confirmed the impact of the $cg$ contribution, from 10 \% up to 45\%
of the direct yield in the case of sea-like $c(x)$.
We also noted that  at larger $P_T$, a significant $\alpha_s^4$ contributions from $cg$ fusion
could also be expected, along with fragmentation-like topology (\cf{fig:cg-frag}). 
Such contribution has been studied by Qiao~\cite{Qiao:2002nd} in 2003 for the Tevatron kinematics
using a DGLAP-like $c$-quark distribution. For the BHPS IC distribution, the  
$P_T$-differential cross section for single $J/\psi$ distribution at large $P_T$ 
and RHIC energy would be expected to show an analogous enhancement as seen at large rapidities in~\cf{fig:fractionpsic}. However, as we showed in 2010~\cite{Lansberg:2010vq}, this occurs 
at significantly higher $P_T$ (see~\cf{fig:plot-dsigdpt-cg}) than where the cross section
can be measured.

In order to  experimentally assess the importance of such $cg$ fusion, 
the measurement of $J/\psi$ in association
with a $D$ meson would be very interesting as noted above. At RHIC or the LHC, 
another accessible observable would the azimuthal correlation 
of $J/\psi + e$ in the central barrels of the ALICE, PHENIX and STAR detectors and
of $J/\psi + \mu$ in the forward region by the PHENIX and the ALICE muon arms. 
At low $P_T$, the key signature for such subprocesses would be the observation a lepton excess
 opposite in azimuthal angle $\phi$ to the detected $J/\psi$. 

In~\cite{Artoisenet:1900zz}, Artoisenet made a brief first survey of the possible impact of the CO contributions to
$pp \to J/\psi +c \bar c$ at the Tevatron at LO ($\alpha_s^4$). At this order, one encounters
the same graphs as those depicted on \cf{fig:gg-Qqq} for the CSM. Those for which the quarkonium is produced
by two heavy-quark lines contribute to the 3 dominant CO transitions\footnote{In~\cite{Artoisenet:1900zz}, $\mopj$ was assumed to zero. For a LO computation, this does not induce any loss of generality.}, namely $\so$, $\sps$, $\pj$. For those where the quarkonium is produced by a single heavy-quark line, \eg\ \cf{fig:OniumQQ-graph7b}, only $\so$ contributes, but they are not part of the leading-$P_T$ contributions. However, for $\so$, there are topologies where 2 gluons fuse into 2 gluons and each fragments into a $Q\bar{Q}$ pair (see \cf{fig:onium-cc-3S18}) .
They are particularly relevant since scaling as $P_T^{-4}$.

With $\mopb=1.06 \times 10^{-2}$ GeV$^3$ --which is admittedly large-- and 
$\mops=2 \times 10^{-2}$ GeV$^3$, he found the $P_T$-differential cross sections 
displayed on \cf{fig:psi-cc-Tevatron}. At low $P_T$, the CS contribution is dominant. 
As expected for a nearly pure $P_T^{-4}$
channel, the $\so$ contribution is the hardest and overshoots the CS contributions
at $P_T^{J/\psi}\simeq 7$~GeV. However, for $\mopb$ on the order of $10^{-3}$ GeV$^3$, 
the CS yield remains dominant certainly until 20 GeV. On the contrary, the $\sps$
contribution always remains more than one order of magnitude below the CS yield, 
with a similar $P_T$ scaling. A first complete NLO NRQCD analysis is eagerly awaited for.

Artoisenet also analysed the polarisation of the $J/\psi$, which he found --without any surprise--
to be transversely polarised as expected for a $\so$ gluon fragmentation channel.
He also analysed~\cite{Artoisenet:1900zz} in detail the angular separation 
between the $J/\psi$ and a charm quark. A simple picture clearly emerges
at large $P_T^{J/\psi}$: in the CSM, one charm is produced near the $J/\psi$, another
recoils on the $J/\psi+c$ system; in the COM, the $J/\psi$ recoils on the charm-quark pair.

As usual for the CS channels, the FD from radially excited states $^3S_1$ 
can be taken into account by constant multiplicative factors : 
$\sim 1.4$ for the $\psi(2S)$ into $J/\psi$,  $\sim 1.1$ for both the $\Upsilon(2S)$ into $\Upsilon(1S)$ and 
$\Upsilon(3S)$ into $\Upsilon(2S)$, provided that one neglects
the mass difference in the decay kinematics.

As for the $P$-wave FD, a dedicated LO study~\cite{Li:2011yc} of $\chi_c+c \bar c$ was carried by 
Li, Ma and Chao in 2011, taking into account both CS and CO contributions, for the
Tevatron and LHC kinematics. One of their motivations was to see whether $\chi_c+c \bar c$
could be a significant source of inclusive $\chi_c$ and could alter the relative
production rate $\sigma_{\chi_{c2}}/\sigma_{\chi_{c1}}$.

They showed that whereas the LO CS contributions indeed include the
quark-fragmentation topology, the approximated scaling of the CS yield
is close to $P_T^{-6}$ for $P_T^{\chi_c}\lesssim 20$ GeV and only
reaches the $P_T^{-4}$ scaling at significantly higher $P_T^{\chi_c}$. This is not the case
of the $\so$ CO contribution --similar to that contribution to $J/\psi+c \bar c$--
which scales like $P_T^{-4}$ even at moderate $P_T^{\chi_c}$. In other words,
the CO contributions exhibit a harder spectrum and are expected to dominate at high enough 
 $P_T^{\chi_c}$. In practice, it takes over the CS yield at $P_T^{\chi_c}\simeq 7$ GeV
with ${\langle\mathcal{O}^{\chi_{cJ}}(\bigl.^3\hspace{-.5mm}P_J^{[8]})\rangle}=(2J+1)\times
2.2  \times 10^{-3}$~GeV$^3$ and 
${\langle\mathcal{O}^{\chi_{cJ}}(\bigl.^3\hspace{-.5mm}P_J^{[1]})\rangle}=
\frac{(2J+1)3 N_C}{2\pi} |R'_P(0)|^2= \frac{(2J+1)3 N_C}{2\pi} \times 0.075$~GeV$^5$.
Finally, let us note that the $P_T$ spectrum of the $J/\psi+c \bar c$
and $\chi_c+c \bar c$ yields is essentially the same up to an overall factor.
This is not surprising since the topologies are the same, except for that depicted
on~\cf{fig:jj-chic-QQ-PT6}. 

As what regards the FD fraction, with the same parameter values and
assuming $P_T^{\chi_c} \simeq P_T^{J/\psi}$, 
$\sum_J d\sigma(\chi_{cJ}+c \bar c)/dP_T\times \Br(\chi_{cJ}\to J/\psi \gamma)$ 
was found to be roughly equal to $d\sigma(J/\psi+c \bar c)/dP_T$ for
$P_T^{J/\psi} \gtrsim 10$~GeV. We note that this statement
however heavily relies on the value of 
${\langle\mathcal{O}^{\chi_{cJ}}(\bigl.^3\hspace{-0.5mm}S_1^{[8]})\rangle}$.

In 2012, LHCb made the first observation~\cite{Aaij:2012dz} of prompt
$J/\psi$ produced in association with another charm hadron ($C$ : $D^0$, $D^+$, $D^+_s$
and $\Lambda_c^+$) with 355 pb$^{-1}$ of data taken at $\sqrt{s}=7$~TeV. Their measurement
covered the range of $P_T^{\psi}$ from 0 up to 12 GeV and for the charmed hadrons 
from 3 up to 12 GeV, both in the rapidity range $2<y<4$.

Using a Tevatron multi-jet event extraction of $\sigma_{\rm eff}$ of 14.5 mb and the pocket formula,
\begin{eqnarray}
\sigma^{\rm DPS}(J/\psi+C)=\frac{\sigma(J/\psi)\sigma(C)}{\sigma_{\rm eff}}.
\end{eqnarray}
they evaluated the expected DPS contribution.  Since the different cases yield 
 similar conclusions, we will restrict the discussion to the most precise measurement, 
\ie\ $J/\psi+D^0$, for which their measured $P_T$ integrated cross section (without 
branching) amounted to $161.0 \pm 3.7 \pm 12.2$ nb with close to 5000 events.

As a comparison, the expected DPS cross section is $146 \pm  39$ nb and that 
from the SPS expectations (LO CS from \MadOnia~\cite{Artoisenet:2007xi} + Pythia or $k_T$ factorisation~\cite{Baranov:2006dh}) 
were found to be significantly smaller on the order of 5 to 15 nb. Even though
NLO corrections may be significant because of the presence of a $P_T$ cut on the
$D$ and not on $J/\psi$, it seems clear that the DPS is dominant. In view of
the previous discussions (di-$\psi$, $J/\psi+\Upsilon$, $J/\psi+Z$ and $J/\psi+W$), 
it may seem normal. At the time, it was the first
of the quarkonium associated-production measurements and it was a surprise.
Except for a visible difference between the $P_T^{\jpsi}$ distribution of the
$J/\psi$ with a $D$ and the inclusive ones (see \cf{dsigdpt-psi-charm-LHCb}), all the kinematical distributions
seem in accordance with a yield strongly dominated by DPS. 

\begin{figure}[hbt!]
\centering
\subfloat{\includegraphics[scale=.33,draft=false]{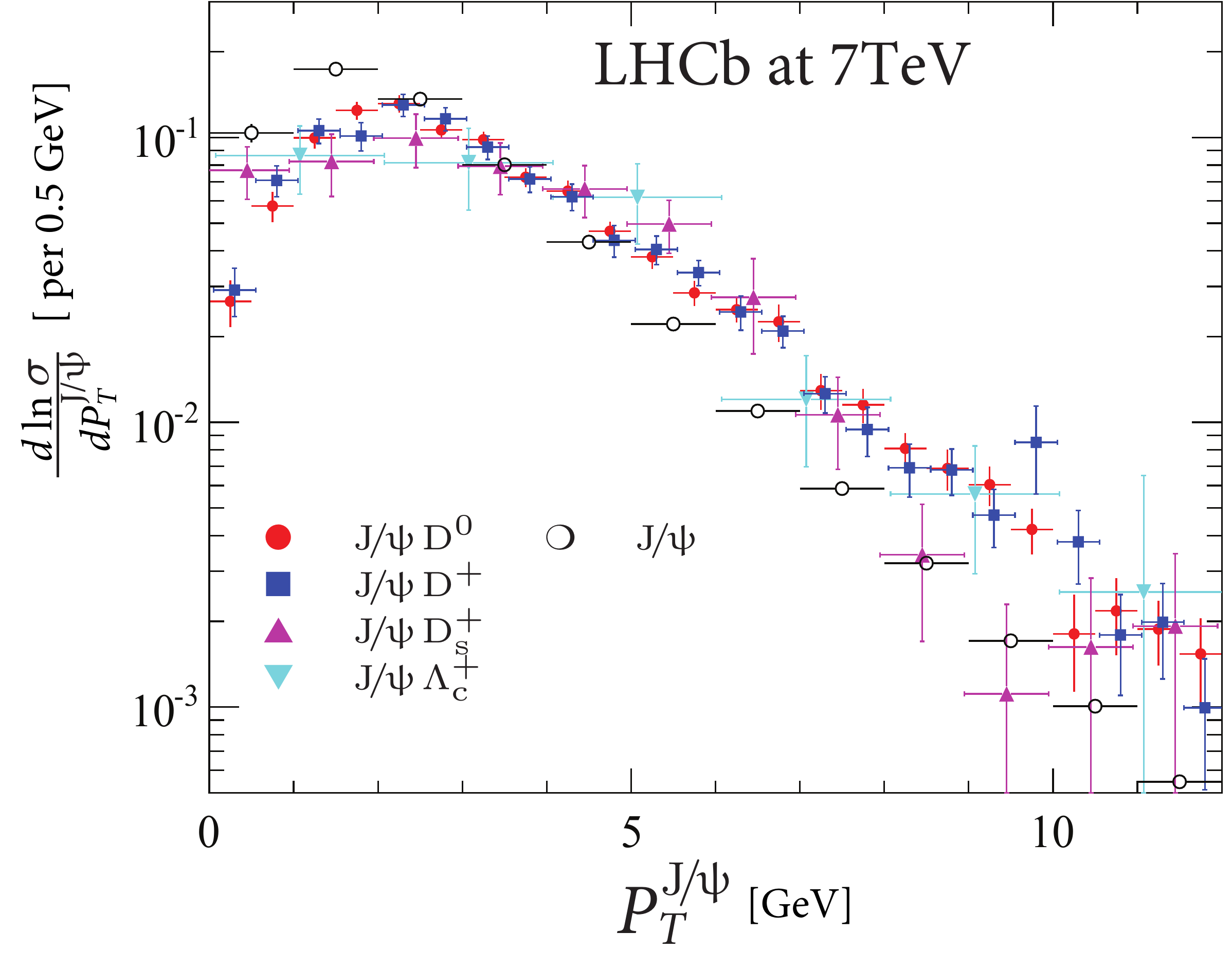}}
\caption{Self-normalised $P_T^{\jpsi}$ differential cross section, $1/\sigma  \times d\sigma/dP_T^{\jpsi}$ for $\jpsi$ produced in association with another charmed hadron compared to the inclusive case (in open black circle) as measured
by LHCb at $\sqrt{s}=7$~TeV. Adapted from~\cite{Aaij:2012dz}.}
\label{dsigdpt-psi-charm-LHCb}
\end{figure}

\paragraph{Production at $B$ factories.}
Associated production of a $J/\psi$ and charm has first been investigated by Clavelli~\cite{Clavelli:1982hp} in 1982. 
Such a channel was then recognised as the potential largest CSM contribution for the CLEO kinematics~\cite{Kiselev:1994pu,Cho:1996cg} before the advent of $B$ factories. 

The whole scene drastically changed when in 2002 Belle reported cross sections significantly larger than these expectations. Once it was found~\cite{Liu:2003jj} out that the COM could not fill the gap, an intense
experimental and theory activity started (see~\eg~\cite{Berezhnoy:2003hz,Liu:2003zr,Qiao:2003ba}). In 2009, thanks to a data sample 30 times larger than in 2002, Belle confirmed its finding and reported~\cite{Pakhlov:2009nj} a prompt cross section of 
\eqs{
\sigma^{\text{ exp.}}(J/\psi+c \bar c)=0.74\pm 0.08^{+0.09}_{-0.08} \text{ pb}.
}
$\alpha_s^3$ (NLO) CS corrections were computed in 2006 by Zhang \etal\ \cite{Zhang:2006ay} and their results were confirmed by Gong \& Wang in 2009~\cite{Gong:2009ng} with a fine analysis of the theoretical uncertainties. $K$ factors on the order of $1.4 \div 1.7$ were found. Relativistic corrections~\cite{He:2007te} were computed as well as QED ISR effects~\cite{Shao:2014rwa}, but were found to be negligible as opposed to the $J/\psi+X_{\text{non } c \bar c}$ case.
Using a scale choice inspired by the BLM procedure~\cite{Brodsky:1982gc}, a CS NLO cross section of 0.73 pb was reported by Gong \& Wang~\cite{Gong:2009ng} which is in perfect agreement with the Belle measurement. We note that this corresponds to a rather low scale choice, $\mu^*=1.6$~GeV, which may lead us to reconsider other scale choices in other quarkonium-production channels.  We also note that the momentum dependence of the NLO CSM~\cite{Gong:2009ng} agrees with that of the data (see~\cf{dsigma_dpstar-ee_psicc-Wang}). 

\begin{figure}[hbt!]
\centering
\subfloat{\includegraphics[scale=.33,draft=false]{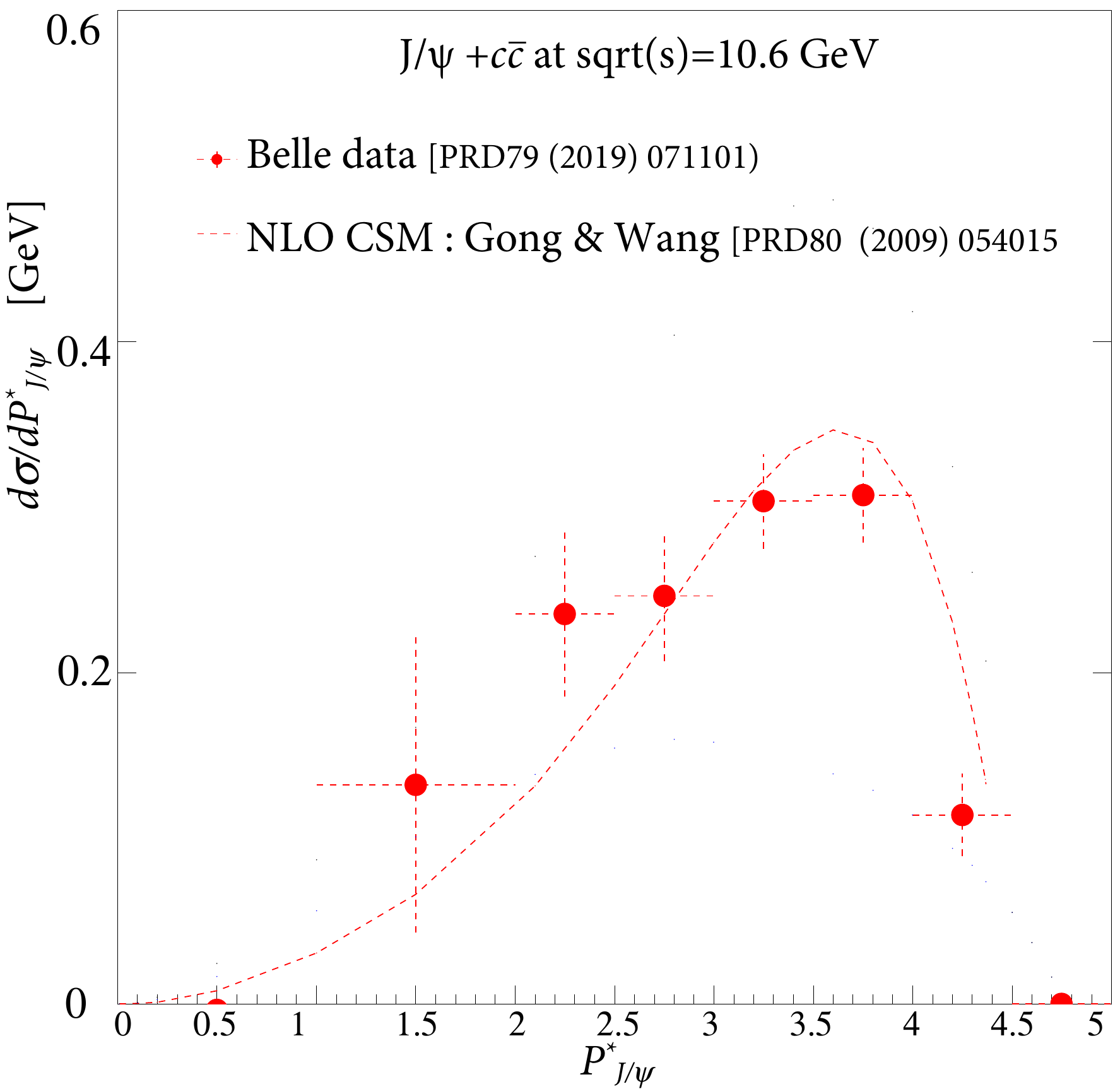}}
\caption{$P^\star_{\jpsi}$ differential cross section for $\jpsi+c\bar c$ production in $e^+e^-$ annihilation at $\sqrt{s}=10.6$~GeV : comparison between the Belle data~\cite{Pakhlov:2009nj} and the NLO CSM predictions~\cite{Gong:2009ng}. Adapted from~\cite{Gong:2009ng}.}
\label{dsigma_dpstar-ee_psicc-Wang}
\end{figure}

As one can imagine the theory uncertainties remain large and it is not clear whether the observed agreement rules out the relevance of the CTs suggested by Nayak \etal\ \cite{Nayak:2007mb,Nayak:2007zb}. Belle-II could certainly contribute to a better understanding of this associated-production channel by looking at the $\psi(2S)$ and the $\chi_c$ produced along with a charm pair. Finally, we note that the only existing CEM evaluation~\cite{Kang:2004zj} points at a cross section as low as 0.05 pb thus 15 times smaller than the Belle value.

\begin{figure}[hbt!]
\centering
\subfloat{\includegraphics[width=0.47\textwidth]{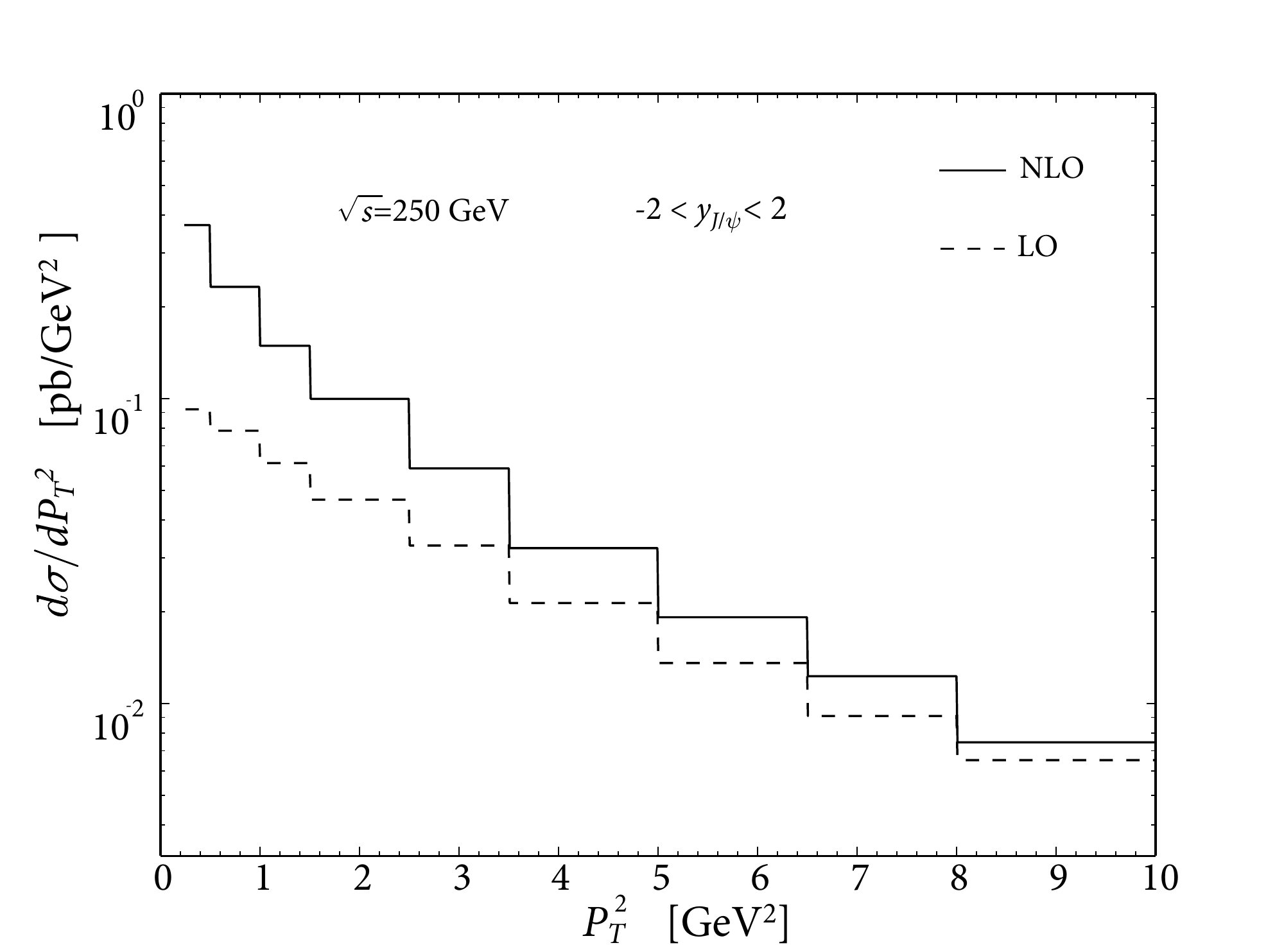}}
\caption{$P^2_T$-differential cross section for (unresolved) $\gamma\gamma\to J/\psi + c\bar{c}$ production predicted by the CSM at LO and NLO at a future CEPC at $\sqrt{s}=250$ GeV (see~\cite{Chen:2016hju} for the theory parameters and the kinematical cuts).
Adapted from~\cite{Chen:2016hju}.
}
\label{fig:ptCEPC-gammagamma-psicc}
\end{figure}

Still in $e^+e^-$ annihilation, associated production of $J/\psi$ and charm has been studied in $\gamma\gamma$ fusion, even up to NLO accuracy. In 2009, Li and Chao~\cite{Li:2009fd} performed a complete study of the possible contributions from $J/\psi + c \bar c$ and found out that it could be the leading CS contribution of all the inclusive channels. As such it gives hope that this channel could be studied on its own.  In 2016, Chen \etal~\cite{Chen:2016hju} advanced the study of this channel with a NLO study and performed predictions for a future Circular Electron-Positron Collider for which they found a $K$ factor of 1.76 (see \cf{fig:ptCEPC-gammagamma-psicc}).

\paragraph{Photoproduction}

The photoproduction of a $J/\psi$ associated with a charm was discussed as early as in 1982 by Berger and Jone~\cite{Berger:1982fh}. At the time, they considered the partonic process $\gamma c \to J/\psi c$ which is directly sensitive to the charm content of the proton. They found out that the scattering amplitudes squared for $\gamma c \to J/\psi c$ and $\gamma g \to J/\psi g$ were similar. As such, it remains a very interesting observable to be studied at an EIC where the c.m.s. energy will be lower than at HERA, thus at higher $x$. 

Similar associated-production channels may also show up at large enough $P_T$ since they include fragmentation topologies
with a hard $P_T^{-4}$ spectrum. It was shown by Godbole \etal\ \cite{Godbole:1995ie} in 1996 that the charm fragmentation contribution was taking over the gluon fragmentation for $P_T> 10$ GeV (see also~\cite{Kniehl:1997gh}). It is understood that charm fragmentation means that another charm pair is produced and such a signature can thus be searched for. We note that, if relevant, CTs~\cite{Nayak:2007mb,Nayak:2007zb} may enhance the signal for configurations where the $J/\psi$ is near another charm quark.

\subsubsection{Bottomonium + bottom} 

Unsurprisingly, the production of bottomonia in association with a $b$ quark
is even less studied. Experimentally, it can only be accessed at hadron colliders with sufficient energies. 
Yet, along with our LO CS study of $J/\psi+c\bar{c}$~\cite{Artoisenet:2007xi}, 
we also performed that of $\Upsilon+b\bar{b}$. Since the computations exactly follows 
the same lines as above, we do not repeat its description here and directly quote
our results, which still constitute the state-of-the-art computation for this process, with the absence of
NLO computations.

Using $|R_{\Upsilon(1S)}(0)|^2=6.48$ GeV$^3$ and $m_b=4.75$ GeV, we found 
that, at $\sqrt{s}=1.96$ TeV and central rapidities, the $P_T$-integrated cross section (times the relevant branching) was on the order of a pb, namely
\eqs{\sigma(\Upsilon +b \bar b)\times {\cal B} (\Upsilon \to \ell^+\ell^-) \simeq  0.5\div 1.5 
\hbox{ pb}}
and at the LHC, at 14 TeV, it was about 10 times larger  (for $|y_\Q|\leq 0.5$): 
\eqs{
\sigma(\Upsilon +b \bar b)\times {\cal B} (\Upsilon \to \ell^+\ell^-) \simeq  5 \div 10
\hbox{ pb}}

As what concerns the $P_T$-differential cross section, it is shown on~\cf{fig:3S1bb}.
Its impact as an NLO correction to single-$\Upsilon$ production is expected to be minor.

\begin{figure}[hbt!]
\centering
\subfloat[]{\includegraphics[width=0.47\textwidth]{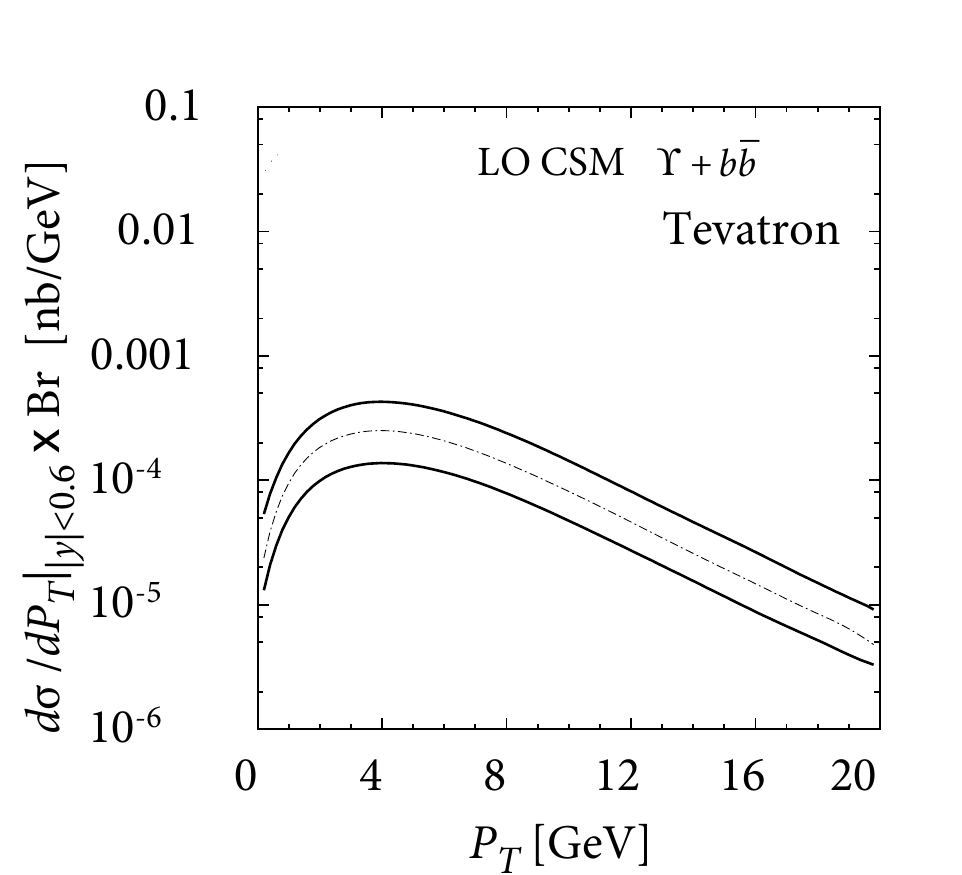}}
\quad
\subfloat[]{\includegraphics[width=0.47\textwidth]{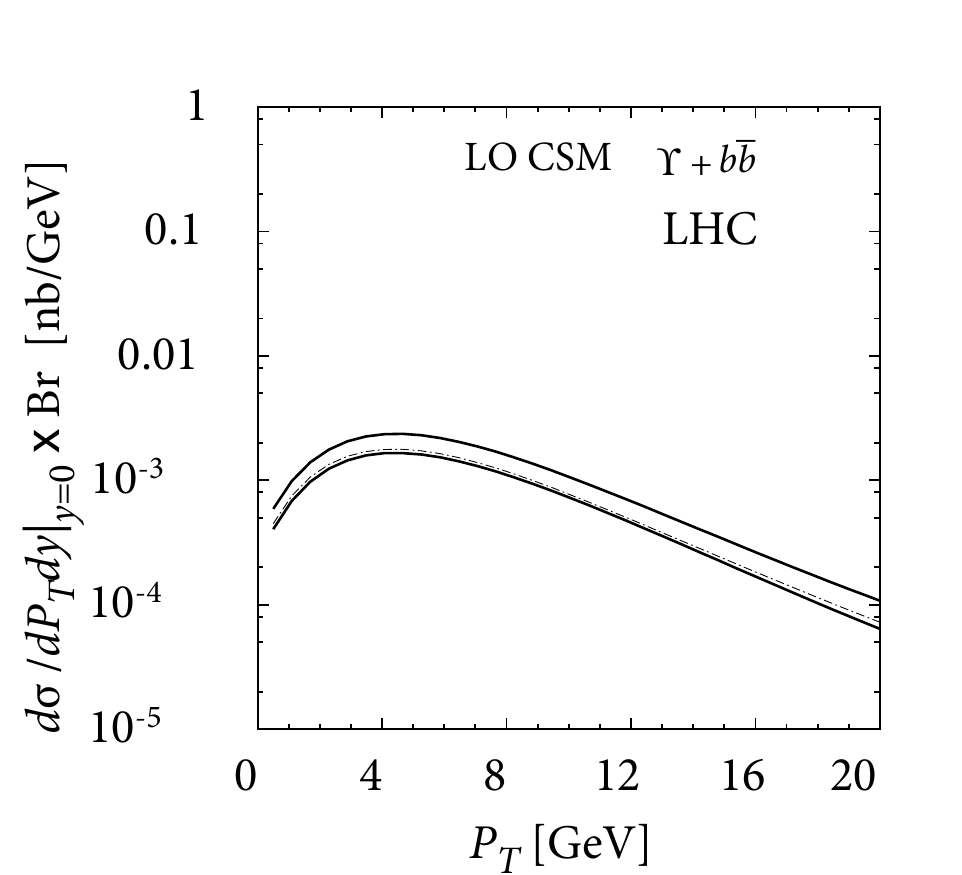}}
\caption{$P_T$-differential cross section for $pp\to \Upsilon + b\bar{b}$ production 
(a) at the Tevatron at $\sqrt{s}=1.96$~TeV and (b) at the LHC at $\sqrt{s}=14$ TeV.
 Adapted from~\cite{Artoisenet:2007xi}.
}
\label{fig:3S1bb}
\end{figure}

The other study that we are aware of is that of Gang~\etal~\cite{Gang:2012js} who 
studied the $gb$ fusion (thus in a five flavour scheme), both for the CS and the CO channels,
into $\Upsilon+b$ and $\chi_b+b$. In principle, taking a $b$ quark as an active parton 
in the proton is meant to account for the resummation of logarithms of $m_b^2/\mu_F^2$
through the evolution of the PDFs at scales $\mu_F$ much larger than $m_b$-- except if one considers
specific contributions with intrinsic bottom~\cite{Pumplin:2005yf,Lyonnet:2015dca}. As such these results should 
mostly be relevant at large $P_T$. However, at this order, the CS contributions
are probably not reliably accounted for since  the $gb \to \Upsilon+b+g$ contributions 
were not considered. 

It however seems  that the computation confirmed the naive expectation, based on the extrapolation of the
$\psi+c+\bar{c}$ case, that the $\Upsilon+b+\bar{b}$ yield should be dominated by CS channels.
The case of $\chi_b+b$ also seems to follow that of $\chi_c+c+\bar{c}$. Given the less precise
determination of the bottomonium LDMEs, their more complex feed-down pattern and
the aforementioned caveat regarding the flavour scheme used, we considered that
further studies, even at LO, are needed to draw stronger conclusions. 

So far, $\Upsilon+\text{bottom}$ has not  been the object of a published
experimental analysis. However, in their $\Upsilon+J/\psi$ analysis~\cite{Abazov:2015fbl}, D0 necessarily
tackled $\Upsilon+ \text{non-prompt } J/\psi$ events which a very reliable proxy
to this channel with an overall branching of $2\%$. Extracting out the corresponding
cross section is nevertheless admittedly a little more complex with a larger background since both muon pairs
cannot be required to emerge from the same vertex.

\subsubsection{Bottomonium + charm} 

Unlike $\Upsilon+\text{bottom}$ associated production, experimental data have already released for
$\Upsilon+\text{charm}$ by LHCb~\cite{Aaij:2015wpa} in 2015. They concluded that their
yield was likely strongly dominated by DPS.
This was expected in the sole theoretical study~\cite{Berezhnoy:2015jga}, where 
it was estimated that the DPS yield was at least one order of magnitude larger
that the SPS yield. 

Soon after the data were published, Likhoded \etal~\cite{Likhoded:2015fdr} carried out a 
more detailed study
on the possibility for a large feed-down of $\chi_b+c+\bar{c}$ which is formally appearing
at one order less in $\alpha_s$ in the CSM, which they assumed to be dominant in analogy
the case of inclusive $\chi_b$ production. They however noted that this channel could contribute
at most 1 or 2 \% of the observed cross section by LHCb.  

Although there are few doubts that the dominance of the DPS on this channel
could be questionned, complete CS and CO computations --even at LO-- would be welcome. 
Let us conclude by mentioning that, under the hypothesis of this dominance, 
LHCb extracted the following value of $\sigma_{\rm eff}$
\eqs{\sigma_{\rm eff}=&19.4 \pm 2.6 \text{(stat)} \pm 1.3 \text{(syst) mb, based on the } \Upsilon(1S)+D^0 \text{ sample,}\\
\sigma_{\rm eff}=&15.2 \pm 3.6 \text{(stat)} \pm 1.5 \text{(syst) mb, based on the } \Upsilon(1S)+D^+ \text{ sample,}
}
in line with their $J/\psi+$ charm analysis discussed above. Such values are admittedly significant smaller than those extracted from the di-$J/\psi$ ATLAS and CMS samples and the $J/\psi+Z\ \&\ W$ ATLAS samples. It is probably a little early to claim for an observation of a violation of $\sigma_{\rm eff}$ universality, but we note that, as far as quarkonium measurements are concerned, those in the forward region hint at larger $\sigma_{\rm eff}$ values, thus smaller DPS contributions, than those in the central rapidity region.

\subsubsection{Charmonium + bottom} 

The case of charmonium + bottom associated production has also seldom been 
studied. In fact, we are only aware of a single study, by 
Gang~\etal~\cite{Gang:2012js} in 2012. Just like the production of a pair of 
$\Upsilon$ and $J/\psi$ such a channel is interesting as the CS contributions
appear at higher orders in $\alpha_s$ than the CO channels and via topologies
which are a priori not favoured. Strangely enough, it shares a number of similarities
with the inclusive $J/\psi$ production case.

Still relying on the sole consideration of the $gb$ fusion processes, they computed the
production cross section for $J/\psi+b$ and $\chi_c+b$. As just mentioned,
the current process looks very similar to the single $J/\psi$ production case, for instance 
as far as the topologies of CO contributions are involved. Hence, 
it is not surprising that they observed a similar pattern
with the dominance of $\so$ at LO. However, owing to the impact of the NLO QCD corrections 
in the relative importance of the different CO contributions, it may be
risky to advance to further conclusions.  

Let us nevertheless add a comment about the CS channels. We have indeed 
seen that, in the case of $J/\psi+W$ where CS channels are 
also suppressed, the QED contributions can become relevant.
Since they are like the $\so$ fragmentation, they even compete with them. 
For $J/\psi+b$, the $b$-quark production does not necessitate the presence of light quarks
from which the photon can easily be emitted. As such,  
the CS channels are probably suppressed for this reaction. 
It would however be expedient to perform a full evaluation including
channels such as $gg \to J/\psi g^\star \to J/\psi + b + \bar{b}$.

On the experimental side, no data on $J/\psi+b$ production have ever been published. 
However, we do believe that they could be at reach with limited efforts. 
An easy way to access this process would indeed be to look a di-$J/\psi$ events where only
one $J/\psi$ is displaced, \ie\ non-prompt. These have certainly already been seen by D0, LHCb, 
CMS and ATLAS.

\subsection{Associated production with (light) hadrons}

As it was clear from the initial discussions of the 3 mostly used production models, these mainly
differ at the level of the quantum numbers of the heavy-quark pair at short distances
and on the likelihood for colour flows or heavy-quark-spin flips to occur
during the hadronisation. This naturally goes along with possible --more or less soft-- 
gluon radiations, which then go along with hadron production. 

The CEM being mostly
phenomenological, not much is known  about the typical amount
of energy released when the pair hadronises.  In the CSM, the situation
is quasi crystal clear as any radiation should be incorporated in the 
hard scattering and treated accordingly. In NRQCD, the IR cut-off $\mu_\Lambda$
 above which any radiation is also incorporated in the hard scattering 
should be the relevant scale used to elaborate a little more on what 
the energy of these soft gluons 
which cannot flip the heavy-quark spin but can rotate its colour is.
It happens that $\mu_\Lambda$ is expected to be larger than
 $\Lambda_{QCD}$ which is the usual other scale used to tell whether a process 
is hard, \ie\ perturbative, or not. Most of the charmonium computations in literature
are for instance done using\footnote{Such an indication is usually very quickly made, 
if not omitted sometimes, since at NLO it does not impact the phenomenology and 
its interpretation, 
except for the $P$-wave case where $\mu_\Lambda$ sets the trade off between CS and CO 
contributions, provided that this distinction makes sense physics wise in this case.} $\mu_\Lambda=m_c$.

Along these lines, the perspective of studying the hadronic activity 
around the quarkonium (see \eg~\cite{Kraan:2008hb}) has always been very appealing.
Yet, the reader should not be confused. Whereas the COM imposes the radiation
of soft gluons in the vicinity of the quarkonium --at least one or two depending
of the CO transition at work--, we have seen in many cases that, in the CSM, 
a similar number of radiations also occur, but in the hard scattering. 
The very simplistic view according to which the CO contributions go along with a higher 
hadronic activity --in general and whatever the other constraints-- as compared to the CS
contributions may be misleading. 
 
Historically, UA1 compared their 
charged-track distributions with Monte Carlo simulations for $J/\psi$
from $b$-hadron decay and $J/\psi$ from  
$\chi_c$ decay~\cite{Albajar:1987ke,Albajar:1990hf}. At the time, the FD from  $\chi_c$ 
 was still expected to be the major source of prompt $J/\psi$. 
Following either the idea of CO transitions or of CS transitions at 
higher-orders, we however  expect now more complex distributions
even for the prompt yield. It is therefore not clear if such methods
are suitable to evaluate the $b$-FD fraction otherwise than with the measurements
of a displaced vertex typical of a $b$-hadron decay as performed by STAR at RHIC~\cite{Adamczyk:2012ey}
by confronting the distribution of events as a function of the azimuthal separation
$\Delta \phi$ between the quarkonium and any other detected hadrons. 
For now, though, the results are compatible between both methods. Let us also note that
in 2013 ZEUS published~\cite{Abramowicz:2012dh} a first study of the momentum flow ``along'' ($P_{\rm along}$) and ``against'' ($P_{\rm against}$) photoproduced $J/\psi$. The event fraction vs $P_{\rm along}$ and $P_{\rm against}$ was compared, at different $J/\psi$ $P_T$, to HERWIG supposed to reproduce the physics of the CSM. A better agreement for the distribution in $P_{\rm along}$ was found than in $P_{\rm against}$. This could tentatively be explained by a better account of the activity related to the hadronisation of the $J/\psi$ than of the short distance scattering. No CEM or COM baselines are however available for further comparisons.

For some of the observables which we discussed above,
the CSM should create isolated quarkonia, in the same way as the isolated photon are defined.
There is however a lack of theoretical studies of the effect of the 
isolation criteria. It is of course more true for the COM and CEM. In the latter, 
one could even wonder if it is possible at all to offer a tractable procedure
to state how an isolation cut could affect the yield.

On the experimental side, two additional sorts of experimental observables
have recently been studied. The first is the ALICE, CMS and STAR analyses of the $J/\psi$ and
$\Upsilon$ production rate as a function of the hadron multiplicity, following
the techniques used in proton-nucleus and nucleus-nucleus collisions
to constrain the geometry of the collisions. The second is the
analysis by LHCb and CMS of the distribution of $J/\psi$ inside jets as a function 
of the fractional energy they carry.  The latter followed NRQCD-based theoretical proposal~\cite{Baumgart:2014upa}
to pin down the impact of the number of gluons involved in the quarkonium
production in a fragmentation-like topologies for the 4 most relevant 
contribution $\ssnew$, $\so$, $\sps$ and $\pj$. In the following, we briefly review both and add some words on the associated production of a quarkonium with jets.

\subsubsection{Production rate as a function of the hadron multiplicity}

With the advent of the LHC, the study of the correlations of quarkonia with charged particles produced in hadronic 
collisions has been proposed to provide new insights into the interplay between hard and soft mechanisms in these reactions~\cite{Porteboeuf:2010dw}. 

In proton-nucleus and nucleus-nucleus collisions, it is a well known fact that quarkonia can 
suffer from final-state interactions and that their production rate can be affected
 accordingly. The comover interaction model~\cite{Capella:1996va,Armesto:1997sa,Armesto:1998rc,Capella:2000zp,Capella:2005cn,Capella:2007zz,Capella:2007jv,Ferreiro:2012rq,Ferreiro:2018wbd} is one of the models which specifically 
addresses this possibility. The key ingredient of this approach is the multiplicity
of the hadrons comoving with the produced quarkonium, namely the hadrons with 
the same rapidity. As such, it is perfectly legitimate to think that such an effect
could arise in very violent proton-proton collisions where the number of hadrons is particularly 
large. In general, it is in fact normal to extend the discussion to wonder whether the
production rate of a quarkonium only depends on its kinematics and that of the initial hadrons, 
or whether other parameters could affect the production rate.

The obvious --but admittedly extreme-- case is that of associated production where the quarkonium
process is so biased that one assumes that other partonic sub-processes are at work. However, as we have seen, the 
production of a $J/\psi$ can also occur quasi independently from that of 
a $\Upsilon$ when a DPS occur in $J/\psi+\Upsilon$ production for instance. 
In such a case, it looks like the bias
of requiring for a $\Upsilon$ in the event has no effect.
That would in fact be misleading to think so. Indeed, the production
of the $J/\psi$ a priori imposes some specific conditions for it to happen, beyond the cost
of producing $\Upsilon$ encoded in $\sigma_\Upsilon$. 
The effective cross section $\sigma_{\rm eff}$ encodes these conditions, \ie\ that another scattering
occurs in the hadron overlapping region.

One can extend the argument to any hadron and 
wonder whether the probability for this hadron production is different
in $J/\psi$ or $\Upsilon$ production events and depends on whether the 
quarkonium $P_T$ is large or not. If this is the case, these 
particle-production cross sections should conversely depend on the 
multiplicity of hadrons. This is what is at stake in such  studies, 
much more than to learn about the 
production mechanisms.

\begin{figure}[hbt!]
\centering
\subfloat[]{\includegraphics[width=0.35\textwidth,draft=false]{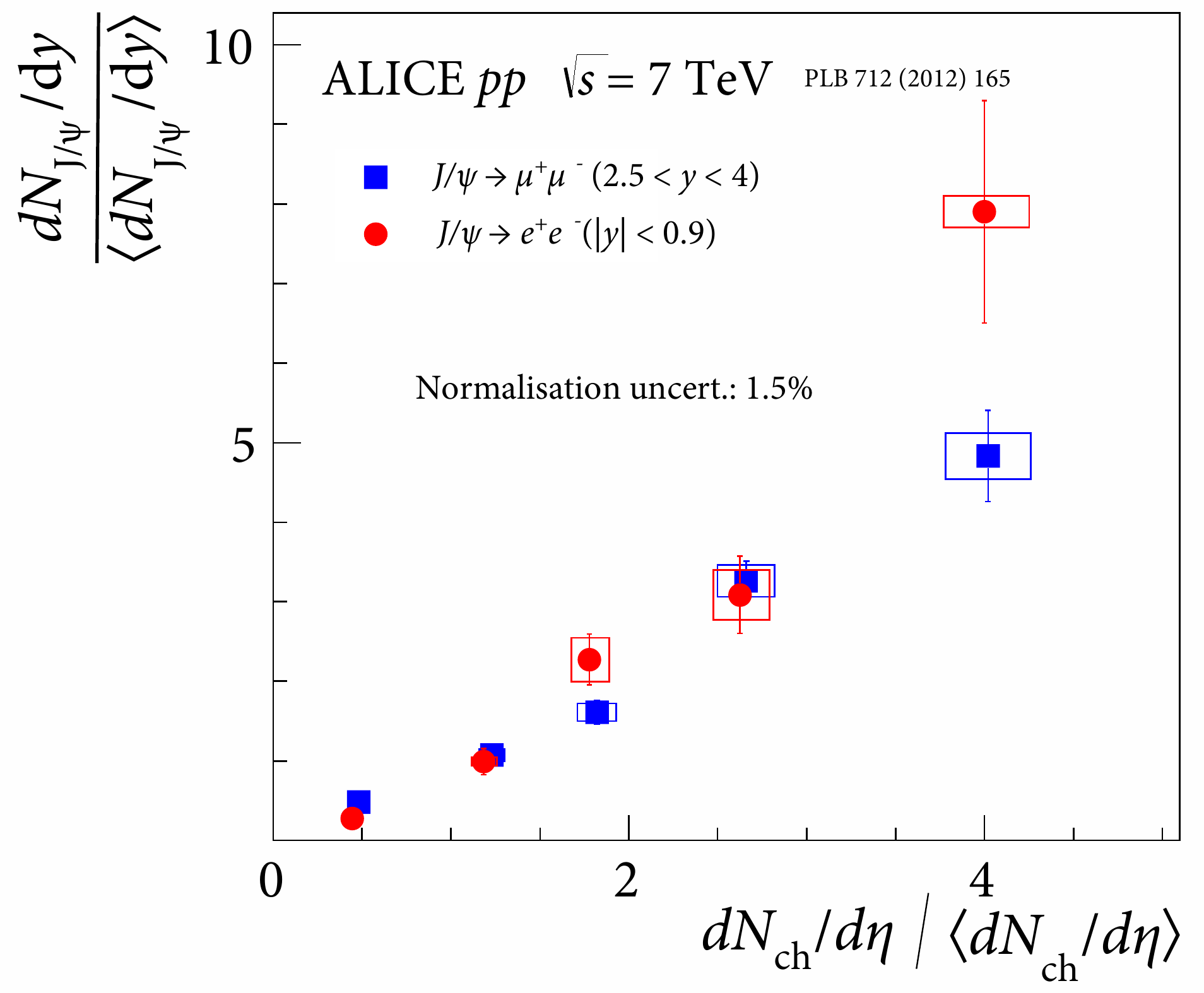}
\label{yields_vs_mult-ALICE-pp-7-TeV}}
\subfloat[]{\includegraphics[width=0.3\textwidth,draft=false]{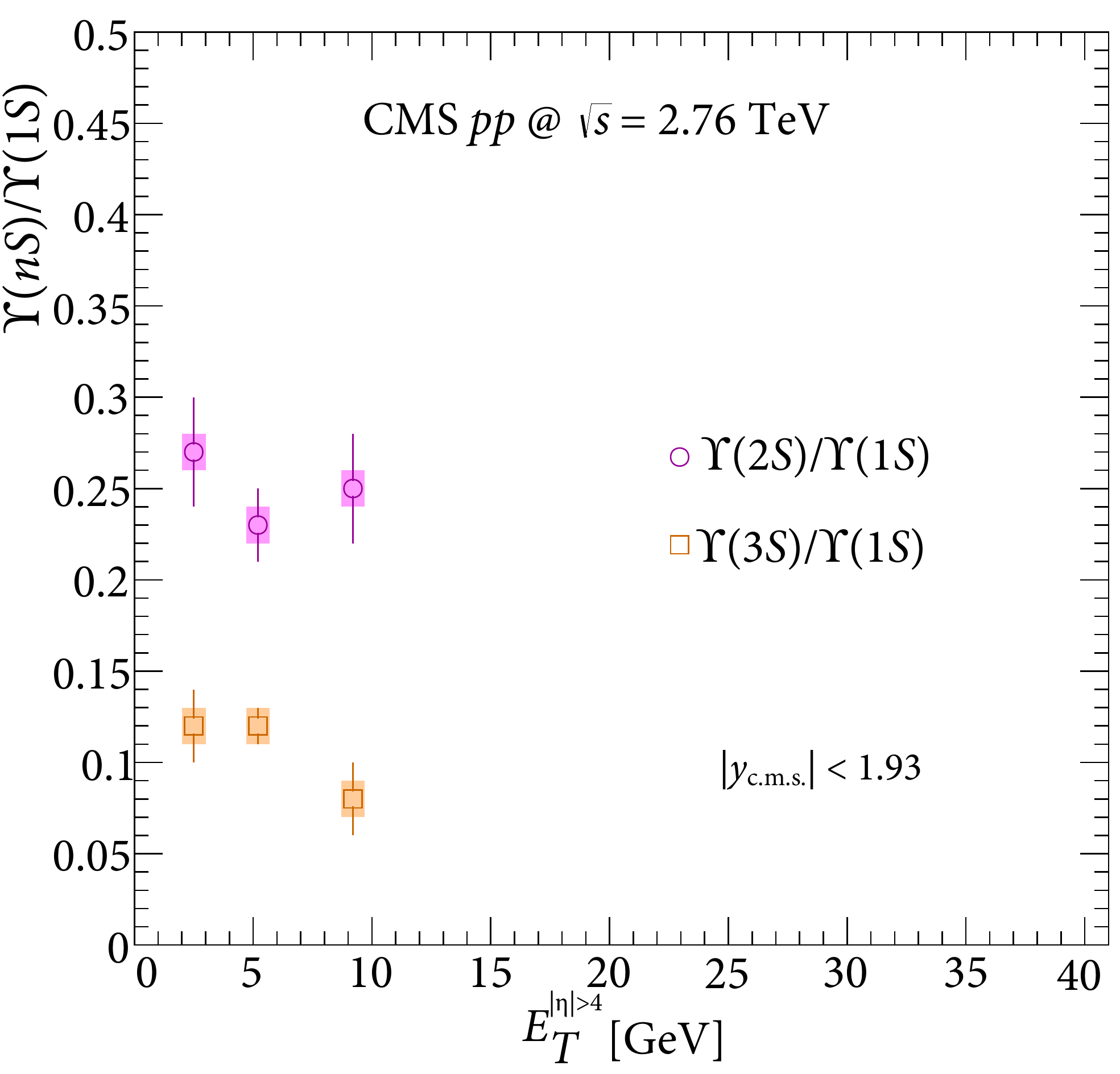}
\label{CMS-Upsilon_vs_ET-ratio-pp}}
\subfloat[]{\includegraphics[width=0.3\textwidth,draft=false]{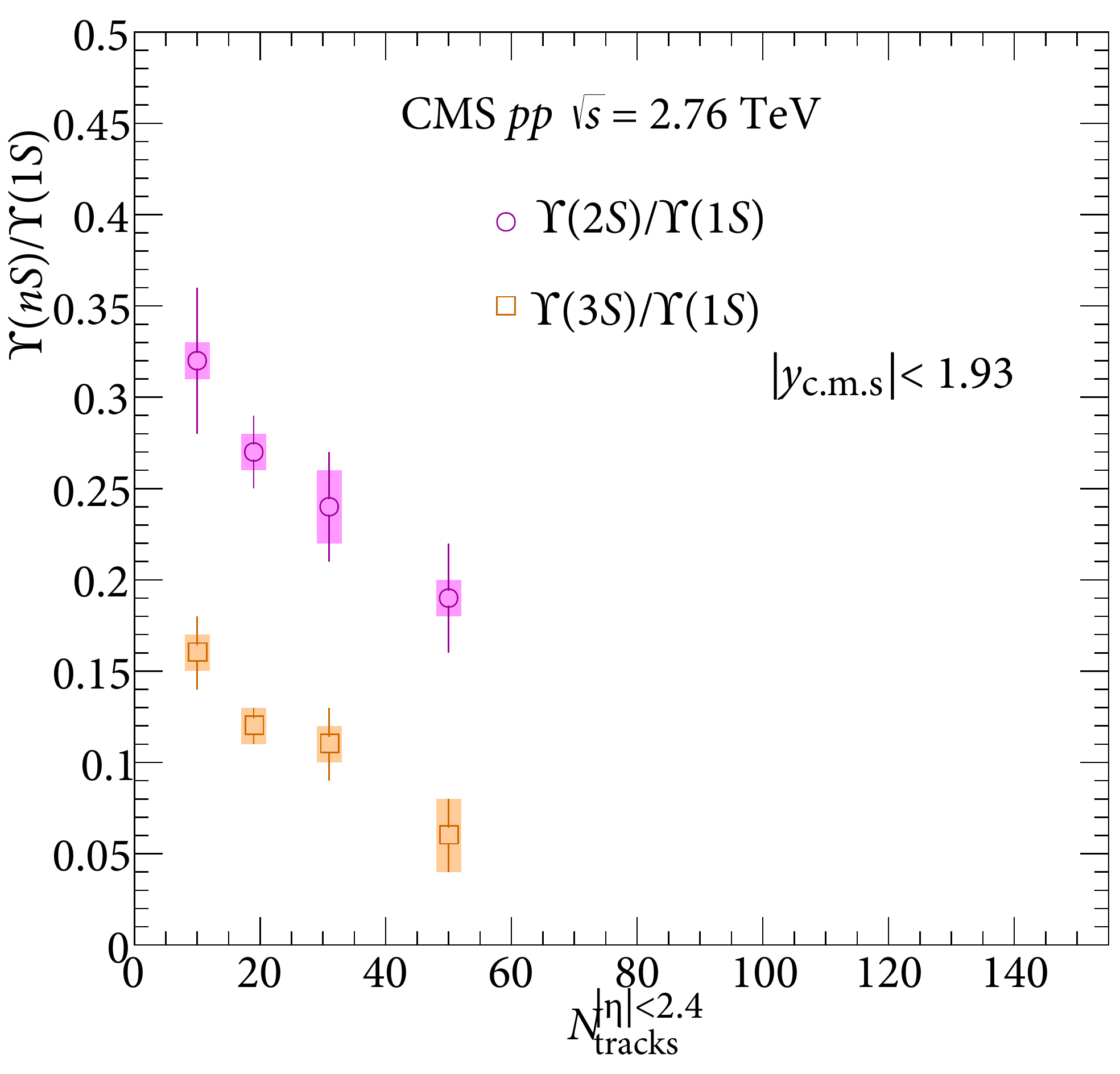}
\label{CMS-Upsilon_vs_Ntracks-ratio-pp}}
\caption{(a) ALICE measurements of $(dN_{\rm ch}/dy)/\langle dN_{\rm ch}/dy\rangle$ in $pp$ collisions at $\sqrt{s} = 7$~TeV~\cite{Abelev:2012rz} via the dimuon-decay channel for $2.5 < y < 4$ and in its dielectron-decay channel for $|y| < 0.9$. CMS measurements the $\Upsilon(nS)$ yield ratio as a function of $E_T$ (b) and $N_{\rm tracks}$ (b) in $pp$ collisions at $\sqrt{s} = 2.76$~TeV. Adapted from (a)~\cite{Abelev:2012rz} and (b-c)~\cite{Chatrchyan:2013nza}.}
\label{fig-onium-hadrons-1}
\end{figure}

At the LHC, ALICE measured in 2012 the relative $J/\psi$ yield, $(dN_\psi/dy)/\langle dN_\psi/dy\rangle$, as a function 
of the relative charged-particle multiplicity,  $(dN_{\rm ch}/dy)/\langle dN_{\rm ch}/dy\rangle$, in $pp$ collisions at $\sqrt{s} = 7$~TeV~\cite{Abelev:2012rz} via the dimuon-decay channel for $2.5 < y < 4$ and via the dielectron-decay channel for $|y| < 0.9$. 
They found similar results in both ranges (see \cf{yields_vs_mult-ALICE-pp-7-TeV}). The relative $J/\psi$ yield linearly increases with the relative charged-particle multiplicity, except maybe for the last point for central rapidities. This increase was interpreted in terms of the hadronic activity accompanying \jpsi production, as well as multiple parton-parton interactions~\cite{Kopeliovich:2013yfa}, or in the percolation scenario~\cite{Ferreiro:2012fb}.

In 2013, CMS performed a similar study of the \ups yields in $pp$ collisions at $\sqrt{s} = 2.76$~TeV~\cite{Chatrchyan:2013nza}. 
The self-normalised cross sections of $\upsa/\langle \upsa \rangle$, $\upsb/\langle \upsb \rangle$ and $\upsc/\langle \upsc \rangle$ at mid-rapidity were found to increase with the relative charged-particle multiplicity. 

To study possible differences among the different $\Upsilon$ states, the ratio of the $\Upsilon(2S)$ and $\Upsilon(3S)$ yields over that of $\Upsilon(1S)$ was analysed as a function of the transverse energy ($E_{\rm T}$) measured in $4.0 < |\eta| < 5.2$ and of the number of charged tracks ($N_{\rm tracks}$) measured in $|\eta| < 2.4$ (see \cf{fig-onium-hadrons-1} (a) \& (b)). These  ratios seemed independent of the event activity as a function of the forward-rapidity $E_{\rm T}$ and of the mid-rapidity $N_{\rm tracks}$, although a possible decrease was observed in the latter case (bearing in mind the experimental uncertainties and the small number of points). Overall, CMS noted that $\Upsilon(1S)$ are produced on average with two extra charged tracks than excited states. Such an effect cannot be described by FD contributions only.  Another interpretation would be the presence of final-state effects, such as those included in the comover interaction model~\cite{Ferreiro:2018wbd}. 

Let us also note that the $\Upsilon$ polarisation was analysed by CMS as a function of 
the multiplicity~\cite{Khachatryan:2016vxr}. Within the experimental uncertainties
of $10 \div 20$ \%, no effect was observed.  Finally, in 2018, STAR performed a first study of $J/\psi$ vs multiplicity at RHIC~\cite{Adam:2018jmp} at $\sqrt{s}=200$~GeV and reported results
qualitatively similar to those of ALICE, although restricted to smaller absolute multiplicities.

\subsubsection{$J/\psi$ in a jet}
\label{sec:psi-in-jet}
In 2017, LHCb reported~\cite{Aaij:2017fak} on the first analysis 
of $J/\psi$ inside jets. In 2014, Baumgart \etal~\cite{Baumgart:2014upa}
had indeed suggested that 
the probability for a jet to contain a $J/\psi$ with a fixed fraction $z$
of its energy $E_{\rm jet}$ was decreasing with $E_{\rm jet}$ for the
CO $\sps$ transition and for $\ssnew$, but
not for the $\so$ transition. As such, this observable could provide new means to disentangle between
the different CO states possibly at work in $J/\psi$  hadroproduction.

LHCb did not exactly carry out this measurement as they only 
analysed the relative transverse-momentum spectrum, referred to as $z_T=P_T^\psi/P_T^{\rm jet}$, of the $J/\psi$ in the 
jets integrated over $P_T^{\rm jet}> 20$~GeV and claimed that
their resulting distribution for the $J/\psi$ was in contradiction with NRQCD + {\sc Pythia},
whereas the nonprompt result agreed with {\sc Pythia}.

However, Bain \etal~\cite{Bain:2017wvk} later showed using both approaches discussed in~\cite{Bain:2016clc}  (the Gluon Fragmentation Improved {\sc Pythia} (GFIP) and the Fragmenting Jet Functions (FJF)) that the LHCb results 
could in fact be accommodated by NRQCD --in fact nearly irrespectively 
of the considered set of channels, thus of the LDME fits (see \cf{plotsGFIP}). The discrepancy initially found by LHCb 
is in fact rather due to the limitation of the possible
reactions accounted for by the LO NRQCD implementation in {\sc Pythia}. 
This also  unfortunately means that the distribution studied by LHCb, 
contrary to the proposed study of the $E_{\rm jet}$ dependence at fixed $z$
is not a very discriminant observable.

\begin{figure}[hbt!]
\centering
\subfloat[]{\includegraphics[width=0.33\textwidth,draft=false]{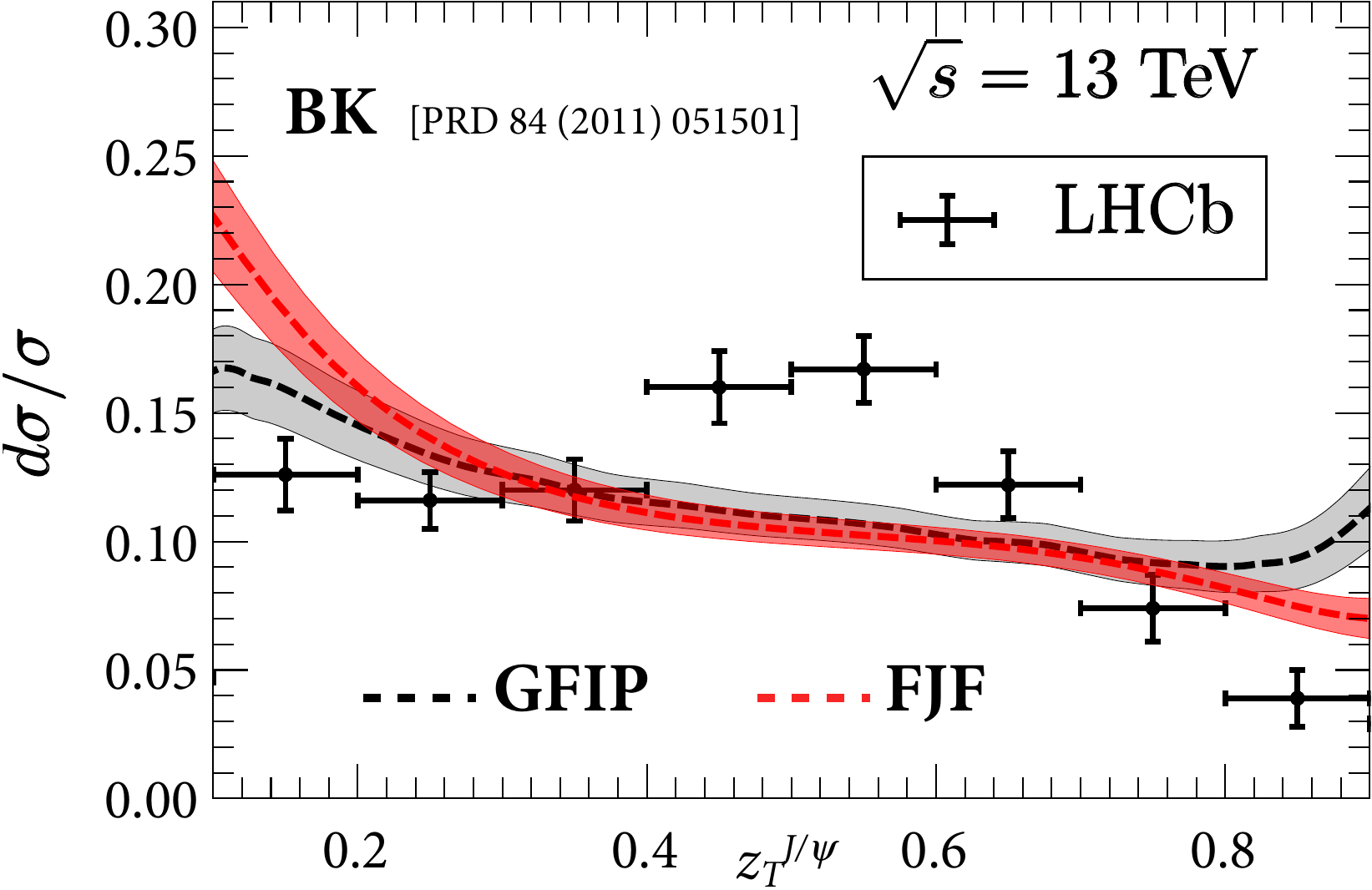}
\label{plotsGFIP-1}}
\subfloat[]{\includegraphics[width=0.33\textwidth,draft=false]{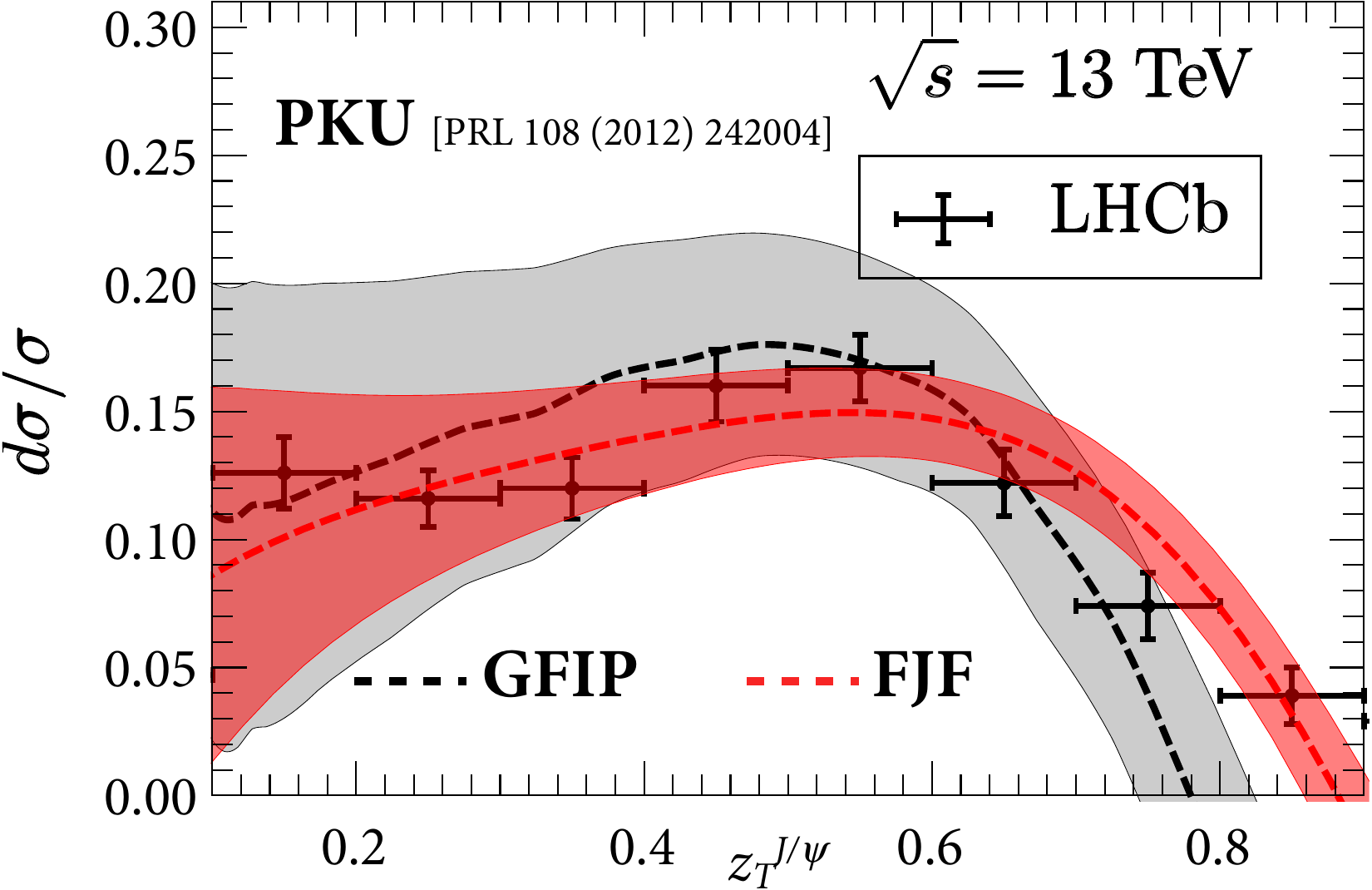}
\label{plotsGFIP-2}}
\subfloat[]{\includegraphics[width=0.33\textwidth,draft=false]{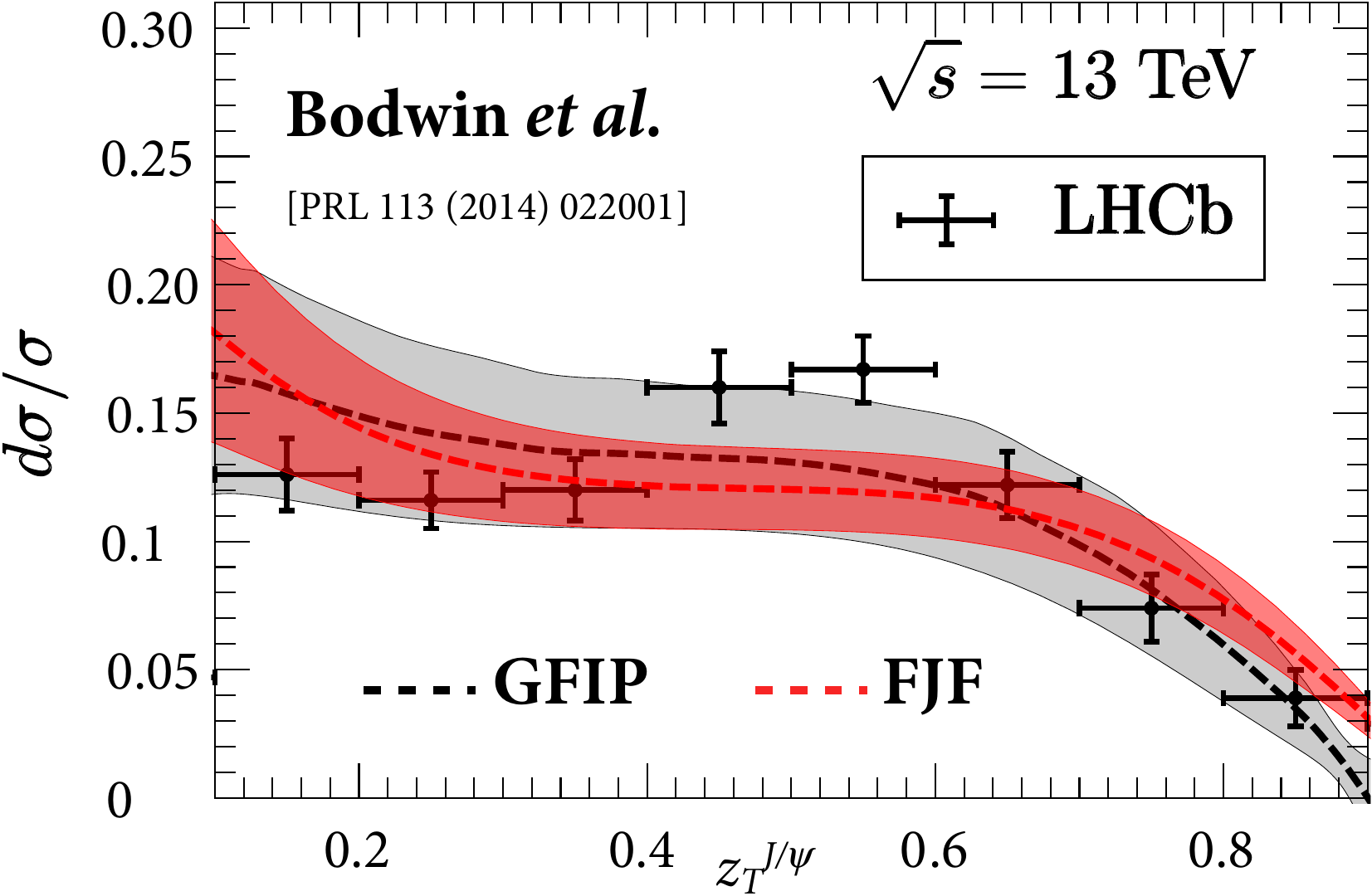}
\label{plotsGFIP-3}}
\caption{Self-normalised $z_T$
distribution of the $J/\psi$ in jets as measured by LHCb compared to the GFIP and FJF predictions using the
  (a) BK~\cite{Butenschoen:2011yh},  (b) PKU~\cite{Chao:2012iv} and (c) Bodwin \etal~\cite{Bodwin:2014gia} LDME NLO fits. Adapted from~\cite{Bain:2017wvk}.}
\label{plotsGFIP}
\end{figure}

We should also note that LHCb released the $z_T$
distribution without normalisation. This is problematic as many
of the quarkonium puzzles bore on normalisation issues. In
particular, a sub-leading contribution which would show a distribution
like that seen by LHCb could mistakenly be considered as 
acceptable. It also does not allow to fully appreciate the observation~\cite{Belyaev:2017lbo} 
that $c \to \psi+c$ fragmentation generates a $z$ distribution in good
agreement with the LHCb data. This observation is indeed intriguing
given that such a fragmentation yields to quasi unpolarised\footnote{Let us also mention recent theory studies of the polarisation of $J/\psi$ inside a jet~\cite{Kang:2017yde,Dai:2017cjq}.} 
$J/\psi$~\cite{Lansberg:2010vq} and could be enhanced by CTs~\cite{Nayak:2007mb,Nayak:2007zb}.

The lack of a normalisation also does not facilitate to assess the relevance of 
the DPS contributions. In the LHCb analysis~\cite{Aaij:2017fak}, the DPS
was probably underestimated as it was evaluated with {\small PYTHIA} 
which would correspond
to $\sigma_{\rm eff}\sim 30$~mb for the production of a $J/\psi$ simultaneously
with a di-jet events, of which one would be the jet within which the $J/\psi$ is found.
In view of the previous discussions, a value of 15 mb, if not below, is more indicated
for hard scatterings in particular those involving quarkonia. In such a case, the
low $z$ region (up to 0.5) may be dominated by DPS events~\cite{Belyaev:2017lbo}.

Finally,
we would like to underline that some of the different features
highlighted between the different CO and CS transitions are related 
to the behaviour of the corresponding fragmentation functions (FFs), 
which are considered to be different between the 
$\ssnew$, $\so$, $\sps$ and $\pj$ states~\cite{Braaten:1993rw,Braaten:1993mp,Braaten:1994kd,Braaten:1995cj}.
These expectations are however based on LO computations.
 Only the FFs of 2 channels are known at NLO, $\so$~\cite{Ma:1995ci,Braaten:2000pc,Feng:2018ulg,Zhang:2018mlo} 
and $\pseudos$~\cite{Artoisenet:2014lpa,Feng:2018ulg,Zhang:2018mlo,Artoisenet:2018dbs}. Very different behaviours
of these FF have been observed at LO and NLO. Hence, one cannot exclude that some of the differences expected for the $J/\psi$ in jets for the different transitions may be washed out when using NLO FFs. Recent theory advances in FF computations with semi-analytical NLO~\cite{Zhang:2018mlo} and analytical LO~\cite{Zhang:2017xoj} results may be very useful for future studies.

Overall, LHCb pioneered a new class of interesting
studies but many caveats remain to be addressed without forgetting that many 
``golden-plated" quarkonium observables have been proposed in the past and were found to
be much more complex --not to say more-- to interpret as soon as NLO computations became available, even
in the cases where expectations for these NLO effect had previously been devised. 

In 2020, the CMS collaboration published 
another study~\cite{Sirunyan:2019vlp} of $J/\psi$ in a jet. Contrary to LHCb, they measured the self-normalised $E_{\rm jet}$ dependence
in bins of $z$ which allowed them to compare their data to FJF predictions for different NRQCD LDME sets. It happens that the distributions
based on the high-$P_T$ fit of Bodwin \etal~\cite{Bodwin:2014gia} are the only ones matching theirs. However, we recall that 
such a fit, with a leading  $\sa$ transition,  badly disagrees with a number of data sets: HERA $J/\psi$ photoproduction, LHC(b) $\eta_c$ hadroproduction, $P_T$-integrated $J/\psi$ hadroproduction and $J/\psi$ production in $e^+e^-$ collisions at $B$ factories. This maybe hints at yet another puzzle in $J/\psi$ production. As discussed above, firmer conclusions nevertheless clearly call for a full NLO analysis based on NLO FFs.

On a more positive side, it is important to understand that, at $z\to 1$, 
the $J/\psi$ identifies with the jet in the LHCb analysis. On the way, this illustrates how much the notion
of the ``fraction of $J/\psi$ in jets'' can be analysis dependent: an isolated $J/\psi$ with high enough $P_T$ is a jet and thus {\it in} a jet whereas the same isolated $J/\psi$ with a $P_T$ lower that the minimum value used to define 
a jet in a given experiment is {\it de facto} not in a jet.
 That being stated, the yield at $z\to 1$ may thus be a proxy
for a measurement of the isolated $J/\psi$, provided that it can properly 
be converted in a cross section. We do think that it should be one of the priority analysis
for the close future, especially if it allows one to isolate the CS contribution. On the
one hand, this would be a test of our understanding of quarkonium production --which would however
require a vigorous and simultaneous theory effort to understand the isolation impact-- and, on the other hand, 
it would help to extend the use of quarkonia as tools for TMD extraction in hadroproduction.

Finally, let us add that the study of a quarkonium {\it in} a charm- and beauty-tagged jet would be another way to access the reactions discussed in section~\ref{sec:onium_HQ} in some specific configurations where 3 heavy quarks are near each other. If they are of the same flavour, the yield could be enhanced due to CTs and is in any case the largest NLO CS contributions at large $\pt$. Such studies should  be attempted especially if they allow one to extend the $\pt$ range of the existing measurements for which the heavy
quarks are tagged by the presence of $b$ and $c$ hadrons and via nonprompt charmonia.

\subsubsection{Associated production of a quarkonium with a jet or with jets}
\label{sec:psi-with-jets}

One of the reasons of the complexity of quarkonium production at finite $P_T$ is
the large impact of QCD corrections associated to the emission of additional hard partons
which recoil on the quarkonium. We have seen in section \ref{ch:dvlpts} how relevant 
such QCD radiations are for CS transitions but also for some CO transitions at increasing $\pt$. On the contrary, 
it does not seem to be the case if the hadronisation occurs like the CEM. 

Naturally, the occurrence of these additional emissions should be visible in the jet multiplicity associated
with high-$\pt$ quarkonium production. In this context, it is particularly surprising that 
there is only a single theory study bearing on the associated production of a quarkonium with a jet and
no experimental study at all. In fact, this theoretical study~\cite{Boussarie:2017oae}
 takes its inspiration
in the study of Mueller-Navelet jets in order to probe the BFKL dynamics, not to probe the quarkonium mechanism.
In this context, we find it useful to make some --qualitative-- statements about the motivations and the feasibility to
study the associated production of a quarkonium with a jet or with jets.

First, at sufficiently high $P_T$, most of the quarkonia should be produced along with a recoiling jet. As such, the rate for such associated channels should be significant and certainly high enough for a number of dedicated measurements with already recorded data. Their precision should likely be better than most of the associated-production studies covered in this section. There should also not be any specific trigger issues for ATLAS and CMS in the presence of rather high $P_T$ dimuons.

Let us outline a couple of possible observables, but there are certainly other ones. First, selecting events with a quarkonium back to back with a jet should enrich the sample in CO vs. CS transitions in particular if it is possible to constrain, in addition, the jet multiplicity to be minimal\footnote{Let us highlight an additional motivation for such observables but rather at mid $P_T$ if rather ``low-$P_T$'' jets can be measured. If, besides selecting such quarkonium + one jet events, it is possible to isolate the quarkonium at mid $P_T$ such as to have instead a dominant
CS yield, which may be done for $\Upsilon$ production with $P_T \simeq M_\Upsilon$,  this would provide an additional observable to study gluon TMDs along the lines of~\cite{Boer:2014lka}. Simply selecting events with a quarkonium back to back with a jet of, say, 50~GeV, as we proposed above, is admittedly very different.}.
The idea would be to look at configurations where the pair has a large invariant masses, $M_{\Q-\rm jet}$. Indeed, the production via $\sb$ will simply occur from topologies like that of \cf{diagram-COM-PT-c} from $gg \to g^\star g$ and will certainly be favoured compared to the possible CS channels. It may also be enhanced with respect to $\sa$ transitions. Clearly, these statements are generic and, given the many golden-plated observables proposed in the history of quarkonium production, some dedicated theoretical evaluations are needed to confirm such expectations and to quantify the achievable sensitivity and the possible improved determination of specific LDMEs for instance.

Second, one could conversely look at the jet multiplicity associated with quarkonium production. A priori, since the leading-$P_T$ contributions to direct $J/\psi$ and $\Upsilon$ production in the CSM are associated with the emission of 3 hard partons, the jet multiplicity should be larger for the contributions from $\ssnew$ than from $\sb$. $\sa$ contributions should be in between. Measurement at different $P_T$ to see whether the multiplicity changes and how would be instructive. 

Third, if the yields are large enough, and since such measurements are likely carried out at large $P_T$
where $P$ waves can more easily be detected, one could compare the jet multiplicity for $J/\psi$ and $\Upsilon$ production to that for $\chi_c$ and $\chi_b$ production. Once again, one would need dedicated theoretical studies to assess how much they would differ and whether experimental measurements would be sensitive enough to see such differences. Yet, observing a similar or a different jet-multiplicity distribution would provide new information of the hard scattering connected to the production of these particles.

There are probably other observables related to jet studies which can be proposed, based on angular correlations, $P_T$ imbalances or rapidity separations for instance. One should however bear in mind the possible DPS contributions. Just like for other associated-production channels, their dedicated study can however provide additional information on $\sigma_{\rm eff.}$. In any case, independently of the above suggestions, any experimental attempt towards quarkonium+jet studies is more than welcome and will certainly be closely followed by an increase of theoretical activity on the topic. Such studies are definitely possible with the 4 LHC detectors.

\section{Conclusions and outlook}

As we have seen all along this review, the physical mechanisms underlying quarkonium production 
in inclusive reactions are not  clearly identified yet -- more than forty years after
the discovery of the first quarkonium. Despite the inflow of high precision data from the LHC with an
unprecedented  kinematical coverage, the progress was not as steady expected.

We have indeed shown in section 2 that the sole measurements of  \jpsi or \upsi production yields
are not sufficient to validate or falsify the predictions of the competing theoretical approaches describing the hadronisation of the quarkonia. Let us mention in particular
the predictions of NRQCD which rely on several non-perturbative parameters which cannot be computed from
first principles. 
In the recent NRQCD analyses, the cross section for the 
direct yields typically depends on 3 of them which appear in specific linear combinations --leaving aside FDs for the moment. If one sticks
to data at rather large $P_T$, which thus excludes most of the other data than hadroproduction, this
precludes their separate determination. This, on the way, highlights the importance of the data selection. 

Until recently, polarisation measurements for the production of the vector quarkonia (\jpsi and \upsi)
were considered to be the ideal means to lift the degeneracy between the experimental constraints. 
In particular, a well-known NRQCD predictions was that, at large $P_T$, both the \jpsi and \upsi
yields should be transversely polarised (in the helicity frame). Not only, this 
observation --although highlighted at many instances as a smoking gun for the CO mechanism
inherited from NRQCD-- was never realised\footnote{The observed polar anisotropies are systematically
close to zero.} but the advent of NLO computations in $\alphaS$ uncovered 
a far more complex situation where polarisation suddenly becomes very subtle to predict.

Since 2009, several groups of physicists have carried out NLO NRQCD studies. As we reviewed it, their
conclusions differ both quantitatively and qualitatively. These diverse interpretations of
the existing data can be traced back to variable data selections motivated by sometimes
contradictory interpretations of the applicability of NRQCD in some given kinematical ranges. This 
critically complicates any scientific procedure aiming at proving or falsifying the relevance
of specific  phenomena encapsulated in NRQCD, in particular the importance of the CO transitions whereby
a heavy-quark pair is produced at short distance in a colourful state and yet hadronises into a colourless meson
after non-perturbative gluon radiations.

Whereas NRQCD clearly allows for a better account of some measurements, in particular
at large $P_T$, it conflicts in an obvious way with a certain number of other observables. Let
us cite the $P_T$-integrated cross sections --or to phrase it differently the total number of produced
quarkonia in a given colliding system, or the production cross section in $e^+e^-$ and $ep$ collisions.
Conversely, the CSM --or equally NRQCD with a limited impact of the CO transitions-- accounts
reasonably well, without any tuned parameter, for these observables. It however seems
to still face clear difficulties to account for the yield at large $P_T$ in spite of the considerable size of the $\alphaS$ corrections.
The third approach, the CEM, also seems unable to catch the subtlety of quarkonium production at high $P_T$
as well as production in $e^+e^-$ collisions. It nevertheless explains without difficulty \eg\ the energy dependence of the $P_T$ cross sections. In short, the current situation does not allow one to call in favour of a specific hadronisation model.

To be exhaustive, we find it important to underline that the present discussion, like that of section 2, bore
on studies based on collinear factorisation. It is however likely that some of the above
conclusions could be quantitatively, and perhaps qualitatively, altered if one rather used
the $k_T$ factorisation which takes into account the growing off-shellness of the initial gluons when
$\sqrt{s}$ increases. As large as these effect may be, we however regret that none of such 
quarkonium studies could so far be carried out beyond the tree level. This {\it de facto}
limits the scope of conclusions which we could drawn regarding the success or the failure of
specific quarkonium hadronisation models used with the $k_T$ factorisation.

In this context, we are convinced that the introduction of new observables is
absolutely essential. On the one hand, these can help better constrain the non-perturbative parameters 
of the aforementioned theoretical approaches, in particular in providing novel constraints
in regions where there is a consensus on the applicability of a given model. A first practical 
example is that of $\eta_c$ production. The LHCb measurement with barely 7 points with limited precision
drastically --and unexpectedly-- tightened the constraints on some NRQCD-based studies, such that
some are now quasi excluded. Other examples are expected in associated-production channels, for some of which predictions are however still carried out at tree level. On the other hand, even without dedicated experimental measurements, some theoretical case studies
clearly helped better understand the structure of the radiative corrections in $\alphaS$ to
single quarkonium production which is far more complex than some associated-production channels.
Another example is that of the associated production of quarkonium along with a photon whose theory studies
taught us that in practice the NRQCD non-perturbative parameters should be positive. Indeed, its cross section
happens to be negative --thus unphysical-- at large $P_T$ when one of these ($\mop0$) is negative.
Section 3 was entirely devoted to the latter class of observables. 

Beside enlarging the spectrum of available observables, it is also necessary that one 
improve both the accuracy and the precision of the evaluation of the perturbative parts of the cross sections.
As such, it would be inadequate to overlook the discussion of the importance of NNLO radiative corrections.
In particular, one must complete the partial NNLO evaluations which we discussed in some details. This is of relevance both at large $P_T$ --where two partial analyses lead to different interpretations-- and to lower $P_T$ to test the convergence of the perturbative series as well as to use quarkonia as tools to learn about (nuclear) PDFs, TMDs as well as the QGP. 

Yet, such partial studies, where the largest contributions to the NNLO yield at large $P_T$ are believed to be 
evaluated through the production of a quarkonium with 2 or 3 partons or jets, should not be overlooked once a full NNLO computation is available. Indeed, these highlight the relevance of studying the jet multiplicity associated with each quarkonium states (pseudo-scalar, vector, tensor, ...) as well as the kinematical distribution of these jets. One should also reach a continuum of understanding between such observables and the associated production with light hadrons, which thus extends to the study of the hadronic activity near the quarkonium. Obviously, different hadronisation mechanisms may lead to a different hadronic activity near the quarkonium. Yet, the effect of harder radiations from the short-distance scattering should not be ignored in such considerations.
Recently, the study of quarkonia in jets has taken over. However, we warn that these are mostly based on expectations using fragmentation functions at LO whose kinematical behaviour is rather extreme. It is essential to advance them to NLO. It is also essential to invest a similar amount of experimental and theoretical efforts in the study of quarkonia {\it outside} jets --thus associated with jets. We regret that not a single theory study exists. In addition, {\it isolated}-quarkonium production cross sections should be measured and predicted in the 3 hadronisation approaches. LHCb and CMS are very close to such a measurement as we discussed.  Going further, one could connect these considerations to the correlations between the production yield and the hadron multiplicity in the event, although such studies would probably tell us more on the initial stages of the collisions (see also the discussion on the DPS below) than on the hadronisation.

Along the same lines, it is absolutely essential that all the associated-production channels be computed at
NLO and accounting for both the CS and CO contributions. To do so, it would be expedient to extend 
to quarkonium production a tool like \MG5aMC which automates the computation of virtually any reactions
involving Standard Model particles at NLO. This would open the path for a truly global analysis of all the constraints
from these new observables. These should thus precisely constrain the non-perturbative physics
involved in quarkonium production to such a level of precision that one could validate
or falsify the 3 most used aforementioned models of quarkonium hadronisation.

\begin{table}[hbt!]\renewcommand{\arraystretch}{1.1}\footnotesize
\begin{tabularx}{\textwidth}{X|p{2.25cm}|p{1.35cm}p{1.25cm}p{1.35cm}|p{4cm}}
Observables & Experiments & CSM & CEM & NRQCD & Interest \\ \hline \hline
$J/\psi+J/\psi$ & LHCb, CMS, ATLAS, D0, NA3 & NLO, NNLO$^\star$ & NLO & LO & \parbox{5.5cm}{\vspace{0.3cm} Test of the CSM; \\ DPS; \\ Gluon TMDs; \vspace{-0.1cm}} ~ \\  \hline
$J/\psi+\psi(2S)$
or $J/\psi+\chi_c$ & -- & LO & NLO & LO & DPS vs SPS;  \\  \hline

$J/\psi+\Upsilon$ & D0 & LO & NLO & LO & \parbox{5.5cm}{\vspace{0.3cm}  Test of the CSM; \\ DPS; \vspace{-0.1cm}} ~ \\ \hline
$\Upsilon+\Upsilon$ & CMS & NLO (?) & NLO & LO &  \parbox{5.5cm}{\vspace{0.3cm}  Test of the CSM; \\ DPS; \\Gluon TMDs;\vspace{-0.1cm}} ~\\ \hline
$J/\psi+$charm & LHCb& LO & -- & LO &  \parbox{5.5cm}{\vspace{0.3cm}  $c\to J/\psi$ fragmentation \& CTs; \\  DPS.\vspace{-0.1cm}} ~ \\ \hline
\parbox{5.5cm}{\vspace{0.3cm}
$J/\psi+$bottom or \\ $J/\psi+$nonprompt $J/\psi$\vspace{-0.1cm}}
& --& -- & -- & LO & \parbox{5.5cm}{\vspace{0.3cm} Test of the COM; \\DPS;\vspace{-0.1cm}} ~\\ \hline
\parbox{5.5cm}{\vspace{0.3cm}
$\Upsilon+$bottom~or \\
 $\Upsilon+$nonprompt $J/\psi$\vspace{-0.1cm}} & -- & LO & -- & LO & \parbox{5.5cm}{\vspace{0.3cm} Test of the CSM/COM; \\DPS; \vspace{-0.1cm}} ~\\ \hline
$\Upsilon+$charm & LHCb & LO & -- & LO &  DPS;\\\hline
$J/\psi+Z$ & ATLAS & NLO & NLO & Partial NLO & \parbox{5.5cm}{\vspace{0.3cm} Test of the CSM/COM; \\ DPS;\vspace{-0.1cm}} ~\\ \hline
$J/\psi+W$ & ATLAS & LO & NLO & NLO (?) & \parbox{5.5cm}{\vspace{0.3cm}Test of the COM; \\DPS;\vspace{-0.1cm}} ~\\ \hline
$\Upsilon+Z$ & -- & NLO & -- &  &\parbox{5.5cm}{\vspace{0.3cm} Test of the CSM/COM; \\DPS;\vspace{-0.1cm}} ~\\ \hline
$\Upsilon+W$ & -- & LO & -- &  &\parbox{5.5cm}{\vspace{0.3cm} Test of the COM;\\ DPS;\vspace{-0.1cm}} ~\\ \hline
$J/\psi$ in jets & LHC, CMS & LO & -- & LO & \parbox{5.5cm}{\vspace{0.3cm}Test of the CSM/COM; \\DPS;\vspace{-0.1cm}} ~\\ \hline
$\Upsilon$ in jets & -- & LO & -- &  LO & \parbox{5.5cm}{\vspace{0.3cm}Test of the CSM/COM;\\ DPS;\vspace{-0.1cm}} ~\\ \hline
Isolated $J/\psi$ & LHC, CMS (?) & LO (?) & -- & -- & Test of the CSM/COM;\\ \hline
Isolated $\Upsilon$ & -- & LO (?) & -- &  -- & Test of the CSM/COM;\\ \hline
$J/\psi+$ jets & -- & -- & -- & -- & Test of the CSM/COM;\\ \hline
$\Upsilon+$ jets & -- & -- & -- &  -- & Test of the CSM/COM;\\ \hline
$J/\psi+$ hadron activity & UA1, STAR & LO (?) & -- & -- &\parbox{5.5cm}{\vspace{0.3cm} Test of the CSM/COM; \\$b$-hadron FD;\vspace{-0.1cm} } ~\\ \hline
$\Upsilon+$ hadron activity & STAR (Prelim.) & LO (?) & -- & -- & Test of the CSM/COM;\\ \hline
$J/\psi$ vs multiplicity & ALICE, CMS, STAR & -- & -- & --& \parbox{5.5cm}{\vspace{0.3cm}Initial stages;\\ final state interactions (?);\vspace{-0.1cm}} ~ \\ \hline
$\Upsilon$ vs multiplicity & CMS& -- & -- & --& \parbox{5.5cm}{\vspace{0.3cm}Initial stages;\\ final state interactions (?);\vspace{-0.1cm}} ~ \\ 
\end{tabularx}
\caption{List of ``new'' observables in quarkonium production with the experimental and theoretical status
of their studies along with the main motivations to consider them.  \label{tab:table-new-obs}}
\end{table}

Yet, as we have seen in section 3, the interpretation of the experimental data of
associated production can, in some cases, be blurred by the possible contribution of 
DPSs. This imposes a systematic experimental effort to take them into account in a data driven
way by assuming their independence in a similar way as some combinatorial backgrounds can be dealt 
with mixed-event techniques. If the independence of both scatterings within a DPS was to become
 debatable also at LHC energies following some future observation, 
it would then be necessary to restrict the study of associated production to kinematical domains
where the DPS contributions can safely be neglected. Such complications inherent to the DPS in associated production could however be turned to our advantage 
as quarkonia then become probes of the multi-gluon dynamics inside the proton by analysing reactions in kinematical regions for which the DPS contributions are believed to be dominant. Already now, the associated production of quarkonia seems to hint at different parton correlations than those measured with jets and $W$/$Z$ data.

A third important aspect of the associated production of quarkonia regards its possible usage to probe the transverse dynamics of gluons inside nucleons. 
This applies to specific cases where (i) the colour flow in the final state is restricted, (ii)
the DPS contributions are negligible and (iii) the $P_T$ of the particle associated with the quarkonium approximately
balances that of the quarkonium. To date, the most promising case is that of a pair of $J/\psi$ at 
small rapidity differences at the LHC. In this case, the CSM contribution is clearly dominant and the DPS contribution becomes small for increasing individual $P_T$. Dedicated experimental studies
are expected soon.

As what concerns the prospects regards topics which could not be treated in this review, we
feel obliged to mention the production of quarkonia in the decay of other particles. Let us cite
the production of charmonia in $b$ hadron or $\Upsilon$ decays, but also more generally the production of quarkonia
in decays of $Z^0$, $W^\pm$ or $H^0$. Indeed, a global understanding of quarkonium production would only be
achieved if it is really inclusive in avoiding to segment, sometimes arbitrarily and --worse-- opportunely, the observables in different classes of variable relevance. Along these lines, it is therefore very important to also draw our attention to the observables beyond those of the collider LHC, \eg\ from Belle-II, COMPASS, the LHC in the fixed-target mode and, in the longer term future, from the US EIC, the FCC or the CEPC-SppC for which very scarce prospects exist in the literature  as what regards the inclusive production case.

\ct{tab:table-new-obs} summarises these conclusions by listing different possible new observables in hadroproduction along with the status of their theoretical and experimental studies and what they are expected to tell us. A similar list can also be made for the 
$\eta_Q$ and $\chi_Q$. In such a case, none are measured and it should be stressed that some motivations based on the expected dominance of one or another mechanism may not be the same. The motivations regarding the test of some mechanism or DPS studies may change depending on what future measurements would reveal. The same applies for TMD studies which require a good control of the DPS and, in most cases, the dominance of the CSM. In presence of a question mark, we invite the reader to refer to the corresponding section in the review for a discussion of possible caveats.

In the other production modes, a measurement of $\psi(2S)+X_{\text{non } c\bar c}$ by Belle-II is absolutely crucial as well as photoproduction studies at large $P_T$ at the US EIC\footnote{We note that $J/\psi$ photoproduction at \eg\ an EIC have recently been investigated in several works~\cite{Mukherjee:2016qxa,Rajesh:2018qks,Bacchetta:2018ivt,Kishore:2018ugo} in order to probe gluon TMDs.}, in particular the measurement of the $J/\psi$ and $\psi(2S)$ cross sections  with the subtraction of $b$ hadron FD.

\clearemptydoublepage

\section*{Acknowledgements}
We are indebted to  P.~Artoisenet, S.J.~Brodsky, B.~Gong, J. Campbell, J.R.~Cudell, Y. Feng, H. Haberzettl, J.~He, Y.L.~Kalinovsky, C. Lorc\'e, F.~Maltoni, M.A. Ozcelik, T.N. Pham, F. Scarpa, H.S. Shao, F. Tramontano, A.~Usachov, J.X.~Wang, N.~Yamanaka, H.F. Zhang, Y.J. Zhang for the fruitful collaborative works from which some of the material presented here resulted.

We thank L.P.~An, A.~Andronic, M.~Anselmino, F. Arl\'eo, R.~Arnaldi, S.P.~Baranov, S.~Barsuk, V.~Belyaev, J.P.~Blaizot, D.~Blaschke
% ...
and less recent discussions, exchanges and comments.

We thank M.A. Ozcelik, H.S. Shao, H.F. Zhang, Y.J.~Zhang for comments on the manuscript.

This work  was partly supported by the French CNRS via the COPIN-IN2P3 agreement, the IN2P3 project "TMD@NLO", the  Franco-Spanish PICS "Excitonium", the project Quarkonium4AFTER of the Franco-Chinese LIA FCPPL and by the Paris-Saclay U. via the P2I Department.


\bibliographystyle{utphys}  
\bibliography{HDR_principal-PR}
\clearemptydoublepage

\end{document}